\documentclass[a5paper, 11pt, oneside]{Thesis}  
\graphicspath{{./Figures/}}  

\usepackage[square, numbers, comma, sort&compress]{natbib}  
\usepackage[T1]{fontenc}
\usepackage{lmodern} 
\usepackage{graphicx}
\usepackage{amssymb,amsmath}
\usepackage{bm}
\usepackage{bbm}
\usepackage{color}
\usepackage{xcolor}
\usepackage{footnote}
\usepackage{epstopdf}
\usepackage{widetext}
\usepackage[Lenny]{fncychap}
\usepackage{xcolor}

\usepackage[toc,page]{appendix}
\usepackage{hyperref}
\usepackage{here} 
\usepackage[applemac]{inputenc}
\usepackage{verbatim}  
\usepackage{vector}  

\usepackage{dsfont}
\hypersetup{urlcolor=RawSienna,colorlinks=true}  

\begin{document}
\frontmatter	  

\title{Electronic correlations in multiorbital systems}
   \authors{
            Jos\'e Mar\'ia Pizarro Blanco
            }
\addresses  {\groupname\\\deptname\\\univname}  
\date {Madrid, Mar 2019}
\subject    {}
\keywords   {}

\maketitle

\newpage{\
\thispagestyle{empty}}
\newpage{\
\thispagestyle{empty}}

\setstretch{1}  

\fancyhead{}  
\rhead{\thepage}  
\lhead{}  

\pagestyle{fancy}  

\clearpage  

\pagestyle{empty}  

\null\vfill

\vfill\vfill\vfill\vfill\vfill\vfill\null
\clearpage  


\acknowledgements{

Para comenzar, me gustar\'{i}a agradecer a mi directora, Leni Bascones, por la oportunidad de realizar esta tesis junto a ella. Gracias por haber tenido tanta paciencia y haberme ense\~{n}ado tanto en estos a\~{n}os. Gracias tambi\'{e}n por haberme dado la oportunidad de viajar e intercambiar conocimientos con otros compa\~{n}eros de profesi\'{o}n en diferentes conferencias y escuelas. Y gracias tambi\'{e}n por haberte preocupado por mi presente y futuro y para que pudiese salir adelante. Aprovecho y tambi\'{e}n te doy las gracias junto a Maria Jos\'{e} Calder\'{o}n y Belen Valenzuela por haberme permitido ayudar a fomentar la Ciencia y la F\'{i}sica de Materiales en diferentes eventos de divulgaci\'{o}n.

Tambi\'{e}n quiero agradecer a mi familia el apoyo durante estos a\~{n}os. A mis hermanos (y sobrinos), a mis t\'{i}os (en especial a mi t\'{i}a Marisa) y dem\'{a}s por preocuparse por m\'{i}. Gracias a mi padre por todo el apoyo que me ha brindado durante a\~{n}os y por la confianza en que lograr\'{i}a ser un buen cient\'{i}fico. Gracias tambi\'{e}n a Enrique y Mila.

Gracias a los compa\~{n}eros y compa\~{n}eras que me han ido acompa\~{n}ando en mi carrera cient\'{i}fica. Gracias a Fer, Teje, a otro Fer, Carol, Gloria, Nadia, Miguel \'{A}ngel, Carlos, Jordi. Gracias tambi\'{e}n a los compa\~{n}eros del ICMM por haber hecho este camino m\'as ameno: M\'{o}nica, Fernando, Sigmund, Bea, \'{A}lvaro, Jos\'{e} Carlos, Jorge, Guillem, Jes\'{u}s y Jordi (otra vez).

I would also like to thank all of the co-workers from Frankfurt: Ying, Sananda, Emanuelle, Hendrik, Karim, Fabian, Vladislav, Steve, Kira, Anand, Thomas. Y por supuesto, gracias a Roser Valent\'{i} por acogerme en su grupo y permitirme aprender tantas cosas.

Por \'{u}ltimo, quiero dar las gracias a Laura por haber estado ah\'{i} durante todos estos a\~{n}os. Muchas gracias por haber estado ah\'{i} todos estos d\'{i}as y por haberme escuchado y animado. Y gracias por haberme ayudado en la escritura y revisi\'on de esta tesis. Sin t\'{i}, esta tesis no habr\'{i}a salido adelante.

}
\clearpage  

\dedicatory{
Para Laura
}
\clearpage
\newpage{\
\thispagestyle{empty}}


\abstract{
\subsection*{Abstract}

The role of electronic correlations in Condensed Matter is at the heart of various important systems, like magnetic materials, superconductors, topological materials, optical lattices, etc. Electronic correlations are those which change the motion of individual electrons when considering the interaction with other electrons in the material. Among the available systems to study electronic correlation effects, in this thesis I focus on unconventional superconductors, specifically in high-$T_c$ iron-based superconductors, and on two-dimensional materials, like the recent magic-angle twisted bilayer graphene or the itinerant ferromagnet $Fe_3GeTe_2$.

In the first chapter, I will briefly review the band theory and Fermi liquid theory for solid systems. In certain situations, the long-range character of the Coulomb interaction can be safely ignored, and short-range Coulomb interaction will result in various interesting behaviors, such as the Mott insulator and the Hund metal, which can change the expectations from band theory. I will constraint to onsite (local) correlations, i.e. those between electrons sitting in the same lattice site. I will also briefly review some of the most important properties of unconventional superconductors and two-dimensional materials.

In the second chapter, I will review the effects of local correlations in multiorbital systems. I will compare with experimental results for high-$T_c$ iron-based superconductors with the expectations given by local correlations, arguing that the iron superconductors are in the Hund metal regime, in which the Hund's coupling plays a major role.

Last chapters are dedicated to the work done during this thesis. I studied the effects of local correlations in various high-$T_c$ iron superconductors, as well as in the magic-angle twisted bilayer graphene. A brief chapter about my ongoing work in $Fe_3GeTe_2$ is presented at the end of the thesis. Through this thesis, I used the Slave-Spin Mean-Field technique to address the local correlations behavior in these multiorbital systems.

In the third chapter, we proposed to search a new family of high-$T_c$ superconductors in the chromium analogues of iron-based superconductors. We argue that, due to the similar strength of electronic correlations, plus a superconducting instability driven by magnetic fluctuations, chromium-based pnictides and chalcogenides could host unconventional superconductivity. This argument is based on the fact that iron-based superconductors can be viewed as electron-doped Mott insulators, where the strength of correlations increases when doping these iron superconductors with holes, and decreases when doping them with electrons. In this picture, chromium pnictides and chalcogenides will be the hole-doped Mott insulator, and we found a similar trend: electronic correlations increase when doping the cromium-based systems with electrons (and decrease when doping with holes).

In the fourth chapter, I studied the strength of local correlations in the quasi-1D two-leg ladder iron-based superconductor $BaFe_2S_3$ for two different pressures. Contrary to other iron-based superconductors, $BaFe_2S_3$ (and related materials) is an insulator. Other authors have pointed out that these quasi-1D systems are Mott insulators. In this chapter, I calculated the strength of local correlations to check the behavior of these systems at $T=0 \, K$. I found a metallic behavior instead, so that we concluded by stating that the insulating behavior could be driven by finite temperature effects. I found a substantial Fermi surface reconstruction due to local correlations, contrary to what happen in other iron-based superconductors. 

In the fifth chapter, I studied the nature of the insulating states in magic-angle twisted bilayer graphene. I implemented the Zeeman effect in the Slave-Spin Mean-Field formalism to address the behavior of the insulating states when varying an onsite magnetic field in the local correlations picture. I found that the behavior is opposite to the experimental evidences. We argued that local correlations by themselves cannot explain the insulating states in magic-angle twisted bilayer graphene. We reviewed the last works done in non-local correlations in other lattices, concluding that the insulating states in magic-angle twisted bilayer graphene could be explained by using the non-local correlations picture.

In the sixth chapter, I obtained the band structure and tight-binding model of the 2D itinerant ferromagnet $Fe_3GeTe_2$. This is a brief chapter about the current status on my work in this system.

A final chapter is devoted to the conclusions and a final overview that can be extracted from this thesis. Various appendices indicate the details of the techniques used during this thesis, as well as some interesting mathematical proofs.

\vfil
{\bf Keywords: } Unconventional superconductivity, strongly correlated electron systems, Mott insulator, multiorbital systems, Hund metal, metal-to-insulator transition, many-body techniques, Slave-Spin Mean Field formalism, local correlations, magic-angle twisted bilayer graphene.
}

\newpage{\
\thispagestyle{empty}}


\resumen{
\subsection*{Resumen}

Las correlaciones electr\'onicas en F\'isica de la Materia Condensada juegan un papel fundamental en varios sistemas importantes, como son los materiales magn\'eticos, los superconductores, los materiales topol\'ogicos, las redes \'opticas, etc. Las correlaciones electr\'onicas son aquellas que modifican el movimiento de electrones individuales cuando se considera su interacci\'on con otros electrones dentro del material. Dentro de los sistemas disponibles para estudiar los efectos de las correlaciones electr\'onicas, en esta tesis voy a centrarme en los superconductores no convencionales, espec\'ificamente en los superconductores de alta $T_c$ basados en hierro, y en sistemas bidimensionales, como el recientemente descubierto grafeno bicapa rotado en \'angulo m\'agico o el ferromagneto itinerante $Fe_3GeTe_2$.

En el primer cap\'itulo, resumir\'e brevemente la teor\'ia de bandas y la teor\'ia del l\'iquido de Fermi para sistemas s\'olidos. En ciertas situaciones, el c\'aracter de largo alcance de la interacci\'on de Coulomb puede ignorarse, y la interacci\'on de Coulomb de corto alcance resulta en varios comportamientos interesantes, tales como el aislante de Mott y el metal de Hund, los cuales cambian las expectativas dadas por la teor\'ia de bandas. Me centrar\'e en las correlaciones en el mismo sitio (locales), es decir, aquellas entre electrones que est\'an localizados en el mismo sitio de la red. Tambi\'en resumir\'e brevemente algunos de las propiedades m\'as importantes de los superconductores no convencionales y de los materiales bidimensionales.

En el segundo cap\'itulo, repasar\'e los efectos de las correlaciones locales en sistemas multi-orbitales. Comparar\'e los resultados experimentales para los superconductores de alta $T_c$ basados en hierro con las expectativas dadas por las correlaciones locales, y argumentar\'e que los superconductores de hierro est\'an en el r\'egimen del metal de Hund, en el cual el acoplo Hund juega un papel esencial.

Los \'ultimos cap\'itulos est\'an dedicados a los trabajos realizados durante esta tesis. Estudi\'e los efectos de las correlaciones locales en varios superconductores de alta $T_c$ de hierro, as\'i como en el grafeno bicapa rotado en \'angulo m\'agico. Al final de la tesis hay un cap\'itulo dedicado a mi trabajo actual en el $Fe_3GeTe_2$. A lo largo de la tesis, he usado la t\'ecnica de Espines Esclavos en Campo Medio para estudiar el comportamiento de las correlaciones locales en estos sistemas multi-orbitales.

En el tercer cap\'itulo, propusimos buscar una nueva familia de superconductores de alta $T_c$ en sistemas de cromo, an\'alogos a los superconductores basados en hierro. Argumentamos que debido a que las correlaciones electr\'onicas tienen una magnitud similar, y asumiendo que la inestabilidad superconductora es debida a fluctuaciones magn\'eticas, los pnicturos y calcogenuros basados en cromo podr\'ian dar lugar a una fase superconductora no convencional. Este argumento est\'a fundamentado en el hecho de que los superconductores basados en hierro pueden verse como aislantes de Mott dopados con electrones en donde el valor de las correlaciones electr\'onicas aumentan cuando se dopan estos superconductores de hierro con huecos, mientras que disminuyen cuando se dopan con electrones. De acuerdo a esta idea, los pnicturos y calcogenuros de cromo se comportan como aislantes de Mott dopados con huecos, y encontramos unas tendencias similares: las correlaciones electr\'onicas aumentan cuando se dopan los sistemas basados en cromo con electrones (y disminuyen cuando se dopan con huecos).

En el cap\'itulo cuarto, estudi\'e el valor de las correlaciones electr\'onicas en el superconductor basado en hierro quasi-1D en escalera de dos patas $BaFe_2S_3$ para dos presiones distintas. Al contrario que en otros superconductores basados en hierro, $BaFe_2S_3$ (y otros materiales relacionados) es un aislante. Otros autores se\~nalaron que estos sistemas quasi-1D son aislantes de Mott. En este cap\'itulo, calcul\'e el valor de las correlaciones electr\'onicas para estudiar el comportamiento de estos sistemas a $T=0\, K$. Encontr\'e un comportamiento met\'alico en vez de aislante, por lo que concluimos que el comportamiento aislante en este sistema podr\'ia ser debido a los efectos de incluir una temperatura finita. Encontr\'e una reconstrucci\'on substancial de la superficie de Fermi debida a las correlaciones electr\'onicas, diferente a lo que sucede en otros superconductores basados en hierro.

En el quinto cap\'itulo, estudi\'e la naturaleza de los estados aislantes en el grafeno bicapa rotado en \'angulo m\'agico. Implement\'e el efecto Zeeman en el formalismo de Espines Esclavos en Campo Medio para estudiar el comportamiento de estos estados aislantes cuando un campo magn\'etico en el mismo sitio var\'ia, considerando solo las correlaciones locales. Encontr\'e que el comportamiento es el opuesto al obtenido experimentalmente. Argumentamos que las correlaciones locales por s\'i solas no pueden explicar los estados aislantes en el grafeno bicapa rotado en \'angulo m\'agico. Examinamos los \'ultimos trabajos hechos en otras redes para las correlaciones no locales, y concluimos que los estados aislantes en el grafeno bicapa rotado en \'angulo m\'agico podr\'ian explicarse teniendo en cuenta las correlaciones no locales.

En el cap\'itulo sexto, obtuve la estructura de bandas y el modelo de enlaces fuertes del ferromagneto itinerante 2D $Fe_3GeTe_2$. \'Este es un cap\'itulo breve sobre el estado actual de mi trabajo en dicho material.

El cap\'itulo final est\'a dedicado a las conclusiones y un resumen final de las ideas que pueden extraerse de esta tesis. Varios ap\'endices indican los detalles de las t\'ecnicas usadas durante esta tesis, as\'i como algunas demostraciones matem\'aticas interesantes.

\vfil
{\bf Palabras clave: } Superconductividad no convencional, sistemas de electrones fuertemente correlacionados, aislante de Mott, metal de Hund, transici\'on metal-aislante, t\'ecnicas de muchos cuerpos, formalismo de espines esclavos en campo medio, correlaciones locales, grafeno bicapa rotado en \'angulo m\'agico.
}

\newpage{\
\thispagestyle{empty}}






\clearpage  












\setstretch{1}  

\pagestyle{fancy}  

    \renewcommand*\contentsname{{\bf Contents}}
\lhead{\emph{ Contents}}  
\tableofcontents  

\newpage{\
\thispagestyle{empty}}

\clearpage

        \renewcommand*\listfigurename{\bf List of Figures}
\lhead{\emph{List of Figures}}  
\listoffigures  


\setstretch{1.5}  
\clearpage  
\lhead{\emph{Abbreviations}}  
\listofsymbols{ll}  
{
\textbf{BZ} & \textbf{B}rillouin \textbf{Z}one \\
\textbf{FT} & \textbf{F}ourier \textbf{T}ransformation \\
\textbf{DOS} & \textbf{D}ensity \textbf{O}f \textbf{S}tates \\
\textbf{DFT} & \textbf{D}ensity \textbf{F}unctional \textbf{T}heory \\
\textbf{LDA} & \textbf{L}ocal \textbf{D}ensity \textbf{A}pproximation \\
\textbf{GGA} & \textbf{G}eneralized \textbf{G}radient \textbf{A}pproximation \\
\textbf{FLT} & \textbf{F}ermi-\textbf{L}iquid \textbf{T}heory \\
\textbf{SCES} & \textbf{S}trongly \textbf{C}orrelated \textbf{E}lectron \textbf{S}ystems \\
\textbf{FeSCs} & Iron-based superconductors \\
\textbf{PM} & \textbf{P}ara\textbf{m}agnetism \\
\textbf{AFM} & \textbf{A}ntiferro\textbf{m}agnetism \\
\textbf{FM} & \textbf{F}erro\textbf{m}agnetism \\
\textbf{NM} & \textbf{N}on-\textbf{M}agnetic \\
\textbf{DMFT} & \textbf{D}ynamical \textbf{M}ean-\textbf{F}ield \textbf{T}heory \\
\textbf{SSMF} & \textbf{S}lave-\textbf{S}pin \textbf{M}ean-\textbf{F}ield \\
\textbf{CDMFT} & \textbf{C}luster \textbf{D}ynamical \textbf{M}ean-\textbf{F}ield \textbf{T}heory \\
\textbf{GA} & \textbf{G}utzwiller \textbf{A}pproximation \\
\textbf{RPA} & \textbf{R}andom \textbf{P}hase \textbf{A}pproximation \\
\textbf{cRPA} & \textbf{c}onstrained \textbf{R}andom \textbf{P}hase \textbf{A}pproximation \\
\textbf{OSMP} & \textbf{O}rbital \textbf{S}elective \textbf{M}ott \textbf{P}hase \\
\textbf{OSMT} & \textbf{O}rbital \textbf{S}elective \textbf{M}ott \textbf{T}ransition \\
\textbf{PES} & \textbf{P}hoto\textbf{E}mission \textbf{S}pectroscopy \\
\textbf{ARPES} & \textbf{A}ngle-\textbf{R}esolved \textbf{P}hoto\textbf{E}mission \textbf{S}pectroscopy \\
\textbf{QOs} & \textbf{Q}uantum \textbf{O}scillations \\
\textbf{MA-TBG} & \textbf{M}agic-\textbf{A}ngle \textbf{T}wisted \textbf{B}ilayer \textbf{G}raphene \\
\textbf{TBG} & \textbf{T}wisted \textbf{B}ilayer \textbf{G}raphene \\
\textbf{TTG} & \textbf{T}wisted \textbf{T}rilayer \textbf{G}raphene \\
\textbf{QCP} & \textbf{Q}uantum \textbf{C}ritical \textbf{P}oint \\
\textbf{hBN} & \textbf{h}exagonal \textbf{B}oron \textbf{N}itride \\
\textbf{TMDC} & \textbf{T}ransition \textbf{M}etal \textbf{D}i\textbf{c}halcogenides \\
\textbf{TMMC} & \textbf{T}ransition \textbf{M}etal \textbf{M}ono\textbf{c}halcogenides \\
\textbf{TMTC} & \textbf{T}ransition \textbf{M}etal \textbf{T}ri\textbf{c}halcogenides \\
\textbf{vdW} & \textbf{v}an-\textbf{d}er-\textbf{W}aals \\
\textbf{BCS} & \textbf{B}ardeen-\textbf{C}ooper-\textbf{S}chrieffer \\
\textbf{MLWF} & \textbf{M}aximally-\textbf{L}ocalized \textbf{W}annier \textbf{F}unctions \\
\textbf{VASP} & \textbf{V}ienna \textbf{a}b-initio \textbf{S}imulation \textbf{P}ackage \\
\textbf{PBE} & \textbf{P}erdew-\textbf{B}urke-\textbf{E}rnzerhof \\
\textbf{fRG} & \textbf{f}unctional \textbf{R}enormalization \textbf{G}roup \\
\textbf{APW} & \textbf{A}ugmented \textbf{P}lane \textbf{W}aves \\
\textbf{LAPW} & \textbf{L}inearized \textbf{A}ugmented \textbf{P}lane \textbf{W}aves
}

\setstretch{1}  

\pagestyle{empty}  

\addtocontents{toc}{\vspace{1em}}  

\newpage{\
\thispagestyle{empty}}

\mainmatter	  
\pagestyle{fancy}  




\chapter{{\bf Introduction}} 
\label{Chap01}
\lhead{\textbf{Chapter \ref{Chap01}}. \emph{Introduction}} 

\section{Motivation of this thesis}
\label{1.1}

A solid material is a system formed by an infinite arrange of atoms ordered in a specific pattern, called the lattice \cite{AshcroftMermin1976}. The material can be obtained by infinitely reproducing the unit cell in the real space, which is the most basic unit arrangement of atoms that describes the system. Each material has a different unit cell, with a different symmetry pattern and/or formed by different atoms. In a first approximation, ions (the atomic nucleus plus the core electrons) in the unit cell are siting motionless in their sites (Born-Oppenheimer approximation), and valence electrons will move through the system according to the band theory of solids.

In band theory, the eigenstates and eigenvalues satisfy the Bloch theorem, which states that electrons move in the average periodic potential created by the motionless atoms. The eigenvalues are also called electronic band energies $E_{k\nu\sigma}$, and they are defined in the $k$-space in the so called first Brillouin zone (BZ), which is the Fourier transformed (FT) unit cell, $\nu$ is the band index and $\sigma$ is the spin index. At the practical level, the electronic band energies (or equivalently the band structure) are obtained by Density Functional Theory (DFT) calculations, explained more in detail later on and in \aref{AppA}. In this framework, the electronic band structure can be extracted from the hybridization of atomic-like orbitals $s$, $p$, $d$, $f$,..., which come from each atom of the unit cell. The number of bands will be determined by the number of atoms in the unit cell $N_{uc}$, the number of orbitals $N$ of each atom (where the total number of orbitals per unit cell can be alternatively defined as $N_{orb}=N_{uc}N$) and the spin degree of freedom for a given system. In DFT calculations, some effects coming from the electron-electron repulsion are included in the average periodic potential in which electrons move.

The number of (valence) electrons per unit cell is given by the total number of valence electrons coming from each atom in the unit cell. Thus, the band energies are filled bottom-up until all the electrons are distributed (following the Pauli exclusion principle) in the band structure. At $T=0\,K$, the last filled band energy is defined as the Fermi energy $\varepsilon_F$. The Fermi energy is defined as the chemical potential at zero temperature, $\varepsilon_F = \mu(T=0\,K)$, where the chemical potential is the change in energy when a new particle is added to the system. At finite and low temperature $T$, electrons distribute in the band energies following the Fermi-Dirac distribution function.

\begin{figure}[h]
   \centering
   \includegraphics[width=0.8\columnwidth]{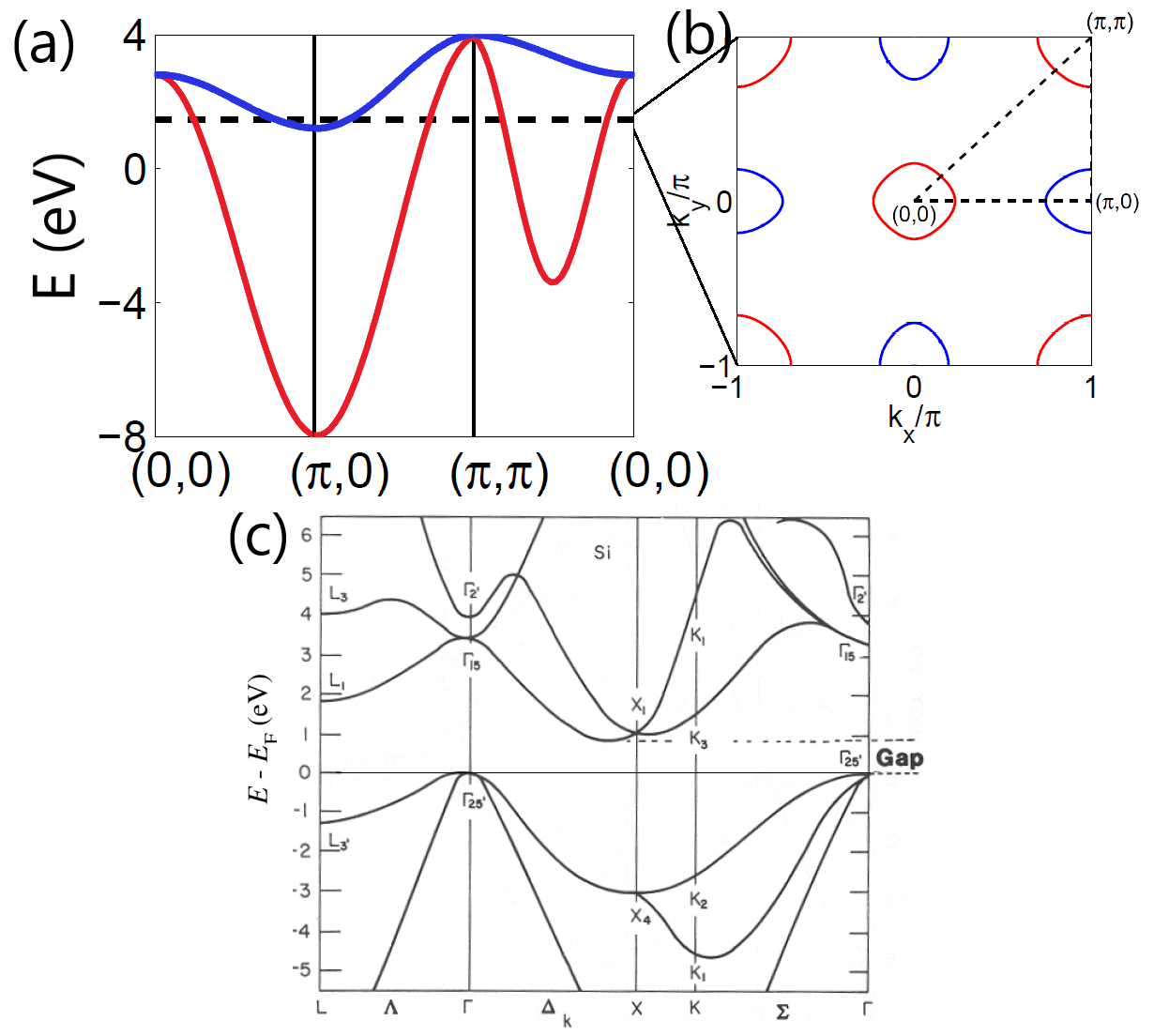}
   \caption[Band structures for a metal and an insulator, and Fermi surface of the metal]{\textbf{(a)} Band structure (dashed black line marks $\varepsilon_F$) and \textbf{(b)} Fermi surface for a model proposed for metallic $LaFeAsO$ in the unfolded BZ (see \sref{1.5.2}). The Fermi surface is formed by various hole- (red lines) and electron-pockets (blue lines). Dashed black line in the Fermi surface marks the $\vec{k}$-path followed to obtain the band structure. Taken and adapted from \cite{RagPRB772008}. \textbf{(c)} Band structure calculation for semiconducting silicon ($\varepsilon_F$ fixed at $0 \, eV$). A (indirect) gap opens at $\varepsilon_F$, where $gap \sim 1 \, eV$. Taken and adapted from \cite{JenPh201954}.}
   \label{fig:1.1}  
\end{figure}

A major success of band theory is the ability to predict a metallic or insulating behavior for a given material. Then, if $\varepsilon_F$ crosses the bands, the system is a metal (\fref{fig:1.1}\textbf{(a)}), while if there is an energy gap between $\varepsilon_F$ and the next available energy, the system is an insulator (\fref{fig:1.1}\textbf{(c)}). Then, for an odd number of electrons per unit cell, the system will be always behave as a metal, whether if it has an even number of electrons per unit cell, \textcolor{black}{band theory predicts that} the system can behave as a metal or an insulator.

When $\varepsilon_F$ crosses the bands, the closed surfaces in the first BZ with band energies $E_{k\nu\sigma}=\varepsilon_F$ are defined as the Fermi surface. The notation hole- or electron-pocket is used to denominate the closed surfaces which have positive or negative curvature for $E_{k\nu\sigma}$ close to $\varepsilon_F$, respectively, see \fref{fig:1.1}\textbf{(b)}. For a given band $E_k$, the electron effective mass $m_{0}$ can be defined as $m_{0} = | \nabla_k^2 E_{k}/\hbar |^{-1}$. This parameter states the variation \textcolor{black}{of the} electron mass when the electron is moving in the periodic potential of a solid.

In band theory, when a new particle (hole or electron) is added to the system, $\varepsilon_F$ shift downwards or upwards and the Fermi surface changes. This shift is said to be a rigid-band shift, i.e. the band structure remains unchanged while $\varepsilon_F$ moves. The parent compound of a material is defined as the system with no external particles added and no external perturbations, such as pressure, magnetic or electric fields.

A priori, the Fermi level shift when new particles are added into the system should not be rigid, due to the fact that the new particle will interact with the other electrons in the material, and hence the band structure should change. Nevertheless, band theory and rigid-band shift work on the basis of the Fermi liquid theory (FLT) for solids \cite{Coleman2015,emergence2017}, see \sref{1.2.2}. FLT justifies the success of band predictions in most of the materials. In this framework, the excitations (i.e. when a new particle is added to the system) are described as single-particle states in the low $T$ and low energy regime. These excitations have a finite lifetime, which in the FLT \textcolor{black}{at low $T$} will be large \textcolor{black}{enough close to the Fermi level} to legitimate the single-particle description. In FLT, the low temperature-dependent electronic properties, such as the resistivity $\rho(T)$, the specific heat $C(T)$ or the susceptibility $\chi(T)$, can be obtained and compared with experimental results.

However, there are some situations in which FLT, and hence the band theory can fail. In these cases, the experimentally observed band structure and electronic properties deviate from the DFT calculations. For example, the observed band structure \textcolor{black}{of some metals} is much narrower than calculated using DFT \cite{LiuPRB922015}, or an expected metallic material behaves as an insulator \cite{PicRMP611989}. These behaviors result from \textcolor{black}{the effects of the} electronic correlations. \textcolor{black}{Including these} electronic correlations will describe how electrons change their motion through the material when they interact with other electrons. Then, further models and approximations beyond DFT are needed to describe such new systems, usually called strongly correlated electron systems (SCES) \cite{Coleman2015,emergence2017,Fazekas1999}. This is the case for the previously commented situations, when the band structure narrows in a metal or when an expected metal behaves as an insulator, the so called Mott insulator \cite{PicRMP611989}.

Among the systems which show increased electronic correlation effects, in this thesis I will focus on high-$T_c$ iron-based superconductors (FeSCs) \cite{emergence2017,BasCRP172016} and 2D materials \cite{RolCSR462017,Schonhoff2017thesis}, specifically in magic-angle twisted bilayer graphene (MA-TBG) and $Fe_3GeTe_2$ (FGT). \textcolor{black}{I will specifically focus on studying such systems when considering the interaction between electrons in the same lattice site, in the so called local approximation (see next section)}. A major question that I would like to answer through this thesis is: \emph{what is the strength of these electronic correlations and how \textcolor{black}{does it} relate with the experimental properties observed?}. This question is crucial in order to understand the origin of various phases, like magnetism, superconductivity, etc, that appear in such systems

The \textcolor{black}{strength} of electronic correlations can be modeled by a parameter called the quasiparticle weight $Z$. \textcolor{black}{Alternatively, in the local approximation}, the effective mass renormalization, or mass enhancement \textcolor{black}{can be defined as} $m^*/m_{0}=1/Z$, where $m_0$ is the non-correlated electron mass obtained from DFT \cite{Coleman2015,emergence2017}. \textcolor{black}{The quasiparticle weight} is a well defined quantity (for a metal) in the FLT and ranges from $0<Z<1$. For $0.7<Z \leq 1$, the \textcolor{black}{general} expectations from band theory are recovered, and the system is said to be weakly correlated. For $Z<1$, the system start to become more and more correlated, progresively entering in the regime which is known as the correlated metal. \textcolor{black}{When} the system becomes the special insulator, called the Mott insulator, \textcolor{black}{$Z=0$}. In this situation, FLT breaks down and further approximations are needed to describe the physics of the system. \textcolor{black}{Due to the relation between $Z$ and $m^*$ found in the local approximation}, $Z$ can be identified as a renormalization pre-factor of the band structure $\sim Z E_{k}$, hence when electronic correlations increase, the bands become narrower. We will later see that this effect translates into a non-rigid shift when the chemical potential varies because \textcolor{black}{$Z$ depends on the electronic filling}. 

The mass renormalization $m^*/m_0 = 1/Z$ gives a physical explanation of what is occurring in the system: at $m^*=m_0$ ($Z=1$), the electron has an effective mass which is equal to the non-correlated one; when correlations increase, $m^*>m_0$ ($Z<1$), the electrons start to become heavier, so their motion is stopped progresively, increasing the resistivity of the material; at $m^* \rightarrow \infty$ ($Z=0$), electrons are infinitively heavy, so they will prefer to stay motionless in the system, and the material becomes an insulator. \textcolor{black}{Note that this description is based on the Brinkman-Rice picture of the Mott transition (see next sections).}

The metal-to-insulator transition described above, also called the Mott transition, is at the heart of various SCES, such as unconventional superconductors, \textcolor{black}{various} two-dimensional (2D) materials, see \sref{1.3}, or for ultra-cold atoms in optical lattices.

\section{Many-body hamiltonian and the concept of correlation}
\label{1.2}

In this section, I will formally introduce the hamiltonian which describes a solid material, also called a many-body hamiltonian, and the concept of electronic correlations. I will follow a similar derivation as the ones done in \cite{AtlandSimons2010,Ferber2012thesis,Schonhoff2017thesis}.

The hamiltonian of a solid material, with $N_{I}$ ions and $N_{el}$ electrons, can be written as:

\begin{equation}
H=\underbrace{- \frac{\hbar}{2M} \sum_{\alpha}^{N_{I}} \nabla_\alpha^2}_{T_{I}} \underbrace{- \frac{\hbar}{2m} \sum_{i}^{N_{el}} \nabla_i^2}_{T_{el}} \underbrace{+ \sum_{\alpha \neq \beta}^{N_I} \frac{Z_\alpha Z_\beta e^2}{|\vec{R}_\alpha - \vec{R}_\beta|}}_{V_{I-I}} \underbrace{- \sum_{\alpha}^{N_{I}} \sum_{i}^{N_{el}} \frac{Z_\alpha e^2}{|\vec{r}_i - \vec{R}_\alpha|}}_{V_{I-el}} \underbrace{+ \sum_{i \neq j}^{N_{el}} \frac{e^2}{|\vec{r}_i - \vec{r}_j|}}_{V_{el-el}}
\label{eq:1.1}
\end{equation}

where atomic units $\varepsilon_0 = 1/4\pi$) have been used. $M$ and $m$ are the free ion and electron masses, $\vec{R}_{\alpha}$ is the ion $\alpha$ position, $\vec{r}_{i}$ is the electron $i$ position and $Z_{\alpha}$ is the atomic number of the $\alpha$ ion. Various kinetic and interaction energy operators have been defined: $T_{I}$ is the ionic kinetic energy, $T_{el}$ is the electronic kinetic energy, $V_{I-I}$ is the ion-ion interaction, $V_{I-el}$ is the ion-electron interaction and $V_{el-el}$ is the electron-electron interaction.

Due to the massive character of ions when compared with electrons ($M\gg m$), ions move much slowly than electrons, so in a first approximation, ions can be treated as sitting motionless, considering $T_I$ as a perturbation $H = H_{BO} + T_{I}$ (Born-Oppenheimer approximation), where:

\begin{equation}
H_{BO} = \underbrace{T_{el}}_{\hat{T}} + \underbrace{V_{I-I} + V_{I-el}}_{\hat{V}_{ext}} + \underbrace{V_{el-el}}_{\hat{V}_{ee}}
\label{eq:1.2}
\end{equation}

The hamiltonian $H_{BO} =\hat{T}+\hat{V}_{ext}+\hat{V}_{ee}$, with $\hat{T}$ being the kinetic energy of the electrons, $\hat{V}_{ext}$ the external potential generated by the ions in which the electrons are moving and $\hat{V}_{ee}$ is the electron-electron interaction. From now on, $T_{I}$ effects will be neglected. 

DFT is the most widely used approximation to solve the many-body hamiltonian \eref{eq:1.2}, see \aref{AppA}. DFT allows to calculate the band structure $E_{k\nu\sigma}$ for different materials. Each atom in the unit cell is considered to contribute with its atomic orbitals. The number of bands $\nu$ equals the number of orbitals per unit cell and spin $2N_{orb}$. When treating $\hat{T}$, the eigenstates are described as a linear combination of single-particle states, hence the electrons will behave almost independently of each other (they are not electronically correlated).  Also, when treating $\hat{V}_{ee}$, a Hartree-like approximation is made, and hence the exchange effects are almost neglected. Thus, the DFT hamiltonian $H_0$ can be written as:

\begin{equation}
H_{0} = \hat{T}^{noncorr} + \hat{V}_{ext} + \hat{V}_{ee}^{H} + \hat{V}_{XC}
\label{eq:1.3}
\end{equation}

where $\hat{T}^{noncorr}$ refers to the non-correlated kinetic energy of the electrons, $\hat{V}_{ee}^{H}$ is the electron-electron interaction in the Hartree approximation and $\hat{V}_{XC}$ encodes the electronic correlation and the exchange effects. In the most common DFT approaches, such as local density approximation (LDA) or generalized gradient approximation (GGA), $\hat{V}_{XC}$ is approximated by different functions, see \aref{AppA} for a brief discussion, and biggest correlation and exchanges effects are neglected. Thus, we can conclude by considering the most common DFT band structure calculations (LDA or GGA) as the weakly correlated limit, and adding new particles to the system will rigidly shift (i.e. the band structure remains unchanged) the Fermi energy $\varepsilon_F$.


\subsection{Tight-binding models}
\label{1.2.1}

The hamiltonian \eref{eq:1.3} can be translated into a tight-binding model \cite{AshcroftMermin1976,AtlandSimons2010}. During this thesis, I will mainly use the second quantization notation. For a tight-binding model in the second quantization notation, $H_0$ is encoded in the hopping integrals $t_{mn}^{ij}$. These hopping integrals describe the possible hopping processes for an electron in an orbital $n$ at the lattice site $j$ which hops to an orbital $m$ in a lattice site $i$. Then, $H_0$ can be written as:

\begin{equation}
H_0 = \sum_{m n} \sum_{ij} \sum_\sigma t_{m n}^{ij} d_{i m \sigma}^\dagger d_{j n \sigma}
\label{eq:1.4}
\end{equation}


where $d_{im\sigma}$ ($d_{im\sigma}^\dagger$) annihilates (creates) an electron in a lattice site $\vec{i}$, in a orbital $m$ and with spin $\sigma$. $\sum_{ij}$ runs for all the neighbors around the considered origin lattice site. $t_{mn}^{ij}$ \textcolor{black}{can be} complex numbers and they usually receive names such as nearest or first neighbor, next-nearest or second neighbor, etc., refering to the distance from the origin. Tight-binding models will only include up to a certain number of neighbor hoppings.

The dispersion relations $\varepsilon_{kmn}$ are defined by the FT of the hopping \textcolor{black}{terms in \eref{eq:1.4}}, as shown in \eref{eq:1.5}, and the band structure $E_{k\nu}$ is obtained by diagonalizing the hamiltonian defined by $H_0=\sum_{kmn\sigma} \varepsilon_{kmn} d_{km\sigma}^\dagger d_{kn\sigma}$. Tight-binding models can be used as simple models to address different physical behaviors, as for example, when considering the 2D square lattice, in a single-orbital system with only nearest neighbors $t$ included: 

\begin{equation}
\varepsilon_{kmn} = \sum_{ij} t_{mn}^{ij} e^{-i \vec{k} \cdot \vec{ij}} \: \rightarrow \: \varepsilon_{k}^{2D \, square} = 2t \left( \cos{k_x} + \cos{k_y} \right)
\label{eq:1.5}
\end{equation}

\begin{figure}[h]
   \centering
   \includegraphics[width=0.8\columnwidth]{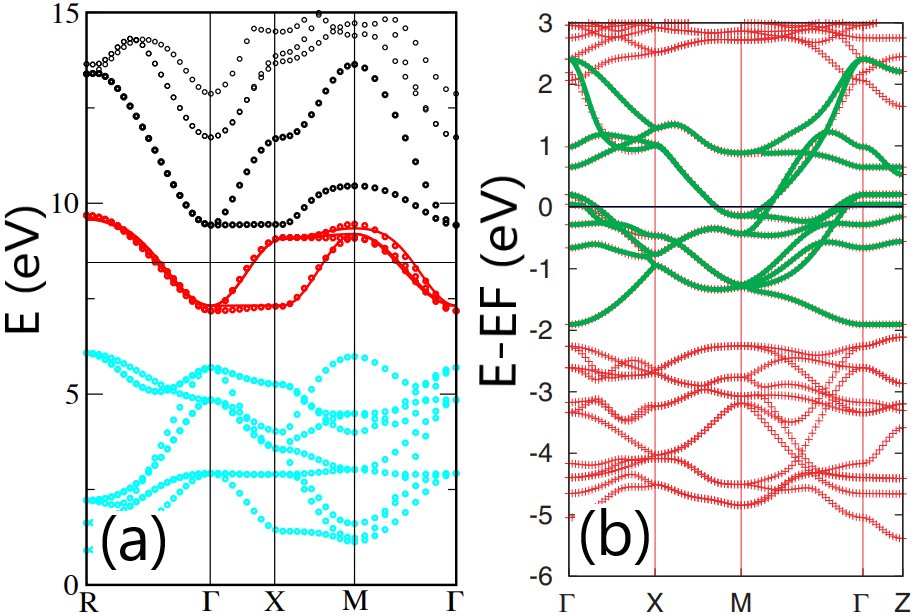}
   \caption[Disentangled and entangled bands, and Wannier parametrization fitting]{\textbf{(a)} Disentangled bands (red points) and Wannier parametrization fitting (red lines) for $SrVO_3$. The perfect fitting is done in a set of bands which are well separated from other untargeted bands. \textbf{(b)} Same for $LaFeAsO$ in the folded BZ, see \sref{1.5.2}, which shows some entanglement between the $Fe$ $d$ bands around the Fermi level and the next bands. The Wannier fitting (green lines) is done in a set of bands which are entangled with other untargeted bands. The fitting shows discrepancies in certain areas (see $E-E_F \sim 2 \, eV$), but the Wannier fitting gives good enough results \textcolor{black}{to address the electronic properties at low energies}. Taken and adapted from \cite{KunCPC1812010}.}
   \label{fig:1.2}  
\end{figure}

Tight-binding models are applied to study real materials. In order to calculate the hopping parameters for a real material, a procedure called Wannier parametrization is used, see \sref{A.2}. In this procedure, a set of targeted band energies $E_{k\nu}$ are fitted to a set of hopping integrals $t_{mn}^{ij}$. The goal of a Wannier parametrization is to be able to reduce the original $\nu$ bands problem, to a smaller set of bands which gives enough information to explain the physics in a given material. Usually, bands around the Fermi level and well separated by a gap from next bands can be targeted to obtain the tight-binding model, see red bands in \fref{fig:1.2}\textbf{(a)}. In some cases, there is not a clear gap between the targeted bands and the next bands (they are said to be entangled), and the procedure will give a less \textcolor{black}{accurate} result, see \fref{fig:1.2}\textbf{(b)}. The orbitals for which $t_{mn}^{ij}$ are obtained are called Wannier orbitals, and they are not necessarily the same as the atomic orbitals. The degree of similarities between Wannier and atomic orbitals depend on how entangled the targeted and untargeted bands are. In most of the cases, and through all this thesis, I will focus on Wannier orbitals which have the same symmetry than the atomic orbitals, hence we can distinguish Wannier orbitals like $s$, $p_x$, $p_y$ and $p_z$, $d_{zx}$, $d_{yz}$, $d_{x^2-y^2}$, $d_{z^2}$ or $d_{xy}$, etc, depending on their quantum angular momentum numbers $l=1,2,\ldots$ and $m_l=-l,\ldots, +l$

\begin{figure}[h]
   \centering
   \includegraphics[width=0.9\columnwidth]{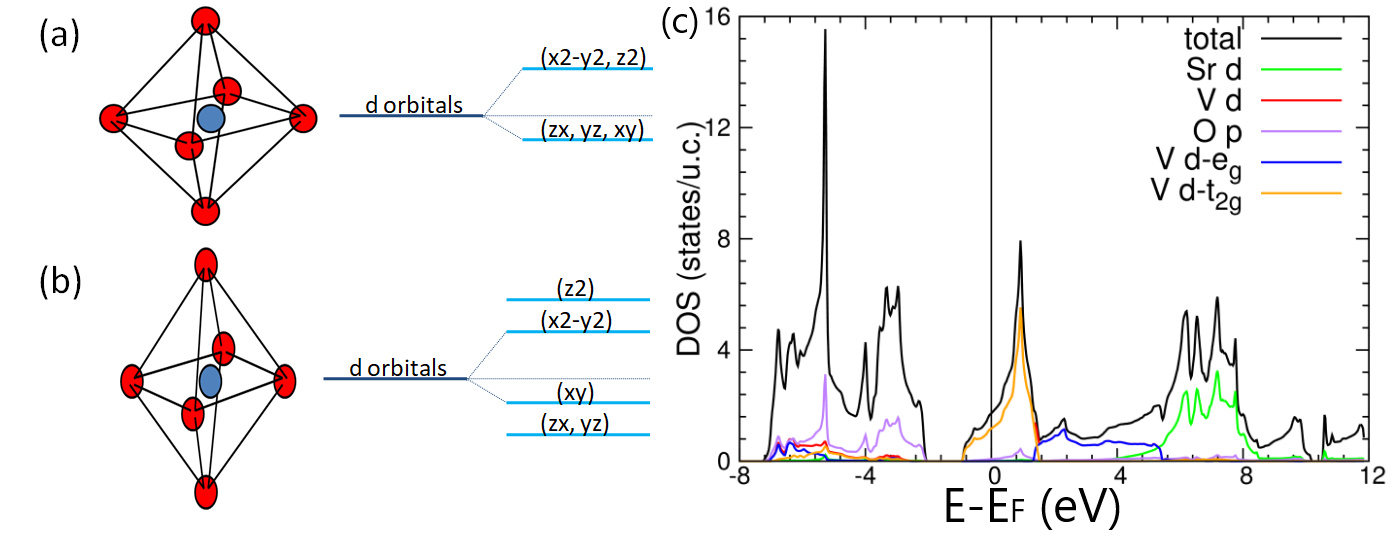}
   \caption[Crystal field splitting in cubic and tetragonal symmetries for perovskite crystal structures, and total and orbital-resolved DOS in $SrVO_3$]{Crystal field splitting in \textbf{(a)} cubic and \textbf{(b)} tetragonal symmetries for perovskite crystal structures. The original $d$ orbitals (marked as a dark blue line) become non-degenerated due to environment and the shape of each orbital (marked as light blue lines). Taken and adapted from \cite{emergence2017}. \textbf{(c)} Total (black lines) and orbital-resolved (colored lines) DOS for $SrVO_3$. $d$ orbitals from $V$ are mostly located from $-1 \, eV$ to $5 \, eV$. Figure obtained by GGA DFT calculations implemented in the \textsc{Wien2k} code.}
   \label{fig:1.3}  
\end{figure}

In principle, the orbitals with the same $l$ and different $m_l$ will be energetically degenerated in each atom. However, due to the environment that surrounds each lattice site in the unit cell, a crystal field spltting $\epsilon_m$ appears, which is $\epsilon_m = t_{mm}^{ij=0}$. i.e. an onsite energy which breaks the $m_l$-orbital degeneracy \cite{Fazekas1999,emergence2017}. The crystal field splitting is an electrostatic interaction between the electrons in each site with their surrounding ions. Different crystal structure symmetries will have different crystal fields. As an example in perovskite structures (see \fref{fig:1.3}), for $d$ orbitals in a cubic symmetry, the crystal field effect breaks the degeneracy between $(d_{zx},d_{yz},d_{xy})$ (also called $t_{2g}$ orbitals) and $(d_{x^2-y^2},d_{z^2})$ (also called $e_{g}$ orbitals), while for $d$-orbitals in a tetragonal symmetry, all the orbitals except $(d_{zx},d_{yz})$ are non-degenerated.

The total bandwidth $W$ of a given band structure is the difference between the minimum and maximum band energies (for example, in \fref{fig:1.1}\textbf{(a)}, $W=12 \, eV$).  It is related with the hopping integrals $t_{mn}^{ij}$, so $W$ approximately states the kinetic energy of the electrons in a given material. In a multiorbital system, each orbital will approximately contribute to the band structure in a certain energy window. This energy window can be identified as an orbital-resolved bandwidth $W_m$. These parameters will be useful when talking about the strength of electronic correlations in the next chapters. In \fref{fig:1.3}\textbf{(c)}, I show the total and orbital-resolved density of states (DOS) for the $SrVO_3$, where $W\approx 18 \, eV$, and $V$ $d$ orbitals are mainly contributing around the Fermi level, with $W_{V \, d} \approx 6 \, eV$, $W_{V \, d \, e_g} \approx 3.5 \, eV$ and $W_{V \, d \, t_{2g}} \approx 2.5 \, eV$. The DOS is defined as the number of accessible states per energy by the electrons of a given material, so it gives the degeneracy of a given band energy $E_{k\nu}$.

\subsection{Fermi liquid theory (FLT)}
\label{1.2.2}

In this section, I will review the FLT for solid state systems, by following \cite{Coleman2015,emergence2017}. I have already mentioned that it is not trivial to see that, when adding a new electron into a material, the band structure remains unchanged while the Fermi level is rigidly shifted. FLT justifies the applicability of this single-particle states approximation for DFT band structure calculations for most of the materials. FLT is a low energy and low temperature theory, in which there is a one-to-one correspondence between the low energy excitations (called quasiparticles) and the real particles of the system. Then, the quasiparticles have the same charge $|e|$ and spin $1/2$ than the real particles. FLT is a phenomenological theory, but it can be justified by perturbative arguments (by using Green's function formalism, see the end of this section), so it is not restricted to weak interactions. FLT fails when dealing with various situations, such as for Mott insulators, Luttinger liquids, strange metal phases, etc.


In FLT, for a non-interacting hamiltonian $H_0$, the interactions between electrons $\hat{U}$ are adiabatically turned on, i.e. the ground state of $H_0$ evolves smoothly with $\hat{U}$, until there is a direct connection with the ground state of the interacting hamiltonian $H_0+\hat{U}$. For magnetic or superconducting phase transitions, the Mott transition, etc, FLT breaks down.

Then, let's assume a many-body system with $N$ particles and the ground states of the non-interacting $|\Psi_{gs}>$ and interacting $|\Psi_{gs}^{*}>$ hamiltonians are connected adiabatically. If a new particle is added to $H_0$, the new state will be $|P>=c_{k\sigma}^\dagger |\Psi_{gs}>$, where $c_{k\sigma}^{\dagger}$ creates a particle in a excited state of the non-interacting hamiltonian $H_0$ with $k$ and spin $\sigma$. When $\hat{U}$ is turned on, the system evolves adiabatically from its initial state $|P>$ to a final state $|QP>=\hat{U} |P>$. Then, quasiparticles can be defined as the elementary excitations of the interacting hamiltonian $|QP>=a_{k\sigma}^\dagger |\Psi_{gs}^{*}>$, where $a_{k\sigma}^{\dagger}$ creates a quasiparticle in a excited state of the interacting hamiltonian $H_0+\hat{U}$ with $k$ and spin $\sigma$, with $a_{k\sigma}^\dagger = \hat{U} c_{k\sigma}^\dagger {\hat{U}}^\dagger$. Thus, the elementary excitations of the interacting system ($H_0+\hat{U}$) are the quasiparticles, and not the real particles.

The electronic particle operator $c_{k\sigma}$ can be written in a perturbative way in terms of $\hat{U}$, not restricted to weak $\hat{U}$ (and dropping out the spin indices $\sigma$ for simplicity):

\begin{equation}
c_{k}^{\dagger} = \sqrt{Z_k} a_{k}^\dagger + \underbrace{\sum_{k_4+k_3=k_2+k} A(k_4,k_3,k_2,k) a_{k_4}^{\dagger}a_{k_3}^{\dagger}a_{k_2}^{}+ \cdots}_{decay \: proccesses} 
\label{eq:1.6}
\end{equation}

$Z_k$ is defined as the quasiparticle weight \textcolor{black}{(mentioned in the previous section)}, and it is given by $Z_{k}=|<\Psi_{gs}^{*}|a_{k}c_k^\dagger|\Psi_{gs}^{*}>|^2>0$, so it is the overlap between particle and quasiparticle states. The quasiparticle weight takes values $0<Z_k<1$, so when $Z_k \approx 1$, the weakly correlated limit is recovered, whether for $Z_k=0$, FLT breaks down. When $Z_k$ decreases, the strength of correlations increases. Then, $Z_k$ can be used to measure the strength of electronic correlations. Higher-order terms in \eref{eq:1.6} describe the decay processes \textcolor{black}{of an electron into several exciations}. \textcolor{black}{Then, there is a part of the electron that cannot be expressed in terms of quasiparticles. This is the so called incoherent part}. As a consequence, the quasiparticles will have a finite lifetime $\tau$.

\textcolor{black}{The single-particle states approximation makes sense if $Z_k$ is a finite number and if the quasiparticles live long enough. It can be shown that in 3D and at low $\omega$ ($|\omega-\mu| \ll \mu$) and low $T$ ($T \ll \mu=k_B T_F$):}

\begin{equation}
\frac{1}{\tau} \propto {(\omega - \mu)}^2 + T^2
\label{eq:1.7}
\end{equation}

where $\omega$ is the frequency of the elementary excitations, $\mu$ is the chemical potential (the change in the total energy when an excitation is created in the system) and $T_F$ is the characteristic temperature of the Fermi liquid metal (Fermi temperature). \textcolor{black}{The quadratic dependencies ensure the long-lived quasiparticles at low $T$ and energies}. The quasiparticle stability requires that $\Gamma / \omega \sim \omega \ll 1$, which is satisfied for 3D\footnote{In 2D $\Gamma / \omega \sim \omega \ln{\omega}$ and stability can be recovered. Interestingly, in 1D systems, $\Gamma / \omega \sim const$, so quasiparticles are not stable, and hence FLT breaks down. This is the situation for the Luttinger liquids.}, see \eref{eq:1.7}. Then, the stability condition states that the decay rate of the quasiparticles have to be much smaller than their characteristic energies.

\begin{figure}[h]
   \centering
   \includegraphics[width=0.8\columnwidth]{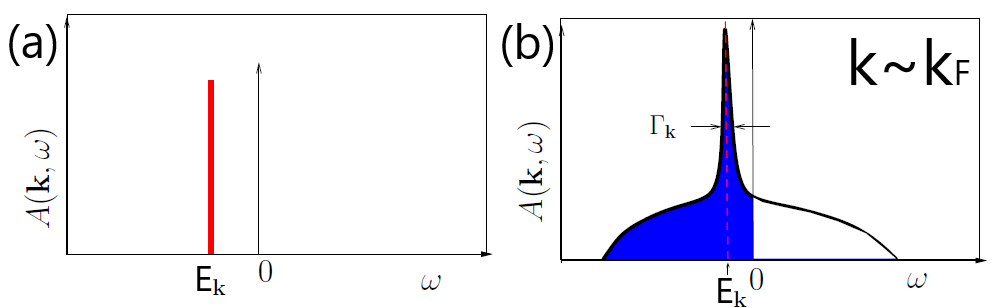}
   \caption[Spectral function for non-interacting and interacting systems]{\textbf{(a)} Spectral function for a non-interacting hamiltonian. In this case, $A(k,\omega)$ is a delta function for $\omega=E_k$. \textbf{(b)} Spectral function for the interacting hamiltonian. The quasiparticle peak can be identified at $\omega=E_k$ with a width $\Gamma$. Apart from the quasiparticle peak, there is a continuum of excitation energies for $\omega>E_k$ and $<E_k$. Taken and adapted from \cite{Coleman2015}.}
   \label{fig:1.4}  
\end{figure}

\textcolor{black}{The quasiparticle weight $Z_k$ is obtained from the excitation spectra. The excitation spectra, or alternatively the spectral function $A(k,\omega)$, is the distribution of excitations which are created when a particle \textcolor{black}{(real electrons)} is added/removed from the system. It can be measured by photoemission spectroscopy (PES). The spectral functions for a non-interacting and interacting system are shown in \fref{fig:1.4}. For a non-interacting system, $A(k,\omega)$ is a delta function. For an interacting system, the excitation spectra is formed by the so called quasiparticle peak (also called the coherent part), plus a continuum of states which are excited by the added/removed particle (the incoherent part). The continuum of states is the contribution of the decay proccesses, see \eref{eq:1.6}. The delta function adquires a width $\Gamma$ related with the decay of a quasiparticle which is not an eigenstate of the interacting hamiltonian, $\Gamma = 1/\tau$.}

The spectral function can be written in terms of the retarded propagator $G(k,\omega)$:

\begin{equation}
A(k,\omega) = - \frac{1}{\pi} Im \, G(k,\omega)
\label{eq:1.8}
\end{equation}

In Green's function formalism, the retarded propagator $G(k,\omega)$ is the probability amplitude for a particle being added/removed from the system, moving through it and ending in a state $(k,\omega)$. Then, $G(k,\omega)$ describes the propagation of a particle when it moves through a material, interacting with other electrons. It is a complex quantity which encodes all the information about the full interacting hailtonian $H_0+\hat{U}$. $G(k,\omega)$ can be written using the Dyson equation:

\begin{equation}
G(k,\omega) = \frac{1}{\omega - \varepsilon_k^0 + \mu - \Sigma(k,\omega)} = \frac{1}{\omega - \varepsilon_k ' + \mu - i \, Im \, \Sigma(k,\omega)}
\label{eq:1.9}
\end{equation}

where $\varepsilon_k^0$ is the dispersion energies of the non-interacting system (see \eref{eq:1.5}), $\varepsilon_k ' = \varepsilon_k^0 + Re \, \Sigma(k,\omega)$, $\mu$ is the chemical potential and $\Sigma (k,\omega)$ is the self-energy. The self-energy is a complex quantity which accounts for the effects \textcolor{black}{of $\hat{U}$} not included in the hamiltonian $H_0$, so it will describe the quantities that we have previously seen for the quasiparticles, $Z_k$ and $\tau$. We will see in the next paragraphs that $Re \, \Sigma(k,\omega)$ is related with $Z_k$ and $Im \, \Sigma(k,\omega)$ with $\tau$.

The poles $\omega_{pole}$ of $G(k,\omega)$ (i.e. the $\omega$ values at which $G(k,\omega)$ diverges) will describe the energies of the excitations of the interacting system, i.e. the quasiparticles. In FLT, the stability conditation states that $\Gamma/\omega = 1/ \tau \omega \ll 1$. By definition $1/\tau = Im \Sigma(k,\omega)$, then \textcolor{black}{the applicability of FLT requires that} $Im \Sigma(k,\omega)/\omega \ll 1$ and $Im \, \Sigma(k,\omega) \ll \varepsilon_k '$. For $T=0 \, K$ and small frequencies (note that $\Sigma(k,\omega) \approx Re\, \Sigma(k,\omega)$), $\Sigma (k,\omega)$ can be Taylor expanded and retain only up to linear-$\omega$ terms \cite{deM1707.03282}:

\begin{equation}
\Sigma(k,\omega) \approx Re \, \Sigma (k,0) + \omega {\left. \frac{\partial Re \, \Sigma(k,\omega)}{\partial \omega} \right|}_{\omega \rightarrow 0}
\label{eq:1.10}
\end{equation}

where the quasiparticle weight $Z_k$ is defined as:

\begin{equation}
Z_k = \frac{1}{1-{\left. \frac{\partial Re \, \Sigma (k,\omega)}{\partial \omega} \right|}_{\omega \rightarrow 0}}
\label{eq:1.11}
\end{equation}

\textcolor{black}{and} $\Sigma (k,\omega) \approx Re \, \Sigma (k,0) + \omega (1 - 1/Z_k)$, so in the FLT the \textcolor{black}{real part of the} self-energy depends linearly on $\omega$ with a factor which depends on the quasiparticle weight $Z_k$. The propagator is redefined as:

\begin{equation}
G(k,\omega) = \frac{Z_k}{\omega - Z_k (\varepsilon_k ' - \mu)}
\label{eq:1.12}
\end{equation}

\eref{eq:1.12} features a pole at $\omega_{pole} = Z_k (\varepsilon_k ' - \mu)$, where $Z_k$ plays the role of weighting this pole (hence the name of $Z_k$). It appears as a pre-factor of $\varepsilon_k '$, so the energies will suffer a renormalization due to this $Z_k$.

Around the Fermi surface $k_F$, $\Sigma (k,\omega)$ can be further expanded to obtain the effective mass renormalization, $({m^*}/{m_0})_k \approx ({1-{\left. \frac{\partial Re \, \Sigma (k,\omega)}/{\partial \omega} \right|}_{k \rightarrow k_F}})/({1-{\left. {\partial Re \, \Sigma (k,\omega)}/{\partial \omega} \right|}_{\omega \rightarrow 0}})$. If $\Sigma(k,\omega)$ is $k$-independent (as I will discuss in \sref{1.3} and during a great part of the thesis), $1-\left. \partial \Sigma ' (k,0) / \partial k \right|_{k \rightarrow k_F} = 1$, and then:

\begin{equation}
\frac{m^*}{m_{0}}=\frac{1}{Z}
\label{eq:1.13}
\end{equation}

In this case, the mass renormalization for $k$-independent self-energy gives a clear physical meaning for $Z$: at $m^*=m_0$ ($Z=1$), the electron has an effective mass which is equal to the non-correlated one, so the system behaves as a non-correlated metal, $m^*<m_0$ ($Z<1$), the electrons become heavier, so the system progresively moves from a weakly correlated metal (\textcolor{black}{$0.7<Z \leq 1$}) to a moderately (\textcolor{black}{$0.3<Z \lesssim 0.7$}), and then a strongly (\textcolor{black}{$0<Z \lesssim 0.3$}) correlated metal; at $m^* \rightarrow \infty$ ($Z=0$), electrons are infinitively heavy, so they will prefer to sit motionless in the material, and the system behaves as an insulator. Now, we will see how to measure the mass renormalization.

Measurable thermodynamic quantities can be extracted by using the Landau's free energy functional $F[\delta \hat{n}_{k}]$, where $\hat{n}_k = a_k^\dagger a_k$ is the quasiparticle number operator. This functional is obtained by an expansion of $H_0+\hat{U}$ around the equilibrium density of quasiparticles, which is a small quantity:

\begin{equation}
F[\delta \hat{n}_k] = F_{gs} + \sum_k \underbrace{\left. \frac{\partial F}{\partial \delta \hat{n}_k} \right|_{\delta \hat{n}_k \rightarrow 0}}_{E_k^*} \delta \hat{n}_k + \sum_{kk'} \underbrace{\left. \frac{\partial^2 F}{\partial \delta \hat{n}_k \partial \delta \hat{n}_{k'}} \right|_{\delta \hat{n}_{k,k'} \rightarrow 0}}_{f_{k,k'}} \delta \hat{n}_k \delta \hat{n}_{k'} + \cdots
\label{eq:1.14}
\end{equation}

\textcolor{black}{where $\delta \hat{n}_k$ is the deviation from the equilibrium density of quasiparticles}, $F_{gs}$ refers to the non-interacting ground state free energy (system without quasiparticles), $E_k^*$ is the single-quasiparticle energy and $f_{k,k'}$ is the residual quasiparticle-quasiparticle interaction (gives the decay processes). They are related with quantities such as the quasiparticle weight $Z_k$, the mass renormalization $m^*/m_0$, the lifetime $\tau$, etc. Here, I will just mention that the electronic excitation terms of various thermodynamics quantities, like the specific heat $C(T)$, the resistivity $\rho(T)$ or the spin susceptibility $\chi^s(T)$, can be obtained as:

\begin{equation}
\left\{ 
\begin{aligned}
C(T) & \sim \gamma T^2 \\
\rho(T) & \sim \rho_0 + AT^2 \\
\chi^{s}(T) & \sim const. = \chi^s_0
\end{aligned}
\right.
\label{eq:1.15}
\end{equation}

The factors $\gamma$, $A$ and $\chi_0^s$ can be obtained from $E_k^*$ and $f_{k,k'}$. They depend on the effective mass renormalization as $\gamma \propto m^*$, $A \propto (m^*)^2$ and $\chi_0^s \propto m^*$, so they increases when electronic correlations increase. \textcolor{black}{Then, the ratio $m^*/m_0 = 1/Z$ can be obtained when comparing the experimental values of $\gamma$, $A$ and $\chi_0^s$ with the results calculated in DFT, e.g. $m^*/m_0 = \gamma^{exp}/\gamma^{DFT}$ (see next chapter)}. When comparing with the experimental results, \eref{eq:1.15} could show other $T$ dependencies, like electron-phonon contributions (for example, $\beta T^4$ term in $C(T)$), or from non-Fermi liquid behaviors. In \sref{3.3.1}, we will further discuss the experimental results for the quantities of \eref{eq:1.15} in FeSCs.

\section{Single-orbital systems}
\label{1.3}

A first approximation to study the physics of \textcolor{black}{many} SCES is the so called Hubbard model. In the spatially localized Wannier orbital basis, \eref{eq:1.3} can be rewritten in second quantization to include strong correlation/interaction effects. In the real space \cite{Ferber2012thesis}:

\begin{equation}
H = \underbrace{\sum_{m n} \sum_{ij} \sum_\sigma t_{m n}^{ij} d_{i m \sigma}^\dagger d_{j n \sigma}}_{weakly correlated \: H_0} + \underbrace{\frac{1}{2} \sum_{m m ' n n '} \sum_{ijkl} \sum_{\sigma \sigma '} U_{m n m ' n '}^{ijkl} d_{i m \sigma}^\dagger d_{j n \sigma '}^{\dagger} d_{l m ' \sigma '} d_{k n ' \sigma}}_{correlated \: \hat{U}}
\label{eq:1.16}
\end{equation}

where $U_{mm'nn'}^{ijkl}$ is the interaction integral between electrons initially with $lm'\sigma '$, and $kn'\sigma$ and which end in $im\sigma$ and $jn\sigma '$. This is the electronic correlations term $\hat{U}$.

In the $k$-space, and taking into account the momentum conservation:

\begin{equation}
H = \sum_{k} \sum_{mn} \sum_{\sigma} \varepsilon_{kmn} d_{km\sigma}^\dagger d_{kn\sigma} + \frac{1}{2}  \sum_{m m ' n n '} \sum_{k k' Q} \sum_{\sigma \sigma '} U_{m n m ' n '}^Q d_{k+Q \, m\sigma}^{\dagger} d_{k'-Q \, n\sigma}^{\dagger} d_{k' \, m ' \sigma' }^{} d_{k \, n' \sigma}^{}
\label{eq:1.17}
\end{equation}

where $Q$ gives the momentum exchange due to the interaction $U^Q$. \eref{eq:1.16} and \eref{eq:1.17} are equivalent. When the $\hat{U}$ effect is strong enough, DFT band structure calculations fail, and further approximations have to be taken into account. 



Two main regimenes can be differentiated when looking into the range of the electronic interaction $\hat{U}$: short- and long-range interactions \cite{Coleman2015}. Like the Coulomb interaction between electrons goes like $\sim 1/r$, it has a long-range character. However, other electrons in the system can screen this Coulomb interaction, reducing its long-range character. In order to have this situation, there has to be a minimum electron density needed to screen this interaction.  When the screening is large enough, the long range character of the Coulomb interaction can be safely ignored, and short-range interactions will dominate the physics of the system.

During this thesis, I will focus on studying short-range interactions, \textcolor{black}{in the limit of} onsite interactions. The onsite interactions are \textcolor{black}{restricted to the} electrons in the same lattice site. \textcolor{black}{These} interactions do not necessarily imply an ordering of the system. In the case of the onsite interactions:

\begin{equation}
U_{m n m ' n '}^{ijkl}  = U_{m n m ' n '} \delta^{ij}\delta^{ik}\delta^{il}
\label{eq:1.18}
\end{equation}

For a single-orbital system, the so called Hubbard model can be derived:

\begin{equation}
H=\sum_{ij,\sigma}(t^{ij}d_{i\sigma}^{\dagger}d_{j\sigma}+h.c.)+U\sum_{i}\widehat{n}_{i\uparrow}\widehat{n}_{i\downarrow}
\label{eq:1.19}
\end{equation}

In this model, $U$ gives the onsite interaction (also called Hubbard interaction) and $\widehat{n}_{i\sigma}=d_{i\sigma}^{\dagger}d_{i\sigma}$ is the occupation number operator. The onsite interaction term gives the repulstion energy between two electrons which are in the same lattice site. We will see in the next section that due to the onsite interaction $U$, double occupied lattice sites (i.e. two electrons with opposite spins in the same lattice site) will be suppressed, as shown in \fref{fig:1.6}.

\textcolor{black}{Within the limit of onsite interactions, two situations can be separated: local and non-local correlations. In the local correlations picture, only correlations between electrons in the same lattice site (also called onsite correlations) are considered, and further neighbor correlations are neglected. In this case, $\Sigma(\omega)$ is $k$-independent, and hence all the observables, like $Z$. In the non-local correlations picture, the correlations between an electron in a site with those in neighbor sites are also included, then $\Sigma(k,\omega)$ is $k$-dependent and also other observables as $Z_k$.}

Although the Hubbard model \eref{eq:1.19} looks quite simple, it can capture the essential physics behind complicated systems, such as the high-$T_c$ cuprates \cite{PicRMP611989} or ultra-cold atoms in optical lattices \cite{GroS3572017}. The most suprising feature of this model is the Mott transition. In various materials, DFT band structure shows a metallic behavior (with an odd number of electrons per unit cell), but the experiments remark their insulating behavior in the paramagnetic (PM) state. This insulating behavior comes from local correlation effects, and it is known as the Mott insulator. We will further explore this insulating state in the next sections.

\subsection{The Mott transition}
\label{1.3.1}

\begin{figure}[h]
   \centering
   \includegraphics[width=0.8\columnwidth]{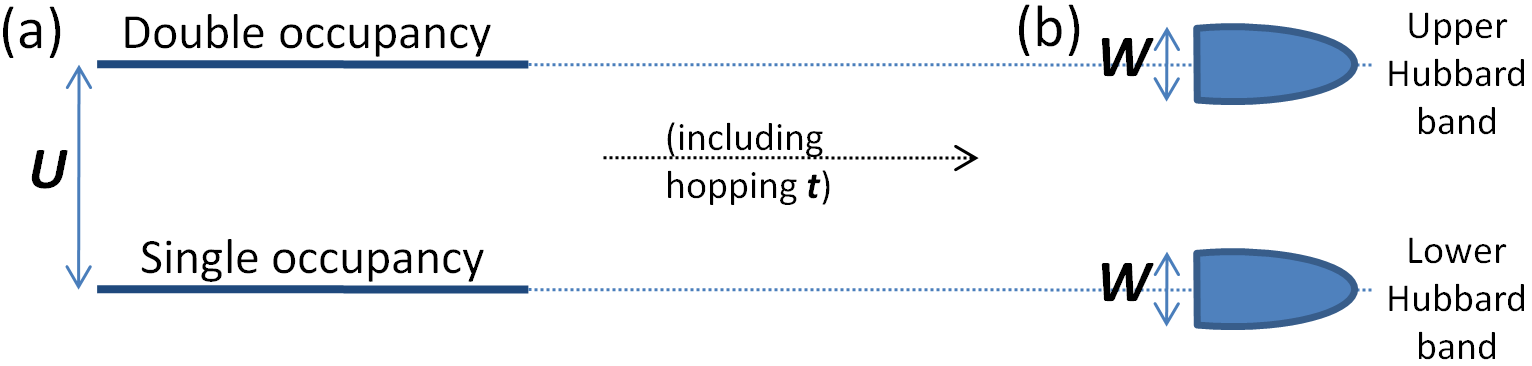}
   \caption[Sketch of the Hubbard bands and its relation with double occupied sites]{\textbf{(a)} Energy levels in the large-$U$ limit (i.e. $t \approx 0$) for an atom, where adding a second electron will cost an energy $U$. These energy levels are labeled by single and double occupancy, so they have a single slot for electrons. \textbf{(b)} When including the hoppings, these energy levels adquire a width of the order of the bandwidth $W$. The dispersive single and double occupancy levels are also called Hubbard bands, which belong to the continuum of states which are not in the quasiparticle state. Adapted from \cite{emergence2017}.}
   \label{fig:1.5}  
\end{figure}

Let's consider a simple model to see the effect of $U$ \cite{Fazekas1999,emergence2017}. Suppose a single-orbital system, where only nearest neighbor hoppings $t$ are included. Consider the large-$U$ limit, with $t \approx 0$. Each lattice site can hold 2 electrons with different spin. Due to the effect of $U$, 2 electrons in the same lattice site will repel each other. Then, adding the second electron in an already occupied lattice site (also called double occupancy) costs an energy $U$, see \fref{fig:1.5}\textbf{(a)}. \textcolor{black}{This results in the non-rigid band shift, i.e. both the chemical potential and the energies change when adding extra particles into the system}. \textcolor{black}{Now, when considering the} hopping $t \neq 0$, the non-degenerated energy levels of \fref{fig:1.5}\textbf{(a)} will adquire a width, as shown in \fref{fig:1.5}\textbf{(b)}. These are the so called Hubbard bands, which belong to the continuum of states which is not in the quasiparticle state, see \sref{1.2.2}. Note that the Hubbard bands represent non-degenerated single and double occupied bands which cannot be explained by the single-particle approximation done in band theory, in constrast to typical band insulators (see \fref{fig:1.1}\textbf{(c)}).

\begin{figure}[h]
   \centering
   \includegraphics[width=0.8\columnwidth]{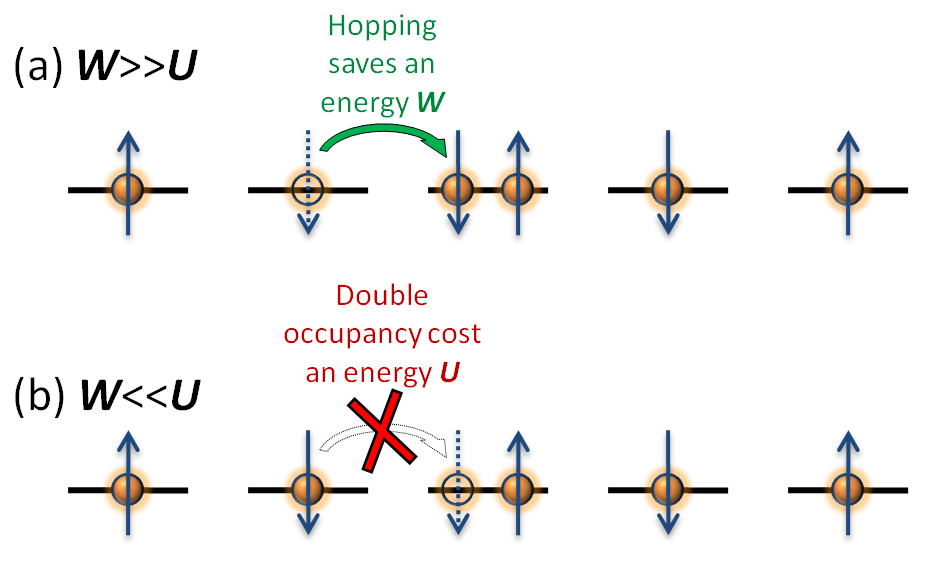}
   \caption[Sketch of the Mott transition in a single-orbital system at half-filling]{Sketch of the Mott transition in a single-orbital system at half-filling, where the effective kinetic energy is identified by the total bandwidth $W$. \textbf{(a)} $W \gg U$, the system behaves as a metal and \textbf{(b)} $W \ll U$, the system becomes a Mott insulator. Adapted from \cite{emergence2017}.}
   \label{fig:1.6}  
\end{figure}

Now, let's assume that the system is at half-filling, i.e. the number of electrons per atom is $n_e=1$. Considering the ratio $U/W$ (which is the ratio between the interaction energy and the effective kinetic energy), with $W$ the bandwidth, two regimenes can be differentiated: when $W\gg U$, \textcolor{black}{the kinetic energy gain is larger than the interaction energy cost for the double occupancy, hence} the electrons will move \textcolor{black}{almost} freely through the lattice (\fref{fig:1.6}\textbf{(a)}), while if $W\ll U$, \textcolor{black}{the interaction energy cost wins over the kinetic energy, so} electrons in the same site repels each other, \textcolor{black}{hence} sitting motionless at each lattice site (\fref{fig:1.6}\textbf{(b)})\footnote{Note that in a model like this one, non-local AFM correlations (note that the spins in \fref{fig:1.6} are distributed anti-parallel from one atom to another, see \sref{1.3.2}) will tend to stabilize the Mott insulating state at an infinitesimal $U \approx 0$, hence the limit $U/W$ does not make sense. However, in the local correlations picture, there is not such magnetic correlations playing any role, and the Mott transition is displaced to $U>0$.}. 

\begin{figure}[h]
   \centering
   \includegraphics[width=0.8\columnwidth]{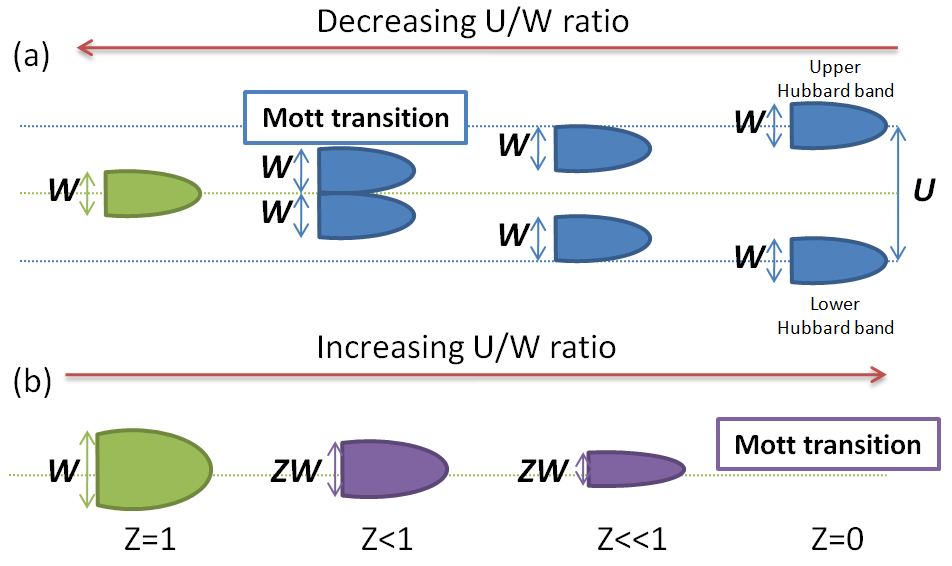}
   \caption[Sketches of the Mott-Hubbard and Brinkman-Rice metal-to-insulator transitions]{Sketches of the \textbf{(a)} Mott-Hubbard and \textbf{(b)} Brinkman-Rice metal-to-insulator transitions. Green dotted line marks the chemical potential $\mu$ when the system is at half-filling $n_e=1$. In the Mott-Hubbard transition, the system starts in the Mott insulating state, and the kinetic energy $W$ is progresively turned on. $U_c$ is defined at the energy at which the upper and lower Hubbard bands merge. In this picture, the metallic phases are not properly described, and $U$ effect is overestimated. In the Brinkman-Rice transition, the system starts in the non-correlated metallic state, and the interaction $U$ is progresively turned on. $U_c$ is defined at the energy at which the quasiparticle peak disappears ($Z=0$). In this picture, the Mott insulating phase cannot be described, and $U$ effect is understimated. Adapted from \cite{emergence2017}.}
   \label{fig:1.7}  
\end{figure}

If only the hopping amplitudes are considered, a metallic solution is found. By introducing the onsite interaction, at a certain value of $U$, \textcolor{black}{the metal-insulator transition can occur without any symmetry breaking. Thus,} the system does not necessarily order magnetically. This is what is called the Mott transition, while the insulator is called Mott insulator. The Mott transition will occur at a critical interaction $U_{c} \simeq W$ and the gap will be of the order of $gap \sim U-W$ \cite{GeoRMP681996}. In \fref{fig:1.7}\textbf{(a)}, I have sketched the Mott transition explained above. Note that this transition will only occur at half-filling, once the system is doped (with holes or electrons) away from half-filling, the system behaves metallically (the chemical potential lies in a partially filled Hubbard band), so that the Mott insulator is only found at half-filling.

This previous picture is what can be alternatively called the Mott-Hubbard transition \cite{Fazekas1999}. It starts from the insulating state, by considering the Hubbard bands and progresively turns on the kinetic energy $W$. \textcolor{black}{In this picture, all the spectral weight goes to the Hubbard bands. Thus, while it captures the essence of the Mott transition, it does not properly describe the metallic state.}

An alternative point of view starts from the metallic side and turns on the interactions $U$ \cite{Fazekas1999,emergence2017}. In this picture, the ground state wavefunction of the interacting system $|\Psi_{gs}^*>$ evolves with a variational parameter $\eta$, which is related with the concentration of double occupied states. At $U=0$, $\eta=1$ and at $U \rightarrow \infty$, $\eta=0$. This variational parameter $\eta$ is also related with the quasiparticle weight $Z$, see \sref{1.2.2}.

This is the so called Brinkman-Rice transition \cite{Fazekas1999}. In \fref{fig:1.7}\textbf{(b)}, I have sketched such transition, where the quasiparticle weight $Z$ (which appears as a renormalization factor of the bandwidth $\sim ZW$) is progresively reduced until it disappears \textcolor{black}{at} the Mott transition at $U_c \simeq 2W$. However, this picture is based on the FLT for solids, and hence it cannot provide further information about the Mott insulator \textcolor{black}{and the Hubbard bands} (e.g., the gap magnitude cannot be obtained), and $U$ effect is understimated.

The quasiparticle weight $Z$ is the order parameter of the Mott transition \cite{emergence2017}. In the local correlations picture, it can be shown that $1/Z = m^*/m_{0}$, see \eref{eq:1.13}, where $m_0$ is the non-correlated effective mass (i.e. obtained from the tight-binding model, see \sref{1.2.1}) and $m^*$ is the correlated effective mass. The quasiparticle weight evolves from $Z=1$ ($m^* = m_0$) in the non-correlated metal to $Z=0$ ($m^* \rightarrow \infty$) in the Mott insulator. When \textcolor{black}{$0.7<Z \leq 1$ ($1 \leq m^*/m_0 < 1.4$)}, the system is said to be a weakly correlated metal, and descriptions based on a DFT band structure calculations and tight-binding models will give good predictions. At \textcolor{black}{$0<Z \lesssim 0.3$ ($3 \lesssim m^*/m_0 < \infty$)} is in a strongly correlated metallic state, and DFT calculations and tight-binding models will leave out the most important local correlations effects. In between, a moderately correlated metal exits with \textcolor{black}{$0.3<Z \lesssim 0.7$ ($1.4 \lesssim m^*/m_0 < 3$)}.

\begin{figure}[h]
   \centering
   \includegraphics[width=0.7\columnwidth]{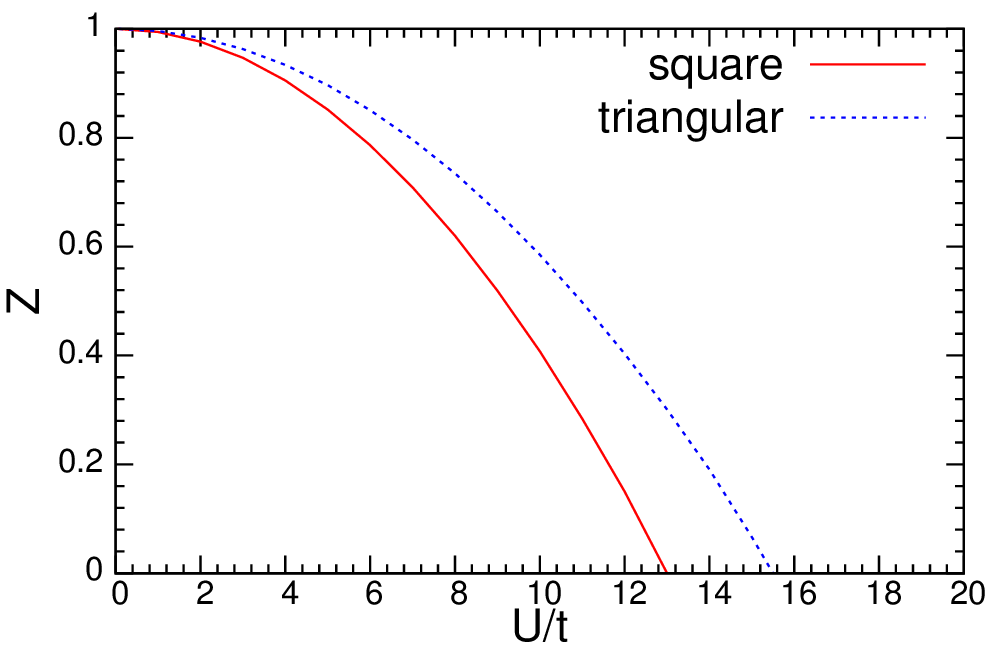}
   \caption[$Z$ versus $U/t$ comparison between 2D square and triangular single-orbital tight-binding models at half-filling]{$Z$ versus $U/t$ comparison between 2D square and triangular single-orbital tight-binding models at half-filling. Only nearest neighbor hoppings $t$ are considered. The triangular lattice is less correlated than the square (it requires a bigger $U_c$ to become a Mott insulator) due to its larger bandwidth $W^{triangular} = 9t > W^{square}=8t$. I have obtained $U_c^{square} \sim 13t$ and $U_c^{triangular} \sim 15.6t$. Figure obtained using SSMF $U(1)$ formalism, see \aref{AppB}.}
   \label{fig:3.1}  
\end{figure}

The behavior explained above for local correlations is robust against any considered tight-binding model, as shown in \fref{fig:3.1} for 2D square and triangular lattices, at half-filling $n_e=1$ and only nearest neighbor hoppings $t$ included, and solved using the slave-spin mean-field (SSMF) formalism, see \sref{1.4.1}, and the detailed discussion in \aref{AppB}. Only the energy scales at which the effects occur will change from one model to another. For \fref{fig:3.1}, the total bandwidth for each lattice is $W^{square}=8t$ and $W^{triangular}=9t$, and hence $U_c^{square} \sim 13t < U_c^{triangular} \sim 15.6t$ (note that $U_c^{square} \sim 1.6W$ and $U_c^{triangular} \sim 1.7W$).

A more complete description requires both Mott-Hubbard and Brinkman-Rice pictures being applied at the same time, and it is provided by numerical calculations using Dynamical Mean-Field Theory (DMFT) \cite{GeoRMP681996}. A description of DMFT calculations is presented in \sref{1.4.1}. \textcolor{black}{In DMFT close to the transition for a given $U/W$ value, two solutions are possible: a metallic one where} the Hubbard bands start to form, with a fully gap ($\sim U-W$) opening at $U_{c \, 1} \sim W$, and \textcolor{black}{an insulating one where} the quasiparticle peak disappears at $U_{c \, 2} \sim 2W$. The Mott transition is then defined as the minimum of the Landau's free energy $F$, which at $T=0 \, K$ coincides with $U_{c \, 2}$, and results in a first order transition (for finite $T$, see \sref{1.3.3}). 

\begin{figure}[h]
   \centering
   \includegraphics[width=0.9\columnwidth]{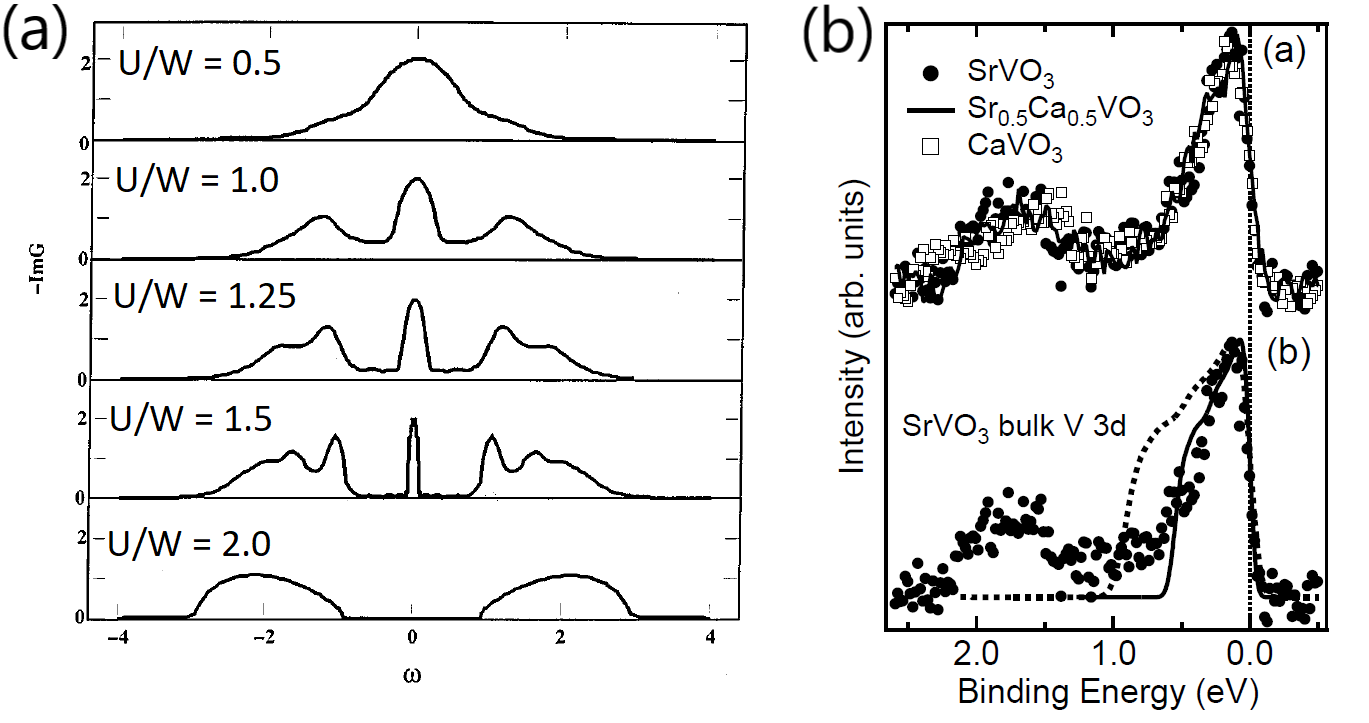}
   \caption[Spectral function in the Mott transition for the Bethe lattice, and photoemission experimental spectra for $SrVO_3$]{\textbf{(a)} DMFT \textcolor{black}{spectral function} for a single orbital \textcolor{black}{in the} Bethe lattice. \textcolor{black}{$\omega$ is in units of the half-bandwidth $D=W/2$}. Modified from \cite{GeoRMP681996} to include the ratio $U/W$. At $U_{c \, 1} \sim W$ a gap ($\sim U-W$) fully opens between the Hubbard bands, and at $U_{c \, 2} \sim 2W$ the quasiparticle peak dissapears, i.e. $Z=0$. At $T=0 \, K$, the Mott transition occurs at $U_c=U_{c \, 2} \sim 2W$. \textbf{(b)} Photoemission spectra of the strontium vanadate $SrVO_3$, where the quasiparticle peak (around $\sim 0.3 \, eV$) and the lower Hubbard band (around $\sim 2 \, eV$) can be clearly seen. Taken from \cite{Sek0206471}.}
   \label{fig:1.8}  
\end{figure}

In \fref{fig:1.8}\textbf{(a)}, \textcolor{black}{the DMFT spectral function} for the Mott transition in the Bethe lattice\footnote{The Bethe lattice is an infinite dimensional hypercubic lattice, where $\varepsilon_k=2t \sum_{d=1}^{\infty} \cos{k_d}$ and it has a semicircular DOS. It is widely used in DMFT calculations due to the fact that this technique is exact in infinite dimensions.} can be seen \cite{GeoRMP681996}. \textcolor{black}{When the ratio $U/W$ increases, the quasiparticle peak at $\omega=0$ narrows, while the Hubbard bands start to form around $\omega \approx \pm 2D$ (where $D=W/2$ is the half-bandwidth), and a gap opens between them, at $U_{c \, 1}$ as previously pointed out. The narrowing of the quasiparticle peak with increasing $U/W$ is related with the decrease (increase) of $Z$ ($m^*/m_0$), hence with the narrowing of the band energies}. In \fref{fig:1.8}\textbf{(b)}, the photoemission spectra, i.e. the spectral function $A(\omega)$, see \sref{1.2.2}) for $SrVO_{3}$ (and $Ca$-doped $SrVO_3)$ clearly shows the coexistance of the quasiparticle peak and the Hubbard bands \cite{Sek0206471}.

\begin{figure}[h]
   \centering
   \includegraphics[width=0.9\columnwidth]{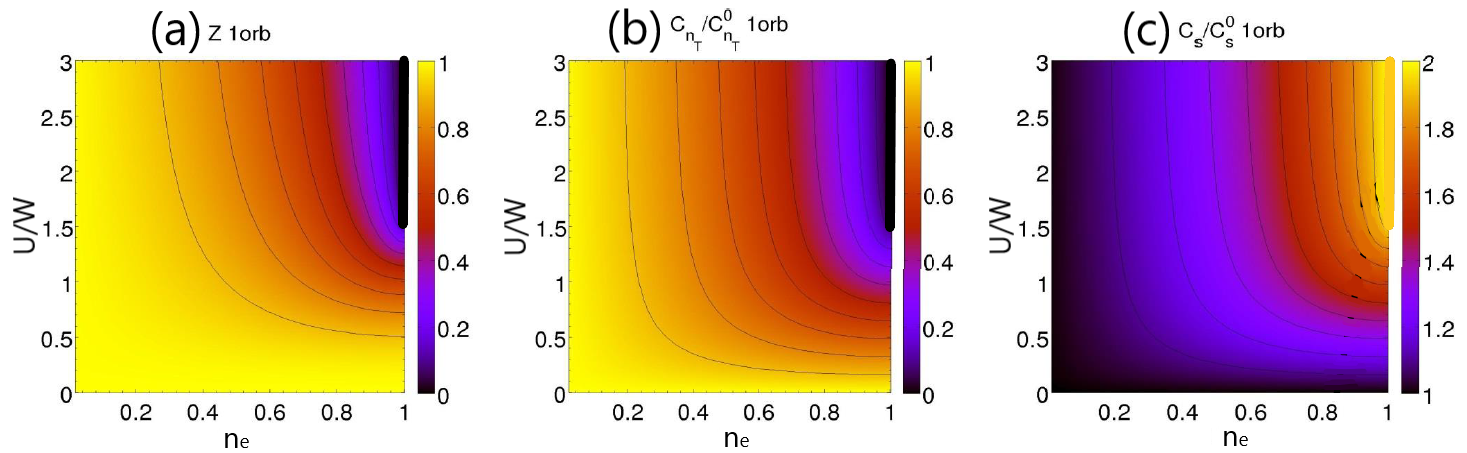}
   \caption[Color maps for $Z$, local charge correlations and local spin correlations in the $U/W$ versus $n_e$ phase diagram.]{Color maps for \textbf{(a)} $Z$, \textbf{(b)} local total charge correlations $C_{n_T}$ and \textbf{(c)} local spin correlations $C_S$ in the $U/W$ versus $n_e$ phase diagram. Solved using a 2D square lattice with only nearest neighbor hoppings in the SSMF $Z_2$ formalism. The Mott transition is marked by a thick line at $n_e=1$. When the Mott transition is reached, $Z=0$, charge fluctuations are suppressed and the atoms become maximally spin polarized. Taken and adapted from \cite{emergence2017}.}
   \label{fig:1.9}  
\end{figure}

An important result from this improved perspective is the fact that $Z$ vanishes at the Mott transition, while double occupancy does not (see next section) \cite{GeoRMP681996}. Thus, $Z$ is the correct order parameter of the Mott transition. \textcolor{black}{In the Brinkman-Rice picture and for a single-orbital system}, when $Z=0$, the local total charge correlations $C_{n_T}$ (which are the fluctuations around the equilibrium charge in each site) are also $0$, hence the Mott insulator implies a localization of the local charge. Another important property is that local spin correlations $C_S$ (which are the fluctuations around the equilibrium spin in each site) are maximum at the Mott insulator, signaling that atoms are in a local high-spin state (each atom is maximally spin polarized). In \fref{fig:1.9}, I present the color maps for $Z$, $C_{n_T}$ and $C_S$ in the $U/W$ versus the number of electrons per atom $n_e$ phase diagram, for the 2D square lattice with only nearest neighbor hoppings $t$ \cite{emergence2017}. These plots are obtained by using the SSMF formalism. Except for half-filling $n_e=1$, the system will behave as a metal. Yellow-red areas in \fref{fig:1.9}\textbf{(a)} mark the weakly and moderately correlated metal, while purple region marks the strongly correlated metal behavior, in which \textcolor{black}{$0<Z \lesssim 0.3$}. In this situation, DFT band structure calculations will fail when predicting the correct band structure, due to the large renormalization.

Calculating $Z$ \textcolor{black}{and when possible, comparing with the experiment}, is then an important first step in any SCES, in order to know the degree of electronic localization, and hence the most suitable space (real space using \eref{eq:1.16}, or $k$-space using \eqref{eq:1.17}) to describe the system.

\subsection{Short-range magnetic correlations for the Mott insulator}
\label{1.3.2}

In the previous sections, we have discussed the Mott transition without the presence of any magnetic (or orbital) tendencies. This emphasizes that the Mott insulator does not necessarily require any magnetic tendency. Nevertheless, short-range magnetic correlations can be promoted. In order to treat magnetic correlations, we have to go beyond local correlations approximation, including neighbor correlations into the calculations (this is the so called non-local correlations picture). In this section, I will briefly discuss the short-range magnetic correlations in the large-$U$ limit, i.e. deep in the Mott insulating phase \cite{Fazekas1999}.

At half-filling, \eref{eq:1.19} in the large-$U$ limit can be rewritten as the Heisenberg hamiltonian:

\begin{equation}
H = J \sum_{ij} \vec{S}_i \cdot \vec{S}_j
\label{eq:1.20}
\end{equation}

where $J = 4t^2/U$ (obtained from second order perturbation theory for the hopping term $t$) is the exchange interaction and $\vec{S}_i$ are the spin operators in each lattice site $i$. Note that $J$ can depend on \textcolor{black}{$i,j$} if second, or higher-order neighbor exchanges are included, \textcolor{black}{resulting in a similar hamiltonian to \eref{eq:1.20}, but placing $J_{ij}$ inside the summatory}. \textcolor{black}{When including a modified hopping hamiltonian in a second step \cite{Fazekas1999}}, it receives the name of $t-J$ model. If $J>0$, antiferromagnetic (AFM) correlations are promoted, and hence the spins in neighboring lattice sites will align antiparallel, while if $J<0$, ferromagnetic (FM) correlations are then promoted, so there is a parallel alignement. Note again that this tendencies do not require a long-range order. 

\textcolor{black}{In the case of single-orbital systems}, AFM correlations are \textcolor{black}{always} promoted. This occurs because a virtual hopping process (as the one described in \fref{fig:1.6}) creates an intermediate empty and doubly occupied state, with an energy $U$. The associated energy gain if the spins of neighboring sites are antiparallel is $\sim -t^2/U$ (calculated from second-order perturbation theory for the hopping), while for parallel spins is zero (due to the Pauli exclusion principle). Then, due to the energy gain, AFM correlations are promoted. Due to this virtual proccesses, double occupancy at the Mott insulator is not strictly zero, as anticipated in the previous section \cite{GeoRMP681996}. \textcolor{black}{Depending on the lattice symmetry, magnetic frustration varies, so the tendency towards the AFM correlations in higher frustrated lattices will decrease, but the character of these correlations will remain AFM}.

\subsection{Temperature effect in local and non-local correlated systems}
\label{1.3.3}

\begin{figure}[h]
   \centering
   \includegraphics[width=0.9\columnwidth]{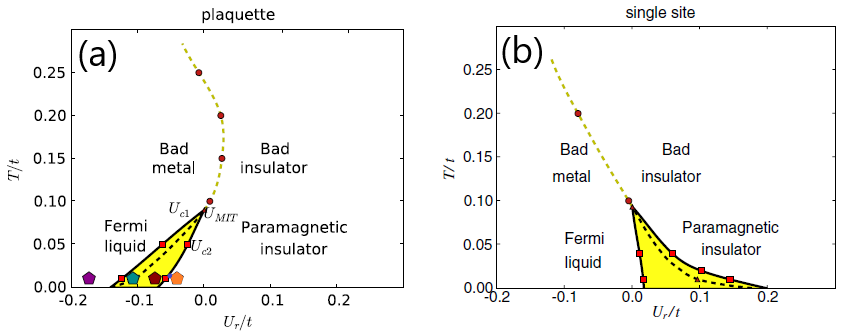}
   \caption[Temperature versus $U$ phase diagram for local and non-local correlations]{$T/t$ versus $U_r/t$ (with $U_r=(U-U_c)/U_c$) phase diagram for \textbf{(a)} non-local and \textbf{(b)} local correlations \textcolor{black}{in the square lattice}. Yellow area (in between $U_{c \, 1}$ and $U_{c \, 2}$ lines) marks the coexistance of both metallic and insulating behaviors. Dashed line inside the yellow region marks the Mott transition $U_c$. In both cases, a critical point appears at $T/t \approx 0.1$. Above $T/t=0.1$, there is a crossover between a bad metal and bad insulator. Orange hexagon marks the pseudogap region. Taken and adapted from \cite{ParPRL1012008}.}
   \label{fig:1.10}  
\end{figure}

In this section, I will review the behavior of the Mott transition in both local and non-local correlation pictures with the temperature $T$, mainly following \cite{ParPRL1012008,VucPRB882013,SchPRB912015}. \textcolor{black}{The behavior of the Mott transition will change with $T$}. \fref{fig:1.10} shows the $T$ vs $U_r=(U-U_c)/U_c$ phase diagram for both local and non-local correlations solved by cellular DMFT (cellDMFT) for a single-orbital at half-filling system in a 2D square lattice. In cellDMFT, non-local correlations between electrons in different real space sites are taken into account. \textcolor{black}{The range of these non-local correlations depends on the size of the cluster used (in \fref{fig:1.10}, it is a $2 \times 2$ cluster in the square lattice)}. The Mott transition at finite $T$ is given by $U_c$, defined as the minimum in the Landau's free energy $F$.

A main feature of non-local correlations when compared with local correlations is that both $U_{c \, 1}$ (\textcolor{black}{in local correlations describes a} fully opened gap between the Hubbard bands) and $U_{c \, 2}$ (\textcolor{black}{in local correlations, the} quasiparticle peak disappears), and hence $U_c$ decrease. In CDMFT calculation for a 2D square lattice at half-filling, $U_c^{nonlocal}\sim 6 t$, while $U_c^{local} \sim 12 t$. $U_c^{nonlocal}$ increases \textcolor{black}{approaching $U_c^{local}$} if the system is more frustrated, i.e. if the magnetic non-local correlations are suppressed. Frustration occurs, for example, \textcolor{black}{when} adding further neighbor hoppings $t'$ to the tight-binding model for the 2D square lattice \eref{eq:1.4}.

In \fref{fig:1.10}, most left-sided line marks $U_{c \, 1}$, while right-sided line marks $U_{c \, 2}$. Yellow area marks the coexistence of both metallic (quasiparticle peak) and insulating (gap separating the Hubbard bands) states, and dashed line marks the Mott transition $U_c$. These three lines end in a second order critical point, and at larger $T$ there exist a crossover between bad metallic and bad insulating behavior (dashed yellow line). At large $T$ and $U$, local correlations dominate the physics behavior. The Mott insulating state is the ground state for both local and non-local correlations pictures \cite{ParPRL1012008}. At lower $T$ and $U$, the behavior is different for both cases. For local correlations, the metallic state is the ground state, whether for non-local correlations, the Mott insulator is the ground state due to AFM short-range correlations \cite{ParPRL1012008}.

Close to critical interaction for the Mott transition in the non-local correlations picture $U_c^{nonlocal}$ (see orange pentagon in \fref{fig:1.10}\textbf{(a)}), the system shows a small gap in $A(k,\omega)$ at the chemical potential, but without the disappearance of the quasiparticle peak, which is splitted. This regime is totally different from local correlations, where once the Mott insulating phase is reached, the Mott gap is fully open ($\sim U-W$) and the quasiparticle peak vanish \cite{GeoRMP681996}. For $U>U_{c}^{nonlocal}$, the gap is fully formed and it recovers the local correlations expectations, with $gap \sim U-W$.


\section{Multiorbital systems}
\label{1.4}

The Hubbard hamiltonian of \eref{eq:1.19} describes a single-orbital system for onsite interactions. In order to generalize this model to the multiorbital case, we have to take into account various considerations \cite{Dagotto2002}. From \eref{eq:1.18}, it can be seen that there are, in principle, $N^4$ different \textcolor{black}{spin-independent} interaction terms $U_{mnm'n'}$. However, not all of them are allowed (due to the Pauli exclusion principle) or independent from each other. For example, $U_{mnm'n'} \equiv U_{m'nmn'} \equiv U_{mn'm'n} \equiv U_{m'n'mn}$. Also, terms which only mix three different orbitals, such as $U_{mmnn'}$, vanish. In conclusion, there are four independent interaction terms:

\begin{equation}
\left\{ 
\begin{aligned}
U_{mm} & = U_{mmmm} \\
U'_{mn} & = U_{mnmn} \\
J_{H \: mn} & = U_{mnnm} \\
J'_{mn} & = U_{mmnn}
\end{aligned}
\right.
\label{eq:1.21}
\end{equation}

where the parameters are defined as \textcolor{black}{(see below for the interacting hamiltonian form and the inclusion of the spin indices)}: $U_{mm}$ is the intraorbital interaction, $U'_{mn}$ is the interorbital interaction for electrons with anti-parallel spins, $U'_{mn} - J_{H \: mn}$ is the interorbital interaction for electrons with parallel spins, $J_{H\: mn}$ is the Hund's coupling and $J'_{mn}$ is the pair-hopping term. The Hund's coupling term comes from two contributions, the interorbital interaction for electrons with anti-parallel spins and the spin-flip term, in which the electrons jump from one atom to another flipping their spin.

Then, the multiorbital system is described by the so called Hubbard-Kanamori hamiltonian:

\begin{equation}
\begin{aligned}
H 	& =  \underbrace{\sum_{ij, mn, \sigma} \left( t_{mn}^{ij} + \delta_{mn} \delta_{ij=0} \epsilon_m \right) d_{im\sigma}^\dagger d_{jn\sigma}}_{H_0} \\
	& \underbrace{+ \sum_{i, m} U_{mm} \hat{n}_{im\uparrow} \hat{n}_{im\downarrow} + \sum_{i, m<n, \sigma} U'_{mn} \hat{n}_{im\sigma} \hat{n}_{in \bar{\sigma}} + \sum_{i, m<n, \sigma} \left( U'_{mn} - J_{H \: mn} \right) \hat{n}_{im\sigma} \hat{n}_{in\sigma}}_{H_{dens}} \\
	& \underbrace{-\sum_{i,m < m'} J_{H \: mn}^{} d_{im\uparrow}^{\dagger} d_{im\downarrow}^{} d_{in\downarrow}^{\dagger} d_{in\uparrow}^{} +  \sum_{i, m \neq n, \sigma \neq \sigma '} \frac{J'_{mn}}{2} d_{im\sigma}^\dagger d_{im \sigma '}^\dagger d_{in' \sigma '} d_{in' \sigma}}_{H_{add}}
\end{aligned}
\label{eq:1.22}
\end{equation}

where $H=H_0 + H_{dens} + H_{add}$, with $H_0$ describing the tight-binding model, $H_{dens}$ being the density-density interaction hamiltonian and $H_{add}$ including both spin-flip and pair-hopping processes. $H=H_0+H_{dens}$ is also known as the Ising hamiltonian, and the essential physical effects \textcolor{black}{discussed in this thesis} are encoded in it. The study of the Ising hamiltonian will be central during this thesis.

In \textcolor{black}{a first approximation}, interactions can be written orbital-independent $U$, $U'$, $J_H$ and $J'$. \textcolor{black}{Even if this approximation looks very crude, it gives good enough results in many materials}. If the \textcolor{black}{system is rotationally invatiant}, the relations $U' = U-2J_H$ and $J'=J_H$ can be invoked \cite{CasPRB181978}. Even if \textcolor{black}{the system is not fully rotationally invariant}, these relations approximately hold for various materials, such as FeSCs. Thus, the effects of local correlations have been reduced to 2 independent parameters, $U$ and $J_H$. For real materials, these parameters can be estimated by using constraint Random Phase Approximation (cRPA) calculations \cite{MiyJPSJ792010}, or by comparing with experiments and taking the most appropiated interaction parameter values \cite{YuRPRB862012,BasCRP172016} (see \cref{Chap4b}).


When solving \eref{eq:1.22} considering only local correlations, new interesting features appear due to the addition of the orbital degree of freedom:

\begin{itemize}

\item The Mott transition occurs not only at half-filling, but also at other integer fillings $n_{e}$, where $n_{e}$ is the number of electrons per atom. 

\item A new interaction energy scale emerges, the Hund's coupling $J_{H}$, and its role in the metallic and Mott insulating states is crucial and depends on the specific $n_{e}/N$ value, where $N$ is the number of orbitals per atom.

\item If the orbitals are non-\textcolor{black}{equivalent}, they will behave differently with $(U,J_{H},n_{e}/N)$, so there is an orbital selectivity of the electronic correlations. This can be translated into an orbital dependent quasiparticle weight $Z_m$.

\end{itemize}

These consequences are more extensively discussed in \cref{Chap03}. Up to now, it is important to note that the appearance of the Hund's coupling, and the phenomenology resulting from it, \textcolor{black}{led the scientific community to introduce} a new type of \textcolor{black}{correlated} metallic state, the Hund metal \cite{emergence2017,WerPRL1012008,HauNJP112009,LiePRB822010,IshPRB812010,HanPRL1042010,YinNM102011,WerNP82012,LanPRB872013,deMPRL1122014,FanPRB922015,deMedici2015,deMedici2017,deM1707.03282,BasCRP172016}. . The interplay and possible connections between Hund metals and Mott insulators are also discussed in \cref{Chap03}. In Hund metals, a bad metallic behavior (i.e. higher resistivity which comes from an increase of the mass enhancements, see \eref{eq:1.15}) is promoted for increasing $J_H$, the atoms are highly (local) spin polarized and there is an orbital decoupling (different orbitals will be decoupled from each other).

Finally, orbital selectivity of the electronic correlations can make the system enters in an orbital selective Mott phase (OSMP), as studied by several authors \cite{deMPRB722005,deMPRL1022009,YuRPRL1102013,YiMPRL1102013,LiuPRB922015}. The OSMP is realised when some $Z_{m}=0$ while others remain finite $Z_{n\neq m}\neq0$, see \sref{3.2.4}. The Hund's coupling promotes orbital decoupling favouring this situation. Nevertheless, in the presence of finite interorbital hoppings $t_{mn}$, the OSMP evolves into a system in which both weakly ($0.7<Z \leq 1$) and strongly correlated ($0 < Z \lesssim 0.3$), and even moderately correlated ($0.3<Z \lesssim 0.7$) electrons coexist in different orbitals at the same time \cite{emergence2017,deMedici2015,deMedici2017,deM1707.03282,BasCRP172016}.

Different experimental probes \cite{BasCRP172016,deMPRL1122014,deM1707.03282,LuDARCMP32012,LanPRB872013,GrePRB842011,LafPRB962017} can be tested to check the effect of the correlations in multiorbital system, similarly to the single-orbital case: angle-resolved photoemission spectroscopy (ARPES), low-T specific heat, optical conductivity, resistivity and magnetic susceptibility measurements, \textcolor{black}{etc.}, see \eref{eq:1.15}. All of these experimental techniques are sensible to the band structure (specially ARPES), and hence to the possible renormalization or deviations from the DFT calculated band structure, hence they are sensible to $(m^*/m_0)_m$. Thus, quasiparticle weights $Z_m$ ($1/Z_m=(m^*/m_0)_m$) can be extracted. In \sref{3.3.1}, we will get into more details when studying FeSCs and the role of electronic correlations in these materials.

In the next section, I will briefly review some of the most widely used techniques to study local correlations in multiorbital systems, which are able to study the the Mott insulator and Hund metal phenomenology in such multiorbital systems.

\subsection{Brief description of local correlations theoretical techniques}
\label{1.4.1}

Various theoretical techniques have been developed to calculate the effect of the local correlations. DMFT \cite{GeoRMP681996} \textcolor{black}{is a technique specially adequate to address the Mott transition and the strength of correlations}. This technique is able to calculate the \textcolor{black}{local} spectral function $A(\omega)$ with the information related with the quasiparticle peak and the continuum of states at the same time. In DMFT, localized electrons in a lattice site interact with a bath of delocalized electrons, thus capturing the real space localized and delocalized character of the electrons in a SCES. \textcolor{black}{This means that DMFT can capture at the same time the role played by the localized electrons, which contribute to the Hubbard bands, and the delocalized ones, which contribute to the quasiparticle peak.}

Other techniques, such as Gutzwiller approximation (GA) \cite{Fazekas1999} or slave-particle mean-field techniques \cite{KotPRL571986,FloPRB702004,deMPRB722005,HasPRB812010,YuRPRB862012}, like slave-boson \cite{KotPRL571986}, slave-rotor \cite{FloPRB702004} or slave-spin \cite{deMPRB722005,HasPRB812010,YuRPRB862012}, are based on the Brinkman-Rice picture of the Mott transition, hence they follow the FLT. They \textcolor{black}{are nevertheless very useful. They} can trace down $Z_m$ when going from the non-correlated metal $Z_m=1$ to the Mott transition, where all $Z_m=0$. These techniques are computationally faster and they are easier to implement. The basic approximation behind GA is the projection out of the double occupied states when the system progresively approaches the Mott transition. The projection out of double occupied states is made by finding the variational parameter $\eta$ for the interacting ground state $|\Psi_{gs}^*>$. For slave-particle mean-field techniques, $Z_m$ appear as multiplying pre-factors of $\varepsilon_{kmn}$ by doing an changing from the original Hilbert basis $\left\{ d_{im\sigma}^\dagger, d_{im\sigma} \right\}$ to a new expanded space, where more degrees of freedom are added to obtain these pre-factors in the hamiltonian.

During this thesis, I will use the slave-spin mean-field (SSMF) formalisms to explore the effects of local correlations in multiorbital systems, specifically in FeSCs and 2D materials. Up to date, $H_{add}$, \textcolor{black}{i.e. the pair-hopping and spin-flip terms, see\eref{eq:1.22}}, is not considered in the calculations, due to the fact that it is unknown how to include it, see \aref{AppB}. In this technique, the original Hilbert space is expanded in a new space with pseudospin operators $\left\{ O_{im\sigma}^\dagger,O_{im\sigma} \right\}$ and auxiliary fermion operators $\left\{ f_{im\sigma}^\dagger,f_{im\sigma} \right\}$. After doing a couple of mean-field approaches, $H_0$ is rewritten for the auxiliary fermion operators $H_f$, while $H_{dens}$ is rewritten for pseudospin operators $H_{PS}$. Both non-interacting $H_0$ and interacting $H_{dens}$ hamiltonians have been then separated for each degree of freedom, $H=H_f + H_{PS}$, where these two hamiltonians are coupled via a set of orbital-dependent parameters, called Lagrange multipliers $\lambda_m$. The hamiltonians $H_f$ and $H_{PS}$, written in \eref{eq:1.23} in the $k$-space, have to be solved self-consistenly. \textcolor{black}{The values of $Z_m$ and $\lambda_m$ depend on the values of the pseudospin operators}. A more extense and detailed discussion, as well as the derivation of \eref{eq:1.23}, can be seen in \aref{AppB}.

\begin{equation}
\left\{
\begin{aligned}
H_f 	& =  \sum_{k,mn,\sigma} \sqrt{Z_{m\sigma}Z_{n\sigma}}\varepsilon_{kmn} f_{km\sigma}^\dagger f_{kn\sigma} + \sum_{k,m,\sigma} \left( \epsilon_m - \mu - \lambda_{m\sigma} + \lambda_{m\sigma}^0 \right) f_{km\sigma}^\dagger f_{km\sigma} \\
H_{PS}	& =  \sum_{m,\sigma} h_{m\sigma} \left( O_{m\sigma}^\dagger + O_{m\sigma} \right) + \sum_{m,\sigma} \lambda_{m\sigma} \left( S_{m\sigma}^z + \frac{1}{2} \right) + H_{dens}^{PS} 
\end{aligned}
\right.
\label{eq:1.23}
\end{equation}

$\lambda_{m\sigma}^0$ is a correction factor which appear to recover the non-interacting limit, and $H_{dens}^{PS}$ is the $H_{dens}$ hamiltonian rewritten for pseudospin operators, see \eref{eq:B.12}. Here, $Z_{m\sigma}$ appear as pre-factors which renormalize the band structure, as already pointed out.

Compared with other slave-particle theories, the SSMF formalisms \cite{deMPRB722005,HasPRB812010,YuRPRB862012} allow to treat multiorbital systems in an economical way, using only $2N$ pseudospin operators, while the slave-boson technique \cite{KotPRL571986} normally increases its number of variables in an exponential way with $N$. The SSMF formalisms permit a treatment of non-degenerated orbitals problem, while slave-rotor technique \cite{FloPRB702004} cannot account for such systems. During the last decade, the SSMF technique has become an useful tool to study \textcolor{black}{the strength of correlations in} multiorbital systems, \textcolor{black}{specially by comparing SSMF with experimental results in FeSCs}, see \cref{Chap03}.

\section{Unconventional superconductors}
\label{1.5}

Superconductors are materials in which the resistivity at a certain critical temperature $T_c$ goes to zero. They also repel the magnetic field (Meissner effect) \cite{Tinkham1996}. In all the superconductors, the electrons are said to be paired, forming the so called Cooper pairs, and moving through the material without resistance. 

Within all the superconductors, two main families can be distinguished: conventional and unconventional superconductors. Conventional superconductors are those systems in which the pairing instability is due to phonons, while for unconventional superconductors is still under debate. According to the Bardeen-Cooper-Schrieffer (BCS) theory for conventional superconductors, electron-phonon coupling is the responsible of this pairing instability. The pairing instability in unconventional superconductors is not yet known, and it is accepted that it does not occur due to phonons. It is believed that magnetic fluctuations play a major role \cite{Coleman2015}. Nowadays, the record $T_c$ belongs to conventional superconductors, with \textcolor{black}{$T_c^{max} \sim 260 \, K$ at huge pressures $\sim 190-200 \, GPa$ in $LaH_x$ \cite{Som1808.07695}}. At ambient pressure, the maximum $T_c$ is found in unconventional superconductors, specifically in a high-$T_c$ cuprate superconductor, $Hg_{0.8} Cu_{0.2} Ba_2 Ca_2 Cu_3 O_{8+\delta}$ with $T_c^{max} \sim 134 \, K$ \cite{BrePCS4062004}.

\textcolor{black}{A key} difference between conventional and unconventional superconductors is the initial theory that explain the non-superconducting compound. In conventional superconductors, the non-superconducting (metallic) phase can be properly described by FLT and DFT band structure calculations. In most unconventional superconductors, the electronic correlations are important in the non-superconducting phase. For example, the parent compound (i.e. the system without doping or external perturbations) of the cuprates \cite{Nor1302.3176} or organic superconductors \cite{KurPRL952005} are Mott insulators, while the parent compound of the heavy-fermion superconductors \cite{Nor1302.3176} or FeSCs \cite{PagNP62010,BasCRP172016} are correlated metals. \textcolor{black}{They also have a tendency to show ordered phases, such as magnetism (see below)}. In this thesis, I will focus on studying the non-superconducting (and not ordered) phase for various unconventional superconductors, specifically in FeSCs, as well as in 2D materials such as MA-TBG, see \sref{1.6.1}.

\begin{figure}[h]
   \centering
   \includegraphics[width=0.9\columnwidth]{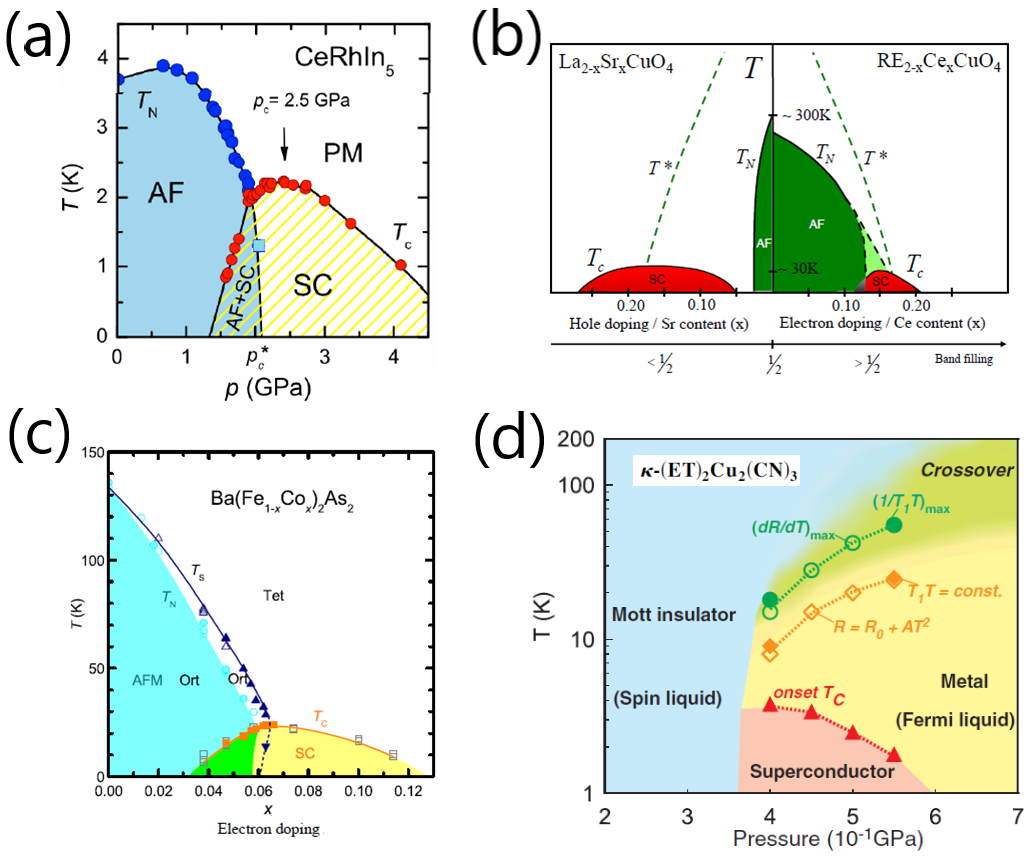}
   \caption[Phase diagrams of unconventional superconductors]{Phase diagrams for compounds belonging to the various families of unconventional superconductors: \textbf{(a)} heavy-fermion superconductors, \textbf{(b)} cuprates, \textbf{(c)} FeSCs and \textbf{(d)} organic superconductors. The superconducting dome usually emerges when an AFM phase is suppressed by doping or pressure. Both phases may coexist. Different metallic and insulating behaviors have been seen in these compounds. For the cuprates, an additional line $T^*$ marks the existence of the pseudogap phase. For FeSCs, there is a tetragonal-to-orthorhombic transition at $T_s$ as a prelude of the AFM phase, which is called the nematic transition. Both the nematic and AFM phases are suppressed with doping or when pressure is applied. In the organic superconductor case, the Mott insulating state can be accompanied by a spin liquid phase, which also suppress when doping or pressure is applied. Taken and adapted from \cite{Nor1302.3176,KurPRL952005}.}
   \label{fig:1.11}  
\end{figure}

Generally, something similar occurs in most of the unconventional superconductors (see \fref{fig:1.11}): the parent (and at ambient pressure) compound develops a magnetic order, normally antiferromagnetic (AFM) or ferromagnetic (FM) (this is not strictly true in all the unconventional superconductors, as for example in $\kappa-(ET)_2Cu_2 (CN)_3$ \cite{KurPRL952005}, as shown in \fref{fig:1.11}\textbf{(d)} or in $FeSe$ \cite{MarCRP172016}), which disappears once doping or pressure is applied. When the magnetic dome disappears, a superconducting dome emerges. More complicated situations may also appear \cite{Nor1302.3176,MarCRP172016}: presence of a pseudogap \textcolor{black}{state}, strange metal behaviors, asymmetry for electron and hole dopings, etc. However, the previous behavior of magnetism being supressed plus superconductivity emerging, seems to be robust feature in most of these systems, so \textcolor{black}{in this introduction, I} will focus the discussion around this general situation, specifically for AFM systems.

By following the phase diagram, it has been proposed that magnetic fluctuations are playing a major role in the superconducting instability \cite{Coleman2015}. An important question emerges when studying the strength of electronic correlations in order to describe the magnetic phase: should the model which describes the magnetism starts from a weakly correlated regime for delocalized electrons or from a strongly correlated regime for localized spins (as described in \sref{1.3.2}) to describe the magnetism \cite{BasCRP172016}? Answering this question is crucial to understand the initial model used to describe the unconventional superconductivity.

\begin{figure}[h]
   \centering
   \includegraphics[width=0.9\columnwidth]{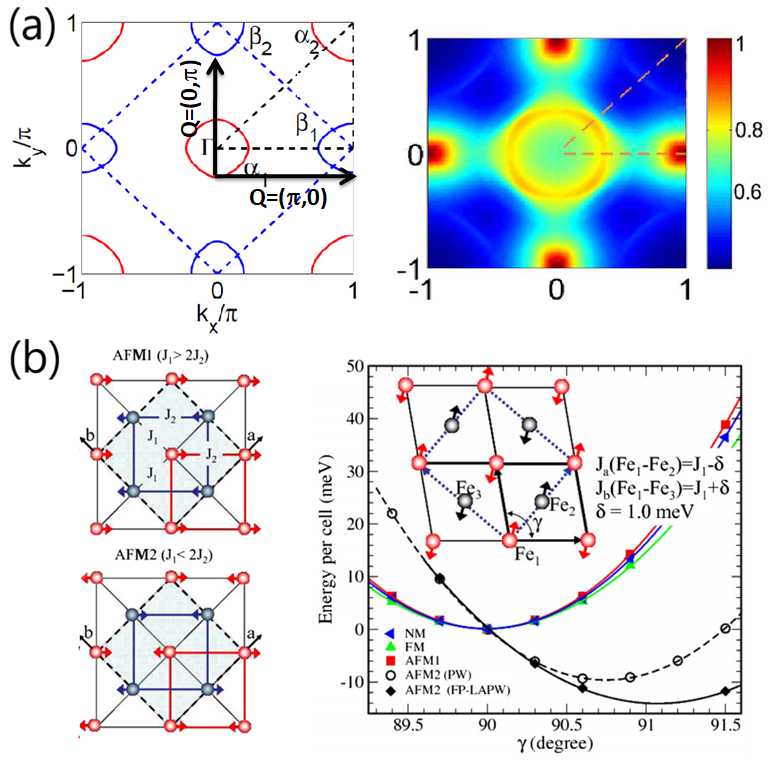}
   \caption[Weakly vs. strongly correlated picture of magnetism in FeSCs]{\textbf{(a)} Weakly correlated picture of magnetism for FeSCs, where the Fermi surface is plotted at the left and $\chi_{spin}$ is plotted at the right. Nesting vector is given by $\vec{Q}=(\pm \pi,0)$ or $(0,\pm \pi)$, signaling a stripe-like AFM order. The scattering originates between $\alpha$ hole pockets $(0,0)$ and $\beta$ electron pockets at $(\pm \pi,0)$ and $(0,\pm \pi)$. Taken and adapted from \cite{RagPRB772008}. \textbf{(b)} Localized picture of magnetism for FeSCs, where the Heisenberg exchanges are plotted in the left and the energy per unit cell in terms of $\gamma$ (where $\gamma$ is the angle defined in the \textsc{Inset}) is plotted in the right. Depending on the ratio $J_1/J_2$, one or another AFM order is promoted, and for $J_2>J_1/2$, AFM2 (checkerboard AFM) is always the ground state for $\gamma>90 \, ^{\circ}$. Taken and adapted from \cite{YilPRL1012008}.}
   \label{fig:1.12}  
\end{figure}

In the weakly correlated limit, for highly delocalized electrons, magnetism is described by a tendency to diverge for the spin susceptibility $\chi_{spin}$ \cite{Fazekas1999}. The spin susceptibility is derived from the Fermi surface, in which nesting features (\textcolor{black}{a portion of the Fermi surface can be superposed on another portion when displaced by a wavevector $\vec{Q}$, called the nesting vector}) promote the magnetic instability. For $\vec{Q}=\vec{0}$, the system shows FM behavior, while for non-zero $\vec{Q}$ it shows different AFM orders. In \fref{fig:1.12}\textbf{(a)}, calculations for $\chi_{spin}$ are shown for the same model as in \fref{fig:1.1}\textbf{(a)}-\textbf{(b)} for $LaFeAsO$, where only the 2D square layers of $Fe$ atoms is considered (see discussion in \sref{1.5.2}). $\vec{Q}=(\pi,0)$ and $(0,\pi)$, which is recognised as a stripe-like AFM order (parallel spin alignement in one axis defined by $\vec{Q}$, while antiparallel alignement in the perpendicular axis), see \sref{1.5.2}.

In the strongly correlated limit (large-$U$ limit), for totally localized spins, magnetism is described by the minimum in the ground state energy obtained from a Heisenberg-like hamiltonian, similar to \eref{eq:1.20} \cite{Fazekas1999}. In \fref{fig:1.12}\textbf{(b)}, Heisenberg model results for first $J_1$ and second $J_2$ neighbor exchanges are presented for the same compound as in \fref{fig:1.12}\textbf{(a)}, $LaFeAsO$. Depending on the ratio $J_1/2J_2$, different AFM order are promoted. For the ground state energy (right part of \fref{fig:1.12}\textbf{(b)}), $J_2>J_1/2$ and $\gamma$ refers to the angle in the vertex of the 2D square lattice of $Fe$ atoms, where for a perfect square $\gamma=90 \, ^{\circ}$, all the magnetic orders are degenerated. For $\gamma > 90 \, ^{\circ}$, the AFM2 order (recognised as the checkerboard AFM order) is always the ground state.

In \cref{Chap03}, we will explore the ongoing work in the moderately to strongly correlated metallic regime \cite{BasCRP172016}.

The superconducting order parameter (related with the supercoducting gap) $\phi$, in absence of spin-orbit coupling, can be decoupled into $\phi = \psi^{orbital} \varphi^{spin}$, where $\psi$ states the orbital symmetry (dictated by $L=0,1,2,\ldots$) and $\varphi$ the spin symmetry. Two different types of superconducting order parameter can be distinguished, spin-singlet, for which the orbital part has $s$-, $d$-, ... wave symmetry, and spin-triplet, for which it has $p$-, $f$-, ... wave symmetry. Spin-singlet superconducting order parameter is promoted by AFM correlations, while spin-triplet is promoted by FM correlations \cite{Coleman2015}.

Spin fluctuation pairing theory is a weakly correlated theory in which spin fluctuations $\chi_{spin}$ enhance the pairing instabitility \cite{HirCRP172016}. The symmetry of the superconducting order parameter is given by the largest eigenvalue of the pairing vertex $\Gamma_{\nu \mu} (\vec{k}, \vec{k}')$, where $\Gamma_{\nu \mu}$ is a scattering matrix (for two bands $\nu$ and $\mu$ which are connected by $\vec{Q}$) which depends on the interaction parameters and on the spin and the charge susceptibilities. The largest eigenvalue of $\Gamma_{\nu \mu}$ gives the largest contribution from spin fluctuations to the pairing instability. 

In the case of strongly correlated theories, exchanges $J_i$ are decoupled in the superconducting pairing channel with a mean-field approach. The superconducting order parameter symmetry will depend on the exchanges $J_{i}$. 

In the next sections, I will very briefly discuss the general phenomenology of heavy-fermion and high-$T_c$ cuprates superconductors, and more in detail about FeSCs.

\subsection{Cuprates \& other unconventional superconductors}
\label{1.5.1}

The heavy-fermion superconductors were discovered in 1978 \cite{StePRL431979}, in $CeCu_{2}Si_{2}$. In these systems \cite{Nor1302.3176}, the $T_{c}$ is usually low (not larger than $\sim 2K$) and the magnetic moment \textcolor{black}{is} high, in contrast with conventional superconductors, which do not show any magnetic moments preceeding the superconducting phase. In some of these compounds \textcolor{black}{the} effective masses can be up to thousands times the one expected from DFT calculations (they are strongly correlated metals), hence the name of the family. The crystal structure of heavy-fermion superconductors is formed by layers of the rare-earth element. In this system, both itinerant (coming from the $f$ orbitals) and localized electrons coexist at the same time. 

There are some heavy-fermion superconductors which show AFM ordering, \textcolor{black}{and} superconductivity emerges \textcolor{black}{when magnetism is suppressed}, such as $CeRh{In}_5$ as shown in \fref{fig:1.3}\textbf{(a)}, while others show a FM order, like $U{Ge}_2$ or $URhGe$. In the latter case, superconductivity is completely inside the FM dome \cite{SaxN4062000}, while for AFM compounds some coexistance between AFM order and superconductivity appears in some compounds \cite{KneCRP122011}, as shown in \fref{fig:1.11}\textbf{(a)}.

\begin{figure}[h] 
   \centering
   \includegraphics[width=0.9\columnwidth]{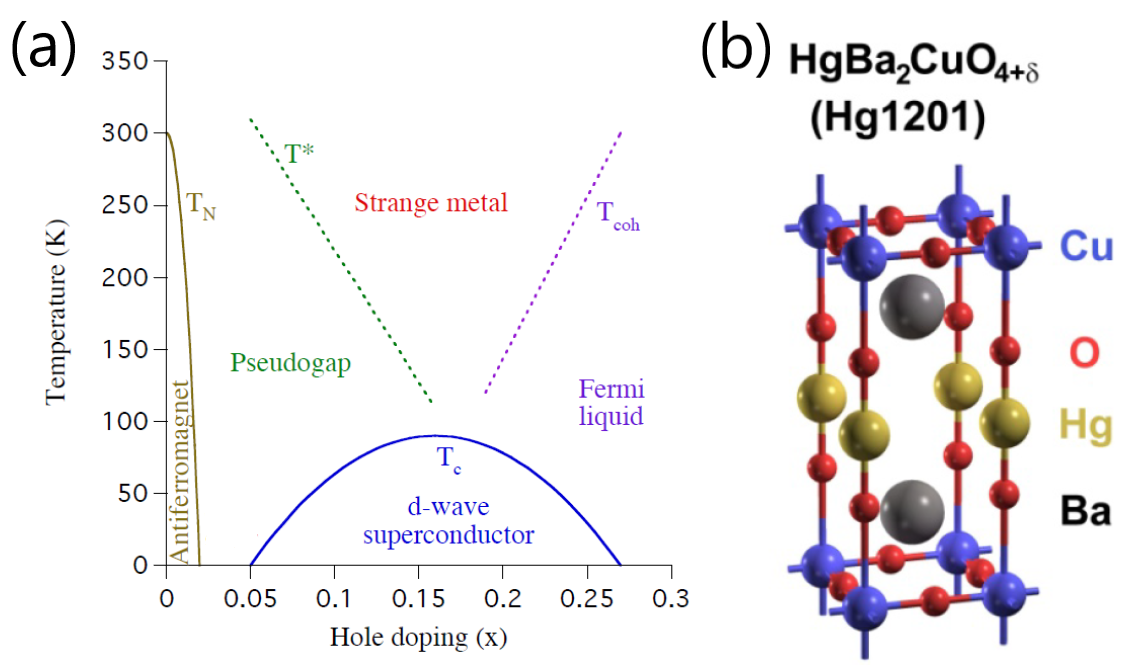}
   \caption[Sketch of the hole-doped phase diagram and crystal structure for high-$T_c$ cuprates superconductors]{\textbf{(a)} Sketch of the hole-doped phase diagram for the cuprates, where various phases can be seen, like AFM checkerboard phase with a characteristic $T_N$, pseudogap state with $T^*$, the unconventional superconducting dome with $T_c$ and the strange metal and Fermi liquid metal \textcolor{black}{states}. Taken and adapted from \cite{Nor1302.3176}. \textbf{(b)} Crystal structure of Hg-1201 cuprate superconductor, where $CuO_2$ layer is known to control the electronic properties of the cuprates. Taken and adapted from \cite{BarPNAS1102013}.}
   \label{fig:1.13}  
\end{figure}

Later in 1986 \cite{BednorzZPBCM641986}, Bednorz and Muller gave birth to a new (and the most famous) family of unconventional superconductors, the high-$T_c$ cuprates. They discovered superconductivity in \textcolor{black}{$Ba_x La_{2-x} Cu O_{4}$} with $T_c \sim 30 \, K$. Rapidly since then, a lot of effort for finding new superconductors in these family rose up $T_c$ to $93 \, K$ in $Y{Ba}_2{Cu}_3{O}_7$ (equivalently called YBCO, or Y-123) \cite{WuMPRL581987}. Up to date, the largest $T_c$ in the cuprates was found in Hg-1223 under pressure, with $T_c \approx 164 \, K$ \cite{ChuN3651993}. The parent compound, \textcolor{black}{i.e. the undoped system (see below)}, of high-$T_c$ cuprates shows a Mott insulating behavior, and a checkerboard AFM phase for $T<T_N \sim 300 \, K$, see \fref{fig:1.13}\textbf{(a)}. The unconventional superconducting dome emerges around the AFM phase for both electron- and hole-doped cuprates, as shown in \fref{fig:1.11}\textbf{(b)}. These systems are still under a huge debate about the physics and different phases that appear in them, as shown in \fref{fig:1.13}\textbf{(a)} for hole-doped cuprates.

The crystal structure of the cuprates is formed by square layers of $CuO_2$ and spacer layers \cite{BarPNAS1102013}. In \fref{fig:1.13}\textbf{(b)}, the crystal structure of Hg-1201 is shown, where the spacer layers are made of $Ba$, $Hg$ and $O$. $CuO_2$ layers control the physics of the cuprates. In \textcolor{black}{the undoped} systems \cite{PicRMP611989}, $Cu$ atoms are filled with $n_e=9$ electrons in 5 $d$ orbitals, where $d_{x^2-y^2}$ remains at half-filling. This $d_{x^2-y^2}$ orbital is well separated from other orbitals, due to crystal field effects. Thus, the system is effectively behaving as a single-orbital system \textcolor{black}{with one electron per site}. Due to its quasi-2D crystal structure, cuprates can be modeled using a 2D square tight-binding model. As expected from the non-interacting limit, the system should behaves as a metal, with the Fermi level crossing the bands. Suprisingly, the undoped compounds are insulators, while at a certain doping, there is a insulator-to-metal transition.

More specifically, $d_{x^2-y^2}$ orbital hybridizes with $p_x$ and $p_y$ orbitals coming from $O$, forming bonding and antibonding orbitals as a linear combination in the $CuO_2$ plane. The $CuO_2$ antibonding orbital will remain at half-filling, \textcolor{black}{then, single-band undoped cuprates can be seen as Mott insulators, as explained in \sref{1.3}}. In fact, it is a charge-transfer (Mott) insulator, because the \textcolor{black}{smallest} direct gap opens between the valence $O$ $p$ bands and the upper Hubbard band ($U$ pushes the lower Hubbard band below the valence $O$ $p$ band). \textcolor{black}{Nevertheless, cuprates are generally referred to as Mott insulators}.

The undoped compound shows an AFM phase with checkerboard pattern $\vec{Q}=(\pi,\pi)$, as anticipated for 2D square lattices in \sref{1.3.2} for Heisenberg-like models \cite{Fazekas1999}. In such a situation, AFM correlations are promoted for $J>0$. The AFM phase is suppressed with doping.

The superconducting order parameter symmetry has been measured to be $d$-wave like, specifically $d_{x^2-y^2}$ \cite{Coleman2015}. It is believed that magnetic \textcolor{black}{correlations are behind} the pairing enhancement. As the undoped compound is a Mott insulator, most of the theories work with strongly correlated models, such as the Heisenberg-like models, like \eref{eq:1.20}. Nevertheless, there are also weakly correlated models applied for the overdoped superconducing dome, i.e. the part which is closer to the Fermi liquid metal.

\begin{figure}[h] 
   \centering
   \includegraphics[width=0.75\columnwidth]{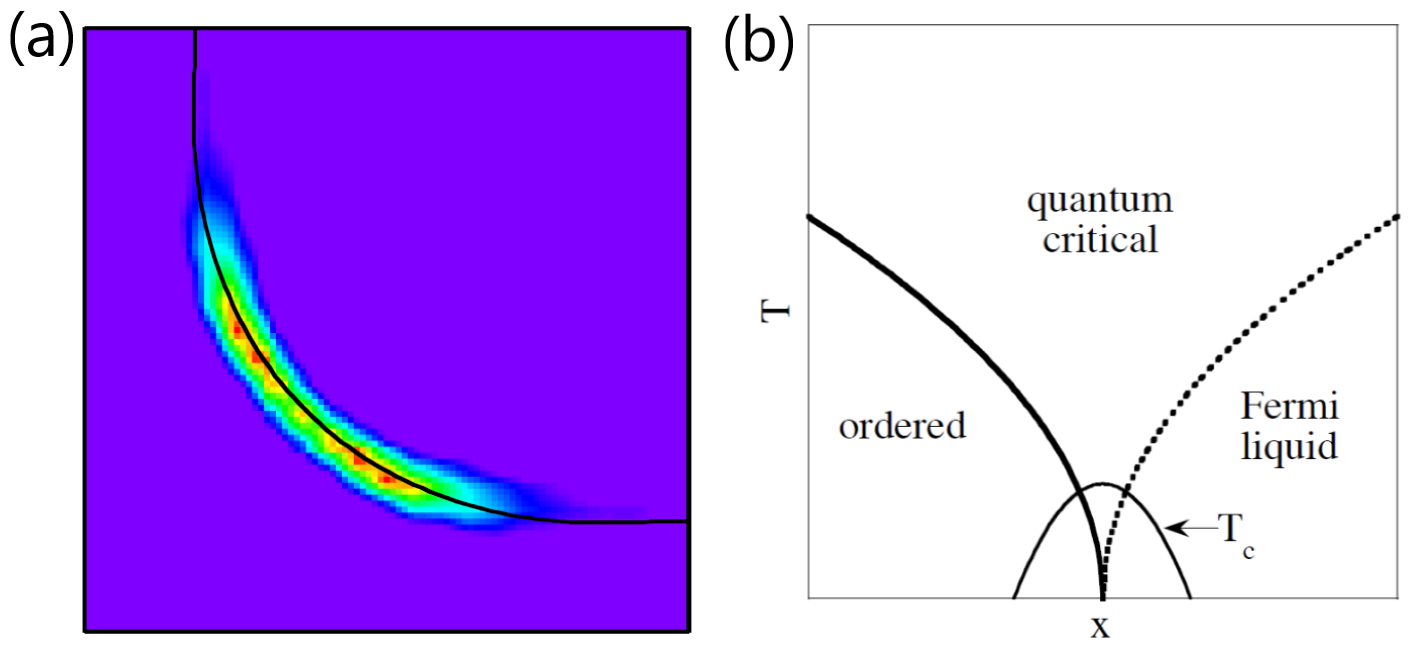}
   \caption[Pseudogap Fermi arcs and QCP phase diagram for cuprates]{\textbf{(a)} Unclosed Fermi arcs appearing in the pseudogap state of cuprates. The Fermi surface of the normal metallic state is plotted in the background. \textbf{(b)} QCP phase diagram for the cuprates. The influence of the QCP at $T=0\,K$ can be seen for $T>0\,K$. Taken and adapted from \cite{Nor1302.3176}.}
   \label{fig:1.14}  
\end{figure}

For the doped compound, at $T^{*}>T_N$ there is a crossover into an exotic pseudogap state \cite{NorPRB762007,TreLTP322006,CivPRL952005}. In this state, the Fermi surface \textcolor{black}{as observed in ARPES}, exhibits an arc of gapless excitations, i.e. it is not closed in the BZ, as shown in \fref{fig:1.14}\textbf{(a)}. The Fermi surface in the weakly correlated metal state (see purple region in \fref{fig:1.13}\textbf{(a)}) is plotted in the background as a black line. There is still a lot of discussion about the origin of the pseudogap state \textcolor{black}{and how it is the complete Fermi surface}. 

The strange metal regime is a metallic \textcolor{black}{state} with a linear-$T$ term in the resistivity. This term does not follow the FLT expectations, see \eref{eq:1.15}. A lot of effort has been put to understand this phase, and its origin it is believed to be related with the presence of a quantum critical point (QCP) inside the superconducting dome \cite{VarPRB612000}. A QCP is a critical point found at $T=0 \, K$ which is driven by quantum fluctuations. The QCP has consequences at $T>0 \, K$, as shown in the phase diagram of \fref{fig:1.14}\textbf{(b)}.


\subsection{Iron-based superconductors}
\label{1.5.2}

In 2008 \cite{KamJACS1302008}, high-$T_c$ superconductivity was found in a new family of compounds, the FeSCs. They discovered $T_c^{max} \sim 26 \, K$, in electron-doped $LaFeAsO$. Different crystal structures have been obtained, but they all share the same $FePn$ or $FeCh$ (with $Pn=As$, $P$ and $Ch=S$, $Se$) layer, where $Fe$ atoms are tetrahedrally coordinated by $Pn$ or $Ch$ atoms \cite{PagNP62010}. Usually, FeSCs are named according to their chemical formula, hence 1111 iron pnictide will refer to $LaFeAsO$ or to a related compound. Among the different compounds, it can be distinguished 1111 (like $LaFeAsO$), 111 (like $LiFeAs$), 122 (like $BaFe_2As_2$), 11 (like $FeSe$), 1144 (like $KCsFe_4As_4$), etc, iron pnictides or chalcogenides. I will study and discuss more in detail the 123 FeSCs compounds in \cref{Chap4b}, which shows various important differences with the other FeSCs compounds. \textcolor{black}{I} will focus \textcolor{black}{in this introduction in} the discussion \textcolor{black}{for} 1111 and 122 FeSCs.

\begin{figure}[h] 
   \centering
   \includegraphics[width=0.8\columnwidth]{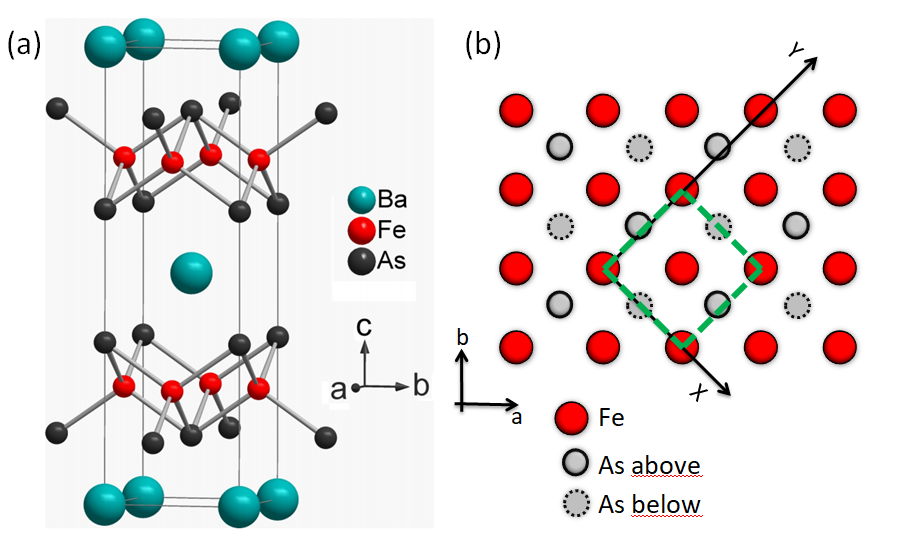}
   \caption[Crystal structure and $Fe$ layer in 122 iron arsenide]{\textbf{(a)} Crystal structure and $Fe$ layer in 122 iron arsenide $BaFe_2As_2$, where $FeAs$ layers are separated by spacer $Ba$. Taken and adapted from \cite{RotPRL1012008}. \textbf{(b)} $FeAs$ layer with unit cell (green dashed line) included with 2 $Fe$ atoms per unit cell. $Fe$ atoms form a 2D square lattice and are tetrahedrally coordinated by $As$ atoms. Adapted from \cite{emergence2017}.}
   \label{fig:1.15}  
\end{figure}

\fref{fig:1.15}\textbf{(a)} shows the crystal structure for the 122 FeSCs, $BaFe_2As_2$ \cite{RotPRL1012008}. In this crystal structure, $FeAs$ (or $FeCh$) layer is sandwiched between spacers (in the case of the figure, the spacers are $Ba$ atoms), similar to what happens in high-$T_c$ cuprates. There are 2 $Fe$ atoms per unit cell, as seen in \fref{fig:1.15}\textbf{(b)} for the $FeAs$ plane. Pnictogen and chalcogen atoms are tetrahedrally coordinating the $Fe$ atoms.

\begin{figure}[h] 
   \centering
   \includegraphics[width=0.9\columnwidth]{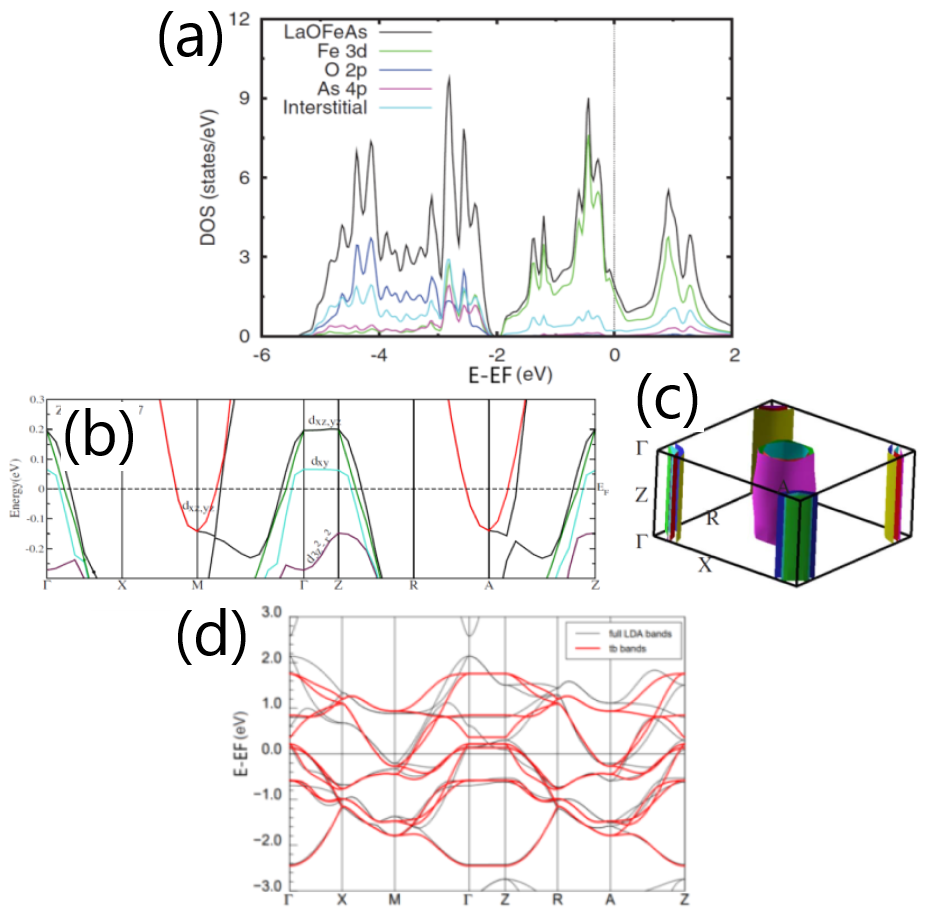}
   \caption[Total and orbital-resolved DOS, band structure around the Fermi level with $d$ orbital character included and 3D Fermi surface for undoped $LaFeAsO$ at $n_e=6$, and 2D Wannier parametrization of the 3D band structure for $FeSe$]{\textbf{(a)} Total (black line) and orbital-resolved (color lines) DOS. $Fe$ $d$ orbitals mostly contribute between $-2$ and $2 \, eV$, with a small contribution from electronic charge in the interstitial region of $Fe$ atoms. \textbf{(b)} DFT band structure around the Fermi level with $d$ orbital character included, and \textbf{(c)} 3D Fermi surface for undoped $LaFeAsO$ at $n_e=6$, remarking the quasi-2D character of FeSCs. Taken and adapted from \cite{VilPRB782008}. \textbf{(d)} 2D Wannier tight-binding fitting (red line) of the 3D band structure (black line) for $FeSe$. Taken and adapted from \cite{EscPRB802009}.}
   \label{fig:1.16}  
\end{figure}

The parent compound of FeSCs is metallic \cite{PagNP62010,BasCRP172016}. The physics in FeSCs is controlled by $Fe$ \textcolor{black}{orbitals}, see \fref{fig:1.16}\textbf{(a)}. In \fref{fig:1.16}\textbf{(a)}, the total and orbital-resolved DOS is plotted for $LaFeAsO$, where the interstitial refers to the electronic charge in between $Fe$ atoms (this effect comes from $As$ $p$ orbitals influence). All the 5 $d$ contributes around the Fermi level \cite{LebPRB752007,VilPRB782008}, with similar crystal field energies, so effective tight-binding models should include, at least, 5 $d$ orbitals \cite{KurPRL1012008,GraNJP112009,CalPRB802009}, see \fref{fig:1.16}\textbf{(b)}, in contrast to single-orbital cuprates. \textcolor{black}{When only taking into account the $Fe$ atoms}, the total number of bands is $10$ ($2$ $Fe$ atoms $\times$ $5$ $d$ orbitals). The parent compound has 6 electrons per $Fe$ atom, $n_e=6$ (which is one extra electron than half-filling, which is defined as $n_e=N=5$, where $N$ is the number of $d$ orbitals per $Fe$ atom). Some authors have argued about the importance of the $Pn$ or $Ch$ atoms, and there are claims that $p$ orbitals should be included in the minimal tight-binding model, giving rise to the $d$, $dp$ and $dpp$ models. In any case, $d$ model is shown to give good enough predictions for most FeSCs, and here \textcolor{black}{I} will focus on \textcolor{black}{the} $d$ model phenomenology. Total bandwidth is estimated $W \sim 4-5 \, eV$ for $d$ models. In the tetragonal phase, $d_{zx}$ and $d_{yz}$ are degenerated, as shown in \fref{fig:1.3}\textbf{(b)}. Fermi surfaces shows a quasi-2D behavior, see \fref{fig:1.16}\textbf{(c)}, where almost cylindrical pockets along $Z$ can be seen. Then, the tight-binding model for FeSCs is usually written for the 2D square layer of $Fe$ atoms. The BZ for the $2$ $Fe$ atoms in the unit cell is also called folded BZ (see next paragraph), and it is formed by two hole pockets centered at $\Gamma=(0,0)$ and two electron pockets at $M=(\pi,\pi)$. In \fref{fig:1.16}\textbf{(d)}, I present the 2D Wannier parametrization of the 3D bands obtained for $FeSe$. Some discrepancies can be seen, but both the bandwidth and the Fermi surface \textcolor{black}{obtained in DFT calculations are} approximately reproduced by the Wannier tight-binding fitting.

\begin{figure}[h] 
   \centering
   \includegraphics[width=1.0\columnwidth]{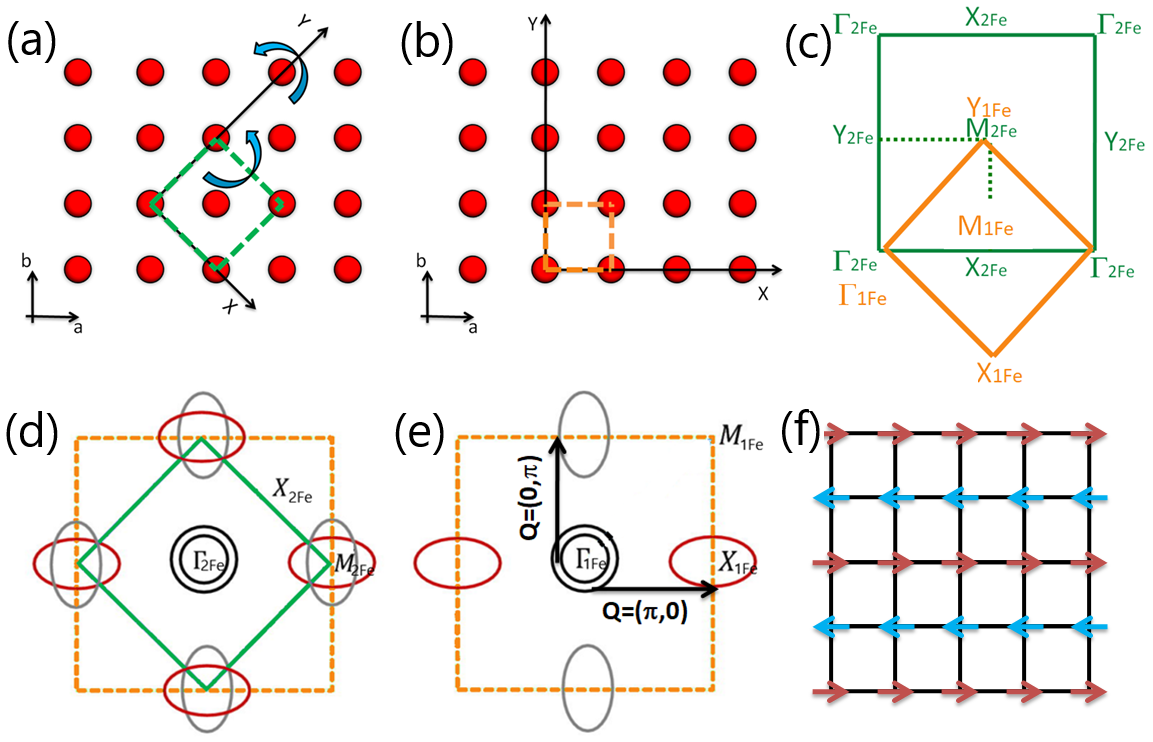}
   \caption[Sketch of the unfolding procedure, folded and unfolded BZs and Fermi surface in both folded and unfolded BZ for FeSCs. Stripe magnetic order in the 2D square lattice of FeSCs]{\textbf{(a)}, \textbf{(b)} Sketch of the unfolding procedure, with unit cell included in both the folded (green dashed lines) and unfolded BZ (orange dashed lines), \textbf{(c)} folded (green square)  and unfolded (orange square) BZs. Adapted from \cite{emergence2017}. 2D Fermi surfaces in both \textbf{(d)} folded and \textbf{(e)} unfolded BZ for FeSCs. Taken and adapted from \cite{HirRPP742011}. I included the nesting vectors $\vec{Q}=(\pi,0)$ and $(0,\pi)$. \textbf{(f)} Stripe-like AFM order in 2D square of $Fe$ atoms. Note that stripe aligned in $X$ or $Y$ directions is equivalent for the tetragonal phase ($Fe$ atoms form a perfect square). Adapted from \cite{emergence2017}}
   \label{fig:1.17}  
\end{figure}

An useful procedure can be done to reduce the difficulty of the problem, the so called unfolding of the BZ \cite{CalPRB802009}. This is a $45 \, ^{\circ}$ rotation plus an unfold in the $y$ direction of the unit cell. By this procedure, the BZ has enlarged, while the unit cell is reduced, with respect to their original shapes in the folded BZ. After this unfolding, the number of atoms in the unit cell is reduced to 1 $Fe$ atom. So instead of having $10$ bands, $5$ bands are obtained for the $Fe$ atom with 5 $d$ orbitals. In \fref{fig:1.17}\textbf{(a)}-\textbf{(b)}, I have sketched this procedure, labeling each folded and unfolded BZ situation with $2Fe$ or $1Fe$, respectively. I also show the resulting BZs in both folded and unfolded cases in \fref{fig:1.17}\textbf{(c)}. The 2D Fermi surface for both cases are shown in \fref{fig:1.17}\textbf{(d)}, \textbf{(e)}, where two hole pockets are centered at $\Gamma$ for both cases, whether two electron pockets appear at $M_{2Fe}$ and $X_{1Fe}/Y_{1Fe}$.

Most of the FeSCs are AFM metals with $T_N \sim 100-150 \, K$ with a stripe-like order $\vec{Q}=(\pi,0)$ and $(0,\pi)$ for the unfolded BZ \cite{BasCRP172016}, as shown in \fref{fig:1.17}\textbf{(f)}. Due to its metallic character, \textcolor{black}{some} approaches are based on a weakly correlated picture of magnetism \cite{RagPRB772008,MazPRL1012008,ChuPRB782008,CveEPL852009}, in which $\chi_{spin}$ is peaked at $\vec{Q}$, as shown in \fref{fig:1.12}\textbf{(a)}, with $\vec{Q}$ being the scattering between $\Gamma$ hole and $X/Y$ electron pockets. On the other hand, there \textcolor{black}{are} calculations in the large-$U$ limit \cite{YilPRL1012008,SiQPRL1012008} for FeSCs with first $J_1$ and second $J_2$ neighbor exchanges, where the $(\pi,0)$ ground state requires $J_2>J_1/2$, as shown in \fref{fig:1.12}\textbf{(b)}. The magnetic moment measured by neutron scattering experiments is $\sim 0.5-1 \, \mu_B$ per $Fe$ atom. From the large-$U$ limit calculations, the expected magnetic moment should be $= 4 \, \mu_B/Fe$, much larger than the observed experimentally. On the other hand, DFT calculations predicted magnetic moments $\sim 1.5-2 \, \mu_B/Fe$, also larger than the obtained experimentally.

ARPES and quantum oscillations (QOs) experiments \cite{LuDARCMP32012} showed the existence of quasiparticles with $m^*/m_0 \sim 2-3$ ($Z \sim 0.3-0.5$) for the parent compound of FeSCs, hence emphasising the importance of electronic correlations. Due to the multiorbital character of FeSCs, it has been argued that both weakly and strongly correlated electrons coexist at the same time in different orbitals. In this situation, magnetic moments at $Fe$ sites appear due to Hund's coupling $J_H$ polarization \cite{FanPRB922015}. I will discuss more in detail these ideas in \cref{Chap03}. 

AFM transition follows a previous tetragonal-to-orthorombic structural transition at $T_s \geq T_N$, also known as nematic transition \cite{MarCRP172016}, which breaks the $x$ and $y$ symmetry, hence the degeneracy between $d_{zx}$ and $d_{yz}$ orbitals. \textcolor{black}{It has an electronic origin}, rather than driven by phonons. Whether it comes from orbital/charge fluctuations or spin fluctuations  is something still under debate. The spin fluctuations seem more plausible due to the fact that magnetism and nematic phases suppresses following the same trend.

\begin{figure}[h] 
   \centering
   \includegraphics[width=0.3\columnwidth]{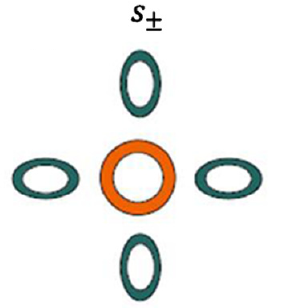}
   \caption[Schematic superconducting order parameter $s_{\pm}$ symmetry obtained for FeSCs]{Schematic superconducting order parameter $s_{\pm}$ symmetry obtained for FeSCs. The different color marks the different sign for the superconducting order parameter. Taken and adapted from \cite{HirCRP172016}.}
   \label{fig:1.18}  
\end{figure}

Due to the metallic character of FeSCs, spin fluctuation pairing theory was proposed to obtain the superconducting order parameter in FeSCs \cite{HirCRP172016}. Magnetic fluctuations with $\vec{Q}=(\pi,0)$ promotes a spin-singlet superconducting order parameter \cite{MazPRL1012008}. Experiments confirmed the spin-singlet superconducting order parameter symmetry. In \fref{fig:1.18}, \textcolor{black}{the $s_{\pm}$-wave symmetry proposed for FeSCs is shown. $s_{\pm}$-wave symmetry is present on most of the FeSCs}.

I conclude by noting that there are still many questions open in FeSCs, in particular for the initial model needed to describe magnetism. In future chapters, I will study local correlations in order to shed light about the strength of such correlations in these systems.

\newpage

\section{2D materials}
\label{1.6} 

\begin{figure}[h]
   \centering
   \includegraphics[width=0.9\columnwidth]{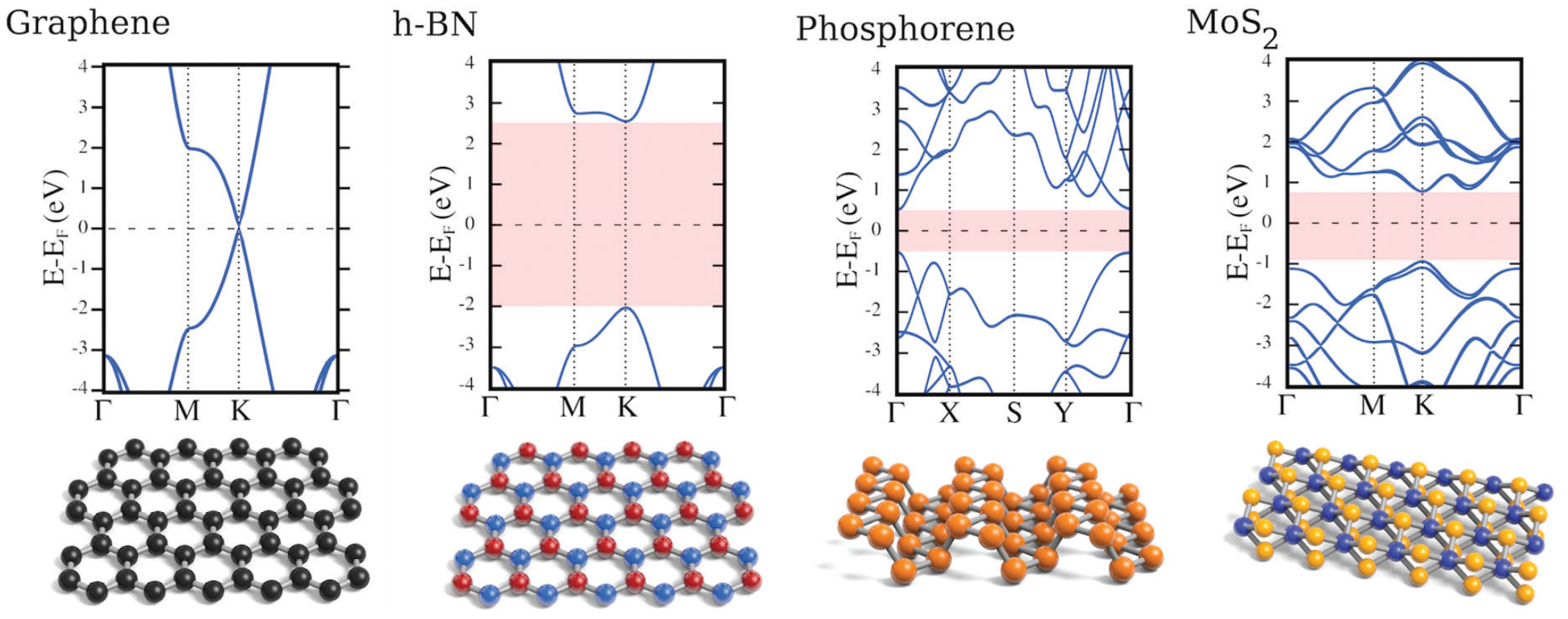}
   \caption[Crystal structure and band structures for different 2D materials: graphene, hBN, phosphorene and $MoS_2$]{Crystal and band structures for various 2D materials: \textbf{(a)} graphene, \textbf{(b)} hBN, \textbf{(c)} phosphorene and \textbf{(d)} $MoS_2$. Graphene shows the famous Dirac points at $K$ due to the hexagonal symmetry of its lattice. The Fermi velocity (in the non-interacting limit) $v_0=\nabla_k E_k=10^{6}\, m\, s^{-1}$ indicates the huge mobility of the electrons in graphene. The other compounds are semiconductors whose gap can be tuned in different ways (see main text). Taken and adapted from \cite{RolCSR462017}.}
   \label{fig:1.19}  
\end{figure}

In this section, I will follow the discussion made in \cite{Schonhoff2017thesis,RolCSR462017} for the role of electronic correlations in 2D materials. In the last decades, 2D materials have received a lot of attention, specially since the discovery of \textcolor{black}{how to produce graphene by exfoliating graphite}. 2D materials can be grouped in: graphene-like systems, hexagonal boron nitride (hBN), phosphorene, transition metal dichalcogenides (TMDCs), such as $MoX_2$, $NbX_2$, $TaX_2$, etc (with $X=S$ or $Se$), transition metal monochalcogenides (TMMCs), like $FeSe$ and $SnS$, and transition metal trichalcogenides (TMTCs), such as $TiS_3$. \textcolor{black}{Other 2D materials, like $Fe_3GeTe_2$, have been recently isolated}. These systems are simple to produce in a 2D version due to the weak van-der-Waals (vdW) forces acting between the layers that form the 3D system, which permit an easy exfoliation of the layers from the bulk crystal, and an easy adaptation to the substrates. Then, it is also simple to build a huge variety of heterostructures to produce different technological devices. In \fref{fig:1.19}, the crystal structure and band structure of various 2D materials, such as graphene, hBN, phosphorene and $MoS_2$, are presented.

\begin{figure}[h]
   \centering
   \includegraphics[width=0.9\columnwidth]{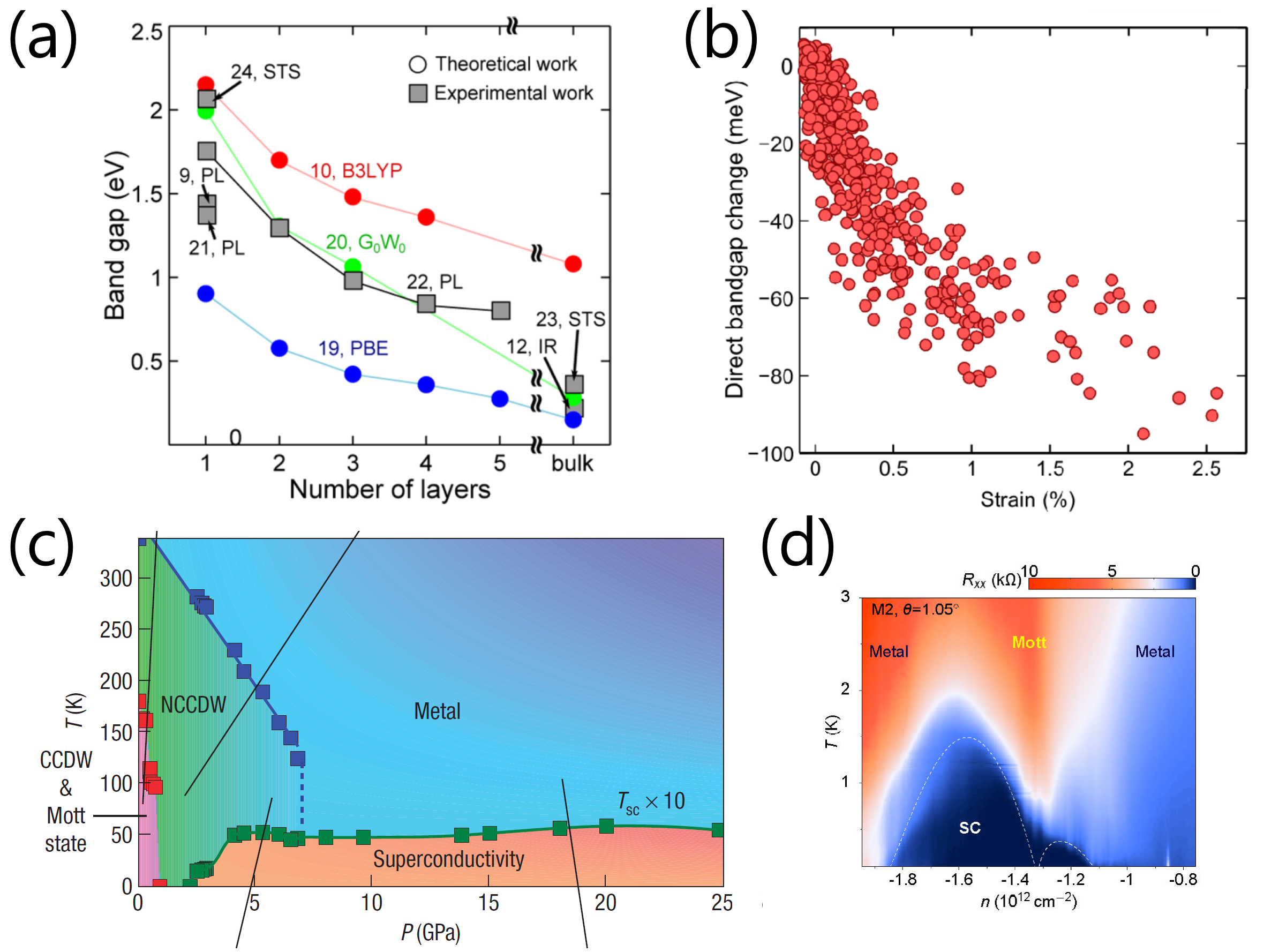}
   \caption[Band gap variation in terms of $N_L$ for phosphorene and strained $MoS_2$, and phase diagrams for $1T-TaS_2$ with pressure and hole-doped MA-TBG]{\textbf{(a)} Band gap versus $N_L$ for phosphorene. Both theoretical and experimental results are included. Band gap shows a exponential decay with $N_L$. Taken from \cite{CasJPCL62015}. \textbf{(b)} Direct band gap verus strain for $MoS_2$. Again, band gap shows a exponential decay with strain. Taken from \cite{CasNL132013}. \textbf{(c)}, \textbf{(d)} Pressure and doping phase diagrams for $1T-TaS_2$ and MA-TBG, respectively. Both compounds show a rich phase diagram, where superconductivity emerges when a Mott insulating state is supressed. Non-invasive doping can be induced in 2D materials by an external gate voltage. Taken and adapted from \cite{SipNM72008,CaoN5562018_sc}.}
   \label{fig:1.20}  
\end{figure}

One of the most important properties of 2D materials is that they are very sensible to changes in the enviroment that surrounds the surface (because the portion of atoms in the surface is comparable to those in the volume). Then, the number of layers $N_L$, different substrates, or dielectric environments, and strain effects will change the electronic properties (in addition to doping or pressure). Moreover, doping can be induced in a non-invasive way by an external gate voltage, rather than by implementation or diffusion as in 3D materials. In \fref{fig:1.20}\textbf{(a)} and \textbf{(b)}, the bandgap evolution for phosphorene \cite{CasJPCL62015} in terms of $N_L$ and for $MoS_2$ in terms of strain \cite{CasNL132013}, respectively, are shown. 

The interaction effects will be important in \textcolor{black}{some of} these systems. In \fref{fig:1.20}\textbf{(c)} and \textbf{(d)}, we can see the phase diagrams of $1T-TaS_2$ when pressure is applied \cite{SipNM72008} and the hole-doped magic-angle twisted bilayer graphene (MA-TBG) \cite{CaoN5562018_sc}. Superconducting domes emerge in both cases when the Mott insulating phase is suppressed, similar to what happens in high-$T_c$ cuprate superconductors.

There are still a lot of questions open for the role of the electronic correlations in 2D materials. In this thesis, I am going to focus on studying the effect of local correlations in MA-TBG, and \textcolor{black}{some properties of} $Fe_3GeTe_2$ (FGT). Here, I will briefly discuss the basic phenomenology of MA-TBG, and in \cref{Chap4c} we will get more into the details after the works by \emph{Cao et al.} were published \cite{CaoN5562018_ins,CaoN5562018_sc}. In \cref{Chap4d}, I will study FGT and the current status of our calculations in this system.

\subsection{Twisted bilayer graphene (TBG)}
\label{1.6.1}

In this section, I will discuss the known phenomenology of twisted bilayer graphene systems \cite{ReiPRB662002,LopPRL992007,SuaPRB822010,BisPNAS1082011,CaoPRL1172016,FanPRB932016,GonPRL1192017,KosPRX82018}, before the discoveries by Pablo Jarillo's group \cite{CaoN5562018_ins,CaoN5562018_sc}.

\textcolor{black}{Graphene is formed by a hexagonal lattice of carbon atoms, in which the unit cell is made of two carbon atoms, usually called $A$ and $B$. The band structure of graphene is formed by two linear dispersions which cross at the so called Dirac points, which corresponds to the $K$ and $K'$ points of the hexagonal BZ (see below) \cite{CasRMP812009}. The region close to each Dirac point is called valley, and these two valleys $K$ and $K'$ are spin-degenerated.}

\begin{figure}[h] 
   \centering
   \includegraphics[width=0.9\columnwidth]{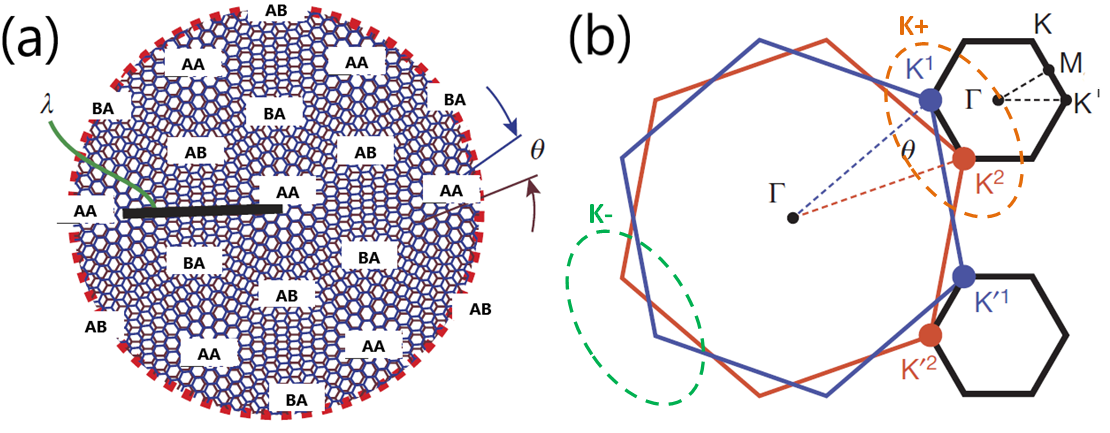}
   \caption[Superlattice and mini-BZ in TBG]{\textbf{(a)} Superlattice and \textbf{(b)} resulting mini-BZ (black hexagons) of TBG at a given rotation angle $\theta$. The size of the mini-BZ is inversely proportional to the size of the superlattice $\lambda$. $AA$, $AB$ and $BA$ regions included as a eye-guide. There are two doubly-degenerated Dirac points located at $K_{\xi}^l$ with $l=1,2$ being the layer index for each $\xi=\pm$ valley. Taken and adapted from \cite{CaoN5562018_ins}.}
   \label{fig:1.21}  
\end{figure}

Twisted graphene systems are heterostructures of $N_L$ graphene layers rotated by a certain set of angles $\theta_{N_L-1}$. In these circumstances, a moir\'e pattern (also called superlattice) emerges, which is described by the alignement of atoms in each graphene layer. For example, in twisted bilayer graphene (TBG), we can distinguish between $AA$ regions, where $A$ atom of the bottom graphene layer is aligned with the $A$ atom of the top layer, $AB$ (and $BA$) regions, where $A$ ($B$) atom of the bottom layer is aligned with $B$ ($A$) atom of the top one, see \fref{fig:1.21}\textbf{(a)}. Similar reasoning will result for twisted trilayer graphene (TTG) \cite{Che1803.01985}. Note that twisted heterostructures might be built of different 2D materials. During this thesis I will focus on TBG.

\fref{fig:1.21}\textbf{(a)} shows the corresponding superlattice that emerges in TBG, where $AA$ regions form a triangular symmetry superlattice and $AB/BA$ regions form a hexagonal symmetry superlattice. The superatomic distance $\lambda$ depends on the rotation angle $\theta$, $\lambda \sim a/(2\sin{\left(\theta/2\right)})$, where $a=2.46 \, \AA$ is the graphene lattice parameter. Due to the emergence of a larger superlattice, the BZ is reduced, which receives the name of mini-BZ, see \fref{fig:1.21}\textbf{(b)}. \textcolor{black}{In the band structure of TBG}, there are two doubly-degenerated Dirac points located at $K_{\xi}^l$ with $l=1,2$ being the layer index for each $\xi=\pm$ valley \cite{BisPNAS1082011,KosPRX82018}. The parent compound of TBG has the Fermi level $\varepsilon_F$ located at these Dirac points (similar to single-layer graphene).

\begin{figure}[h] 
   \centering
   \includegraphics[width=0.9\columnwidth]{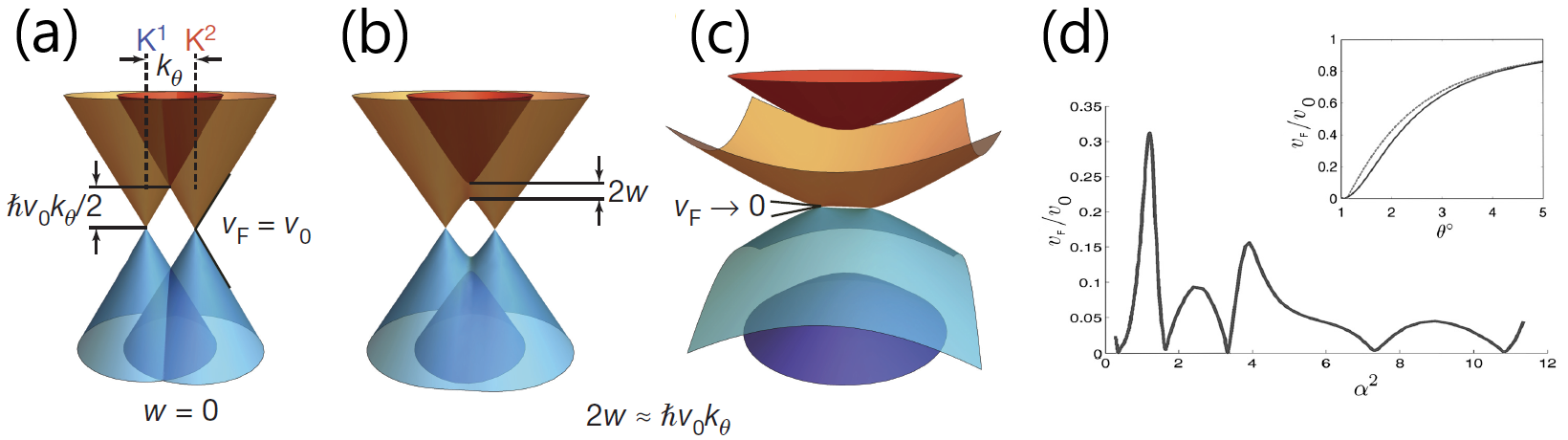}
   \caption[Band evolution and renormalized Fermi velocity of TBG in terms of the rotation angle $\theta$]{\textbf{(a)}-\textbf{(c)} Sketch of the band evolution for TBG. For $\theta>6 \, ^{\circ}$, two independent graphene layers are obtained, as shown in \textbf{(a)}. For $\theta<6 \, ^{\circ}$, hybridization gaps of size $\approx 2w$ emerge between the bands around the Fermi level and the higher in energy bands. For even smaller angles $\theta<2 \, ^{\circ}$, a set of incommensurate magic angles appears at which $v_F \rightarrow 0$, and flat bands emerge. Taken and adapted from \cite{CaoN5562018_ins}. \textbf{(d)} $v_F/v_0$ in terms of $\alpha^2 \sim \theta^{-2}$, showing the cyclic behavior of $v_F/v_0$. The first magic angle appears at $\theta_{1}^{magic} \approx 1.05 \, ^{\circ}$. \textsc{Inset}: $v_F/v_0$ versus $\theta$, where it can be seen the dependence with the rotation angle, for $\theta>6\, ^{\circ}$, $v_F \sim v_0$, and both graphene layers behave as isolated one from another. Taken from \cite{BisPNAS1082011}.}
   \label{fig:1.22}  
\end{figure}

An interesting behavior appears for the band structure in terms of the rotation angle $\theta$, see \fref{fig:1.22}. Two different regimes can be found: for $\theta > 6 \, ^{\circ}$, both graphene layers behave independently of each other, while for smaller angles $\theta<6 \, ^{\circ}$, both layers start to interact. The layer interaction is identified by a hybridization hopping, which will produce gaps, called the superlattice gaps, in the band structure. Assuming that there is no corrugations (out-of-plane distortions in the interlayer space in such a way that the distance is the widest in $AA$ regions and the narrowest in $AB$/$BA$ regions between the two layers) this behavior can be understood by an angle-dependent hybridization hopping $w(\theta)$ between the layers, where \textcolor{black}{around the Dirac points \cite{BisPNAS1082011}}:

\begin{equation}
w(\theta) \approx \left\{
\begin{aligned}
	& 0 \qquad & if \, \theta>6 \, ^{\circ}\\
	& \frac{\hbar v_0 k_{\theta}}{2} \qquad & if \, \theta<6 \, ^{\circ}\\
\end{aligned}
\right.
\label{eq:1.24}
\end{equation}

With $k_\theta \propto \sin{\theta}$ and $v_0 = 10^6 ms^{-1}$ the Fermi velocity of non-interacting single-layer graphene (proportional to $m_0$). In the hybridized case, the hybridization gaps appear between the bands around the Fermi level and the higher-in-energy bands, see \fref{fig:1.22}\textbf{(b)}. For a set of slightly incommensurate angles below $2 \, ^{\circ}$, the bandwidth $W$ of the bands around the Fermi level strongly reduces down to a few dozens of $meV$, see \fref{fig:1.22}\textbf{(c)}. In this situation, the renormalized (due to the interacting layers) Fermi velocity of TBG is $v_F \rightarrow 0$. The angles at which these flat bands emerge are called the magic angles $\theta^{magic}$, see \fref{fig:1.22}\textbf{(d)}. The first magic angle occurs at $\theta_1^{magic} \approx 1.05 \, ^{\circ}$.

\begin{figure}[h] 
   \centering
   \includegraphics[width=0.9\columnwidth]{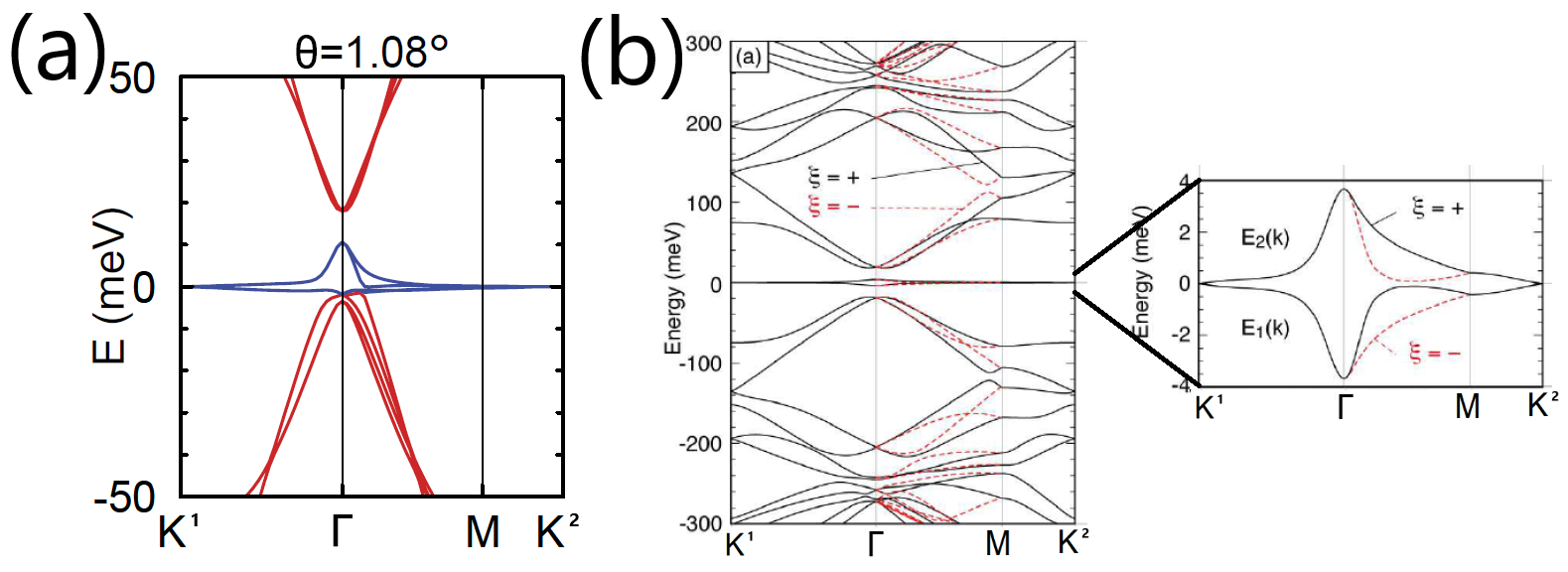}
   \caption[Flat bands at the first magic angle for non-corrugated and corrugated low-energy continuum model for MA-TBG]{Flat bands at the first magic angle for \textbf{(a)} non-corrugated and \textbf{(b)} corrugated effects included in the low-energy continuum model. The flat bands around $K^l$ with $l=1,2$ show a parabolic dispersion, and reach extreme values at $\Gamma$. Dirac points at $K^l$ are doubly-valley degenerated. In total, there are $4$ bands forming the flat bands. Taken and adapted from \cite{CaoN5562018_ins,KosPRX82018}.}
   \label{fig:1.23}  
\end{figure}

\textcolor{black}{A popular} approach \textcolor{black}{to describe the band structure of TBG}, based on a low-energy model (also called low-energy continuum model), \textcolor{black}{is used} to describe the flat bands dispersion around the Dirac points \cite{BisPNAS1082011,KosPRX82018}. It can be shown that the intervalley hopping parameters will decay exponentially, hence there is \textcolor{black}{an approximate} $\xi$ valley degeneracy for $K_{\xi}^l$ Dirac points. Adding the spin degree of freedom, the TBG non-interacting hamiltonian can be build by $4 \times 4$ block matrices, with non-zero diagonal $4 \times 4$ blocks $H_{\xi}$, and zero off-diagonal blocks. The derivation of \textcolor{black}{the non-interacting hamiltonian for the other valley} $H_{-}$ can be done by changing $\vec{k} \rightarrow - \vec{k}$ in the equations of $H_{+}$ (see below). $H_{\xi}$ is written in the lattice sites basis \textcolor{black}{of each layer}, $\left\{ A_1,B_1,A_2,B_2 \right\}$:

\begin{equation}
H_{0} \approx \begin{pmatrix}
H_{+}	& 	0 \\
0	& 	H_{-} \\
\end{pmatrix} \: \rightarrow \:
H_{\xi} \approx \begin{pmatrix}
h^1	& 	h^{\theta \: \dagger} \\
h^{\theta}	& 	h^2 \\
\end{pmatrix} 
\label{eq:1.25}
\end{equation}

where $h^l$ (with $l=1,2$) are the intralayer hamiltonians and $h^\theta$ is the interlayer hamiltonian. These terms can be written as:

\begin{equation}
\begin{aligned}
h^l 	& = -\hbar v_0 \left( \vec{k} - \vec{K}_\xi^l \right) \cdot \left( \xi \sigma_x, \sigma_y \right) \\
h^\theta	& = \begin{pmatrix}
				h_{A_2A_1}	& 	h_{A_2B_1} \\
				h_{B_2A_1}	& 	h_{B_2B_1} \\
			\end{pmatrix} 
\end{aligned}
\label{eq:1.26}
\end{equation}

where $\sigma_x$ and $\sigma_y$ are the $2 \times 2$ Pauli spin matrices. $h_{ij}$ terms depend on the angle-dependent interlayer hopping $w(\theta)$. In fact, it can be proved there are two angle-dependent interlayer hoppings $w$ and $w'$ contributing to the diagonal and off-diagonal elements of $h^\theta$, respectively. $w' \neq w$ accounts for corrugation effects in TBG \cite{KosPRX82018}. If the two graphene layers do not have any interlayer corrugation, then $w' \neq w$. In \fref{fig:1.23}, the band structure \textcolor{black}{close to the Dirac points is shown} for both cases at the first magic angle are shown. Note that for $w'=w$, there is no superlattice gap \textcolor{black}{(the gap between the flat bands and the next bands)} in the \textcolor{black}{valence ($E<\varepsilon_F$)} part, but for $w' \neq w$ \textcolor{black}{both} superlattice gaps are recovered, hence remarking the importance of corrugation effects in MA-TBG. The resulting flat bands show a \textcolor{black}{very flat} dispersion \textcolor{black}{around} $K^l$, and reach extreme values at $\Gamma$. Note that the Fermi level $\varepsilon_F$ is located at the Dirac points $K^l$. In total, there are 4 bands forming the flat bands.

In \cref{Chap4c}, \textcolor{black}{I} will discuss more extensively the current status of MA-TBG experimental and theoretical advances.

\section{Organization of this thesis}
\label{1.7}


In \cref{Chap03}, I will present \textcolor{black}{an introduction to the electronic correlations in} multiorbital systems by considering \textcolor{black}{only} the effects of local correlations. First, I will focus \textcolor{black}{in the case of} equivalent orbitals systems, and I will review the behaviour of $Z_m$ in terms of $U$, $J_H$ and $n_e/N$. A new regime called the Hund metal is found for a certain region of the $(U,J_H,n_e/N)$ space. This metallic state shows sizable electronic correlations ($Z_m$ values are low), the atoms are maximally spin polarized and there is an orbital decoupling behavior. Secondly, I will study the effect of local correlations in real materials, with special focus on FeSCs. The experimental features to identify the local correlation effects in real materials are also presented \textcolor{black}{at} this point.

\textbf{Chapters \ref{Chap4a}}, \textbf{\ref{Chap4b}}, \textbf{\ref{Chap4c}} and \textbf{\ref{Chap4d}} are devoted to present the contributions done during the thesis.

In \cref{Chap4a} \cite{Pizarro1}, I will show the results obtained for the chromium pnictide analogue of the FeSCs, \textcolor{black}{where $Cr$ atoms are considered instead of $Fe$ ones}. The main idea in this work is to present a plausible new unconventional \textcolor{black}{superconducting family} based on chromium instead of iron, where the main difference is the electronic filling of each transition metal atom ($n_e=6$ electrons in $5$ $d$ orbitals in $Fe$, and $n_e=4$ in $5$ $d$ orbitals in $Cr$). \textcolor{black}{We proposed that, for a similar crystal and band structure, and a similar strength of the electronic correlations, $Cr$-based materials could host unconventional superconductivity}. First, \textcolor{black}{I} will check if the behavior of increasing electronic correlations when moving towards half-filling \textcolor{black}{($n_e=5$ electrons in $5$ $d$ orbitals)} in a multiorbital system (see \cref{Chap03}) occurs for a $5$ $d$ orbitals tight-binding model (proposed for \textcolor{black}{FeSCs}) from both $n_e=6$ and $n_e=4$. \textcolor{black}{As an example, we considered $LaCrAsO$}. Then, DFT band structure calculations, and a Slater-Koster tight-binding model are obained for $LaCrAsO$. I will study the effects of local correlations by using SSMF $Z_2$ formulation. The magnetic and superconducting instabilities are studied using the renormalized band structure and by using multiorbital Random Phase Approximation (RPA). \textcolor{black}{We} will conclude by \textcolor{black}{stating} that, due to the presence of similar values for $Z_m$, \textcolor{black}{and AFM tendencies}, a new family of unconventional superconductors might be found in $Cr$ pnictides and chalcogenides.

In \cref{Chap4b} \cite{Pizarro2}, I will study the effect of local correlations in the quasi-one dimensional 123 FeSC, $BaFe_2S_3$. The \textcolor{black}{differences} of this material \textcolor{black}{with respect to other FeSCs} is that \textcolor{black}{the crystal structure is quasi-1D and} the parent compound is an insulator. The superconductivity emerges when applying pressure at the same time that the metallicity is recovered. I use the tight-binding \textcolor{black}{proposed for this material} \cite{AriPRB922015} for zero pressure and the pressure around which superconductivity emerges. I will calculate local correlations by using SSMF $U(1)$ formulation. At zero pressure, the system shows \textcolor{black}{very} strong correlations ($Z_m \sim 0.03-0.06$) for $d_{zx}$- and $d_{yz}$-like orbitals, \textcolor{black}{much stronger than in other quasi-2D FeSCs}. For the pressure at which superconductivity emerges, $Z_m$ values are close to those of other FeSCs. I will show that, contrary to other FeSCs, the Fermi surface suffers a huge reconstruction due to local correlations. This reconstruction will affect the typical weakly correlated approaches, which are based on the Fermi surface topology.

In \cref{Chap4c} \cite{Pizarro3}, I will study the nature of the insulating states found in MA-TBG. I will consider a hexagonal lattice symmetry with two orbitals and nearest neighbor hoppings ($t \sim 2 meV$) tight-binding model, in order to mimic the four bands obtained for the flat bands. A discussion about the current status of experimental and theoretical calculations in MA-TBG is presented. I will check if \textcolor{black}{the experimental observations \cite{CaoN5562018_ins} can be explained in terms of a Mott insulating state including} only local correlations, \textcolor{black}{in particular}: increasing the external magnetic field $B$ and the temperature $T$ promotes metallicity, and the gap found $\sim 0.3 \, meV$ is two orders of magnitude smaller than the bandwidth $\sim 10 \, meV$. \textcolor{black}{For this work, I have implemented the Zeeman term} in the SSMF $U(1)$ formalism and study the effect of $B$ in the $Z_{m\sigma}$. I will conclude by stating that local correlations cannot explain the experimental features. \textcolor{black}{We} will propose to go beyond local correlations, \textcolor{black}{including} non-local correlations to explain the behavior found in this system.

In \cref{Chap4d}, I will present my current calculations in the 2D room-temperature itinerant ferromagnet $Fe_3GeTe_2$ (FGT). I will comment the recent results in bulk and \textcolor{black}{low dimensional} FGT, and the importance of finding a room-temperature ferromagnet free of rare-earth elements. \textcolor{black}{I will perform} DFT calculations \textcolor{black}{for this material}, and I will compare the band structure for bulk and single-layer FGT. I will then derive the Wannier tight-binding model for single-layer FGT, including $Fe$ $d$, $Ge$ $p$ and $Te$ $p$ orbitals.

\textcolor{black}{The final \cref{Chap8} is devoted to the conclusions extracted from this thesis.}

\textcolor{black}{In the Apendices section, \aref{AppA} gives a brief introduction of DFT and Wannier parametrization calculations, \aref{AppB} gets into the physical details and mathematical derivations of SSMF, and \aref{AppC} gives some useful mathematical proofs.}



\chapter{{\bf Local electronic correlations in multiorbital systems: the Hund metal}} 
\label{Chap03}
\lhead{Chapter \ref{Chap03}. \emph{Local correlations in multiorbital systems: Hund metal}} 

\section{Introduction}
\label{3.1}

In this chapter, I summarize the most important results for local correlations in multiorbital systems. I will follow the discussions in \cite{emergence2017,WerPRL1012008,HauNJP112009,LiePRB822010,IshPRB812010,HanPRL1042010,YinNM102011,WerNP82012,LanPRB872013,TerPRB872013,HarPRL1112013,deMPRL1122014,NakSC42014,FanPRB922015,deMedici2015,EilPRL1162016,HarPRB942016,deMedici2017,deM1707.03282,BasCRP172016}. But before starting the discussion for multiorbital systems, I am going to summarize some results from the previous chapter for the effects of local correlations in single-orbital systems, see \sref{1.3}. 

I have already presented the Hubbard hamiltonian in \eref{eq:1.19} for \textcolor{black}{onsite} interactions and single-orbital systems \cite{HubPRSLA2761963,HubPRSLA2771964,HubPRSLA2811964,HubPRSLA2851965}. The local correlations physics which control this hamiltonian will depend on the ratio $U/W$, with $U$ being the onsite Hubbard interaction (related with the double occupancy interaction energy cost) and $W$ the bandwidth (related with the kinetic energy gain of the electrons when they hop through the system). At half-filling (one electron per atom, i.e. $n_e=1$), the system can undergo a metal-to-insulator transition, called the Mott transition, at a critical interaction $U_c$ \textcolor{black}{at which the electrons become localized at the atomic sites}. The $U_c$ value \textcolor{black}{depends only weakly} on the lattice symmetry, and it ranges $U_c \sim 1.5-2 W$ at $T=0 \, K$ \cite{GeoRMP681996}. The insulating state is called the Mott insulator. Out of half-filling ($n_e>1$ or $n_e<1$), the system \textcolor{black}{is metallic} for all $U$.

The quasiparticle weight $Z$ is the order parameter \textcolor{black}{for} the Mott transition \cite{emergence2017}. In the local correlations picture, it can be shown that $1/Z = m^*/m_{0}$, see \eref{eq:1.13}, where $m_0$ is the non-correlated effective mass (obtained from the tight-binding model) and $m^*$ is the correlated effective mass. The quasiparticle weight evolves from $Z=1$ ($m^* = m_0$) in the non-correlated metal to $Z=0$ ($m^* \rightarrow \infty$) in the Mott insulator. When \textcolor{black}{$0.7 < Z \leq 1$ ($1 \leq m^*/m_0 < 1.4$)}, the system is said to be a weakly correlated metal, and descriptions based on a DFT band structure calculations and tight-binding models will give good predictions. At \textcolor{black}{$0 < Z \lesssim 0.3$ ($3 \lesssim m^*/m_0 < \infty$)}, the system is a strongly correlated metal, and DFT calculations and tight-binding models will leave out the most important local correlations effects. In between, a moderately correlated metal exits with \textcolor{black}{$0.3 < Z \lesssim 0.7$ ($1.4 \lesssim m^*/m_0 < 3$)}.

In the Brinkman-Rice picture of the Mott tansition\footnote{During this chapter and the next ones, we will focus in this picture, as already pointed out in \cref{Chap01}.} \cite{Coleman2015,emergence2017}, $Z$ acts as a renormalization pre-factor of the band energy, hence when $Z$ is reduced the bandwidth $W$ will also shrink, see \fref{fig:1.7}\textbf{(b)}. When varying the filling of the system (for example, by doping), this will produce a non-rigid band shift of the Fermi level $\varepsilon_F$, in contrast with what happens in the non-correlated limit, where $\varepsilon_F$ is rigidly shifted when the filling varies.



In single-orbital systems, the Mott insulating state is characterized by a vanishing quasiparticle weight $Z=0$ and a suppression of local charge correlations $C_{n_T}$, see \fref{fig:1.9}. Due to the clear relation between $Z$ and $C_{n_T}$, a low $Z$ value signals the charge localization in each atom, with a low $C_{n_T}$ value. We will see that in multiorbital systems, this equivalence is not straightforward.

In order to study local correlations effects in multiorbital systems, I will discuss the solutions for \eref{eq:1.21}, mainly using slave-spin mean-field (SSMF) formalisms, which are briefly described in \sref{1.4.1}, and more in detail in \aref{AppB}.

In the SSMF framework, $H_{add}$ is neglected, see \eref{eq:1.22}. Up to date, it is not known how to properly include it in SSMF calculations. Also, I will assume: $(i)$ orbital-independent interactions $U$, $U'$, $J_H$ and $J'$ and $(ii)$ rotational invariance relations are invoked $U'=U-2J_H$ and $J'=J_H$ (even if the orbitals are not fully spherical) \cite{CasPRB181978}. $U$ and $J_H$ are positive (see \cref{Chap4c} for negative $J_H$ discussion), and the maximum value $J_H/U=1/3$, which occurrs to maintain the interaction $(U-3J_H)$ (see \eref{eq:1.22}) repulsive. In SSMF, the two coupled hamiltonians of \eref{eq:1.23} have to be solved self-consistenly, as described in \sref{B.5}. Then, for spin-degenerated systems, the orbital-dependent quasiparticle weights $Z_m$ can be obtained in terms of the interaction parameters $U$ and $J_H$, as well as in terms of the number of electrons per atom $n_e$ and the number of orbitals per atom $N$. Equivalently, I will use the total filling per atom, orbital and spin $x=n_e/2N$ defined between $x \in [0,1]$, where $x=0$ is the fully-emptied, $x=1$ is the totally-filled and $x=1/2$ is the half-filled system.

In a multiorbital system, we can define other useful quantities, such as the orbital filling per spin $n_{m\sigma}$ ($n_{m\sigma} \in [0,1]$, with $n_{m\sigma}=0$ being the emptied orbital and $n_{m\sigma}=1$ the totally occupied orbital), and the orbital-resolved bandwidth $W_m$, which is an estimation of the energy window in which each orbital contributes to the band structure, see \fref{fig:1.3}\textbf{(c)}.

\textcolor{black}{The description of SSMF for the Mott transition captures the Brinkman-Rice picture}, see \fref{fig:1.7}\textbf{(b)} \cite{Coleman2015,emergence2017}. The quasiparticle weights evolve from the weakly correlated metal, where all $Z_m \approx 0.8-1$, to the Mott insulating state where all $Z_m = 0$. When adding the orbital degree of freedom, the strongly correlated metal state where all $Z_m \sim 0.1-0.5$ can be also found, as in the case of a single-orbital system (see \sref{1.3}). But then, new additional cases can appear, where weakly (\textcolor{black}{$0.7 < Z_m \leq 1$}), moderately (\textcolor{black}{$0.3 < Z_n \lesssim 0.7$}) and strongly (\textcolor{black}{$0 < Z_p \lesssim 0.3$}) correlated orbitals coexist at the same time. For example, when some $Z_m =0$ but others are $Z_{n \neq m} \neq 0$, the system is in the so called orbital selective Mott phase (OSMP) \cite{deMPRB722005,deMPRL1022009,YuRPRL1102013,YiMPRL1102013,LiuPRB922015}. I will comment later on that iron-based superconductors (FeSCs) can be seen as systems where weakly, moderately and strongly correlated orbitals coexist \cite{emergence2017,deMedici2015,deMedici2017,deM1707.03282,BasCRP172016}.

The band energy is renormalized by these $Z_m$ factors (or equivalently, the orbital mass enhancements $(m^*/m_0)_m$), until the bands dissappear at the Mott transition, see \fref{fig:1.7}\textbf{(b)}. Thus, once the Mott insulator is reached, SSMF cannot provide further information.  As already mentioned, this technique will allow for a \textcolor{black}{faster} treatment of local correlations effects, where \textcolor{black}{hundreds} of points can be obtained without increasing too much the time spent by the calculations. At $T=0 \, K$ and for equivalent orbitals systems, SSMF gives good results when comparing with other more sofisticated techniques, such as Dynamical Mean-Field theory (DMFT) (see \textcolor{black}{comparison in} \fref{fig:3.4}) \cite{FanPRB922015}. In the case of real materials, like FeSCs, SSMF gives also good predictions when comparing with the experimental results \cite{BasCRP172016,deMPRL1122014,LuDARCMP32012,GrePRB842011,LafPRB962017}.

In \sref{3.2}, I will study the most important features of multiorbital systems by exploring the equivalent orbitals systems when varying $U$, $J_H$, $n_e$ and $N$ \cite{emergence2017,FanPRB922015,deM1707.03282}. I will explain in detail the Hund metal phenomenology, as well as its relation with Mott insulator physics. I will briefly review the orbital selective Mott transition (OSMT) for non-equivalent orbitals systems \cite{deMPRB722005,deMPRL1022009}. In \sref{3.3}, I will focus on the theoretical and experimental results obtained in FeSCs \cite{BasCRP172016,deMPRL1122014,LuDARCMP32012,LanPRB872013,GrePRB842011,LafPRB962017}. I will show that FeSCs can be seen as multiorbital systems in which weakly, moderately and strongly correlated orbitals coexist at the same time. In \sref{3.4}, a brief summary of the chapter is given.

\section{Equivalent orbitals systems}
\label{3.2}

Local correlations effects are robust from one lattice to another, and the details of the dispersion relations (like the lattice symmetry or the number of hoppings included) are not very important when studying this phenomenology in a first approximation. The parameters which control the effects of local correlations are the total bandwidth $W$, the interaction energy scales $U$ and $J_H$, the number of electrons per atom $n_e$ and the number of orbitals per atom $N$. Thus, the results obtained for simple toy models, as the ones presented here, can be used to give a first approximation for the phenomenology present in many real materials, such as FeSCs. In this section, I will review the results on equivalent orbitals systems (see next paragraph) and the local correlations phenomenology when varying $U$, $J_H$, $n_e$ and $N$.

In a system of $N$ equivalent orbitals, these orbitals have the same dispersion relation and zero crystal field splittings. Except otherwise indicated, I will consider a 2D square lattice symmetry, in which only intraorbital nearest neighbor hoppings $t$ are included, then only intraorbital dispersion relations will be non-zero $\varepsilon_{km} = -2t(\cos{k_x}+\cos{k_y})$. All the results are presented in units of $t$ or $W=8t$ (or equivalently, half-bandwidth $D=W/2$). In some cases (as in \fref{fig:3.4}), Bethe lattice is considered, which is an infinite-dimensional lattice with $\varepsilon_{km} = -2t\sum_{d=1}^{\infty} \cos{k_d}$ and a semicircular DOS ($W=4t$). For equivalent orbitals, all the $Z_m$ are reduced to the same $Z$ value. I will restrict the discussion to $T=0 \, K$.

For a multiorbital system, the kinetic energy gain $\widetilde{W}$ when the electrons hop through the system is not simply the total bandwidth $W$, as in the single-orbital case. Due to the $N$-orbital degeneracy, there will be $N$ degenerated bands. Thus, the electrons have $N$ degenerated channels to hop through the system. Then, the kinetic energy gain $\widetilde{W}$ will increase with $N$ and $W$ \cite{GunPRB541996}. We will see later on that, a finite $J_H$ will promote the breaking of the ground state degeneracy, and hence an effective reduction of $\widetilde{W}$ to $\sim W$. 

In terms of the total filling $x=n_e/2N$, the interaction energy cost of a hopping process is defined as the atomic gap $\Delta^x$, see \eref{eq:3.1}.

\begin{equation}
\Delta^x = E(n_e+1) - E(n_e-1) - 2E(n_e)
\label{eq:3.1}
\end{equation}

where $E(n_e)$ is the energy of an atom for a certain number of electrons per atom $n_e$. $\Delta^x$ depends on $U$, $J_H$, $N$ and integer $n_e$ values. The atomic gap is a stability energy, i.e. it states how stable an electronic configuration of electrons in a atom is in the large-$U$ limit ($t \approx 0$).

For $J_H=0$, $\Delta^x = U$ for any $N$ and any integer filling $n_e$. Then, the Mott insulating state does not only appear for half-filling ($n_e=N$), but also for other integer fillings $n_e=1,2,\ldots N$. 

The critical interaction for the Mott transition $U_c^x$ at a certain filling $x$ will depend on the competition between this interaction energy cost $\Delta^x$ and the kinetic energy gain $\widetilde{W}$. Thus, for multiorbital systems, the local electronic correlations effects are controlled by the ratio $\Delta^x/\widetilde{W}$, which reduces to $U/W$ in the single-orbital situation. When $\Delta^x$ increases, $U_c^x$ decreases (due to the enhancement of the ratio $\Delta^x/\widetilde{W}$). $U_c^x$ will increase if $N$ increases when including the effect of $\widetilde{W}$. \textcolor{black}{In \cite{FloPRB662002}, the authors demonstrated that, at half-filling, $U_c$ increases linearly with $N$}:

\begin{equation}
U_c = 8 |\sum_k \varepsilon_k <f_k^\dagger f_k^{}>| N
\label{eq:3.2}
\end{equation}

\begin{figure}[h]
   \centering
   \includegraphics[width=0.8\columnwidth]{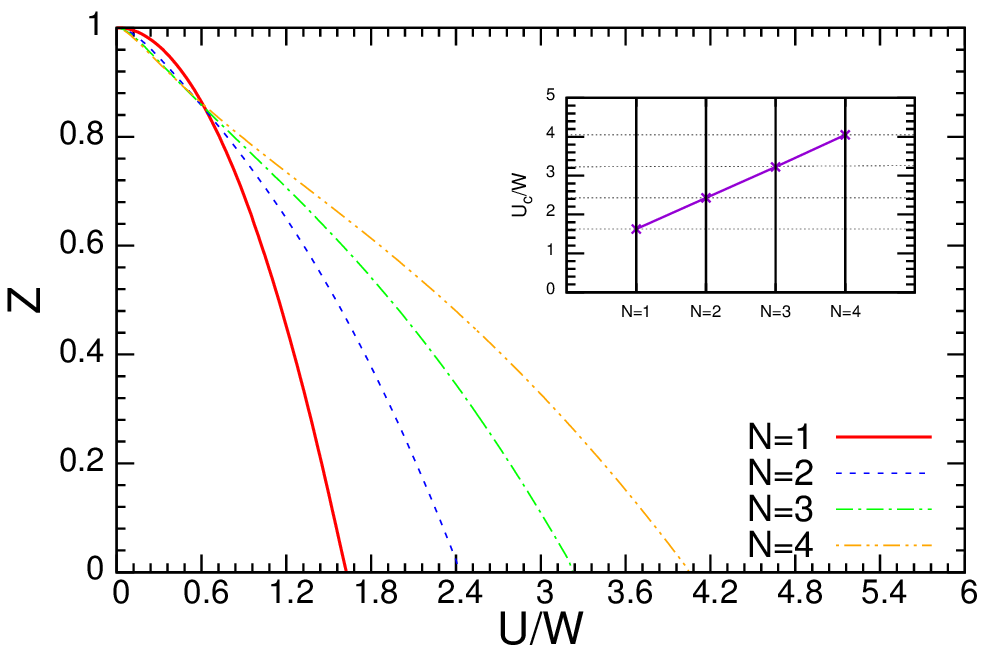}
   \caption[SSMF calculations of Z versus $U$ for $J_H=0$ and different $N$, inset: $U_c$ versus $N$]{SSMF calculations of $Z$ versus $U/W$ for different systems with $N$ equivalent orbitals at half-filling and $J_H=0$. \textsc{Inset}: $U_c/W$ for different $N$ values (extracted from $Z$ vs. $U/W$ curves). $U_c$ increases linearly with $N$, \textcolor{black}{as expected from} \eref{eq:3.2}. Adapted from \cite{deMPRB722005}.}
   \label{fig:3.2}  
\end{figure}

\textcolor{black}{In \fref{fig:3.2}, I show the SSMF calculations for $Z$ in terms of the ratio $U/W$ for the square lattice and various $N$-equivalent orbitals cases. The $U_c$ versus $N$ behavior of \eref{eq:3.2} can be seen clearly in the \textsc{Inset} of the figure.}

\newpage

\subsection{The effect of the Hund's coupling}
\label{3.2.1}

Now, I will consider $J_H \neq 0$ and study its effects on $\Delta^x(J_H)$. In \eref{eq:3.3}, the interaction energy cost $\Delta^x(J_H>0)$ is estimated, where Mott insulating states can appear, not only for half-filled systems, but also for any integer filling $n_e=1, 2, \ldots, N-1$, same as in the $J_H=0$ situation. 

\begin{equation}
\Delta^x \, (J_H > 0) = \left\{
\begin{aligned}
& = U + (N - 1) J_H \quad & n_e=N \\
& = U - 3 J_H \quad & other \, integer \, n_e
\end{aligned}
\right.
\label{eq:3.3}
\end{equation}

\begin{figure}[h]
   \centering
   \includegraphics[width=0.8\columnwidth]{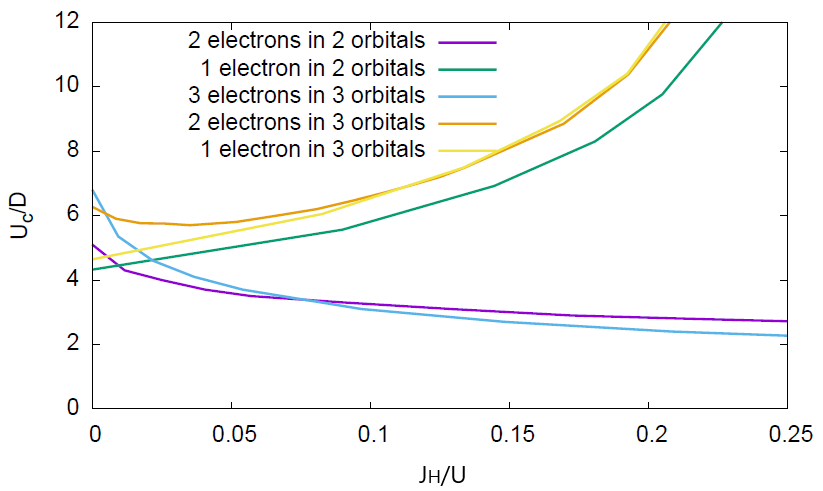}
   \caption[SSMF calculations of $U_c/D$ versus $J_H/U$ for $N=2$ and $N=3$ at different integer fillings]{SSMF calculations of $U_c/D$ (where $D=W/2$ is the half-bandwidth) in terms of $J_H/U$ for various integer fillings $n_e$ and $N=2,3$. At half-filling (blue and purple lines in the plot), $U_c$ becomes smaller when $J_H$ increases, following \eref{eq:3.3}. At $n_e=1$ (green and yellow lines), $U_c$ trend in terms of $J_H$ also follows \eref{eq:3.3}, where $U_c$ increases when $J_H$ increases. See main text for the discussion of the case $N=3$ and $n_e=2$ (orange line). Taken from \cite{deM1707.03282}.}
   \label{fig:3.3}  
\end{figure}

$J_H$ favors the Mott insulator for half-filled systems, whether disfavors it for other integer filling systems. In \fref{fig:3.3}, $U_c$ in terms of $J_H$ is presented, for various integer fillings $n_e$ and $N=2$ and $N=3$ equivalent orbitals systems. For half-filling ($n_e=N$) and one-electron systems ($n_e=1$), the behavior of $U_c$ in terms of $J_H$ is well captured by \eref{eq:3.3}, \textcolor{black}{except for small $J_H$ (see previous section)}. For $n_e=1$, $U_c$ increases if $J_H$ also increases (Mott insulator is disfavored by $J_H$), and for $n_e=N$, $U_c$ decreases if $J_H$ increases (Mott insulator is favored by $J_H$).

\begin{figure}[h]
   \centering
   \includegraphics[width=0.9\columnwidth]{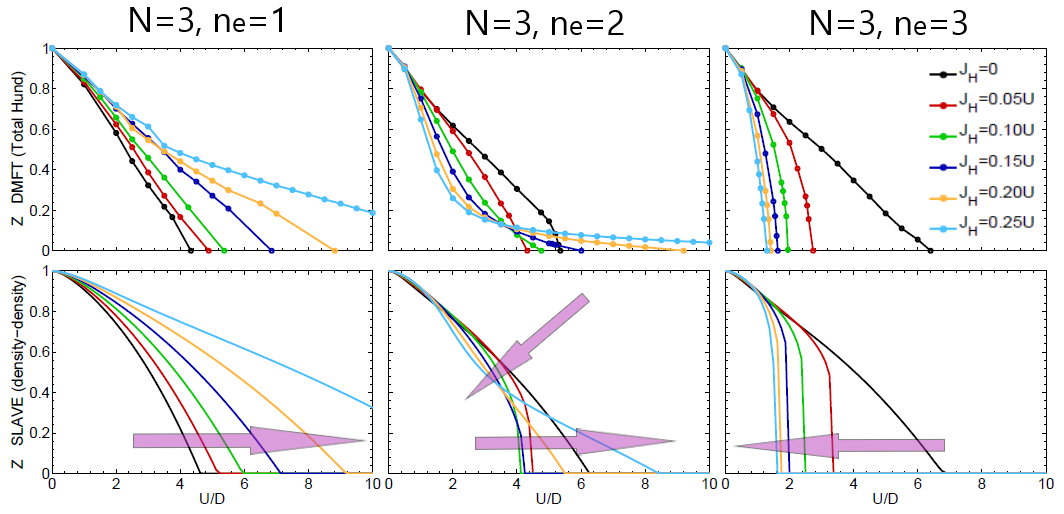}
   \caption[$Z$ versus $U/D$ for different integer fillings and $J_H/U$ in the $N=3$ case]{$Z$ in terms of $U/D$ for different integer fillings and different $J_H/U$ values, in the Bethe lattice for $N=3$. DMFT (upper row) and SSMF (lower row) calculations are presented. DMFT calculations are done using the full hamiltonian of \eref{eq:1.22}, with the pair-hopping and spin-flip terms included, while in SSMF those terms are neglected. Each column stands for different integer fillings $n_e=1,2,3$. Big violet arrows are marked in order to follow the increasing $J_H/U$, where for $n_e=1$ when $J_H$ increases, $U_c$ increases and for $n_e=N=3$ when $J_H$ increases, $U_c$ decreases. For $n_e=2$, there is a non-monotonic behavior, where for increasing $J_H$, first $U_c$ decreases, and then it starts to increase. Taken from and adapted from \cite{FanPRB922015}.}
   \label{fig:3.4}  
\end{figure}

However, for $N=3$ in $n_e=2$ (orange line in \fref{fig:3.3}), the behavior of the system is controled by $\Delta^x (J_H>0)$ only at large $J_H$, where Hund's coupling disfavors Mott insulating behavior. As shown in \fref{fig:3.4} (for DMFT and SSMF calculations\footnote{As anticipated, a very good agreement between DMFT and SSMF can be seen, specially for $n_e=1$ and $n_e=3$ situations, with $Z$ being slightly overstimated in SSMF due to its Brinkman-Rice character (larger $Z$ is translated into an understimation of $U$ effects). At $n_e=2$, the differences are more evident, where $Z$ suppression is weaker in SSMF than in DMFT. In any case, the behavior is well captured by SSMF when compared with DMFT} in a Bethe lattice), for small $J_H$, Hund's coupling favors the Mott insulator. This non-monotonic behavior is general to all the other cases for $1<n_e<N$ and $N>2$.

\begin{figure}[h]
   \centering
   \includegraphics[width=0.9\columnwidth]{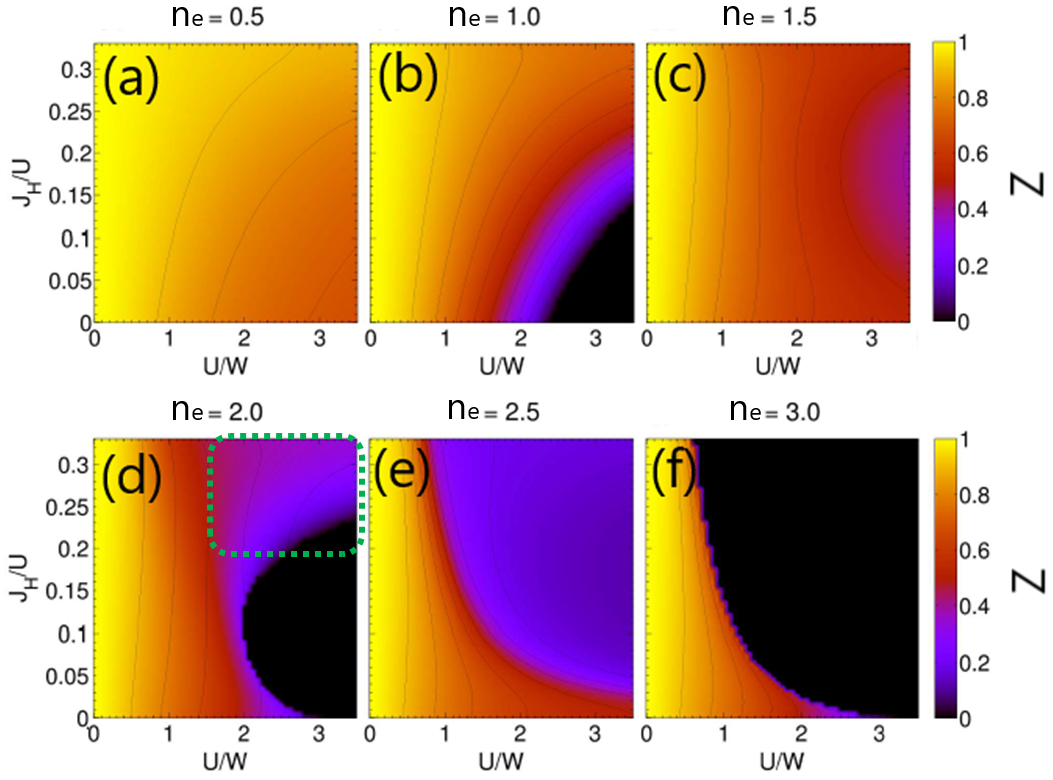}
   \caption[SSMF calculations of $Z$ color map for $J_H/U$ versus $U/W$ phase diagram for $N=3$ at different fillings $n_e$]{SSMF calculations of $Z$ color map for the $J_H/U$ versus $U/W$ phase diagram, for $N=3$ at different fillings $n_e$. Integer and non-integer fillings are included, in order to show that the appearance of the Mott insulator (black region) is restricted to integer fillings. Colored areas follow the next scheme: yellow region for the weakly correlated metal ($Z \sim 0.8-1$), purple region for the strongly correlated metal ($Z \sim 0.1-0.5$) and black region for the Mott insulator ($Z=0$). Dark red regions describe the moderately correlated metal ($Z\sim 0.6-0.7$). Purple region surrounded by a green dashed square for \textbf{(d)} $n_e=2$ can be identified as the Hund metallic state. Other purple regions are not necessarily describing a Hund metal, see \sref{3.2.2}. This correlated metal state follow the same dependence as the Mott insulator with $U$ and $J_H$. Taken and adapted from \cite{FanPRB922015}.}
   \label{fig:3.5}  
\end{figure}

In \fref{fig:3.5}, the $Z$ color map for $J_H/U$ versus $U/W$ is presented for $N=3$. Here, the non-monotonic behavior can be more easily identified in \fref{fig:3.5}\textbf{(d)} for $n_e=2$ (i.e. the purple region surrounded by a green dashed square). It can be seen that this correlated metallic state (with low $Z$ values) follows the same dependence on the interaction energy scales $U$ and $J_H$ as the Mott insulator. In \sref{3.2.3}, I will show that this correlated metallic state and the Mott insulator follows the same dependence with the total filling $n_e$. This correlated metal is what is known as the \textbf{Hund metal} \cite{YinNM102011}. The Hund metal can also appear for other situations between single-electron and half-filling systems ($1<n_e<N$) \textcolor{black}{with} $N>2$. Note that other purple regions are not necessarily describing the Hund metal regime, due to the additional properties that I will discuss in the next section.

\begin{figure}[h]
   \centering
   \includegraphics[width=0.9\columnwidth]{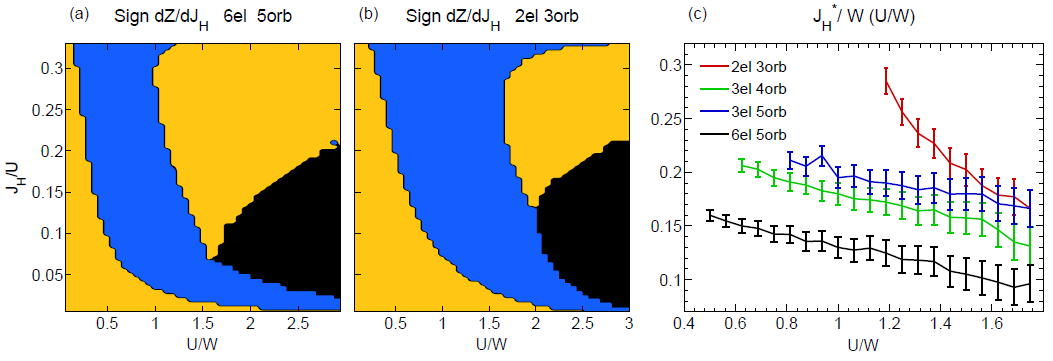}
   \caption[SSMF calculations of $dZ/d{J_{H}}$ maps and $J_H^{*}/W$ curves for different $n_e/N$ cases, marking the crossover regions between uncorrelated and Hund metals]{SSMF calculations of $dZ/d{J_{H}}$ maps for \textbf{(a)} $N=5$, $n_e=6$ and \textbf{(b)} $N=3$, $n_e=2$. Blue area marks $dZ/d{J_{H}}<0$, yellow areas mark the metallic states with $dZ/d{J_{H}} \geq 0$ and black area marks the Mott insulating phase. Note that the blue region below the Mott insulating phase is just a signal of a decreasing $Z$ value, and does not signal the crosssover to the Hund metal regime. \textbf{(c)} $J_H^{*}/W$ curves for different $n_e/N$ cases. $J_H^{*}/W$ is smaller if $x$ is smaller, except when comparing $n_e/N=2/3$ and $n_e/N=3/5$. Taken from \cite{FanPRB922015}.}
   \label{fig:3.6}  
\end{figure}

In \fref{fig:3.6}, $dZ/d{J_{H}}$ values for the plane $J_H/U$ versus $U/W$ and \textbf{(a)} $N=5$, $n_e=6$ and \textbf{(b)} $N=3$, $n_e=2$ are shown. \textcolor{black}{Depending on the value of the parameters $(U,J_H,n_e)$, $Z$ can increase or decrease with $J_H$. In \fref{fig:3.6}\textbf{(a)} and \textbf{(b)}, this is signaled by the positive (yellow regions) or negative (blue region) sign of the slope $dZ/dJ_H$}.

Between the weakly correlated metal and the Hund metal, there is a crossover. Its width depends on the lattice, $U$, $J_H$ and $n_e/N$. \textcolor{black}{In \cite{FanPRB922015}, the authors identified the crossover by a characteristic $J_H^*$ value}. This crossover between the weakly correlated metal and the Hund metal was located around the most prominent suppression of $Z$ with $J_H$ ($dZ/d{J_{H}}<0$). In \fref{fig:3.6}\textbf{(c)}, several $J_H^{*}/W$ are presented for different $n_e/N$ situations. $J_H^{*}/U$ decreases if $x$ decreases, except when comparing the cases $n_e/N=2/3$ ($x \sim 0.33$) and $n_e/N=3/5$ ($x=0.3$).

\newpage

\subsection{Local spin and charge correlations}
\label{3.2.2}

In the previous section, I have pointed out that there is a region of the $(U, \, J_H)$ space in which a specific correlated metallic state, the Hund metal, can be found \cite{emergence2017,WerPRL1012008,HauNJP112009,LiePRB822010,IshPRB812010,HanPRL1042010,YinNM102011,WerNP82012,LanPRB872013,deMPRL1122014,FanPRB922015,deMedici2015,deMedici2017,deM1707.03282,BasCRP172016}. In this section, I will review the origin and properties of the Hund metal by following the discussion in \cite{FanPRB922015}.

The local spin $C_S$ correlations can be defined as:

\begin{equation}
C_S = <S^2> - {<S>}^2
\label{eq:3.4}
\end{equation}

where $S=\sum_{m} (\widehat{n}_{m\uparrow} - \widehat{n}_{m\downarrow})$, and $\widehat{n}_{m\sigma}=d_{m\sigma}^\dagger d_{m\sigma}$ are the orbital and spin occupation number operators. For a spin-degenerate system, $<S>=0$ and $<\widehat{n}_{m\uparrow}>=<\widehat{n}_{m\downarrow}> \equiv <\widehat{n}_{m}>$, and for equivalent orbitals $<\widehat{n}_{m}>=n_e/2N=x$ for all $m$.

\begin{figure}[h]
   \centering
   \includegraphics[width=0.9\columnwidth]{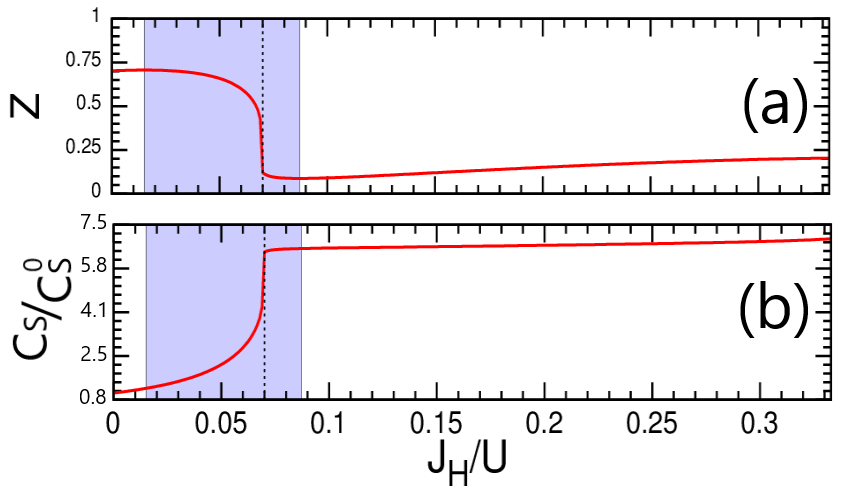}
   \caption[SSMF calculations of $Z$ and local spin correlations in terms of $J_H/U$ for $N=5$ and $n_e=6$ at $U/W=1.5$]{SSMF calculations of \textbf{(a)} $Z$ and \textbf{(b)} local spin correlations \textcolor{black}{$C_S/C_S^0$} in terms of $J_H/U$ for $N=5$ and $n_e=6$ at $U/W=1.5$. Blue shaded area marks the negative slope $dZ/dJ_H<0$ region, and dashed line marks the crossover $J_H^*/U=0.07$. At $J_H^*$, $Z$ drops to a finite value and \textcolor{black}{$C_S/C_S^0$} get enhanced. Once the system enters in the Hund metal regime ($J_H > J_H^*$), $Z$ \textcolor{black}{increases slightly} and \textcolor{black}{$C_S/C_S^0$} saturates \textcolor{black}{to the local moment in the atom ($\approx 6-7$)}. There is a clear relation between \textcolor{black}{the drop of} $Z$ and \textcolor{black}{the enhancement of $C_S/C_S^0$ at the Hund metal crossover}.}
   \label{fig:3.7}  
\end{figure}

$C_S$ refers to the local interorbital spin correlations. \textcolor{black}{The ratio $C_S/C_S^0$ (where $C_S^0$ is the non-interacting value)} describes the fluctuations around the spin configuration in each atom, \textcolor{black}{hence it is directly related with the local moment in each atom}. Then, if \textcolor{black}{$C_S/C_S^0$} is in a low value, the spins are aligned anti-parallel in each orbital, while if it is in a large value, the spins are aligned parallel in each orbital. In \fref{fig:3.7}, I have plotted both $Z$ and \textcolor{black}{$C_S/C_S^0$} when varying $J_H/U$. The figure is obtained for $N=5$ orbitals at $n_e=6$ and $U/W=1.5$. The $U/W$ value is selected in order to see the crossover $J_H^*$ (dashed line) and the Hund metal regime $J_H>J_H^*$. The blue shaded region marks $dZ/dJ_H<0$, same as in \fref{fig:3.6}.

At $J_H^*$, $Z$ suffers a large drop, while $C_S$ get enhanced. Once the system is in the Hund metal regime $J_H>J_H^*$, $Z$ has an almost constant value while $C_S$ saturates to a high value. Deep in the Hund metal $J_H \gg J_H^*$, $Z$ increases slighly. The saturation of $C_S$ indicates that the atoms are in a local high-spin configuration (the spins in different orbitals will be parallel to each other). \textcolor{black}{This is nothing but the Hund's rules being satisfied locally in each atom. Like the spins have to be parallel, correlations are induced between electrons, so that $Z$ decays}.

\begin{figure}[h]
   \centering
   \includegraphics[width=0.9\columnwidth]{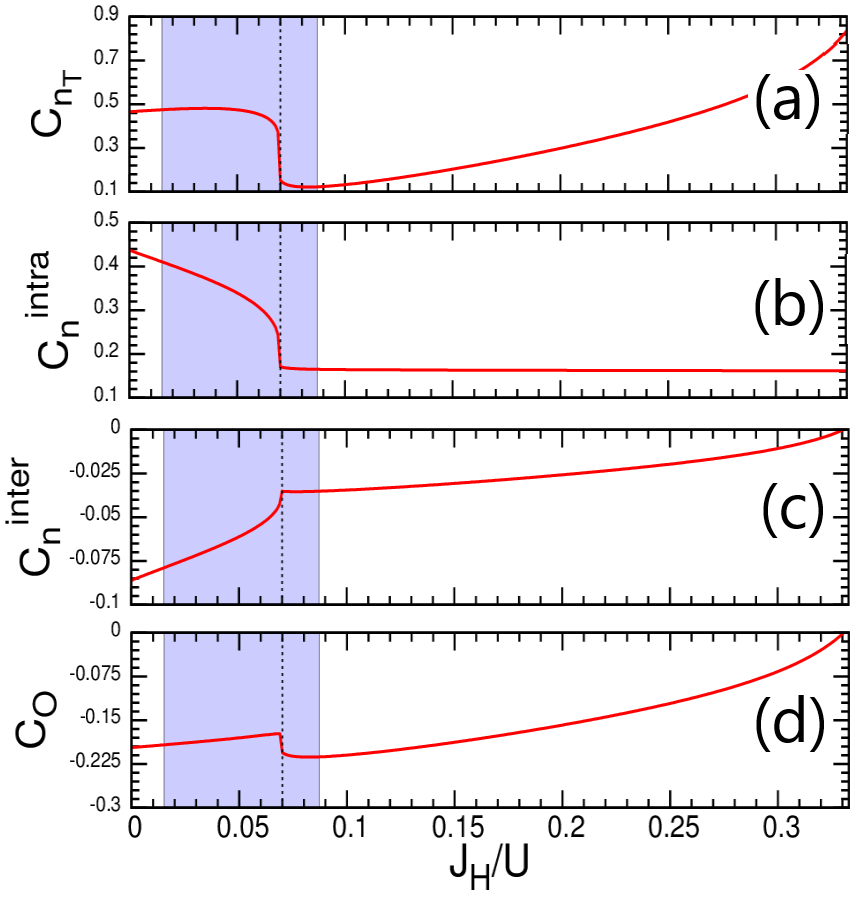}
   \caption[SSMF calculations of local charge correlations, and orbital coupling in terms of $J_H/U$ for $N=5$ and $n_e=6$ at $U/W=1.5$]{SSMF calculations of \textbf{(a)} local total charge correlations $C_{n_T}$, \textbf{(b)} intraorbital charge correlations $C_{n}^{intra}$, \textbf{(c)} interorbital charge correlations $C_{n}^{inter}$ and \textbf{(d)} orbital coupling $C_O$ in terms of $J_H/U$ for $N=5$ and $n_e=6$ at $U/W=1.5$. Blue shaded area marks the negative slope $dZ/dJ_H<0$ region, and dashed line marks the crossover $J_H^*/U=0.07$. At $J_H^*$, $C_{n_T}$ suffers a drop due to the effect of $C_n^{intra}$ (whether the contribution from $C_n^{inter}$ is not important in this point). Once the system enters in the Hund metal regime ($J_H > J_H^*$), $C_{n_T}$ increases due to the contribution from $C_n^{inter}$. In this regime, $C_n^{intra}$ saturates and $C_n^{inter}$ gets strongly suppressed. $C_O$ approaching $0$ signals the orbital decoupling.}
   \label{fig:3.8}  
\end{figure}

\textcolor{black}{Another aspect of the Hund metal is related with the} local total charge correlations $C_{n_T}$, \textcolor{black}{which} can be defined as:

\begin{equation}
C_{n_T} = <n_T^2> - {<n_T>}^2
\label{eq:3.5}
\end{equation}

where $n_T=\sum_{m} (\widehat{n}_{m\uparrow} + \widehat{n}_{m\downarrow})$. $C_{n_T}$ describes how delocalized is the charge, then for a low value, the charge is localized, while for a large value, the charge becomes more delocalized through the system. For a spin- and orbital-degenerated systems, $C_{n_T}$ can be alternatively written as:

\begin{equation}
C_{n_T} = N ( C_n^{intra} + (N-1) C_n^{inter} )
\label{eq:3.6}
\end{equation}

where $C_n^{intra}=<\widehat{n}_m^2> - {<\widehat{n}_m>}^2$ is the intraorbital charge correlations (fluctuations around the average charge in a given orbital) and $C_n^{inter}=<\widehat{n}_m \widehat{n}_n> - <\widehat{n}_m><\widehat{n}_n>$ is the interorbital charge correlations (fluctuations around the non-correlated value for the charge between \textcolor{black}{two different} orbitals, \textcolor{black}{$m$ and $n$}). $C_n^{intra}$ is related with the double occupancy in a given orbital, while $C_n^{inter}$ states how correlated are two different orbitals. $C_{n}^{intra}$ is zero or positive (in the non-correlated limit has a finite value), while $C_n^{inter}$ is zero or negative (in the non-correlated limit goes to zero). Thus, both of them contribute with different sign to $C_{n_T}$. In the Mott insulator, $C_{n_T}$ vanishes, because $C_n^{intra}$ and $C_n^{inter}$ cancel each other. We can also define the orbital coupling $C_O=C_n^{inter}/C_n^{intra}$ as the correlation of an electron when it jumps between different orbitals. In the case of $C_O=0$, the orbitals are decoupled of each other. In \fref{fig:3.8}, I show these correlations in terms of $J_H/U$ for the same system as in \fref{fig:3.7}.

\begin{figure}[h]
   \centering
   \includegraphics[width=0.9\columnwidth]{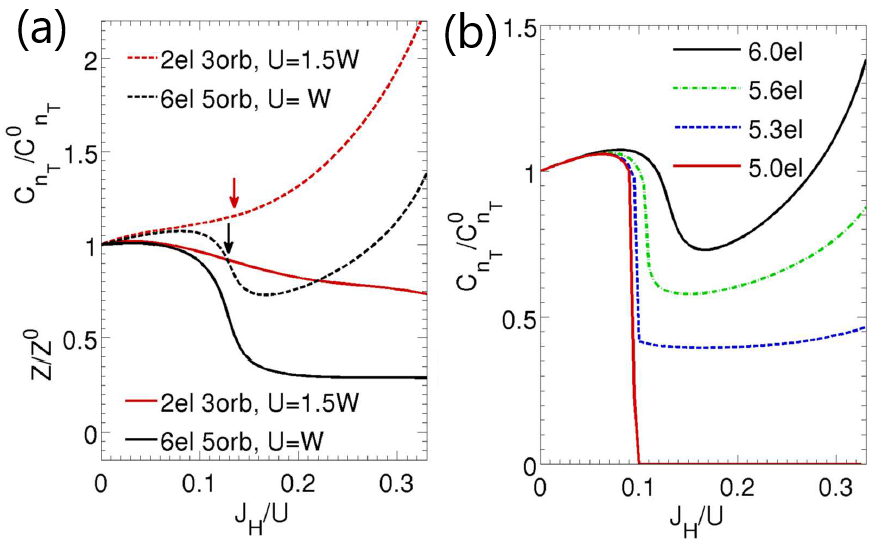}
   \caption[$C_{n_T}/C_{n_T}^0$ versus $J_H/U$ for $N=5$ and $N=3$ and different $n_e$ and $U/W$ values]{\textbf{(a)} $C_{n_T}/C_{n_T}^0$ (dotted lines) and $Z/Z^0$ (continuos lines) versus $J_H/U$ comparison between $n_e/N=2/3$ (red lines) and $5/6$ (black lines) cases. Note that for $n_e/N=2/3$, there is no suppression \textcolor{black}{at the entrance of the Hund metal regime} (arrows mark $J_H^*/U$). This is due to the fact that $C_n^{inter}$ dominates over the whole interaction parameters. For $n_e/N=5/6$ there is a suppression, due to the competition between $C_n^{intra}$ and $C_n^{inter}$. A clear independent behavior can be seen between $Z$ and $C_{n_T}$. \textbf{(b)} $C_{n_T}/C_{n_T}^0$ versus $J_H/U$ for $N=5$ and various $n_e$ values. When the system approaches half-filling, the suppression of the local charge correlations becomes more prominent, until it wins when the Mott insulator is realised at half-filling. Taken from \cite{FanPRB922015}.}
   \label{fig:3.9}  
\end{figure}

At $J_H^*$, $C_{n_T}$ gets suppressed, and then it is enhanced in the Hund metal regime ($J_H>J_H^*$). This enhancement signals an increase of metallicity in the system. However, this is done in spite of a low $Z$ value (see \fref{fig:3.7}\textbf{(a)}). This situation is totally different from single-orbital systems, in which the low $Z$ value implies charge localization in each atom (i.e. low $C_{n_T}$ value). Thus, for systems with sizable electronic correlations and Hund metal phenomenology, the system shows an increased metallic behavior. The \textcolor{black}{different} behavior between $Z$ and $C_{n_T}$ can be seen in \fref{fig:3.9}\textbf{(a)} for $n_e/N=2/3$ and $6/5$, and $U/W=1.5$ and $1$, respectively. Then, a low $Z$ value does not \textcolor{black}{necessarily} imply charge localization for a multiorbital system.

$C_{n}^{intra}$ and $C_{n}^{inter}$ compete to give the behavior of $C_{n_T}$, with both being suppressed with $J_H$. $C_{n}^{intra}$ mainly contributes to the initial drop in $C_{n_T}$ at $J_H^*$. $C_{n}^{inter}$ governs the enhancement of $C_{n_T}$ in the Hund metal regime ($J_H>J_H^*$).

The orbital coupling $C_O$ follows the opposite tendency of $C_{n_T}$ with $J_H$. Thus, at $J_H^*$ there is a small enhancement of the orbital coupling, and in the Hund metal regime ($J_H>J_H^*$), $C_O$ tends to $0$, signaling an orbital decoupled behavior. This orbital decoupling is driven by the suppression of $C_n^{inter}$ in the Hund metal regime.

For $N=3$ and $n_e=2$, total charge correlations $C_{n_T}$ does not show the initial drop, as appeared in other cases, as shown in \fref{fig:3.9}\textbf{(a)} for $U/W=1.5$ (red lines). This indicates that in such situations, $C_n^{inter}$ dominates the behavior of the charge correlations for all $J_H$, while $C_n^{intra}$ \textcolor{black}{gives a smaller} contribution \textcolor{black}{to the $J_H$-dependence}. \textcolor{black}{I} will discuss this situation in the next paragraphs.

The enhancement of total charge correlations is suppressed when the system approaches to half-filling. \textcolor{black}{In \fref{fig:3.9}\textbf{(b)}, the evolution of $C_{n_T}/C_{n_T}^0$ (where $C_{n_T}^0$ is the $U=J_H=0$ value) in terms of the filling can be seen, for $U/W=1$ and when approaching $n_e=N=5$}.

Now, I will discuss the fact that the local spin polarization given by $C_S$ (\fref{fig:3.7}\textbf{(b)}) is responsible of the Hund metal phenomenology, hence driving the drop in $Z$ and the orbital decoupling behavior. I will follow the energetic arguments discussed in \cite{FanPRB922015}. 

\begin{figure}[h]
   \centering
   \includegraphics[width=1.0\columnwidth]{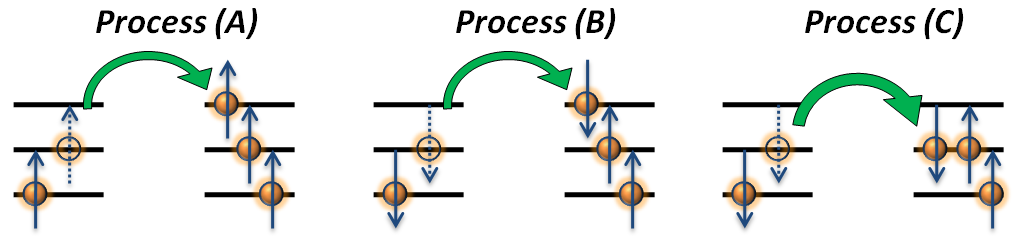}
   \caption[Virtual hopping processes for a multiorbital system with finite Hund's coupling $J_H$ close to the Hund metal regime]{Virtual hopping processes for a $N=3$ system with finite Hund's coupling $J_H$ around the Hund metal regime. \textbf{(A)} both sites have all the orbitals with parallel spin alignement, and one electron jumps to an empty orbital, \textbf{(B)} both sites have all the orbitals with anti-parallel spin alignement, and one electron jumps to an empty orbital, and \textbf{(C)} both sites have all the orbitals with anti-parallel spin alignement, and one electron jumps to an occupied orbital.}
   \label{fig:3.10}  
\end{figure}

Let's consider two sites with $N$ orbitals and $n_e \leq N$ (particle-hole symmetry is assumed, so similar arguments can be applied to $n_e>N$) and the large-$U$ limit. In the correlated metallic state, each atom is spin polarized due to $J_H$, so the possible virtual hopping processes are: $(A)$ both \textcolor{black}{atoms have the same spin alignement}, and one electron jumps to an empty orbital ($\Delta_{\uparrow \uparrow}$), $(B)$ both \textcolor{black}{atoms have the opposite spin alignement}, and one electron jumps to an empty orbital ($\Delta_{\uparrow \downarrow}^{inter}$) and $(C)$ both \textcolor{black}{atoms have the opposite spin alignement}, and one electron jumps to an occupied orbital ($\Delta_{\uparrow \downarrow}^{intra}$). $\Delta$ gives the interaction cost for these virtual hopping processes. These virtual hopping processes are depicted in \fref{fig:3.10}. 

\begin{equation}
\left\{
\begin{aligned}
Process \, (A) & \, \rightarrow \, \Delta_{\uparrow \uparrow} = U - 3 J_H \\
Process \, (B) & \, \rightarrow \, \Delta_{\uparrow \downarrow}^{inter} = U - (n_e - 3) J_H \\
Process \, (C) & \, \rightarrow \, \Delta_{\uparrow \downarrow}^{intra} = U - (n_e - 1) J_H 
\end{aligned}
\right.
\label{eq:3.6}
\end{equation}

At half-filling ($n_e=N$) \textcolor{black}{there is one electron per orbital, so} processes $(A)$ and $(B)$ are blocked, \textcolor{black}{hence} $(C)$ controls $U_c$ behavior ($J_H$ promotes Mott insulating behavior). When $n_e < N$ \textcolor{black}{(for $N<7$)} and $J_H$ is large, process $(A)$ \textcolor{black}{has the lowest energy, so it} controls the Mott transition, \textcolor{black}{hence} $J_H$ disfavors Mott insulating behavior (see \fref{fig:3.3} for large $J_H/U$ values at $n_e \neq N$). \textcolor{black}{Process $(A)$ is promoted by $J_H$, and enhances metallicity}.

At $J_H^*$, the entrance in the Hund metal regime emerges to avoid the process $(C)$. Process $(C)$ is suppressed by $J_H$ for $n_e>1$. As shown in \eref{eq:3.3}, suppression of process $(C)$ is related with the Mott insulating phase at half-filling ($n_e=N$), and this suppression strongly reduces $C_n^{intra}$. Process $(B)$ is suppressed by $J_H$ for $n_e>3$ and promoted for $n_e<3$. Then, systems with $n_e=2$ will behave differently than $n_e>2$ ones (note the absence of the drop for $C_{n_T}$ at $J_H^*/U$ in \fref{fig:3.9}\textbf{(a)}). In this $n_e=2$ situation, $C_n^{inter}$ dominates the behavior of $C_{n_T}$ for all $J_H/U$ (and not only for large $J_H>J_H^*$ values).

At $J_H^*$, the suppression of process $(C)$ (which is accompanied by a large reduction of $C_n^{intra}$) indirectly suppress the process $(B)$ (reducing $C_n^{inter}$). In the Hubbard-Kanamori hamiltonian of \eref{eq:1.22}, local anti-parallel or parallel configurations in different orbitals of the same atom will have an effective interaction cost $U-2J_H$ and $U-3J_H$, respectively. Then, at finite $J_H$, anti-parallel configurations in each atom will cost more energy, so they become less frequent, enhancing $C_S$, and then reducing $C_n^{inter}$. 

This mechanism is behind the orbital decoupling promoted by Hund's coupling (due to the reduction of $C_n^{inter}$ when $(B)$ is suppressed). If $J_H$ further increases ($J_H>J_H^*$), the orbital decoupling is enhanced (the orbital coupling $C_O$ is reduced). At the maximum value $J_H/U=1/3$, $C_n^{inter} = 0$, \textcolor{black}{hence} the orbitals are totally decoupled ($C_O = 0$).

Then, at $J_H^*$, the reduction of $C_n^{intra}$, and the indirect reduction of $C_n^{inter}$, comes from an effective reduction of local anti-parallel configurations. \textcolor{black}{So} the enhancement of $C_S$ drives the low $Z$ values and the orbital decoupling mechanism in the Hund metal regime.

\textcolor{black}{In summary, in the Hund metal, the atoms are in a high spin state due to the Hund's coupling $J_H$, and hopping processes $(C)$ involving intraorbital double occupancy are suppressed. Such suppression reduces $Z$. However, because the hopping processes with parallel spin to an empty orbital are promoted by $J_H$, the system is metallic.}




\subsection{Relation between Mott insulators and Hund metals}
\label{3.2.3}

In the previous section, I have discussed the fact that the enhancement of the local spin polarization in the Hund metal, reduces the intraorbital double occupancy ($C_n^{intra}$), and indirectly reduces $C_n^{inter}$. The suppression of the intraorbital double occupancy is directly related with the Mott transition at half-filling.

\begin{figure}[h]
   \centering
   \includegraphics[width=0.9\columnwidth]{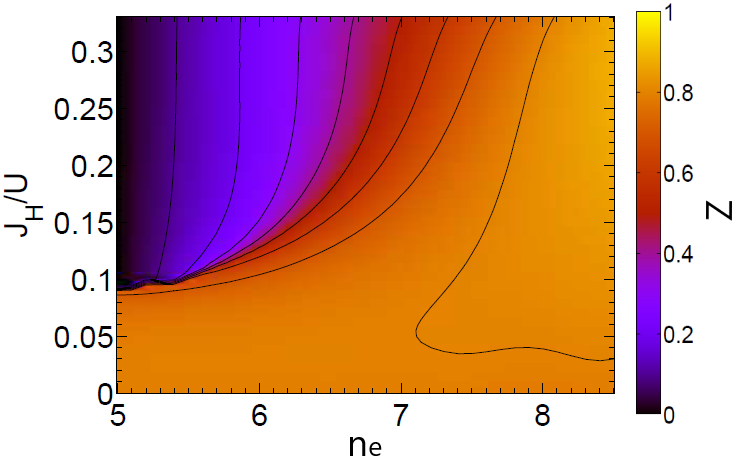}
   \caption[$Z$ color map for the phase diagram $J_H/U$ versus filling $n_e$ for $N=5$ at $U/W=1$]{$Z$ color map for the phase diagram $J_H/U$ versus number of electrons per atom $n_e$ for $N=5$ at $U/W=1$. A clear relation can be seen between Mott insulator at half-filling and the correlated metallic regions. This can be more easily identified by following the $Z=0.1,0.2, \ldots$ lines included in the phase diagram. Taken from \cite{FanPRB922015}.}
   \label{fig:3.11}  
\end{figure}

In \fref{fig:3.11}, $Z$ color map for the $J_H/U$ versus $n_e$ phase diagram of $N=5$ for $U/W=1$ is presented. There is a clear relation between the Mott insulator at half-filling and the correlated metallic region (Hund metal). This relation can be more easily identified by looking at the $Z=0.1,0.2, \ldots$ lines included in the figure, taken for each $0.1$ step. When approaching the half-filled Mott insulator, the strength of correlations increase.

In general, for any $n_e/N$ and \textcolor{black}{almost all} $J_H \neq 0$ \textcolor{black}{(larger than a certan threshold)}, the system becomes more correlated when approaching half-filling \cite{IshPRB812010,LiePRB822010,WerNP82012,BasPRB862012,TerPRB872013,HarPRL1112013,NakSC42014,CalPRB902014,deMPRL1122014,EilPRL1162016,HarPRB942016,BasCRP172016}. For $J_H =0$, the situation is different, as seen in \fref{fig:3.3}. In this situation, the orbital fluctuations stabilize the metallic state at half-filling \cite{RozPRB551997,FloPRB702004}, disfavoring the Mott insulator. Then, at $J_H=0$ and half-filling, $U_c$ will increase. In \cref{Chap4c}, I will discuss more in detail this situation for $N=2$ orbitals system.

The relation between the Hund metal and the Mott insulator is at the heart of the ideas that I will discuss in the following sections and, specially in \cref{Chap4a}, where I will use this link between both Mott insulator physics and Hund metal physics to propose a new high-$T_c$ superconducting family.

\subsection{Hund's coupling and orbital-selective Mott transition (OSMT)}
\label{3.2.4}

In this section, I will briefly review an important consquence of the Hund's coupling related with the orbital decoupling. This is the already mentioned orbital-selective Mott transition (OSMT) \cite{deMPRB722005,deMPRL1022009,YuRPRL1102013,YiMPRL1102013,LiuPRB922015}. 

Due to the Hund's coupling $J_H$, an orbital decoupling behavior is promoted (see \sref{3.2.2}). For a \textcolor{black}{system with} non-equivalent orbitals, this behavior can lead to the OSMT, i.e. some of the $Z_m=0$ while others $Z_{n \neq m}$ remain finite \cite{deMPRB722005,deMPRL1022009,YuRPRL1102013,YiMPRL1102013,LiuPRB922015}. When orbital degeneracy is broken and orbital decoupling behavior is acting on the system, each non-degenerated $Z_m$ will behave differently in terms of $U$, $J_H$ and $n_e$. This property is called orbital differentiation. The equivalence between different orbitals can be broken in \textcolor{black}{three} distinct manners: $(1)$ by changing the hopping parameters for different orbitals, hence there will be different dispersion relations for each orbital $\varepsilon_{km}$ \cite{deMPRB722005}, $(2)$ by introducing a crystal field splitting, so there will be different $\epsilon_m$ parameters for each orbital \cite{deMPRL1022009}, \textcolor{black}{or $(3)$ by considering orbital-dependent interacting parameters. I discuss the first two cases below. The third case goes along similar lines}. Note that I will only consider intraorbital hoppings $t_{mm}^{ij}$ and neglect interorbital ones $t_{mn}^{ij}$.

\begin{figure}[h]
   \centering
   \includegraphics[width=0.9\columnwidth]{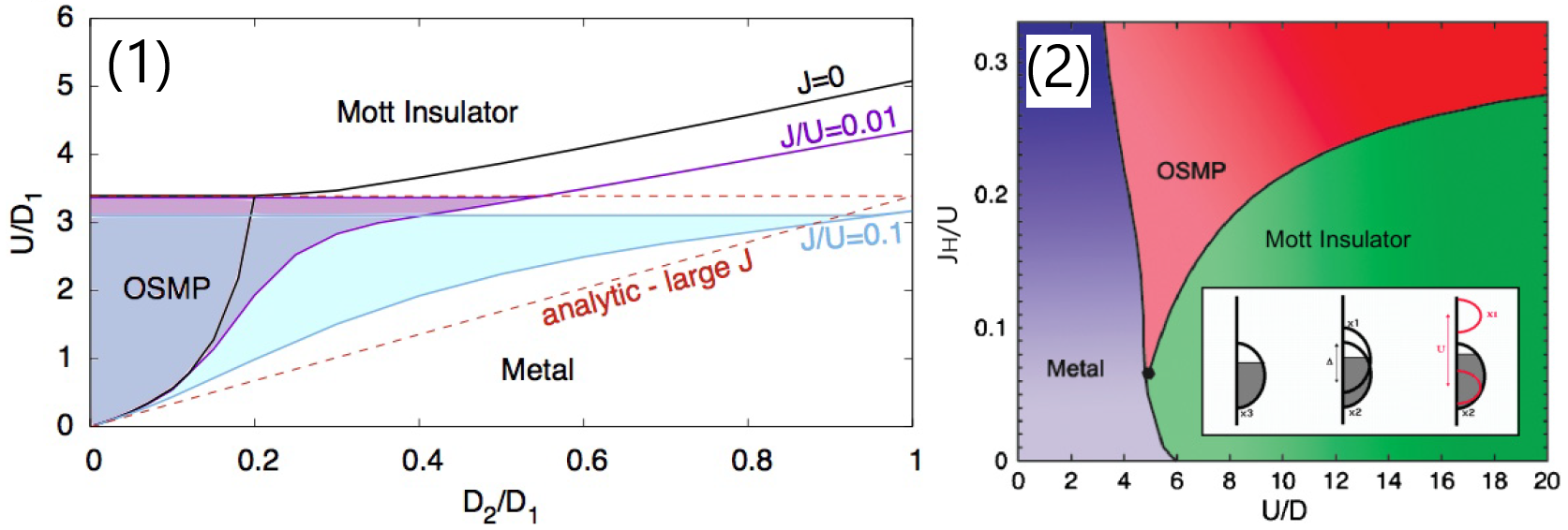}
   \caption[OSMT for different bandwidths for $N=2$ at half-filling and different crystal field splittings for $n_e/N=3/4$]{\textbf{(1)} $U/D_1$ versus $D_2/D_1$ phase diagram for $N=2$ at half-filling, where metallic, OSMP and Mott insulator phases are shown. Various $J_H/U=0,0.01$ and $0.1$ values are presented. Once $J_H/U$ increases, OSMP enlarges, due to the promotion by $J_H$ of the orbital decoupling mechanism. Large-$J_H/U$ limit included as a dashed red line. Taken from \cite{deMPRB722005}. \textbf{(2)} $J_H/U$ versus $U/D$ phase diagram for $n_e/N=4/3$, where one of the orbitals has a finite crystal field different to the other two orbitals. \textcolor{black}{The orbital filling per spin is fixed to $n_{3\sigma} = 0.5$ and $n_{1\sigma}=n_{2\sigma}=0.75$}. \textsc{Inset}: Sketch for the OSMT, where black bands refer to the two equivalent orbitals, whether red band is the one splitted by a finite crystal field $\epsilon_3 \neq \epsilon_1=\epsilon_2$. Taken from \cite{deMPRL1022009}.}
   \label{fig:3.12}  
\end{figure}

The first case can be equivalently recognised with an orbital-dependent bandwidth $W_m$. In \fref{fig:3.12}\textbf{(a)}, the phase diagram $U/D_1$ versus $D_2/D_1$ (where $D_m=W_m/2$) for $N=2$ at half-filling and different $J_H/U$ values is presented. For $J_H=0$, a really large $D_2/D_1 \approx 0.2$ is needed to obtain a OSMP, whether if $J_H/U$ (and hence the orbital decoupling) is increased, this OSMP widens. Here, the orbital which takes $Z_m =0$ is the one with the lowest $W_m$.

In \fref{fig:3.12}\textbf{(b)}, the phase diagram $J_H/D$ versus $U/D$ for $N=3$ at $n_e=4$ is shown, when one of the orbitals has a finite crystal field splitting $\Delta$ with respect to the other two. \textcolor{black}{The orbital filling per spin was fixed to $n_{3\sigma} = 0.5$ and $n_{1\sigma}=n_{2\sigma}=0.75$}. As appreciated, this phase diagram looks really similar to the one in \fref{fig:3.5}\textbf{(d)}, showing the important role played by Hund's coupling in this area. Here, the orbital which takes $Z_m=0$ is the one with an orbital filling $n_{m\sigma}$ closer to half-filling.

As a summary, Hund's coupling $J_H$ promotes a local spin polarization, which results in an orbital decoupling behavior. For non-equivalent orbitals, $J_H$ can drive the system into an OSMT. Any hybridization between orbitals, like an interorbital hopping $t_{mn}^{ij}$, will act against the appearance of a OSMT. In this situation, any finite interorbital hybridization will turn \textcolor{black}{(depending on the strength of the orbital differentiation, i.e. different $Z_m$ values)} the OSMT into a state with weakly (\textcolor{black}{$0.7<Z_m \leq 1$}), moderately (\textcolor{black}{$0.3<Z_n \lesssim 0.7$}) and strongly (\textcolor{black}{$0<Z_p \lesssim 0.3$}) correlated orbitals coexisting at the same time \cite{emergence2017,deMedici2015,deMedici2017,deM1707.03282,BasCRP172016}. 

The most correlated orbitals will be determined by the smaller $W_m$ (due to an increase of the ratio $\Delta^x/W_m$) and the closeness to half-filling of the orbital filling per spin $n_{m\sigma}$.

\section{Hund metals and FeSCs}
\label{3.3}

In \cref{Chap01}, the discussion about the strength of electronic correlations in FeSCs was presented, arguing about a weakly \cite{RagPRB772008,MazPRL1012008,ChuPRB782008,CveEPL852009} or a strongly \cite{YilPRL1012008,SiQPRL1012008} correlated approach to study FeSCs. In this section, I will show that in FeSCs, weakly, moderately and strongly correlated electrons coexist at the same time, with the Hund metal phenomenology playing a major role.

FeSCs are multiorbital systems for which the parent compound (i.e. the system without doping or external perturbations) has $n_e=6$ electrons in 5 $d$ orbitals in a tetragonal symmetry. Non-zero interorbital hoppings $t_{mn}^{ij}$ will turn a possible OSMT into a metallic state with weakly ($Z_m \sim 0.8-1$), moderately ($Z_n \sim 0.6-0.7$) and strongly ($Z_p \sim 0.1-0.5$) correlated orbitals.

In this section, I present a review on the role of local correlations in real materials, specifically on FeSCs  \cite{BasCRP172016,deMPRL1122014,LuDARCMP32012,LanPRB872013,GrePRB842011,LafPRB962017}. First, I will briefly review the theoretical calculations for various FeSCs, such as $LaFeAsO$, $Ba{Fe}_2{As}_2$ and $FeSe$. Then, in \sref{3.3.1}, I will present the experimental evidences which support that FeSCs are \textcolor{black}{close to or in} the Hund metal regime.

\begin{figure}[h]
   \centering
   \includegraphics[width=0.9\columnwidth]{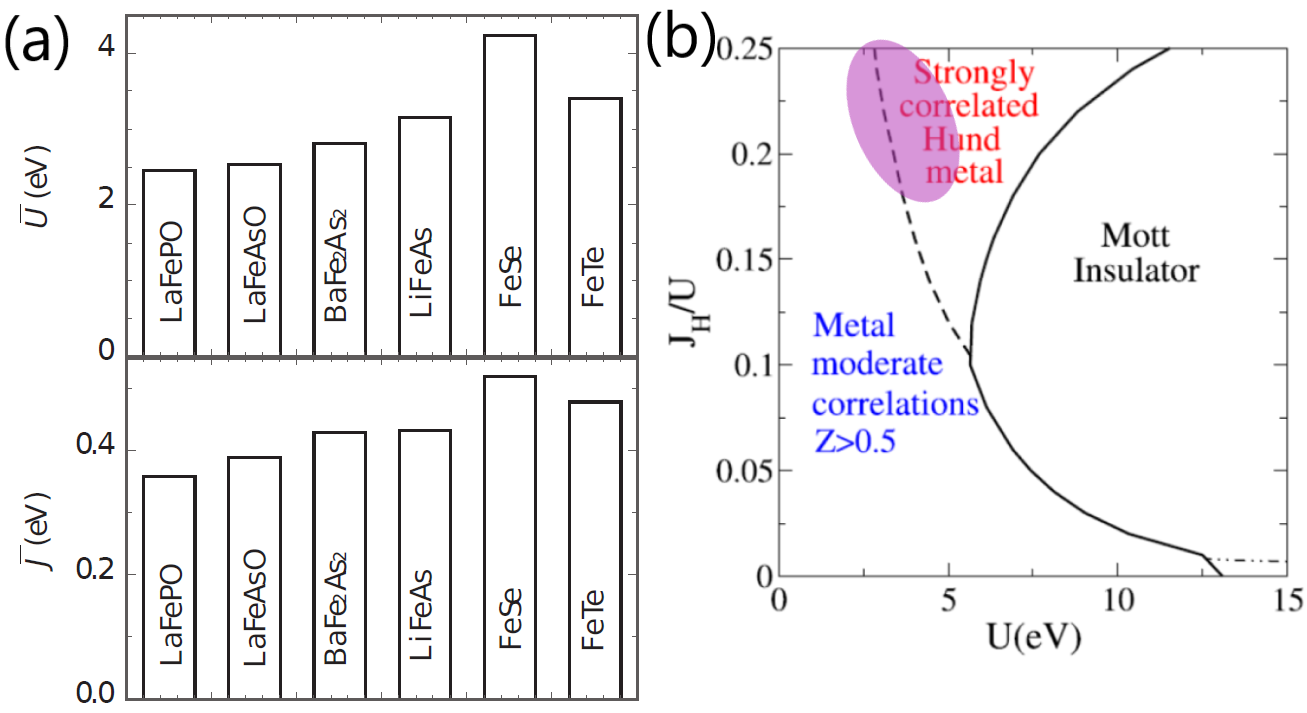}
   \caption[Orbital-averaged interaction parameters for various FeSCs and $J_H/U$ versus $U$ phase diagram for undoped $LaFeAsO$]{\textbf{(a)} Orbital-averaged interaction parameters for various FeSCs. $U$ ranges from $2.5$ to $4 \, eV$ and $J_H$ from $0.35$ to $0.5 \, eV$. Relation $U' = U-2J_H$ is approximately satisfied. Taken and adapted from \cite{MiyJPSJ792010}. \textbf{(b)} SSMF $U(1)$ $J_H/U$ versus $U$ phase diagram for $LaFeAsO$ at $n_e=6$ electrons. Shaded circle marks an approximate region for locating FeSCs, where the area is estimated when comparing SSMF calculations with experimental $Z_m$ values from ARPES and QOs. Taken and adapted from \cite{YuRPRB862012,BasCRP172016}.}
   \label{fig:3.13}  
\end{figure}

In \fref{fig:3.13}\textbf{(a)}, a table with the orbital-averaged interactions for various FeSCs calculated using constrained random phase approximation (cRPA) \cite{MiyJPSJ792010} is shown. The authors obtained $U \approx 2.5-4 \, eV$ and $J_H \approx 0.35-0.5 \, eV$, with 1111 ($LaFePnO$ with $Pn=P,As$) and 122 ($BaFe_2As_2$) families being the less correlated and 11 ($FeCh$ with $Ch=Se,Te$) and 111 ($LiFeAs$) the most correlated ones. The differences between the orbital-averaged ($U$ and $J_H$) and the orbital-dependent interactions ($U_{mm}$ and $J_{H \: mn}$) oscillate between $1\%$ and $35\%$, so that $U$ and $J_H$ can be seen as a good approximation. The relation $U'=U-2J_H$ is approximately satisfied, where the differences with respect to its orbital-dependent values $U'_{mn}$ are $\approx 5\%$.

In \fref{fig:3.13}\textbf{(b)}, the SSMF $U(1)$ phase diagram obtained for undoped ($n_e=6$) $LaFeAsO$ in the unfolded Brillouin zone (BZ), i.e. in a 5 $d$ orbitals tight-binding model, is shown \cite{YuRPRB862012,BasCRP172016}. Note that SSMF formalism tends to underestimate the effects of the real interaction values, so that the $U$ and $J_H$ are not straighforwardly extracted from \fref{fig:3.13}\textbf{(a)}, and they should be a bit larger in SSMF. Shaded circle marks an estimated region where most of the FeSCs can be localized. This shaded area is obtained when comparing $Z_m$ values from theoretical with experimental results from orbital-resolved band structure measurements, such as ARPES and quantum oscillations (QOs) \cite{LuDARCMP32012}. Both experimental techniques are sensible to the real band structure, and when comparing with DFT band structure calculations, $Z_m$ can be obtained. Here, FeSCs are located around the Hund metal crossover and in the Hund metal regime, so it is expected that they show the properties explained in \sref{3.2}.

\begin{figure}[h]
   \centering
   \includegraphics[width=0.8\columnwidth]{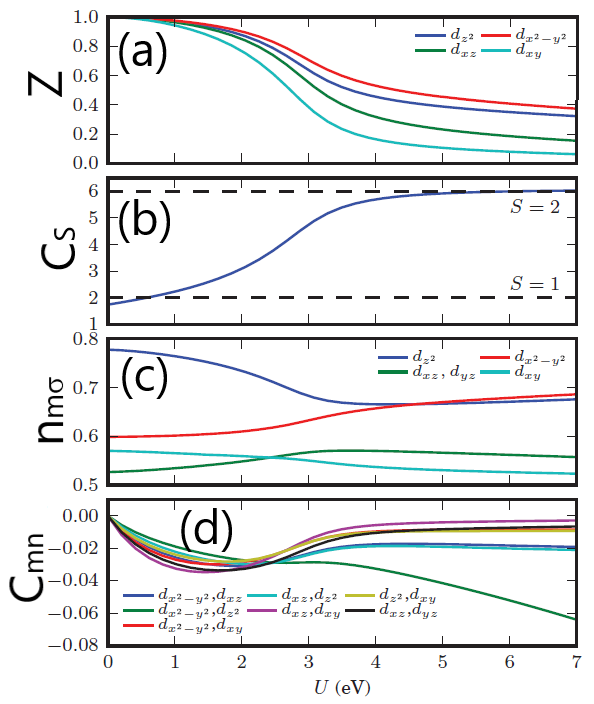}
   \caption[DFT + Gutzwiller approximation calculations for $FeSe$ in a 5 $d$ orbitals tight-binding model at $n_e=6$ and $J_H/U=0.224$]{DFT + Gutzwiller approximation calculations for $FeSe$ in a 5 $d$ orbitals tight-binding model at $n_e=6$ and $J_H/U=0.224$. \textbf{(a)} Orbital quasiparticle weights $Z_m$, \textbf{(b)} total spin correlations $C_S$, \textbf{(c)} orbital fillings per spin $n_{m\sigma}$ and \textbf{(d)} interorbital charge correlations $C_{mn} \equiv C_{n}^{inter}$. The strength of local correlations for each orbital is stated by the closeness to half-filling ($n_{m\sigma}=1/2$) of the orbital fillings $n_{m\sigma}$. Spin correlations are enhanced and interorbital charge correlations are suppressed (except for $C_{x^2-y^2,z^2}$, due to their less correlated behavior). Taken and adapted from \cite{LanPRB872013}.}
   \label{fig:3.14}  
\end{figure}

In \fref{fig:3.14}, DFT + Gutzwiller approximation (GA) calculation (see \sref{1.4.1}) in a 5 $d$ orbitals tight-binding model for undoped ($n_e=6$) $FeSe$ at $J_H/U=0.224$ is presented \cite{LanPRB872013}. As shown, all the features mentioned in \sref{3.2} are well reproduced in this system. The most correlated orbital (i.e., with the lowest $Z_m$) is $xy$, followed by $xz$ (which is degenerated with $yz$) and $z^2$, and the less correlated orbital is $x^2-y^2$. In the Hund metal regime ($U>3 \, eV$), this trend follows the closeness to half-filling of each orbital filling per spin $n_{m\sigma}$, plotted in \fref{fig:3.14}\textbf{(c)}, rather than depend on the total filling ($n_e=6$), which signals an orbital decoupling behavior. It can be seen that, even for a large $U=7 \, eV$, the system is not yet in the Mott insulating state, due to the effect of $J_H/U=0.224$, disfavoring the appearance of the Mott insulator. The enhancement of spin correlations $C_S$ and orbital decoupling (with the interorbital charge correlations $C_{mn}$ approaching $0$) can be clearly seen in \fref{fig:3.14}\textbf{(b)} and \textbf{(d)}, except for interorbital correlations $C_{x^2-y^2, z^2}$, due to their less correlated behavior in which the considered $J_H$ was not able to produce an orbital decoupling.

\begin{figure}[h]
   \centering
   \includegraphics[width=0.8\columnwidth]{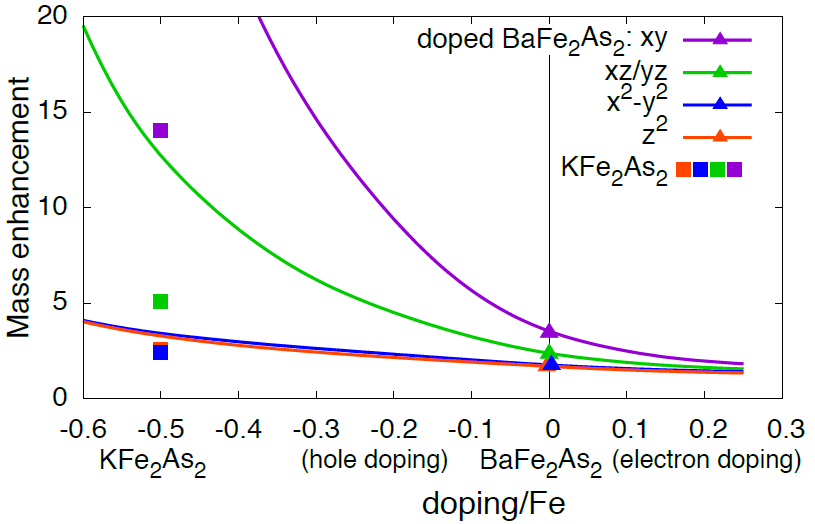}
   \caption[SSMF $Z_2$ calculations of mass enhancements for $BaFe_2As_2$, and hole-doped $Ba_{1-x}K_xFe_2As_2$]{SSMF $Z_2$ calculations of a 5 $d$ orbitals tight-binding model for $Ba{Fe}_2{As}_2$ and hole-doped $Ba_{1-x}K_xFe_2As_2$. Correlations increase when half-filling is approached. Each orbital behaves differently in terms of the filling, as an indication of the orbital selectivity of local correlations. Colored squares describe a tight-binding model for $K{Fe}_2{As}_2$ (instead of using the one for $Ba{Fe}_2{As}_2$ and doped it to $n_e=5.5$). Taken and adapted from \cite{deMPRL1122014}.}
   \label{fig:3.15}  
\end{figure}

In \fref{fig:3.15}, SSMF $Z_2$ calculations for $Ba{Fe}_2{As}_2$ at $U=2.7 \, eV$ and $J_H/U=0.25$ when varying the filling $n_e$ of the system, are shown \cite{deMPRL1122014}. The $0$ in the $x$-axis represents the undoped $BaFe_2As_2$ with $n_e=6$. Negative $x$ represents hole doping, while positive represents electron doping. As far as $K{Fe}_2{As}_2$ (with $n_e=5.5$) is approached with hole doping, the strength of correlations increases, giving larger mass enhancement factors $(m^{*}/m_{0})_m$ (i.e. lower $Z_m$ values) due to the influence of the Mott insulator at half-filling $n_e=5$, as discussed in \sref{3.2.3}. Colored squares mark the solution for a tight-binding model for $K{Fe}_2{As}_2$, instead of using the one for $Ba{Fe}_2{As}_2$ and doped it up to $n_e=5.5$. This is done in order to reproduce the real band structure of $KFe_2As_2$. Differences in the scale of the strength of local correlations are found, but the increase of the strength towards half-filling is still present.

\subsection{Experimental evidences for Hund metal behavior in FeSCs}
\label{3.3.1}

Now, I will comment the experimental evidences which support that the FeSCs are indeed in a Hund metal regime (or at least, in the crossover) due to local correlations \cite{BasCRP172016,deMPRL1122014,deM1707.03282,LuDARCMP32012,LanPRB872013,GrePRB842011,LafPRB962017}. 

\begin{figure}[h]
   \centering
   \includegraphics[width=1.0\columnwidth]{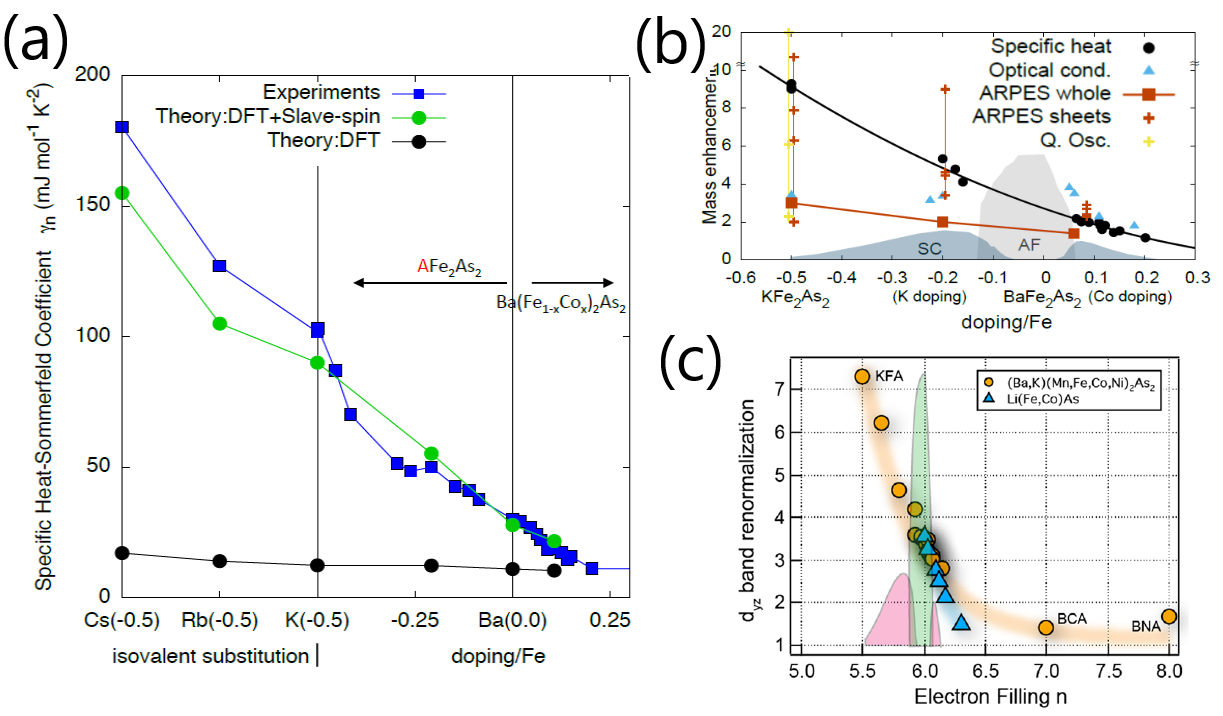}
   \caption[Low-$T$ specific heat measurements, and comparison with DFT and SSMF calculations, and mass enhancement factors for different experimental probes for various 122 FeSCs compounds]{\textbf{(a)} Low-$T$ specific-heat in terms of the total filling, where the Sommerfeld coefficient $\gamma$ is shown for various 122 FeSCs. A comparison with DFT and SSMF $Z_2$ calculations is shown. A really good agreement can be seen between SSMF and experimental results. Taken from \cite{HarPRB942016,deMedici2017,deM1707.03282}. \textbf{(b)} Mass enhancement factors $(m^*/m_{0})_m$ for different experimental probes when varying the total filling for various 122 FeSCs. Each experimental probe is sensible to different orbitals with respect to their behavior with local correlations. Specific heat is sensible to the most strongly correlated orbitals. Optical conductivity and ARPES whole are sensible to weakly correlated orbitals. ARPES and QO are orbital-resolved measurements. In the background, the phase diagram for $Ba{Fe}_2{As}_2$ is shown as a eyeguide to locate the different phases of this FeSC. The strength of correlations clearly increases when doping with holes, i.e. when approaching half-filling (left side of the $x$ axis in the figure), and decreases with electron doping. Taken and adapted from \cite{deMPRL1122014,deM1707.03282}. \textbf{(c)} ARPES measurements for the renormalization of the $d_{yz}$ orbital in terms of the electronic filling $n_e$ of the 122 FeSCs, $BaFe_2As_2$. $KFA$, $BCA$ and $BNA$ labels stand for the isostructural compounds, $KFe_2As_2$ (with $n_e=5.5$), $BaCo_2As_2$ (with $n_e=7$) and $BaNi_2As_2$ (with $n_e=8$), respectively. Taken from \cite{YiMnpjQM22017}.}
   \label{fig:3.16}  
\end{figure}

In \fref{fig:3.16}\textbf{(a)}, the experimental Sommerfeld coefficient $\gamma$ (i.e. the linear term of the low-$T$ specific heat $C(T) \sim \gamma T^2$) for various 122 FeSCs (from $n_e=6$ to $n_e=5.5$ and $n_e=6.25$) and the comparison with DFT calculations are shown. SSMF $Z_2$ (labeled with DFT + Slave-spin in the legend) calculations are also shown \cite{HarPRB942016,deMedici2017,deM1707.03282}. The Sommerfeld coefficient was already introduced in \eref{eq:1.15}, and depends on the mass enhancements $(m^*/m_0)_m$ for a multiorbital system as:

\begin{equation}
\gamma^* = \frac{\pi^2 k_B^2 N^*(\mu)}{3} \rightarrow \frac{\gamma^*}{\gamma_{0}} = \sum_{m} (\frac{m^*}{m_{0}})_m = \sum_{m} \frac{1}{Z_m}
\label{eq:3.8}
\end{equation}

where $N^*(\mu)$ is the DOS at the chemical potential for the correlated system which is proportional to the mass enhancements in the multiorbital system $(m^*/m_0)_m = 1/Z_m$, and $\gamma_0$ is the Sommerfeld coefficient in the non-correlated limit. When comparing experiments and theoretical results, the deviation with respect to DFT calculations to see the effect of local correlations is used, as for example $\gamma^* / \gamma_{0}$. The strength of correlations increases when approaching and decreases when moving away from half-filling. A good agreement can be seen between SSMF and experimental results, signaling the importance of local correlations in these systems.

$\gamma$ will be mainly controled by the most correlated orbitals, i.e. those with the largest mass enhancement factors $(m^*/m_0)_m$ (lowest $Z_m$ values). On the other hand, other measurements are sensible to the less correlated orbitals, such as the Drude weight $D$ obtained from optical conductivity measurements ($D/D_{0} = \sum_m (m_0/m^*)_m$, where $D_0$ is the non-correlated Drude weight), or the renormalization of the whole bandstructure measured by ARPES (labeled as ARPES whole in \fref{fig:3.16}\textbf{(b)}). Finally, ARPES and QOs are orbital-resolved measurements, so they are sensible to $(m^*/m_0)_m$ for each orbital. \fref{fig:3.16}\textbf{(b)} shows a compendium of the different experimental probes in the 122 FeSCs $BaFe_2As_2$ family \cite{deMPRL1122014,deM1707.03282}. In the background, an sketch of the phase diagram for $Ba{Fe}_2{As}_2$ can be seen, to locate the different phases in this material. The strength of correlations clearly increases when approaching half-filling (left side of the $x$ axis in the figure), as expected from SSMF calculations (see \fref{fig:3.15}). In \fref{fig:3.16}\textbf{(c)}, I show the ARPES measurements for the renormalization of the $d_{yz}$ orbital in terms of the electronic filling $n_e$ of the 122 FeSCs, $BaFe_2As_2$ \cite{YiMnpjQM22017}. $KFA$, $BCA$ and $BNA$ labels stand for the isostructural compounds, $KFe_2As_2$ (with $n_e=5.5$), $BaCo_2As_2$ (with $n_e=7$) and $BaNi_2As_2$ (with $n_e=8$), respectively. 

An important consequence can be extracted from these experimental results \cite{deM1707.03282,BasCRP172016}: different experimental probes will be sensible to different weakly or strongly correlated orbitals. All the measurements together show that both weakly and strongly correlated orbitals coexist at the same time. This can be more clearly seen in orbital-resolved measurements, such as ARPES or QOs \cite{YiMNC62015,YiMnpjQM22017}.

\begin{figure}[h]
   \centering
   \includegraphics[width=0.6\columnwidth]{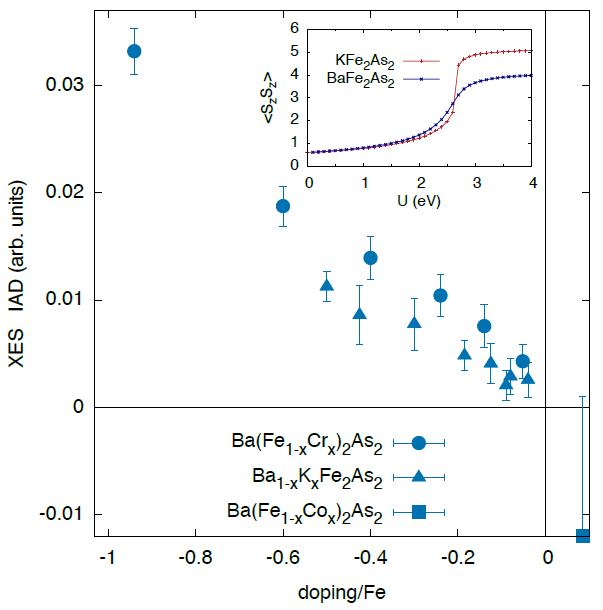}
   \caption[XES measurements and total spin correlations SSMF calculations for ${Ba}_{1-x}K_x$ ${Fe}_2{As}_2$]{XES IAD values for ${Ba}_{1-x}{K}_x{Fe}_2{As}_2$. XES IAD values are proportional the local moment, i.e. to the local spin correlations, in a non-ordered system. \textsc{Inset}: local spin correlations for $Ba{Fe}_2{As}_2$ and $K{Fe}_2{As}_2$ systems calculated using SSMF $Z_2$. No difference is seen for the weakly correlated regime ($U \leq 1 \, eV$), and only being in deep in the Hund metal regime ($U \geq 2.5 \, eV$) would reconcile the experimental results with the theoretical expectations. Taken from \cite{deM1707.03282}.}
   \label{fig:3.17}  
\end{figure}

\fref{fig:3.17} shows the experimental results for X-ray emission spectroscopy (XES) in hole-doped ${Ba}_{1-x}{K}_x{Fe}_2{As}_2$ systems \cite{deM1707.03282,GrePRB842011,LafPRB962017}. This experimental probe is sensible to the local magnetic moment (XES IAD value is proportional to the local moment) in a system which is not necessarily ordered. \textsc{Inset} shows the total spin correlations $C_S$ calculated with SSMF for $Ba{Fe}_2{As}_2$ ($n_e=6$) and $K{Fe}_2{As}_2$ ($n_e=5$) in terms of $U$. In the weakly correlated regime ($U \leq 1 \, eV$), local spin correlations are the same for both compounds as shown in the \textsc{inset}, totally different to what happens in the experimental results, where local moment increases when approaching half-filling. Experiment agrees well with theory for $U\geq 2.5 \, eV$, where a different magnitude of local spin correlations for both fillings can be found. This indicates that the compounds are in (or at least in the crossover to) the Hund metal.

All of these measurements sustain the idea that FeSCs are \textcolor{black}{close to or in} the Hund metal regime with the phenomenology explained in \sref{3.2}. The ideas explained here can be used for other materials and establish a basis to check whether local correlations are playing an important role in the system under study.

\section{Conclusions}
\label{3.4}

Along this chapter, I have presented the phenomenology for multiorbital systems and the role of local correlations \cite{emergence2017,WerPRL1012008,HauNJP112009,LiePRB822010,IshPRB812010,HanPRL1042010,YinNM102011,WerNP82012,LanPRB872013,TerPRB872013,HarPRL1112013,deMPRL1122014,NakSC42014,FanPRB922015,deMedici2015,EilPRL1162016,HarPRB942016,deMedici2017,deM1707.03282,BasCRP172016}.

In a multiorbital system, the Mott insulating state will appear, not only at half-filling, but also for any integer fillings $n_e=1,2,\ldots N-1$ \cite{FanPRB922015,deM1707.03282}. The critical interaction for the Mott transition $U_c^x$ and its behavior with $J_H$ depend on the filling $n_e$. For single-occupied systems ($n_e=1$), $U_c$ increases when $J_H$ increases, while the opposite happens for half-filled systems ($n_e=N$).

A new correlated metallic \textcolor{black}{state}, the Hund metal \cite{emergence2017,WerPRL1012008,HauNJP112009,LiePRB822010,IshPRB812010,HanPRL1042010,YinNM102011,WerNP82012,LanPRB872013,deMPRL1122014,FanPRB922015,deMedici2015,deMedici2017,deM1707.03282,BasCRP172016} is found for a certain filling region $1<n_e<N$. This Hund metal regime shows a correlated metallic behavior which is promoted for large $J_H$ values, so that the Mott insulating state is disfavored when $J_H$ increases (similar to what happens in $n_e=1$ systems). 

In the Hund metal regime, the atoms are \textcolor{black}{highly} spin polarized (Hund's rule is satisfied). As a consequence, $Z_m$ are in a low value and there is an orbital decoupling behavior \cite{FanPRB922015,deM1707.03282,BasCRP172016}. Between the weakly correlated metal and the Hund metal there is a crossover defined by a characteristic $J_H^*$, whose \textcolor{black}{value} depends on the lattice symmetry and $n_e/N$.

The enhancement of local spin correlations in the Hund metal regime drives also an orbital decoupling behavior \cite{FanPRB922015,BasCRP172016}. This comes from the fact that there is an effective reduction of local anti-parallel spin configurations, while parallel configurations are promoted. The reduction of anti-parallel spin configurations results in a reduction of $C_n^{intra}$, and an indirect reduction of $C_n^{inter}$, which translates in the orbital decoupling. 

The suppression of $C_n^{intra}$ is directly related with the Mott transition at half-filling, hence providing a link between the Hund metal regime at $1<n_e<N$ and the Mott insulating state at half-filling $n_e=N$ \cite{IshPRB812010,LiePRB822010,WerNP82012,BasPRB862012,TerPRB872013,HarPRL1112013,NakSC42014,CalPRB902014,deMPRL1122014,EilPRL1162016,HarPRB942016,BasCRP172016}. .

In the case of non-equivalent orbitals, the orbital decoupling promoted by $J_H$ can lead to an OSMT \cite{deMPRB722005,deMPRL1022009,YuRPRL1102013,YiMPRL1102013,LiuPRB922015}. For any finite interorbital hopping $t_{mn}$, this OSMT is suppressed, and the system has weakly, moderately and strongly correlated orbitals coexisting at the same time. Different orbitals will have different correlation strengths (orbital differentiation).

In this framework, various theoretical and experimental results for FeSCs (low-$T$ specific heat, ARPES renormalized band structure, XES for the local moment, etc.) \cite{BasCRP172016,deMPRL1122014,LuDARCMP32012,LanPRB872013,GrePRB842011,LafPRB962017} sustain the idea that these systems can be seen as being close to or in the Hund metal regime.

%

\chapter{{\bf Strong correlations and the search for high-$T_c$ superconductivity in chromium pnictides and chalcogenides\footnote{Results presented in this Chapter published in \cite{Pizarro1}}}} 
\label{Chap4a}
\lhead{Chapter \ref{Chap4a}. \emph{Search for high-$T_c$ superconductivity in chromium pnictides}} 

\begin{small}

In \cref{Chap03}, I have mentioned a connection between the Mott insulating state at half-filling and the Hund metal regime for multiorbital systems. This connection could serve as a link between the two most well-known families of high-$T_c$ superconductors, the cuprates, in which the undoped system (with one $d$ orbital at half-filling) is a Mott insulator, and the iron-based superconductors (FeSCs), in which the system (where the total number of electrons per $Fe$ atom is $n_e=6$ in $5$ $d$ orbitals) is close to or in the Hund metal regime. Both the Mott insulator and Hund metal phenomenology appear due to local correlations. In 2010, \emph{Ishida and Liebsch} \cite{IshPRB812010} showed that FeSCs can be seen as electron-doped \textcolor{black}{systems from the Mott insulator at half-filling ($n_e=5$ in $5$ $d$ orbitals)}, instead of being a completely different family when compared with the cuprates. In this chapter, we explore more deeply this idea by going into the other side of the Mott insulator at half-filling, i.e. by studying the hole-doped Mott insulator systems.

Here, we propose to search for high-$T_c$ superconductivity in a new material family based on chromium instead of iron. We expect that the strength of electronic correlations \textcolor{black}{evolves in the hole-doped side ($n_e<5$) in a similar way as in the electron-doped side ($n_e>5$)}. Then, we \textcolor{black}{assume} that an optimal degree of the electronic correlations is needed in order to obtain high-$T_c$ superconductivity, \textcolor{black}{and focus on the holpe doping level at which the strength of correlations is closest to the one found in FeSCs}. We will assume that this superconducting phase is driven by AFM fluctuations, as it seems to occurr in most FeSCs, \textcolor{black}{and discuss the superconducting instability}. As an example, we study the 1111 $LaCrAsO$ chromium pnictide, the analogue of the iron superconductor $LaFeAsO$. $LaCrAsO$ \cite{ParIC522013} has the same crystal structure as its iron counterpart (see \fref{fig:1.15}), with $Cr$ atoms forming a 2D square lattice and tetrahedrally coordinated by $As$ atoms. $LaCrAsO$ shows a DFT band structure \textcolor{black}{similar to other 1111 transition metal pnictides, like $LaFeAsO$}, with all the 5 $d$ orbitals of $Cr$ mainly contributing to the bands around the Fermi level, with a total bandwidth $W \sim 4 \, eV$ (in $LaFeAsO$ is \textcolor{black}{$W \sim 4.5-5 \, eV$). Experimentally, this compound shows an antiferromagnetic (AFM) phase with $(\pi,\pi)$ ordering (checkerboard AFM), which can be suppressed by doping \cite{ParIC522013}}. The main difference in undoped $LaCrAsO$ with respect to undoped $LaFeAsO$ is the number of electrons per site, which is $n_e=4$ in $LaCrAsO$, while $n_e=6$ in $LaFeAsO$.

In this chapter, I will first check if a generic $5$ $d$ orbitals model for FeSCs (in particular for $LaFeAsO$) follows the idea of increasing the strength of local correlations when approaching half-filling $n_e=5$ from both sides, i.e. I will study the mass enhancement factors $m_{m}^{*}/m_0 \sim 1/Z_m$ when varying $n_e$ from $4$ and from $6$ to $n_e=5$. 

After demonstrating that the strength of local correlations increases (i.e. $m_m^*/m_0$ increases) when moving towards half-filling from the electron- and the hole-doped part, I will study the case of chromium systems, using $LaCrAsO$ as an example. I will present the results obtained by my co-workers to calculate the DFT band structure for the undoped $LaCrAsO$ ($n_e=4$), and the $5$ $d$ orbitals tight-binding model using the same Slater-Koster fitting procedure as in the case of $LaFeAsO$. Then, I will calculate the mass enhancement factors $m_m^*/m_0$ for $LaCrAsO$ when varying the total filling from $n_e=4$ to $n_e=5$, i.e. when approaching half-filling. I found that the strength of local correlations is similar to the undoped $LaFeAsO$ for $LaCrAsO$ at $n_e=4.5$. \textcolor{black}{We expect that this behavior will hold for any other $Cr$ pnictide or chalcogenide.}

To study \textcolor{black}{the strength of} local correlations, I will use the SSMF $Z_2$ formalism. $U$ and $J_H$ will be fixed for both $LaFeAsO$ and $LaCrAsO$. In the case of $LaFeAsO$, $U=3 \, eV$ and $J_H/U=0.25$ are the values for which the mass enhancement factors $m_m^*/m_0$ obtained by SSMF equal the ones observed experimentally in angle-resolved photoemission spectroscopy (ARPES) and quantum oscillations (QOs). For $LaCrAsO$, we will \textcolor{black}{assume} that, due to the equivalency of both systems, $U$ and $J_H$ will take the same values as in $LaFeAsO$.

$LaCrAsO$ shows a metallic behavior, as well as $LaFeAsO$, for realistic interaction parameters $U$ and $J_H$. Then, we assumed that the spin fluctuation theory to study the superconducting instability is applicable, but instead of using the weakly correlated DFT band structure, we used the renormalized one as a better approach for the real band structure. First, my co-workers calculated the spin susceptibility by means of multiorbital random phase approximation (RPA) in the renormalized band structure. They obtained that the AFM with nesting vector $\vec{Q}=(\pi,\pi)$ (checkerboard-like) is the ground state of $LaCrAsO$, as expected from experiments \cite{ParIC522013}. A secondary peak at $\vec{Q}_2=(\pi/2,0)$ (and $(0,\pi/2)$ due to the fact that the system is in the tetragonal phase) is found. At $n_e=4.5$, my co-workers also found that the superconducting instability is driven by this secondary peak $(\pi/2,0)$, rather than by the nesting vector responsible of the AFM order. They argued that this enhanced response at $\vec{Q}_2$ is due to the nearness of a Lifshitz transition. Finally, they found that the superconducting order parameter symmetry is $d_{xy}$.

We conclude by stating that it seems plausible to find a new family of high-$T_c$ superconductors in the chromium analogue of $LaFeAsO$, $LaCrAsO$. If this behavior is found in such a system, it will open a new route to search for new high-$T_c$ superconductors based on the Mott doped scenario. 

\end{small}

\newpage

\section{Introduction \& Motivation}
\label{4.1}

In previous chapters, I have summarized several important properties present in high-$T_c$ cuprates superconductors and FeSCs, as well as explained the phenomenology of local correlations in single- and multiorbital systems, which results in the emergence of Mott insulator and Hund metal behaviors.

As shown in \cref{Chap01}, in most cuprates and FeSCs, and other unconventional superconductors, the parent compound shows a magnetic phase, usually AFM. Once the system is doped with electrons or holes (or pressure is applied in some cases), the AFM phase is suppressed and the unconventional superconducting dome emerges. This behavior is quite robust in unconventional superconductors. It is believed that the magnetic phase suppression is a key ingredient to find unconventional superconductivity

\textcolor{black}{Nevertheless, there is an important difference between cuprates and FeSCs. The parent magnetic phase of the cuprates is Mott insulating. There has been an intense debate about the role of the Mott physics in high-$T_c$ superconductivity. On the other hand, the parent magnetic phase of FeSCs is metallic. This led to the quastion whether Mott insulator physics was necessary for obtaining high-$T_c$ superconductivity.}

A new point of view came out in 2010, when \emph{Ishida and Liebsch} \cite{IshPRB812010} found, by using DMFT, that that the strength of correlations in FeSCs are stronger when moving from the undoped $LaFeAsO$, with $n_e=6$ in 5 $d$ orbitals, towards half-filling $n_e=5$. Based on these results, they proposed to see FeSCs as electron-doped Mott insulators, making a possible connection between cuprates and FeSCs. \textcolor{black}{Interestingly, $n_e=6$ corresponds to $1.2$ electrons per orbital on average, close to the filling of the single relevant orbital at optimal doping ($\sim 1.16$) in the cuprates \cite{RamS62322015}}. A lot of experiments \cite{TerPRB872013,HarPRL1112013,NakSC42014,EilPRL1162016,HarPRB942016} and theoretical \cite{LiePRB822010,WernerNP82012,BasPRB862012,MisPRL1082012,deMPRL1122014,CalPRB902014} results have confirmed this Mott doped scenario, see \cref{Chap03}, and for a review \cite{BasCRP172016}. \textcolor{black}{This suggests that} not only a magnetic phase suppression is needed as a prelude of the \textcolor{black}{high-$T_c$} superconducting dome, but also an optimal degree of electronic correlations.

By taking into account the optimal degree of electronic correlations and their increase when approaching half-filled systems, we propose to search for a new family of high-$T_c$ superconductors similar to FeSCs \cite{Pizarro1}. Considering that FeSCs can be seen as electron-doped Mott insulators, we looked into the other side of the half-filled Mott insulator, i.e. for the hole-doped Mott systems. By considering a similar system as the FeSCs, we propose to search for unconventional superconductivity in the $Cr$ pnictides and chalcogenides. For the same crystal structure of quasi-2D FeSCs, only the total number of electrons per atom in the $d$ orbitals will vary when changing $Fe$ atoms by $Cr$ ones, where for FeSCs is $n_e=6$ in the $5$ $d$ orbitals, while for $Cr$ systems is $n_e=4$ (in this situation, the half-filled systems are the $Mn$ pnictides and chalcogenides, where $n_e=5$ in $5$ $d$ orbitals). If unconventional superconductivity is found in these chromium-based systems\footnote{Note that I will not talk about the quasi-1D ${Cr}_3{As}_3$ systems \cite{BaoPRX52015}, but rather focus on the quasi-2D chromium pnictides, with crystal structures equivalent to the ones for FeSCs.}, the connection between cuprates and FeSCs, and the role of the electronic correlations would became clear. This could open a new route to search for new high-$T_c$ superconductors.

In the \textcolor{black}{doped} Mott scenario previously described, there is an explanation of why other transition metal (TM) pnictides and chalcogenides have failed when searching for high-$T_c$ superconductivity. For $TM=Ni$, $Pd$ or $Pt$ the system has $n_e/N=8/5$, while for $Co$, $Ir$ or $Rh$ it has $n_e/N=7/5$. In these situations, the distance from the Mott insulator at $n_e=N=5$ is too large, and hence, the strength of correlations is too small to host high-$T_c$ superconductivity.

In the next section, I will give a brief summary of the current experiments and theoretical calculations for $Mn$ and $Cr$ pnictides and chalcogenides to give a context in the actual status on the study of such systems.

\subsection{Preliminary results in $Mn$ and $Cr$ pnictides and chalcogenides}
\label{4.1.1}

\begin{figure}[h]
   \centering
   \includegraphics[width=0.8\columnwidth]{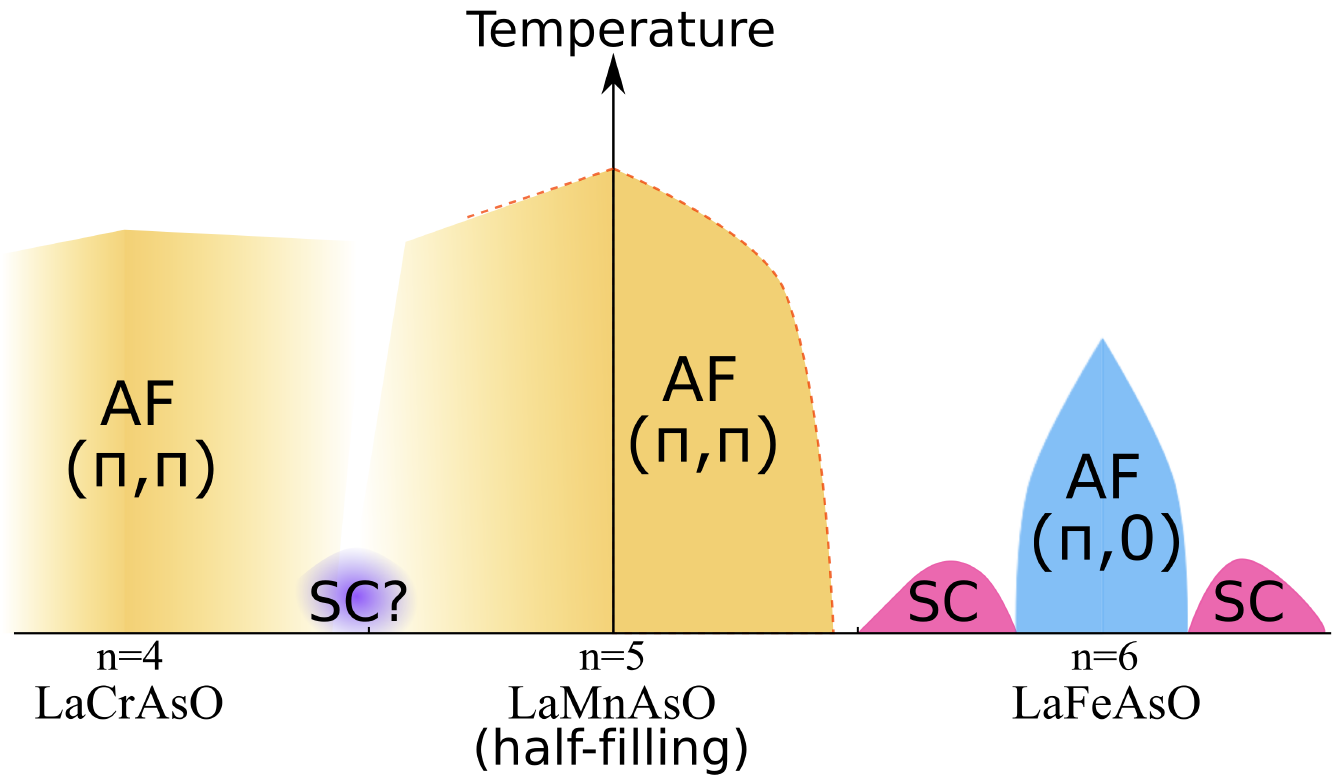}
   \caption[Sketch of the proposed unifying phase diagram for $LaTMAsO$ with $TM=Fe$, $Mn$ and $Cr$, and expected superconducting dome in electron-doped $Cr$ pnictides]{Sketch of the proposed unifying phase diagram for $LaTMAsO$ \textcolor{black}{and related compounds}, with $TM=Fe$ ($n_e=6$), $Mn$ ($n_e=5$) and $Cr$ ($n_e=4$). The different phases are reproduced from the existing literature. By comparing with cuprates and FeSCs, AFM phase in $LaCrAsO$ has to be suppress around $n_e=4.5$ in order to the superconducting dome to appear. If high-$T_c$ superconductivity is found in chromium pnictides or chalcogenides, the connection between high-$T_c$ cuprates and FeSCs will be made. This could open a route to discover new high-$T_c$ superconductors based on the Mott doped scenario.}
   \label{fig:4.1}  
\end{figure}

In this section, I summarize the available results obtained in the literature for $Mn$- and $Cr$-based compounds. I will conclude by arriving to the sketch of a proposed unified phase diagram for $LaTMAsO$, where $TM=Fe$, $Mn$ and $Cr$, \fref{fig:4.1}. 

As a brief reminder of the previous chapters, the undoped FeSCs are AFM $(\pi,0)$ (stripe AFM ordering, with spins in $x$ direction aligned parallel, while antiparallel in the $y$ direction) metals with magnetic moments $\mu \sim 0.5-1 \, \mu_B/Fe$ and N\'eel temperatures $T_N \sim 100-150 \, K$. The AFM phase can be suppressed with doping or pressure, and unconventional superconducting domes emerge, with high critical temperatures $T_c \sim 20-50 \, K$, see \fref{fig:1.11}\textbf{(c)} for the electron-doped phase diagram of $BaFe_2As_2$. From the theoretical point of view, FeSCs are multiorbital systems \cite{LebPRB752007,VilPRB782008}, where the usual tight-binding models have $5$ $d$ orbitals coming from the $Fe$ atoms in the unfolded Brillouin zone (BZ), which it is identified as the $1Fe$ atom unit cell, see \sref{1.5.2} \cite{KurPRL1012008,GraNJP112009,CalPRB802009}. The undoped compound has $n_e=6$ electrons per $Fe$ atom. In \cref{Chap03}, I have shown that FeSCs present sizable electronic correlations.

$Mn$ pnictides and chalcogenides have been extensively studied and no-superconducting dome has been found up to date \cite{SimPRB842011,SatPRB842011,SimPNAS1092012,GuoSC32013,SapSC32013,Bel1307.6404,LamPRB872013,McN1402.6312,DonJAP1152014,WilPRB902014,CalPRB892014,GuDSC42014,ZhaPRB912015,UelPRL1142015,PanPRB922015,McNPRB922015,McGPRB932016,ZhaPRB932016,ZinPRB942016}. These systems are AFM insulators, with large values for $\mu \sim 3-4 \, \mu_B/Mn$ and $T_N \sim 300-600 \, K$. The AFM phase is checkerboard inplane $(\pi,\pi)$, and it may vary between $C$- (FM order of $Mn$ layers in the $z$ direction) or $G$-type (AFM order of $Mn$ layers in the $z$ direction). In the case of $LaMnPO$ \cite{SatPRB842011}, when pressure is applied, the system first undergoes an insulator-to-metal transition (at around $15-20 \, GPa$), and then a second transition where AFM is suppressed (at around $25-30 \, GPa$). In $Ba{Mn}_2{As}_2$ \cite{SatPRB842011}, an insulator-to-metal transition can be driven when it is doped with holes (with $K$) at $x \approx 0.016$, while the AFM is not suppressed, so it is robust against doping. We assume that the electronic correlations are too large to make $Mn$ pnictides and chalcogenides able to host unconventional superconductivity.

Different $Cr$ pnictides and chalcogenides have been studied, with only few articles coming out during the last decade \cite{SinPRB792009,DinSSC1492009,HuSJPCC1142010,HuS1011.2576,MarPRB832011,ParIC522013,ParPRB892014,JiaPRB922015}. There are studies on $Ba{Cr}_2{As}_2$ \cite{SinPRB792009}, $Eu{Cr}_2{As}_2$ \cite{ParPRB892014}, $LN CrAsO$ (with $LN=La$, $Ce$, $Pr$ and $Nd$) \cite{ParIC522013}, ${Sr}_2{Cr}_3{As}_2{O}_2$ \cite{JiaPRB922015} and $CrSe$ \cite{DinSSC1492009}, and also mixed systems with layers of $CrFe$ \cite{HuSJPCC1142010,HuS1011.2576,MarPRB832011}. Here, I briefly review the most important features in these compounds.

$Ba{Cr}_2{As}_2$ \cite{SinPRB792009} is an AFM $(\pi,\pi)$ metal. There is no signal for the magnetic ordering temperature $T_N$ (up to $200 \, K$) in the specific heat and the resistivity, so there is no data available for $T_N$ or other transitions. In \cite{SinPRB792009}, the authors performed DFT calculations for the magnetic ground state, obtaining $C$-type AFM $(\pi,\pi)$ with $\mu = 2.01 \, \mu_B/Cr$. 

$LNCrAsO$ with $LN=La$, $Ce$, $Pr$ and $Nd$ \cite{ParIC522013} are AFM $(\pi,\pi)$ metals up to $300 \, K$. No data for $T_N$ is available. From DFT calculations, it was concluded that the $C$-type AFM $(\pi,\pi)$ is the most stable order. The computed magnetic moment is $\mu \sim 2.6 \, \mu_B/Cr$. In \cite{ParIC522013}, the authors doped $LaCrAsO$ with $F$ (holes) and $Mn$ (electrons), obtaining an increase/decrease of the resistitivy for $F$/$Mn$-doped compounds, with a metal-to-insulator transition found at $La{Cr}_{0.8}{Mn}_{0.2}AsO$. $LaCrAsO$ doped with $F$/$Mn$ corresponds to a hole/electron doping, respectively.

$CrSe$ \cite{DinSSC1492009} has not be synthesized yet, but it is expected that it will be stable and crystallize in the same crystal structure as $FeSe$. In \cite{DinSSC1492009}, the authors performed DFT calculations and show that both $CrSe$ and $MnSe$ present an itinerant FM ground state.

As an example, \textcolor{black}{later in this chapter,} we will pick $LaCrAsO$ as our system to study the effects of electronic correlations in $Cr$ pnictides and chalcogenides. Due to the robustness of the local correlations effects, the results explained here can be applied for any other similar $Cr$ pnictides and chalcogenides. Using $LaCrAsO$ will be also useful when making a direct analogy with the 1111 FeSCs, $LaFeAsO$.

\section{Results \& Discussion}
\label{4.2}

Here, I present the results that I have obtained and published in \cite{Pizarro1}, altogether with the ones obtained by my co-workers. In \sref{4.2.1}, I use a 5 $d$ Slater-Koster tight-binding model proposed for FeSCs, in particular for $LaFeAsO$ \cite{CalPRB802009}, to demonstrate that the increase of the strength of local correlations when approaching half-filling is a robust feature from both the electron- ($n_e>5$) and the hole-doped ($n_e<5$) part. However, the applicability to a real material of this tight-binding model in regions where $n_e$ is much different from $n_e=6$ is not realistic. So in \sref{4.2.2}, I show the DFT band structure calculation performed by my co-workers for $LaCrAsO$, and the tight-binding model obtained by performing a similar Slater-Koster fitting procedure as it was made for $LaFeAsO$ \cite{CalPRB802009}. In \sref{4.2.3}, I calculate the mass enhancement factors $m_{m}^{*}/m_0 \sim 1/Z_m$ for $LaCrAsO$ when varying the filling $4<n_e<5$, showing that the phenomenology of strength of correlations getting increase towards half-filling \textcolor{black}{is maintained}. Local correlations calculations are done both in \sref{4.2.1} and \sref{4.2.3} using SSMF $Z_2$ formalism. Finally, in \sref{4.2.4}, I show the calculations done by my co-workers on the spin susceptibility and the superconducting order parameter in the renormalized band structure. Multiorbital RPA is used with focus on singlet solutions, due to the expected AFM order as the ground state for $LaCrAsO$. We conclude by proposing that this system could host high-$T_c$ superconductivity in a similar fashion than occurrs in FeSCs.

\subsection{Electronic correlations for FeSCs around half-filling}
\label{4.2.1}

In this section, I will use the Slater-Koster tight-binding model proposed for $LaFeAsO$ in \cite{CalPRB802009} to study the strength of local correlations when the total filling varies and approaches half-filling from $n_e=6$ and $n_e=4$. In the Slater-Koster framework \cite{SlaPR941954}, the fitting to the DFT band structure \textcolor{black}{using} a tight-binding model is done in the $FeAs$ 2D layer by writting each hopping parameter $t_{mn}^{ij}$ as functions of various overlap integrals and the angle $\alpha$. $\alpha$ is defined as the angle between the $As$ atoms and the $Fe$ plane, where for a regular tetrahedron $\alpha = 35.3 \, ^{\circ}$. The overlap integrals describe the direct overlap between $d$ orbitals in neighbor $Fe$ atoms (denoted by $(dd\sigma)_{i}$, $(dd\pi)_i$ and $(dd\delta)_i$ where $i=1,2$ are the first and second neighbors, respectively) and the indirect overlap between $Fe$ $d$ and $As$ $p$ orbitals (denoted by $pd\sigma$ and $pd\pi$). 

The $t_{mn}^{ij}$ expressions  in terms of these indirect overlaps are computed in second order perturbation theory, neglecting the crystal field splitting in between $Fe$ and $As$ orbitals and in between the different $As$ orbitals. The matrix elements $t_{mn}^{ij}$ depend on the difference $|\epsilon_p - \epsilon_d|$. As an example, the intraorbital first neighbor hopping for $xy$ in the $X$ direction is written as:

\begin{equation}
t_{xy,xy}^{x} = \underbrace{\frac{1}{|\epsilon_p - \epsilon_d|} \left( - \frac{3}{2} pd\sigma^2 - 2 pd\pi^2 + 2 \sqrt{3} pd\sigma pd\pi \right)}_{indirect \, overlap} \underbrace{\cos^4{\alpha} \sin^2{\alpha}}_{angle \, term} + \underbrace{(dd\pi)_1}_{direct \, overlap}
\label{eq:4.1}
\end{equation}

Similar expressions can be obtained for all the other first and second neighbor hoppings $t_{mn}^{ij}$.

\begin{figure}[h]
   \centering
   \includegraphics[width=0.9\columnwidth]{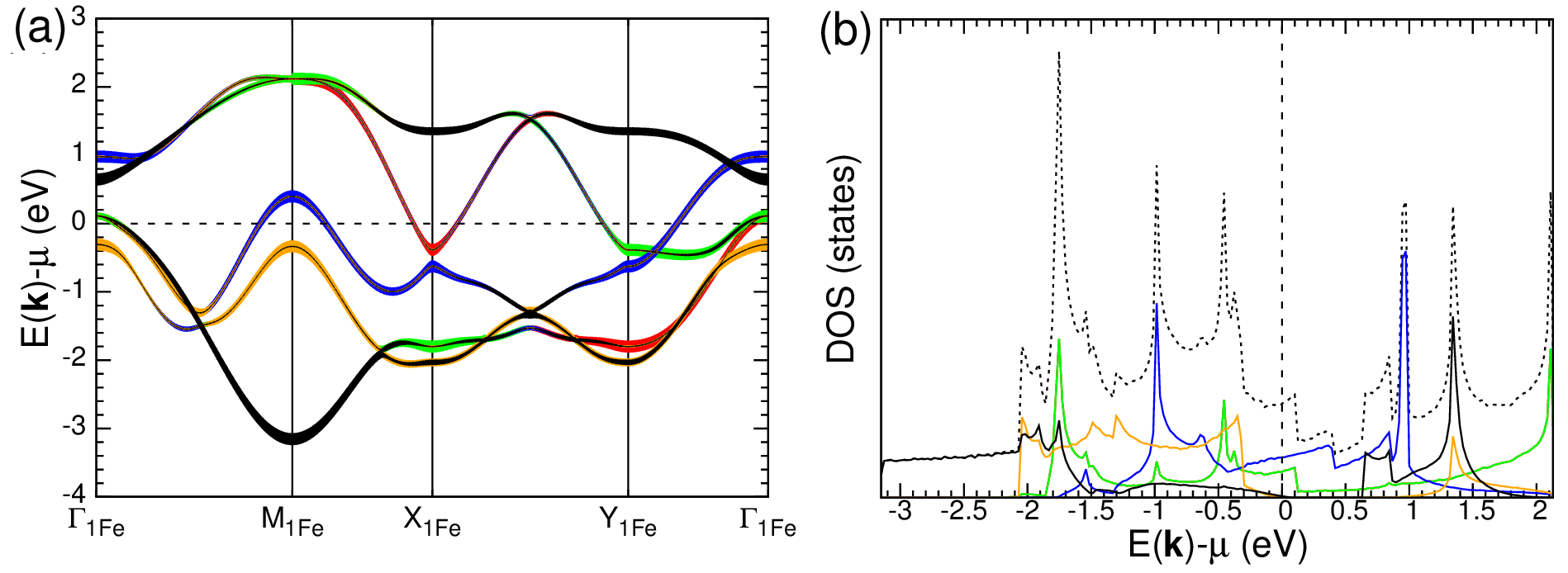}
   \caption[Slater-Koster tight-binding model for $LaFeAsO$ with $n_e=6$ and $\alpha = 35.3 \, ^{\circ}$, with orbital character included]{Slater-Koster tight-binding model for $LaFeAsO$ with $n_e=6$ and $\alpha = 35.3 \, ^{\circ}$ in the unfolded BZ, where \textbf{(a)} shows the band structure with the orbital character included and \textbf{(b)} the total DOS (dotted line) and the orbital-resolved DOS. Color code for the orbitals: red for $yz$, green for $zx$, blue for $xy$, orange for $3z^2-r^2$ and black for $x^2-y^2$. Figures obtained from the Slater-Koster tight-binding model of \cite{CalPRB802009}.}
   \label{fig:4.2}  
\end{figure}

The hopping matrix elements in the tight-binding model depend on the direct and indirect overlaps, and on the angle $\alpha$. A crystal field splitting term $\epsilon_m$ between the $Fe$ $d$ orbitals is also included in the model. These parameters can be tuned to give a similar band structure to the one obtained in DFT calculations. The $d$ orbitals are defined in the unfolded BZ, with $zx$ and $yz$ being degenerated. In \fref{fig:4.2}\textbf{(a)},\textbf{(b)}, the Slater-Koster tight-binding band structure and the DOS, with the orbital character included can be seen (color code in the caption of the figure). Overlap integrals and crystal field splittings are those considered in \cite{CalPRB802009}. This Slater-Koster model compares well with the DFT band structure (see \fref{fig:1.16}), giving a good description of the band energies and orbital content. This model was previously used in the context of FeSCs to study their magnetic properties, the optical conductivity, the Raman spectra and the anisotropy present in the FeSCs \cite{BasCRP172016}.

All the $d$ orbitals contribute for the band structure around the Fermi level. The total bandwidth is $W \sim 5.3 \, eV$, while the individual orbital bandwidths are approximately $W_{zx,yz} \approx 3.75 \, eV$, \textcolor{black}{$W_{xy} \approx 2 \, eV$}, $W_{3z^2-r^2} \approx 3.5 \, eV$ and $W_{x^2-y^2} \approx 4.5 \, eV$.  The smallest orbital bandwidth is $W_{xy}$, whether the largest is the one of $x^2-y^2$. A more extense discussion about this Slater-Koster tight-binding model can be found in \cite{CalPRB802009}.

\begin{figure}[h]
   \centering
   \includegraphics[width=0.7\columnwidth]{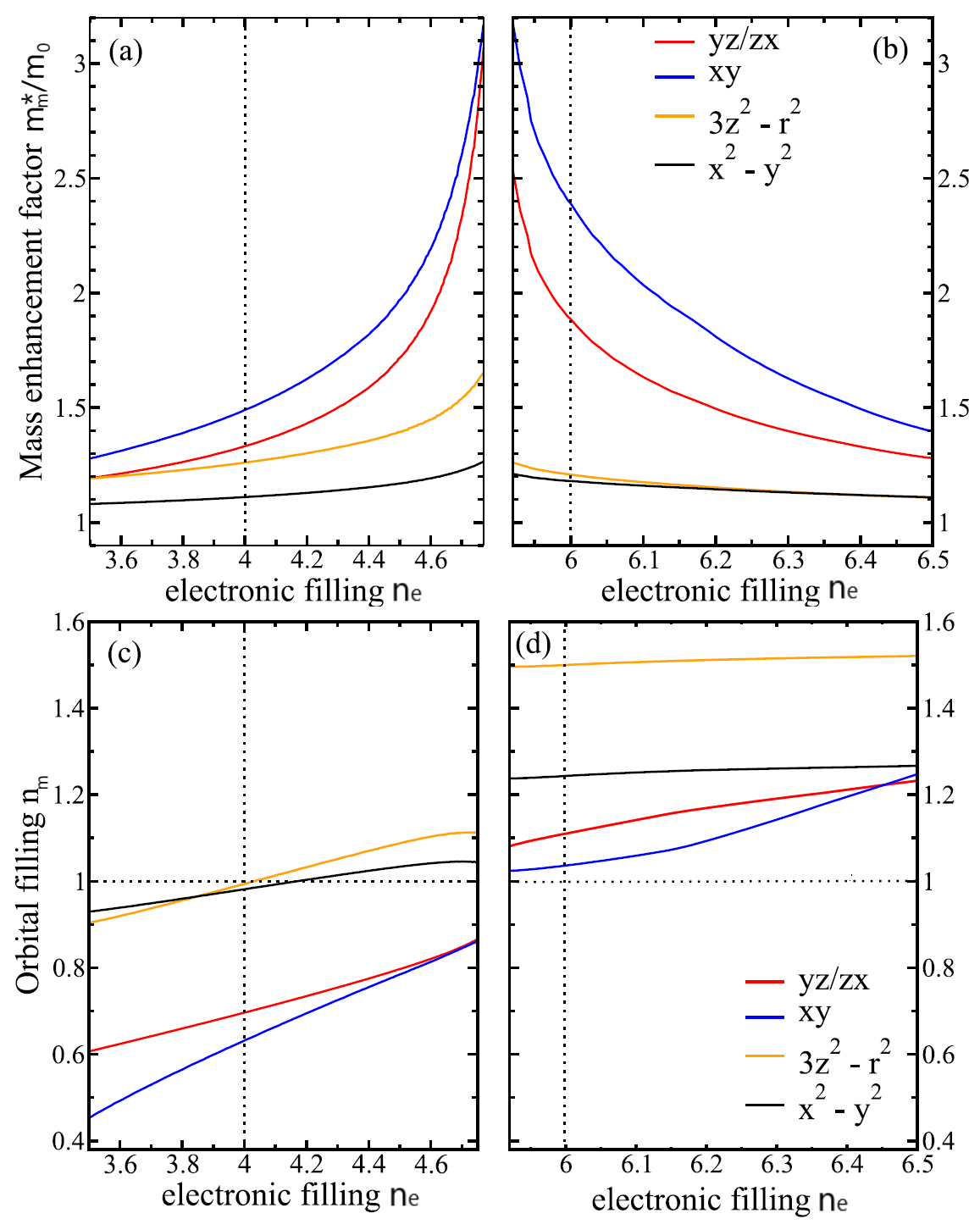}
   \caption[Orbital mass enhancement factors and orbital fillings for the Slater-Koster model of $LaFeAsO$ in terms of $n_e$ and at $U=3 \, eV$ and $J_H/U=0.25$]{\textbf{(a)}, \textbf{(b)} Orbital mass enhancement factors $m_{m}^{*}/m_0$ and \textbf{(c)},\textbf{(d)} orbital fillings $n_m$ for the Slater-Koster model of $LaFeAsO$ in terms of $n_e$ and at $U=3 \, eV$ and $J_H/U=0.25$. Dotted line marks the undoped $n_e=4$ or $n_e=6$ compound, and the half-filling line ($n_m = 1$). Color code for the orbitals: red for $yz \equiv zx$, blue for $xy$, orange for $3z^2-r^2$ and black for $x^2-y^2$.}
   \label{fig:4.3}  
\end{figure}

I now apply the SSMF $Z_2$ formalism to obtain the mass enhancement factors \textcolor{black}{$m_m^*/m_0=1/Z_m$} in terms of the number of electrons per $Fe$ atom $n_e$, keeping $U=3 \, eV$ and $J_H/U=0.25$ as constants. These are the interaction values for which the mass enhancement factors $m_m^*/m_0$ obtained by SSMF equal the ones observed experimentally in ARPES and QOs for undoped $LaFeAsO$ \cite{LuDARCMP32012}. I will vary the total filling $4<n_e<6$. The enhancement of $m_m^*/m_0$ signals the increase on the strength of correlations. In \fref{fig:4.3}, I present the mass enhancement factors $m_m^*/m_0$ and the orbital filligns $n_m$ in terms of the number of electrons $n_e$. The orbital fillings are defined as $n_m = 2 n_{m\sigma}$ where $n_{m\sigma}$ is the orbital filling per spin, hence $n_m \in [0,2]$ with $n_m=1$ defined \textcolor{black}{as} half-filling.

The strength of correlations of all orbitals increases when approaching half-filling $n_e=5$ from both sides, as expected from the Mott doped scenario. The most correlated orbitals, \textcolor{black}{i.e. larger $m_m^*/m_0$}, are $xy$ and $zx/yz$ , while $3z^2-r^2$ and $x^2-y^2$ are the less correlated ones. Interestingly, in the hole-doped part (around $n_e=4$), the orbitals \textcolor{black}{closer to} half-filling ($n_m=1$) \textcolor{black}{are not} the most correlated ones, due to the effect of the effective kinetic energies given by the orbital bandwidths $W_m$, see \cref{Chap03}.

The situation when comparing $3z^2-r^2$ and $zx/yz$ might be troublesome when looking to the lowest $W_m$ and the closeness to half-filling of each $n_m$. In this case, the most correlated orbitals are $xz/yz$, even if they have a larger $W_m$ and they are further away from half-filling. This is due to the fact that at the considered interaction values $U$ and $J_H$, the system is in the crossover to the Hund metal regime, and the orbitals are still coupled. For larger $U$, the system will show an orbital decoupling behavior, and hence the strength of correlations will follow the lowest $W_m$ and the closest to half-filling $n_m$ expectations, as explained in \cref{Chap03}.

\begin{figure}[h]
   \centering
   \includegraphics[width=0.8\columnwidth]{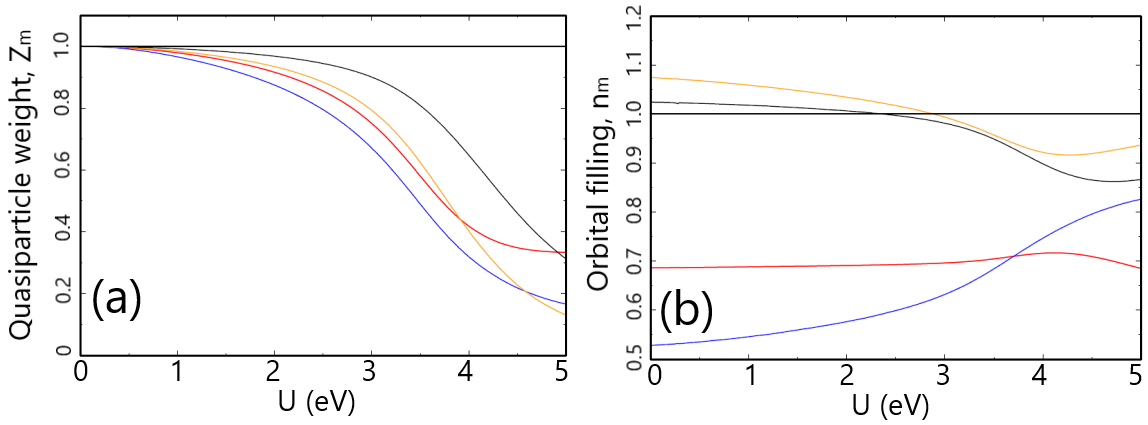}
   \caption[Orbital quasiparticle weights and orbital fillings for the Slater-Koster model of $LaFeAsO$ in terms of $U$ and at $n_e=4$ and $J_H/U=0.25$]{\textbf{(a)} Orbital quasiparticle weights $Z_m$ and \textbf{(b)} orbital fillings $n_m$ for the Slater-Koster model of $LaFeAsO$ in terms of $U$ and at $n_e=4$ and $J_H/U=0.25$. Deep in the Hund metal regime ($U > 4 \, eV$), $Z_m$ evolution follows the lowest $W_m$ and the closeness to half-filling behavior. Same color code as in \fref{fig:4.3}.}
   \label{fig:4.4}  
\end{figure}

In \fref{fig:4.4}, I have plotted the $Z_m$ and $n_m$ evolution for $n_e=4$ and $J_H/U=0.25$ in terms of $U$. Once the system is deep in the Hund metal regime ($U>4.5 \, eV$), $Z_m$ follows the expectations for the lowest $W_m$ and the closest to half-filling behavior. \textcolor{black}{Above $U>4\, eV$, $3z^2-r^2$ is more correlated than $zx$, $yz$, and for} $U>4.5 \, eV$, $3z^2-r^2$ becomes the most correlated orbital, due to its closeness to half-filling.

Thus, I have demonstrated that the Mott doped scenario is valid when approaching from both sides of half-filling for a generic tight-binding model for FeSCs. However, the applicability of this tight-binding model for $n_e=4$ is not realistic. Then, in the next section I will present the DFT band structure and the Slater-Koster tight-binding model for $Cr$-based systems, specifically for $LaCrAsO$.

\newpage

\subsection{DFT band structure and tight-binding model for $LaCrAsO$}
\label{4.2.2}

\begin{figure}[h]
   \centering
   \includegraphics[width=0.85\columnwidth]{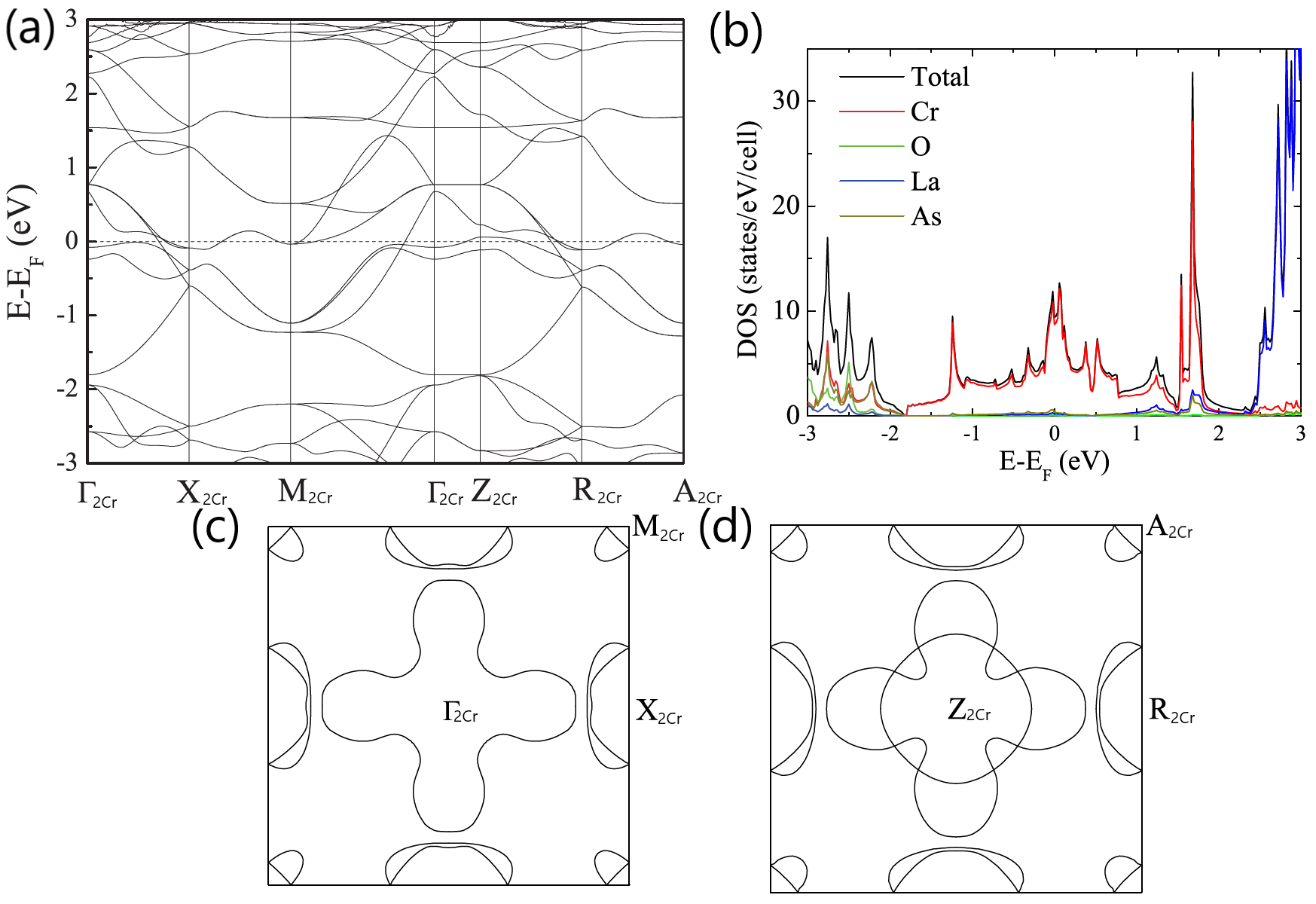}
   \caption[DFT calculations obtained by VASP code (band structure, DOS and Fermi surfaces) for tetragonal $P4/nmm$ $LaCrAsO$ in the folded BZ]{DFT calculations obtained by VASP code for tetragonal $P4/nmm$ $LaCrAsO$, where the crystal structure is fixed to the experimental parameters ($a=b=4.0412 \, \AA$, $c=8.9863 \, \AA$). \textbf{(a)} Band structure, \textbf{(b)} DOS with the atomic character included and \textbf{(c)} $k_z=0$and \textbf{(d)} $k_z=\pi$ Fermi surfaces. Figures given in the folded BZ. See main text for the discussion.}
   \label{fig:4.5}  
\end{figure}

In this section, I will show the DFT band structure calculations done by my co-workers, using the Vienna ab-initio Simulation Package (VASP), with the GGA exchange-correlation potential, see \aref{AppA}. The crystal structure of $LaCrAsO$ is tetragonal (space group $P4/nmm$), same as in $LaFeAsO$, see \fref{fig:1.15}, with crystal parameters fixed at the experimental values ($a=b=4.0412 \, \AA$, $c=8.9863 \, \AA$). The DFT band structure is given in the folded BZ, i.e. in the $2Cr$ atoms unit cell. In \fref{fig:4.5}, I show the DFT band structure, the DOS with the atomic character included and the Fermi surfaces at $k_z=0$ and $k_z=\pi$.


The band structure for $Cr$ atoms extends between $-2 \, eV$ and $2.5 \, eV$ ($W \approx 4.5 \, eV$), with $Cr$ $d$ orbitals mostly contributing in this energy window (and a negligible contribution from $La$ presents around $\sim 1.5 \, eV$). The band structure is quite similar to those obtained for $Mn$ and $Fe$ isostructural compounds. The Fermi surface is quasi-2D. Thus, $LaCrAsO$ can be modeled as a quasi-2D system where $CrAs$ layers control the physics in it. The Fermi surface is composed of a flower-shaped hole pocket at $\Gamma_{2Cr}$, electron pockets at $X_{2Cr}/Y_{2Cr}$ and shallow electron pockets at $M_{2Cr}$. Note that this topology will change when the unfolding of the BZ is performed (see \sref{1.5.2}). The folded and unfolded BZ will be identified with the subscript $2Cr$ and $1Cr$, as a reference to the number of atoms per unit cell.

\begin{figure}[h]
   \centering
   \includegraphics[width=0.9\columnwidth]{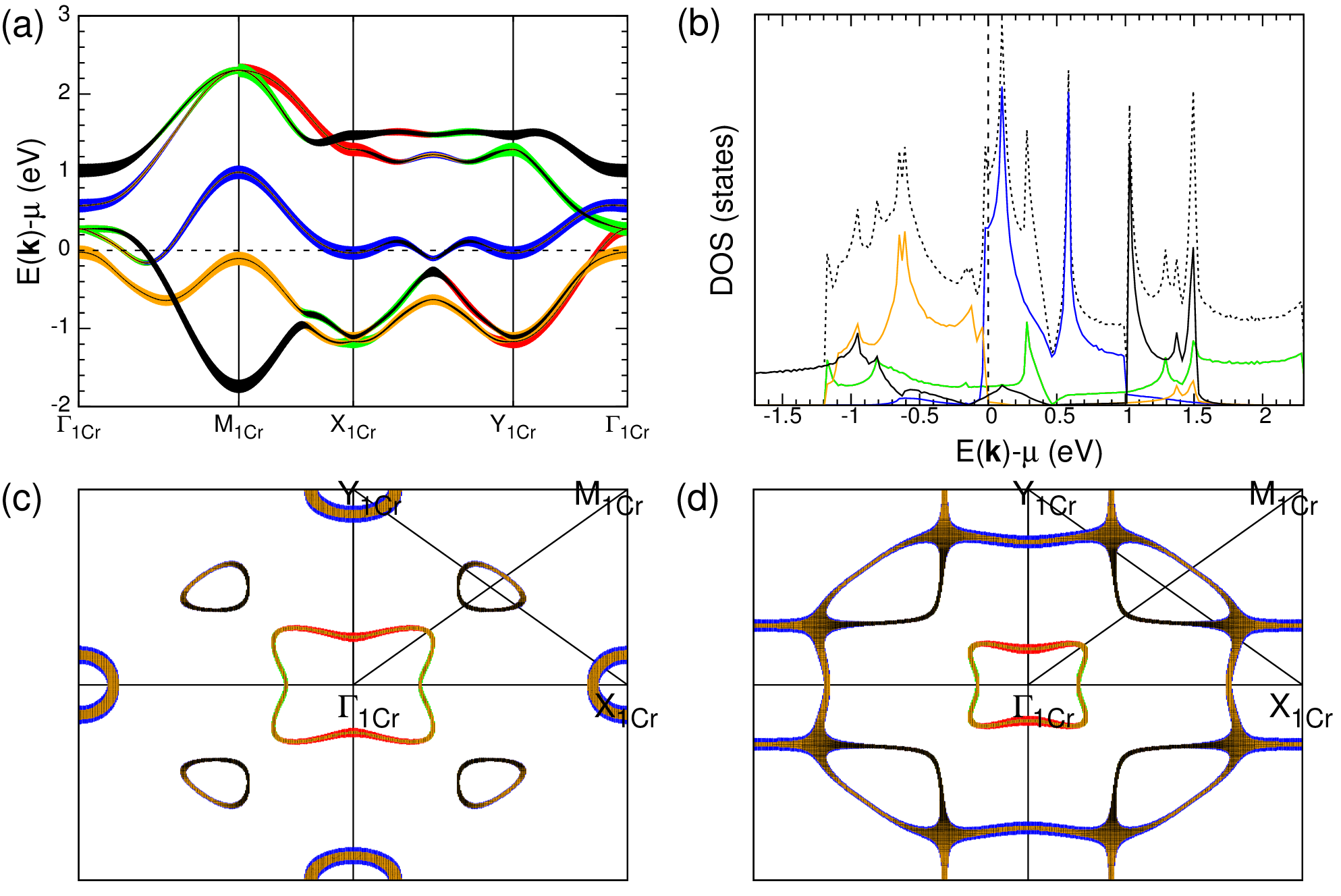}
   \caption[Slater-Koster tight-binding model for $LaCrAsO$ with $n_e=4$ and $\alpha = 35.26 \, ^{\circ}$, with orbital character included. Fermi surfaces at $n_e=4$ and $n_e=4.5$ with orbital character included]{Slater-Koster tight-binding model for $LaCrAsO$ with $\alpha = 35.3 \, ^{\circ}$ in the unfolded BZ, where \textbf{(a)} shows the band structure with the orbital character included and \textbf{(b)} the total DOS (dotted line) and the orbital-resolved DOS, and \textbf{(c)} $n_e=4$ and \textbf{(d)} $n_e=4.5$ show the Fermi surfaces with orbital character included. At $n_e=4.5$, there is a Lifshitz transition between the electron pockets at $X_{1Cr}/Y_{1Cr}$ and $(\pi/2,\pi/2)$. Color code for the orbitals: red for $yz$, green for $zx$, blue for $xy$, orange for $3z^2-r^2$ and black for $x^2-y^2$.}
   \label{fig:4.6}  
\end{figure}

For the tight-binding model, the same 2D Slater-Koster fitting than for $LaFeAsO$ \cite{CalPRB802009} was used, with some changes. $\alpha = 35.26 \, ^{\circ}$, the crystal fields splittings are set to $\epsilon_{zx,yz}=0$, $\epsilon_{xy}=-0.3$, $\epsilon_{3z^2-r^2}=-0.9$ and $\epsilon_{x^2-y^2}=-0.48$, the overlap integrals are set to $pd{\sigma}=1.12$, $pd{\pi}=-0.79$, $dd\sigma_1=-0.42$, $dd\pi_1=0.36$, $dd\delta_1=-0.12$ and $dd\sigma_2=-0.024$, and $\epsilon_d - \epsilon_p = 3$, all in $eV$ units. The quantity $0.2$ is substracted from the first-nearest neighbor hoppings $t_{yz,yz}^{i,i+1}$ and symmetry-related ones. The tight-binding band structure and the Fermi surface are well reproduced, see \fref{fig:4.6}. Note that these calculations are done in the unfolded BZ.

The Slater-Koster tight-binding model for $LaCrAsO$ has a total bandwidth $W \approx 4 \, eV$ , where the most striking feature is that $xy$ is much more concentrated in a smaller energy window than in $LaFeAsO$ (see \fref{fig:4.2}). The orbital bandwidths can be estimated as $W_{zx,yz} \approx 3.5 \, eV$, $W_{xy} \approx 2 \, eV$, $W_{3z^2-r^2} \approx 2.5 \, eV$ and $W_{3z^2-r^2} \approx 3 \, eV$. In \fref{fig:4.6}\textbf{(c)} and \textbf{(d)}, I show the Fermi surfaces for $n_e=4$ and $n_e=4.5$, respectively. For $n_e=4$, a central $\Gamma_{1Cr}$ hole-pocket (a bit distorted flower-shaped) with mainly $zx$ and $yz$ character, and electron pockets at $X_{1Cr}$ and $Y_{1Cr}$ with mainly $xy$ character and other electron pockets centered at $ \approx (\pi/2,\pi/2)$ with $3z^2-r^2$ and $x^2-y^2$ character are obtained. Both the DFT and the Slater-Koster model calculations compares \textcolor{black}{reasonably} well.

\subsection{Electronic correlations in $LaCrAsO$}
\label{4.2.3}

Once the tight-binding is obtained, I can calculate the mass enhancement factors $m_m^*/m_0$ in $LaCrAsO$ when varying $n_e$. I apply the SSMF $Z_2$ formalism \cite{deMPRB722005,HasPRB812010} fixing $U$ and $J_H/U$ values, see \aref{AppB}. There are not available experimental results for $Z_m$ to compare with and obtain $U$ and $J_H$ values. Thus, I will use the same $U=3 \, eV$ and $J_H/U=0.25$ values as in $LaFeAsO$ \cite{BasCRP172016}, by considering the similarities between both compounds. I will vary $n_e$ from $3.5$ to $4.5$ to obtain $m_m^*/m_0$, and I will also calculate the renormalized band structures and Fermi surfaces.

\begin{figure}[h]
   \centering
   \includegraphics[width=0.9\columnwidth]{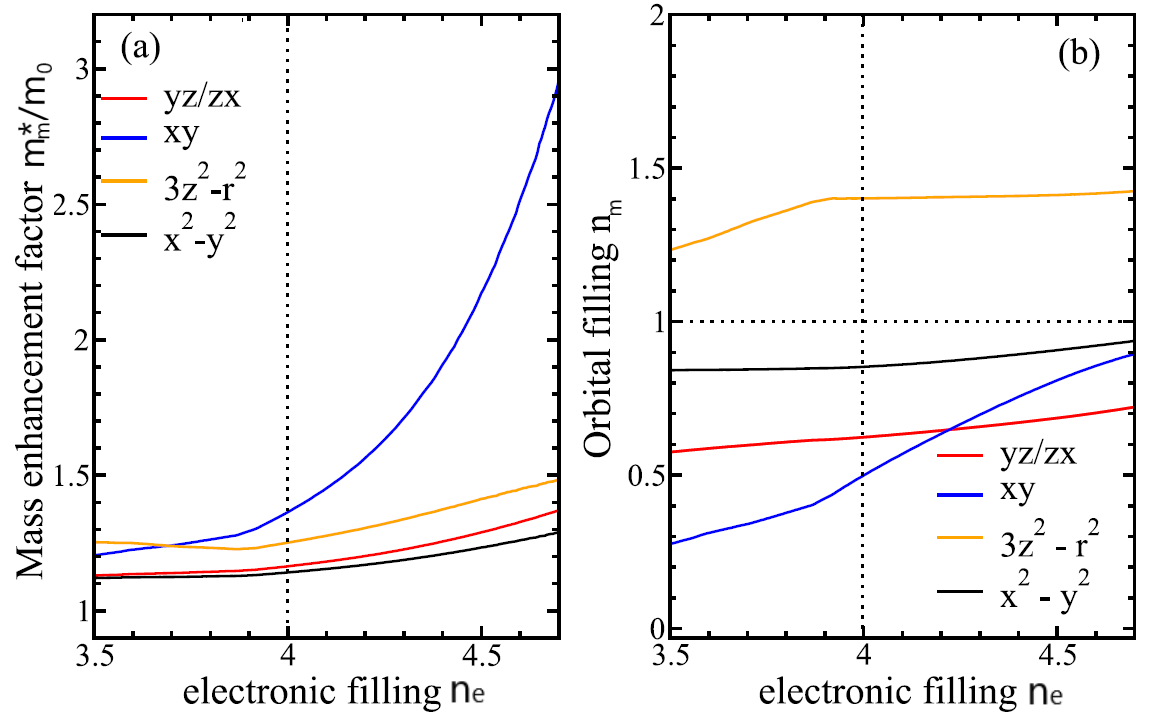}
   \caption[Orbital mass enhancement factors and orbital fillings for $LaCrAsO$ in terms of $n_e$ for $U=3 \, eV$ and $J_H/U=0.25$]{\textbf{(a)} Orbital mass enhancement factors $m_m^{*}/m_0$ and \textbf{(b)} orbital fillings $n_m$ in terms of $n_e$ for the tight-binding model for $LaCrAsO$ explained in \sref{4.2.2}, and $U=3 \, eV$ and $J_H/U=0.25$. Color code included in the legend of the figure.}
   \label{fig:4.7}  
\end{figure}

\fref{fig:4.7} shows the mass enhancement factors $m_m^*/m_0$ and orbital fillings $n_m$ for $LaCrAsO$ when varying $n_e$. I obtained the similar trend as in $LaFeAsO$ (\fref{fig:4.3}). \textcolor{black}{$LaCrAsO$ at $n_e=4$} is less correlated \textcolor{black}{than $LaFeAsO$ at $n_e=6$}, but at larger filling, the correlations get of the order of \textcolor{black}{undoped} $LaFeAsO$. For $n_e \sim 4.5$, the strength of correlations in $LaCrAsO$ is similar to those of undoped ($n_e=6$) $LaFeAsO$. $xy$ is the most correlated orbital, due to its reduced bandwidth $W_{xy}$. In $LaCrAsO$ the order between different orbitals for the strength of correlations changes when compared with $LaFeAsO$. This reorganization will come from the different tight-binding model used (hence the different $W_m$ and $n_m$ values).

The dependence of the strength of correlations when doping $LaCrAsO$ towards half-filling is similar to the one found in FeSCs, \textcolor{black}{so} it behaves as a hole-doped Mott insulator with an increase of the strength of electronic correlations when approaching half-filling. Even if $xy$ \textcolor{black}{orbital filling} is the \textcolor{black}{furthest} away from half-filling at $n_e=4$, it is the most correlated orbital due to its very small $W_{xy}$.

Local correlations will alter the Fermi surface topology in a non-negligible way. $\Gamma_{1Cr}$ hole pocket becomes square-like for all the fillings $4<n_e<4.5$, while $X_{1Cr}/Y_{1Cr}$ electron pockets become really shallow for $n_e<4.3$ (see next section). In the next section, I will explain my co-workers calculations for the spin susceptibility $\chi_{spin}^{RPA}$ and the superconducting order parameter $g(\vec{k})$, by means of multiorbital RPA \cite{TakPRB692004,KubPRB752007,GraNJP112009} in the spin fluctuation framerwork \cite{HirRPP742011,HirCRP172016}. \textcolor{black}{Like in FeSCs, the} applicability of the spin fluctuation theory is still under a debate \cite{BasCRP172016}, as discussed in \cref{Chap01}. In these calculations, the band structure and Fermi surface used to calculate the spin fluctuations is renormalized using the values calculated in \fref{fig:4.7}, as a better approach to the real band structure and Fermi surfaces of $LaCrAsO$.

\textcolor{black}{Due to the strength of electronic correlations, we expect that the superconducting dome is found in the range $n_e \sim 4.4-4.7$. In the next section, we considered that the superconducting emerges around $n_e=4.5$, and study the magnetic and superconducting instability at this filling (as well as around it).}

\subsection{Spin susceptibility and superconducting order parameter for $LaCrAsO$}
\label{4.2.4}


To search for the magnetic and superconducting instabilities, the system is considered a metal with a renormalized band structure. In this situation, spin fluctuation theory \cite{HirRPP742011,HirCRP172016} (i.e. superconductivity driven by spin fluctuations) can be applied for the renormalized band structure, as it will be a better approximation for the real band structure and Fermi surfaces for $LaCrAsO$. In this section, I will show the calculations done by my co-workers when applying multiorbital RPA \cite{TakPRB692004,KubPRB752007,GraNJP112009} to calculate the spin susceptibility $\chi_{spin}^{RPA}$ and the superconducting order parameter $g(\vec{k})$.

Only singlet solutions are considered, due to the expected AFM $(\pi,\pi)$ ground state for $LaCrAsO$ \cite{ParIC522013}. The magnetic ground state will be recognised by a divergence in the spin susceptibility $\chi_{spin}^{RPA}$. The interaction parameter $U$ assumed in RPA is not the same as in other techniques, such as in SSMF or DMFT, due to the weakly correlated character of RPA (see \fref{fig:3.13}). The interaction values in RPA are labeled as $U_{RPA}$. As explained in \cref{Chap01}, the correct symmetry for the superconducting order parameter is given by the largest eigenvalue of the pairing interaction vertex $\Gamma_{\nu \mu} (\vec{k},\vec{k}')$, with $\Gamma$ being the scattering matrix between two $k$ points in the BZ and for bands $\nu$ and $\mu$. Thus, the largest eigenvalue will give the most important scattering process for the pairing instability. The pairing vertex $\Gamma_{\nu \mu}$ is a matrix which depends on the interaction parameters, and on the spin and charge susceptibilities.

\begin{figure}[h]
   \centering
   \includegraphics[width=0.9\columnwidth]{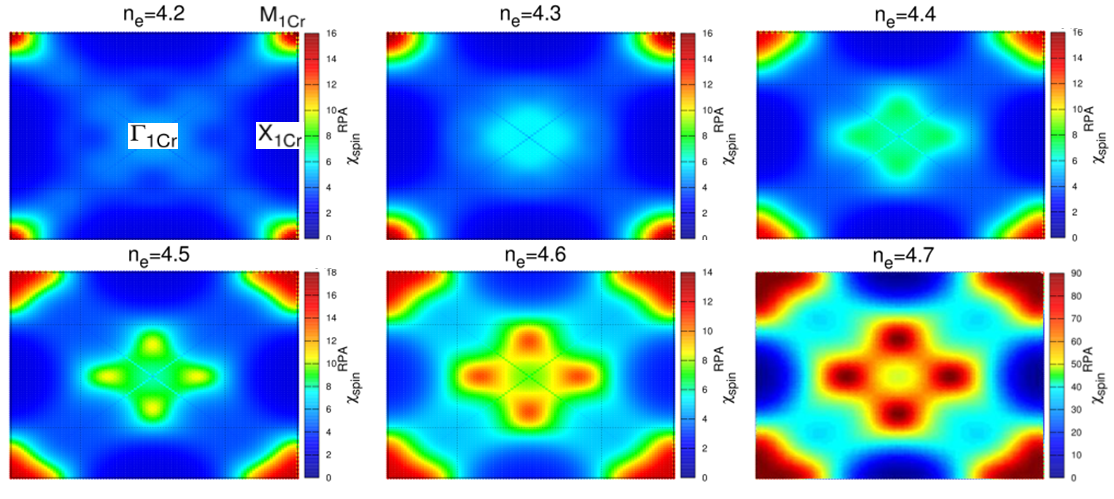}
   \caption[Spin susceptibility $\chi_{spin}^{RPA}$ calculated by multiorbital RPA in the renormalized Fermi surface of doped $LaCrAsO$ at various $n_e$]{Spin susceptibility $\chi_{spin}^{RPA}$ calculated by multiorbital RPA in \textcolor{black}{doped} $LaCrAsO$ at various $n_e$. $\chi_{spin}^{RPA}$ peaks are colored by red, while negligible $\chi_{spin}^{RPA} \sim 0$ regions are shown in blue. Figures taken for $U_{RPA}<U_{RPA}^c$. $\chi_{spin}^{RPA}$ diverges at $U_{RPA}^c$ for $\vec{Q}=(\pm \pi,\pm \pi)$, signaling a AFM $(\pi,\pi)$ ground state, as expected from experiments. For $n_e>4.4$, secondary peaks at $\vec{Q}_2=(\pm \pi/2,0)$ (and $(0,\pm \pi/2)$) appear. At $n_e=4$, $\chi_{spin}^{RPA}$ peaks at $\vec{Q}' \approx (\pm 3\pi/4,0)$ (and $(0,\pm 3\pi/4)$). \textcolor{black}{The electron} pockets at $X_{1Cr}/Y_{1Cr}$ \textcolor{black}{play a major role} in the formation of the AFM $(\pi,\pi)$ ground state for $LaCrAsO$.}
   \label{fig:4.8}  
\end{figure}

\fref{fig:4.8} shows the \textcolor{black}{$n_e \geq 4.2$} evolution of the spin susceptibility $\chi_{spin}^{RPA}$ in the unfolded BZ for the renormalized band structure. Red color marks a peak in $\chi_{spin}^{RPA}$. Figures are taken for $U_{RPA}$ slightly smaller than the critical interaction $U_{RPA}^c$ at which $\chi_{spin}^{RPA}$ diverges. The relative height of the peaks depends on $n_e$ and $U_{RPA}$. $\chi_{spin}^{RPA}$ diverges for $\vec{Q}=(\pm \pi,\pm \pi)$ at $U_{RPA}^c$, signaling a checkerboard AFM ground state, as obtained in experiments. \textcolor{black}{The electron pockets at $X_{1Cr}/Y_{1Cr}$ play a major role in the formation of this AFM $(\pi,\pi)$ ground state for $LaCrAsO$.}

\begin{figure}[h]
   \centering
   \includegraphics[width=0.9\columnwidth]{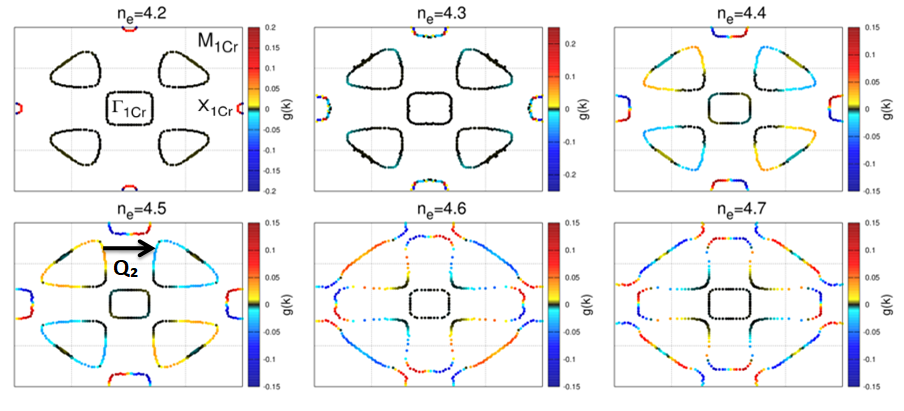}
   \caption[Superconducting order parameter $g(\vec{k})$ calculated by multiorbital RPA in the renormalized Fermi surface of $LaCrAsO$ at various $n_e$]{Superconducting order parameter $g(\vec{k})$ calculated by multiorbital RPA in the renormalized Fermi surface of $LaCrAsO$ at various $n_e$. Local correlations modified the Fermi surface, with $\Gamma_{1Cr}$ hole pocket becoming square-like for all the fillings $n_e<5$, and $X_{1Cr}/Y_{1Cr}$ electron pockets becoming really shallow for $n_e<4.3$. Except for $n_e<4.3$, the maximum values of $g(\vec{k})$ occur in the tips of $X_{1Cr}$ and $Y_{1Cr}$ electron pockets.  $\vec{Q}_2$ (and not $\vec{Q}$) is the responsible of the leading superconducting order parameter symmetry. This occur due to an enhanced scattering response for the tips of $X_{1Cr}/Y_{1Cr}$ electron pockets which are close to the Lifshitz transition. At $n_e = 4.5$, the superconducting order parameter symmetry is $d_{xy}$.}
   \label{fig:4.9}  
\end{figure}

\textcolor{black}{In electron-doped $LaCrAsO$}, there are secondary peaks appearing at $\vec{Q}_2 \approx (\pm \pi/2,0)$ (and $(0,\pm \pi/2)$) \textcolor{black}{for $n_e \geq 4.4$}, where \fref{fig:4.9} at $n_e=4.5$ shows this scattering vector $\vec{Q}_2$. The presence of \textcolor{black}{these} secondary peaks signal the competing magnetic instabilities. Note that a Lifshitz transition happens between electron pockets at $X_{1Cr}/Y{1Cr}$ and at $(\pm \pi/2,\pm \pi/2)$ \textcolor{black}{occurs} at $n_e \sim 4.5-4.6$, see \fref{fig:4.9}.

In \fref{fig:4.9}, I show the superconducing order parameter $g(\vec{k})$ for the same $n_e$ values in the renormalized Fermi surface as in \fref{fig:4.8} and interaction parameters $U_{RPA}$. Red and blue colors marks the different sign of $g(\vec{k})$.

$g(\vec{k})$ shows a complex evolution with $n_e$. For interactions $U_{RPA} < U_{RPA}^{c}$, the largest eigenvalue has a $d_{xy}$ symmetry ($g(\vec{k})$ changes sign at the $x$ and $y$ axis), where the maximum amplitude of $g(\vec{k})$ is found in the tips of the $X_{1Cr}$ and $Y_{1Cr}$ electron pockets. Suprisingly, the leading nesting vector for the superconducting instability is given by $\vec{Q}_2$, and not $\vec{Q}$, at which $\chi_{spin}^{RPA}$ diverges for $U_{RPA}^{c}$ (it was expected to give a major contribution). The enhanced response at $\vec{Q}_2$ originates between the tips of the electron pockets at $X_{1Cr}$ and $Y_{1Cr}$ (I signaled $\vec{Q}_2$ in \fref{fig:4.9} at $n_e=4.5$ for clarification purposes), due to the proximity of the Lifshitz transition.

The discussion above for $\chi_{spin}^{RPA}$ and $g(\vec{k})$ was made for the renormalized Fermi surfaces, which show really shallow electron pockets at $X_{1Cr}$ and $Y_{1Cr}$. For other fittings, other interaction parameters $U$ and $J_H$ (and then, different renormalizations $m_m^*/m_0$) and possibly other $Cr$ related compounds, the preferred superconducting symmetry might vary, \textcolor{black}{but it will most probably be $d$-wave due to the peak in $\chi_{spin}^{RPA}$ at $\vec{Q}$. Finding} the superconducting instability due to spin fluctuations is possible in these systems, and the details \textcolor{black}{will only} vary the \textcolor{black}{specific $d$-wave} superconducting order parameter symmetry, but not the possibility of finding a solution for $g(\vec{k})$.

\section{Summary}
\label{4.3}

I conclude this chapter by making a summary of the obtained results:

\begin{itemize}

\item Considering the Mott doped scenario \cite{IshPRB812010}, we proposed to search for a new family of unconventional superconductors, based on chromium instead of iron. For the same crystal structure, only the total number of electrons per atom in the $d$ orbitals will vary when changing $Fe$ atoms by $Cr$ ones, where for FeSCs is $n_e=6$ in the $5$ $d$ orbitals, while for $Cr$ systems is $n_e=4$ (in this situation, the half-filled systems are the $Mn$ pnictides and chalcogenides, where $n_e=5$ in $5$ $d$ orbitals). As an example, we have chosen $LaCrAsO$ to study the evolution of the strength of local correlations, and also the leading AFM and superconducting order. We expect that the results obtained here could be applied in any $Cr$ related pnictide or chalcogenide. $LaCrAsO$ is an isostructural compound of $LaFeAsO$ \cite{ParIC522013}, so the crystal structure and crystal parameters are similar, and the main difference relies in the total filling $n_e$, with $Cr$-based compounds describing the hole-doped Mott insulating part.

\item In \sref{4.2.1}, I demonstrated that the strength of correlations increases when approaching to half-filling from both electron- and hole-doped Mott insulators. I used a Slater-Koster tight-binding model proposed for FeSCs, in particular for $LaFeAsO$ \cite{CalPRB802009}, keeping $U=3 \, eV$ and $J_H/U=0.25$ as constants. These are the interaction values for which the mass enhancement factors $m_m^*/m_0$ obtained by SSMF equal the ones observed experimentally in ARPES and QOs for undoped $LaFeAsO$ \cite{LuDARCMP32012}. I \textcolor{black}{varied} $4<n_e<6$. I demonstrated that the Mott doped scenario still holds for the hole-doped part.

\item However, using a tight-binding model for $LaFeAsO$ in the $n_e=4$ region does not seem too realistic, then a realistic tight-binding model to describe $LaCrAsO$ is needed. In \sref{4.2.2}, my co-workers calculated the ab-initio DFT band structure and the Slater-Koster tight-binding model for $LaCrAsO$. The band structure is quite similar to $LaFeAsO$ (with $W_{LaCrAsO} \leq W_{LaFeAsO}$), but the topology of the Fermi surface has changed. In $LaCrAsO$, the $xy$ is concentrated in a smaller energy window ($W_{xy} \approx 2 \, eV$). The Fermi surface consists in a hole-pocket centered at $\Gamma_{1Cr}$ and electron pockets centered at $X_{1Cr}/Y_{1Cr}$ and $\approx (\pm \pi/2, \pm \pi/2)$. The electron pockets $X_{1Cr}/Y_{1Cr}$ and $(\pm \pi/2, \pm \pi/2)$ are near of a Lifshitz transition at $n_e \sim 4.5$.

\item In \sref{4.2.3}, I calculated the evolution of the strength of correlations with the total filling for $LaCrAsO$ in $3.5<n_e<4.5$. The values $m_m^*/m_0$ show a similar trend as in $LaFeAsO$ when approaching half-filling. The strength of correlations \textcolor{black}{is smaller in electron-doped $LaCrAsO$ at $n_e<5$ than for hole-doped $LaFeAsO$ at $n_e>5$.} Around $n_e \sim 4.5$, $LaCrAsO$ has similar values for $m_m^*/m_0$ as for undoped $LaFeAsO$. \textcolor{black}{Due to this similarity, we assumed that this would be the doping at which superconductivity is more likely to appear.}

\item In \sref{4.2.4}, my co-workers calculated the spin susceptibility and the superconducting order parameter using multiorbital RPA, in the framework of the spin fluctuation theory. They used the renormalized band structure as a better approximation for the real band structure. For $U_{RPA}^c$, the peaks at $\vec{Q}$ diverge, signaling the $(\pi,\pi)$ AFM ground state as found in experiments \cite{ParIC522013}. Other AFM instabilities are present, like $\vec{Q}_2=(\pi/2,0)$ and $(0,\pi/2)$. These peaks at $\vec{Q}$ and $\vec{Q}_2$ are interaction and filling dependent. My co-workers argued that $X_{1Cr}/Y_{1Cr}$ electron pockets are controlling the magnetic and superconducting instabilities.

\item Interestingly, for $U_{RPA}<U_{RPA}^c$ at $n_e=4.5$, the largest eigenvalue for the superconducting order parameter is given by $\vec{Q}_2$ (and not by the AFM nesting vector $\vec{Q}$) and it has a $d_{xy}$ symmetry. This enhanced response at $\vec{Q}_2$ originates between the tips of the electron pockets at $X_{1Cr}$ and $Y_{1Cr}$. This enhancement appears due to the proximity of the Lifshitz transition at $n_e \sim 4.5-4.6$. The symmetry $d_{xy}$ found for $LaCrAsO$ might vary for a different tight-binding model or a different choice of interacting parameters $U$ and $J_H$ (hence the renormalized Fermi surface will change). However, we expect that a solution for the superconducting order parameter can be still found for $n_e \sim 4.4-4.7$, \textcolor{black}{maintaining the $d$-wave character due to the peak in $\chi_{spin}^{RPA}$ at $\vec{Q}$.}

\end{itemize}

I would like to conclude by giving a few comments on other related articles that came out after the publication in arXiv of the present work \cite{EdePRB952017,FilPRB952017,NayPNAS1142017,RicPRB952017,Jis1807.03041}. Simultaneously, another article \cite{EdePRB952017} proposing $Ba{Cr}_2 {As}_2$ as a new high-$T_c$ superconductor was published on arXiv. This article motivates the search by following the same ideas that we proposed, and the results presented in it are consistent with ours. They focused on the idea that an enhancement of the electronic compressibility (defined by $\kappa_{el} = dn/d\mu$) will drive a charge instability, hence a possible superconducting instability. A few experimental \cite{FilPRB952017,NayPNAS1142017,RicPRB952017} and theoretical \cite{Jis1807.03041} articles about $Ba{Cr}_2 {As}_2$ have been published afterwards. They pointed out a remarkable reduction of the effect of local correlations in this compound, due to the influence of the larger hybridization of $Cr$ $d$ orbitals with $As$ $p$ orbitals, as compared with $Ba{Fe}_2 {As}_2$. \textcolor{black}{We have also found the same trend.}


\chapter{{\bf Strong electronic correlations and Fermi surface reconstruction in the quasi-one dimensional iron superconductor $BaFe_2 S_3$\footnote{Results presented in this Chapter published in \cite{Pizarro2}}}} 
\label{Chap4b}
\lhead{Chapter \ref{Chap4b}. \emph{Strong correlations and Fermi surface reconstruction in $BaFe_2S_3$}} 

\begin{small}

In \cref{Chap01}, I have emphasized that iron-based superconductors (FeSCs) are usually metals, in contrast with cuprates which are Mott insulators. \textcolor{black}{Recently, in 2015, a new} family of FeSCs \textcolor{black}{was discovered, showing} insulating behavior which persists for $T>T_N$, i.e. for $T$ larger than the antiferromagnetic (AFM) ordering temperature $T_N$. By applying pressure, there is an insulator-to-metal transition, the AFM order is suppressed and superconductivity emerges, all around the same pressure. These systems are the 123 two-leg ladder family of iron chalcogenides. \textcolor{black}{The peculiarity of these 123 FeSCs is that} the crystal structure is similar to other FeSCs (see \fref{fig:1.15}), but with every third column of $Fe$ atoms in the $Fe$ layer is absent, showing a quasi-1D crystal structure. $Ba{Fe}_2 {S}_3$ shows a $T_c^{max} \sim 24 \, K$ for $P_c \sim 11 \, GPa$ \cite{TakNM142015,YamPRL1152015} and $Ba{Fe}_2 {Se}_3$ shows a $T_c^{max} \sim 11 \, K$ for $P_c \sim 12.7 \, GPa$ \cite{YinPRB952017}. \textcolor{black}{It} was argued by some authors \cite{TakNM142015,YamPRL1152015} that these systems are Mott insulators.

\textcolor{black}{In the 123 FeSCs, the filling per $Fe$ atom is the same as in other FeSCs families, i.e. $n_e=6$ electrons per $Fe$ atom. Then, by following the arguments of previous chapters, we expect that the system is a metal, rather than a Mott insulator (Hund's coupling tends to disfavor the appearance of the Mott insulating state)}. Here, I will calculate the effect of local correlations at $T=0\, K$ for $Ba{Fe}_2 {S}_3$ to clarify the origin of the insulating \textcolor{black}{state}. I will consider the 5 $d$ orbitals tight-binding model, proposed by \cite{AriPRB922015} using Wannier parametrization of the Density Functional Theory (DFT) band structure obtained for two pressures, $0 \, GPa$ and $12.4 \, GPa$, \textcolor{black}{where at the latter pressure, the system is superconducting}. I will apply the slave-spin mean-field (SSMF) $U(1)$ formalism for both pressures. I will show that, for realistic $U$ and $J_H$, the system is still metallic at both pressures for $T=0 \, K$. At $0 \, GPa$, \textcolor{black}{the strength of correlations for the orbitals which mostly contribute to the band structure around the Fermi level is very strong, so that} we \textcolor{black}{will} argue that the insulating behavior in a local correlations picture could come from temperature \textcolor{black}{effects} (see \fref{fig:1.10}\textbf{(b)}). At $12.4 \, GPa$, the strength of correlations shows similar values as in other FeSCs, as shown in \cref{Chap03}. The Fermi surfaces suffer a huge reconstruction due to local correlations effects, contrary to what happens in other quasi-2D FeSCs.

In this chapter, I will first briefly review the tight-binding model presented for $Ba{Fe}_2 {S}_3$ at $0 \, GPa$ and $12.4 \, GPa$ \cite{AriPRB922015}. I will show that the band structures around the Fermi level and Fermi surfaces for both pressures show a 3D character, in contrast with \textcolor{black}{the quasi-1D structure and with} other FeSCs which show a quasi-2D behavior. Due to the substraction of every third column of $Fe$ atoms, there is a breaking of degeneracy of the $x$ and $y$ directions, hence $d_{zx}$ and $d_{yz}$ are not degenerated anymore. All 5 $d$ non-degenerated orbitals contribute to the band structure. In $BaFe_2S_3$, the unit cell has $4$ equivalent $Fe$ atoms, hence the tight-binding hamiltonian in the $k$-space is a $20 \times 20$ matrix ($5 \, d$ orbitals $\times$ $4 \, Fe$ atoms), substantially increasing the complexity of the problem. I have checked the impossibility of performing an unfolding of the Brillouin zone (BZ) in these systems (\textcolor{black}{while} it was possible for other quasi-2D FeSCs, see \sref{1.5.2}), due to various off-plane hoppings which are not negligible.

In the $4$ $Fe$ atoms unit cell, there are non-zero off-diagonal crystal field energies $\epsilon_{mn}$ with $m\neq n$. In order to \textcolor{black}{simplify} the SSMF calculations, \textcolor{black}{we rewrite} the tight binding hamiltonian to retain only diagonal crystal field splittings $\epsilon_m$, see \aref{AppB}. The original tight-binding model is written in a new orbital basis (labeled by ${w_m}$ with $m=zx,yz,xy,x^2-y^2$ and $3z^2-r^2$), in which only diagonal crystal field energies persist. I will comment the effects of writing the tight-binding hamiltonian in this new orbital basis. The main effect is that $w_{zx}$ and $w_{yz}$ orbitals mostly contribute to the bands around the Fermi level, hence highly reducing the orbital bandwidths $W_{w_m}$ for these two orbitals.

Then, I will calculate $Z_{w_m}$ in the new orbital basis by using SSMF $U(1)$ formalism, setting $n_e=6$ and discussing the correct $U$ and $J_H$ values for $Ba{Fe}_2S_3$ at both pressures \cite{BasCRP172016,AriPRB922015}. I obtained finite \textcolor{black}{quasiparticle weights $Z_{w_m} \neq 0$} for all orbitals, at both pressures and for realistic values of $U$ and $J_H$. \textcolor{black}{These values range between $Z_{w_m} \sim 0.03-0.06$}, so the system is a correlated metal at $T=0 \, K$. However, $Z_{w_{zx}}$ and $Z_{w_{yz}}$ at $0 \, GPa$ are very small, and we will argue that the system can be driven to an insulating behavior by a finite $T$. \textcolor{black}{Applying pressure will increase the total bandwidth, hence the system is expected to be less correlated (due to the increase of the kinetic energy gain $\widetilde{W}$) at $12.4 \, GPa$}. For $12.4 \, GPa$, the strength of correlations, \textcolor{black}{in fact, decreases and becomes} similar to other FeSCs. \textcolor{black}{This finding may} help to link the optimal degree of correlations plus an AFM order being suppresed with the appearance of high-$T_c$ superconductivity, an idea pointed out in previous chapters.

Finally, I will calculate the renormalized band structure for both pressures. I will show that there is a non-negligible Fermi surface reconstruction for both pressures, in contrast with other FeSCs, in which the Fermi surface is almost unaffected by local correlations. This reconstruction will alter the results extracted from Fermi surface-based theories for magnetic and superconducting instabilities.

\end{small}

\newpage

\section{Introduction \& Motivation}
\label{5.1}

Recently, in 2015 \cite{TakNM142015,YamPRL1152015}, and later in 2017 \cite{YinPRB952017}, a new family of FeSCs has been discovered in the 123 iron chalcogenides, for \textcolor{black}{$Ba{Fe}_2 {S}_3$ and $Ba{Fe}_2 {Se}_3$}, respectively. The undoped \textcolor{black}{compounds} at ambient pressure are AFM insulators, \textcolor{black}{and} the insulating behavior persist for $T>T_N$, with $T_N \sim 240-255 \, K$, $\mu \sim 2.8 \, \mu_B/Fe$ for $Ba{Fe}_2 {Se}_3$ \cite{CarPRB842011,MedJETPL952012,LuoPRB872013} and $T_N \sim 100-120 \, K$, $\mu \sim 1.20 \, \mu_B/Fe$ for $Ba{Fe}_2 {S}_3$ \cite{TakNM142015,YamPRL1152015,ChiPRL1172016,WanPRB952017_123}. $BaFe_2S_3$ shows closer $T_N$ and $\mu$ values to other quasi-2D FeSCs, see \sref{1.5.2}. The AFM ordering is different in both compounds, see next paragraphs.

\begin{figure}[h]
   \centering
   \includegraphics[width=0.9\columnwidth]{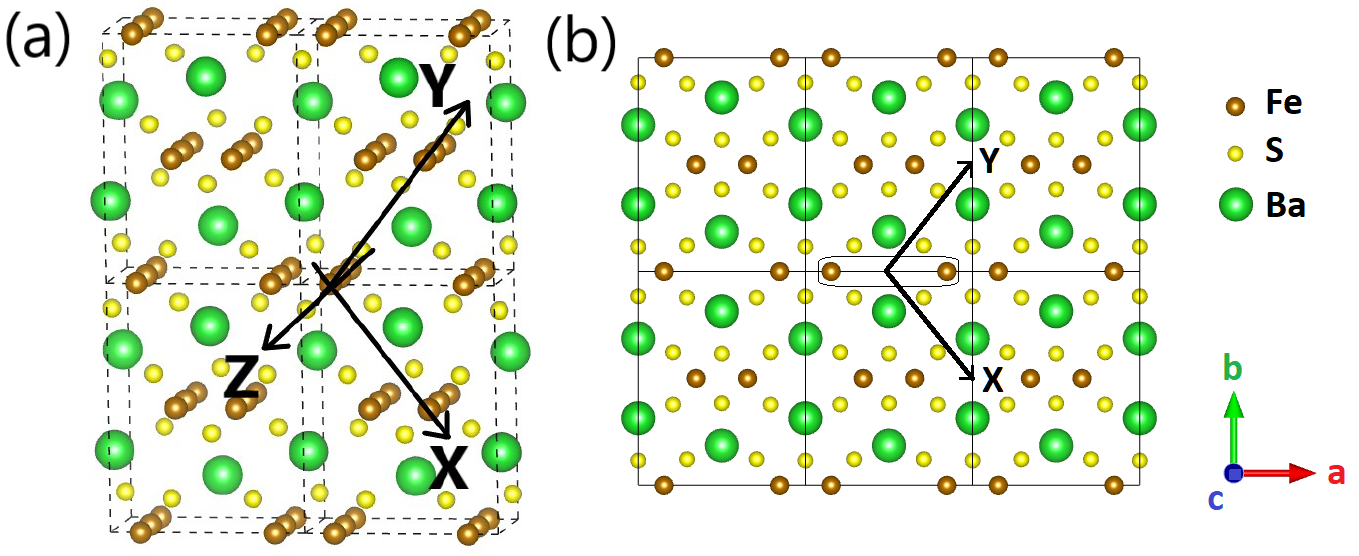}
   \caption[Crystal structure of the two-leg ladder FeSC $Ba{Fe}_2{S}_3$ where real space and crystallographic axes are included]{\textbf{(a)} General view and \textbf{(b)} perpendicular to the ladder view of  the crystal structure $CmCm$ for $Ba{Fe}_2S_3$. Two-leg ladders can be visualized like the square $Fe$ layers of quasi-2D FeSCs, but substracting every third column of $Fe$ atoms. Real space axes $X$, $Y$ and $Z$ and crystallographic axes $a$, $b$ and $c$ are marked, where $Z$ and $c$ coincides, and $XY$ and $ab$ run in different directions. When obtaining the tight-binding model, the orbitals will be oriented along the crystallographic axes $abc$, while the hopping amplitudes are written in the real space axes $XYZ$. Note also the difference with quasi-2D FeSCs, where the Fe layer is in the $ab$ plane, while here it is in the $ac$ plane.}
   \label{fig:5.1}  
\end{figure}

123 FeSCs have a different crystal structure \textcolor{black}{than} other quasi-2D FeSCs. 123 FeSCs are formed by two-leg ladders of iron atoms, where $Fe$ atoms are tetrahedrally coordinated with the chalcogen atoms $Se$ or $S$ (like in quasi-2D FeSCs). This can be identified as a quasi-1D crystal structure. The two-leg ladders structure is similar to the quasi-2D layers of $Fe$, but substracting every third column of $Fe$ atoms, then effectively reducing the dimensionality, see \fref{fig:5.1} for $Ba{Fe}_2{S}_3$. 


\begin{figure}[h]
   \centering
   \includegraphics[width=0.8\columnwidth]{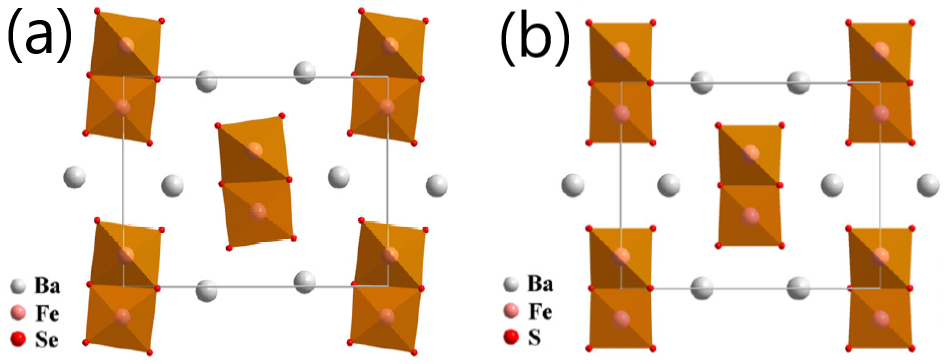}
   \caption[Crystal structures for $Ba{Fe}_2{Se}_3$ and $Ba{Fe}_2{S}_3$]{Crystal structures for \textbf{(a)} $Ba{Fe}_2{Se}_3$ ($Pnma$) and \textbf{(b)} $Ba{Fe}_2{S}_3$ ($CmCm$). The two-leg ladders run perpendicular to the paper. The difference between both structures is the tilting of the two-leg ladders for the $Pnma$ crystal structure, which is also believed to be related with the magnetoelastic coupling in $Ba{Fe}_2{Se}_3$ and the different AFM ordering in both crystal structure symmetries. Taken and adapted from \cite{Svi1808.03952}.}
   \label{fig:5.2}  
\end{figure}

In \fref{fig:5.2}, I compare the two different crystal structures present in 123 FeSCs. $Ba{Fe}_2 {Se}_3$ has $Pnma$ crystal structure symmetry, while $Ba{Fe}_2 {S}_3$ has $CmCm$ crystal structure symmetry. The difference relies in the tilting of the two-leg ladders.

The AFM order is different in both compounds. \textcolor{black}{$BaFe_2Se_3$ shows a} Block AFM order, \textcolor{black}{where} the spins are parallel for each $2 \times 2$ arrange of $Fe$ atoms in the two-leg ladder, and they oriented out-of-plane of the two-leg ladders. In $Ba{Fe}_2 {S}_3$, the AFM order is the typical stripe AFM, present in other FeSCs (see \sref{1.5.2}), with spins oriented parallel in the $a$ direction of the two-leg ladder and antiparallel in the $c$ direction \cite{TakNM142015,SuzPRB922015}. In $Ba{Fe}_2 {Se}_3$, there is a sizable magnetoelastic coupling (i.e. \textcolor{black}{the crystal structure parameters} are sensible to changes in the \textcolor{black}{magnetism}) \cite{CarPRB842011,NamPRB852012,CarPRB852012,YinPRB952017}, which is believed to be related with the formation of the Block AFM order.

\begin{figure}[h]
   \centering
   \includegraphics[width=0.9\columnwidth]{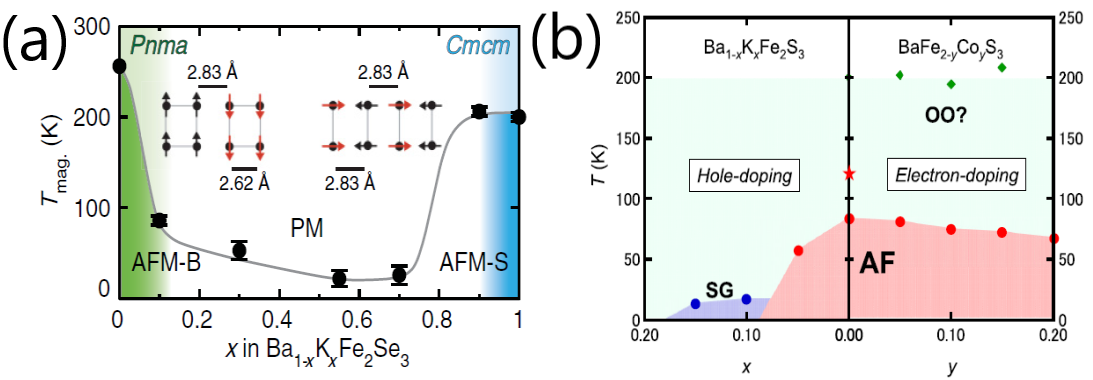}
   \caption[Doping phase diagrams for $Ba{Fe}_2{Se}_3$ and $Ba{Fe}_2{S}_3$]{Doping phase diagrams for \textbf{(a)} ${Ba}_{1-x}K_x{Fe}_2{Se}_3$ (hole-doped) and \textbf{(b)} ${Ba}_{1-x}K_x{Fe}_{2-y}{Co}_y{S}_3$ (both hole- and electron-doped). For $Ba{Fe}_2{Se}_3$, the Block AFM order and the $Pnma$ crystal structure symmetry is progresively replaced by the stripe AFM order and the $CmCm$ crystal structure symmetry when $K{Fe}_2{Se}_3$ is reached. For $Ba{Fe}_2{S}_3$, the stripe AFM order is suppressed for the hole-doped part, while it remains robust for the electron-doped part. A spin-glass phase appears at $x\sim0.1$. The authors \cite{HirPRB922015} argued about a possible orbital ordering at $T_{OO}$ as the origin of a kink in the resistivity for $BaFe_{2-y}Co_y S_3$, but yet it is not fully established and further studied would be required. Taken and adapted from \cite{CarPRB852012,HirPRB922015}.}
   \label{fig:5.3}  
\end{figure}

Different dopings on $Ba{Fe}_2 {Se}_3$ and $Ba{Fe}_2 {S}_3$ have been studied \cite{CarPRB852012,HirPRB922015}, see \fref{fig:5.3}. For hole-doped ${Ba}_{1-x}K_x{Fe}_2 {Se}_3$ \cite{CarPRB852012}, the Block AFM phase gets suppressed, and there are structural (from $Pnma$ to $CmCm$) and magnetic (from Block AFM to stripe AFM) transitions when approaching $K{Fe}_2{Se}_3$ ($n_e=5.5$ per $Fe$ atom). For hole-doped ${Ba}_{1-x}K_x{Fe}_2 {S}_3$, $CX$-AFM order (spins aligned parallel along the $x$ axis, and antiparallel along the $y$ axis, see \fref{fig:5.3}\textbf{(a)}) gets suppressed. For electron-doped $Ba {Fe}_{2-y}{Co}_y S_3$ \cite{HirPRB922015}, stripe AFM phase is more robust and no sign of suppression is seen up to $y \sim 0.2$. A resistivity anomaly at $\approx 200 \, K$ have been seing for $BaFe_{2-y}Co_y S_3$ \cite{HirPRB922015}. The authors argued about a possible orbital ordering (OO) as the origin of this anomaly, but yet it is not fully established and further studied would be required. No superconducting transition have been found when doping these systems.

\begin{figure}[h]
   \centering
   \includegraphics[width=0.9\columnwidth]{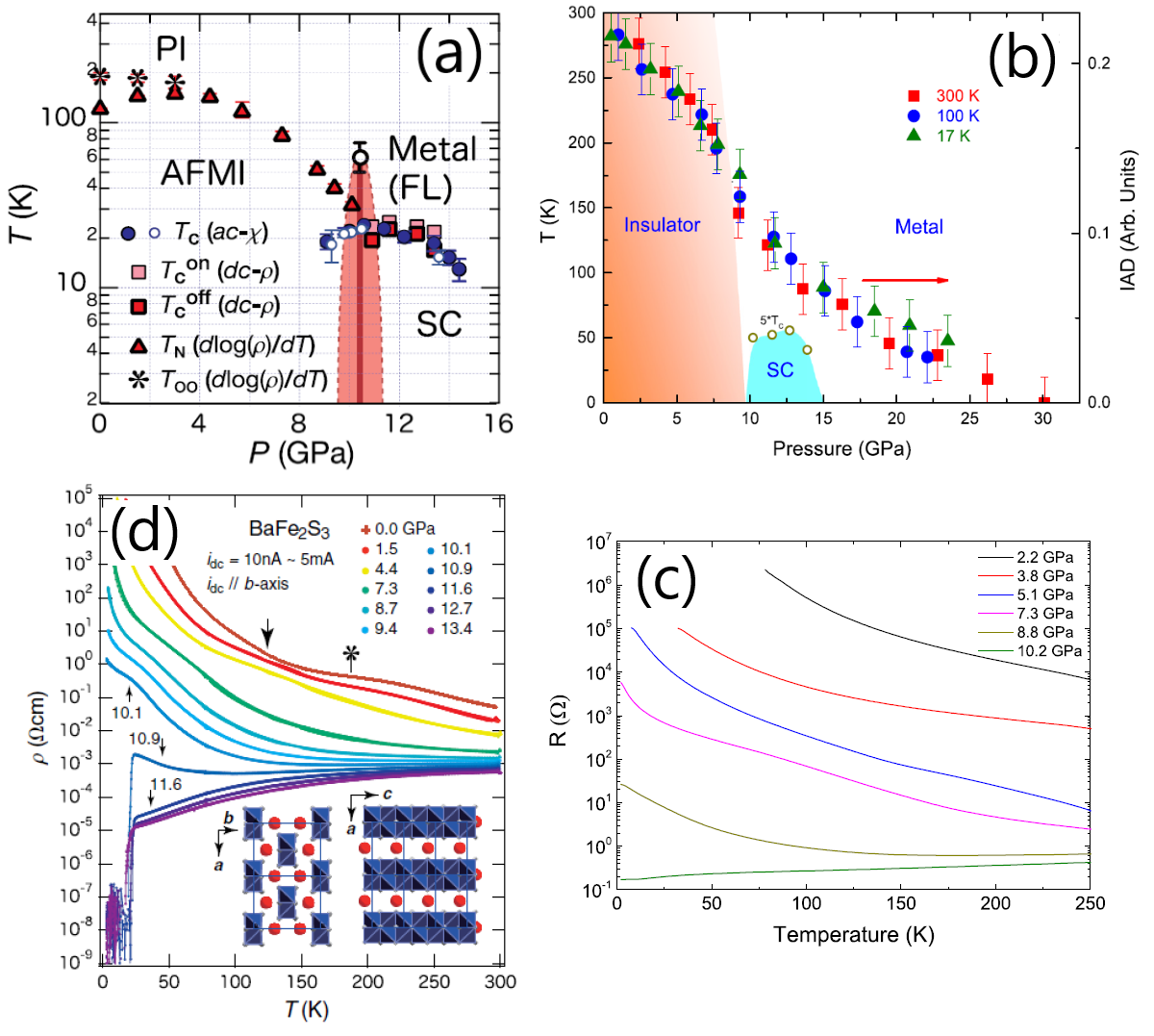}
   \caption[Pressure phase diagram and resistivity versus temperature curves for $Ba{Fe}_2{Se}_3$ and $Ba{Fe}_2{S}_3$]{Pressure phase diagram for \textbf{(a)} $Ba{Fe}_2{S}_3$ and \textbf{(b)} $Ba{Fe}_2{Se}_3$, where the insulating (both AFM and paramagnetic, PI), the metallic and superconducting phases are marked. $Ba{Fe}_2 {Se}_3$ shows a $T_c^{max} \sim 11 \, K$ for $P_c \sim 12.7 \, GPa$, and $Ba{Fe}_2 {S}_3$ shows a $T_c^{max} \sim 24 \, K$ for $P_c \sim 11 \, GPa$. Stars in the phase diagram of $Ba{Fe}_2 {S}_3$ signal a kink in the resistivity. \textbf{(c)}, \textbf{(d)} Resistivity versus temperature curves for $BaFe_2S_3$ and $BaFe_2Se_3$ at various pressures. The insulating behavior persist for $T>T_N$ ($\rho(T)$ decreases with $T$) in both systems. Taken from \cite{YamPRL1152015,YinPRB952017}.}
   \label{fig:5.4}  
\end{figure}

When pressure is applied for both $Ba{Fe}_2 {Se}_3$ and $Ba{Fe}_2 {S}_3$ \cite{TakNM142015,YamPRL1152015,YinPRB952017}, the AFM phase is suppressed, there is an insulator-to-metal transition and superconductivity emerges, all around the same $P_c$, see \fref{fig:5.4}. $Ba{Fe}_2 {Se}_3$ shows a $T_c^{max} \sim 11 \, K$ for $P_c \sim 12.7 \, GPa$, and $Ba{Fe}_2 {S}_3$ shows a $T_c^{max} \sim 24 \, K$ for $P_c \sim 11 \, GPa$. Interestingly, the insulating behavior for ambient pressure 123 FeSCs, persists for $T>T_N$ \cite{NamPRB852012,DuFPRB852012,TakNM142015,YamPRL1152015,YinPRB952017}.

$Fe$ two-leg ladders control the band structure around the Fermi level, with a total number of electrons per $Fe$ atom $n_e=6$, like in other FeSCs. The reduced dimensionality will decrease the kinetic energy gain $\widetilde{W}$ of the electrons, hence increasing the ratio $\Delta^x/\widetilde{W}$, see \sref{3.2}, and increasing the electronic correlations effects. This has led to some authors \cite{TakNM142015,YamPRL1152015} to \textcolor{black}{claim} that these systems are Mott insulators, like high-$T_c$ cuprates superconductors. Nevertheless, in previous chapters, I have shown that in Hund metals (thus in FeSCs) the Mott insulating behavior is disfavored by $J_H$, so $\Delta^x/\tilde{W}$ should be strongly reduced, by the effective reduction of the dimensionality, in order to be able to drive the system into a Mott insulating state, see \fref{fig:3.13}\textbf{(b)}. This situation does not seem plausible according to the band structure calculations (see next paragraph).

In \cite{AriPRB922015}, the band structure and tight-binding models were calculated for $Ba{Fe}_2 {S}_3$, finding that all 5 $d$ orbitals of $Fe$ atoms mainly contribute to the band structure around the Fermi level with a total bandwidth $W \sim 4 \, eV$ (see \sref{5.2.1}). Also in \cite{AriPRB922015}, constraint random phase approximation (cRPA) calculations showed that the realistic $U$ and $J_H$ values for $BaFe_2S_3$ are similar to $LiFeAs$, i.e. larger than in $Ba{Fe}_2 {As}_2$ and smaller than in $FeSe$, \textcolor{black}{therefore, we expect that} $BaFe_2S_3$ \textcolor{black}{is close to or} in the Hund metal regime (see violet shaded area in \fref{fig:3.13}\textbf{(b)}). 

DFT calculations can reproduce the correct AFM but with a larger magnetic moment $\mu_{DFT} \sim 2.1 \, \mu_B/Fe$ \cite{SuzPRB922015,ZhaPRB972018}, as occurs in other FeSCs, see \sref{1.5.2}. From the experimental point of view, in $Ba{Fe}_2S_3$, the magnetic moment is $\sim 1.20 \, \mu_B/Fe$, much smaller than in a Mott insulator where the saturation gives $4 \, \mu_B/Fe$ \cite{ChiPRL1172016}, see \sref{1.5.2}. Photoemission and X-ray experiments in $BaFe_2S_3$ suggest the coexistance of local and itinerant electrons at the same time in this system \cite{OotPRB912015,TakPRB962017}.

\textcolor{black}{So far, the theoretical analysis in 123 FeSCs on the role of electronic correlations have used models with two or three $Fe$ $d$ orbitals \cite{LuoPRB872013,PatPRB942016,LiSPRB942016}. Due to the hybridization between the $5$ $d$ orbitals, the orbital bandwidths and fillings depend highly on the number of orbitals included in the initial model. Thus, the two and three orbitals models cannot properly address the strength of the electronic correlations for each orbital. Furthermore, in the case of two orbitals models \cite{LuoPRB872013,PatPRB942016}, no Hund metal physics is found (see \cref{Chap03}). The three orbital model \cite{LiSPRB942016} cannot reproduce neither the correct band structure nor the correct quasiparticle weights for each orbital in the 123 FeSCs (see next sections). Another important failure of these two and three orbitals models is that the authors assumed a 1D two leg ladder to reproduce the band structure. In $BaFe_2S_3$, the band structure shows a non-negligible 3D dispersion \cite{AriPRB922015}, hence questioning the validity of these previous works. To address the strength of electronic correlations in this family, a suitable technique for a model containing the $5$ $Fe$ $d$ orbitals is required}.

All of these results point to the importance of electronic correlations for a $5$ $d$ orbitals model of 123 FeSCs, and also to the need of clarification about the insulating phase in these multiorbital systems. From the perspective worked on previous chapters, it does not seem plausible to find a Mott insulator at $T = 0\, K$, as assumed by other authors.

\section{Results \& Discussion}
\label{5.2}

Now, I will present the results that I have obtained and published in \cite{Pizarro2}. In \sref{5.2.1}, I will briefly review the tight-binding model obtained by \cite{AriPRB922015} for $Ba{Fe}_2 {S}_3$ at two different pressures $0 \, GPa$ and $12.4 \, GPa$. $Fe$ $d$ orbitals are mainly contributing to the band structure around the Fermi level, with $n_e=6$ in each $Fe$ atom. I will discuss the results of this model in the folded BZ, i.e. with $4$ $Fe$ atoms in the unit cell. Then, I performed a \textcolor{black}{change of basis} from the non-diagonal crystal field original orbitals to a new orbital basis with diagonal crystal field energies (labeled by $w_m$). \textcolor{black}{This change of basis is done in order to simplify the local correlations calculations, see \aref{AppB}}. In \sref{5.2.2}, I calculated $Z_{w_m}$ using SSMF $U(1)$ formalism at $T=0 \, K$. We have estimated $U$ and $J_H$ for the different pressure values, according to the comments made in \cite{AriPRB922015} when using cRPA, and taking into account the cRPA and SSMF comparison, as discussed in \fref{fig:3.13}. I will show that the system at $T=0 \, K$ and both pressures is still in the Hund metal regime, and no Mott (or even orbital-selective Mott transition) is found. \textcolor{black}{However, at $0 \, GPa$, the quasiparticle weights $Z_{w_m}$ of the orbitals mostly contributing to the band structure around the Fermi level are very small, $\sim 0.03-0.06$. Thus, we argue that a finite $T$ (see \fref{fig:1.10}\textbf{(b)}) the incoherence of the system increases, hence} the insulating nature of $Ba{Fe}_2 {S}_3$ could be reconciled \textcolor{black}{with} the local correlations picture. Finally, in \sref{5.2.3}, I calculate the renormalized band structure, showing the sizable reconstruction suffered by the Fermi surfaces at both pressures. The consequences of this reconstruction for Fermi surface instability theories are discussed. The Fermi surface reconstruction will change the expectations for Fermi surface-based magnetic and superconducting instabilities.

\subsection{Tight-binding model for $Ba{Fe}_2 {S}_3$ at $0 \, GPa$ and $12.4 \, GPa$}
\label{5.2.1}

Here, I reproduce the tight-binding model derived by \cite{AriPRB922015} for $Ba{Fe}_2 {S}_3$ at $0\, GPa$ and $12.4 \, GPa$. This tight-binding model is obtained by doing a Wannier parametrization for the $Fe$ $d$ orbitals. For $12.4 \, GPa$, crystal structural parameters shrinks around $92-96 \, \%$ of those at $0\, GPa$. 

In $Ba{Fe}_2 {S}_3$, the unit cell has $4$ $Fe$ atoms. Thus, the total number of orbitals in the unit cell is $N_{orb}=20$ ($5 \, d$ orbitals $\times$ $4 \, Fe$ atoms). In the unit cell there is only 1 inequivalent $Fe$ atom, so there are 5 $d$ fourly-degenerated orbitals. I have checked that the unfolding of the BZ (as done in other FeSCs) is not possible in this system. Thus, we have to work in the folded BZ, with $4$ $Fe$ atoms in the unit cell, hence with all the 20 $d$-like bands. 

The real (and $k$-) space axes $X$, $Y$ and $Z$ are oriented as shown in \fref{fig:5.1}, and the hopping parameters are written using these axes. On the other hand, $d$ orbitals are oriented along the crystalographic axes $a$, $b$ and $c$. The crystalographic axes (shown in \fref{5.1}\textbf{(b)}) are \textcolor{black}{chosen differently} from other FeSCs, where 2D square of $Fe$ atoms is in the $ab$ plane and $c$ is the offplane axis. In $Ba{Fe}_2 {S}_3$, the two-leg ladder is oriented along $c$, in the plane $ac$, whether $b$ stands for the offplane axis. Then, the $d$ orbitals are not \textcolor{black}{defined in the same way as} in $BaFe_2S_3$ as in other quasi-2D FeSCs\footnote{For example, in $Ba{Fe}_2 {S}_3$, $zx$ orbital will be equivalent to $xy$ orbital in other quasi-2D FeSCs.}. Due to the \textcolor{black}{breaking of the degeneracy} of $x$ and $y$ directions in the two-leg ladder plane, ${xz}$ and ${yz}$ orbitals are not degenerate in 123 FeSCs. Also, it can be seen that $(0,0,k_z) \rightarrow (\pi,0,k_z)$ and $(0,0,k_z) \rightarrow (0,\pi,k_z)$ directions are equivalent, while $(0,0,k_z) \rightarrow (\pi,\pi,k_z)$ and $(0,0,k_z) \rightarrow (\pi,- \pi,k_z)$ are not, where $k_z \in [-\pi,\pi]$.


\begin{figure}[h]
   \centering
   \includegraphics[width=0.9\columnwidth]{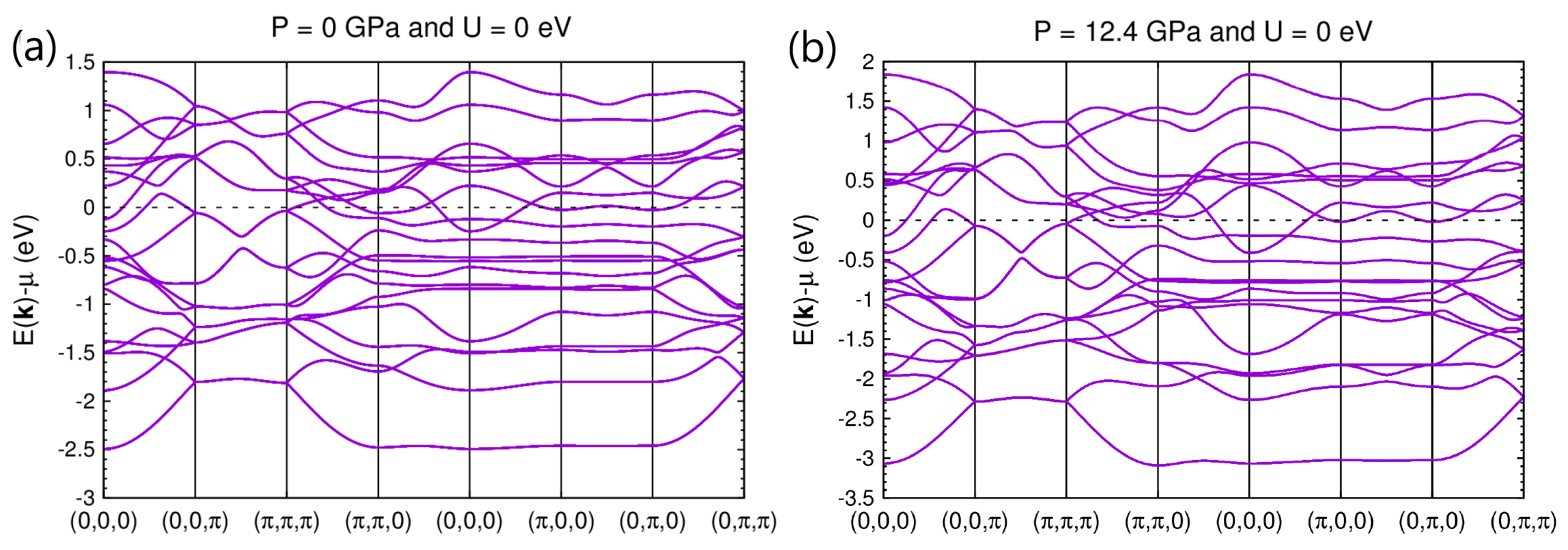}
   \caption[Band structures and zooms around the Fermi level at $0 \, GPa$ and $12.4 \, GPa$ for non-interacting $Ba{Fe}_2S_3$]{Band structures for \textbf{(a)} $0 \, GPa$ and \textbf{(b)} $12.4 \, GPa$ in the non-interacting limit ($U=0 \, eV$). Bandwidth is of the order of other quasi-2D FeSCs, with $W^{0 \, GPa} \sim 4 \, eV$ and $W^{12.4 \, GPa} \sim 5 \, eV$. The band structure close to the Fermi level shows a marked 3D character, besides the system shows a quasi-1D crystal structure. Reproduced from \cite{Pizarro2} using the parameters proposed in \cite{AriPRB922015}.}
   \label{fig:5.5}  
\end{figure}

\fref{fig:5.5}\textbf{(a)} and \textbf{(b)} show the band structure for both pressures, where the total bandwidth is $W^{0 \, GPa} \sim 4 \, eV$ and $W^{12.4 \, GPa} \sim 5 \, eV$. \textcolor{black}{Close to the Fermi level, the} band structure have a singular 3D character (see the non-negligible dispersion when varying $k_x$ or $k_y$), \textcolor{black}{on spite of the} quasi-1D crystal structure.

\begin{figure}[h]
   \centering
   \includegraphics[width=1.0\columnwidth]{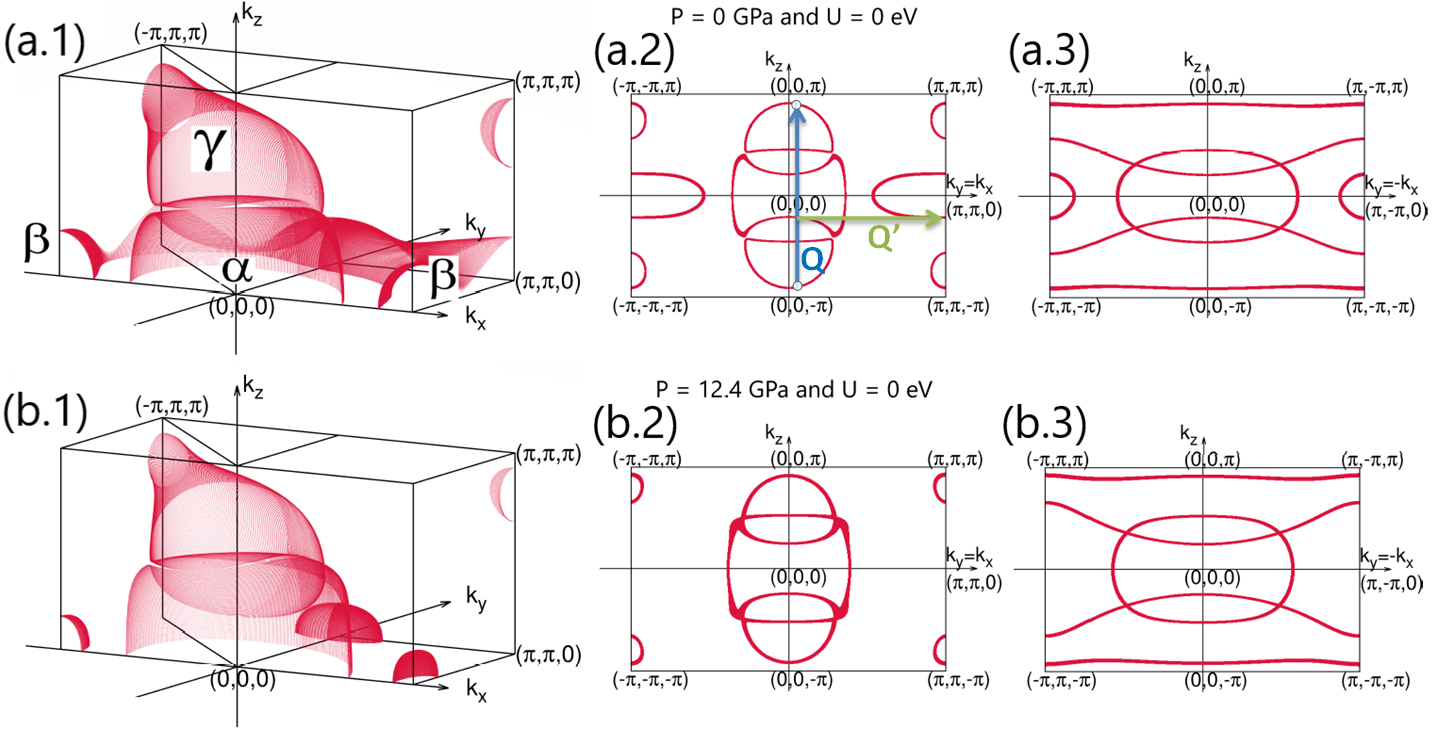}
   \caption[3D Fermi surfaces and cuts along two inequivalent directions for $0 \, GPa$ and $12.4 \, GPa$ for the non-interacting $Ba{Fe}_2S_3$]{\textbf{(a.1)} $0 \, GPa$ and \textbf{(b.1)} $12.4 \, GPa$ 3D Fermi surfaces in the reduced BZ. The Fermi surfaces consist on different pockets, which can be identified as $\alpha$, $\beta$ and $\gamma$ (see main text). \textbf{(a.2)} $0 \, GPa$ and \textbf{(b.2)} $12.4 \, GPa$ Fermi surface cuts along $k_z$-$k_x-k_y$ plane, and \textbf{(a.3)} $0 \, GPa$ and \textbf{(b.3)} $12.4 \, GPa$ Fermi surface cuts along $k_z$-$k_x+k_y$ plane, in the full BZ. Figures taken at $n_e=6$ and for the non-interacting limit. $\vec{Q}$ intra-ladder AFM nesting vector \cite{SuzPRB922015} connects the top and bottom $\gamma$ pockets, while $\vec{Q}'$ inter-ladder AFM nesting vector connects $\alpha$ and $\beta$ pockets. $\vec{Q}$ and $\vec{Q}'$ included as blue and green lines, respectively.}
   \label{fig:5.7}  
\end{figure}

In \fref{fig:5.7}, I present the Fermi surfaces for both pressures and undoped ($n_e=6$) $BaFe_2S_3$, in the reduced BZ (the full BZ is just the repetition of this one-quarter reduced BZ). The top row refers to $0 \, GPa$ and the bottom row to $12.4 \, GPa$. I included the Fermi surface \textbf{(a.2)}, \textbf{(b.2)} cuts along $k_z$-$(k_x-k_y)$ plane, and \textbf{(a.3)}, \textbf{(b.3)} cuts along $kz$-$(k_x+k_y)$ plane, both in the full BZ. The Fermi surfaces consist on several pockets with a 3D character. At $0 \, GPa$, the Fermi surface has an electron pocket $\alpha$ at $\Gamma$, surrounded by a large electron pocket $\beta$, and a couple of hole pockets $\gamma$ around $(0,0, \pm 3\pi/4)$ and directed along $(k_x, -k_x,\pm 3\pi/4)$. With pressure, the size of the pockets change, specially for the electron pocket $\beta$ which shrinks significatively, dissappearing around the corners $(\pm \pi, \pm \pi, 0))$.

From a weakly correlated picture \textcolor{black}{of} magnetism, the nesting vector appears around a slightly incommensurate $\vec{Q} \sim (0,0,2\pi)$ (see blue vector in \fref{fig:5.7}\textbf{(a.2)} and \textbf{(b.2)}), and it \textcolor{black}{could give some contribution (see below) for} the intra-ladder stripe AFM order \cite{SuzPRB922015} (in the unfolded BZ, $\vec{Q}$ would transform into $\vec{Q}^* \sim (0,0,\pi)$). The nesting vector which \textcolor{black}{would} produce the inter-ladder stripe AFM order is given by $\vec{Q}' = (\pm \pi, \pm \pi, k_z)$ (see green vector in \fref{fig:5.7}\textbf{(a.2)} and \textbf{(b.2)}). $\vec{Q}$ \textcolor{black}{would not be} suppressed by pressure, but $\vec{Q}'$ \textcolor{black}{will}, so the experimental suppression of AFM when applying pressure could by explained as a decrease of the inter-ladder scattering. Nevertheless, \textcolor{black}{at $0 \, GPa$, the system is an insulator, and I will show later on that the orbitals controlling the Fermi surface are very strongly correlated, so} weakly correlated theories are highly in doubt to be applicable for the ambient pressure compound.

The tight-binding model of \cite{AriPRB922015} is written in the non-diagonal (crystal field energies) orbitals basis, hence there exist non-zero $\epsilon_{m \neq n}$, \textcolor{black}{i.e. different orbitals have onsite mixing terms}. \textcolor{black}{To simplify} the SSMF $U(1)$ \textcolor{black}{calculations} (see \aref{AppB}), the tight-binding hamiltonian \textcolor{black}{is} rewritten in a new \textcolor{black}{basis in which the onsite terms are diagonal}. This is done by taking the matrix of crystal field energies, diagonalizing it, and rewriting $H_0$ (where $H_0$ is the tight-binding hamiltonian, see \eref{eq:1.4}) in the new orbital basis (defined by the eigenvectors obtained in the crystal field energies matrix diagonalization).

\begin{figure}[h]
   \centering
   \includegraphics[width=0.9\columnwidth]{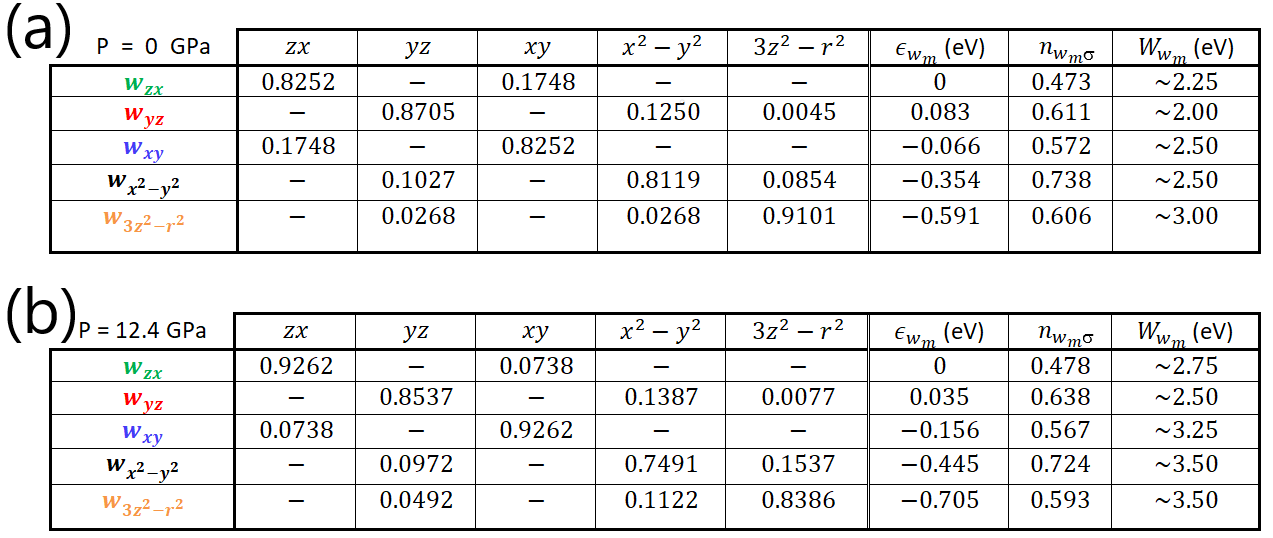}
   \caption[Table for the orbital weights of the change of basis from the old orbital basis to the new one for $0 \, GPa$ and $12.4 \, GPa$, and values of the new crystal fields, orbital fillings per spin and estimated orbital bandwidths]{Table for the basis change from the old orbital basis to the new one for \textbf{(a)} $0 \, GPa$ and \textbf{(b)} $12.4 \, GPa$, and values of the new crystal field splittings, orbital fillings per spin and orbital bandwidths. The name of the new orbitals is given by looking at the largest contribution of the old ones when changing the basis. }
   \label{fig:5.6}  
\end{figure}

The new orbitals are labeled as $w_m$ with $m=zx,yz,xy,3z^2-r^2$ and $x^2-y^2$, and they are a linear combination of the old orbitals. The subscripts stand for the largest contribution to these new orbitals from the old ones. \fref{fig:5.6} show the tables for $0\, GPa$ and $12.4 \, GPa$, with the contributions of each original orbital for the new orbitals, as well as the new crystal field splittings $\epsilon_{w_m}$, the orbital filllings per spin $n_{w_m\sigma}$ ($n_{w_m\sigma}=0.5$ is defined as half-filling) and the estimated orbital bandwidths $W_{w_m}$. $\epsilon_{w_m}$ are of the same order, so all new orbitals are playing an important role in the band structure. Applying pressure does not vary significantly the orbital filling, but it increases $W_{w_m}$. Then, we expect that the strength of local correlations will be reduced by pressure for all the orbitals, due to an increase of the kinetic energy gain.

\begin{figure}[h]
   \centering
   \includegraphics[width=0.9\columnwidth]{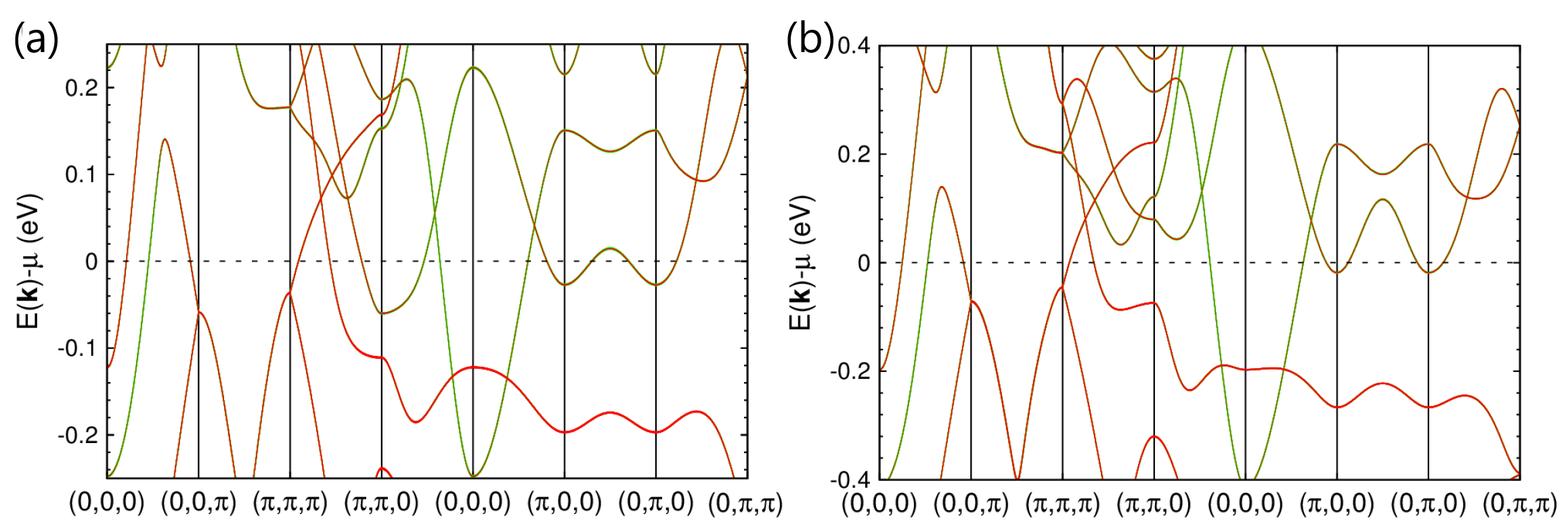}
   \caption[Zooms of the band structure around the Fermi level at $0 \, GPa$ and $12.4 \, GPa$ for non-interacting $Ba{Fe}_2S_3$]{Zooms of the non-interacting band structure around the Fermi level for \textbf{(c)} $0 \, GPa$ and \textbf{(d)} $12.4 \, GPa$. New orbitals $w_m$ (see main text) majority contribution included, where $w_{zx}$ (green) and $w_{yz}$ (red) contribute to the band structure around the Fermi level.}
   \label{fig:5.5a}  
\end{figure}

\textcolor{black}{\fref{fig:5.5a}\textbf{(a)} and \textbf{(b)} show zooms of the band structure around the Fermi level for $0 \, GPa$ and $12.4 \, GPa$, with the majority orbital (in the new orbital basis) contribution included}. Red and green colors refer to the contribution of $w_{yz}$ and $w_{zx}$ orbitals, respectively. \textcolor{black}{As it can be seen, both orbitals dominate the band structure around the Fermi level, hence they also dominate the Fermi surfaces}.

\newpage

\subsection{Local electronic correlations in $Ba{Fe}_2 {S}_3$}
\label{5.2.2}

Now, I apply the SSMF $U(1)$ formalism to calculate the local correlations for $n_e=6$ in the new orbital basis $w_m$. In the local correlations approach, each $Fe$ site of the unit cell is equivalent to other sites, hence the interacting hamiltonian $H_{dens}^{PS}$ of \eref{eq:1.23} only affects 5 $w_{m}$ orbitals, and only 5 different quasiparticle weights $Z_{w_m}$ have to be calculated. Note that I considered $H_{dens}^{PS}$ written in the new orbital basis, and not in the old one. If $H_{dens}^{PS}$ would have been written in the old orbital basis, and then change to the new one, some non-density-density terms would have appeared (similar to the pair-hopping and spin-flip $H_{add}$ terms). Nevertheless, these terms would have been neglected in the SSMF $U(1)$ formulation, because, \textcolor{black}{as mentioned in previous chapters, it is not known how to treat} non-density-density terms.


In \cite{AriPRB922015}, the authors calculated $U$ and $J_H$ interaction parameters by means of cRPA in $Ba{Fe}_2 {S}_3$, locating it (at $0 \, GPa$) close to $LiFeAs$, i.e. between $Ba{Fe}_2 {As}_2$ and $FeSe$. For these compounds, SSMF gives good results when comparing with angle-resolved photoemission spectroscopy (ARPES) experiments \cite{BasCRP172016} for $J_H/U=0.25$, $U_{Ba{Fe}_2 {As}_2} = 2.7 \, eV$ \cite{deMPRL1122014} and $U_{FeSe}=3 \, eV$ \cite{FanPRB952017}. I will fix $J_H/U=0.25$ and vary $U$. In \sref{5.2.3} when calculating the renormalized band structure, we will assume that $U_{Ba{Fe}_2 {S}_3}^{0 \, GPa}=2.9 \, eV$. Also in \cite{AriPRB922015}, the authors mentioned that the interaction parameters are reduced by a $6-7 \%$ when applying pressure. Thus, we will also consider $U_{Ba{Fe}_2 {S}_3}^{12.4 \, GPa}=2.75 \, eV$.

\begin{figure}[h]
   \centering
   \includegraphics[width=0.9\columnwidth]{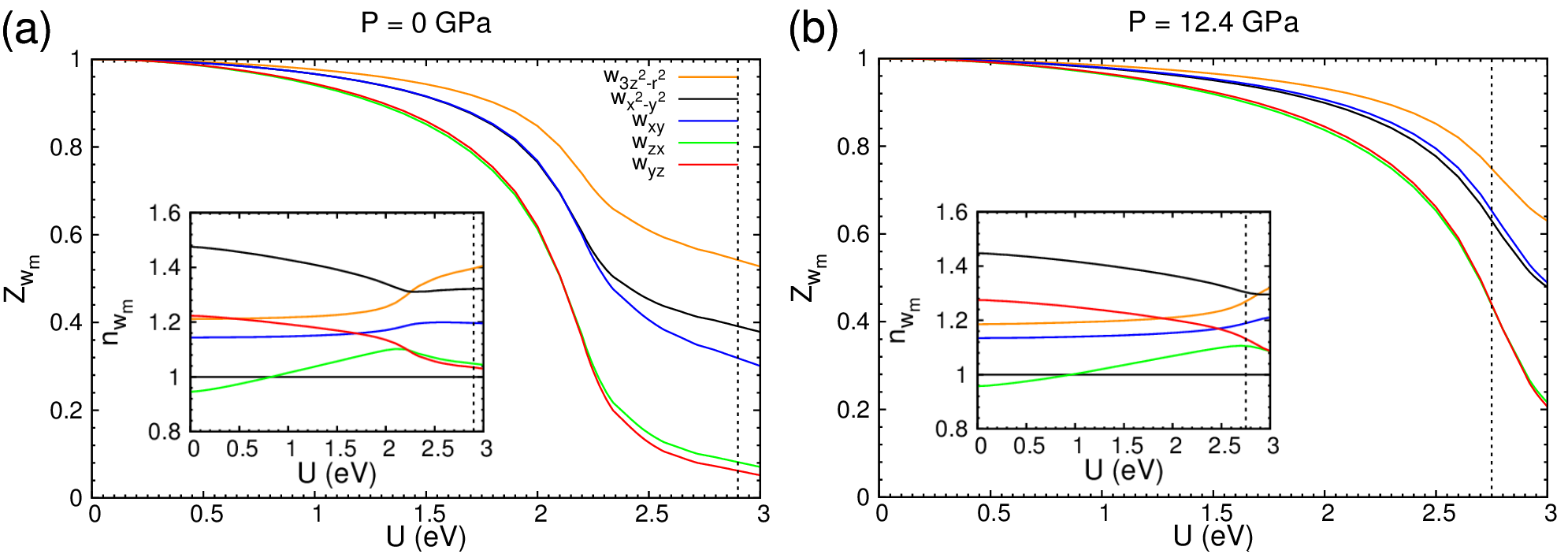}
   \caption[SSMF $U(1)$ results for $Z_{w_m}$ and $n_{w_m}$ at $n_e=6$ and $J_H/U=0.25$, for $Ba{Fe}_2S_3$ at $0 \, GPa$ and $12.4 \, GPa$]{$Z_{w_m}$ in terms of $U$ at \textbf{(a)} $0 \, GPa$ and \textbf{(b)} $12.4 \, GPa$, using SSMF $U(1)$ formalism at $n_e=6$ and $J_H/U=0.25$. \textsc{Inset:} same evolution for the orbital fillings $n_{w_m}$ in terms of $U$, where half-filling ($n_{w_m}=1$) is marked as a straight black line. Dotted lines mark the estimated interaction $U^P$ for $Ba{Fe}_{2} {S}_3$ at both pressures, $U^{0 \, GPa}=2.9 \, eV$ and $U^{12.4 \, GPa}=2.75 \, eV$. For both pressures, the strength of correlations of different orbitals is related with the lowest $W_{w_m}$ and the closeness to half-filling. $12.4 \, GPa$ shows a less correlated behavior (as anticipated), and while the $0 \, GPa$ compound is in the Hund metal regime, $12.4 \, GPa$ is located in the crossover region. The crossover region is found at $U_{0 \, GPa}^* \sim 2.1 \, eV$ and $U_{12.4 \, GPa}^* \sim 2.6-2.7 \, eV$. For the estimated $U^P$, the system is still metallic at both pressures. The color code for each orbital is included in the legend.}
   \label{fig:5.8}  
\end{figure}

\fref{fig:5.8} shows the results for $Z_{w_m}$ and $n_{w_m}$ for both pressures, when $U$ varies at $n_e=6$ and $J_H/U=0.25$. Dotted line marks the $U^{P}$ estimated in the previous paragraph for $0 \, GPa$ and $12.4 \, GPa$. The crossover from the weakly correlated to the Hund metal is located around $U_{0 \, GPa}^* \sim 2.1 \, eV$ and $U_{12.4 \, GPa}^* \sim 2.6-2.7 \, eV$, so the system at $0 \, GPa$ is located in the Hund metal regime and the system at $12.4 \, GPa$ in the crossover. \textcolor{black}{As anticipated, at $12.4 \, GPa$, there is a decrease of the strength of local correlations when comparing with $0 \, GPa$, so the Hund metal regime moves to larger $U$ values}. For both cases, $w_{zx}$ and $w_{yz}$ are the most correlated orbitals, followed by $w_{xy}$ and $w_{x^2-y^2}$, and finally the less correlated orbital is $w_{3z^2-r^2}$. This behavior can be traced down to the lowest $W_{w_m}$ and the closeness to half-filling of each orbital filling $n_{w_m}$ (note that $n_{w_m}=1$ is defined as half-filling), as already pointed out in previous chapters.

From our calculations at $T= 0 \, K$, no Mott insulating (or even orbital-selective Mott insulating) behavior is found, so the system is still metallic at both pressures. At $0 \, GPa$, the strength of correlations is considerably larger for $w_{zx}$ and $w_{yz}$ (with $Z_{w_m}<0.1$), while it is more similar to quasi-2D FeSCs for other orbitals (with $Z_{w_m} \sim 0.3-0.6$). \textcolor{black}{The strongest correlated orbitals ($w_{zx}$ and $w_{yz}$) are those contributing to the band structure around the Fermi level. This would have important implications when studying the magnetic and superconducting instabilities (see below)}. Interestingly, at $12.4 \, GPa$, the strength of correlations becomes similar to those found in quasi-2D FeSCs for all the orbitals, in accordance with the ideas of needing an optimal degree of electronic correlations to host unconventional superconductivity (see \cref{Chap4a}).

Note that the SSMF $U(1)$ calculations are performed for $T=0 \, K$. As shown in \sref{1.3.3} for the local correlations behavior with temperature, \textcolor{black}{when $T$ increases, incoherence also increases, hence reducing $Z$ , and promoting an insulating behavior. Due to the extremely low values of the quasiparticle weights for $w_{zx}$ and $w_{yz}$ ($Z_{w_m} \sim 0.03-0.06$), we expect that $T$ could lead to the disappearance of the quasiparticle peak and the opening of a gap at the Fermi level}. Thus, we expect that calculations for finite $T$ at $0 \, GPa$ would be able to drive the system to an insulating state.

\subsection{Band renormalization and Fermi surface reconstruction by local correlations}
\label{5.2.3}

In other quasi-2D FeSCs, local correlations do not almost change the Fermi surface because different pockets renormalize in a similar way. Thus, a weakly correlated picture of magnetism in quasi-2D FeSCs will be mostly unaffected by including local correlations for the Fermi surface. In some cases (like in 111 FeSCs), local correlations will mainly alter the pockets with $xy$ character ($yz$ and $zx$ ones are renormalized without nematicity), see for example \cite{YiMnpjQM22017}, hence the expectations from a weakly correlated picture of magnetism and superconductivity will vary.

\begin{figure}[h]
   \centering
   \includegraphics[width=0.9\columnwidth]{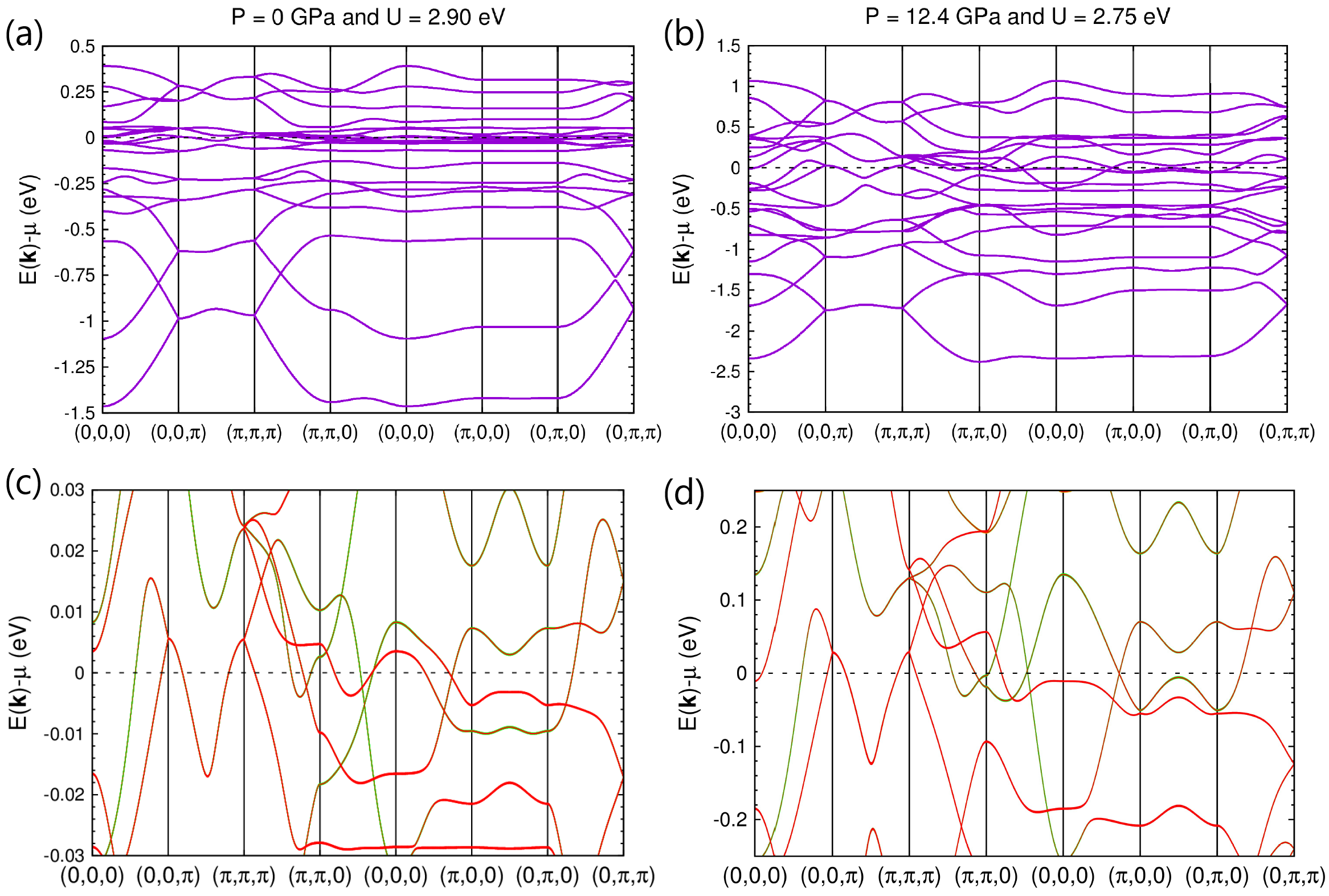}
   \caption[Band structure and zooms around the Fermi level at $0 \, GPa$ and $12.4 \, GPa$ for the interacting $Ba{Fe}_2S_3$]{Band structures for \textbf{(a)} $0 \, GPa$ at $U=2.9 \, eV$ and \textbf{(b)} $12.4 \, GPa$ at $U=2.75 \, eV$, with $J_H/U=0.25$. Bandwidths are reduced with respect to their non-interacting value (see \fref{fig:5.5}) due to the local correlations effect, $W^{0 \, GPa} \sim 2 \, eV$ and $W^{12.4 \, GPa} \sim 3.5 \, eV$. The color code is the same as in previous figures. Around the Fermi level, the bands suffer a considerable change, which can be traced back to a upwards shift for $w_{yz}$-like bands and a downwards shift for $w_{zx}$-like bands. This large reconstruction appears due to the fact that $w_{zx}$ and $w_{yz}$ are the most correlated orbitals.}
   \label{fig:5.9}  
\end{figure}

In \fref{fig:5.9}, I show the renormalized band structure for both pressures, as well as the zoom around the Fermi level. The interactions reduce the total bandwidth around $\sim 50 \%$ for $0 \, GPa$ and $\sim 30 \%$ for $12.4 \, GPa$, to $W^{0 \, GPa} \sim 2 \, eV$ and $W^{12.4 \, GPa} \sim 3.5 \, eV$.

\begin{figure}[h]
   \centering
   \includegraphics[width=1.0\columnwidth]{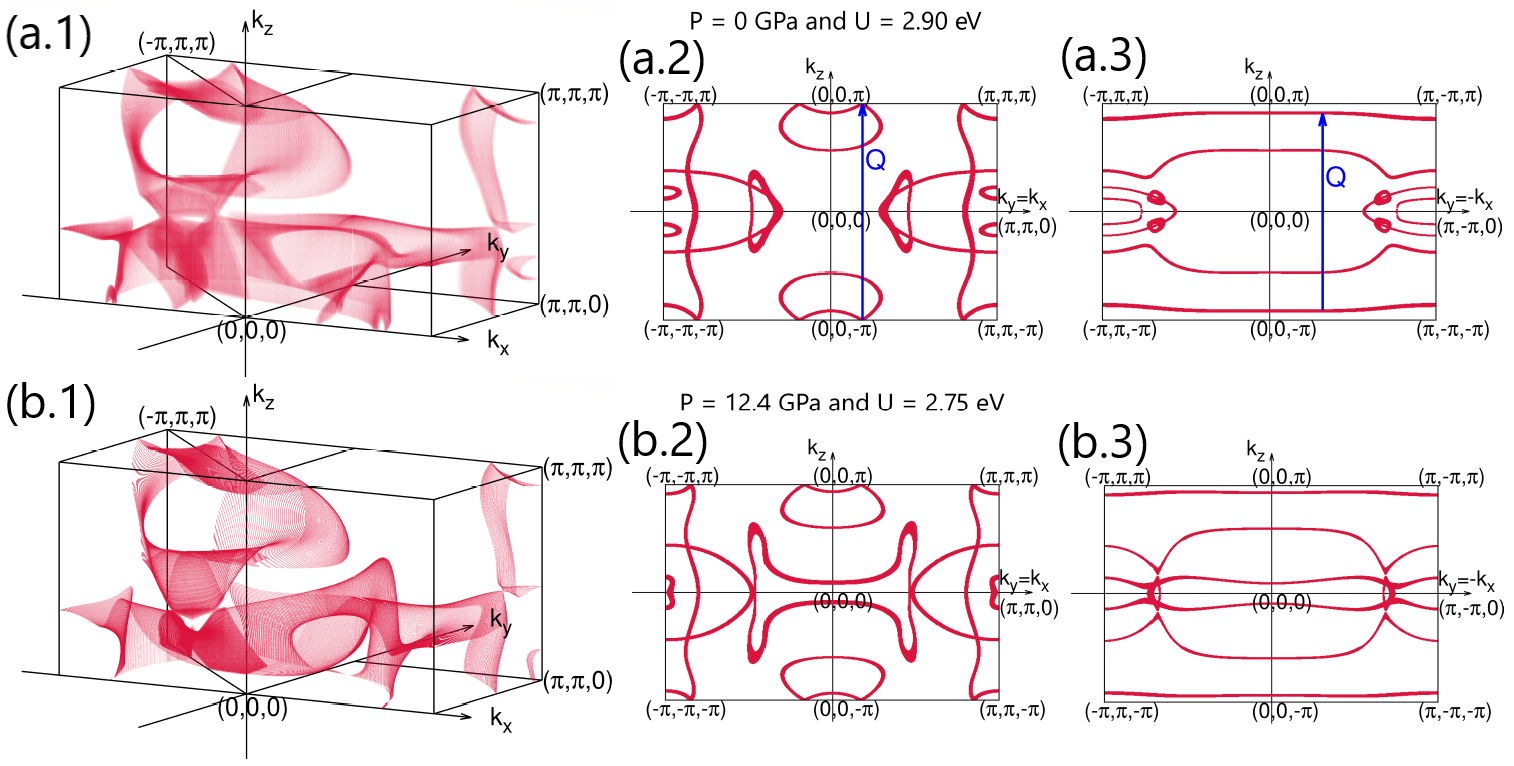}
   \caption[3D Fermi surfaces and cuts along two inequivalent directions for $0 \, GPa$ and $12.4 \, GPa$ for the interacting $Ba{Fe}_2S_3$]{\textbf{(a.1)} $0 \, GPa$ and \textbf{(b.1)} $12.4 \, GPa$ 3D Fermi surfaces in the reduced BZ. The Fermi surfaces evolve, with some pockets dissapearing and new ones appearing (see main text). \textbf{(a.2)} $0 \, GPa$ and \textbf{(b.2)} $12.4 \, GPa$ Fermi surface cuts along $k_z$-$k_x-k_y$ plane, and \textbf{(a.3)} $0 \, GPa$ and \textbf{(b.3)} $12.4 \, GPa$ Fermi surface cuts along $k_z$-$k_x+k_y$ plane. Figures taken at $n_e=6$ and $U=2.9 \, eV$ for $0 \, GPa$ and at $U=2.75 \, eV$ for $12.4 \, GPa$. The Fermi surfaces suffer a sizable reconstruction, which will affect the predictions for Fermi surface instability based theories. Intra-ladder AFM nesting vector $\vec{Q}$ included, where the scattering gets enhanced by the cut of the $\gamma$ pockets with the $k_z$ axis, which is not present in the $U=0 \, eV$ case. }
   \label{fig:5.10}  
\end{figure}

Interestingly, the most strongly correlated orbitals $w_{zx}$ and $w_{yz}$ (with $Z_{w_m} \sim 0.03-0.06$) are those who mainly contribute to the band structure around the Fermi level, in a range of a few dozens and hundreds of $meV$. Local correlations alter the bands, where the most striking feature is the upwards/downwards shift of the $w_{yz}$/$w_{zx}$-like bands. Thus, there is a sizable Fermi surface reconstruction in $Ba {Fe}_2 S_3$ due to local correlations. This can be more clearly seen in \fref{fig:5.10}, where I present the Fermi surfaces (as well as the same cuts as in \fref{fig:5.7}) for $0\, GPa$, $U = 2.9 \, eV$ and $12.4 \, GPa$, $U=2.75 \, eV$.

When including the effect of local correlations, the $\gamma$ hole pockets change their shape and cut the border $k_z= \pm \pi$, and the $\beta$ electron pocket grows and new smaller hole pockets appear close to $(\pm \pi,\pm \pi,0)$. At $0 \, GPa$, the $\alpha$ electron pocket dissapears and evolves in triangular-shaped electron pockets around $(\pm \pi/2, \pm \pi/2,0)$. At $12.4 \, GPa$, the $\alpha$ electron pocket evolves into a H-shaped electron pocket. Some of these pockets (for example, those around $(\pi,\pi,0)$) are extremely shallow, specially for $0 \, GPa$, where a small change on the Fermi level will make these pockets dissappear, see \fref{fig:5.9}\textbf{(c)} and \textbf{(d)}.

An important consequence of the Fermi surface reconstruction is \textcolor{black}{how the nesting changes}. When interactions are turned on, there is an enhancement of the scattering with the intra-ladder AFM nesting vector $\vec{Q}=(0,0,2 \pi)$ (marked in \fref{fig:5.10}\textbf{(a.2)} and  \textbf{(a.3)}) between the $\gamma$ hole pockets with respect to the non-interacting case, see \fref{fig:5.7}. This is due to the fact that in the interacting case, $\gamma$ hole pockets cut the $k_z=\pm \pi$ border, while in the non-interacting case, they do not. 

Thus, in a Fermi surface-based magnetic instability, this intra-ladder AFM susceptibility will be enhanced. Nevertheless, \textcolor{black}{at $0 \, GPa$, the strength of correlations for the orbitals controlling the Fermi surface ($w_{zx}$ and $w_{yz}$) is very strong, and} a weakly correlated picture of AFM \textcolor{black}{does not seem} a good starting point. In the case of $12.4 \, GPa$, where correlations are of the order of quasi-2D FeSCs, a weakly correlated picture to study the superconducting instability could maybe be applied. An interesting issue which requires further work is whether this weakly correlated picture for superconductivity is invalidated by the non-weakly correlated picture of magnetism or not. If so, the Fermi surface reconstruction will affect the superconducting instability calculation.

\section{Summary}
\label{5.3}

I conclude this chapter by making a summary of the obtained results:

\begin{itemize}

\item In 2015 and 2017, high-$T_c$ superconductivity was found in a new family of FeSCs, the 123 iron chalcogenides \cite{TakNM142015,YamPRL1152015,YinPRB952017}. Superconductivity in these systems appears when applying pressure. The undoped at ambient pressure compound is an insulator with an AFM order. The insulating behavior persists for $T>T_N$. There is an insulator-to-metal transition around the pressure at which superconductivity emerges and AFM is suppressed. 123 FeSCs form quasi-1D structures of two-leg ladders of $Fe$ atoms, which are tetrahedrally coordinated with the chalcogen atoms. \textcolor{black}{Part of the community claim} that these systems are Mott insulators \cite{TakNM142015,YamPRL1152015}.

\item In \sref{5.2.1}, I have \textcolor{black}{reproduced} the tight-binding model developed in \cite{AriPRB922015} for $Ba{Fe}_2S_3$ at two pressures, $0 \, GPa$ and $12.4 \, GPa$. The tight-binding model is given for $4$ $Fe$ atoms in the unit cell, and considering the 5 $d$ orbitals of each $Fe$ atom, which are controlling the band structure around the Fermi level. As expected, pressure increases $W$, so it will promote metallic behavior in the presence of electronic correlations. Even if the system shows a quasi-1D crystal structure, the bands and the Fermi surfaces have a non-negligible 3D character. \textcolor{black}{I rewrote the tight-binding hamiltonian in the diagonal onsite energy basis, where the new orbitals are labeled by $w_m$. In this new orbital basis, $w_{zx}$ and $w_{yz}$ are concentrated around the Fermi level, giving the largest contribution for the Fermi surfaces.}

\item In \sref{5.2.2}, I calculated $Z_{w_m}$ and $n_{w_m}$ for both pressures and varying $U$, using SSMF $U(1)$ formalism at $n_e=6$. We \textcolor{black}{use} $J_H/U=0.25$, and $U^{0 \, GPa}=2.9 \, eV$ and $U^{12.4 \, GPa}=2.75 \, eV$. When pressure is applied, the strength of correlations decreases for all the orbitals, as anticipated. For both pressures, the system is still metallic. \textcolor{black}{At} $0 \, GPa$, compound is located in the Hund metal regime, whether \textcolor{black}{at} $12.4 \, GPa$, it is in the crossover region, defined by $U_{0\, GPa}^* \sim 2.1 \, eV$ and $U_{12.4\, GPa}^* \sim 2.6-2.7 \, eV$. At $0 \, GPa$, $w_{zx}$ and $w_{yz}$ are very strongly correlated, with $Z_{w_m}\sim 0.03-0.06$, while the other 3 orbitals are moderately and strongly correlated, with $Z_{w_m} \sim 0.3-0.6$. At $12.4 \, GPa$, all the orbitals $w_m$ are moderately and strongly correlated, with $Z_{w_m} \sim 0.2-0.6$, similar to \textcolor{black}{other} quasi-2D FeSCs.

\item The system is metallic at both pressures. However, SSMF $U(1)$ calculations are made at $T=0 \, K$. \textcolor{black}{Including} $T$ effects in the local correlations picture \textcolor{black}{increases the incoherence, hence promoting an} insulating behavior. Thus, \textcolor{black}{at $0 \, GPa$)}, we conclude by stating that the insulating phase in 123 FeSCs \textcolor{black}{could be} driven by $T$ due to the low values of $Z_{w_m}$ \textcolor{black}{for the orbitals which control the band structure around the Fermi level.}

\item In \sref{5.2.3}, I calculated the effect of local correlations in the band structure and the Fermi surfaces at $0 \, GPa$ and $12.4 \, GPa$. The total bandwidths are reduced $\sim \%50$ for $0 \, GPa$ and $\sim \% 30$ for $12.4 \, GPa$. Fermi surfaces suffer a sizable reconstruction, which can be traced back to the shift downwards/upwards of the $w_{zx}$/$w_{yz}$-like bands. $\gamma$ hole pockets change their shape and cut $k_z=\pm \pi$. When including local correlations, there is an enhancement of the intra-ladder AFM nesting vector $\vec{Q}=(0,0,2\pi)$ with respect to the non-interacting case. This enhancement is due to the cut of $\gamma$ pockets with $k_z=\pm \pi$. Nevertheless, at $0 \, GPa$, the Fermi surface will be absent (due to the insulating nature), and any explanation of the AFM \textcolor{black}{in terms of a weakly correlated picture is highly in doubt}. At $12.4 \, GPa$, the strength of correlations becomes similar to other FeSCs. An interesting issue which requires further work is whether the weakly correlated picture for superconductivity is invalidated by the non-weakly correlated picture of magnetism or not. The Fermi surface reconstruction will be important and will affect the possible instabilities, \textcolor{black}{opposite to what usually happens in most quasi-2D FeSCs.}

\end{itemize}


\chapter{{\bf The nature of correlations in the insulating states of twisted bilayer graphene\footnote{Results presented in this Chapter published in \cite{Pizarro3}}}} 
\label{Chap4c}
\lhead{Chapter \ref{Chap4c}. \emph{Nature of correlations in the insulating states of MA-TBG}} 

\begin{small}

During the APS Meeting 2018, Pablo Jarillo-Herrero presented the recent results of his group in the magic-angle twisted bilayer graphene (MA-TBG) \cite{pablojarillo_talkAPS2018}. The results were published in \textcolor{black}{two articles in} Nature \cite{CaoN5562018_ins,CaoN5562018_sc}. The authors have mainly discovered that: $(1)$ electron- and hole-doped MA-TBG show an unexpected insulating behavior for a certain commensurate fillings \cite{CaoN5562018_ins} and $(2)$ around the insulating state of the hole-doped MA-TBG, a couple of superconducting domes emerge with $T_c \sim 1-2 \, K$ \cite{CaoN5562018_sc}. These results open a new area for studying unconventional superconductivity, where the possibilities of creating magic-angle twisted systems and unveling the origin of unconventional superconductivity in a non-invasive and sistematic way are enormous.

In \sref{1.6.1}, I have already introduced the phenomenology previously known for MA-TBG \cite{ReiPRB662002,LopPRL992007,SuaPRB822010,BisPNAS1082011,CaoPRL1172016,FanPRB932016,GonPRL1192017}. In twisted bilayer graphene (TBG), a moir\'e pattern emerges, as shown in \fref{fig:1.21}. In the moir\'e pattern, or alternatively called the superlattice, $AA$ \textcolor{black}{and $AB/BA$} regions form triangular \textcolor{black}{and hexagonal patterns, respectively}. For \textcolor{black}{$\theta < 2 \, ^{\circ}$} and due to the interlayer hybridization $w(\theta)$, at a certain (magic) angles the bands around the Fermi level become very flat, with \textcolor{black}{theoretically obtained} bandwidths $W \sim 10 \, meV$ (see \fref{fig:1.23} for the low-energy continuum model \textcolor{black}{results}), with Dirac points at $K$ and \textcolor{black}{$K'$} \cite{BisPNAS1082011}. Between the flat bands and the next bands, there are two superlattice gaps $\sim 40 \, meV$, \textcolor{black}{obtained by transport measurements} \cite{CaoPRL1172016,CaoN5562018_ins,CaoN5562018_sc,Yan1808.07865}. Around these $K$ points, the electronic density is mostly located in $AA$ regions \cite{FanPRB932016}.

In MA-TBG, there are a total of four spin-degenerated flat bands, in which there is a valley degeneracy, see \sref{1.6.1} \cite{LopPRL992007,BisPNAS1082011,YuaPRB982018,PoHPRX82018,KanPRX82018}. In 2D systems, doping can be induced in a non-invasive way via an external gate voltage \cite{CaoN5562018_ins,CaoN5562018_sc,CaoPRL1172016,Yan1808.07865}. \textcolor{black}{The unexpected insulating states in MA-TBG appear at a doping of $2$ electrons and $2$ holes with respect to the charge neutrality point (CNP), which is defined as zero filling. Alternatively to this notation, here I will define half-filling of the flat bands at the CNP.} At the CNP, the system shows the typical Dirac points of the graphene hexagonal lattice, with no gap opening \textcolor{black}{and a semimetallic behavior seen} in \textcolor{black}{transport} experiments. Due to the fact that in MA-TBG the bandwidth is strongly reduced for the flat bands (hence, $U/W$ ratio will largely increase), \textcolor{black}{as well as that these unexpected insulating states appear at commensurate fillings}, the authors \cite{CaoN5562018_ins,CaoN5562018_sc} argued that the insulating states in MA-TBG are, indeed, Mott insulating states.

During the next months, there \textcolor{black}{has been} a lot of activity in the scientific community to model and discuss the phenomenology of MA-TBG. Here, I present the results that we have published \cite{Pizarro3} about \textcolor{black}{our analysis of} the nature of the insulating states in electron- and hole-doped MA-TBG. \textcolor{black}{Our goal is to check whether the insulating states in MA-TBG are compatible with Mott physics.} We \textcolor{black}{will} assume that the electronic interaction in MA-TBG is short-ranged due to the screening effect of the external gate voltage, \textcolor{black}{hence we will be able to use an effective Hubbard model for the flat bands}. \textcolor{black}{We \textcolor{black}{will focus on} the analysis} of three experimental results which will not be explained when only local correlations\textcolor{black}{, i.e. correlations between electrons in the same lattice \textcolor{black}{site (see \cref{Chap01})}} are considered: $(i)$ the insulating behavior \textcolor{black}{disappears} when temperature increases, $(ii)$ the insulating behavior also dissapears when the magnetic field increases and $(iii)$ the gap size ($\sim 0.3 \, meV$) is two orders of magnitude smaller than \textcolor{black}{the expected gap} $U-W \sim 15-25 \, meV$. We will conclude by proposing that \textcolor{black}{the Mott physics can explain the insulating states of MA-TBG where} non-local correlations \textcolor{black}{(correlations between electrons in different lattice sites)} are playing a major role in MA-TBG, rather than be \textcolor{black}{an effect of the} local correlations.

In this chapter, I first discuss the tight-binding model used for MA-TBG. \textcolor{black}{I will introduce} the triangular and hexagonal symmetries discussed when modeling MA-TBG \cite{YuaPRB982018,PoHPRX82018,KanPRX82018,KosPRX82018,Rad1805.05294}. We chose the hexagonal \textcolor{black}{moir\'e superlattice} as the best initial approximation for MA-TBG. We consider 2 \textcolor{black}{equivalent} orbitals, and only nearest neighbor intraorbital hoppings $t \approx 2 \, meV$ (in a hexagonal lattice, the total bandwidth is $W=6t \approx 12 \, meV$). In this model, the $A$ and $B$ sites of the hexagonal lattice represent the $AB$ and $BA$ regions of MA-TBG. I write all the energy scales in units of the hopping $t$

I will calculate the \textcolor{black}{critical interaction for the Mott transition by using the} SSMF $U(1)$ \textcolor{black}{formalism} for $N=2$ equivalent orbitals \textcolor{black}{in a honeycomb lattice. Similar results were already studied for other models in} \cite{RozPRB551997,HanPRB581998,FloPRB702004,deMPRB832011,YuaPRB982018}. For multiorbital systems, Mott insulating states appear at certain commensurate fillings, which in the case of $N=2$ is for a number of electrons per site \textcolor{black}{$n_e=1$ (quarter filling), $n_e=N=2$ (half-filling) and $n_e=3$ (three-quarter filling)}. In order to reproduce the experimental findings, \textcolor{black}{i.e.} at half-filling \textcolor{black}{(the CNP of MA-TBG where the Dirac points are located)} the system shows a metallic behavior, and at quarter \textcolor{black}{and three-quarter} filling \textcolor{black}{of the flat bands} the system is an insulator, I will show that $J_H$ should be small, \textcolor{black}{so I will take} $J_H =0$ as an initial guess for its value in MA-TBG.

Then, I will calculate \textcolor{black}{the effect of a Zeeman field in the Mott transition} in the SSMF $U(1)$ formalism by including a spin-dependent onsite energy shift, to study the insulating states of MA-TBG response with the magnetic field. Breaking spin-degeneracy is a novel implementation in SSMF techniques. I will show that the opposite tendency with respect to the experiments is found: when the Zeeman magnetic field increases, the system becomes more insulating (and not metallic). \textcolor{black}{I will argue that the metallic state promoted by temperature $T$} is also at odds with the theoretical results in local correlations, as shown in \sref{1.3.3} for single \textcolor{black}{and multiorbital} systems in which insulating behavior is promoted when $T$ increases \cite{GeoRMP681996,VucPRB882013,Fazekas1999}. For $N=2$ orbitals, this phenomenology holds \cite{RozPRB551997}. Also, the small size of the gap \textcolor{black}{could not} be explained when considering only local correlations \cite{DodPRB982018,Liu1804.10009,GeoRMP681996}, in which \textcolor{black}{the gap ($\sim U-W$) is expected to be two orders of magnitude smaller than the one obtained in transport measurements ($\sim 0.3 \, meV$), see \sref{1.3.1}.}

Thus, \textcolor{black}{assuming that the Mott physics is at the heart of the insulating states in MA-TBG}, we \textcolor{black}{will show} that by only considering local correlations the experimental results cannot be explained, and hence it is necessary to go beyond local correlations, and consider non-local correlations effects. \textcolor{black}{Based on previous results in other lattices \cite{MaiRMP772005,PotPRL932004,BiePRL942005,KitPRB792009,NomPRB892014,NomPRB912015,ParPRL1012008,SchPRB912015,LiePRB872013,LiQPRB922015,MomPRB581998,KimPRB962017,DeFPRB982018,ZhoPRB932016,CorPRX22012,JakPRB932016,CorPRL1072011}, we argued that the} experimental results could be explained by including \textcolor{black}{these} non-local phenomenology. Nevertheless, further calculations are needed to study the insulating states in MA-TBG. For non-local correlations, the tight-binding model details are important, and hence these calculations would required a good enough tight-binding model for MA-TBG, something which is yet not available

\end{small}

\newpage

\section{Introduction \& Motivation}
\label{6.1}

On \textcolor{black}{March} 2018, Pablo Jarillo's group \textcolor{black}{reported} \cite{CaoN5562018_ins} that MA-TBG shows unexpected insulating states at a certain commensurate electron and hole dopings \textcolor{black}{for the moir\'e superlattice (and not for the lattice of $C$ atoms, see below)}. Pablo Jarillo's group also discovered \cite{CaoN5562018_sc} that, around the insulating state for hole-doped MA-TBG, two superconducting domes with $T_c \sim 1-2 \, K$ emerge, in a similar fashion as in high-$T_c$ cuprates superconductors, see \sref{1.5}. Since then, a lot of scientific effort has been put by the community to understand the physics playing a role in MA-TBG. \textcolor{black}{The emergence of insulating states at commensurate fillings suggests the possibility to describe the system using an} effective tight-binding model for the moir\'e superlattice. However, \textcolor{black}{this} tight-binding model \textcolor{black}{for the moir\'e superlattice} is still difficult to establish \cite{YuaPRB982018,PoHPRX82018,KanPRX82018,KosPRX82018}. In this section, I am going to explain various important results from Pablo Jarillo's group \cite{CaoN5562018_ins,CaoN5562018_sc}, as well as giving a brief summary of the works done after these discoveries about the tight-binding model \textcolor{black}{for the moir\'e superlattice} \cite{YuaPRB982018,PoHPRX82018,KanPRX82018,KosPRX82018,Rad1805.05294}. \textcolor{black}{At the end of this chapter, I summarize} the results presented in this work, as well as making a brief comment of the recent work and discussion about the tight-binding model in MA-TBG \cite{PoH1808.02482}.

In \sref{1.6.1}, I have explained the basic phenomenology known about TBG \cite{ReiPRB662002,LopPRL992007,SuaPRB822010,BisPNAS1082011,CaoPRL1172016,FanPRB932016,GonPRL1192017} before Pablo Jarillo's group discoveries. In TBG, a moir\'e pattern appears for a twisting angle $\theta$, see \fref{fig:1.21}. The size of this moir\'e pattern $\lambda$ varies with $\theta$ and takes values of the order of $10-20 \, nm$ \textcolor{black}{for $\theta \sim 1-1.5 \, ^{\circ}$}. In the moir\'e pattern, or alternatively the superlattice, $AA$ regions form a triangular superlattice, while $AB$ and $BA$ form a hexagonal superlattice. \textcolor{black}{This} notation is used to refer to the alignement between bottom and top \textcolor{black}{sublattice of the graphene layers}. For \textcolor{black}{$\theta < 2 \, ^{\circ}$} and due to the interlayer hybridization $w(\theta) \propto \sin{\theta} $, at a certain (magic) angles the bands around the Fermi level become \textcolor{black}{very} flat (see \fref{fig:1.23}), with Dirac points at $K$ and \textcolor{black}{$K'$}. The set of angles at which the bands around the \textcolor{black}{CNP} become very flat appears due to a vanishing Fermi velocity $v_F =0$, hence $v_F$ is a cyclic function of $\theta$, as shown in \fref{fig:1.22}\textbf{(d)}. The flat bands disperse between the maxima and minima located at $\Gamma$, and between them and the next bands, there are a couple of superlattice gaps which appear due to the hybridization of the two graphene layers \cite{BisPNAS1082011,KosPRX82018,CaoPRL1172016}, and also due to corrugation effects in the interface of the two graphene layers \cite{KosPRX82018} (see \sref{1.6.1} and discussion about $w \neq w'$). Around the $K$ points, the electronic density is located in the $AA$ regions \cite{FanPRB932016}.

\begin{figure}[h]
   \centering
   \includegraphics[width=0.8\columnwidth]{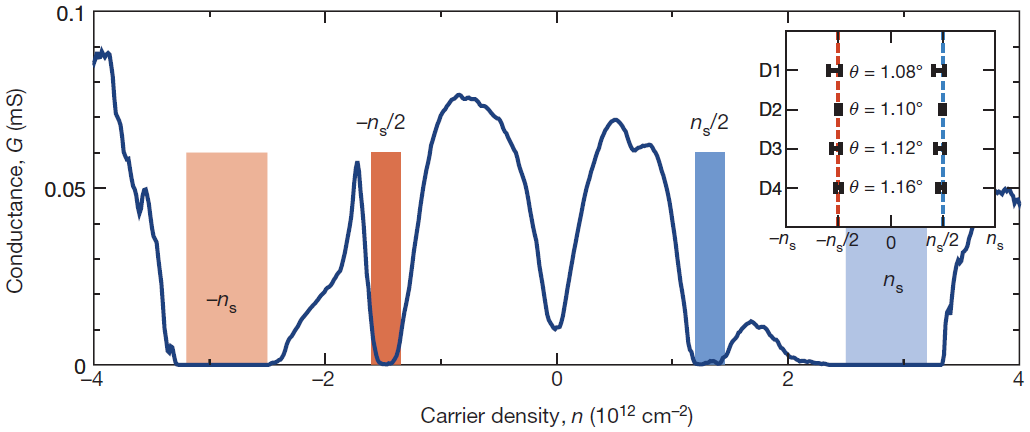}
   \caption[Conductance versus external gate voltage induced doping for MA-TBG]{Conductance versus external gate voltage induced doping for MA-TBG with $\theta=1.08 \, ^{\circ}$ measured at $T=0.3 \, K$. Flat bands develop between $\pm n_s \approx 2.7 \times 10^{12} \, cm^{-2}$, which are identified as the superlattice gaps $\approx 40 \, meV$. At half-filling ($n_s=0$) a V-shaped conductance is found, typical for Dirac points. \textcolor{black}{Insulating} states are found at $\pm n_s/2$. \textcolor{black}{During this chapter, I will use} the alternative notation in which superlattice gaps develop at $x=0$ (light red area) and $x=1$ (light blue area), half-filling is $x=1/2$ (V-shaped conductance), and insulating states appear at commensurate fillings $x=1/4$ (dark red area) and $x=3/4$ (dark blue area). \textsc{Inset:} Locations of \textcolor{black}{the unexpected} insulating states for several rotation angles $\theta$. Taken from \cite{CaoN5562018_ins}.}
   \label{fig:6.1}  
\end{figure}

In \cite{CaoN5562018_ins}, Pablo Jarillo's group have discovered that two gaps ($\approx 0.3 \, meV$) open at a certain commensurate fillings per moir\'e unit cell, \textcolor{black}{$2$ electrons and $2$ holes with respect to the CNP. Here, I will use an alternative notation for the moir\'e superlattice (see below),} they appear at $n_{e}^{moire}=2$ and $6$ electrons \textcolor{black}{with respect to the bottom of the flat bands}, see \fref{fig:6.1} for the conductance curves at $\theta_{1}^{magic} =1.08 \, {^\circ}$ (these two insulating states are marked by a darker red and blue regions, with the label $\pm n_s/2$). The superlattice gaps ($\approx 40 \, meV$), \textcolor{black}{i.e. the gaps associated to the total \textcolor{black}{emptying} ($n_e^{moire}=0$) and total \textcolor{black}{filling} ($n_e^{moire}=8$) of the flat bands,} can be also seen in \fref{fig:6.1} (marked by light red and blue regions and labeled as $\pm n_s$).

\textcolor{black}{According to a hexagonal symmetry,} I will alternatively use the number of electrons per moir\'e site ($n_e=n_e^{moire}/2$), where $n_e= 2$ is half-filling (and the total filling per site, spin and orbital is $x=n_e/2N=1/2$), $n_e = 1$ is quarter filling ($x=1/4$) and $n_e = 3$ is three-quarter filling ($x=3/4$). Due to the small value $W\sim 10 \, meV$ ($U/W$ ratio largely increases), \textcolor{black}{the authors \cite{CaoN5562018_ins,CaoN5562018_sc} argued} that these unexpected insulating states in MA-TBG are in fact Mott insulating states. Interestingly, at half-filling \textcolor{black}{(CNP of MA-TBG where the Dirac points are located)} there is no insulating behavior, and a V-shaped conductance \textcolor{black}{typical for semimetals} is seen \cite{CaoN5562018_ins}.

\begin{figure}[h]
   \centering
   \includegraphics[width=0.8\columnwidth]{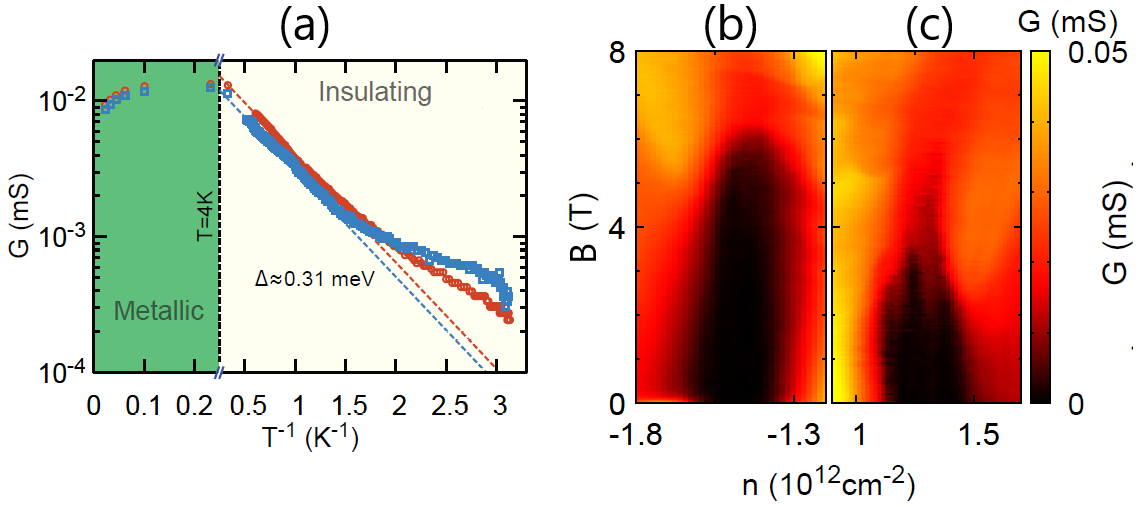}
   \caption[Insulator-to-metal transition for insulating states at quarter and three-quarter fillings when increasing $T$ and increasing the Zeeman magnetic field in MA-TBG]{Insulator-to-metal transition for \textcolor{black}{the unexpected} insulating states \textcolor{black}{(see dark red and dark blue regions in \fref{fig:6.1})} at \textcolor{black}{quarter} (red line) and \textcolor{black}{three-quarter} (blue line) fillings when \textbf{(a)} increasing $T$ and increasing Zeeman magnetic field $B$ for \textbf{(b)} \textcolor{black}{quarter} and \textbf{(c)} \textcolor{black}{three-quarter filling} in MA-TBG. Both transitions appear when the thermal ($k_B T$) and magnetic ($\mu_B B$) excitation energies are of the same order of the $gap\sim 0.3 \, meV$. Taken and adapted from \cite{CaoN5562018_ins}.}
   \label{fig:6.2}  
\end{figure}

The insulating states at quarter and three-quarter filling behave similarly with respect to $T$ and with respect to an external magnetic field $B$\textcolor{black}{, see below} \cite{CaoN5562018_ins}. The magnetic field behavior is the same for both perpendicular and parallel (with respect to the superlattice) directions, hence in \cite{CaoN5562018_ins}, the authors concluded that it comes from a Zeeman magnetic field effect. In both cases, there is an insulator-to-metal transition at $T \approx 4 \, K$ ($\sim 0.3 \, meV$) and $B \approx 5-6 \, T$ ($\sim 0.3 \, meV$). These excitation energies are of the same order of the gap size $\sim 0.3 \, meV$\textcolor{black}{, which is determined by thermal-activated fitting to transport measurements}. In \fref{fig:6.2}, both experimental results are shown. \textcolor{black}{I will argue} in \sref{6.2} that the $T$ and $B$ behaviors signal a major role played by non-local correlations, rather than being a local correlations effect, as usually expected for a Mott insulator.

\begin{figure}[h]
   \centering
   \includegraphics[width=0.8\columnwidth]{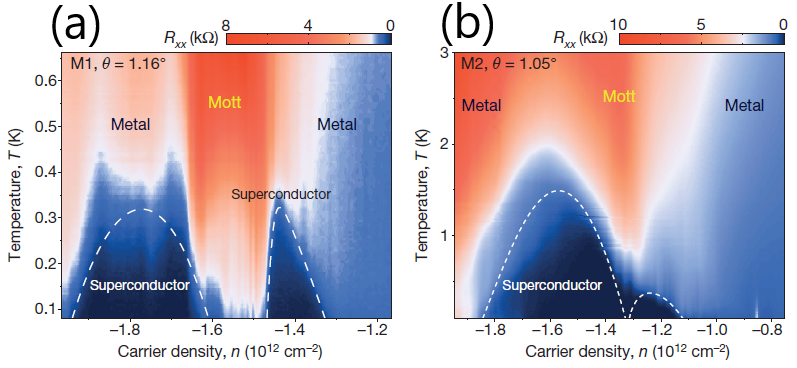}
   \caption[Color map of the resistance for the temperature phase diagram and hole-doped MA-TBG in terms of external gate voltage induced doping at $\theta=1.16 \, ^{\circ}$ and $1.05 \, ^{\circ}$]{Color map of the resistance for the temperature phase diagram and hole-doped MA-TBG in terms of external gate voltage induced doping at \textbf{(a)} $\theta=1.16 \, ^{\circ}$ and \textbf{(b)} $1.05 \, ^{\circ}$. \textcolor{black}{The} insulating phase is colored in red, superconductivity is colored in dark blue. The metallic states are marked in light blue and light red. Superconductivity appears around the insulating state, similar to cuprates and other unconventional superconductors. It appears at a record carrier density, with $T_c^{max} \approx 1.7 \, K$ at $\approx -1.5 \times 10^{12} {cm}^{-2}$. Taken from \cite{CaoN5562018_sc}.}
   \label{fig:6.3}  
\end{figure}

Pablo Jarillo's group \cite{CaoN5562018_sc} also found that, around the quarter filling insulating state, a couple of superconducting domes emerge, see \fref{fig:6.3} for two different twisting angles $\theta$ around $\theta_1^{magic}$. Around the same carrier density at which superconductivity emerges, there is an insulator-to-metal transition. This behavior resembles the one obtained in cuprates and other unconventional superconductors, see \sref{1.5}. MA-TBG superconducting domes show a record low carrier densities (previously held by interfaces $LaAlO_3/SrTiO_3$ with $1.5 \times 10^{13} \, {cm}^{-2}$ \cite{SaiNRM22016}), with $T_c^{max} \approx 1.7 \, K$ at $\approx -1.5 \times 10^{12} {cm}^{-2}$.

\begin{figure}[h]
   \centering
   \includegraphics[width=0.85\columnwidth]{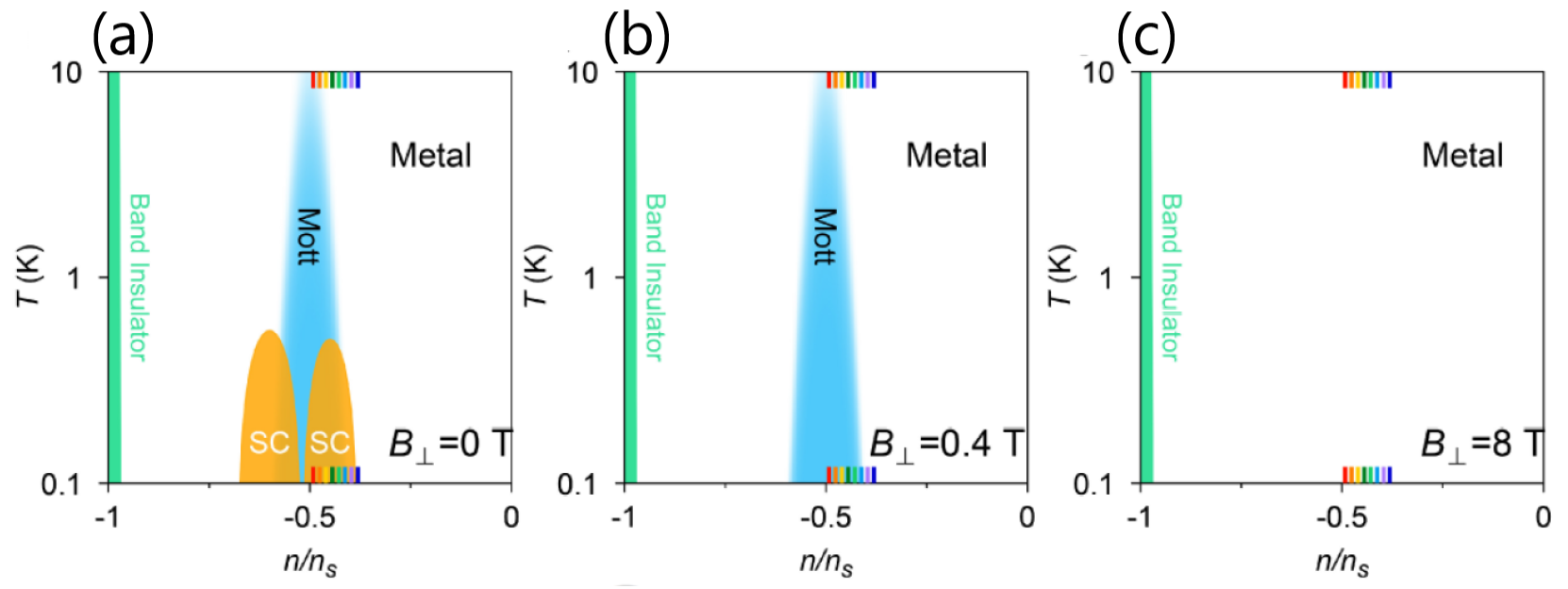}
   \caption[Sketch of the phase diagram for hole-doped MA-TBG in terms of perpendicular magnetic field]{\textbf{(a)}-\textbf{(c)} Sketch of the phase diagram for hole-doped MA-TBG in terms of perpendicular magnetic field. At $B_{\perp}=0$, \textcolor{black}{the} insulating state appears at $x=1/4$, and also a couple of unconventional superconducting domes around \textcolor{black}{it}. At $B_{\perp}=0.4 \, T$ ($\approx 20 \, \mu eV$), superconductivity is suppressed, and at $B_{\perp}=8 \, T$ ($\approx 0.4 \, meV$), \textcolor{black}{the} insulating state dissappears. Taken from \cite{CaoN5562018_sc}.}
   \label{fig:6.4}  
\end{figure}

Superconductivity is supressed by a perpendicular magnetic field $B_{\bot}\approx 70 \, mT$ ($\approx 4 \, \mu eV$) \cite{CaoN5562018_sc}. \fref{fig:6.4} shows the schematic evolution of the phase diagram in terms of the perpendicular magnetic field. \textcolor{black}{As mentioned}, there is a \textcolor{black}{first} suppression of the superconducting domes at \textcolor{black}{$70 \, mT$}, and then an insulator-to-metal transition at \textcolor{black}{$6 \, T$}.

\subsection{Works published after \emph{Y. Cao et al.} discoveries}
\label{6.1.1}

During the months after the discovery of Pablo Jarillo's group presentation and publications \cite{pablojarillo_talkAPS2018,CaoN5562018_ins,CaoN5562018_sc}, a lot of works have been published in arXiv about studying MA-TBG. Here, I will briefly review the ones related with the initial tight-binding model for MA-TBG, focusing in the works \cite{YuaPRB982018,PoHPRX82018,KanPRX82018,KosPRX82018}. \textcolor{black}{These works were uploaded to arXiv prior or simultaneously to our work.}

In \sref{1.6.1}, I have discussed the low-energy continuum model \cite{LopPRL992007,BisPNAS1082011,KosPRX82018}, in which the flat bands around the Fermi level \textcolor{black}{compare well with calculations from tight-binding models in the lattice of $C$ atoms}. Each layer will contribute with its own dispersions $h^l$, where $l=1,2$ is the layer index, and for $\theta<6 \, ^{\circ}$, it can be shown that there is an interlayer hybridization $w(\theta)$ contribution. In the mini-BZ of MA-TBG, there are four bands, with \textcolor{black}{two} Dirac points located at \textcolor{black}{$K$ and $K'$ for each valley $\xi=\pm$ \cite{BisPNAS1082011}, see \sref{1.6.1}}. The low-energy continuum hamiltonian is block diagonal, where each non-zero diagonal block is a $4 \times 4$ (including the \textcolor{black}{layer and} spin degree of freedom) matrix $H_\xi$, see \eref{eq:1.24}. Then, the flat bands consist on two bands per spin and valley (\textcolor{black}{eight} bands if we take \textcolor{black}{into account explicitly} the \textcolor{black}{spin and} valley degrees of freedom). 

\textcolor{black}{Due} to corrugation effects \cite{KosPRX82018}, superlattice gaps open between the flat bands and the \textcolor{black}{higher-in-energy} bands, see \fref{fig:1.23}\textbf{(b)}. In order to be able to study the effects of electronic correlations, a correct and minimal tight-binding model is needed for \textcolor{black}{the moir\'e superlattice of MA-TBG}. Due to the opening of superlattice gaps, it is reasonable to try to obtain the tight-binding model describing only the flat bands. The number of orbitals $N$ which has to be taken to describe the flat bands coincides with the number of bands per spin and per valley (note the valley and spin degeneracy), hence flat bands are described by $N=2$ orbitals per spin and valley \cite{YuaPRB982018,PoHPRX82018,KanPRX82018,KosPRX82018}.

Two different tight-binding models \textcolor{black}{were} proposed for \textcolor{black}{the moir\'e superlattice: the first one consider $s$- and off-plane $p$-like orbitals centered at each $AA$ site of the triangular superlattice \cite{EfiPRB982018,XuCPRL1212018,GuoPRB872018}, while the second one consider two in-plane $p$-like orbitals centered at $AB$ and $BA$ sites of the hexagonal superlattice \cite{YuaPRB982018,PoHPRX82018,KanPRX82018,KosPRX82018}}. Due to the fact that the electronic density is mainly located in $AA$ regions around $K$ points \cite{FanPRB932016}, it seems reasonable to start from a triangular superlattice model. However, according to the symmetry analysis done in \cite{YuaPRB982018,KosPRX82018,PoHPRX82018,KanPRX82018}, the two orbitals should be centered at the $AB$ and $BA$ regions, forming a hexagonal superlattice.  The basic idea of these arguments is that the flat bands  have to fulfill certain symmetries at the principal points of the mini-BZ, $\Gamma$, $K$ and $M$. It turns out that the triangular \textcolor{black}{superlattice} model does not correctly \textcolor{black}{capture the symmetries of the flat bands}, \textcolor{black}{like} for example, \textcolor{black}{it} cannot reproduce the Dirac points at $K$ \textcolor{black}{and $K'$}. Then, \textcolor{black}{it was proposed that} the initial tight-binding model for the flat bands have to be written in a hexagonal superlattice with $N=2$ in-plane $p$-like orbitals.

\begin{figure}[h]
   \centering
   \includegraphics[width=0.8\columnwidth]{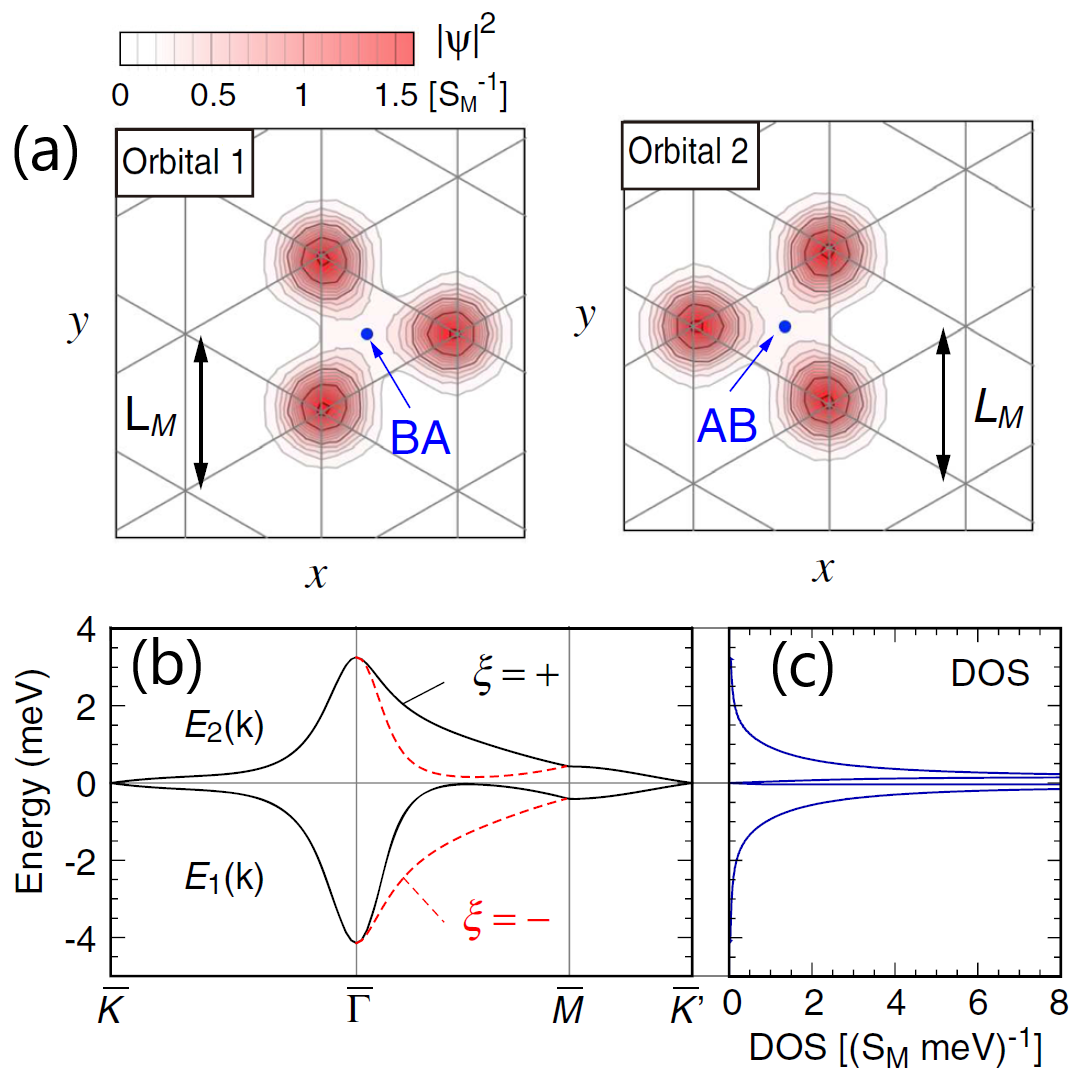}
   \caption[Squared amplitudes for the two in-plane $p$-like orbitals in the hexagonal superlattice formed by $AB/BA$ regions for MA-TBG, and resulting tight-binding flat bands and density of states when including up to five neighbor hoppings]{\textbf{(a)} Squared amplitudes $| \psi |^2$ for the two $p$ orbitals (labeled as $1$ and $2$), centered in the hexagonal superlattice formed by $AB/BA$ regions. The maximum amplitude is located at the $AA$ regions. \textbf{(b)} Tight-binding model \textcolor{black}{obtained from the fitting} (black dashed line) using five neighbor hoppings for the two in-plane $p$-like orbitals hexagonal superlattice model, where there is a good enough agreement with the low-energy continuum model flat bands (green continuous line). \textcolor{black}{\textbf{(c)} Total DOS for the low-energy continuum model of MA-TBG, which shows the typical Dirac point at the CNP, and also two very large van-Hove singularities emerge \textcolor{black}{because of the $M$ saddle point}. The quarter and three-quarter fillings are located at a larger energy than the location of this van-Hove peaks}. Taken and adapted from \cite{KosPRX82018}.}
   \label{fig:6.5}  
\end{figure}

\textcolor{black}{When obtaining the Wannier parametrization in the hexagonal superlattice model}, the maximum amplitude of the in-plane $p$-like orbitals is located in the $AA$ regions, as shown in \fref{fig:6.5}\textbf{(a)}. This result agrees with the fact that the electronic density is located at $AA$ regions around the $K$ points. The resulting tight-binding model, \textcolor{black}{as well as the density of the states (DOS) of the flat bands in the low-energy continuum model} \cite{KosPRX82018} are shown in \fref{fig:6.5}\textbf{(b)} and \textbf{(c)}. \textcolor{black}{Up to five neighbor hoppings are included in this tight-binding model}. Note that in this hexagonal superlattice model, first neighbor \textcolor{black}{hopping} refers to an interorbital hopping, second neighbor \textcolor{black}{hopping} to an intraorbital one, \textcolor{black}{etc}. The total bandwidth is $W \sim 8 \, meV$. A \textcolor{black}{very} good agreement can be seen between \textcolor{black}{the} $N=2$ \textcolor{black}{$p$-like} orbitals tight-binding model (marked as a black dashed line) and the low-energy continuum model (green continuous line) for the flat bands. \textcolor{black}{The total DOS shows the typical Dirac point at the CNP, and also two very large van-Hove singularities emerge \textcolor{black}{because of the $M$ saddle point}. The quarter and three-quarter fillings are located at a larger energy than the location of this van-Hove peaks \cite{KosPRX82018}. The low-energy continuum model shows particle-hole symmetry, while in the real material this symmetry is broken \cite{MooPRB872013}.}

\section{Results \& Discussion}
\label{6.2}

Now, I will present the results that \textcolor{black}{we} have obtained and published in \cite{Pizarro3}. In \sref{6.2.1}, I will first discuss the tight-binding model \textcolor{black}{that we have used. We} chose $N=2$ equivalent orbitals per site in a hexagonal lattice with only intraorbital hoppings $t \sim 2 \, meV$ ($W = 6t \sim 12 \, meV$) model. \textcolor{black}{We have assumed} that long-range interactions are screened in MA-TBG due to the external gate voltage (which induces the doping in MA-TBG) in the experiments of Pablo Jarillo's group \cite{CaoN5562018_ins,CaoN5562018_sc}. \textcolor{black}{This is because the metallic gate is placed at a distance similar to the moir\'e superlattice constant}. Then, I will \textcolor{black}{present} the local correlations calculations in this $N=2$ equivalent orbitals model \cite{RozPRB551997,HanPRB581998,FloPRB702004,deMPRB832011,YuaPRB982018}, using the SSMF $U(1)$ formalism at half-filling $n_e=N=2$ ($x=1/2$) and quarter filling $n_e=1$ ($x=1/4$). I will study the critical interactions for the Mott transition $U_c/t$ in terms of $J_H/U$. I will \textcolor{black}{show} that, in order to reproduce the experimental findings (at quarter filling, the system behaves as an insulator, while at half-filling, it shows metallic behavior), a small $J_H$ value has to be considered. I will consider then $J_H =0$ as a first approximation for the real $J_H$ value in MA-TBG. In \sref{6.2.2}, I implement the Zeeman effect in the SSMF $U(1)$ formalism. I will \textcolor{black}{also show that the evolution of the insulating states in a magnetic field $H$ as observed experimentally ($H$ promotes metallic behavior) cannot be explained if only local correlations are considered}. I will also remind that, \textcolor{black}{when only onsite correlations are \textcolor{black}{taken} into account}, $T$ promotes an insulating behavior \cite{GeoRMP681996,VucPRB882013,Fazekas1999,RozPRB551997} and the gap should be of the order of $\sim U-W$ \textcolor{black}{(where estimations give $U \sim 20-25 \, meV$ and $W \sim 10 \, meV$)} \cite{DodPRB982018,Liu1804.10009,GeoRMP681996}, also at odds with the experimental results. \textcolor{black}{Assuming that the Mott physics still plays a major role in the insulating states of MA-TBG, these theoretical results point to a non-local correlations origin of these insulating states}. In \sref{6.2.3}, I will briefly review the results obtained for non-local correlations in various single- and two-orbital systems \cite{MaiRMP772005,PotPRL932004,BiePRL942005,KitPRB792009,NomPRB892014,NomPRB912015,ParPRL1012008,SchPRB912015,LiePRB872013,LiQPRB922015,MomPRB581998,KimPRB962017,DeFPRB982018,ZhoPRB932016,CorPRX22012,JakPRB932016,CorPRL1072011}. Then, we conclude by \textcolor{black}{proposing that the nature of the insulating states in MA-TBG could be explained by Mott physics if non-local correlations are explictly taken into account.}

\subsection{Local correlations in 2-orbital tight-binding model for hexagonal symmetry MA-TBG}
\label{6.2.1}

In order to study the \textcolor{black}{effects of} local correlations in MA-TBG, we \textcolor{black}{have used a} tight-binding model by considering $N=2$ equivalent orbitals in a hexagonal lattice with only intraorbital hoppings $t\approx 2 \, meV$ included, \textcolor{black}{aimed to qualitatively reproduce the flat bands of MA-TBG}. As mentioned in \cref{Chap01}, the details of the tight-binding model will not affect the local correlations physics (like for example, the behavior of the critical interaction for the Mott transition $U_c$ in terms of $J_H$), and only the energy scales at which the effects occurr will change \textcolor{black}{slightly with the inclusion of further neighbor hoppings}. In this hexagonal lattice model, $A$ and $B$ sites represent the $AB$ and $BA$ regions in MA-TBG, and the two orbitals are located in each $A$ and $B$ site in order to reproduce the \textcolor{black}{valley degree of freedom of the} four \textcolor{black}{flat} bands. The total bandwidth $W$ is $6t \approx 12 \, meV$. For simplicity, I will usually write all the energy scales in units of $t$.

\begin{figure}[h]
   \centering
   \includegraphics[width=0.9\columnwidth]{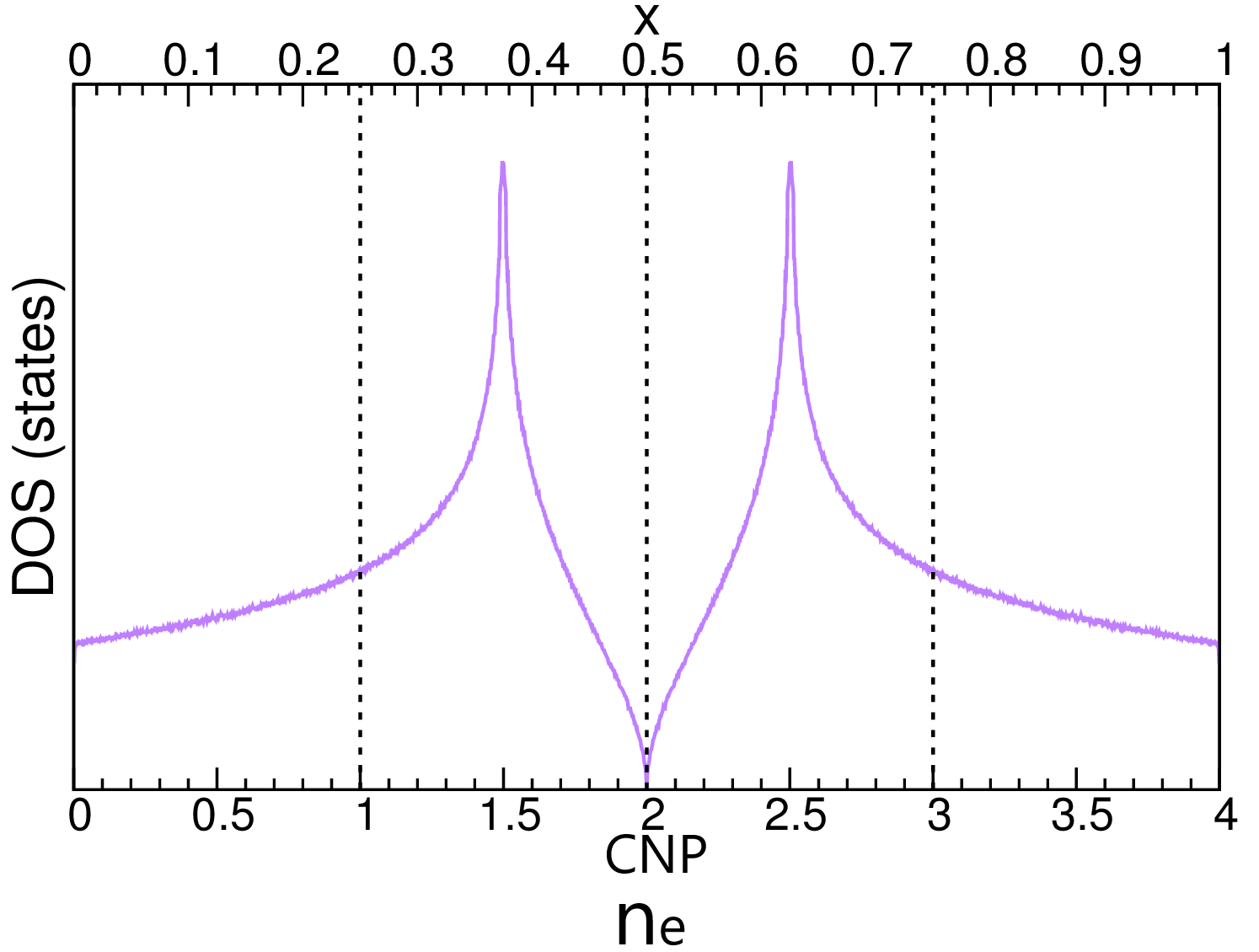}
   \caption[DOS for the $N=2$ equivalent orbitals hexagonal lattice with only nearest neighbor hoppings included]{DOS for the $N=2$ equivalent orbitals hexagonal lattice with only nearest neighbor hoppings included, as described in \eref{eq:6.1}. \textcolor{black}{Our model is particle-hole \textcolor{black}{symmetric} with respect to half-filling $n_e=2$ ($x=1/2$). Vanishing DOS can be seen at the CNP, as expected for Dirac points. The van-Hove singularities are located $n_e \sim 1.5$ and $\sim 2.5$, in between quarter (and three-quarter) filling and half-filling, \textcolor{black}{quantitatively} similar to the low-energy continuum model \cite{KosPRX82018}.}}
   \label{fig:6.6}  
\end{figure}

The tight-binding hamiltonian $H_0$ (note that the dispersion relations are $2 \times 2$ matrices which are the same for both degenerated orbitals) for a hexagonal lattice with intraorbital nearest neighbor hoppings $t$ (in the hexagonal first Brillouin zone) is:

\begin{equation}
H_0 \begin{pmatrix}
0	& 	\varepsilon_k^{AB} \\
\left(\varepsilon_k^{AB}\right)^\dagger	& 	0 \\
\end{pmatrix} \: \rightarrow \:
\varepsilon_k^{AB} = - t - 2 t \cos{\left(\frac{k_x}{2}\right)} e^{i \sqrt{3}k_y/2}
\begin{pmatrix}
1	& 	0 \\
0	& 	1 \\
\end{pmatrix}
\label{eq:6.1}
\end{equation}

where the electrons jump between $A$ and $B$ sites of the hexagonal lattice and between the same orbitals. In \fref{fig:6.6}, I have plotted the DOS in terms of the number of electrons per site $n_e$ and the filling per orbital, spin and site $x=n_e/2N$. \textcolor{black}{This model shows a particle-hole symmetry around the CNP, but we expect that the results when including only onsite correlations will not change qualitatively}. The Dirac points \textcolor{black}{are located} at half-filling, and the van-Hove singularities are located in the middle point of quarter (and three-quarter) and half-filling. \textcolor{black}{Compared with the low-energy continuum model of \cite{KosPRX82018}, our model would be at least qualitatively similar, where also the van-Hove peaks are located in between quarter and three-quarter filling and the CNP}.

In principle, due to the 2D character of MA-TBG, long-range interactions will become more important because of the reduced screening, see \sref{1.6}. In Pablo Jarillo's group experiments \cite{CaoN5562018_ins,CaoN5562018_sc}, MA-TBG is sandwiched between hexagonal boron-nitride (hBN), whose thichness is $\approx 10-30 \, nm$, of the order of the superlattice size ($\lambda \sim 13 \, nm$) for MA-TBG. The doping is induced by an external gate voltage, and due to the similar thickness of the hBN layers with respect to the superlattice distance, we \textcolor{black}{believe} that this external electric field is effectively screening the long-range character of the Coulomb interaction, hence disminishing the long-range interaction effects, in a first approximation. We then assume that the interaction effects will \textcolor{black}{be restricted to electrons which are in the same superlattice site}. I will use the SSMF $U(1)$ formalism to calculate the effect of local correlations, see \aref{AppB}.

\begin{figure}[h]
   \centering
   \includegraphics[width=0.9\columnwidth]{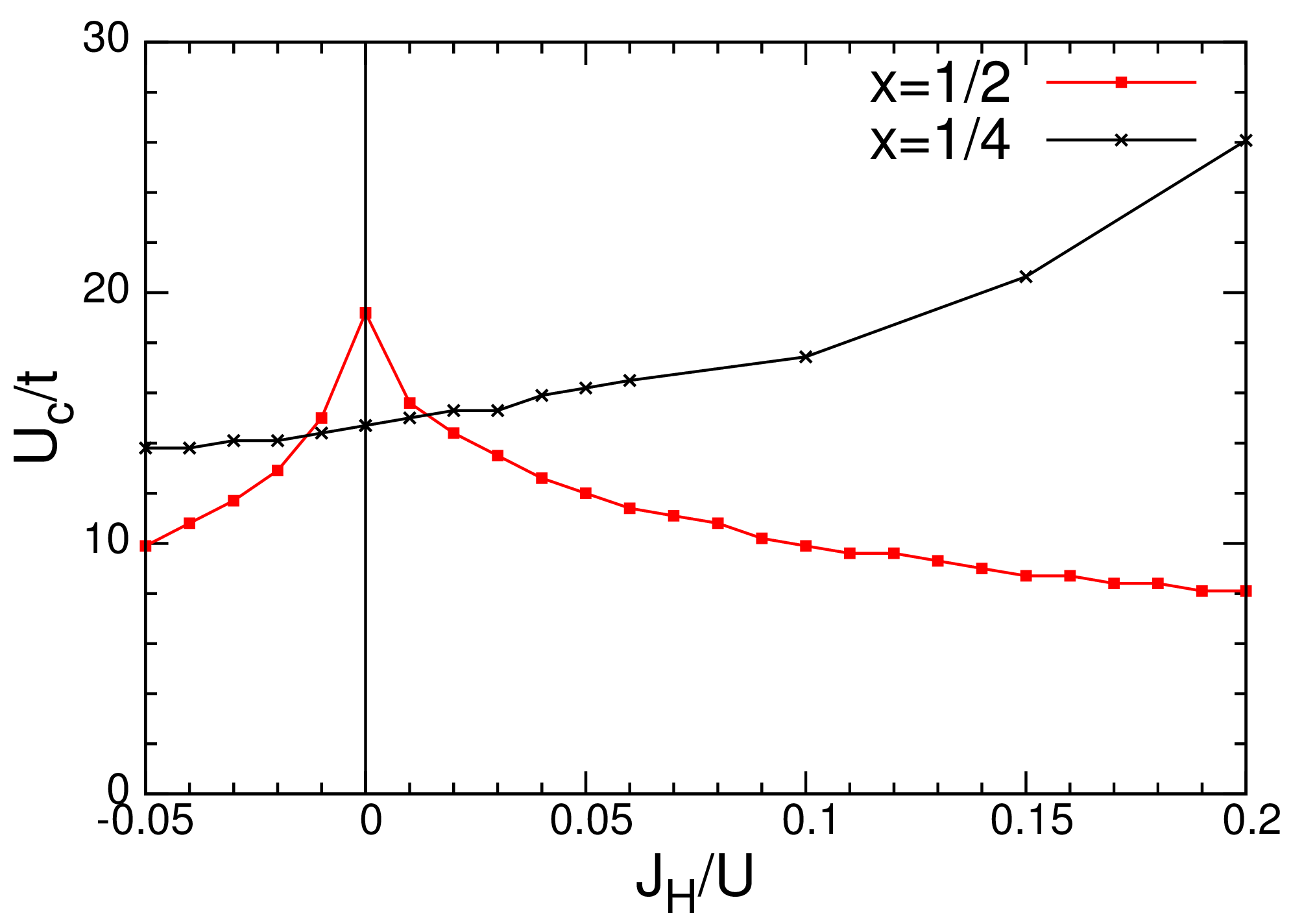}
   \caption[$U_c/t$ versus $J_H/U$ for $x=1/2$ and $x=1/4$ in the hexagonal tight-binding model]{$U_c/t$ versus $J_H/U$ for $x=1/2$ (red line) and $x=1/4$ (black line) in the hexagonal tight-binding model proposed \eref{eq:6.1}. $U_c$ behavior follows \eref{eq:3.3} and \eref{eq:6.2}. We found that metallicity at $x=1/2$ and a Mott insulating state at $x=1/4$ occurr in a small range $-0.01<J_H/U<0.01$.}
   \label{fig:6.7}  
\end{figure}

In \fref{fig:6.7}, I reproduce the critical interaction for the Mott transition $U_c/t$ versus $J_H/U$ \cite{RozPRB551997,HanPRB581998,FloPRB702004,deMPRB832011,YuaPRB982018}, by using SSMF $U(1)$ formalism at $x=1/2$ (red line) and $x=1/4$ (black line).

\textcolor{black}{For $J_H \leq 0$, $U_c$ behavior follows the evolution already explained in \cref{Chap03}}. The region $J_H/U <0$ is \textcolor{black}{included because} some authors \cite{NomJPCM282016} argued that, once the effect of phonons (with frequencies $\omega \sim 200 \, meV$) is included in MA-TBG, $J_H$ may take an effective small negative value. This region might be of interest when applying non-local correlations techniques. For $J_H < 0$, the interaction energy cost $\Delta^x$ can be obtained as done in \eref{eq:3.3} for $J_H>0$, so in the large-$U$ limit:

\begin{equation}
\Delta^x \, (J_H < 0) = \left\{
\begin{aligned}
& = U + 5 J_H \quad & n_e=N \\
& = U \quad & other \, integer \, n_e
\end{aligned}
\right.
\label{eq:6.2}
\end{equation}

In a multiorbital system, the ratio between the interaction energy cost and the kinetic energy gain is given by $\Delta^x/\widetilde{W}$, see \cref{Chap03}, and this ratio controls the critical interaction $U_c^x$ at a given filling $x$. So, when $\Delta^x$ increases, $U_c^x$ will decrease. Then, for the $J_H<0$ region, at half-filling $n_e=N=2$, $U_c$ decreases if $|J_H|$ increases (Mott insulating behavior promoted), and at quarter filling $n_e=1$, $U_c$ is unaffected by $J_H<0$.

As seen in \fref{fig:6.7}, there is a small region of $J_H/U$ in which a Mott insulating state can be obtained for $x=1/4$ and a metallic state for $x=1/2$ (same as in the experimental results, see \fref{fig:6.1}), i.e. $U_c^{x=1/2}>U_c^{x=1/4}$. This region occurs in the small range $-0.01<J_H/U<0.01$. This region appears due to the promotion of metallic behavior in the case of $x=1/2$ (thus, increasing $U_c^x$), but not for $x=1/4$ \cite{RozPRB551997,HanPRB581998,FloPRB702004,deMPRB832011,YuaPRB982018}. \textcolor{black}{I will consider from now on $J_H=0$ as an approximation of its value in MA-TBG. In the case of the hexagonal lattice at $J_H=0$, $U_c^{x=1/4}=14.7t$ ($\sim 30 \, meV$) and $U_c^{x=1/2}=19.2t$  ($\sim 40 \, meV$), and also $U_c^{x=1/2}/U_c^{x=1/4} \sim 1.28$}. In other lattices, such as the square lattice, $U_c^{x=1/2}/U_c^{x=1/4} \sim 1.13$ \cite{RozPRB551997,FloPRB702004,deMPRB832011}. \textcolor{black}{This difference emerges due to the vanishing DOS in the hexagonal lattice at half-filling, in contrast with the van-Hove peak in the square lattice at half-filling}. The relation $U_c^{x=1/2}/U_c^{x=1/4} > 1$ \textcolor{black}{at zero $J_H$} holds for any lattice symmetry in the local correlations case for $J_H=0$.

\begin{figure}[h]
   \centering
   \includegraphics[width=0.9\columnwidth]{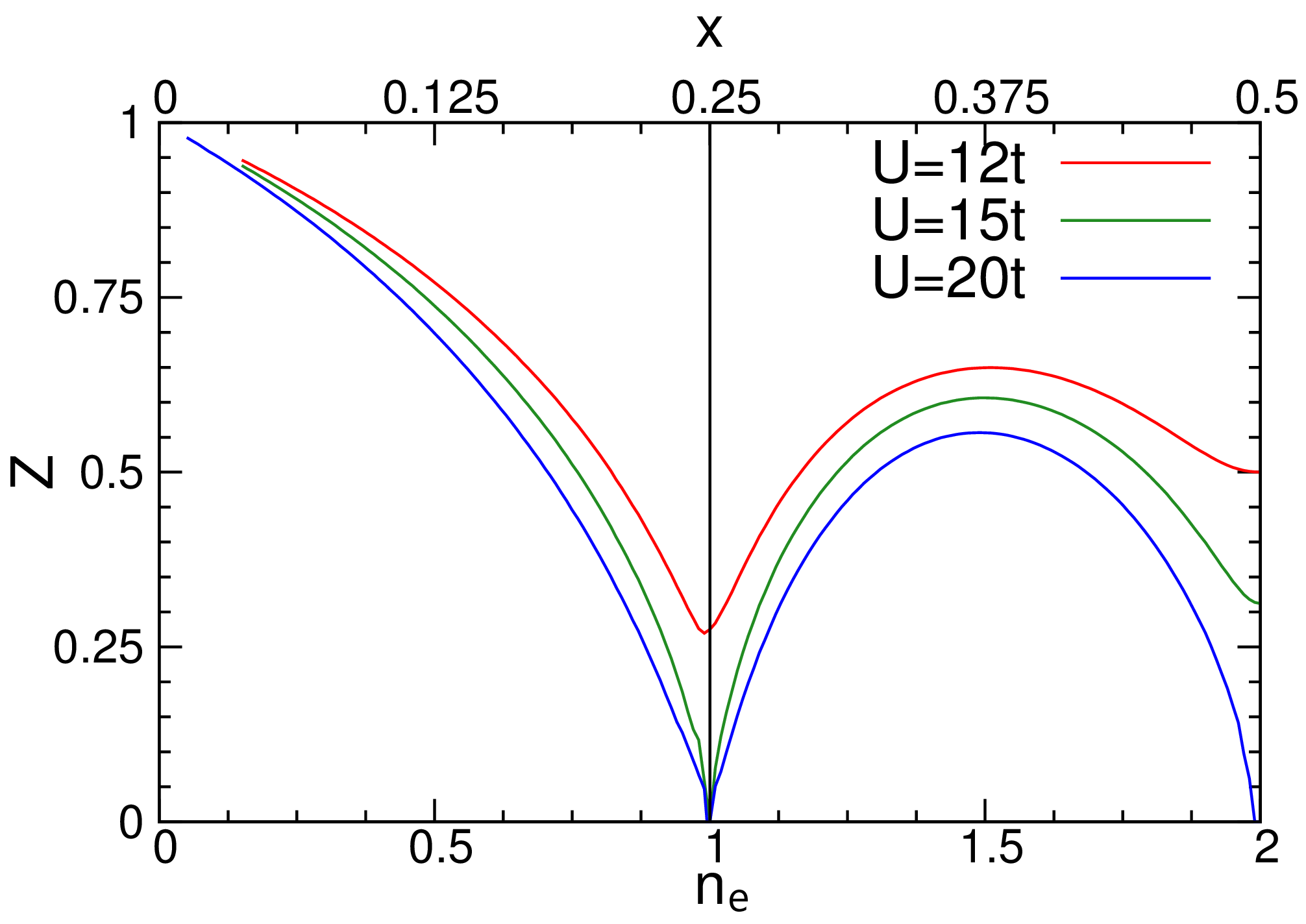}
   \caption[$Z$ versus $n_e$ and $x$ at $U=12t$, $15t$ and $20t$ values and $J_H=0$  in the hexagonal tight-binding model]{$Z$ versus filling curves for $U=12t$ (red line), $15t$ (green line) and $20t$ (blue line), and for $J_H=0$. The experimental results are consistent for $U_c^{x=1/4}<U<U_c^{x=1/2}$, where $U_c^{x=1/4}=14.7t$ and $U_c^{x=1/2}=19.2t$. \textcolor{black}{Correlations are sizable at all dopings, specially at half-filling (where $Z \leq 0.3$), which would produce a sizable effect in the renormalization of the Fermi velocity $v_F$ of MA-TBG. The strength of correlations is asymmetric with respect to the insulating state at $x=1/4$, due to the asymmetry of the DOS around this filling, and also due to the proximity of the Mott insulating state at $x=1/2$ (see \cref{Chap03}). The van-Hove singularity at $x=0.375$ does not have a significant effect.}}
   \label{fig:6.8}  
\end{figure}

In \fref{fig:6.8}, I show $Z$ versus $n_e$ and $x=n_e/2N$ for different $U$ values at $J_H=0$ \cite{RozPRB551997,HanPRB581998,FloPRB702004,deMPRB832011,YuaPRB982018}. \textcolor{black}{Due to the particle-hole symmetry, $Z$ is symmetric with respect to half-filling $n_e=2$}. For interactions $U_c^{x=1/4}<U<U_c^{x=1/2}$, there is a finite $Z$ value (i.e. metallic behavior) at $x=1/2$ and $Z=0$ (Mott insulator behavior) at $x=1/4$. \textcolor{black}{Nevertheless, correlations are sizable at all dopings, specially at half-filling (where $Z \leq 0.3$), which would produce a sizable effect in the renormalization of the Fermi velocity $v_F$ of MA-TBG. The strength of correlations is asymmetric with respect to the insulating state at $x=1/4$, due to the asymmetry of the DOS around this filling, and also due to the proximity of the Mott insulating state at $x=1/2$ (see \cref{Chap03}). The van-Hove singularity at $x=0.375$ does not have a significant effect.}


\subsection{Local correlations and Zeeman effect in MA-TBG}
\label{6.2.2}

I will now study \textcolor{black}{the effect that a Zeeman magnetic field has on the insulating states at} quarter filling ($x=1/4$). In order to do the calculations, \textcolor{black}{I have implemented the Zeeman field term in the SSMF $U(1)$ formalism, as detailed in \aref{AppC}, by adding the following term to the Hubbard-Kanamori hamiltonian of \eref{eq:1.22}:}

\begin{equation}
H_{Zeeman} = H \sum_{im} (\hat{n}_{im\uparrow}^d - \hat{n}_{im\downarrow}^d)
\label{eq:6.3}
\end{equation}

where $\hat{n}_{im\sigma}^{d} = d_{im\sigma}^\dagger d_{im\sigma}^{}$. \textcolor{black}{Through this section, I will check how the critical interaction for the Mott transition $U_c$ evolves in terms of the Zeeman field $H$.} 

The Zeeman field $H$ breaks the spin degeneracy. \textcolor{black}{Here} the majority spin band \textcolor{black}{is labeled with} $\downarrow$ and the minority spin with $\uparrow$. This term is a spin-dependent onsite energy shift, so the majority spin band will move downwards, \textcolor{black}{being} progresively filled when $H$ increases, and the minority spin band will move upwards, \textcolor{black}{being} emptied when $H$ increases. When the minority spin band is empty, its quasiparticle weight $Z_{\uparrow}$ cannot be defined anymore\footnote{Note that the condition to define the quasiparticles in the Fermi liquid theory (FLT) is that the corresponding bands are crossed by the Fermi level, see \sref{1.2.2}.}. Then, $H$ will produce a spin polarization which is enhanced by increasing interactions. \textcolor{black}{This enhancement appears when the majority spin band progresively approaches half-filling, thus following the ideas pointed out in previous chapters. Then, \textcolor{black}{the filling} per spin $n_{m\sigma}$ of the} majority spin band evolves from $n_{m\downarrow} = 0.25$ (quarter filling) to $0.5$, \textcolor{black}{and} the minority spin band is emptied.




\begin{figure}[h]
   \centering
   \includegraphics[width=0.8\columnwidth]{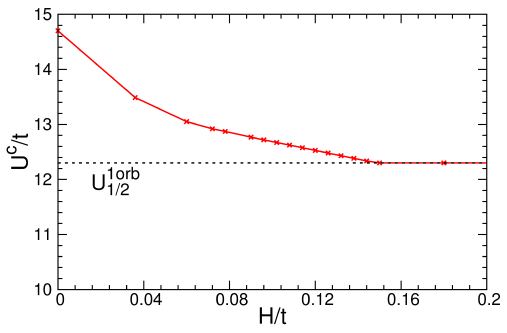}
   \caption[Critical interaction for the Mott transition in terms of the Zeeman field for the \textcolor{black}{honeycomb} lattice at quarter filling]{\textbf{(a)} Critical interaction for the Mott transition $U_c/t$ versus Zeeman field $H/t$ for the \textcolor{black}{honeycomb} lattice model at quarter filling $x=1/4$. At $H_c \sim 0.15t$, the system fully polarizes, saturating to an effective single-orbital at half-filling system, in which the critical interaction is given by $U_{1/2}^{1orb}$.}
   \label{fig:6.9}  
\end{figure}

$H$ promotes insulating behavior (when $H$ increases, $\Delta^{x=1/4}$ decreases, and hence $U_c^{x=1/4}$ increases), rather than promoting metallicity as expected from the experiments \cite{CaoN5562018_ins}. At $J_H=0$, $H$ \textcolor{black}{reduces the degeneracy of the ground state, hence it reduces the effective kinetic energy $\tilde{W}$}, see \sref{3.2.2}. Then, $U_c^{x=1/4}$ will be reduced with respect to its zero Zeeman field value $U_c^{x=1/4}(H=0)$. In \fref{fig:6.9}, I show the results for $U_c/t$  in terms of the Zeeman field $H/t$ at quarter filling $x=1/4$.

\begin{figure}[h]
   \centering
   \includegraphics[width=0.9\columnwidth]{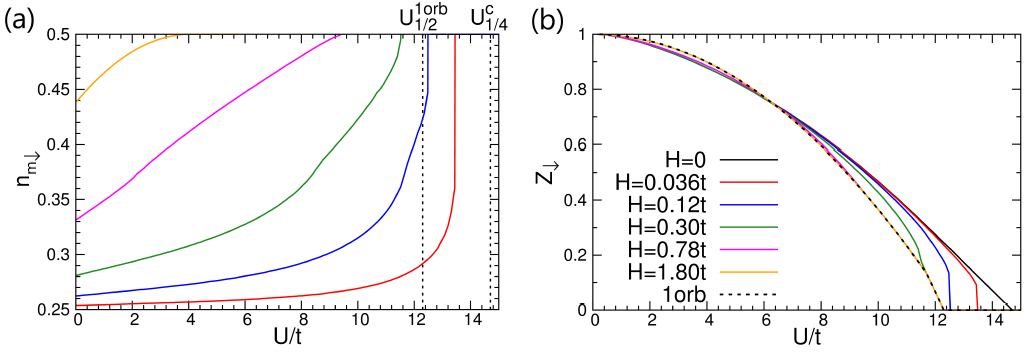}
   \caption[Majority spin filling and quasiparticle weight for various values of the Zeeman field for the hexagonal lattice at quarter filling]{\textbf{(a)} Majority spin filling $n_{m\downarrow}$ and \textbf{(b)} quasiparticle weight $Z_{\downarrow}$ curves versus $U/t$ at various Zeeman field $H/t$ values. $H/t$ values included in the legend. I included the $Z$ evolution in terms of $U/t$ for the single-orbital system at half-filling (labeled by "1orb"). The system fully polarizes at a $H \sim 0.15t$ ($\sim 0.3 \, meV$). For weak Zeeman field, $H \leq 0.15t$, the system jumps into the Mott insulator at $U_c>U_{1/2}^{1orb}$, while for strong Zeeman field, $H>0.15t$, it follows the single-orbital at half-filling behavior.}
   \label{fig:6.10}  
\end{figure}

Once the orbitals are completely spin polarized (i.e. $n_{m\downarrow}=0.5$ and $n_{m\uparrow}=0$), I have found that the system becomes equivalent to a single-orbital model at half-filling with a critical interaction $U_{1/2}^{1orb}<U_c^{x=1/4}(H=0)$. Two regimes can be distinguished depending on the $H/t$ value: weak Zeeman field $H\leq 0.15t$ and strong Zeeman field $H>0.15t$. As shown in \fref{fig:6.10}, in the strong Zeeman field regime, the system saturates to $U_{1/2}^{1orb}$, following the $Z$ evolution of the single-orbital system at half-filling down to the Mott insulating phase. In the weak Zeeman field regime, once the system fully polarizes, it jumps into the Mott insulator at $U>U_{1/2}^{1orb}$.

Then, I have shown that the Zeeman field at quarter filling in the local correlations picture will promote insulating behavior ($U_c^{x=1/4}$ decreases when $H$ increases) \textcolor{black}{as it reduces the degeneracy of the ground state}. This is precisely the opposite behavior than the \textcolor{black}{one found in} the experiments \cite{CaoN5562018_ins}.

Besides ours, other works \cite{DodPRB982018,Liu1804.10009} also remarked the fact that the gap ($\approx 0.3 \, meV$) of the Mott insulating phases is in contradiction with the expected gap from a Mott insulator in the local correlations picture. In a single-orbital at half-filling system \cite{GeoRMP681996}, \textcolor{black}{the Mott transition takes place at $U_c \sim 1.5-2 W$ and the gap is $\sim U-W \sim 15-25 \, meV$}. In multiorbital systems, the gap \textcolor{black}{\textcolor{black}{is of the} same order of magnitude} \cite{DodPRB982018,Liu1804.10009}. 

In \sref{1.3.3}, I have shown that in local correlations for a single-orbital at half-fillling system, the temperature $T$ promotes an insulating behavior \cite{GeoRMP681996,VucPRB882013,Fazekas1999}, rather than metallicity (see \fref{fig:1.10}\textbf{(b)}). \textcolor{black}{In} multiorbital systems the trend is similar \cite{RozPRB551997}. Then, \textcolor{black}{when} including only local correlations, the experimental behaviors for the Mott insulating states observed at quarter and three-quarter fillings in MA-TBG \cite{CaoN5562018_ins} \textcolor{black}{cannot be explained (a metallic state is promoted when increasing $T$)}. \textcolor{black}{In next section, we will argue that these insulating states could be theoretically explained by Mott physics if non-local correlations are taken into account explicitly.}

\subsection{Non-local correlations as a possible explanation}
\label{6.2.3}

\textcolor{black}{When non-local correlations are included in the description}, correlations between different adjacent sites are also taken into account \cite{MaiRMP772005,PotPRL932004,BiePRL942005,KitPRB792009,NomPRB892014,NomPRB912015,ParPRL1012008,SchPRB912015,LiePRB872013,LiQPRB922015,MomPRB581998,KimPRB962017,DeFPRB982018,ZhoPRB932016,CorPRX22012,JakPRB932016,CorPRL1072011}. Note that the interactions in a non-local correlations picture are still between electrons in the same lattice site. Inter-site magnetic correlations will promote a short-range ordering of the spins at different lattice sites. Inter-site orbital correlations will promote the location of electrons in specific orbitals from one lattice site to another. \textcolor{black}{These inter-site correlations can produce long-range order. However, even if there is not long-range order, the short-range correlations which are established (e.g. spin singlets if the correlations have antiferromagnetic character) can result in a Mott-like insulating behavior.}

Including the effect of non-local correlations could explain the experimental behaviors of the unexpected insulating states in MA-TBG. In contrast with local correlations physics, in which the details of the tight-binding model are not too important when studying the physical trends, for the non-local correlations effects, the specific tight-binding model will influence the behavior of the system. In MA-TBG, the tight-binding model is not yet fully established and understood \cite{YuaPRB982018,KosPRX82018,PoHPRX82018,KanPRX82018,PoH1808.02482}. Nevertheless, there are qualitative conclusions which are expected to be valid regardless of the model.

Here, I will briefly review the effect that including non-local correlations has on the basis of known results in other lattice models, mainly the 2D square lattice \cite{KitPRB792009,ParPRL1012008,SchPRB912015,LiePRB872013,LiQPRB922015}. These results were obtained by means of cellular dynamical mean-field theory (cellDMF), dynamical cluster approximation (DCA), variational cluster approximation (VCA) and dynamical vertex approximation (D$\Gamma$A). Due to the computational cost of these techniques, and specially for multiorbital systems, only a few results are available. Up to date, there is not available any study for the $N=2$ orbitals hexagonal lattice in the existing literature. This section would prepare the path for a future calculation in MA-TBG.

In \cref{Chap01} and for the single-orbital 2D square lattice at half-filling, I have explained that there are two critical interactions $U_{c \, 1}$ and $U_{c \, 2}$ (with $U_{c \, 1}<U_{c \, 2}$), where the first interaction $U_{c \, 1}$ is related with the formation of the Hubbard bands and the emergence of a gap between them, and the second $U_{c \, 2}$ is related with the dissapearance of the quasiparticle peak \cite{GeoRMP681996}, see \fref{fig:1.8}\textbf{(a)}. For a finite $T$, the Mott transition $U_c$ is defined in between this two critical interaction values, see \sref{1.3.3}. 

\begin{figure}[h]
   \centering
   \includegraphics[width=0.9\columnwidth]{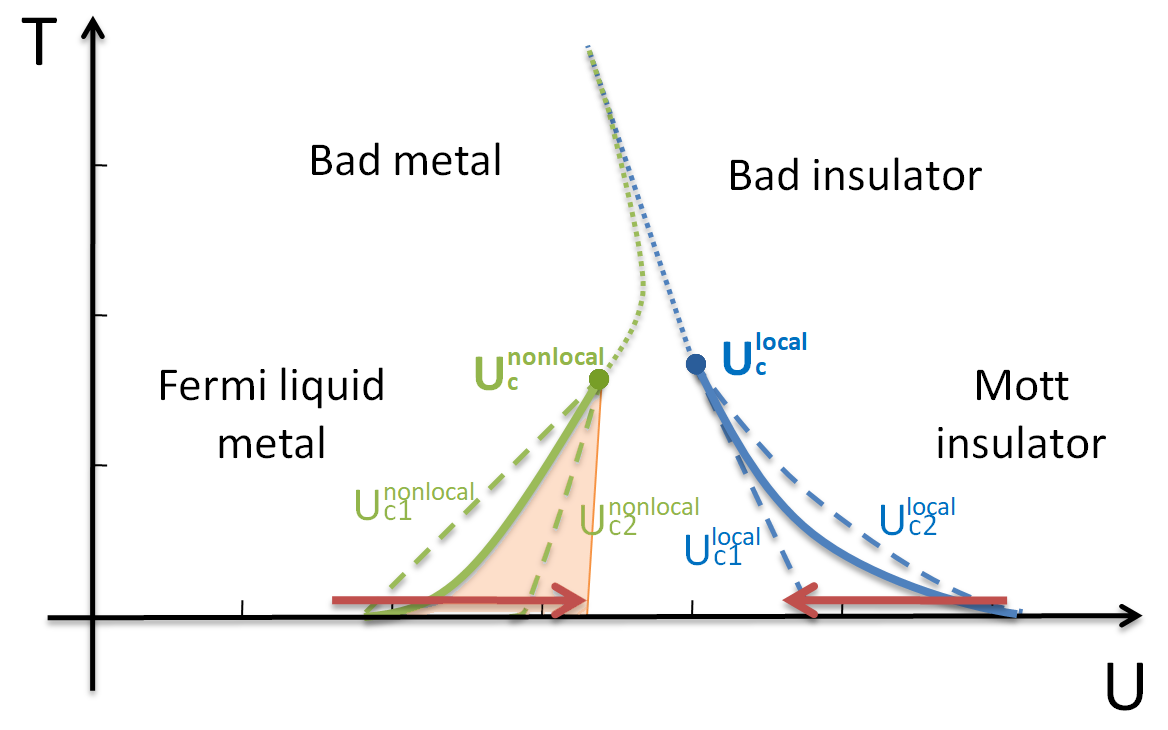}
   \caption[Sketch of the phase diagram $T$ vs $U$ for a single-orbital 2D square lattice at half-filling, including local and non-local correlations results]{Sketch of the phase diagram $T$ vs $U$ for a single-orbital 2D square lattice at half-filling, including local (blue lines) and non-local (green lines) correlations results. $U_c^{local}$ and $U_c^{nonlocal}$ are plotted as continuous lines, while $U_{c \, 1}$ (Hubbard bands start to form and gap opens) and $U_{c \, 2}$ (quasiparticle peak dissapears) for both cases as dashed lines. For $U<U_c$, a metallic behavior is found, while for $U>U_c$ the system is a Mott insulator. These lines end in a critical point marked by a filled circle. The dotted lines mark the crossover between a bad metal and bad insulating behavior at larger $T$. At large $T$ and $U$, local correlations give good results. The orange shaded region marks the area in which a small gap is opened (lower than the expectation for local correlations $\sim U-W$). For $J_H \leq 0$, Zeeman field $H$ will shift the Mott transition to lower (larger) $U$ values for local (non-local) correlations, see red arrows. Sketch made by using the results in \cite{GeoRMP681996,VucPRB882013,Fazekas1999,ParPRL1012008,SchPRB912015,Pizarro3}.}
   \label{fig:6.11}  
\end{figure}

In the presence of non-local correlations, both $U_{c \, 1}$ and $U_{c \, 2}$ are shifted towards smaller interaction values, maintaining the relation $U_{c \, 1}<U_{c \, 2}$ \cite{KitPRB792009,ParPRL1012008,SchPRB912015,LiePRB872013,LiQPRB922015}. At $T=0 \, K$, $U_c^{local}$ coincides with $U_{c \, 2}$, while for non-local correlations, $U_{c}^{nonlocal}$ coincides with $U_{c \, 1}$ \cite{ParPRL1012008,SchPRB912015}. In \fref{fig:6.11}, I show an unified sketch of the phase diagrams presented in \fref{fig:1.10} and obtained using the results in \cite{GeoRMP681996,VucPRB882013,Fazekas1999,ParPRL1012008,SchPRB912015}, as well as the results obtained in this chapter. 

Thus, non-local correlations will shift the Mott transition to a smaller $U_c^{nonlocal}<U_c^{local}$ for all $n_e$ fillings and $N$ orbitals \cite{KitPRB792009,ParPRL1012008,SchPRB912015,LiePRB872013,LiQPRB922015}. This is due to the interplay between local charge correlations $C_{n_T}$ and inter-site magnetic and orbital correlations, see \sref{1.3.2} for the case of inter-site magnetic correlations in the single-orbital 2D square lattice at half-filling. As seen in \fref{fig:6.11}, at large $T$ and $U$, local correlations results are recovered, and the system is not sensible to the non-local correlations effects. Close to $U_c^{nonlocal}$ (orange shaded region in \fref{fig:6.11}), the Zeeman field behavior, the temperature behavior \cite{ParPRL1012008,SchPRB912015} and the gap size \cite{ParPRL1012008} will be controlled by the inter-site correlations.

For multiorbital systems, depending on $J_H$ value, AFM or FM inter-site correlations are promoted. For $J_H>0$, inter-site FM and antiferro-orbital (i.e. electrons will sit in different orbitals with parallel spins from one site to another) correlations are promoted. For $J_H \leq 0$, inter-site AFM and ferro-orbital (i.e. electrons will sit in the same orbital with antiparallel spins from one site to another) correlations are promoted \cite{KitPRB792009,MomPRB581998}. A sufficiently large Zeeman magnetic field could suppress the AFM inter-site correlations and promote metallicity, as obtained in the experiment. I signaled the increasing Zeeman magnetic field $H$ with red arrows in \fref{fig:6.11}. At $J_H=0$, when $H$ increases, $U_c^{local}$ is pushed to lower values (insulating behavior is promoted), as shown in \sref{6.2.2}, and $U_c^{nonlocal}$ is pushed to larger values (metallicity is promoted) if $J_H \leq 0$.

This behavior will qualitatively hold for a tight-binding model without magnetic frustration. For frustrated lattices, the non-local critical interaction $U_c^{nonlocal}$ approaches the local one, hence the effects of the non-local correlations will be reduced. In the limit of an infinite dimensional lattice (Bethe lattice) completely frustrated, the local correlations limit is recovered. For the non-frustrated honeycomb model used in this work, we expect that the phenomenology explained here will qualitatively hold.

\textcolor{black}{Due} to the experimental behavior found in MA-TBG with respect to the Zeeman field and temperature, we expect that the system shows the phenomenology described above for $J_H \leq 0$, so that inter-site AFM correlations will be controlling the behavior of this system.

I have already commented the resulting $T$ behavior in a single-orbital 2D square lattice at half-filling \cite{GeoRMP681996,VucPRB882013,Fazekas1999,ParPRL1012008,SchPRB912015}, see \sref{1.3.3}. Inter-site AFM correlations are suppressed when increasing $T$ in square and hexagonal single-orbital lattices at half-filling \cite{ParPRL1012008,SchPRB912015}. Close to $U_c^{nonlocal}$ (orange shaded region in \fref{fig:6.11}), at low $T$ the system is insulating and at larger $T$ is metallic, opposite to what happens in local correlations. The suppression of the inter-site AFM correlations when increasing $T$ have been also obtained for $N=2$ orbitals 2D square lattice at quarter filling $x=1/4$ and $J_H=0$ \cite{KitPRB792009}. Then, \textcolor{black}{we expect that} the experimental $T$ behavior could be also explained by non-local correlations physics.

\begin{figure}[h]
   \centering
   \includegraphics[width=0.8\columnwidth]{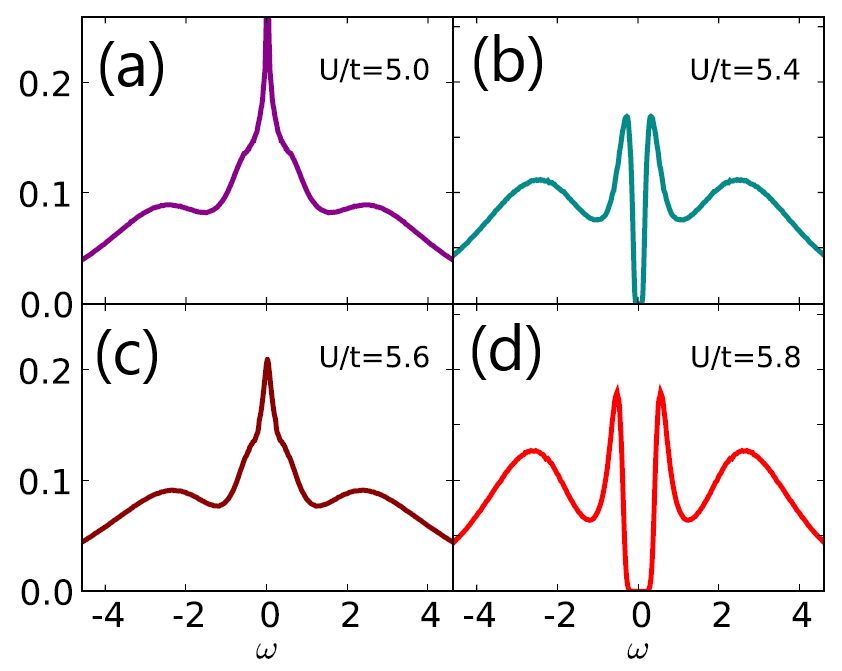}
   \caption[CellDMFT spectral function at $(\pi,0)$ for 2D square lattice at half-filling and various interactions around the non-local critical interaction for the Mott transition]{CellDMFT spectral functions at $(\pi,0)$ and $T = 0.01t$ (well below the end critical point in \fref{fig:6.11}) for the 2D square lattice at half-filling and various interactions around the non-local critical interaction for the Mott transition. \textbf{(a)} At the metallic state ($U<U_{c \, 1}^{nonlocal}$), the spectral function shows the typical quasiparticle peak at the Fermi level ($\omega =0$) with a sizable renormalization $Z \sim 0.4$. In the coexistance region ($U_{c \, 1}^{nonlocal}<U<U_{c \, 2}^{nonlocal}$), a \textbf{(b)} small gap starts to develop and \textbf{(c)} the quasiparticle peak shows a reduced weight $Z<0.4$. \textbf{(d)} At the insulating state ($U>U_{c \, 2}^{nonlocal}$), but for $U<U_{c \, 1}^{local}$, a small gap ($\sim 0.2 t$) is found. Taken and adapted from \cite{ParPRL1012008}.}
   \label{fig:6.12}  
\end{figure}

The small gap found in experiments might be \textcolor{black}{also reconciled with the one in a Mott insulator if non-local correlations are included}. For non-local correlations, a small gap opens in the quasiparticle peak close to $U_c^{nonlocal}$ \cite{ParPRL1012008}, which is signaled by an orange shaded area. For larger $U \gg U_c^{nonlocal}$, the gap size will converge to the local correlation expected one $\sim U-W \sim 15-25 \, meV$ \cite{DodPRB982018,Liu1804.10009,GeoRMP681996}. The evolution of the spectral function at $(\pi,0)$ obtained by cellDMFT can be seen in \fref{fig:6.12} for various points of the phase diagram of \fref{fig:6.11}, for the metallic state ($U<U_{c \, 1}^{nonlocal}$), for the coexistance region ($U_{c \, 1}^{nonlocal}<U<U_{c \, 2}^{nonlocal}$) and for the insulating state ($U>U_{c \, 2}^{nonlocal}$) for $U<U_{c \, 1}^{local}$. At the metallic state ($U<U_{c \, 1}^{nonlocal}$), the spectral function shows the typical quasiparticle peak at the Fermi level ($\omega =0$). In the coexistance region ($U_{c \, 1}^{nonlocal}<U<U_{c \, 2}^{nonlocal}$), a small gap starts to develop, and the quasiparticle peak shows a reduced weight. At the insulating state ($U>U_{c \, 2}^{nonlocal}$), but for $U<U_{c \, 1}^{local}$, a small gap ($\sim 0.2 t$) is found

Finally, another important effect of non-local correlations is that the ratio $U_c^{x=1/2}/U_c^{x=1/4}$ for multiorbital systems will be altered \cite{KitPRB792009,NomPRB892014}. In order to reproduce the experimental results, the ratio $U_c^{x=1/2}/U_c^{x=1/4}$ has to be larger than $1$. Specific calculations should address whether the inequality $U_c^{x=1/2,nonlocal}/U_c^{x=1/4,nonlocal}<1$ is obtained for $J_H \leq 0$ in MA-TBG in order to reproduce the experimental behavior \cite{CaoN5562018_ins}.

Note that these results are obtained for different tight-binding models that the one which would describe MA-TBG. However, it is interesting to note that non-local correlations could reconcile the experimental results with the theoretical expectations. In \fref{fig:6.11}, I point to the possible location of MA-TBG in the orange shaded area. In this region, a metallic behavior promoted by $T$ and the Zeeman magnetic field $H$, as well as the small gap size are obtained.

\section{Summary}
\label{6.3}

I conclude this chapter by making a summary of the obtained results:

\begin{itemize}

\item On April 2018, unexpected insulating states at commensurate fillings \textcolor{black}{of the moir\'e superlattice} \cite{CaoN5562018_ins}, and superconducting domes around the insulating state for the hole-doped MA-TBG \cite{CaoN5562018_sc} were found, with $T_c^{max} \approx 1.7 \, K$ at $\approx -1.5 \times 10^{12} {cm}^{-2}$. In this system, the band structure around the Fermi level is formed by four degenerated flat bands, with the total bandwidth $W \sim 10 \, meV$ \cite{LopPRL992007,BisPNAS1082011,YuaPRB982018,PoHPRX82018,KanPRX82018}. The flattening of the bands around the Fermi level comes from the hybridization between the twisted graphene layers. The insulating states of MA-TBG show an interesting behavior: the gap is $\sim 0.3 \, meV$, two orders of magnitude smaller than $W$, and they are suppressed by $T$ and by a Zeeman magnetic field $H$, at the same energy scales as the size of the gap.

\item The correct tight-binding model for MA-TBG is still under discussion \cite{YuaPRB982018,PoHPRX82018,KanPRX82018,KosPRX82018,EfiPRB982018,XuCPRL1212018,GuoPRB872018}. At the time when we made this work, there \textcolor{black}{was} a consensus about the fact that the correct lattice symmetry should show the Dirac points at $K$ \textcolor{black}{and $K'$}, and that two in-plane $p$-like orbitals should be centered at the $AB$ and $BA$ regions \cite{YuaPRB982018,PoHPRX82018,KanPRX82018,KosPRX82018}. These orbitals show a maximum amplitude centered at $AA$ regions, in agreement with the previously known fact that the electronic density around $K$ is mainly located in $AA$ regions \cite{FanPRB932016}. Then, we have adopted a simple hexagonal lattice tight-binding model with only intraorbital nearest neighbor hoppings $t \approx 2 \, meV$ (then, $W=6t\approx 12 \, meV$) in order to mimic the correct tight-binding model for MA-TBG. Our aim is studying local correlations effects in MA-TBG. In local correlations picture, the tight-binding model details are not too important, hence the trends that I have obtained will be robust and applicable to MA-TBG.

\item We argued that the external gate voltage responsible for the induced doping, also screens the long-range character of the Coulomb interaction. Hence, we concluded that \textcolor{black}{onsite} interactions physics are playing a major role in MA-TBG. Whether local or non-local correlations are responsible of these effects is something that we explored in this work. We clarified that local correlations cannot explain the experiments that are found in MA-TBG, but non-local correlations could in principle reconcile experimental and theoretical results.

\item In \sref{6.2.1}, I have reviewed the results obtained in $N=2$ orbital systems \cite{RozPRB551997,HanPRB581998,FloPRB702004,deMPRB832011,YuaPRB982018}. I solved the hexagonal lattice tight-binding model with SSMF $U(1)$ formalism. I extended the $U_c$ versus $J_H$ curves to include a negative $J_H<0$ region, which may appear due to the effect of phonons with $\omega \sim 200 \, meV$ \cite{NomJPCM282016}. I found that, in order to obtain a Mott insulator at $x=1/4$ and a metal at $x=1/2$ like in the experiment (i.e. $U_c^{x=1/4}=14.7t \sim 30 \, meV<U_c^{x=1/2}=19.2t \sim 40 \, meV$), $J_H$ should lie between $-0.01<J_H/U<0.01$. In the following, we assumed $J_H=0$ as a good approximation for the real $J_H$ value in MA-TBG. I found that $U_c^{x=1/2}/U_c^{x=1/4}\sim 1.28$ (the experimental behavior is reproduced if $U_c^{x=1/2}/U_c^{x=1/4} > 1$). Even if the system at $x=1/2$ is not a Mott insulator, the electronic correlations are sizable, giving $Z\approx 0.3$.

\item In \sref{6.2.2}, I studied the Zeeman effect for the \textcolor{black}{same} $N=2$ orbitals hexagonal lattice model. I implemented it in SSMF $U(1)$ \textcolor{black}{via} a spin-dependent onsite energy. I considered $J_H=0$ and $x=1/4$. By estimating $\Delta^{x=1/4}$, I found that \textcolor{black}{the Zeeman field $H$ does not promote metallicity}. In contrast, $U_c^{x=1/4}$ will be reduced with respect to $U_c^{x=1/4}(H=0)$, and Mott insulating behavior is promoted by the Zeeman field $H$. This behavior is opposite to the \textcolor{black}{one observed in the} experiments.

\item For strong Zeeman field ($H>0.15t$), the system will be fully spin-polarized (i.e. $n_{\downarrow}=0.5$ and $n_{\uparrow}=0$) at $U<U_{1/2}^{1orb}$, and \textcolor{black}{at larger interactions, it will} effectively behave as a single-orbital at half-filling system. $U_c^{x=1/4}$ saturates to $U_{1/2}^{1orb}$ at $0.15$. For weak Zeeman field ($0<H<0.15t$), the system jumps to the Mott insulating state at $U_c>U_{1/2}^{1orb}$, \textcolor{black}{value at which} the system fully polarizes.

\item In the local correlations picture, when $T$ increases, the Mott insulating state is promoted \cite{GeoRMP681996,VucPRB882013,Fazekas1999,RozPRB551997}, see \sref{1.3.3}, opposite to what occurs in the experiment. The expected gap for local correlations is $gap\sim U-W \sim 15-25 \, meV$ \cite{DodPRB982018,Liu1804.10009,GeoRMP681996}, much larger than the obtained experimentally. All of these behaviors \textcolor{black}{suggest} that local correlations cannot explain the phenomenology of MA-TBG. Then, in \sref{6.2.3}, we proposed to go beyond local correlations, by including the effects of non-local correlations in order to explain the phenomenology of MA-TBG.

\item I made a brief review on the present results for non-local correlations in single- and multiorbital systems \cite{MaiRMP772005,PotPRL932004,BiePRL942005,KitPRB792009,NomPRB892014,NomPRB912015,ParPRL1012008,SchPRB912015,LiePRB872013,LiQPRB922015,MomPRB581998,KimPRB962017,DeFPRB982018,ZhoPRB932016,CorPRX22012,JakPRB932016,CorPRL1072011}. \textcolor{black}{When non-local correlations are included, all} the critical interactions $U_c$, are pushed to lower values \cite{KitPRB792009,ParPRL1012008,SchPRB912015,LiePRB872013,LiQPRB922015}, due to the interplay between local charge correlations $C_{n_T}$ and inter-site magnetic and orbital correlations. At large $T$ and $U$, local correlations will control the physics of the system \cite{GeoRMP681996,VucPRB882013,Fazekas1999,ParPRL1012008,SchPRB912015}. Close to $U_c^{nonlocal}$ and for $J_H\leq 0$, AFM and ferro-orbital inter-site correlations are promoted \cite{KitPRB792009,MomPRB581998}. In this case, \textcolor{black}{a Zeeman field will suppress} AFM inter-site correlations. Also, $T$ \cite{ParPRL1012008,SchPRB912015} and the small gap \cite{ParPRL1012008} results can be reconcile with the experiment in this region. On the other hand, the needed relation $U_c^{x=1/2}/U_c^{x=1/4} > 1$ is not known \cite{KitPRB792009,NomPRB892014}. But non-local correlations results are sensible to the correct tight-binding model, and hence for MA-TBG this factor may change.

\end{itemize}

Now, I would like to mention the current status of the discussion about the tight-binding model for MA-TBG. In \sref{1.6.1} and \sref{6.1.1}, I have mentioned that there is a $\xi$ valley degeneracy, due to the fact that the Dirac points are uncoupled, which results in four flat bands per spin. Around $K$, the electronic density is mainly located at the $AA$ regions. \textcolor{black}{In} order to fit the flat bands, MA-TBG \textcolor{black}{was} modeled as a $N=2$ orbitals system. Due to symmetry arguments \cite{YuaPRB982018,PoHPRX82018,KanPRX82018}, the two orbitals have an in-plane $p$-like character and they are centered at the $AB$ and $BA$ regions, forming a hexagonal superlattice. The maximum amplitude of these orbitals is centered at $AA$ regions \textcolor{black}{at $K$}.


\textcolor{black}{However, there is still a debate about the applicability of this model. It was discussed that this model suffers a} topological obstruction \cite{PoHPRX82018} (related with the valley symmetry at $K_\xi^l$ Dirac points) when trying to construct a minimal model only for the flat bands. Some authors \cite{KanPRX82018} have argued that such topological obstruction can be overcomed by considering that both valleys are not perfectly uncoupled, and hence there is not a perfect valley \textcolor{black}{degeneracy}. 

These two problems point to the necessity of including more than only the four flat bands to model MA-TBG. Ten and six bands models (valley degeneracy considered) were proposed in \cite{PoH1808.02482}. The ten bands model fits the two flat bands and the four bands above and below. There are different orbitals centered at different regions of the superlattice. In this model, there are three $p$-like orbitals centered at the $AA$ regions forming a triangular superlattice, four $p$-like orbitals centered at the $AB$ and $BA$ regions forming a hexagonal superlattice and one $s$-like orbital centered at the middle point between $AB$ and $BA$ regions, forming a Kagome superlattice. In total, there are 10 orbitals (3 $p$ $\times$ 1 site+ 4 $p$ $\times$ 2 sites + 1 $s$ $\times$ 3 sites) which will reproduce the 10 bands. The \textcolor{black}{simplified} six bands model fits the two flat bands and the four bands below \textcolor{black}{the,}, where only the 3 $p$-like orbitals in the triangular superlattice and the $s$-like orbital in the Kagome superlattice are retained. These models looks promising, \textcolor{black}{but yet there is not a consensus in the community about them.}


\chapter{{\bf Wannier tight-binding model for the single-layer $Fe_3GeTe_2$}} 
\label{Chap4d}
\lhead{Chapter \ref{Chap4d}. \emph{Model and study of local correlations in $Fe_3GeTe_2$}} 

\begin{small}

Very recently, few layers of an iron-based material, $Fe_3GeTe_2$ (FGT), were obtained (by exfoliating the bulk FGT), reaching the monolayer limit. \textcolor{black}{FGT is the first metallic ferromagnet which has been isolated in the monolayer form \cite{DenN5632018,FeiNM172018}, and it is free of rare-earth elements}. In this material, the ferromagnetic temperature in the bulk is $T_{F}\sim 220 \, K$. When exfoliating the bulk system, $T_F$ decreases until $T_F \sim 20 \, K$ in monolayer FGT. This critical temperature $T_F$ can be tuned with doping via an external voltage, reaching room temperature ferromagnetism for the trilayer FGT at an electron carrier density doping of $\sim 10^{14} \, cm^{-2}$. 

Bulk FGT was already experimentally and theoretically studied \cite{DeiEJIC20062006,CheJPSJ822013,ZhuPRB932016,Jie2DM42017,ZhaSA42018}. On the other hand, few layer FGT is theoretically less known.

ARPES and low-$T$ specific heat measurements in bulk FGT \cite{ZhaSA42018} reported the presence of sizable electronic correlations. Here, we want to study the possible similarities between bulk and monolayer FGT, and other iron chalcogenides, with an special focus on the possible role of the Hund's coupling in FGT. \textcolor{black}{In order to do this, the first step is to calculate the band structure and Wannier tight-binding model of FGT.}

In this chapter, I will present the calculations that I have performed to calculate the DFT band structure and the Wannier tight-binding model for FGT. I will study both the bulk and the monolayer FGT. Lastly, I will give some final comments. This chapter could serve as a first step to the study the ferromagnetism and other electronic properties in FGT.

\end{small}

\newpage

\section{Introduction \& Motivation}
\label{7.1}

\begin{figure}[h]
   \centering
   \includegraphics[width=0.9\columnwidth]{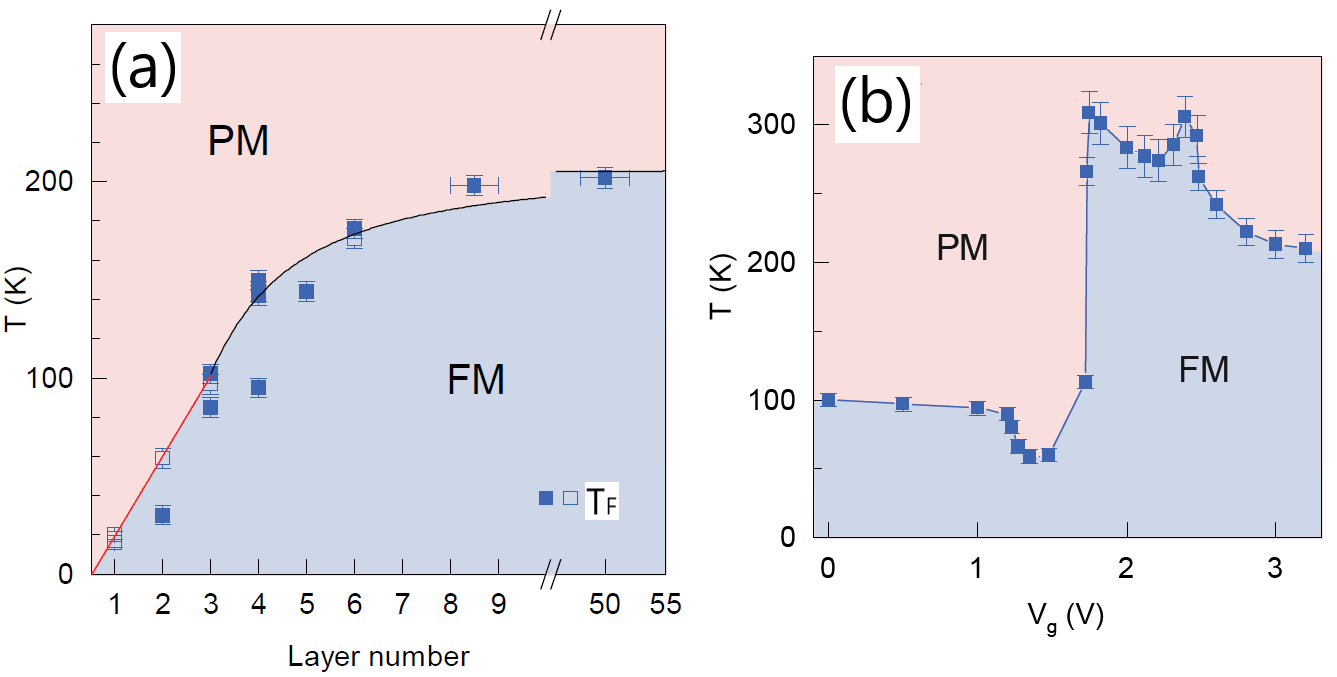}
   \caption[Temperature versus layer number phase diagram for FGT and versus ionic external voltage for trilayer FGT]{\textbf{(a)} Temperature $T$ versus layer number phase diagram, where $T_F$ evolves from $\sim 220 \, K$ in the bulk FGT (identified as $50-55$ layers) down to $\sim 20 \, K$ in monolayer FGT. \textbf{(b)} $T$ versus ionic external voltage $V_g$ (which acts as an effective electron doping) phase diagram for trilayer FGT. Due to the itinerant character of FGT, it is expected that this phase diagram follows the evolution of the total DOS of FGT. Taken and adapted from \cite{DenN5632018}.}
   \label{fig:7.1}  
\end{figure}

FGT is a van der Waals crystal, so it can be easily exfoliated to obtain atomically thin layers. \textcolor{black}{It was recently discovered that monolayer FGT is the first monolayer metallic ferromagnetic system \cite{DenN5632018,FeiNM172018}, and it is free of rare-earth elements.} When exfoliating FGT, the critical temperature of the itinerant ferromagnetic phase $T_F$ evolves from $T_F \sim 220 \, K$ in the bulk to $T_F \sim 20 \, K$ in the monolayer, as shown in \fref{fig:7.1}\textbf{(a)}. A large intrinsic magnetocrystalline anisotropy in the monolayer FGT lifts out the restriction imposed by the Mermin-Wagner theorem\footnote{This theorem \cite{MerPRL171966} states that in an isotropic system with dimensions lower than three, thermal fluctuations suppress any possible long-range order.}, hence the 2D long-range ferromagnetic order can be formed. When applying an external ionic voltage which effectively dopes with electrons the system, $T_F$ changes \textcolor{black}{non-monotonously}, and it can reach room temperature ($T_F \sim 300 \, K$) in the case of trilayer FGT for an electron carrier density $\sim 10^{14} \, cm^{-2}$ \cite{DenN5632018}, see \fref{fig:7.1}\textbf{(b)}.

\begin{figure}[h]
   \centering
   \includegraphics[width=0.85\columnwidth]{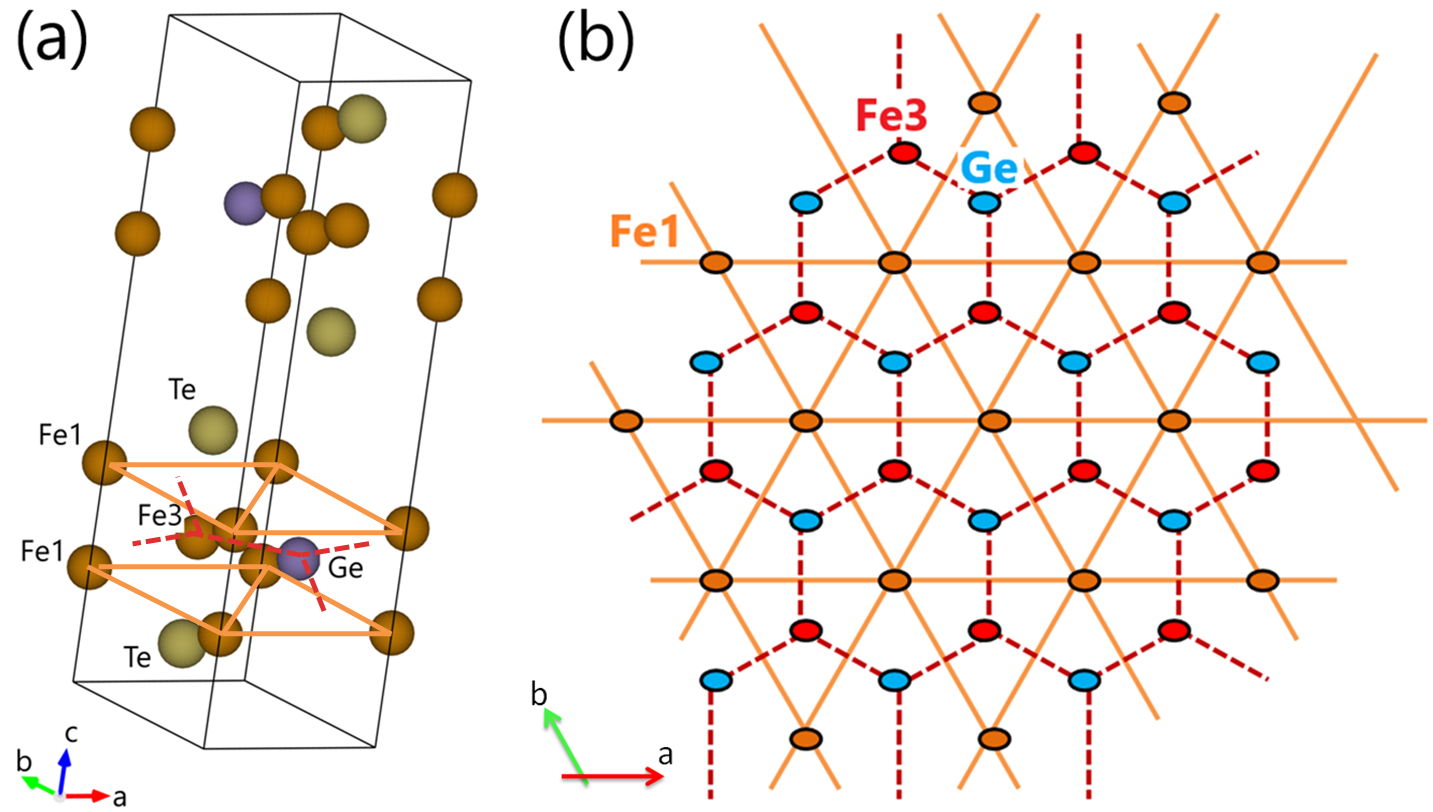}
   \caption[3D and top view of the crystal structure of bulk FGT]{\textbf{(a)} Hexagonal crystal structure (space group $P63/mmc$) of bulk FGT where the different atomic species are recognised, and $Fe1$ and $Fe3+Ge$ planes are marked. $c$ axis directed along adjancent layers of FGT. \textbf{(b)} Perpendicular view of FGT, along the $ab$ plane. $Fe1$ atoms form a triangular lattice which sandwiches the hexagonal lattice formed by $Fe3$ and $Ge$.}
   \label{fig:7.2}  
\end{figure}

$Fe_3GeTe_2$ has a hexagonal symmetry (space group $P63/mmc$), see \fref{fig:7.2}. The crystal structure parameter along the $c$ axis is quite large ($c=16.333 \AA$) when compared with the in-plane parameters ($a=b=3.991 \AA$), hence indicating a quasi-2D crystal structure. In the unit cell, $2$ $Fe$ atoms are equivalent (labeled by $Fe1$), while the third one (labeled by $Fe3$) is not. $Fe1$ atoms form a triangular lattice which sandwiches the $Fe3$ and $Ge$ atoms forming an hexagonal lattice, see \fref{fig:7.2}\textbf{(b)}. $2$ $Te$ spacers atoms are located between adjacent $Fe1-Fe3-Ge$ layers. A single stack of $2Fe1$, $Fe3-Ge$ and $2Te$ atoms forms the monolayer.

Due to the itinerant character of FGT, a weakly correlated Stoner criteria \cite{Fazekas1999} \textcolor{black}{could be possible applied. In the Stoner criteria for ferromagnetism,} the peaks in the total density of states (DOS) give the tendency towards ferromagnetism. Hence, the total DOS for FGT should give the same evolution as the one shown in \fref{fig:7.1}\textbf{(b)} when doping with electrons. On the other hand, ARPES and low-$T$ specific heat measurements \cite{ZhaSA42018} signaled the presence of strong correlations, by measuring large mass enhancement factors.

So far, only calculations for bulk FGT have been performed \cite{ZhuPRB932016,Jie2DM42017,ZhaSA42018}. These calculations show that the band structure around the Fermi level consists on several flat bands of mainly $Fe$ $d$ orbitals character. It urges to clarify the nature of this itinerant ferromagnetic phase, as well as its relation with the number of layers reduction, by studying the band structure of low dimensional FGT in the non-magnetic state. It is also interesting to study the possible similarities with other $Fe$-based materials, such as $FeSe$ \cite{BohJPCM302017} or $FeGe$ \cite{GutPRL1182017}, and clarify the role of the Hund's coupling in FGT.

In this chapter, I calculate the band structure, DOS and Fermi surfaces for bulk and monolayer FGT, as well as the Wannier parametrization for monolayer FGT. I will also compare with experimental expectations. This chapter could serve as an initial step to study the origin of ferromagnetism and other electronic properties in FGT and related systems.

\section{Results \& Discussion}
\label{7.2}

In this section, I will present the calculations that I have performed on bulk and monolayer FGT. I will calculate the band structure, DOS and Fermi surfaces for both situations, and the Wannier tight-binding model for monolayer FGT by using the DFT \textsc{Wien2k} and \textsc{Wannier90} codes, as explained in \aref{AppA}. In \sref{7.2.1}, I obtain the band structure for bulk and monolayer FGT, and the DOS and Fermi surface for monolayer FGT. In \sref{7.2.2}, I calculate \textcolor{black}{a} Wannier tight-binding model for the monolayer FGT.

\subsection{Band structure, DOS and Fermi surfaces of $Fe_3GeTe_2$}
\label{7.2.1}

In this section, I will show the DFT \textsc{Wien2k} calculations for bulk and monolayer $Fe_3GeTe_2$. The calculations were run with a $k$-mesh of $28 \times 28 \times 2$ for bulk FGT and $23 \times 23 \times 1$ for monolayer FGT, where this notation refers to the grid in the $k_x \times k_y \times k_z$ directions of the Brillouin zone (BZ).

I calculate the total number of electrons per atom for the electronic shells in bulk and monolayer FGT for those shells which will mainly contribute to the band structure around the Fermi level, i.e. the valence electronic shells. In both cases, $Fe1$ and $Fe3$ atoms have a total number of electrons per atom in the $d$ electronic shells of $n_e^{Fe1} \approx 6.07$ and $n_e^{Fe3} \approx 6.12$. Thus, FGT shows the same number of electrons per $Fe$ atom as in the case of iron-based superconductors (FeSCs), see \sref{1.5.2}, which could be important when calculating the effects of the electronic correlations. In the remaining atoms, the $p$ shells have a total number of electrons per atom of $n_e^{Ge} \approx 0.95$ and $n_e^{Te} \approx 1.99$. 

\begin{figure}[h]
   \centering
   \includegraphics[width=0.9\columnwidth]{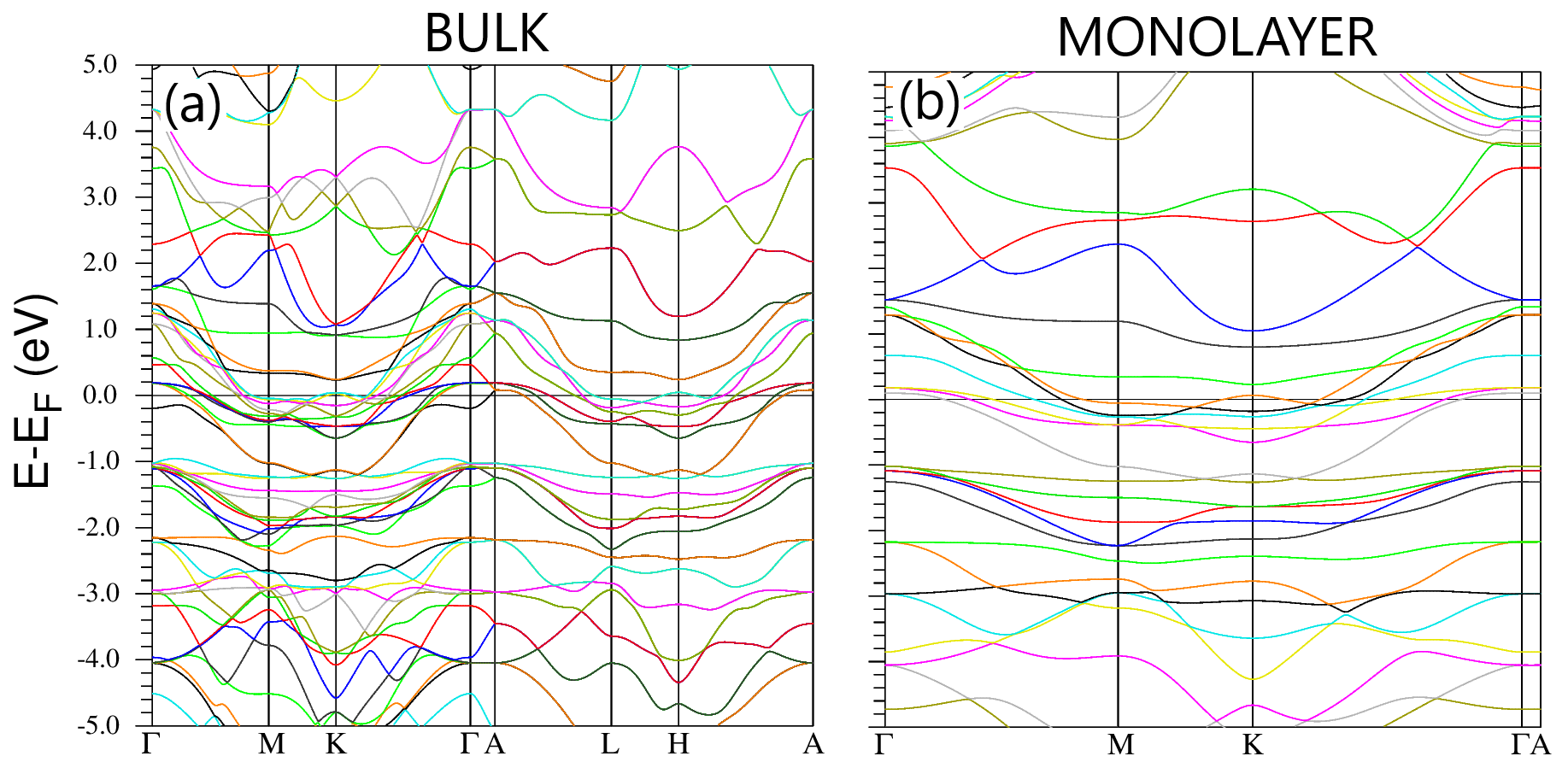}
   \caption[DFT band structure for bulk and monolayer FGT]{DFT band structure of \textbf{(a)} bulk and \textbf{(b)} monolayer FGT. Flat bands with a predominant $Fe1$ and $Fe3$ $d$ orbitals character dominate the band structure around the Fermi level. In the case of bulk FGT, a doubling in the number of bands appears due to the folded BZ considered in the calculations. The principal directions $\Gamma$, $M$ and $K$ for $k_z=0$ (while $A$, $L$ and $H$ for $k_z=\pi$) are defined in the hexagonal BZ.}
   \label{fig:7.3}  
\end{figure}

In \fref{fig:7.3}, I show the band structure for both bulk and monolayer FGT. 

\begin{figure}[h]
   \centering
   \includegraphics[width=0.4\columnwidth]{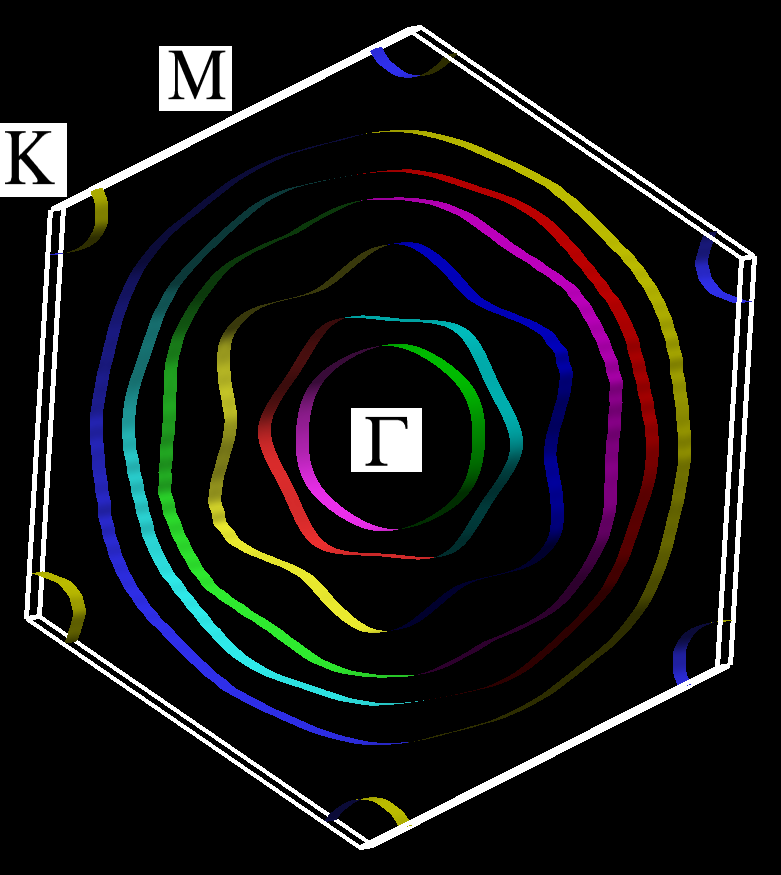}
   \caption[Fermi surfaces of monolayer FGT in the hexagonal BZ]{Fermi surfaces of monolayer FGT in the hexagonal BZ. 6 hole pockets surround the $\Gamma$ point, with different shapes, while small electron pockets are located in the $K$ points.}
   \label{fig:7.4}  
\end{figure}

As already mentioned, a set of weakly dispersive bands dominate the band structure around the Fermi level. The number of bands in bulk FGT for $k_z=0$ is doubled with respect to monolayer FGT, due to the \textcolor{black}{presence of two monolayers in the unit cell, see \fref{fig:7.2}\textbf{(a)}}. \textcolor{black}{Except for the doubling of bands, the band structure is very similar} in both cases, with some differences, apart from the folded/unfolded character of the BZ for bulk/monolayer FGT. This result remarks the quasi-2D character that bulk FGT already has. $6$ hole pockets can be seen surrounding $\Gamma$, and small electron pockets around $K$ points. I plotted the Fermi surfaces for monolayer FGT (\textcolor{black}{in bulk FGT, twice the Fermi surfaces appear}) in \fref{fig:7.4}, where the mentioned hole and electron pockets can be seen.

\begin{figure}[H]
   \centering
   \includegraphics[width=0.9\columnwidth]{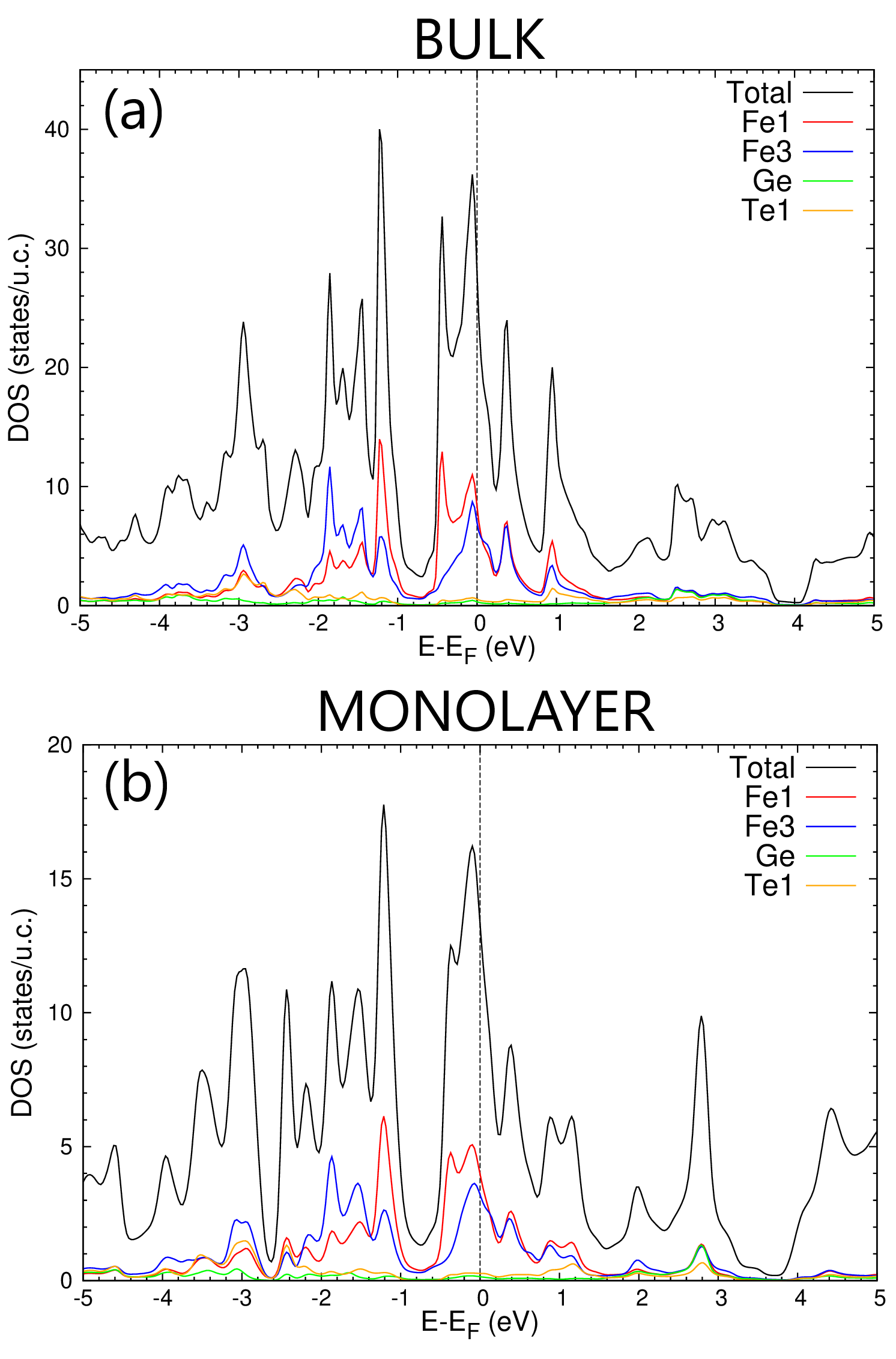}
   \caption[Total and contributions from each atom DOS for bulk and monolayer FGT]{Total (black lines) and contributions from each atom (color lines) DOS for \textbf{(a)} bulk and \textbf{(b)} monolayer FGT. \textcolor{black}{Note that I have not included the contribution from $Fe2$ and $Te2$ due to their degeneracy with $Fe1$ and $Te1$, respectively}. $Fe1$ and $Fe3$ $d$ orbitals contribute to the band structure around the Fermi level between $-2$ and $2 \, eV$. $Fe1$ $d$ orbitals mostly contribute to the flat bands around the Fermi level, while the major contribution from $Fe3$ $d$ orbitals appear around $-2$ and $0 \, eV$. $Ge$ $p$ orbitals mostly contribute around $2.5-3 \, eV$. $Te$ $p$ orbitals mostly contribute around $-2.5$ and $-3 \, eV$.}
   \label{fig:7.5}  
\end{figure}

The band structure extends between $-6$ and $4 \, eV$. The bands around the Fermi level between $-4$ and $3 \, eV$ have mostly $Fe1$ and $Fe3$ $d$ orbitals character, as shown in the DOS plot of \fref{fig:7.5}, for both bulk and monolayer FGT. The bulk FGT DOS compares very well with previous calculations \cite{ZhuPRB932016}. The most important feature in both bulk and monolayer FGT is the very large DOS at and close to the Fermi level. The bandwidth for $Fe1$ and $Fe3$ $d$ orbitals is $W \sim 3-4 \, eV$. $Fe1$ $d$ orbitals mostly contribute to the flat bands around the Fermi level, while the major contribution from $Fe3$ $d$ appears around $-2$ and $0 \, eV$. An important contribution from $Ge$ $p$ orbitals appear around $2.5-3 \, eV$, while $Te$ $p$ orbitals contribute around $-2.5 \, eV$ and $-3 \, eV$. In the case of monolayer FGT, two depletions can be seen around $-0.7 \, eV$ and $-1.6 \, eV$, whether in bulk FGT the first depletion appears, while the second one does not. In monolayer FGT, there is an splitting of the DOS around $1 \, eV$.

\begin{figure}[H]
   \centering
  \includegraphics[width=0.9\columnwidth]{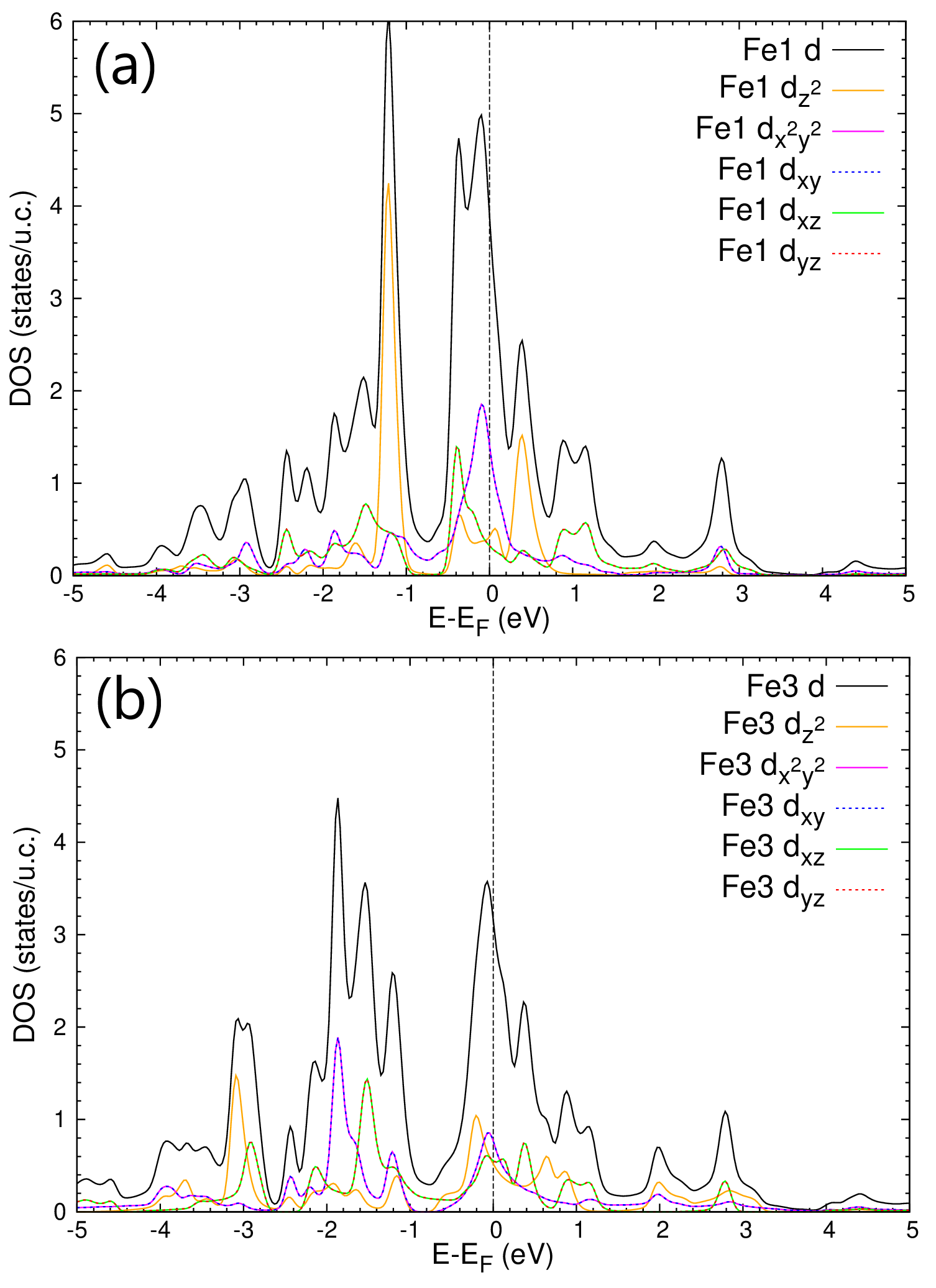}
   \caption[Orbital DOS for monolayer FGT for $Fe1$ and $Fe3$ $d$ orbitals]{Orbital DOS for monolayer FGT for \textbf{(a)} $Fe1$ and \textbf{(b)} $Fe3$ $d$ orbitals around the Fermi level. All the $d$ orbitals contribute to the band structure around the Fermi level. \textcolor{black}{$d_{xy}$ (blue dashed line) and $d_{x^2-y^2}$ (magenta line) are degenerated, as well as $d_{xz}$ (green line) and $d_{yz}$ (red dashed line)}.}
   \label{fig:7.6}  
\end{figure}

For the chosen \textcolor{black}{non-orthogonal} orbital axis $a$ and $b$, \textcolor{black}{see \fref{fig:7.2}}, $d_{xz}$ and $d_{yz}$ are equivalent, as well as a $d_{x^2-y^2}$ and $d_{xy}$\footnote{Note that the orbitals in FGT are defined differently than in FeSCs. Here, the lobes of the $d$ orbitals form $120 \, ^{\circ}$, while in FeSCs these lobes are orthogonal.}. This holds for both $Fe1$ and $Fe3$ atoms. In \fref{fig:7.6}, I plot the orbital resolved DOS for $Fe1$ and $Fe3$ $d$ orbitals in the monolayer FGT. In the next section, I will calculate that, in order to obtain a Wannier tight-binding model for monolayer FGT, it is also necessary to include the $p$ orbitals coming from $Ge$ and $Te$, due to their sizable hybridization with $Fe$ $d$ orbitals, which results in a non-negligible contribution to the bands around $\pm 2.5 \, eV$.

\newpage

\subsection{Wannier tight-binding model for monolayer $Fe_3GeTe_2$}
\label{7.2.2}

In order to fit the previous band structure, a tight-binding model is proposed here considering the $Fe1$ $d$, $Fe3$ $d$, $Ge$ $p$ and $Te$ $p$ orbitals, as well as a minimal one using only $Fe1$ and $Fe3$ $d$ orbitals. The Wannier fitting was done using the projector method explained in \aref{AppA}, as implemented in the \textsc{Wannier90} code. The energy window to obtain the $ddpp$ is selected between $-6$ and $4 \, eV$, with a total of $24$ bands being fitted. For the $dd$ model, the energy window is between $-3$ and $1.5 \, eV$.

\begin{figure}[h]
   \centering
   \includegraphics[width=0.9\columnwidth]{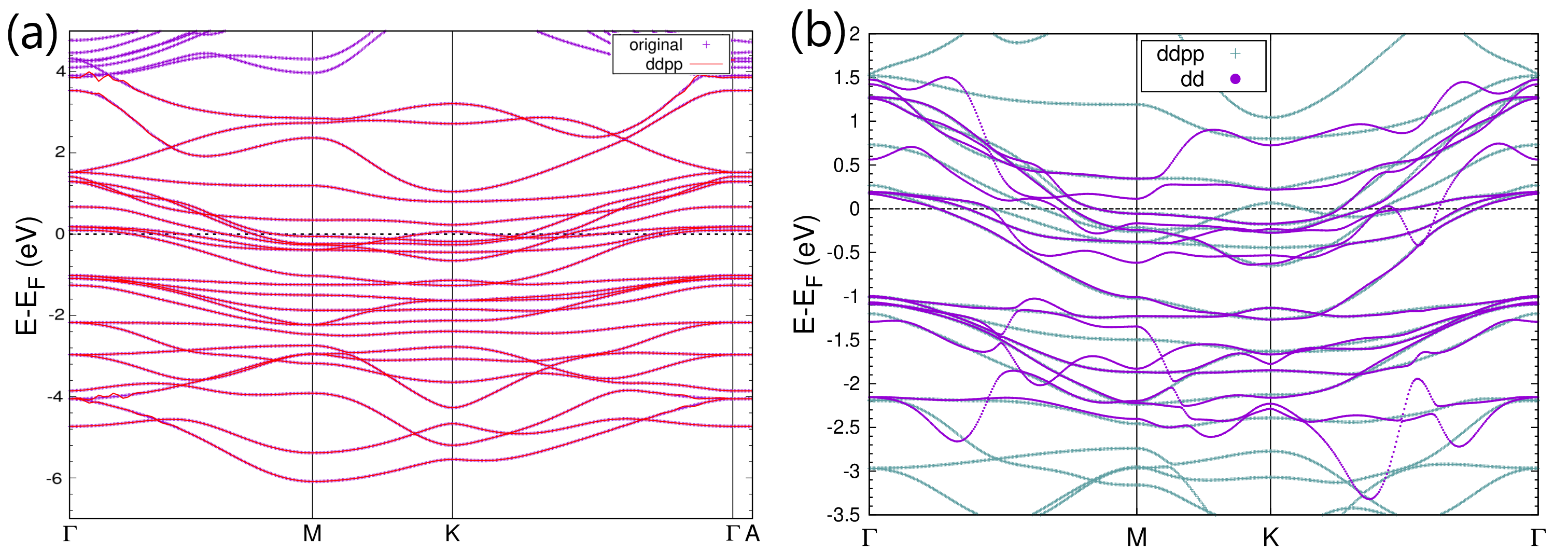}
   \caption[Wannier $dd$ and $ddpp$ tight-binding models, and comparison of $ddpp$ model with DFT band structure for monolayer FGT]{\textbf{(a)} DFT band structure (purple crosses) and $ddpp$ model (red line) for monolayer FGT. The agreement between both cases is very good. \textbf{(b)} Wannier $ddpp$ (green crosses) and $dd$ (purple points) tight-binding model comparison. The $dd$ model shows large deviations from the original band structure (here represented by the $ddpp$ model), specially for the bands close to $-2$ and $1.5 \, eV$. Calculations done using the projector method as \textcolor{black}{explained} in the \textsc{Wannier90} code, see \aref{AppA}.}
   \label{fig:7.7}  
\end{figure}

Due to the entanglement present between the bands with mainly $Fe$ $d$ orbitals character and with $Ge$ and $Te$ $p$ orbitals character, the tight-binding will include all of them. I will call this tight-binding the $ddpp$ model. In \fref{fig:7.7}\textbf{(a)}, I show the band structure (red line) obtained by the Wannier tight-binding $ddpp$ model as compared with the DFT band structure (purple crosses). The Wannier fitting shows a very good agreement with the DFT band structure. I tried to reduce the number of orbitals needed for the tight-binding model. In the case of a Wannier parametrization using only $Fe1$ and $Fe3$ $d$ orbitals (called $dd$ model), the fitting shows very large deviations from the original band structure, see \fref{fig:7.7}\textbf{(b)} for the comparison between both $dd$ and $ddpp$ model.

Due to the similar values of the bandwidth of the $Fe$ $d$ orbitals $\sim 3-4 \, eV$ and the filling per $Fe$ atom ($n_e^{Fe} \approx 6$) for FGT with respect to other FeSCs, we expect that Hund metal phenomenology might play a role in this system. It seems natural to think that Hund's coupling and atomic spin polarization could play a major role in the ferromagnetic phase of FGT. However, for FGT, the crystal symmetry and band structure differs from other quasi-2D FeSCs, see previous chapters. In FGT, the hybridization between $Fe$ $d$ and $p$ orbitals is larger than in most of the FeSCs. \textcolor{black}{The study of electronic correlations will be very complicated when using the $ddpp$ model. A more appealing model will only include $Ge$ $p$ orbitals (due to the fact that the minimal $dd$ model do not correctly reproduce the band structure), in what I will call the $ddp$ model. Yet further work is needed along this path}.

\section{Summary}
\label{7.3}

I conclude this chapter by making a summary of the obtained results:

\begin{itemize}

\item Very recently, the rare-earth free compound $Fe_3GeTe_2$ was \textcolor{black}{exfoliated and became the first monolayer itinerant ferromagnet} \cite{DenN5632018,FeiNM172018}. When the \textcolor{black}{number of layers} is reduced, the critical temperature $T_F$ evolves from $\sim 220 \, K$ in the bulk FGT to $\sim 20 \, K$ in monolayer FGT. $T_F$ can be further tuned by effectively doping with electrons via an external voltage \cite{DenN5632018,FeiNM172018}. In trilayer FGT, $T_F$ rises up to room temperature ($\sim 300 \, K$). The study of the band structure for \textcolor{black}{FGT is an important first step} to be done. I calculated here the band structure, DOS, Fermi surfaces and a minimal Wannier tight-binding model for the monolayer FGT by using DFT \textsc{Wien2k} and \textsc{Wannier90} codes.

\item In \sref{7.2.1}, I presented the DFT calculations on the band structure, DOS and Fermi surfaces for bulk and monolayer FGT. The number of electrons per $Fe$ atom is the same as in the FeSCs ($n_e \approx 6$) studied in previous chapters. A set of flat bands dominate the band structure around the Fermi level with mainly $Fe1$ and $Fe3$ $d$ orbitals character. The band structure of bulk and monolayer FGT is quite similar, signaling the quasi-2D character of bulk FGT. The bandwidth for $Fe1$ and $Fe3$ $d$ orbitals is $W \sim 3-4 \, eV$. $Fe1$ $d$ orbitals mostly contribute to the flat bands around the Fermi level, while the major contribution from $Fe3$ $d$ appears around $-2$ and $0 \, eV$. An important contribution from $Ge$ $p$ orbitals appear around $2.5-3 \, eV$, while $Te$ $p$ orbitals contribute around $-2.5 \, eV$ and $-3 \, eV$. An interesting difference between bulk and monolayer FGT is a splitting that emerges around $1 \, eV$ in the DOS.

\item In \sref{7.2.2}, I obtained the Wannier tight-binding model for monolayer FGT. I derived the $dd$ (only $Fe1$ and $Fe3$ $d$ orbitals included) and the $ddpp$ model for monolayer FGT. Even though $Fe1$ and $Fe3$ $d$ orbitals dominate the bands around the Fermi level, the entanglement with $Ge$ and $Te$ $p$ orbitals is very important, hence the $dd$ model shows large discrepancies with respect to the original DFT band structure. The $ddpp$ model is fitted for an energy window between $-6$ and $4 \, eV$, and it shows a good agreement with the DFT band structure.  

\end{itemize}

This brief chapter was a summary of my recent work on FGT. Some calculations and analysis have to be performed yet, meanwhile I have shown here the actual status of this work.



\chapter{{\bf Conclusions}} 
\label{Chap8}
\lhead{Chapter \ref{Chap8}. \emph{Conclusions}} 

During this thesis, I have studied the effects of electronic correlations in various systems. I have focused on studying the local correlations effects in multiorbital systems, and the phenomenology that appears in such multiorbital systems in contrast with single-orbital systems when including the orbital degrees of freedom. I have studied these effects in real materials, specially in iron-based superconductors (FeSCs) and related materials, such as $LaCrAsO$, see \cref{Chap4a}, and $BaFe_2S_3$, see \cref{Chap4b}, and in 2D materials, specifically on magic-angle twisted bilayer graphene (MA-TBG), see \cref{Chap4c}. I have also obtained the DFT band structure and Wannier tight-binding model for the 2D material $Fe_3GeTe_2$ (FGT), see \cref{Chap4d}. \textcolor{black}{Understanding the band structure serves as a first step to study the electronic correlations.}


After reviewing some generic concepts of correlations in \cref{Chap01}, in \cref{Chap03} I have introduced the phenomenology of local correlations in multiorbital systems. In multiorbital systems, the ratio between the interaction energy cost $\Delta^x$ and the kinetic energy gain $\widetilde{W}$ depends on the number of orbitals per atom $N$, the number of electrons per atom $n_e$ and the Hund's coupling $J_H$. The Mott insulator is not restricted to half-filled systems ($n_e=N$), but \textcolor{black}{it can also happen at} other integer fillings ($n_e=1,2,\ldots, N-1$). When $J_H$ increases, the critical interaction for the Mott transition $U_c$ decreases (increases) for half-filling $n_e=N$ (single-electron $n_e=1$) systems. For $N>2$, $1<n_e<N$ and \textcolor{black}{intermediate to large $J_H$}, a correlated metallic state emerges, called the Hund metal. In the Hund metal regime, the strength of local correlations is large, the atoms are highly and locally spin polarized, and there is orbital decoupling. The Hund metal phenomenology is driven by the enhancement of local spin correlations, \textcolor{black}{which occurs} due to a decrease of anti-parallel spin configurations in each atom. \textcolor{black}{Then, there is a} atomic spin polarization \textcolor{black}{promoted} by $J_H$. The suppression of the anti-parallel spin configurations is related with the Mott insulator at half-filling, hence a link exists between the Mott insulator physics and the Hund metal physics, \textcolor{black}{besides the difference of being in an insulating or metallic state}. In the case of non-equivalent orbitals systems, $J_H$ promotes orbital \textcolor{black}{differentiation}, hence an orbital selective Mott transition (OSMT) might occur. Various theoretical and experimental results for FeSCs (low-$T$ specific heat, angle-resolved photoemission spectroscopy (ARPES) renormalized band structure, X-ray emission spectroscopy (XES) for the local moment, etc.) sustain the idea that these systems can be seen as being close to or in the Hund metal regime.

\newpage
The main results of the thesis can be summarized as follows:

\begin{itemize}

\item In \cref{Chap4a} \cite{Pizarro1}, we have proposed to search for a new family of high-$T_c$ superconductors similar to the FeSCs, but based on chromium instead of iron. Our argument is based on the idea of having an optimal degree of electronic correlations, as it occurrs in most of the unconventional superconductors. We \textcolor{black}{proposed} that $Cr$ pnictides and chalcogenides \textcolor{black}{could} host high-$T_c$ superconductivity, similar to FeSCs. The main difference between $Cr$-based systems and FeSCs is the total number of electrons per atom, which is $n_e=4$ for $Cr$-based materials and $n_e=6$ for FeSCs. The strength of local correlations increases when approaching half-filling ($n_e=5$) from $n_e=6$ for FeSCs, then it has been argued that FeSCs can be seen as electron-doped Mott insulators, hence linking both high-$T_c$ cuprates and FeSCs physics. I have shown that the similar trend is obtained when moving from the hole-doped Mott part $n_e=4$ towards half-filling. We have chosen $LaCrAsO$ as an example to study the evolution of local correlations, and to study the magnetic and superconducting instabilities in a Fermi surface instability picture. I have shown that the strength of correlations in $LaCrAsO$ is similar to FeSCs when $n_e=4.5$, hence we proposed to search for unconventional superconductivity in this region. My co-workers showed that the \textcolor{black}{most plausible} superconducting order parameter has a \textcolor{black}{$d$-wave symmetry. In the band structure studied, $d_{xy}$ is favored. Nevertheless, we have pointed out that} the electron pockets responsible for the superconductivity are shallow, and the expectations may change if these shallow pockets are not present, \textcolor{black}{to other symmetry, most probably $d_{x^2-y^2}$}.

\item In \cref{Chap4b} \cite{Pizarro2}, I have studied the \textcolor{black}{strength of} correlations in the quasi-1D two-leg ladder FeSC, $BaFe_2S_3$. \textcolor{black}{At ambient pressure, the undoped} $BaFe_2S_3$ shows an insulating behavior which persists for $T>T_N$. Various authors have argued that these systems are Mott insulators. \textcolor{black}{When considering only local correlations in multiorbital systems, in the Hund metal regime (in which $BaFe_2S_3$ seems to lie, see below), this statement does not seem very plausible, see \fref{fig:3.13}}. The aim of this work \textcolor{black}{was} to clarify the nature of the insulating states in these materials. I have used the tight-binding model proposed by \emph{Arita et al.} \cite{AriPRB922015} for two pressures $0 \, GPa$ and $12.4 \, GPa$, which considers all the 5 $Fe$ $d$ orbitals on each of the $4$ $Fe$ atoms of the unit cell. In this model, the band structure of $BaFe_2S_3$ shows a 3D dispersion, even if the system is initially seen as a quasi-1D material. I have studied the \textcolor{black}{strength of correlations} for both pressures at $T=0 \, K$. I have obtained a metallic behavior for both pressures. We have argued that \textcolor{black}{temperature $T$ could reconcile the local correlations picture with the experimental behavior, due to the fact that $T$ favors a more incoherent system, hence it promotes an insulating behavior.} I have also shown that, at $12.4 \, GPa$ (which is the pressure at which superconductivity emerges), the strength of correlations is similar to other FeSCs. I have obtained a remarkable \textcolor{black}{reconstruction of the Fermi surface, which can modify} the expectations \textcolor{black}{from} Fermi surface-based instability theories. We argued that \textcolor{black}{due to this reconstruction}, the intra-ladder AFM scattering vector is enhanced. At $0 \, GPa$, the system shows an insulating behavior, and hence it does not have Fermi surface, so any Fermi surface-based instability picture is highly in doubt to be applicable. At $12.4 \, GPa$, the Fermi surface modifications will be important when elucidating the Fermi surface instabilities, such as the superconductivity order parameter.

\item In \cref{Chap4c} \cite{Pizarro3}, I studied the nature of the insulating states found by Pablo Jarillo's group in MA-TBG \cite{CaoN5562018_ins,CaoN5562018_sc}. In MA-TBG, unexpected insulating states appear \textcolor{black}{when the charge neutrality point (CNP) is doped with $2$ electrons or $2$ holes}. Various authors have argued that these insulating behaviors are Mott insulating states, but the insulating nature is still unknown. We assumed that MA-TBG can be describe by an effective model for the moir\'e superlattice, and that flat bands are responsible for its correlated behavior. We also assumed that the interactions are Hubbard-like, and studied three experimental features of these insulating states to check whether local correlations could explain their behavior: when increasing an external magnetic field or the temperature $T$, there is a transition to a metallic state, and the gap size ($\sim 0.3 \, meV$) of these insulating states is two orders of magnitude smaller than the expected bandwidth ($W \sim 10 \, meV$). I have shown that all these behaviors cannot be explained by considering only local correlations. We then considered the change in the phenomenology which appears in Mott states when non-local correlations are important. Based on the phenomenology driven by non-local correlations in related models, we argued that if these non-local correlations are included, the experimental observations could be reconciled with the expectations for Mott-like insulating states.

\item In \cref{Chap4d}, I have presented a DFT calculation and the Wannier tight-binding model for the rare-earth free 2D material FGT. This system shows an itinerant ferromagnetic (FM) phase which persists down to the monolayer FGT. The critical temperature $T_F$ can be tuned by effective electron doping via an external voltage. Trilayer FGT shows room temperature itinerant ferromagnetism at an density electron doping of $\sim 10^{14} \, cm^{-2}$. \textcolor{black}{Due to the itinerant character of the FM phase, a peak in the density of states (DOS) could explain the origin of this phase} (following a Stoner criteria). On the other hand, ARPES and low-$T$ specific heat measurements pointed to the existance of strong correlations. Similarities between FGT and FeSCs (like the total filling per $Fe$ atom, or the bandwidth of the $Fe$ $d$ orbitals) suggest that the Hund's coupling could be playing a role in the ferromagnetic phase of FGT. As a fisrt step to study the electronic correlations, I calculated via DFT \textsc{Wien2k} and \textsc{Wannier90} codes the band structure, DOS and Fermi surfaces for bulk and monolayer FGT, as well as the tight-binding model for monolayer FGT. \textcolor{black}{I showed that the most important feature of the DOS for bulk and monolayer FGT is the very large peaks at and around the Fermi level}. I also showed that the contribution from $Ge$ and $Te$ $p$ orbitals is not negligible. \textcolor{black}{I calculated the so called $ddpp$ model, as well as a minimal version including only $Fe$ $d$ orbitals, the so called $dd$ model. However, the $dd$ model shows large discrepancies with respect to the original band structure}. Yet, further work is needed to elucidate the origin of this ferromagnetic phase \textcolor{black}{and the role of electronic correlations in few layers} FGT.

\end{itemize}

I expect that the local correlations in multiorbital systems phenomenology explained in this thesis may serve as a background for future theoretical calculations and the interpretation of experimental results in real materials. I \textcolor{black}{believe} that having an optimal degree of electronic correlations is an important ingredient to find new unconventional superconductors, and hence understanding the phenomenology of such electronic correlations is essential to understand the unconventional superconductivity. 

\chapter{{\bf Conclusiones}} 
\label{Chap8sp}
\lhead{Chapter \ref{Chap8sp}. \emph{Conclusiones}} 

A lo largo de esta tesis, he estudiado los efectos de las correlaciones electr\'onicas en varios sistemas. Me he centrado en estudiar los efectos de las correlaciones locales en sistemas multi-orbitales, y en la fenomenolog\'ia que aparecer en dichos sistemas multi-orbitales en comparaci\'on con sistemas de un solo orbital cuando se incluyen los grados de libertad orbitales. He estudiado estos efectos en materiales reales, especialmente en superconductores basados en hierro (FeSCs) y materiales relacionados, tales como $LaCrAsO$, ver \crefsp{Chap4a}, y $BaFe_2S_3$, ver \crefsp{Chap4b}, y en materiales 2D, espec\'ificamente en el grafeno bicapa rotado en \'angulo m\'agico (MA-TBG), ver \crefsp{Chap4c}. Tambi\'en he obtenido la estructura de bandas DFT y el modelo de enlaces-fuertes Wannier para el material 2D $Fe_3GeTe_2$ (FGT), ver \crefsp{Chap4d}. Entender la estructura de bandas sirve como primer paso para estudiar las correlaciones electr\'onicas.

Despu\'es de revisar algunos conceptos gen\'ericos de las correlaciones en el \crefsp{Chap01}, en el \crefsp{Chap03} he introducido la fenomenolog\'ia de las correlaciones locales en sistemas multi-orbitales. En los sistemas multi-orbitales, el ratio entre el coste de energ\'ia de interacci\'on $\Delta^x$ y la ganancia de energ\'ia cin\'etica $\widetilde{W}$ depende del n\'umero de orbitales por \'atomo $N$, el n\'umero de electrones por \'atomo $n_e$ y el acoplo Hund $J_H$. El aislante de Mott no est\'a restringido a sistemas a llenado mitad ($n_e=N$), si no que tambi\'en puede aparecer a otros llenados enteros ($n_e=1,2, \ldots, N-1$). Cuando $J_H$ aumenta, la interacci\'on cr\'itica para la transici\'on de Mott $U_c$ disminuye (aumenta) para sistemas a llenado mitad $n_e=N$ (a llenado de un solo electr\'on $n_e=1$). Para $N>2$, $1<n_e<N$ y valores intermedios y largos de $J_H$, un estado met\'alico correlacionado surge, llamado metal de Hund. En el r\'egimen del metal de Hund, la fuerza de las correlaciones locales es grande, los \'atomos est\'an localmente en un estado de esp\'in grande, y hay un desacople orbital. La fenomenolog\'ia del metal de Hund es manejada por el aumento de las correlaciones de esp\'in locales, que ocurre debido a la disminuci\'on de las configuraciones de esp\'in anti-paralelas en cada \'atomo. Por tanto, hay una polarizaci\'on at\'omica de esp\'in promovida por $J_H$. La supresi\'on de las configuraciones de esp\'in anti-paralelas est\'a relacionada con el aislante de Mott a llenado mitad, por lo tanto, esto marca un enlace entre la f\'isica del aislante de Mott y la del metal de Hund, a pesar de la diferencia de ser un estado aislante o met\'alico. En el caso de sistemas de orbitales no equivalentes, $J_H$ promueve la diferenciaci\'on orbital, por lo que una transici\'on de Mott selectiva en orbitales puede ocurrir. Varios resultados te\'oricos y experimentales para los FeSCs (calor espec\'ifico a baja $T$, espectroscop\'ia de fotoemisi\'on resuelta en \'angulo (ARPES) para la estructura de bandas renormalizada, espectroscop\'ia de emisi\'on de rayos X (XES) para el momento local, etc.) sostienen la idea de que estos sistemas pueden verse como estando cerca o en el r\'egimen del metal de Hund.

Los principales resultados de esta tesis pueden resumirse como sigue:

\begin{itemize}

\item En el \crefsp{Chap4a} \cite{Pizarro1}, hemos propuesto buscar una nueva familia de superconductores de alta $T_c$ similar a los FeSCs pero basados en cromo en vez de en hierro. Nuestro argumento est\'a basado en la idea de necesitar un grado \'optimo de correlaciones electr\'onicas, tal y como ocurre en la mayor\'a de superconductores no-convencionales. Hemos propuesto que los pnicturos y calcogenuros de $Cr$ podr\'ian alojar superconductividad de alta $T_c$, similar a los FeSCs. La diferencia m\'as importante entre los sistemas basados en $Cr$ y los FeSCs es el n\'umero total de electrones por \'atomo, el cual es $n_e=4$ en los materiales basados en $Cr$ y $n_e=6$ en los FeSCs. La fuerza de las correlaciones locales aumenta cuando nos acercamos al llenado mitad ($n_e=5$) desde $n_e=6$ para los FeSCs, por lo que se argument\'o que los FeSCs pueden verse como aislantes de Mott dopados con electrones, as\'i enlazando la f\'isica de los cupratos de alta $T_c$ y de los FeSCs. He demostrado que una tendencia similar se obtiene cuando nos movemos desde el aislante de Mott dopado con huecos $n_e=4$ hacia llenado mitad. Hemos elegido $LaCrAsO$ como ejemplo para estudiar la evoluci\'on de las correlaciones locales, y para estudiar las inestabilidades magn\'eticas y superconductoras en una descripci\'on basada en inestabilidades de la superficie de Fermi. He demostrado que la fuerza de las correlaciones en $LaCrAsO$ es similar a los FeSCs cuando $n_e=4.5$, por lo que hemos propuesto buscar superconductividad no-convencional en esta regi\'on. Mis colaboradores demostraron que el par\'ametro de orden superconductor m\'as probable tiene simetr\'ia de onda $d$. Para la estructura de bandas estudiada, $d_{xy}$ se favorece. No obstante, hemos se\~nalado que los \textit{pockets} de electrones responsables de la superconductividad son poco profundos, y las expectativas pueden cambiar si estos \textit{pockets} de electrones no est\'an presentes, dando lugar a otra simetr\'ia, la m\'as probable $d_{x^2-y^2}$.

\item En el \crefsp{Chap4b} \cite{Pizarro2}, he estudiado la fuerza de las correlaciones electr\'onicas en el FeSCs quasi-1D en escalera de dos patas, $BaFe_2S_4$. A presi\'on ambiente, el $BaFe_2S_3$ no dopado muestra un comportamiento aislante, el cual persiste para $T>T_N$. Varios autores han argumentado que estos sistemas son aislantes de Mott. Cuando se consideran \'unicamente las correlaciones locales en sistemas multi-orbitales, en el r\'egimen de Hund (en el cual $BaFe_2S_3$ parece estar situado, ver m\'as abajo), esta afirmaci\'on no parece muy acertada, ver \frefsp{fig:3.13}. El objetivo de este trabajo fue clarificar la naturaleza de los estados aislantes en estos materiales. He usado el modelo de enlaces-fuertes propuesto por \emph{Arita et al.} \cite{AriPRB922015} para dos presiones $0 \, GPa$ y $12.4 \, GPa$, el cual considera los $5$ orbitales $d$ del $Fe$ en cada uno de los $4$ \'atomos de $Fe$ de la celda unidad. En este modelo, la estructura de bandas del $BaFe_2S_3$ muestra una dispersi\'on 3D, incluso aunque el sistema fuera inicialmente visto como un material quasi-1D. He estudiado la fuerza de las correlaciones para ambas presiones a $T=0 \, K$. He obtenido un comportamiento met\'alico para ambas presiones. Hemos argumentado que el efecto de la temperatura $T$ podr\'ia reconciliar las expectativas de las correlaciones locales con el comportamiento experimental, debido al hecho de que $T$ favorece un sistema m\'as incoherente, por lo que promueve el comportamiento aislante. Tambi\'en he mostrado que, a $12.4 \, GPa$ (que es la presi\'on a la que la superconductividad surge), la fuerza de las correlaciones es similar a otros FeSCs. He obtenido una reconstrucci\'on notable de la superficie de Fermi, lo cual puede cambiar las expectativas de teor\'ias basadas en inestabilidades de la superficie de Fermi. Hemos argumentado que debido a esta reconstrucci\'on, el vector de dispersi\'on AFM entre escaleras se ve aumentado. A $0\, GPa$, el sistema muestra un estado aislante, por lo que no tiene superficie de Fermi, as\'i que cualquier descripci\'on basada en inestabilidades de la superficie de Fermi est\'a altamente en duda de ser aplicable. A $12.4 \, GPa$, las modificaciones en la superficie de Fermi ser\'an importante a la hora de estudiar las inestabilidades de la superficie de Fermi, tales como el par\'ametro de orden superconductor.

\item En el \crefsp{Chap4c} \cite{Pizarro3}, he estudiado la naturaleza de los estados aislantes encontrados en el MA-TBG por el grupo de Pablo Jarillo \cite{CaoN5562018_ins,CaoN5562018_sc}. En el MA-TBG, unos estados aislantes inesperados aparecen cuando el punto de carga neutral (CNP) es dopado con $2$ electrones o $2$ huecos. Varios autores argumentaron que estos comportamientos aislantes corresponden a estados aislantes de Mott, pero la naturaleza de dichos estados es todav\'ia desconocida. Hemos asumido que el MA-TBG puede describirse con un modelo efectivo para la superred de moir\'e, y que las bandas planas son responsables de su comportamiento correlacionado. Tambi\'en, hemos asumido que las interacciones son de tipo Hubbard, y hemos estudiado tres resultados experimentales de estos estados aislantes para comprobar si las correlaciones locales podr\'ian explicarlos: cuando un campo magn\'etico externo o la temperatura $T$ aumentan, hay una transici\'on a un estado met\'alico, y el tama\~no del gap ($\sim 0.3 \, meV$) de estos estados aislantes es dos \'ordenes de magnitud m\'as peque\~no que la anchura de banda esperada ($W \sim 10 \, meV$). He demostrado que todos estos comportamientos no pueden explicarse considerando \'unicamente las correlaciones locales. Consideramos m\'as tarde el cambio en la fenomenolog\'ia que aparece en los estados de Mott cuando las correlaciones no-locales son importantes. Basados en la fenomenolog\'ia debida a las correlaciones no-locales en modelos relacionados, hemos argumentado que si estas correlaciones no-locales son incluidas, las observaciones experimentales podr\'ian reconciliarse con las expectativas para los estados aislantes de tipo Mott.

\item En el \crefsp{Chap4d}, he presentado los c\'alculos DFT y el modelo de enlaces-fuertes Wannier para el material 2D libre de tierras raras FGT. Este sistema muestra una fase ferromagn\'etica (FM) itinerante que persiste hasta el sistema monocapa FGT. La temperatura cr\'tica $T_F$ puede modificarse por el dopaje efectivo de electrones via un voltaje externo. La tricapa FGT muestra ferromagnetismo itinerante a temperatura ambiente a un dopaje de densidad de electrones de $\sim 10^{14} \, cm^{-2}$. Debido al car\'acter itinerante de la fase FM, los picos en la densidad de estados (DOS) podr\'ian explicar el origen de esta fase (siguiendo el criterio de Stoner). Por otro lado, medidas ARPES y de calor espec\'ifico a baja $T$ se\~nalan la existencia de correlaciones fuertes. Similitudes entre FGT y los FeSCs (como el llenado total por \'atomo de $Fe$, o la anchura de banda de los orbitales $d$ del $Fe$) sugieren que el acoplo Hund podr\'ia estar jugando un papel en la fase ferromagn\'etica del FGT. Como un primer paso para estudiar las correlaciones electr\'onicas, he calculado via los c\'odigos de DFT \textsc{Wien2k} y \textsc{Wannier90}, la estructura de bandas, DOS y las superficies de Fermi del material en volumen y monocapa FGT, as\'i como el modelo de enlaces-fuertes para la monocapa FGT. He mostrado que la propiedad m\'as importante en el material en volumen y monocapa FGT es la aparici\'on de grandes picos en la DOS alrededor del nivel de Fermi. Tambi\'en he mostrado que la contribuci\'on de los orbitales $p$ del $Ge$ y $Te$ no son despreciables. He calculado el modelo $ddpp$, as\'i como una versi\'on m\'as sencilla incluyendo solo los orbitales $d$ del $Fe$, el modelo $dd$. Sin embargo, el modelo $dd$ muestra grandes discrepancias con respecto a la estructura de bandas original. No obstante, se necesita m\'as trabajo para elucidar el origen de la fase ferromagn\'etica y el papel de las correlaciones electr\'onicas en pocas capas de FGT.

\end{itemize}

Espero que la fenomenolog\'ia sobre las correlaciones locales en sistemas multi-orbitales explicada en esta tesis pueda servir como antecedente para futuros estudios te\'oricos y la interpretaci\'on de resultados experimentales en materiales reales. Creo que tener un grado \'optimo de correlaciones electr\'onicas es un ingrediente importante para encontrar nuevos superconductores no-convencionales, por lo que entender la fenomenolog\'ia de dichas correlaciones electr\'onicas es esencial para entender la superconductividad no-convencional.




\addtocontents{toc}{\vspace{1em}} 

\begin{appendices}
\chapter{\bf Density Functional Theory and Wannier parametrization}
\label{AppA} 
\lhead{Appendix \ref{AppA}: DFT and Wannier parametrization}

In this appendix, I will briefly present the Density Functional Theory (DFT) for band structure calculations, following the overview given in \cite{Ferber2012thesis,Schonhoff2017thesis}, and the Wannier parametrization in order to calculate the hopping integrals $t^{ij}_{mn}$ and crystal field splittings $\epsilon_m$ for any given material, by following the manual of \textsc{Wannier90} code \cite{MosCPC1852014}.

\textcolor{black}{DFT and Wannier parametrization calculations were used through all this thesis. Personally, I used them in \cref{Chap4d} when calculating the DFT band structure and the Wannier tight-binding model for $Fe_3GeTe_2$ (FGT).}

\section{DFT framework}
\label{A.1}

DFT was originally developed by Hohenberg, Kohn and Sham in 1964 and 1965 \cite{HohPRB1361964,KohPRB1401965}. This theory has become very useful in these days to calculate the electronic properties of solids, like band structure, band-gaps, optical responses, etc. It is based on two main theorems called the Hohenberg-Kohn theorems \cite{HohPRB1361964}, which state:

\begin{enumerate}

\item The ground state properties of a system of particles are uniquely determined by the electronic density $\rho(\vec{r})$. This means that the energy is a functional of the electronic density $E[\rho(\vec{r})]$.

\item The correct ground state electronic density minimizes the energy functional, so the ground state energy is $E_0 = E[\rho_0 (\vec{r})] \leq E[\rho (\vec{r})]$

\end{enumerate}

Then, DFT uses the electronic density instead of the wavefunctions to determine the ground state properties. In this situation, the original Schr\"odinger problem of \eref{eq:1.1} for solving $N$ differential equations greatly reduces to solve the equation for the electronic density in terms of $\vec{r}$. Due to the fact that DFT calculates the ground state properties, it tends to have problems with \textcolor{black}{excitation} properties, such as the band gap of semiconductors.

In the Born-Oppenheimer approximation (where ions will sit motionless in each lattice site because they are more massive than electrons, $M \ll m$), the equation to be solved for $E[\rho (\vec{r})]$ is given by \eref{A.1}. 

\begin{equation}
E [\rho] = T[\rho] + V[\rho] + U[\rho]
\label{eq:A.1}
\end{equation}

\textcolor{black}{where $V[\rho]$ is a material-dependent (and known) quantity. In this equation, $T$ refers to the kinetic energy of electrons,  $V$ to the potential created by the ions in which the electrons are moving and $U$ to the electron-electron interaction (see also \eref{eq:1.2}). $U[\rho]$ can be written as a Hartree potential $U_{H} [\rho]$ plus an exchange-like interaction term $E_{X}[\rho]$:}

\begin{equation}
U [\rho] \approx \underbrace{\frac{1}{2} \int \frac{\rho (\vec{r}) \rho (\vec{r}')}{| \vec{r}-\vec{r}' |} d^3 \vec{r} d^3 \vec{r} '}_{U_H [\rho]} + E_{X} [\rho]
\label{eq:A.2}
\end{equation}

\subsection{Kohn-Sham equations}
\label{A.1.1}

The form of $T[\rho]$ will depend on further approximations. \textcolor{black}{In most DFT approximations}, the single-particle states approximation is used, $\rho(\vec{r}) = \sum_{i=1}^{N} f_i |\psi_{i}(\vec{r})|^2$ where $f_i$ is the occupation number, $\psi_{i}(\vec{r})$ are the so called Kohn-Sham orbitals \textcolor{black}{and $N$ is the number of these Kohn-Sham orbitals}. Thus, $T[\rho] \approx T_{hom}[\rho] + E_C[\rho]$, where $T_{hom}[\rho]$ is the kinetic energy of the non-correlated homogeneous electron gas (a well-known quantity) and $E_C[\rho]$ is the kinetic correlation energy left out when $\rho$ is \textcolor{black}{approximated by an expression in terms of single-particle states} \cite{CepPRL451980,PerPRB451992}. 

After these considerations, \eref{eq:A.1} can be rewritten as:

\begin{equation}
E [\rho(\vec{r})] = T_{hom}[\rho(\vec{r})] + \int v_{ext}(\vec{r}) \rho (\vec{r}) d^3 \vec{r} + \frac{1}{2} \int \frac{\rho (\vec{r}) \rho (\vec{r}')}{| \vec{r}-\vec{r}' |} d^3 \vec{r} d^3 \vec{r} ' + E_{XC} [\rho(\vec{r})]
\label{eq:A.3}
\end{equation}

with $E_{XC}[\rho] = E_{X}[\rho] + E_{C}[\rho]$. The form of $E_{XC}[\rho]$ is given by the different approximations used, as in local density approximation (LDA), generalized gradient approximation (GGA), etc. The exchange-correlation potential is $V_{XC}(\vec{r}) = \delta E_{XC}[\rho(\vec{r})] / \delta \rho(\vec{r})$. In LDA $V_{XC}$ is approximated as a local potential. It only depends on the electronic density, which is a constant function of $\vec{r}$, hence $V_{XC}^{LDA}(\rho)$. In GGA \cite{BecPRA381988}, $V_{XC}$ further depends on the first derivative of $\rho$ with respect to $\vec{r}$, hence $V_{XC}^{LDA}(\rho, \nabla \rho)$. 

With the aid of the Kohn-Sham orbitals, the Kohn-Sham equation can be derived:

\begin{equation}
\left[- \frac{1}{2} \nabla^2 + v_{ext} + U_{H} + V_{XC} \right] \psi_i = \varepsilon_i \psi_i
\label{eq:A.4}
\end{equation}

Note that this equation is not equivalent to the original many-body hamiltonian of \eref{eq:1.2}. Only in the ideal case that $V_{XC}$ is exactly treated (which is not solvable), the full correlations and interactions effects will be included. In the most widely used approximations, LDA and GGA, these correlations and interactions are almost neglected, and further models are needed to solve the problem.

\subsection{Obtaining the band structure using DFT}
\label{A.1.2}

In order to obtain the band structure of a given material, Kohn-Sham orbitals have to be expanded in a suitable basis of functions $\left\{ \phi_{k\nu} \right\}$, then $\psi_i = \sum_\nu c_{k\nu}^i \phi_{k\nu}$. There are different approximations to calculate this basis, like using augmented plane waves (APW) or linearized augmented plane waves (LAPW). \textcolor{black}{The difference between APW and LAPW lies in the expansion used to write $\phi_{k\nu}$. In the case of LAPW, both the energy dependence of the radial part and its derivative are considered, while for APW only the energy dependence for the radial part is considered (see below)}. These approximations exploit the fact that close to the atomic nuclei, the ionic potential in a solid is similar to the one of an isolated atom, whether far from the nuclei, the potential varies weakly and electrons move freely. This is the so called muffin-tin spheres (MTS) approximation. Inside an sphere around the atom, the electron has atomic-like functions \textcolor{black}{(radial and spherical harmonics functions)}, whereas in the interstitial region is described by free plane waves-like functions. 

These basis are useful for valence states, while core and semi-core states are not so well treated. This problem can be improved by using the LAPW,LO basis. In this basis, band energies $\varepsilon_{k \nu}$ and Bloch states $\phi_{k \nu} (\vec{r})$ are calculated using \eref{eq:A.5} (note that $\vec{r}$ is defined inside the MTS).

\begin{equation}
\phi_{lm \, \alpha}^{LAPW,LO} (\vec{r}) = \left[ A_{l \, m}^{\alpha, \, LO}  u_{l}^{\alpha}(r,E_{1,l}^{\alpha}) + B_{l \, m}^{\alpha, \, LO}  \dot{u}_{l}^{\alpha}(r,E_{1,l}^{\alpha}) + C_{l \, m}^{\alpha, \, LO}  u_{l}^{\alpha}(r,E_{2,l}^{\alpha})  \right] Y_l^m (\theta,\varphi)
\label{eq:A.5}
\end{equation}

where $\alpha$ is the atom index, $l$ and $m$ are the angular quantum numbers associated with the angular solution $Y_l^m(\theta,\varphi)$ of \eref{eq:A.4}, $A$, $B$ and $C$ are expansion coefficients to be determined self-consisenly, $u_l^\alpha$ is the radial part solution of each atom \textcolor{black}{and its derivative $\dot{u}_l^{\alpha}$ with respect to the atomic energy $E_{1,l}^\alpha$}. Outside the MTS, $\phi_{lm \, \alpha}$ can be written as free plane waves.

Different codes are available to be used for the calculation of the band structure by means of DFT. In this thesis, I will use the \textsc{Wien2k} codes results.

\section{Wannier parametrization}
\label{A.2}

The Wannier parametrization is a method in which the hopping integrals $t_{mn}^{ij}$ (as well as crystal field energies $\epsilon_m$) are calculated by fitting a set of specific orbitals to the band structure around the Fermi level. This method is based on the idea that the orbitals which mostly contribute to form the bands around the Fermi level are those which control the physics in the system.

Wannier functions \cite{WanPR521937} $w_{m}^{\alpha} (\vec{r}-\vec{R})$ are localized functions centered at lattice sites $\vec{R}$ can be obtained from Fourier transforming the Bloch states $\phi_{k \nu}(\vec{r})$ (i.e. free plane waves):

\begin{equation}
w_{m}^{\alpha}(\vec{r}-\vec{R}) = \sum_k e^{-i \vec{k} \cdot \vec{R}} \sum_{\nu \in W} P_{km\nu}^{\alpha} \phi_{k\nu} (\vec{r})
\label{eq:A.6}
\end{equation}

where $m$ is the orbital index, $\nu$ is the band index and $\nu \in W$ refers to the energy window $W$ taken to fit the band structure to a set of Wannier orbitals. $P_{km\nu}^{\alpha}$ are the projector matrices, which can be obtained from the coefficients $A$, $B$ and $C$.

In order to calculate $t_{mn}^{ij}$, let's consider the DFT hamiltonian $H_0$ given in \eref{eq:A.4}. This hamiltonian follows the Schr\"odinger equation $H_0 \phi_{k\nu} = E_{k\nu} \phi_{k\nu}$, where $E_{k\nu}$ are the band energies. Then, in the basis of Bloch states $\left\{ \phi_{k\nu}\right\}$, $H_0$ is diagonal:

\begin{equation}
\phi_{k\nu}^\dagger(\vec{r})  H_0 \phi_{k\nu}(\vec{r})  = E_{k\nu} \mathbb{I}
\label{eq:A.7}
\end{equation}

where $\mathbb{I}$ is the identity matrix.

In the Wannier orbital basis $\left\{ w_m^{\alpha} (\vec{r}) \right\}$, $H_0$ is not diagonal and its elements are given by $t_{mn}^{ij}$, or by the dispersion relations $\varepsilon_{kmn}$ if the hoppings are Fourier transformed (FT) into the $k$-space. In the $k$-space:

\begin{equation}
w_{km}^{\alpha}(\vec{r}) = \sum_{\nu \in W} P_{km\nu}^{\alpha} \phi_{k\nu} (\vec{r})
\label{eq:A.8}
\end{equation}

Then, in the Wannier orbitals basis:

\begin{equation}
w_{km}^{\alpha \: \dagger}(\vec{r})  H_0 w_{km}^\alpha(\vec{r})  = P_{km\nu}^{\alpha \: \dagger}  \left( E_{k\nu} \mathbb{I} \right) P_{km\nu}^{\alpha}
\label{eq:A.9}
\end{equation}

So that, the hopping integrals $t_{mn}^{ij}$ and crystal field splitting $\epsilon_m$ can be numerically obtained as (assuming that Wannier functions are centered at $\vec{R}=\vec{0}$):

\begin{equation}
\left\{
\begin{aligned}
t_{mn}^{ij}	& = \sum_k <w_{km}(\vec{i}) | H_0 | w_{kn}(\vec{j}) > e^{-i \vec{k} \cdot \vec{ij}} \\
\epsilon_{m}	& = \sum_k <w_{km}(\vec{i}) | H_0 | w_{km}(\vec{i}) >
\end{aligned}
\right.
\label{eq:A.10}
\end{equation}

The most widely used code to calculate $t_{mn}^{ij}$ is \textsc{Wannier90} \cite{MosCPC1852014}. Both the projection method explained above and the so called maximally localized Wannier functions (MLWFs) procedure are implemented in this code. MLWFs follows the same scheme as the projection method, but it includes the maximal localization of the Wannier functions $w_{km}(\vec{r})$ \cite{MarPRB561997}, hence reducing the spatial spread of the Wannier orbitals in order to ensure the real space localization of these Wannier orbitals.


\chapter{\bf Slave-Spin Mean Field formalism}
\label{AppB} 
\lhead{Appendix \ref{AppB}: SSMF formalism}

In this appendix, an extense discussion about slave-spin mean-field (SSMF) formalisms (both $Z_2$ and $U(1)$) can be found. I will use the articles where both formalisms were presented \cite{deMPRB722005,HasPRB812010,YuRPRB862012}, as well as my personal notes and mathematical proofs. A comparison between $Z_2$ and $U(1)$ formulations is also presented. The solving procedure for a SSMF calculation is explained at the end of this appendix.

\section{SSMF background}
\label{B.1}

SSMF formalism is a framework well suited up to study the effect of local correlations in a multiorbital system \textcolor{black}{via a self-consistent} calculation of the quasiparticle weights $Z_{m\sigma}$ for different $m$ orbitals (with $m=1,2... N$) and $\sigma$ spin (with $\sigma=+,-$ or $\sigma=\uparrow,\downarrow$). \textcolor{black}{In the so called single-site approximation}, SSMF is constrained to onsite correlations. Usually, the local paramagnetic (PM) solution is obtained, hence $Z_m$ will be spin indepentent. The exception can be seen in \cref{Chap4c}, where I implemented the calculation of $Z_{m\sigma}$ for the Zeeman effect, in which spin degeneracy is broken by an onsite shift which depends on the Zeeman magnetic field.

SSMF techniques follow the Brinkman-Rice picture for the Mott transition \cite{Fazekas1999,Coleman2015,emergence2017}, in which $Z_m$ is traced down from $Z_m=1$ in the non-correlated metal to $0$ in the Mott insulator when varying $U$. Thus, SSMF is expected to slightly understimate the effect of local correlations. For a multiorbital problem, SSMF permits a calculation of $Z_m$ in terms of $U$, $J_H$, the number of electrons per atom $n_e$ and the number of orbitals per atom $N$. Once Mott insulator is reached with all $Z_m=0$, SSMF cannot provide further information about the system.

Compared with other slave-particle theories, the slave-spin mean field (SSMF) (both in $Z_2$ and $U(1)$ formulations) allows to treat multiorbital systems in an economical way, using $2N$ \textit{pseudospin-}$1/2$ variables, while slave-boson technique \cite{KotPRL571986} normally increases its number of variables in an exponential way with $N$. SSMF formalisms allow a treatment of non-equivalent orbitals system, while slave-rotor technique \cite{FloPRB702004} cannot account for it. Compared with other more sofisticated numerical techniques (such as DMFT \cite{GeoRMP681996}), SSMF formalisms are faster and easier to implement, and the results are quite good when comparing with such techniques and with experimental results (see discussion and comparison in \cref{Chap03}).

\eref{eq:1.22} will be the initial hamiltonian to work with. \textcolor{black}{For simplicity}, interactions \textcolor{black}{are assumed to be} orbital-independent, rotational invariant relations $U'=U-2J_H$ and $J'=J_H$ are used and $H_{add}$ is neglected. $H_{add}$ is difficult to treat in SSMF, due to the fact that it mixes pseudospin-1/2 and auxiliary fermion operators; in \cite{deMPRB722005}, the authors tried to include $H_{add}$ by introducing some operators which actually mix the physical states with the unphysical ones (see text). \textcolor{black}{Later on, the authors realized that such an approximation was not giving good results \cite{lucapriv}}. Nevertheless, the physics discussed in this thesis is mainly contained in $H_{dens}$. Thus, we will study the so called Ising hamiltonian $H=H_0+H_{dens}$, where $H_0$ describes the tight-binding model and $H_{dens}$ the density-density interaction terms.

SSMF formalisms are, considering the change of representation ($Z_2$ or $U(1)$), a \textcolor{black}{change of basis} from the initial Hilbert space defined by the occupation numbers of the electrons $\mathcal{H}_e \equiv \{ |n^d=0>,|n^d=1> \}$, with $n^d=d^\dagger d$ (dropping subindices for simplicity) to another expanded Hilbert space in which the occupation number operator $n^d$ is identified by a pseudospin-$1/2$ operator $S^z$. This trick is made in order to obtain the pre-factors of the dispersion relations $\varepsilon_{kmn}$ which will precisely be the quasiparticle weights $Z_m$.

Auxiliary fermion operators $n^f$ have to be defined in order to fulfill the anti-commutation rules, so that the new Hilbert space is defined by $\mathcal{H}_{SSMF} \equiv \{|n^f,S^z>\}$. By performing this \textcolor{black}{change of basis}, a constraint has to be imposed in order to project out the unphysical states obtained by this basis change, as described in \eref{eq:B.1}.

\begin{equation}
\left.
\begin{aligned}
	& physical \rightarrow \left\{
	\begin{aligned}
		|n^d=0> = |n^f=0,S^z=-1/2>\\
		|n^d=1> = |n^f=1,S^z=+1/2>\\
	\end{aligned}
	\right. \\
	& unphysical \rightarrow \left\{
	\begin{aligned}
		|\varnothing>	= |n^f=1,S^z=-1/2>\\
		|\varnothing> = |n^f=0,S^z=+1/2>
	\end{aligned}
	\right.
\end{aligned}
\right.
\label{eq:B.1}
\end{equation}

In SSMF, the pseudospin operators carry the charge of the real electrons, while the auxiliary fermions carry the spin. Thus, the metallic phase corresponds to the ordered pseudospins, \textcolor{black}{i.e. with a finite $<O> \neq 0$ value, where $O$ is a pseudospin operator}. The constraint has the form:

\begin{equation}
\widehat{n}_{}^{d} = d_{}^{\dagger} d_{}^{} \equiv \widehat{n}_{}^{f} = f_{}^{\dagger} f_{}^{} = S_{}^{z} + \frac{1}{2}
\label{eq:B.2}
\end{equation}

Generalizing to site $i$, orbital $m$ and spin $\sigma$ indices:

\begin{equation}
\widehat{n}_{im\sigma}^{d} = d_{im\sigma}^{\dagger} d_{im\sigma}^{} \equiv \widehat{n}_{im\sigma}^{f} = f_{im\sigma}^{\dagger} f_{im\sigma}^{} = S_{im\sigma}^{z} + \frac{1}{2}
\label{eq:B.3}
\end{equation}

This constraint is taken into account in $H=H_0+H_{dens}$ by introducing a set of parameters, the time-dependent Lagrange multipliers $\lambda_{im\sigma} (\tau)$ in the following way:

\begin{equation}
H=H_0^{} + H_{dens}^{} + \sum_{im\sigma} \lambda_{im\sigma}^{} (\tau) \left( S_{im\sigma}^{z} + \frac{1}{2} - \widehat{n}_{im\sigma}^{f} \right)
\label{eq:B.4}
\end{equation}

A set of $2N$ pseudospin-$1/2$ and auxiliary fermion operators is being introduced in each lattice site $i$. If the constraint is fully treated as in \eref{eq:B.4}, the solution would be exact. However, a full treatment is not solvable, and static Lagrange multipliers $\lambda_{im\sigma}$ are imposed by time-averaging the constraint:

\begin{equation}
\lambda_{im\sigma} = <\lambda_{im\sigma}(\tau)>_{time}
\label{eq:B.5}
\end{equation}

In order to express the non-interacting and the interacting parts of the hamiltonian of \eref{eq:B.4}, an appropriate representation of the operators $d_{im\sigma}$ and $d_{im\sigma}^{\dagger}$ in this new Hilbert space $\mathcal{H}_{SSMF}$ has to be choosen. An initial try could be introducing the pseudospin operators $S^{\pm}$:

\begin{equation}
\left\{
\begin{aligned}
d_{}^{} & = f_{}^{} S_{}^{-}\\
d_{}^{\dagger} & = f_{}^{\dagger} S_{}^{+}
\end{aligned}
\right.
\label{eq:B.6}
\end{equation}

However, applying further mean-field approximations will end on a $Z \neq 1$ at the non-correlated limit ($U=J_H=0$), which is an unphysical solution. 

I will explain now the $Z_2$ \cite{deMPRB722005,HasPRB812010} and the $U(1)$ \cite{YuRPRB862012} SSMF formulations, which are able to propose an appropiate representation of $d_{im\sigma}^\dagger$ and $d_{im\sigma}$ in the new Hilbert space $\mathcal{H}_{SSMF}$ in order to outcome the $Z \neq 1$ at $U=J_H=0$ problem.


\section{SSMF $Z_2$ formulation}
\label{B.2}

Following the ideas of \cite{deMPRB722005,HasPRB812010}, the change for the electron operators can be chosen to be:

\begin{equation}
\left\{
\begin{aligned}
d_{}^{} & = f_{}^{} O_{}^{}\\
d_{}^{\dagger} & = f_{}^{\dagger} O_{}^{\dagger}
\end{aligned}
\right. \ \
where \ \ O_{}=\begin{pmatrix}
0	& 	c_{} \\
1	& 	0 \\
\end{pmatrix} 
\label{eq:B.7}
\end{equation}

where $O$ is \textcolor{black}{a} pseudospin operator\footnote{$O$ is the generalization of the operator $2S^x$, which at the beginning of the technique development was constrained to be used only for half-filling systems \cite{deMPRB722005}.} and $c_{}$ is an arbitrary complex number. Note that $dim \ O_{m\sigma} = 2^{2N}$.

The choice of $c_{im\sigma}$ is a gauge \textcolor{black}{freedom} of the $Z_2$ formalism. It states that different pseudospin operators act in the same way in the physical states but differently in the unphysical ones; if the constraint is treated exactly, the difference of the $c_{im\sigma}$ actuation doesn't affect the problem. \textcolor{black}{Like the full treatment of the constraint is not possible, so} the problem has to be approximated \textcolor{black}{by being solved} with an appropriate $c_{im\sigma}$.

\textcolor{black}{In the single-site approximation, $c_{m\sigma}$ are real numbers. To recover the non-correlated limit $U=J_H=0$:}

\begin{equation}
c_{m\sigma} = \frac{1}{\sqrt{n_{m\sigma}(1-n_{m\sigma})}} - 1
\label{eq:B.8}
\end{equation}

where $n_{m\sigma}$ is the filling per atom, spin $\sigma$ and orbital $m$ ($n_{m\sigma} \in \left[ 0,1 \right]$ with $n_{m\sigma}=0.5$ defined as half-filling). The derivation of this formula can be seen in \aref{AppC}.

\textcolor{black}{Now, $H_0$ and $H_{dens}$ (see \eref{eq:B.4}) have to be written in the new Hilbert space $\mathcal{H}_{SSMF}$. $H_{dens}$ is written \textcolor{black}{in terms of} pseudospin operators \textcolor{black}{only}. Then (note that $H_{dens}=H_{dens}^{PS}-K_{PS}$):}

\begin{equation}
H_{dens}^{PS} = \frac{U-U'}{2} \sum_{im} \left( \sum_\sigma \tilde{S}_{im\sigma}^{z} \right) ^2 + \frac{U'}{2} \left( \sum_{im\sigma} \tilde{S}_{im\sigma}^{z} \right) ^2 - \frac{J_H}{2} \sum_{i\sigma} \left( \sum_m \tilde{S}_{im\sigma}^{z} \right) ^2
\label{eq:B.12}
\end{equation}

\textcolor{black}{where the constant $K_{PS}$ is a term which does not affect the mean-field calculations (note that $\widehat{n}_{im\sigma} \equiv S_{im\sigma}^{z} + \frac{1}{2} = \tilde{S}_{im\sigma}^{z}$):}

\begin{equation}
K_{PS} = \frac{U-J_{H}}{2} \sum_{im\sigma} \tilde{S}_{im\sigma}^{z \ 2}
\label{eq:B.11}
\end{equation}

\textcolor{black}{To obtain \eref{eq:B.12}, the following expressions have to be considered:}

\begin{equation}
\small{
\left\{
\begin{aligned}
{\left( \sum_{m\sigma} S_{im\sigma}^{z} \right)}^2 & = \underbrace{\sum_{m\sigma} S_{im\sigma}^{z \ 2}}_{constant} + 2\sum_{m} S_{im\uparrow}^{z} S_{im\downarrow}^{z} + 2\sum_{\substack{\sigma \\ m < m'}} S_{im\sigma}^{z} S_{im'\bar \sigma}^{z} + 2\sum_{\substack{\sigma \\ m < m'}} S_{im\sigma}^{z} S_{im'\sigma}^{z} \\
\sum_{m} {\left( \sum_{\sigma} S_{im\sigma}^{z} \right)}^2 & = \underbrace{\sum_{m\sigma} S_{im\sigma}^{z \ 2}}_{constant} + 2\sum_{m} S_{im\uparrow}^{z} S_{im\downarrow}^{z} \\
\sum_{\sigma} {\left( \sum_{m} S_{im\sigma}^{z} \right)}^2 & = \underbrace{\sum_{m\sigma} S_{im\sigma}^{z \ 2}}_{constant} + 2\sum_{\substack{\sigma \\ m < m'}} S_{im\sigma}^{z} S_{im'\sigma}^{z}
\end{aligned}
\right.}
\label{eq:B.10}
\end{equation}

On the other hand, $H_0$ can be written as:

\begin{equation}
H_0 = \sum_{i \neq j \, mm'\sigma} \left( t_{mm'}^{ij} f_{im\sigma}^{\dagger} O_{im\sigma}^{\dagger} f_{jm'\sigma}^{} O_{jm'\sigma}^{} + h.c. \right) + \sum_{im\sigma} \left( \epsilon_{m}^{} - \mu_{}^{} \right) f_{im\sigma}^{\dagger} f_{im\sigma}^{}
\label{eq:B.13}
\end{equation}

\textcolor{black}{In a first mean-field approximation, it is assumed that the auxiliary fermions, which carry the spin, and the pseudospins, which carry the charge, are uncorrelated. Taking into account that $O_{im\sigma}^\dagger$ and $f_{jm'\sigma}^{}$ commute, the tight-binding term can be written as:}

\begin{equation}
\begin{aligned}
f_{im\sigma}^{\dagger} f_{jm'\sigma}^{} O_{im\sigma}^{\dagger} O_{jm'\sigma}^{} \approx & f_{im\sigma}^{\dagger} f_{jm'\sigma}^{} <O_{im\sigma}^{\dagger} O_{jm'\sigma}^{}> + O_{im\sigma}^{\dagger} O_{jm'\sigma}^{} <f_{im\sigma}^{\dagger} f_{jm'\sigma}^{}> \\
& - <f_{im\sigma}^{\dagger} f_{jm'\sigma}^{}><O_{im\sigma}^{\dagger} O_{jm'\sigma}^{}>
\end{aligned}
\label{eq:B.14}
\end{equation}

Now, a second mean-field approximation will be applied. \textcolor{black}{It assumes that operators in different sites are uncorrelated.} This second mean-field gives rise to the single-site approximation. \textcolor{black}{Then, for $i \neq j$}:

\begin{equation}
\begin{aligned}
O_{im\sigma}^{\dagger} O_{jm'\sigma}^{} \approx & \ O_{im\sigma}^{\dagger} <O_{jm'\sigma}^{}> + O_{jm'\sigma}^{} <O_{im\sigma}^{\dagger}> - <O_{im\sigma}^{\dagger}><O_{jm'\sigma}^{}>\\
\end{aligned}
\label{eq:B.15}
\end{equation}

Using this last expression, it can be shown that the expected value $<O_{im\sigma}^{\dagger} O_{jm'\sigma}^{}>$ factorizes in another quantity $<O_{im\sigma}><O_{jm'\sigma}>$. In the single-site approximation, the site indices $i$ and $j$ can be dropped out from the pseudospin operators. Then, as shown in \aref{AppC}, the quasiparticle weight $Z_{m\sigma}$ is given by:

\begin{equation}
\begin{aligned}
Z_{m\sigma} & = <O_{m\sigma}^{\dagger} O_{m\sigma}^{}>\\
		& = <O_{m\sigma}>^2
\end{aligned}
\label{eq:B.15aa}
\end{equation}

Then, the first mean-field approximation of \eref{eq:B.14} can be rewritten as:

\begin{equation}
\small{
\begin{aligned}
f_{im\sigma}^{\dagger} f_{jm'\sigma}^{} O_{m\sigma}^{\dagger} O_{m'\sigma}^{} \approx & \sqrt{Z_{m\sigma}}\sqrt{Z_{m'\sigma}} f_{im\sigma}^{\dagger} f_{jm'\sigma}^{} +<f_{im\sigma}^{\dagger} f_{jm'\sigma}^{}> \left( \sqrt{Z_{m'\sigma}} O_{m\sigma}^{\dagger} + \sqrt{Z_{m\sigma}} O_{m'\sigma}^{} \right) \\
& - \sqrt{Z_{m\sigma}}\sqrt{Z_{m'\sigma}} <f_{im\sigma}^{\dagger} f_{jm'\sigma}^{}> \\
\end{aligned}}
\label{eq:B.16}
\end{equation}

where the last term is a constant which does not affect the mean-field calculations. $H_0$ can be written as (note that $t_{mm'}^{ij} \equiv t_{m'm}^{ij}$):

\begin{equation}
\begin{aligned}
H_0 & = \sum_{ijmm'\sigma} \sqrt{Z_{m\sigma}}\sqrt{Z_{m'\sigma}} \ t_{mm'}^{ij}\left( f_{im\sigma}^{\dagger} f_{jm'\sigma}^{} + h.c. \right) + \sum_{im\sigma} \left( \epsilon_m - \mu \right) f_{im\sigma}^{\dagger} f_{im\sigma}^{} \\
& + \sum_{ijmm'\sigma} t_{mm'}^{ij} \sqrt{Z_{m'\sigma}} \left[ O_{m\sigma}^{\dagger} \left( <f_{im\sigma}^{\dagger} f_{jm'\sigma}^{}> + <f_{jm'\sigma}^{\dagger} f_{im\sigma}^{}> \right) + h.c. \right] - K_f
\end{aligned}
\label{eq:B.17}
\end{equation}

where:

\begin{equation}
K_{f} = \sum_{ijmm'\sigma} t_{mm'}^{ij} \sqrt{Z_m\sigma} \sqrt{Z_{m'\sigma}} \left( <f_{im\sigma}^{\dagger} f_{jm'\sigma}^{}> + <f_{jm'\sigma}^{\dagger} f_{im\sigma}^{}> \right)
\label{eq:B.18}
\end{equation}

Finally, using the Fourier transformation (FT) of the auxiliary fermion operators:

\begin{equation}
f_{im\sigma}^{\dagger} = \sum_k e^{-i\vec{k}\vec{i}} f_{km\sigma}^{\dagger}
\label{eq:B.19}
\end{equation}

And the dispersion relations $\varepsilon_{kmm'}$ are defined as in \eref{eq:1.5}:

\begin{equation}
\small{ \sum_{ijmm'\sigma} t_{mm'}^{ij} \left( f_{im\sigma}^{\dagger} f_{jm'\sigma}^{} + h.c. \right) = \sum_{kmm'\sigma} \varepsilon_{kmm'}^{} f_{km\sigma}^{\dagger} f_{km'\sigma}^{}}
\label{eq:B.20}
\end{equation}

The non-correlated hamiltonian $H_0$ can be written as the sum of 2 terms $H_{PS}^0$ and $H_{f}$:

\begin{equation}
\begin{aligned}
H_0 & = \underbrace{\sum_{kmm'\sigma} \sqrt{Z_{m\sigma}} \sqrt{Z_{m'\sigma}} \varepsilon_{kmm'}^{} f_{km\sigma}^{\dagger} f_{km'\sigma}^{} + \sum_{km\sigma} \left( \epsilon_m - \mu \right) f_{km\sigma}^{\dagger} f_{km\sigma}^{}}_{H_f} \\ 
& + \underbrace{\sum_{m\sigma} h_{m\sigma}^{} \left( O_{m\sigma}^{\dagger} + h.c. \right)}_{H_{PS}^0} - K_f
\end{aligned}
\label{eq:B.21}
\end{equation}

where the effective Weiss fields $h_{m\sigma}$ are defined as (note also that $h_{m\sigma}^{\dagger} = h_{m\sigma}^{}$ in the single-site approximation):

\begin{equation}
h_{m\sigma} = \sum_{m'} \sqrt{Z_{m'\sigma}} \sum_k \varepsilon_{kmm'} <f_{km\sigma}^{\dagger} f_{km'\sigma}^{}>
\label{eq:B.22}
\end{equation}

The initial hamiltonian $H$ in the new basis $\mathcal{H}_{SSMF}$ is rewritten as a sum of 2 coupled (via the Lagrange multipliers $\lambda_{m\sigma}$) terms $H = H_{f} + H_{PS}$, where:

\begin{equation}
\small{
\left\{
\begin{aligned}
H_f 		& = \sum_{kmm'\sigma} \sqrt{Z_{m\sigma}} \sqrt{Z_{m'\sigma}} \varepsilon_{kmm'}^{} f_{km\sigma}^{\dagger} f_{km'\sigma}^{} + \sum_{km\sigma} \left( \epsilon_{m}^{} - \mu_{}^{} -\lambda_{m\sigma}^{} + \lambda_{m\sigma}^{0}  \right) f_{km\sigma}^{\dagger} f_{km\sigma}^{} - K_f \\
H_{PS}	& = \sum_{m\sigma} h_{m\sigma}^{} \left( O_{m\sigma}^{\dagger} + h.c. \right) + \sum_{m\sigma} \lambda_{m\sigma}^{} \left( S_{m\sigma}^{z} + \frac{1}{2} \right) + H_{dens}^{PS} - K_{PS}
\end{aligned}
\right.}
\label{eq:B.23}
\end{equation}

Then, these two hamiltonians \eref{eq:B.23} have to be solved self-consistenly, taking into account the definitions \eref{eq:B.15aa} and \eref{eq:B.22} and the condition given by averaging the constraint in the single-site approximation:

\begin{equation}
 n_{m\sigma} = <S_{m\sigma}^{z}> + \frac{1}{2}
\label{eq:B.25}
\end{equation}

In \eref{eq:B.23}, I have added by hand the onsite energies $\lambda_{m\sigma}^{0}$. In order to recover the tight-binding solution (i.e., when $U=J_H=0$, all $Z_{m\sigma}=1$), the noon-zero Lagrange multipliers $\lambda_{m\sigma}$ that are going to appear when the system is out of half-filling have to be removed. The expression for these corrections $\lambda_{m\sigma}^{0}$ is found to be equal to the non-zero $\lambda_{m\sigma}$ that will appear at $U=J_H=0$, see \aref{AppC}:

\begin{equation}
\lambda_{m\sigma}^{0} = -2 \frac{n_{m\sigma}^0 - 1/2}{n_{m\sigma}^0 (1-n_{m\sigma}^0)} h_{m\sigma}^0
\label{eq:B.26}
\end{equation}

\textcolor{black}{where $n_{m\sigma}^0$ and $h_{m\sigma}^0$ are the non-interacting values of the orbital fillings $n_{m\sigma}$ and the Weiss fields $h_{m\sigma}$.}

\section{SSMF $U(1)$ formulation}
\label{B.3}

Another approach to SSMF is the $U(1)$ formalism, as derived in \cite{YuRPRB862012}. The new pseudospin operators $o_{im\sigma}$ \textcolor{black}{(different from the pseudospin operators $O_{im\sigma}$ defined previously)} are defined as dressed operators:

\begin{equation}
\left\{
\begin{aligned}
d_{im\sigma}^{} & = f_{im\sigma}^{} o_{im\sigma}^{}\\
d_{im\sigma}^{\dagger} & = f_{im\sigma}^{\dagger} o_{im\sigma}^{\dagger}
\end{aligned}
\right. \ \
where \ \ o_{im\sigma}=P_{im\sigma}^{+} S_{im\sigma}^{+} P_{im\sigma}^{-}
\label{eq:B.27}
\end{equation}

\begin{equation}
o_{im\sigma}^{\dagger} = P_{im\sigma}^{+} S_{im\sigma}^{+} P_{im\sigma}^{-}
\label{eq:B.27}
\end{equation}

where:

\begin{equation}
P_{im\sigma}^{\pm} = \frac{1}{\sqrt{1/2 \pm S_{im\sigma}^{z}}}
\label{eq:B.28}
\end{equation}

These operators $o_{im\sigma}$ will play the same role as $O_{im\sigma}$ in the SSMF $Z_2$ formulation. The same \eref{eq:B.23}-\eref{eq:B.25} are obtained by substituting $O_{m\sigma}$ by $o_{m\sigma}$ (note that I have not added any correction $\lambda_{m\sigma}^{0}$ at this point):

\begin{equation}
\small{
\left\{
\begin{aligned}
H_f 		& = \sum_{kmm'\sigma} \sqrt{Z_{m\sigma}} \sqrt{Z_{m'\sigma}} \varepsilon_{kmm'}^{} f_{km\sigma}^{\dagger} f_{km'\sigma}^{} + \sum_{km\sigma} \left( \epsilon_{m}^{} - \mu_{}^{} -\lambda_{m\sigma}^{}  \right) f_{km\sigma}^{\dagger} f_{km\sigma}^{} - K_f \\
H_{PS}	& = \sum_{m\sigma} h_{m\sigma}^{} \left( o_{m\sigma}^{\dagger} + h.c. \right) + \sum_{m\sigma} \lambda_{m\sigma}^{} \left( S_{m\sigma}^{z} + \frac{1}{2} \right) + H_{dens}^{PS} - K_{PS}
\end{aligned}
\right.}
\label{eq:B.29}
\end{equation}

where:

\begin{equation}
\left\{
\begin{aligned}
& Z_{m\sigma} = <o_{m\sigma}>^2 \\
& h_{m\sigma} = \sum_{m'} <o_{m'\sigma}> \sum_k \varepsilon_{kmm'} <f_{km\sigma}^{\dagger} f_{km'\sigma}^{}>
\end{aligned}
\right.
\label{eq:B.30}
\end{equation}

Now, as an additional step, a Taylor expansion of $o$ and $o^{\dagger}$ around $\hat{A} - <\hat{A}>$ (where $\hat{A} = S^z, S^{+}$) can be done, keeping up the linear terms of $\hat{A} - <\hat{A}>$ (see \aref{AppC}). This is possible if the system is not close to the fully-filled or fully-emptied situations: 

\begin{equation}
o_{m\sigma}^{\dagger} \approx O_{m\sigma}^{\dagger} + <O_{m\sigma}^{\dagger}> \eta_{m\sigma} \left[ 2S_{m\sigma}^{z} - \left( 2n_{m\sigma}^f - 1 \right) \right]
\label{eq:B.31}
\end{equation}

where (note that $<o_{m\sigma}> = <O_{m\sigma}>$):

\begin{equation}
\left\{
\begin{aligned}
& O_{m\sigma}^{\dagger} = <P_{m\sigma}^{+}> S_{m\sigma}^{+} <P_{m\sigma}^{-}> \\
& \eta_{m\sigma} = \frac{1}{2} \frac{n_{m\sigma} - 1/2}{n_{m\sigma} \left( 1 - n_{m\sigma} \right)}
\end{aligned}
\right.
\label{eq:B.32}
\end{equation}

\textcolor{black}{The operators $O_{m\sigma}$ are named so due to the similarities (see below) with the $Z_2$ pseudospin operators.} This is a remarkable result: the expression for the pseudospin operators $O_{m\sigma}$ in terms of the so called gauge $c_{m\sigma}$ has been recovered:

\begin{equation}
<P_{m\sigma}^{+}><P_{m\sigma}^{-}> = \frac{1}{\sqrt{\left( <S_{m\sigma}^z> + 1/2 \right) \left( - <S_{m\sigma}^z> + 1/2 \right)}}
\label{eq:B.33}
\end{equation}

So that, $c_{m\sigma}$ are obtained as:

\begin{equation}
\rightarrow <P_{m\sigma}^{+}><P_{m\sigma}^{-}> = \frac{1}{\sqrt{n_{m\sigma} \left( 1-n_{m\sigma} \right)}} = 1 + c_{m\sigma}
\label{eq:B.34}
\end{equation}

By doing this Taylor expansion, an expression which resembles the \eref{eq:B.23} in the SSMF $Z_2$ formalism is obtained:

\begin{equation}
\small{
\left\{
\begin{aligned}
H_f 		& = \sum_{kmm'\sigma} \sqrt{Z_{m\sigma}} \sqrt{Z_{m'\sigma}} \varepsilon_{kmm'}^{} f_{km\sigma}^{\dagger} f_{km'\sigma}^{} + \sum_{km\sigma} \left( \epsilon_{m}^{} - \mu_{}^{} -\lambda_{m\sigma}^{} + \lambda_{m\sigma}^{0}  \right) f_{km\sigma}^{\dagger} f_{km\sigma}^{} - K_f \\
H_{PS}	& = \sum_{m\sigma} h_{m\sigma}^{} \left( O_{m\sigma}^{\dagger} + h.c. \right) + \sum_{m\sigma} \lambda_{m\sigma}^{} \left( S_{m\sigma}^{z} + \frac{1}{2} \right) + H_{dens}^{PS} - K_{PS}
\end{aligned}
\right.}
\label{eq:B.35}
\end{equation}

with:

\begin{equation}
\left\{
\begin{aligned}
K_f 		& = \sum_{m\sigma} \sqrt{Z_{m\sigma}} h_{m\sigma} \\
K_{PS}	& = \frac{U-J_{H}}{2} \sum_{m\sigma} \tilde{S}_{m\sigma}^{z \ 2} + \sum_{m\sigma} \lambda_{m\sigma}^{0} n_{m\sigma}
\end{aligned}
\right.
\label{eq:B.36}
\end{equation}

Interestingly, the expression of the correction \textcolor{black}{$\lambda_{m\sigma}^0$} is similar to the $Z_2$ representation, except for the $\sqrt{Z_{m\sigma}}$ pre-factor which appears in the $U(1)$ formalism:

\begin{equation}
\lambda_{m\sigma}^{0} = -2 \sqrt{Z_{m\sigma}} \frac{n_{m\sigma} - 1/2}{n_{m\sigma} (1-n_{m\sigma})} h_{m\sigma}
\label{eq:B.37}
\end{equation}

\textcolor{black}{In this case, $\lambda_{m\sigma}^0$\footnote{Note that the superindex $0$ here is maintained to make a connection with $Z_2$ SSMF formalism, and not to a fixed $U=0$ value.} depends on the interaction via $Z_{m\sigma}$, $n_{m\sigma}$ and $h_{m\sigma}$, opposite to what happens in the SSMF $Z_2$ formalism.}

\section{Comparison between SSMF $Z_2$ and $U(1)$ formalisms}
\label{B.4}

As it can be seen when comparing \eref{eq:B.23} with \eref{eq:B.35}, and \eref{eq:B.26} with \eref{eq:B.37}, both representations looks really similar.

\begin{figure}[h]
   \centering
   \includegraphics[width=1.0\columnwidth]{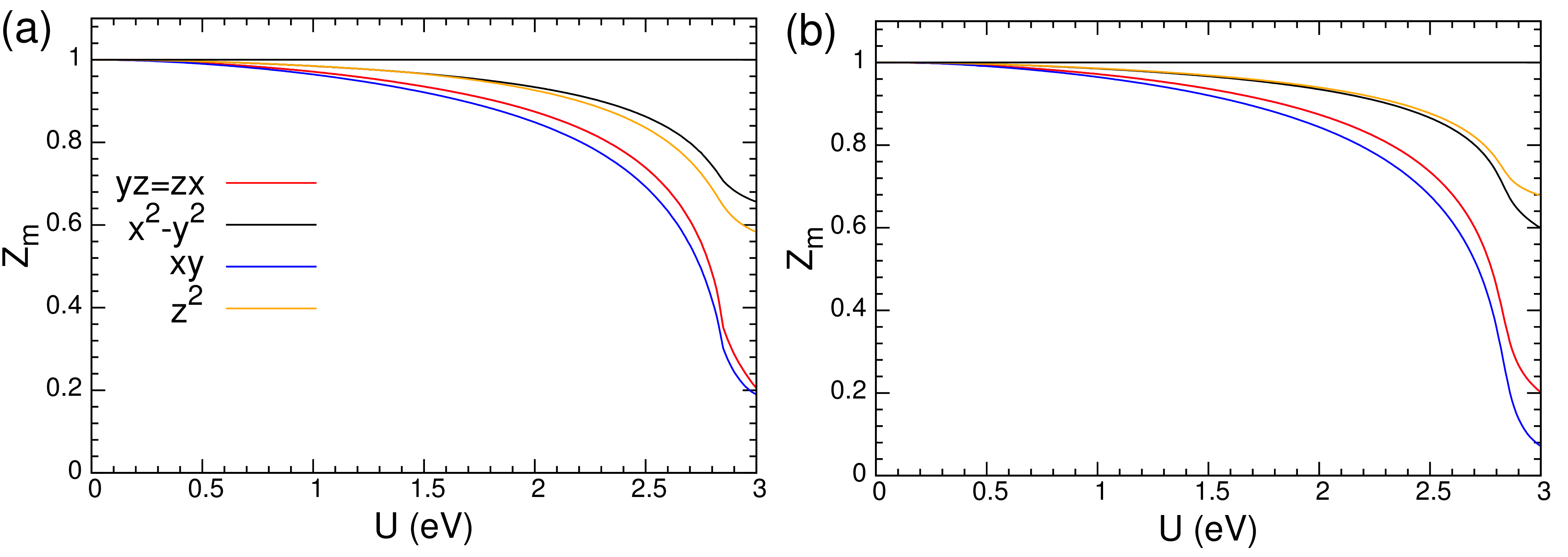}
   \caption[Comparison between SSMF $Z_2$ and $U(1)$ formalisms for undoped $LaFeAsO$ 5 $d$ orbitals tight-binding model at $J_H/U=0.25$]{$Z_m$ in terms of $U$ for $J_H=0.25U$ for the \emph{Graser et al.} \cite{GraNJP112009} 5 $d$ orbitals tight-binding model for $LaFeAsO$, where the filling per atom is $n_e = 6$. \textbf{(a)} $U(1)$ formulation calculation and \textbf{(b)} $Z_2$ formulation calculation with $\tilde{c}_m$ varying with $U$ following \eref{eq:B.8}.}
   \label{fig:B.1}  
\end{figure}

In fact, in the single-site approximation, both formulations give the same \textcolor{black}{expression} for the Weiss term in $H_{PS}$, $O_{m\sigma}^{\dagger} + O_{m\sigma}^{}$. Also, both formulations give the same results for equivalent orbitals systems, \textcolor{black}{due to the equivalent parameters $Z$, $\lambda$ and $\lambda^0$ which vary during the whole process}. The gauge parameters $\tilde{c}_{m\sigma}=1+c_{m\sigma}$ appear in both formulations inside the pseudospin operators $O_{m\sigma}$. $Z_2$ gauges are obtained by calculating it in the non-interacting limit $U=J_H=0$, while for $U(1)$ is obtained for all $U$ values. If $\tilde{c}_{m\sigma}$ are fixed to the value at $U=0$ in the $Z_2$ formulation, the system goes to the unphysical $Z>1$ situation with $U>0$ (see \sref{C.1}). This can be overcomed by varying $\tilde{c}_{m\sigma}$ with $U,J_H$, \fref{fig:B.1}\textbf{(b)}.

The corrections $\lambda_{m\sigma}^{0}$ are different in both techniques, although the final expressions looks similar. $\lambda_{m\sigma}^0$ arise from a Taylor expansion (far from the cases around full-filled and full-emptied fillings) in the $U(1)$ formalism, while they have to be included by hand in the $Z_2$ formalism. Also, the pre-factor $Z_{m\sigma}$ \textcolor{black}{and the interaction-dependent} \eref{eq:B.37} for $U(1)$ formalism makes the corrections being renormalized during the whole proccess of varying $U$ and $J_H$, while the values of \eref{eq:B.26} have not this pre-factor and they are fixed at the non-interacting value $U=J_H=0$. 

The summary of these conclusions can be seen in the \tref{tab:B.1}. In \fref{fig:B.1}, a comparison between different cases can be seen for a 5 $d$ orbitals tight-binding proposed by \emph{Graser et al.} \cite{GraNJP112009} for undoped ($n_e=6$) $LaFeAsO$ at a fixed $J_H/U=0.25$. In this case, both formalisms give \textcolor{black}{similar} results, except for the order of $Z_{z^2}$ and $Z_{x^2-y^2}$ at $U>0$. \textcolor{black}{Compared with DMFT results in the same tight-binding model \cite{IshPRB812010}, the SSMF $U(1)$ formalism compares better than the SSMF $Z_2$ one.}

\begin{table}[H]
\begin{center}
\renewcommand*{\arraystretch}{2}
\begin{tabular}{|c|c|}
\hline
$Z_2$ formalism & $U(1)$ formalism \\
\hline \hline
$O_{m\sigma}^{} = S_{m\sigma}^{-} + c_{m\sigma}^{} S_{m\sigma}^{+}$ & $O_{m\sigma}^{} = <P_{m\sigma}^{+}> S_{m\sigma}^{-} <P_{m\sigma}^{-}>$ \\ 
$O_{m\sigma}^{\dagger} = c_{m\sigma}^{} S_{m\sigma}^{-} + S_{m\sigma}^{+}$ & $O_{m\sigma}^{\dagger} = <P_{m\sigma}^{+}> S_{m\sigma}^{+} <P_{m\sigma}^{-}>$ \\ \hline
\small{$\tilde{c}_{m\sigma}^{} = 1 + c_{m\sigma}^{} = \frac{1}{\sqrt{n_{m\sigma}(1-n_{m\sigma})}} $} & \small{$\tilde{c}_{m\sigma} = <P_{m\sigma}^{+}> <P_{m\sigma}^{-}> = \frac{1}{\sqrt{n_{m\sigma}(1-n_{m\sigma})}}$} \\ \hline
$\lambda_{m\sigma}^{0} = - 2 |h_{m\sigma}| (n_{m\sigma} - \frac{1}{2}) \tilde{c}_{m\sigma}$ & $\lambda_{m\sigma}^{0} = - 2 Z_{m\sigma} |h_{m\sigma}| (n_{m\sigma} - \frac{1}{2}) \tilde{c}_{m\sigma}$ \\ \hline
\end{tabular}
\caption{Comparison between various quantities in the $Z_2$ and $U(1)$ formulations.}
\label{tab:B.1}
\end{center}
\end{table}

\section{How to solve the SSMF formalism}
\label{B.5}

In this section, I will briefly explain how to solve both SSMF $Z_2$ and $U(1)$ formalisms.

\begin{figure}[h]
   \centering
   \includegraphics[width=0.9\columnwidth]{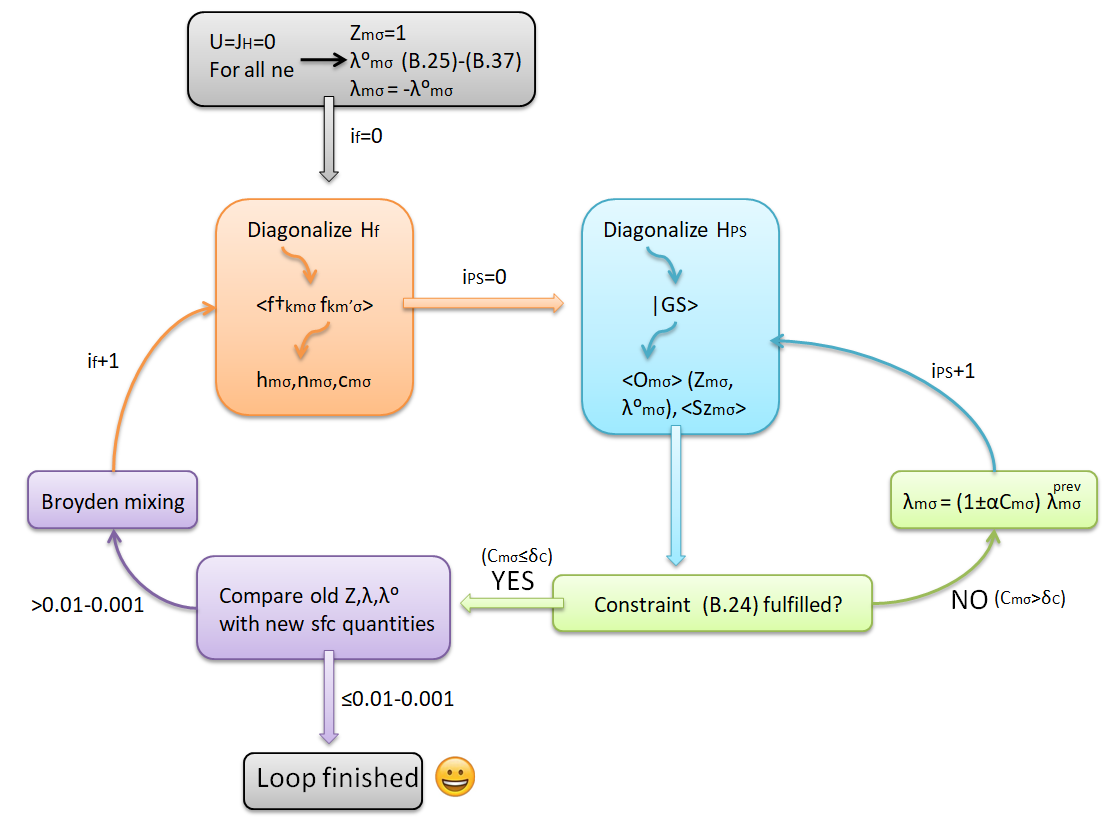}
   \caption[Sketch of the SSMF loop]{Sketch of the SSMF loop for $U(1)$ formalism ($Z_2$ formalism is solved by not updating $\lambda_{m\sigma}^0$ in the loop). $C_{m\sigma}=n_{m\sigma}-(<S_{m\sigma}^z>+1/2)$ has to go to $0$ in order to fulfill the constraint. $\delta_C$ is a dynamical convergence criteria which goes from $10^{-3}$ down to $10^{-6}$ according to $\delta_C^{i_f+1}=\delta_C^{i_f}/(i_f+1)$. Apart from the convergence of $Z_{m\sigma}$, $\lambda_{m\sigma}$ and $\lambda_{m\sigma}^0$, the constraint has to be also fulfilled below $10^{-6}$ at the end of the SSMF loop. See details in main text.}
   \label{fig:B.2a}  
\end{figure}

The two hamiltonians $H_f$ and $H_{PS}$ (see previous sections) have to be solved self-consistenly by fulfilling the constraint (see \eref{eq:B.25}), and by self-consistenly calculating the parameters $Z_{m\sigma}$, $\lambda_{m\sigma}$ and, in the case of the $U(1)$ formalism, also $\lambda_{m\sigma}^0$ (see \eref{eq:B.15aa}, \eqref{eq:B.26} or \eref{eq:B.37}). The SSMF loop procedure can be seen in \fref{fig:B.2a}.

The initial step is done for $U=J_H=0$, where all $Z_{m\sigma}=1$, and $\lambda_{m\sigma}$, $c_{m\sigma}$ and $\lambda_{m\sigma}^0$ can be calculated, see previous equations. $\lambda_{m\sigma}$ are set to be the same as $\lambda_{m\sigma}^0$ (\eref{eq:B.26} or \eref{eq:B.37}) but with opposite sign. Then, we can proceed with the following steps:

\begin{itemize}

\item In the first iteration, $Z_{m\sigma}$, $\lambda_{m\sigma}$ and $\lambda_{m\sigma}^0$ are considered as those obtained for the previous iteration step. When initializing new values of $(U,J_H,n_e)$, the initial $Z_{m\sigma}$, $\lambda_{m\sigma}$ and $\lambda_{m\sigma}^0$ are set to those of the previous $(U,J_H,n_e)$.

\item $H_f$ is diagonalized. The orbital fillings per spin $n_{m\sigma}$ and Weiss fields $h_{m\sigma}$ are calculated, see \eref{eq:B.22} or \eref{eq:B.30}. Using $n_{m\sigma}$, the gauge $c_{m\sigma}$ is also calculated, see \eref{eq:B.8}.

\item Then, the hamiltonian $H_{PS}$ is diagonalized, using the obtained $h_{m\sigma}$ and $c_{m\sigma}$. The expected values $<O_{m\sigma}>$ and $<S_{m\sigma}^z>$ are calculated for the ground state $|GS>$ of $H_{PS}$.

\item New values of $Z_{m\sigma}$ (and $\lambda_{m\sigma}^0$ in $U(1)$ formalism) are calculated, see \eref{eq:B.15aa}.

\item Then, the constraint $C_{m\sigma}=n_{m\sigma}-(<S_{m\sigma}^z>+1/2)$ is checked.

\item If the constraint is not fulfilled below a certain criteria ($|C_{m\sigma}|>\delta_C$), we have to calculate a new $\lambda_{m\sigma}=(1 \pm \alpha C_{m\sigma})\lambda_{m\sigma}^{prev}$, where $\lambda_{m\sigma}^{prev}$ is the Lagrange multiplier from the previous iteration and $\alpha = 0.01-0.5$ depending on the problem\footnote{A smaller $\alpha$ gives a slower evolution, but also smoother, so in situations where the interorbital coupling is important, $\alpha$ should be reduced. The best recipe to chose $\alpha$ is by trying and failing}.  The sign of $\alpha$ will depend on how the constraint evolves in each iteration step $i_{PS}$ (if $C_{m\sigma}$ grows from the previous iteration, then $\alpha$ changes sign). $\delta_C$ is a dynamical quantity which evolves between $\delta_C^{initial}=10^{-3}$ down to $10^{-6}$ with the number of iterations $i_f$, $\delta_C^{i_f+1}=\delta_C^{i_f}/i_f$.

\item Then, the new $\lambda_{m\sigma}$ is introduced in $H_{PS}$. This hamiltonian is diagonalized, and previous step repeats until the constraint is fulfilled.

\item If the constraint is fulfilled ($C_{m\sigma} \leq \delta_C$), the initial and the new calculated $Z_{m\sigma}$, $\lambda_{m\sigma}$ and $\lambda_{m\sigma}^0$ are compared. If the differences are below a certain criteria ($|X^{old} - X^{new}| < 0.01-0.001$ with $X=Z$, $\lambda$ and $\lambda^0$\footnote{A relative criteria $|X^{old} - X^{new}|/|X^{old}|$ can be used instead of the absolute error presented in the main text. This can be done to give a better convergence for $X$, but taking into account that $X$ should not be close enough to $0$.}), and $C_{m\sigma} \leq 10^{-6}$, then the calculation at the corresponding $(U,J_H,n_e)$ is finished. If not, a numerical method, the Broyden scheme \cite{broyden} is applied to mix the initial and new quantities (this mixing varies between $10-50 \%$ for the new variable $X^{new}$ depending on the problem). In this case, $H_f$ is diagonalized again, and the previous steps are repeated.

\end{itemize}

For a given set of parameters $(U,J_H,n_e)$, the run can take from few minutes to various hours. This time will depend on: the number of orbitals and possible degeneracies, the number of $k$ points used to diagonalize $H_f$, precision criterias for the constraint and the quantities $Z$, $\lambda$ and $\lambda^0$. Close to the Hund metal crossover or the Mott insulating transition, the run has to be carefully made, with a better precision and lower mixing quantities (e.g. $\alpha$), hence around this regions the run will take more time to finish.

The biggest system that was treated during this thesis, $BaFe_2S_3$ (where tight-binding model is built on $20$ $d$ orbitals), see \cref{Chap4b}, lasted around three-four weeks to finish. The number of $U$ points taken for fixed $J_H$ and $n_e$ were $\approx 100$. For smallest systems, like $N$ equivalent orbitals systems, SSMF can run from a couple of minutes ($N=1$), up to a few hours ($N=5$) for $\approx 100$ points. For usual $5$ $d$ orbitals models in FeSCs, SSMF can take from 2 to 10 days to finish, also for $\approx 100$ points.



\chapter{\bf Mathematical proofs}
\label{AppC} 
\lhead{Appendix \ref{AppC}: Mathematical proofs}

In this appendix, I am going to present some mathematical proofs which have been pointed out in previous chapters and appendices.

\section{Gauge and correction expressions in SSMF $Z_2$ formalism}
\label{C.1}

\textcolor{black}{Here, I will proof the expressions \eref{eq:B.8} and \eref{eq:B.26} obtained for the gauge $c_{m\sigma}$ and the correction $\lambda_{m\sigma}^0$ in the SSMF $Z_2$ formalism}. I will consider the single-orbital case, for simplicity. I begin by imposing that the non-correlated limit is well reproduced, i.e. when $U=J_H=0$, $Z=1$ for any giving filling per atom and spin $n$.

Taking into account \eref{eq:B.23} for $H_{PS}$:

\begin{equation}
H_{PS} = h \left(O^{\dagger} + h.c. \right) + \lambda \left( S^z + \frac{1}{2} \right)
\label{eq:C.1}
\end{equation}

\textcolor{black}{where $O$ and $S^z$ are $2 \times 2$ spin matrices}. Diagonalizing:

\begin{equation}
|H_{PS} - E \mathbb{I}_{}| = \left| \begin{matrix}
\lambda-E			& 	a \\
a								& 	-E \\
\end{matrix} \right|=0
\label{eq:C.2}
\end{equation}

where $a=h(1+c)$ (see \eref{eq:B.7}) and the ground state is defined as:

\begin{equation}
\begin{aligned}
E_{GS} &= \frac{\lambda}{2} - \sqrt{\frac{\lambda^2}{4}+a^2} = \frac{\lambda}{2} - R\\
|GS> &= \frac{1}{N} \begin{pmatrix}
\frac{\lambda}{2}-R \\
a
\end{pmatrix}
\end{aligned}
\label{eq:C.3}
\end{equation}

Now, the normalization factor $N$ has to be calculated:

\begin{equation}
\begin{aligned}
& \small{<GS|GS>=\frac{1}{N^2} \left[ \left( \frac{\lambda}{2}-R \right)^2 + a^2 \right] = 1 \Longrightarrow} \\
& \small{N^2 = 2 R^2 - \lambda R \rightarrow N = \sqrt{2R(R-\frac{\lambda}{2})}}
\end{aligned}
\label{eq:C.4}
\end{equation}

\textcolor{black}{with $R$ defined in \eref{eq:C.3}. Using this definition, the} expectation values of $S^z$ and $O$ can be calculated as:

\begin{equation}
\small{
\left\{
\begin{aligned}
<S^z> 	& = <GS|S^z|GS> = \frac{1}{2N^2} \left[ \left( \frac{\lambda}{2} - R \right) - a^2 \right] = - \frac{\lambda}{4R} \\
<O> 	& = <GS|O|GS> = \frac{1}{N^2} a \left( 1 + c \right) \left( \frac{\lambda}{2} - R \right) = -\frac{a(1+c)}{2R}
\end{aligned}
\right.}
\label{eq:C.5}
\end{equation}

Now, applying the condition for the constraint given by \eref{eq:B.25}, I obtain an analytical expression for $\lambda$:

\begin{equation}
n-\frac{1}{2}=<S^z>=-\frac{\lambda}{4R}
\label{eq:C.6}
\end{equation}

Note that $\lambda=0$ if the system is at half-filling, i.e. $n=1/2$. In the non-correlated limit:

\begin{equation}
Z=1=<O>^2=\frac{a^2(1+c)^2}{4R^2}
\label{eq:C.7}
\end{equation}

Now, I square \eref{eq:C.6}, then divide both \eref{eq:C.6} $\div$ \eref{eq:C.7}, restoring the expression of $R$, and obtaining:

\begin{equation}
\frac{\lambda^2}{a^2} = 4 \left( 1+c \right)^2 \left( n-\frac{1}{2} \right)^2
\label{eq:C.8}
\end{equation}

The expression extracted from \eref{eq:C.8} for $\lambda$ will be used to obtain the correction $\lambda^0$ in \eref{eq:B.26}. This is due to the fact that $\lambda$ and $\lambda^0$ has to be equal (with a different sign) at $U=J_H=0$ in order to cancel each other and reproduce the non-correlated limit, see \aref{AppB}. Otherwise (generalizing to the multiorbital case), $\lambda_m \neq 0$ out of half-filling, resulting in unphysical additional crystal field splittings at $U=J_H=0$.

\textcolor{black}{Substituting the value of $R$ in \eref{eq:C.3}, and using \eref{eq:C.8},} the expression for the gauge $c$ \textcolor{black}{can be obtained} in terms of the filling $n$, which is the expression that I wanted to demonstrate:

\begin{equation}
\begin{aligned}
&\small{\left( 1+c \right)^2 = \frac{\lambda^2}{a^2} + 4 = 4 \left( 1+c \right)^2 \left( n-\frac{1}{2} \right)^2 + 4 \Longrightarrow} \\
&\small{\left( 1+c \right)^2 = \frac{1}{n(1-n)} \rightarrow c = \frac{1}{\sqrt{n(1-n)}} - 1}
\end{aligned}
\label{eq:C.9}
\end{equation}

In general, for a given number of orbitals $m$ it can be shown, by using this procedure, that the results for $c_{m\sigma}$ and $\lambda_{m\sigma}^0$ for a given orbital filling per atom, spin and orbital $n_{m\sigma}$ are:

\begin{equation}
\left\{
\begin{aligned}
c_{m\sigma} & = \frac{1}{\sqrt{n_{m\sigma}(1-n_{m\sigma})}} - 1 \\
\lambda_{m\sigma}^0 & = -2 |h_{m\sigma}| (n_{m\sigma}-\frac{1}{2}) (1+c_{m\sigma})
\end{aligned}
\right.
\label{eq:C.10}
\end{equation}

\begin{figure}[h]
   \centering
   \includegraphics[width=0.8\columnwidth]{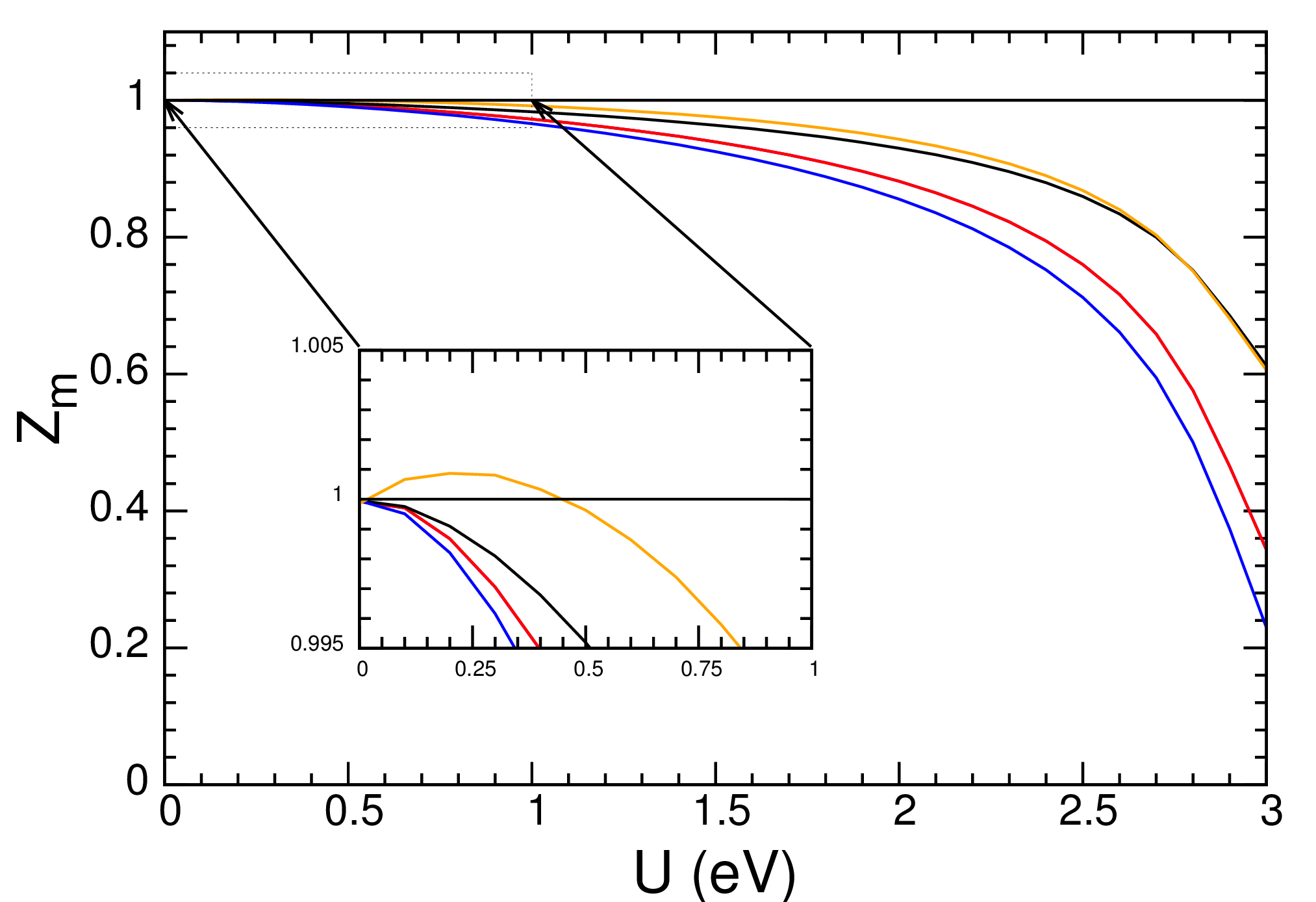}
   \caption[SSMF $Z_2$ formalism for undoped $LaFeAsO$ 5 $d$ orbitals tight-binding model at $J_H/U=0.25$ and fixed gauge $c_{m\sigma}$ at the non-interacting value]{$Z_m$ in terms of $U$ for $J_H=0.25U$ for the \emph{Graser et al.} \cite{GraNJP112009} 5 $d$ orbitals tight-binding model for $LaFeAsO$, where the filling per atom is $n_e = 6$. $Z_2$ formulation with $\tilde{c}_m$ fixed to $U=J_H=0$ value, \textsc{Inset} shows a zoom into the unphysical solution $Z>1$ for $z^2$.}
   \label{fig:C.1}  
\end{figure}

As shown here, the expressions in \eref{eq:C.10} are obtained for the non-correlated limit, $U=J_H=0$. There is no justification to consider that this quantities should vary with non-zero $U$ and $J_H$. However, not doing so in the case of the gauge will produce an unphysical $Z>1$ for $U>0$, as shown in \fref{fig:C.1}.

During my thesis, I found the aforementioed analytical expression for $\lambda_{m\sigma}^0$. It coincides with previous calculations in which $\lambda_{m\sigma}^0$ was self-consistenly calculated in a first step, hence the analytical formula permits to save an important time when initializing the SSMF calculations.

\section{Taylor expansion in $U(1)$ SSMF formulation}
\label{C.2}

Let's see now how to obtain the \eref{eq:B.31} - \eref{eq:B.34}. I am going to leave out all the indices (note that this derivation is the same if we include such indices). Then:

\begin{equation}
o^\dagger = P^{+} S^{+} P^{-}
\label{eq:C.11}
\end{equation}

By taking the expression of $P^{\pm}$ and applying a typical mean-field approach $S^z = <S^z> + \delta S^z \longleftrightarrow \delta S^z = S^z - <S^z>$, I obtain:

\begin{equation}
\begin{aligned}
P^{\pm} & = \frac{1}{\sqrt{1/2 \pm S^{z}}} = \frac{1}{\sqrt{1/2 \pm (<S^{z}> + \delta S^z)}} \\
	& = \frac{1}{\sqrt{1/2 \pm <S^{z}>}} \frac{1}{\sqrt{1 \pm \frac{\delta S^{z}}{1/2 \pm <S^z>}}}
\end{aligned}
\label{eq:C.12}
\end{equation}

If the system is far from the fully-filled or fully-emptied situations, $\frac{1}{2} \pm <S^z> \sim \pm 1$, then $\frac{\delta S^z}{1/2 \pm S^z} << 1$ (because $\delta S^z << 1$). Now, a Taylor expansion can be performed:

\begin{equation}
\frac{1}{\sqrt{1 \pm x}} \approx 1 \mp \frac{1}{2} x + \frac{3}{8} x^2 - \cdots
\label{eq:C.13}
\end{equation}

Keeping up to linear terms:

\begin{equation}
\frac{1}{\sqrt{1 \pm \frac{\delta S^z}{\frac{1}{2} \pm <S^z>}}} \approx 1 \mp \frac{1}{2} \frac{\delta S^z}{1/2 \ \pm <S^z>}
\label{eq:C.14}
\end{equation}

An important result is obtained for $<P^{\pm}>$:

\begin{equation}
\begin{aligned}
<P^{\pm}> & = \langle \frac{1}{\sqrt{1/2 \pm <S^{z}>}} \frac{1}{\sqrt{1 \pm \frac{\delta S^{z}}{1/2 \pm <S^z>}}} \rangle \\
	& = \frac{1}{\sqrt{1/2 \pm <S^{z}>}} \langle \left( 1 \mp \frac{1}{2} \frac{\delta S^z}{1/2 \ \pm <S^z>} \right) \rangle \\
	& = \frac{1}{\sqrt{1/2 \pm <S^{z}>}} \mp \frac{1}{\sqrt{1/2 \pm <S^{z}>}} \frac{1}{2} \frac{<\delta S^z>}{1/2 \ \pm <S^z>} \\
	& = \frac{1}{\sqrt{1/2 \pm <S^{z}>}}
\end{aligned}
\label{eq:C.15}
\end{equation}

where I have used that $<\delta S^z> = \langle S^z - <S^z> \rangle = 0$. Also, note that, by using \eref{eq:B.33}:

\begin{equation}
\left\{
\begin{aligned}
<P^{+}> & = \frac{1}{\sqrt{n}} \\
<P^{-}> & = \frac{1}{\sqrt{1-n}}
\end{aligned}
\right.
\label{eq:C.16}
\end{equation}

Then (considering that ${\delta S^{z}}^2 \approx 0$ and $\delta S^z \delta S^{+} \approx 0$):

\begin{equation}
\small{
\begin{aligned}
o^\dagger & \approx <P^{+}> \left( 1 - \frac{1}{2} <P^{+}>^2 \delta S^z \right) S^{+} <P^{-}> \left( 1 + \frac{1}{2} <P^{-}>^2 \delta S^z \right) \\
	& = <P^{+}> S^{+} <P^{-}> + <P^{+}> S^{+} <P^{-}> \frac{1}{2} \delta S^z \left( - <P^{+}>^2 + <P^{-}>^2 \right) \\
	& = <P^{+}> S^{+} <P^{-}> + <P^{+}> <S^{+}> <P^{-}> \frac{1}{2} \left( S^z - <S^z> \right) \left( - \frac{1}{n} + \frac{1}{1-n} \right) \\
	& = O^\dagger + <O^\dagger> \eta \left[ 2S^z - \left( 2n - 1 \right) \right]
\end{aligned}}
\label{eq:C.17}
\end{equation}

where I have defined:

\begin{equation}
\left\{
\begin{aligned}
O^\dagger & = <P^{+}> S^{+} <P^{-}> \\
\eta & = \frac{1}{2} \frac{n-1/2}{n \left( 1-n \right)}
\end{aligned}
\right.
\label{eq:C.18}
\end{equation}

And the expressions from \eref{eq:B.31} to \eref{eq:B.34} are obtained.


\section{Implementation of the Zeeman field effect in SSMF}
\label{C.3}

In this section, I will show the implementation of the Zeeman field term presented in \cref{Chap4c}, in the SSMF formalism. I include the Zeeman magnetic field term in the Hubbard-Kanamori hamiltonian of \eref{eq:1.22}:

\begin{equation}
H=H_0 + H_{dens} + H_{add} + \underbrace{H\sum_{im} \left( d_{im\uparrow}^{\dagger}d_{im\uparrow}^{} - d_{im\downarrow}^{\dagger}d_{im\downarrow}^{} \right)}_{H_{Zeeman}}
\label{eq:C.27}
\end{equation}

where $H_0$ is the tight-binding, $H_{dens}$ is the density-density interaction part and $H_{add}$ encodes the pair-hopping and spin-flip terms. The Zeeman field $H$ breaks the spin degeneracy, moving one spin $\uparrow$ upwards and the other one $\downarrow$ downwards. The majority spin band is labeled with $\downarrow$, and the minority spin band with $\uparrow$.

In the SSMF formalism, $H_{Zeeman}$ is written in the basis of auxiliary fermion operators:

\begin{equation}
H_{Zeeman} = H \sum_{im} \left( \hat{n}_{im\uparrow}^f - \hat{n}_{im\downarrow}^f \right)
\label{eq:C.28}
\end{equation}

where $\hat{n}_{im\sigma}^f = f_{im\sigma}^\dagger f_{im\sigma}$. Then, $H_{Zeeman}$ has to be added in the $H_f$ of \eref{eq:B.23} or \eref{eq:B.35} as a spin-dependent onsite energy shift. In this situation, all the parameters $Z_{m\sigma}$, $\lambda_{m\sigma}$, etc. will depend explicitly on the spin index $\sigma$.

The implementation of the Zeeman field effect on SSMF is constraint to a specific regime. The discussion of next paragraphs works for equivalent orbitals systems which are out of half-filling and for small $J_H$. In \cref{Chap4c}, I precisely study this regime for a $N=2$ equivalent orbitals system at quarter filling and $J_H=0$.

I will consider that the total filling of the system is less than half-filling, and drop out the orbital index. Thus, when the Zeeman field $H$ increases, the majority spin band is progressively filling, while the minority spin band is getting emptied (approaching $0$ filling). This poses a problem when $n_{\sigma}$ approaches $0$ (band totally emptied) or to $1$ (band totally filled). In this situation, $Z_{\sigma}$ cannot be defined anymore for the band which is evolving to the totally filled/emptied situation, see \sref{1.2.2}. By following the numerical evolution of $h_{\sigma}$ and $c_{\sigma}$ for increasing $H$, it can be shown that when $n_{\sigma} \rightarrow 0$ or $1$, $a_{\sigma}=h_{\sigma}(1+c_{\sigma}) \rightarrow 0$, and $H_{PS}$ is a block diagonal matrix. In the case of a single-orbital system:

\begin{equation}
H_{PS} = \left( \begin{matrix}
\lambda_\uparrow+\lambda_\downarrow+U			& 	a_\downarrow	&	0	&	0 \\
a_\downarrow			& 	\lambda_\uparrow	&	0	&	0 \\
0			& 	0	&	\lambda_\downarrow	&	a_\downarrow \\
0			& 	0	&	a_\downarrow	&	0
\end{matrix} \right)
\label{eq:C.29}
\end{equation}

Interestingly, the ground state $|GS>$ of \eref{eq:C.29}, only depends on the parameters of the spin band $\downarrow$ (bottom right $2 \times 2$ block), i.e. the band which is not getting totally emptied. A similar demonstration can be done for $N=2$ equivalent orbitals (not shown).


I will then proceed with the implementation of $H_{Zeeman}$ in the SSMF calculations as follows: once one of the spin bands is totally emptied or filled, both spin channels are decoupled by neglecting the evolution of the spin band which is getting totally emptied/filled. This is done by simply re-formulating $H_f$ and $H_{PS}$ with respect to the spin band which remains in a finite filling. This spin decoupling has to be done in order to overcome possible oscillations in the filling of the totally empty/fill spin band, hence in all the other SSMF parameters.

\section{Propagators in SSMF}
\label{C.4}

Let's consider a single-orbital system in the local PM metal state. By taking the operators representation of \eref{eq:B.7} and the mean-field treatment, and the time-ordered propagators in the real space are:

\begin{equation}
G^d (ij,t) = G^f (ij,t) G^{PS} (ij,t)
\label{eq:C.19}
\end{equation}

where the superindices accounts for the different operators, $ij$ is the relative position of the electron, $t$ is the time and:

\begin{equation}
\left\{
\begin{aligned}
G^d (ij,t) & = -i{<T d_{i\sigma} (t) d_{j\sigma}^\dagger (0)>}^d \\
G^f (ij,t) & = -i{<T f_{i\sigma} (t) f_{j\sigma}^\dagger (0)>}^f \\
G^{PS} (ij,t) & = {<T O_{i\sigma} (t) O_{j\sigma}^\dagger (0)>}^{PS} 
\end{aligned}
\right.
\label{eq:C.20}
\end{equation}

These expressions can be Fourier transformed into $G^d(k,\omega)$, where $k$ is the reciprocal space vector and $\omega$ is the frequency. $G^d(k,\omega)$ is found as the convolution of $G^f (k,\omega)$ and $G^{PS} (k,\omega)$:

\begin{equation}
G^d (k,\omega) = \int \frac{d\omega '}{2\pi} \frac{d^3 \vec{k} '}{8\pi^3} G^f (k ',\omega ') G^{PS} (k-k ',\omega - \omega ')
\label{eq:C.21}
\end{equation}

After the second mean-field of \eref{eq:B.15}, it can be seen that pseudospins are long-range ordered with $<O_{i\sigma}^\dagger><O_{j\sigma}> \neq 0$ in the metallic phase. Then, at low frequencies $\omega$, as required in FLT, $G^{PS} (k,\omega)$ can be divided into a small and large $k$ and $\omega$ contributions:

\begin{equation}
G^{PS} (k,\omega) = Z_k \delta (k) \delta (\omega) + G^{PS, \: incoh} (k,\omega)
\label{eq:C.22}
\end{equation}

where the parameter $Z_k$ is defined as the FT of $<O_{i\sigma}^\dagger><O_{j\sigma}>$, and $G^{PS, \: incoh} (k,\omega)$ as the incoherent part, see \sref{1.2.2}. In a first approximation, $G^{PS, \: incoh} (k,\omega)$ can be neglected and:

\begin{equation}
G^d (k,\omega) = Z_k G^f (k,\omega)
\label{eq:C.23}
\end{equation}

$H_f$ describes a non-interacting hamiltonian with a renormalized band structure. Hence, the Dyson equation for $G^f (k,\omega)$ can be written (note that $\Sigma^f (k,\omega) = 0$ in a non-interacting hamiltonian):

\begin{equation}
G^f (k,\omega) = \frac{1}{\omega - Z_k \varepsilon_k + \mu + \lambda_k - \lambda_k^0}
\label{eq:C.24}
\end{equation}

where $\lambda_k$ and $\lambda_k^0$ are the FT Lagrange multipliers and correction energy shifts. Taking into account \eref{eq:C.23} and \eref{eq:C.24}:

\begin{equation}
G^d (k,\omega) = \frac{Z_k}{\omega - Z_k \varepsilon_k + \mu + \lambda_k - \lambda_k^0}
\label{eq:C.25}
\end{equation}

Using the definition \eref{eq:1.9} for $G^d (k,\omega)$, the self-energy $\Sigma(k,\omega)$ of the interacting hamiltonian $H$ can be obtained as:

\begin{equation}
\Sigma(k,\omega) = \underbrace{\mu - \frac{\mu+\lambda_k - \lambda_k^0}{Z_k}}_{Re \, \Sigma(k,0)} + \omega \left( 1 - \frac{1}{Z_k} \right)
\label{eq:C.26}
\end{equation}

Hence, the linear-$\omega$ dependence of the $\Sigma(k,\omega)$ in the FLT is recovered if $Z_k$ is defined as the FT of $<O_{i\sigma}^\dagger><O_{j\sigma}>$, as anticipated. In the single-site approximation for a PM system, this condition reduces to $<O>^2=Z$.


\end{appendices}
\addtocontents{toc}{\vspace{1em}} 

\clearpage  





\addtotoc{List of publications}  
\publications{
\addtocontents{toc}{\vspace{1em}}  

\begin{enumerate}

\item Strong correlations and the search for high-$T_c$ superconductivity in chromium pnictides and chalcogenides\\
{\bf J. M. Pizarro}, M. J. Calder\'{o}n, J. Liu, M. C. Mu\~noz and E. Bascones\\
\href{https://journals.aps.org/prb/abstract/10.1103/PhysRevB.95.075115/}{ Phys. Rev. B {\bf 95}, 075115 (2017)}\\

 \item Strong electronic correlations and Fermi surface reconstruction in the quasi-one dimensional iron superconductor $Ba{Fe}_2S_{3}$\\
 {\bf J. M. Pizarro} and Elena Bascones\\
\href{https://journals.aps.org/prmaterials/abstract/10.1103/PhysRevMaterials.3.014801}{Phys. Rev. Materials {\bf 3}, 014801 (2018)}

 \item The nature of correlations in the insulating states of twisted bilayer graphene\\
{\bf J. M. Pizarro}, M. J. Calder\'{o}n and E. Bascones\\
\href{https://arxiv.org/abs/1805.07303/}{arXiv 1805.07303 (2018)}

\end{enumerate}
}
\clearpage  
\renewcommand\bibname{{\bf References}} 
\addtocontents{toc}{\vspace{0em}}  
\backmatter
\label{References}
\lhead{{ References}}  
\bibliographystyle{unsrtnat}  
\bibliography{thesisbib}  

\begin{thebibliography}{226}
\providecommand{\natexlab}[1]{#1}
\providecommand{\url}[1]{\texttt{#1}}
\expandafter\ifx\csname urlstyle\endcsname\relax
  \providecommand{\doi}[1]{doi: #1}\else
  \providecommand{\doi}{doi: \begingroup \urlstyle{rm}\Url}\fi

\bibitem[Ashcroft and Mermin(1976)]{AshcroftMermin1976}
Neil~W. Ashcroft and N.~Mermin.
\newblock \emph{Solid {$\text{S}$}tate {$\text{P}$}hysics}.
\newblock Cengage Learning, 1976.
\newblock ISBN 978-0030839931.
\newblock URL
  \url{https://cengage.com.au/product/title/solid-state-physics/isbn/9780030839931}.

\bibitem[Raghu et~al.(2008)Raghu, Qi, Liu, Scalapino, and Zhang]{RagPRB772008}
S.~Raghu, Xiao-Liang Qi, Chao-Xing Liu, D.~J. Scalapino, and Shou-Cheng Zhang.
\newblock Minimal two-band model of the superconducting iron oxypnictides.
\newblock \emph{Phys. Rev. B}, 77:\penalty0 220503, 2008.
\newblock \doi{10.1103/PhysRevB.77.220503}.
\newblock URL \url{https://link.aps.org/doi/10.1103/PhysRevB.77.220503}.

\bibitem[Jenkins(1954)]{JenPh201954}
D.P. Jenkins.
\newblock The electronic band structure of silicon.
\newblock \emph{Physica}, 20\penalty0 (7):\penalty0 967 -- 970, 1954.
\newblock \doi{https://doi.org/10.1016/S0031-8914(54)80206-0}.

\bibitem[Coleman(2015)]{Coleman2015}
Piers Coleman.
\newblock \emph{Introduction to Many-Body Physics}.
\newblock Cambridge University Press, 2015.
\newblock \doi{10.1017/CBO9781139020916}.

\bibitem[Bascones et~al.(2017)Bascones, Calder\'on, Aguado, Valenzuela,
  Cortijo, and Rold\'an]{emergence2017}
E.~Bascones, M.J. Calder\'on, R.~Aguado, B.~Valenzuela, A.~Cortijo, and
  R.~Rold\'an.
\newblock Emergence of quantum phases in novel materials.
\newblock Teor\'ia y Simulaci\'on de Materiales, Instituto de Ciencia de
  Materiales de Madrid, ICMM-CSIC, Sep 2017.

\bibitem[Liu et~al.(2015)Liu, Yi, Zhang, Hu, Yu, Zhu, He, Chen, Hashimoto,
  Moore, Mo, Hussain, Si, Mao, Lu, and Shen]{LiuPRB922015}
Z.~K. Liu, M.~Yi, Y.~Zhang, J.~Hu, R.~Yu, J.-X. Zhu, R.-H. He, Y.~L. Chen,
  M.~Hashimoto, R.~G. Moore, S.-K. Mo, Z.~Hussain, Q.~Si, Z.~Q. Mao, D.~H. Lu,
  and Z.-X. Shen.
\newblock Experimental observation of incoherent-coherent crossover and
  orbital-dependent band renormalization in iron chalcogenide superconductors.
\newblock \emph{Phys. Rev. B}, 92:\penalty0 235138, 2015.
\newblock \doi{10.1103/PhysRevB.92.235138}.
\newblock URL \url{https://link.aps.org/doi/10.1103/PhysRevB.92.235138}.

\bibitem[Pickett(1989)]{PicRMP611989}
Warren~E. Pickett.
\newblock Electronic structure of the high-temperature oxide superconductors.
\newblock \emph{Rev. Mod. Phys.}, 61:\penalty0 433--512, 1989.
\newblock \doi{10.1103/RevModPhys.61.433}.
\newblock URL \url{https://link.aps.org/doi/10.1103/RevModPhys.61.433}.

\bibitem[Fazekas(1999)]{Fazekas1999}
Patrik Fazekas.
\newblock \emph{Vol. 5: Lecture Notes on Electron Correlation and Magnetism}.
\newblock WORLD SCIENTIFIC, 1999.
\newblock ISBN 978-981-02-2474-5.
\newblock URL \url{https://www.worldscientific.com/doi/abs/10.1142/2945}.

\bibitem[Bascones et~al.(2016)Bascones, Valenzuela, and
  Calder\'on]{BasCRP172016}
Elena Bascones, Bel\'en Valenzuela, and Mar\'ia~Jos\'e Calder\'on.
\newblock Magnetic interactions in iron superconductors: A review.
\newblock \emph{Comptes Rendus Physique}, 17\penalty0 (1):\penalty0 36 -- 59,
  2016.
\newblock \doi{https://doi.org/10.1016/j.crhy.2015.05.004}.
\newblock URL
  \url{http://www.sciencedirect.com/science/article/pii/S1631070515000924}.

\bibitem[Rold\'an et~al.(2017)Rold\'an, Chirolli, Prada, Silva-Guill\'en,
  San-Jose, and Guinea]{RolCSR462017}
Rafael Rold\'an, Luca Chirolli, Elsa Prada, Jose~Angel Silva-Guill\'en, Pablo
  San-Jose, and Francisco Guinea.
\newblock Theory of 2{$\text{D}$} crystals: graphene and beyond.
\newblock \emph{Chem. Soc. Rev.}, 46:\penalty0 4387--4399, 2017.
\newblock \doi{10.1039/C7CS00210F}.
\newblock URL \url{http://dx.doi.org/10.1039/C7CS00210F}.

\bibitem[Sch\"onhoff(2017)]{Schonhoff2017thesis}
Gunnar Sch\"onhoff.
\newblock \emph{Coulomb interaction and phonons in doped semiconducting and
  metallic two-dimensional materials}.
\newblock PhD thesis, Universit\"at Bremen, 2017.

\bibitem[Atland and Simons(2010)]{AtlandSimons2010}
Alexander Atland and Ben~D. Simons.
\newblock \emph{Condensed {$\text{M}$}atter {$\text{F}$}ield
  {$\text{T}$}heory}.
\newblock Cambridge University Press, 2010.
\newblock ISBN 9780521769754.
\newblock URL
  \url{https://www.cambridge.org/de/academic/subjects/physics/condensed-matter-physics-nanoscience-and-mesoscopic-physics/condensed-matter-field-theory-2nd-edition?format=HB&isbn=9780521769754}.

\bibitem[Ferber(2012)]{Ferber2012thesis}
Johannes Ferber.
\newblock \emph{Density Functional Theory and Dynamical Mean Field Theory:
  Applications to Correlated Materials}.
\newblock PhD thesis, Johann Wolfgang Goethe-Universit\"at, Frankfurt am Main,
  2012.

\bibitem[Kune$\check{s}$ et~al.(2010)Kune$\check{s}$, Arita, Wissgott, Toschi,
  Ikeda, and Held]{KunCPC1812010}
Jan Kune$\check{s}$, Ryotaro Arita, Philipp Wissgott, Alessandro Toschi,
  Hiroaki Ikeda, and Karsten Held.
\newblock Wien2wannier: From linearized augmented plane waves to maximally
  localized {$\text{W}$}annier functions.
\newblock \emph{Computer Physics Communications}, 181\penalty0 (11):\penalty0
  1888 -- 1895, 2010.
\newblock \doi{https://doi.org/10.1016/j.cpc.2010.08.005}.
\newblock URL
  \url{http://www.sciencedirect.com/science/article/pii/S0010465510002948}.

\bibitem[{de' Medici}(2017)]{deM1707.03282}
L.~{de' Medici}.
\newblock Hund's metals, explained.
\newblock \emph{ArXiv e-prints}, 2017.
\newblock URL \url{https://arxiv.org/abs/1707.03282}.

\bibitem[Gross and Bloch(2017)]{GroS3572017}
Christian Gross and Immanuel Bloch.
\newblock Quantum simulations with ultracold atoms in optical lattices.
\newblock \emph{Science}, 357\penalty0 (6355):\penalty0 995--1001, 2017.
\newblock \doi{10.1126/science.aal3837}.
\newblock URL \url{http://science.sciencemag.org/content/357/6355/995}.

\bibitem[Georges et~al.(1996)Georges, Kotliar, Krauth, and
  Rozenberg]{GeoRMP681996}
Antoine Georges, Gabriel Kotliar, Werner Krauth, and Marcelo~J. Rozenberg.
\newblock Dynamical mean-field theory of strongly correlated fermion systems
  and the limit of infinite dimensions.
\newblock \emph{Rev. Mod. Phys.}, 68:\penalty0 13--125, 1996.
\newblock \doi{10.1103/RevModPhys.68.13}.
\newblock URL \url{https://link.aps.org/doi/10.1103/RevModPhys.68.13}.

\bibitem[{Sekiyama} et~al.(2002){Sekiyama}, {Fujiwara}, {Imada}, {Eisaki},
  {Uchida}, {Takegahara}, {Harima}, {Saitoh}, and {Suga}]{Sek0206471}
A.~{Sekiyama}, H.~{Fujiwara}, S.~{Imada}, H.~{Eisaki}, S.~I. {Uchida},
  K.~{Takegahara}, H.~{Harima}, Y.~{Saitoh}, and S.~{Suga}.
\newblock Genuine electronic states of vanadium perovskites revealed by
  high-energy photoemission.
\newblock \emph{eprint arXiv:cond-mat/0206471}, 2002.
\newblock URL \url{https://arxiv.org/abs/cond-mat/0206471v2}.

\bibitem[Park et~al.(2008)Park, Haule, and Kotliar]{ParPRL1012008}
H.~Park, K.~Haule, and G.~Kotliar.
\newblock Cluster {$\text{D}$}ynamical {$\text{M}$}ean {$\text{F}$}ield
  {$\text{T}$}heory of the {$\text{M}$}ott transition.
\newblock \emph{Phys. Rev. Lett.}, 101:\penalty0 186403, 2008.
\newblock \doi{10.1103/PhysRevLett.101.186403}.
\newblock URL \url{https://link.aps.org/doi/10.1103/PhysRevLett.101.186403}.

\bibitem[Vu\ifmmode \check{c}\else \v{c}\fi{}i\ifmmode \check{c}\else
  \v{c}\fi{}evi\ifmmode~\acute{c}\else \'{c}\fi{} et~al.(2013)Vu\ifmmode
  \check{c}\else \v{c}\fi{}i\ifmmode \check{c}\else
  \v{c}\fi{}evi\ifmmode~\acute{c}\else \'{c}\fi{}, Terletska,
  Tanaskovi\ifmmode~\acute{c}\else \'{c}\fi{}, and
  Dobrosavljevi\ifmmode~\acute{c}\else \'{c}\fi{}]{VucPRB882013}
J.~Vu\ifmmode \check{c}\else \v{c}\fi{}i\ifmmode \check{c}\else
  \v{c}\fi{}evi\ifmmode~\acute{c}\else \'{c}\fi{}, H.~Terletska,
  D.~Tanaskovi\ifmmode~\acute{c}\else \'{c}\fi{}, and
  V.~Dobrosavljevi\ifmmode~\acute{c}\else \'{c}\fi{}.
\newblock Finite-temperature crossover and the quantum {$\text{W}$}idom line
  near the {$\text{M}$}ott transition.
\newblock \emph{Phys. Rev. B}, 88:\penalty0 075143, 2013.
\newblock \doi{10.1103/PhysRevB.88.075143}.
\newblock URL \url{https://link.aps.org/doi/10.1103/PhysRevB.88.075143}.

\bibitem[Sch\"afer et~al.(2015)Sch\"afer, Geles, Rost, Rohringer, Arrigoni,
  Held, Bl\"umer, Aichhorn, and Toschi]{SchPRB912015}
T.~Sch\"afer, F.~Geles, D.~Rost, G.~Rohringer, E.~Arrigoni, K.~Held,
  N.~Bl\"umer, M.~Aichhorn, and A.~Toschi.
\newblock Fate of the false {$\text{M}$}ott-{$\text{H}$}ubbard transition in
  two dimensions.
\newblock \emph{Phys. Rev. B}, 91:\penalty0 125109, 2015.
\newblock \doi{10.1103/PhysRevB.91.125109}.
\newblock URL \url{https://link.aps.org/doi/10.1103/PhysRevB.91.125109}.

\bibitem[Dagotto(2002)]{Dagotto2002}
Elbio Dagotto.
\newblock \emph{Nanoscale phase separation and colossal magnetoresistance}.
\newblock Springer-Verlag, 2002.
\newblock ISBN 978-3-662-05244-0.
\newblock URL \url{https://www.springer.com/de/book/9783540432456}.

\bibitem[Castellani et~al.(1978)Castellani, Natoli, and
  Ranninger]{CasPRB181978}
C.~Castellani, C.~R. Natoli, and J.~Ranninger.
\newblock Metal-insulator transition in pure and cr-doped
  ${\mathrm{v}}_{2}$${\mathrm{o}}_{3}$.
\newblock \emph{Phys. Rev. B}, 18:\penalty0 5001--5013, 1978.
\newblock URL \url{https://link.aps.org/doi/10.1103/PhysRevB.18.5001}.

\bibitem[Miyake et~al.(2010)Miyake, Nakamura, Arita, and Imada]{MiyJPSJ792010}
Takashi Miyake, Kazuma Nakamura, Ryotaro Arita, and Masatoshi Imada.
\newblock Comparison of ab initio low-energy models for lafepo, lafeaso,
  bafe2as2, lifeas, fese, and fete: Electron correlation and covalency.
\newblock \emph{Journal of the Physical Society of Japan}, 79\penalty0
  (4):\penalty0 044705, 2010.
\newblock \doi{10.1143/JPSJ.79.044705}.
\newblock URL \url{https://journals.jps.jp/doi/abs/10.1143/JPSJ.79.044705}.

\bibitem[Yu and Si(2012)]{YuRPRB862012}
Rong Yu and Qimiao Si.
\newblock {$U(1)$} slave-spin theory and its application to {$\text{M}$}ott
  transition in a multiorbital model for iron pnictides.
\newblock \emph{Phys. Rev. B}, 86:\penalty0 085104, 2012.
\newblock \doi{10.1103/PhysRevB.86.085104}.
\newblock URL \url{https://link.aps.org/doi/10.1103/PhysRevB.86.085104}.

\bibitem[Werner et~al.(2008)Werner, Gull, Troyer, and Millis]{WerPRL1012008}
Philipp Werner, Emanuel Gull, Matthias Troyer, and Andrew~J. Millis.
\newblock Spin freezing transition and non-{$\text{F}$}ermi-liquid self-energy
  in a three-orbital model.
\newblock \emph{Phys. Rev. Lett.}, 101:\penalty0 166405, 2008.
\newblock \doi{10.1103/PhysRevLett.101.166405}.
\newblock URL \url{https://link.aps.org/doi/10.1103/PhysRevLett.101.166405}.

\bibitem[Haule and Kotliar(2009)]{HauNJP112009}
K~Haule and G~Kotliar.
\newblock Coherence–incoherence crossover in the normal state of iron
  oxypnictides and importance of {$\text{H}$}und's rule coupling.
\newblock \emph{New Journal of Physics}, 11\penalty0 (2):\penalty0 025021,
  2009.
\newblock URL \url{http://stacks.iop.org/1367-2630/11/i=2/a=025021}.

\bibitem[Liebsch and Ishida(2010)]{LiePRB822010}
Ansgar Liebsch and Hiroshi Ishida.
\newblock Correlation-induced spin freezing transition in {$FeSe$}: A dynamical
  mean field study.
\newblock \emph{Phys. Rev. B}, 82:\penalty0 155106, 2010.
\newblock \doi{10.1103/PhysRevB.82.155106}.
\newblock URL \url{https://link.aps.org/doi/10.1103/PhysRevB.82.155106}.

\bibitem[Ishida and Liebsch(2010)]{IshPRB812010}
Hiroshi Ishida and Ansgar Liebsch.
\newblock Fermi-liquid, non-fermi-liquid, and {$\text{M}$}ott phases in iron
  pnictides and cuprates.
\newblock \emph{Phys. Rev. B}, 81:\penalty0 054513, 2010.
\newblock \doi{10.1103/PhysRevB.81.054513}.
\newblock URL \url{https://link.aps.org/doi/10.1103/PhysRevB.81.054513}.

\bibitem[Hansmann et~al.(2010)Hansmann, Arita, Toschi, Sakai, Sangiovanni, and
  Held]{HanPRL1042010}
P.~Hansmann, R.~Arita, A.~Toschi, S.~Sakai, G.~Sangiovanni, and K.~Held.
\newblock Dichotomy between large local and small ordered magnetic moments in
  iron-based superconductors.
\newblock \emph{Phys. Rev. Lett.}, 104:\penalty0 197002, 2010.
\newblock \doi{10.1103/PhysRevLett.104.197002}.
\newblock URL \url{https://link.aps.org/doi/10.1103/PhysRevLett.104.197002}.

\bibitem[Yin et~al.(2011)Yin, Haule, and Kotliar]{YinNM102011}
Z.~P. Yin, K.~Haule, and G.~Kotliar.
\newblock Kinetic frustration and the nature of the magnetic and paramagnetic
  states in iron pnictides and iron chalcogenides.
\newblock \emph{Nature Materials}, 10:\penalty0 932, 2011.
\newblock URL \url{http://dx.doi.org/10.1038/nmat3120}.

\bibitem[Werner et~al.(2012{\natexlab{a}})Werner, Casula, Miyake, Aryasetiawan,
  Millis, and Biermann]{WerNP82012}
Philipp Werner, Michele Casula, Takashi Miyake, Ferdi Aryasetiawan, Andrew~J.
  Millis, and Silke Biermann.
\newblock Satellites and large doping and temperature dependence of electronic
  properties in hole-doped {$BaFe_2As_2$}.
\newblock \emph{Nature Physics}, 8:\penalty0 331, 2012{\natexlab{a}}.
\newblock URL \url{http://dx.doi.org/10.1038/nphys2250}.

\bibitem[Lanat\`a et~al.(2013)Lanat\`a, Strand, Giovannetti, Hellsing, de'
  Medici, and Capone]{LanPRB872013}
Nicola Lanat\`a, Hugo U.~R. Strand, Gianluca Giovannetti, Bo~Hellsing, Luca de'
  Medici, and Massimo Capone.
\newblock Orbital selectivity in {$\text{H}$}und's metals: The iron
  chalcogenides.
\newblock \emph{Phys. Rev. B}, 87:\penalty0 045122, 2013.
\newblock \doi{10.1103/PhysRevB.87.045122}.
\newblock URL \url{https://link.aps.org/doi/10.1103/PhysRevB.87.045122}.

\bibitem[de' Medici et~al.(2014)de' Medici, Giovannetti, and
  Capone]{deMPRL1122014}
Luca de' Medici, Gianluca Giovannetti, and Massimo Capone.
\newblock Selective {$\text{M}$}ott physics as a key to iron superconductors.
\newblock \emph{Phys. Rev. Lett.}, 112:\penalty0 177001, 2014.
\newblock \doi{10.1103/PhysRevLett.112.177001}.
\newblock URL \url{https://link.aps.org/doi/10.1103/PhysRevLett.112.177001}.

\bibitem[Fanfarillo and Bascones(2015)]{FanPRB922015}
L.~Fanfarillo and E.~Bascones.
\newblock Electronic correlations in {$\text{H}$}und metals.
\newblock \emph{Phys. Rev. B}, 92:\penalty0 075136, 2015.
\newblock \doi{10.1103/PhysRevB.92.075136}.
\newblock URL \url{https://link.aps.org/doi/10.1103/PhysRevB.92.075136}.

\bibitem[de' Medici(2015)]{deMedici2015}
Luca de' Medici.
\newblock \emph{Weak and {$\text{S}$}trong Correlations in {$\text{Fe}$}
  Superconductors}.
\newblock Springer International Publishing, 2015.
\newblock ISBN 978-3-319-11254-1.
\newblock \doi{10.1007/978-3-319-11254-1_11}.
\newblock URL \url{https://doi.org/10.1007/978-3-319-11254-1_11}.

\bibitem[de' Medici and Capone(2017)]{deMedici2017}
Luca de' Medici and Massimo Capone.
\newblock \emph{Modeling {$\text{M}$}any-{$\text{B}$}ody Physics with
  {$\text{S}$}lave-{$\text{S}$}pin {$\text{M}$}ean-{$\text{F}$}Ield:
  {$\text{M}$}ott and {$\text{H}$}und's Physics in
  {$\text{Fe}$}-Superconductors}.
\newblock Springer International Publishing, 2017.
\newblock ISBN 978-3-319-56117-2.
\newblock \doi{10.1007/978-3-319-56117-2_4}.
\newblock URL \url{https://doi.org/10.1007/978-3-319-56117-2_4}.

\bibitem[de'Medici et~al.(2005)de'Medici, Georges, and Biermann]{deMPRB722005}
L.~de'Medici, A.~Georges, and S.~Biermann.
\newblock Orbital-selective {$\text{M}$}ott transition in multiband systems:
  Slave-spin representation and dynamical mean-field theory.
\newblock \emph{Phys. Rev. B}, 72:\penalty0 205124, 2005.
\newblock \doi{10.1103/PhysRevB.72.205124}.
\newblock URL \url{https://link.aps.org/doi/10.1103/PhysRevB.72.205124}.

\bibitem[de' Medici et~al.(2009)de' Medici, Hassan, Capone, and
  Dai]{deMPRL1022009}
Luca de' Medici, S.~R. Hassan, Massimo Capone, and Xi~Dai.
\newblock Orbital-{$\text{S}$}elective {$\text{M}$}ott {$\text{T}$}ransition
  out of band degeneracy lifting.
\newblock \emph{Phys. Rev. Lett.}, 102:\penalty0 126401, 2009.
\newblock \doi{10.1103/PhysRevLett.102.126401}.
\newblock URL \url{https://link.aps.org/doi/10.1103/PhysRevLett.102.126401}.

\bibitem[Yu and Si(2013)]{YuRPRL1102013}
Rong Yu and Qimiao Si.
\newblock Orbital-{$\text{S}$}elective {$\text{M}$}ott {$\text{P}$}hase in
  multiorbital models for alkaline iron selenides
  {${\mathbf{K}}_{1\ensuremath{-}x}{\mathrm{Fe}}_{2\ensuremath{-}y}{\mathrm{Se}}_{2}$}.
\newblock \emph{Phys. Rev. Lett.}, 110:\penalty0 146402, 2013.
\newblock \doi{10.1103/PhysRevLett.110.146402}.
\newblock URL \url{https://link.aps.org/doi/10.1103/PhysRevLett.110.146402}.

\bibitem[Yi et~al.(2013)Yi, Lu, Yu, Riggs, Chu, Lv, Liu, Lu, Cui, Hashimoto,
  Mo, Hussain, Chu, Fisher, Si, and Shen]{YiMPRL1102013}
M.~Yi, D.~H. Lu, R.~Yu, S.~C. Riggs, J.-H. Chu, B.~Lv, Z.~K. Liu, M.~Lu, Y.-T.
  Cui, M.~Hashimoto, S.-K. Mo, Z.~Hussain, C.~W. Chu, I.~R. Fisher, Q.~Si, and
  Z.-X. Shen.
\newblock Observation of temperature-induced crossover to an
  {$\text{O}$}rbital-{$\text{S}$}elective {$\text{M}$}ott {$\text{P}$}hase in
  {${\mathrm{A}}_{x}{\mathrm{Fe}}_{2\mathrm{\text{\ensuremath{-}}}y}{\mathrm{Se}}_{2}$}
  ({$A\mathbf{=}\mathrm{K}$}, {$\text{Rb}$}) superconductors.
\newblock \emph{Phys. Rev. Lett.}, 110:\penalty0 067003, 2013.
\newblock \doi{10.1103/PhysRevLett.110.067003}.
\newblock URL \url{https://link.aps.org/doi/10.1103/PhysRevLett.110.067003}.

\bibitem[Lu et~al.(2012)Lu, Vishik, Z, Chen, Moore, and Shen]{LuDARCMP32012}
Donghui Lu, Inna~M. Vishik, Ming Z, Yulin Chen, Rob~G. Moore, and Zhi-Xun Shen.
\newblock Angle-{$\text{R}$}esolved {$\text{P}$}hotoemission studies of quantum
  materials.
\newblock \emph{Annual Review of Condensed Matter Physics}, 3\penalty0
  (1):\penalty0 129--167, 2012.
\newblock \doi{10.1146/annurev-conmatphys-020911-125027}.
\newblock URL \url{https://doi.org/10.1146/annurev-conmatphys-020911-125027}.

\bibitem[Gretarsson et~al.(2011)Gretarsson, Lupascu, Kim, Casa, Gog, Wu,
  Julian, Xu, Wen, Gu, Yuan, Chen, Wang, Khim, Kim, Ishikado, Jarrige, Shamoto,
  Chu, Fisher, and Kim]{GrePRB842011}
H.~Gretarsson, A.~Lupascu, Jungho Kim, D.~Casa, T.~Gog, W.~Wu, S.~R. Julian,
  Z.~J. Xu, J.~S. Wen, G.~D. Gu, R.~H. Yuan, Z.~G. Chen, N.-L. Wang, S.~Khim,
  K.~H. Kim, M.~Ishikado, I.~Jarrige, S.~Shamoto, J.-H. Chu, I.~R. Fisher, and
  Young-June Kim.
\newblock Revealing the dual nature of magnetism in iron pnictides and iron
  chalcogenides using x-ray emission spectroscopy.
\newblock \emph{Phys. Rev. B}, 84:\penalty0 100509, 2011.
\newblock \doi{10.1103/PhysRevB.84.100509}.
\newblock URL \url{https://link.aps.org/doi/10.1103/PhysRevB.84.100509}.

\bibitem[Lafuerza et~al.(2017)Lafuerza, Gretarsson, Hardy, Wolf, Meingast,
  Giovannetti, Capone, Sefat, Kim, Glatzel, and de' Medici]{LafPRB962017}
S.~Lafuerza, H.~Gretarsson, F.~Hardy, T.~Wolf, C.~Meingast, G.~Giovannetti,
  M.~Capone, A.~S. Sefat, Y.-J. Kim, P.~Glatzel, and L.~de' Medici.
\newblock Evidence of {$\text{M}$}ott physics in iron pnictides from x-ray
  spectroscopy.
\newblock \emph{Phys. Rev. B}, 96:\penalty0 045133, 2017.
\newblock \doi{10.1103/PhysRevB.96.045133}.
\newblock URL \url{https://link.aps.org/doi/10.1103/PhysRevB.96.045133}.

\bibitem[Kotliar and Ruckenstein(1986)]{KotPRL571986}
Gabriel Kotliar and Andrei~E. Ruckenstein.
\newblock New functional integral approach to strongly correlated fermi
  systems: The {$\text{G}$}utzwiller approximation as a saddle point.
\newblock \emph{Phys. Rev. Lett.}, 57:\penalty0 1362--1365, 1986.
\newblock \doi{10.1103/PhysRevLett.57.1362}.
\newblock URL \url{https://link.aps.org/doi/10.1103/PhysRevLett.57.1362}.

\bibitem[Florens and Georges(2004)]{FloPRB702004}
Serge Florens and Antoine Georges.
\newblock Slave-rotor mean-field theories of strongly correlated systems and
  the {$\text{M}$}ott transition in finite dimensions.
\newblock \emph{Phys. Rev. B}, 70:\penalty0 035114, 2004.
\newblock \doi{10.1103/PhysRevB.70.035114}.
\newblock URL \url{https://link.aps.org/doi/10.1103/PhysRevB.70.035114}.

\bibitem[Hassan and de' Medici(2010)]{HasPRB812010}
S.~R. Hassan and L.~de' Medici.
\newblock Slave spins away from half filling: Cluster mean-field theory of the
  {$\text{H}$}ubbard and extended {$\text{H}$}ubbard models.
\newblock \emph{Phys. Rev. B}, 81:\penalty0 035106, 2010.
\newblock \doi{10.1103/PhysRevB.81.035106}.
\newblock URL \url{https://link.aps.org/doi/10.1103/PhysRevB.81.035106}.

\bibitem[Tinkham(1996)]{Tinkham1996}
Michael Tinkham.
\newblock \emph{Introduction to superconductivity}.
\newblock McGraw-Hill, Inc., 1996.
\newblock ISBN 0-07-064878-6.

\bibitem[{Somayazulu} et~al.(2018){Somayazulu}, {Ahart}, {Mishra}, {Geballe},
  {Baldini}, {Meng}, {Struzhkin}, and {Hemley}]{Som1808.07695}
Maddury {Somayazulu}, Muhtar {Ahart}, Ajay~K {Mishra}, Zachary~M. {Geballe},
  Maria {Baldini}, Yue {Meng}, Viktor~V. {Struzhkin}, and Russell~J. {Hemley}.
\newblock Evidence for superconductivity above 260 {$\text{K}$} in lanthanum
  superhydride at megabar pressures.
\newblock \emph{ArXiv e-prints}, 2018.
\newblock URL \url{https://arxiv.org/abs/1808.07695}.

\bibitem[Breitzke et~al.(2004)Breitzke, Eremin, Manske, Antipov, and
  L\"uders]{BrePCS4062004}
H.~Breitzke, I.~Eremin, D.~Manske, E.V. Antipov, and K.~L\"uders.
\newblock Formation of magnetic moments in the cuprate superconductor
  {$Hg0.8Cu0.2Ba2Ca2Cu3O8+\delta$} below {$\text{Tc}$} seen by {$\text{NQR}$}.
\newblock \emph{Physica C: Superconductivity}, 406\penalty0 (1):\penalty0 27 --
  36, 2004.
\newblock \doi{https://doi.org/10.1016/j.physc.2004.02.183}.
\newblock URL
  \url{http://www.sciencedirect.com/science/article/pii/S0921453404004988}.

\bibitem[{Norman}(2013)]{Nor1302.3176}
M.R. {Norman}.
\newblock {Unconventional Superconductivity}.
\newblock \emph{ArXiv e-prints}, 2013.
\newblock URL \url{https://arxiv.org/abs/1302.3176}.

\bibitem[Kurosaki et~al.(2005)Kurosaki, Shimizu, Miyagawa, Kanoda, and
  Saito]{KurPRL952005}
Y.~Kurosaki, Y.~Shimizu, K.~Miyagawa, K.~Kanoda, and G.~Saito.
\newblock Mott transition from a spin liquid to a {$\text{F}$}ermi liquid in
  the spin-frustrated organic conductor
  {$\ensuremath{\kappa}\mathrm{\text{\ensuremath{-}}}(\mathrm{ET}{)}_{2}{\mathrm{Cu}}_{2}(\mathrm{CN}{)}_{3}$}.
\newblock \emph{Phys. Rev. Lett.}, 95:\penalty0 177001, 2005.
\newblock \doi{10.1103/PhysRevLett.95.177001}.
\newblock URL \url{https://link.aps.org/doi/10.1103/PhysRevLett.95.177001}.

\bibitem[Paglione and Greene(2010)]{PagNP62010}
Johnpierre Paglione and Richard~L. Greene.
\newblock High-temperature superconductivity in iron-based materials.
\newblock \emph{Nature Physics}, 6:\penalty0 645, 2010.
\newblock URL \url{http://dx.doi.org/10.1038/nphys1759}.

\bibitem[Martinelli et~al.(2016)Martinelli, Bernardini, and
  Massidda]{MarCRP172016}
Alberto Martinelli, Fabio Bernardini, and Sandro Massidda.
\newblock The phase diagrams of iron-based superconductors: Theory and
  experiments.
\newblock \emph{Comptes Rendus Physique}, 17\penalty0 (1):\penalty0 5 -- 35,
  2016.
\newblock \doi{https://doi.org/10.1016/j.crhy.2015.06.001}.
\newblock URL
  \url{http://www.sciencedirect.com/science/article/pii/S1631070515001267}.

\bibitem[Yildirim(2008)]{YilPRL1012008}
T.~Yildirim.
\newblock Origin of the 150 {$\text{K}$} anomaly in {$\text{LaFeAsO}$}:
  Competing antiferromagnetic interactions, frustration, and a structural phase
  transition.
\newblock \emph{Phys. Rev. Lett.}, 101:\penalty0 057010, 2008.
\newblock \doi{10.1103/PhysRevLett.101.057010}.
\newblock URL \url{https://link.aps.org/doi/10.1103/PhysRevLett.101.057010}.

\bibitem[Hirschfeld(2016)]{HirCRP172016}
Peter~J. Hirschfeld.
\newblock Using gap symmetry and structure to reveal the pairing mechanism in
  {$Fe$}-based superconductors.
\newblock \emph{Comptes Rendus Physique}, 17\penalty0 (1):\penalty0 197 -- 231,
  2016.
\newblock \doi{https://doi.org/10.1016/j.crhy.2015.10.002}.
\newblock URL
  \url{http://www.sciencedirect.com/science/article/pii/S1631070515001693}.

\bibitem[Steglich et~al.(1979)Steglich, Aarts, Bredl, Lieke, Meschede, Franz,
  and Sch\"afer]{StePRL431979}
F.~Steglich, J.~Aarts, C.~D. Bredl, W.~Lieke, D.~Meschede, W.~Franz, and
  H.~Sch\"afer.
\newblock Superconductivity in the presence of strong {$\text{P}$}auli
  paramagnetism: {$\text{Ce}{\mathrm{Cu}}_{2}$${\mathrm{Si}}_{2}$}.
\newblock \emph{Phys. Rev. Lett.}, 43:\penalty0 1892--1896, 1979.
\newblock \doi{10.1103/PhysRevLett.43.1892}.
\newblock URL \url{https://link.aps.org/doi/10.1103/PhysRevLett.43.1892}.

\bibitem[Saxena et~al.(2000)Saxena, Agarwal, Ahilan, Grosche, Haselwimmer,
  Steiner, Pugh, Walker, Julian, Monthoux, Lonzarich, Huxley, Sheikin,
  Braithwaite, and Flouquet]{SaxN4062000}
S.~S. Saxena, P.~Agarwal, K.~Ahilan, F.~M. Grosche, R.~K.~W. Haselwimmer, M.~J.
  Steiner, E.~Pugh, I.~R. Walker, S.~R. Julian, P.~Monthoux, G.~G. Lonzarich,
  A.~Huxley, I.~Sheikin, D.~Braithwaite, and J.~Flouquet.
\newblock Superconductivity on the border of itinerant-electron ferromagnetism
  in {$\text{UGe2}$}.
\newblock \emph{Nature}, 406:\penalty0 587, 2000.
\newblock URL \url{http://dx.doi.org/10.1038/35020500}.

\bibitem[Knebel et~al.(2011)Knebel, Aoki, and Flouquet]{KneCRP122011}
Georg Knebel, Dai Aoki, and Jacques Flouquet.
\newblock Antiferromagnetism and superconductivity in cerium based
  heavy-fermion compounds.
\newblock \emph{Comptes Rendus Physique}, 12\penalty0 (5):\penalty0 542 -- 566,
  2011.
\newblock \doi{https://doi.org/10.1016/j.crhy.2011.05.002}.
\newblock URL
  \url{http://www.sciencedirect.com/science/article/pii/S1631070511001204}.

\bibitem[Bednorz and M{\"u}ller(1986)]{BednorzZPBCM641986}
J.~G. Bednorz and K.~A. M{\"u}ller.
\newblock Possible high {$\text{Tc}$} superconductivity in the
  {$\text{Ba-La-Cu-O}$}ba−la−cu−o system.
\newblock \emph{Zeitschrift f{\"u}r Physik B Condensed Matter}, 64\penalty0
  (2):\penalty0 189--193, 1986.
\newblock \doi{10.1007/BF01303701}.
\newblock URL \url{https://doi.org/10.1007/BF01303701}.

\bibitem[Wu et~al.(1987)Wu, Ashburn, Torng, Hor, Meng, Gao, Huang, Wang, and
  Chu]{WuMPRL581987}
M.~K. Wu, J.~R. Ashburn, C.~J. Torng, P.~H. Hor, R.~L. Meng, L.~Gao, Z.~J.
  Huang, Y.~Q. Wang, and C.~W. Chu.
\newblock Superconductivity at 93 {$\text{K}$} in a new mixed-phase
  {$\text{Y-Ba-Cu-O}$} compound system at ambient pressure.
\newblock \emph{Phys. Rev. Lett.}, 58:\penalty0 908--910, 1987.
\newblock \doi{10.1103/PhysRevLett.58.908}.
\newblock URL \url{https://link.aps.org/doi/10.1103/PhysRevLett.58.908}.

\bibitem[Chu et~al.(1993)Chu, Gao, Chen, Huang, Meng, and Xue]{ChuN3651993}
C.~W. Chu, L.~Gao, F.~Chen, Z.~J. Huang, R.~L. Meng, and Y.~Y. Xue.
\newblock Superconductivity above 150 {$\text{K}$} in
  {$\text{HgBa2Ca2Cu3O8+d}$} at high pressures.
\newblock \emph{Nature}, 365:\penalty0 323, 1993.
\newblock URL \url{http://dx.doi.org/10.1038/365323a0}.

\bibitem[Bari{\v s}i{\'c} et~al.(2013)Bari{\v s}i{\'c}, Chan, Li, Yu, Zhao,
  Dressel, Smontara, and Greven]{BarPNAS1102013}
Neven Bari{\v s}i{\'c}, Mun~K. Chan, Yuan Li, Guichuan Yu, Xudong Zhao, Martin
  Dressel, Ana Smontara, and Martin Greven.
\newblock Universal sheet resistance and revised phase diagram of the cuprate
  high-temperature superconductors.
\newblock \emph{Proceedings of the National Academy of Sciences}, 110\penalty0
  (30):\penalty0 12235--12240, 2013.
\newblock \doi{10.1073/pnas.1301989110}.
\newblock URL \url{http://www.pnas.org/content/110/30/12235}.

\bibitem[Norman et~al.(2007)Norman, Kanigel, Randeria, Chatterjee, and
  Campuzano]{NorPRB762007}
M.~R. Norman, A.~Kanigel, M.~Randeria, U.~Chatterjee, and J.~C. Campuzano.
\newblock Modeling the {$\text{F}$}ermi arc in underdoped cuprates.
\newblock \emph{Phys. Rev. B}, 76:\penalty0 174501, 2007.
\newblock \doi{10.1103/PhysRevB.76.174501}.
\newblock URL \url{https://link.aps.org/doi/10.1103/PhysRevB.76.174501}.

\bibitem[Tremblay et~al.(2006)Tremblay, Kyung, and S\'en\'echal]{TreLTP322006}
A.-M.~S. Tremblay, B.~Kyung, and D.~S\'en\'echal.
\newblock Pseudogap and high-temperature superconductivity from weak to strong
  coupling. {$\text{T}$}owards a quantitative theory (review article).
\newblock \emph{Low Temperature Physics}, 32\penalty0 (4):\penalty0 424--451,
  2006.
\newblock \doi{10.1063/1.2199446}.
\newblock URL \url{https://doi.org/10.1063/1.2199446}.

\bibitem[Civelli et~al.(2005)Civelli, Capone, Kancharla, Parcollet, and
  Kotliar]{CivPRL952005}
M.~Civelli, M.~Capone, S.~S. Kancharla, O.~Parcollet, and G.~Kotliar.
\newblock Dynamical breakup of the {$\text{F}$}ermi surface in a doped
  {$\text{M}$}ott insulator.
\newblock \emph{Phys. Rev. Lett.}, 95:\penalty0 106402, 2005.
\newblock \doi{10.1103/PhysRevLett.95.106402}.
\newblock URL \url{https://link.aps.org/doi/10.1103/PhysRevLett.95.106402}.

\bibitem[Varma(2000)]{VarPRB612000}
C.~M. Varma.
\newblock Proposal for an experiment to test a theory of high-temperature
  superconductors.
\newblock \emph{Phys. Rev. B}, 61:\penalty0 R3804--R3807, 2000.
\newblock \doi{10.1103/PhysRevB.61.R3804}.
\newblock URL \url{https://link.aps.org/doi/10.1103/PhysRevB.61.R3804}.

\bibitem[Kamihara et~al.(2008)Kamihara, Watanabe, Hirano, and
  Hosono]{KamJACS1302008}
Yoichi Kamihara, Takumi Watanabe, Masahiro Hirano, and Hideo Hosono.
\newblock Iron-based layered superconductor {$La[O1-xFx]FeAs$} (x =
  0.05−0.12) with {$T_c = 26 K$}.
\newblock \emph{Journal of the American Chemical Society}, 130\penalty0
  (11):\penalty0 3296--3297, 2008.
\newblock \doi{10.1021/ja800073m}.
\newblock URL \url{https://doi.org/10.1021/ja800073m}.

\bibitem[Rotter et~al.(2008)Rotter, Tegel, and Johrendt]{RotPRL1012008}
Marianne Rotter, Marcus Tegel, and Dirk Johrendt.
\newblock Superconductivity at 38 {$\text{K}$} in the iron arsenide
  {$({\mathrm{Ba}}_{1\ensuremath{-}x}{\mathrm{K}}_{x}){\mathrm{Fe}}_{2}{\mathrm{As}}_{2}$}.
\newblock \emph{Phys. Rev. Lett.}, 101:\penalty0 107006, 2008.
\newblock \doi{10.1103/PhysRevLett.101.107006}.
\newblock URL \url{https://link.aps.org/doi/10.1103/PhysRevLett.101.107006}.

\bibitem[Leb\`egue(2007)]{LebPRB752007}
S.~Leb\`egue.
\newblock Electronic structure and properties of the fermi surface of the
  superconductor {$LaOFeP$}.
\newblock \emph{Phys. Rev. B}, 75:\penalty0 035110, 2007.
\newblock \doi{10.1103/PhysRevB.75.035110}.
\newblock URL \url{https://link.aps.org/doi/10.1103/PhysRevB.75.035110}.

\bibitem[Vildosola et~al.(2008)Vildosola, Pourovskii, Arita, Biermann, and
  Georges]{VilPRB782008}
Ver\'onica Vildosola, Leonid Pourovskii, Ryotaro Arita, Silke Biermann, and
  Antoine Georges.
\newblock Bandwidth and fermi surface of iron oxypnictides: Covalency and
  sensitivity to structural changes.
\newblock \emph{Phys. Rev. B}, 78:\penalty0 064518, 2008.
\newblock \doi{10.1103/PhysRevB.78.064518}.
\newblock URL \url{https://link.aps.org/doi/10.1103/PhysRevB.78.064518}.

\bibitem[Kuroki et~al.(2008)Kuroki, Onari, Arita, Usui, Tanaka, Kontani, and
  Aoki]{KurPRL1012008}
Kazuhiko Kuroki, Seiichiro Onari, Ryotaro Arita, Hidetomo Usui, Yukio Tanaka,
  Hiroshi Kontani, and Hideo Aoki.
\newblock Unconventional pairing originating from the disconnected fermi
  surfaces of superconducting
  {${\mathrm{LaFeAsO}}_{1\ensuremath{-}x}{\mathrm{F}}_{x}$}.
\newblock \emph{Phys. Rev. Lett.}, 101:\penalty0 087004, 2008.
\newblock \doi{10.1103/PhysRevLett.101.087004}.
\newblock URL \url{https://link.aps.org/doi/10.1103/PhysRevLett.101.087004}.

\bibitem[Graser et~al.(2009)Graser, Maier, Hirschfeld, and
  Scalapino]{GraNJP112009}
S~Graser, T~A Maier, P~J Hirschfeld, and D~J Scalapino.
\newblock Near-degeneracy of several pairing channels in multiorbital models
  for the {$Fe$} pnictides.
\newblock \emph{New Journal of Physics}, 11\penalty0 (2):\penalty0 025016,
  2009.
\newblock URL \url{http://stacks.iop.org/1367-2630/11/i=2/a=025016}.

\bibitem[Calder\'on et~al.(2009)Calder\'on, Valenzuela, and
  Bascones]{CalPRB802009}
M.~J. Calder\'on, B.~Valenzuela, and E.~Bascones.
\newblock Tight-binding model for iron pnictides.
\newblock \emph{Phys. Rev. B}, 80:\penalty0 094531, 2009.
\newblock \doi{10.1103/PhysRevB.80.094531}.
\newblock URL \url{https://link.aps.org/doi/10.1103/PhysRevB.80.094531}.

\bibitem[Eschrig and Koepernik(2009)]{EscPRB802009}
Helmut Eschrig and Klaus Koepernik.
\newblock Tight-binding models for the iron-based superconductors.
\newblock \emph{Phys. Rev. B}, 80:\penalty0 104503, 2009.
\newblock \doi{10.1103/PhysRevB.80.104503}.
\newblock URL \url{https://link.aps.org/doi/10.1103/PhysRevB.80.104503}.

\bibitem[Hirschfeld et~al.(2011)Hirschfeld, Korshunov, and Mazin]{HirRPP742011}
P~J Hirschfeld, M~M Korshunov, and I~I Mazin.
\newblock Gap symmetry and structure of {$Fe$}-based superconductors.
\newblock \emph{Reports on Progress in Physics}, 74\penalty0 (12):\penalty0
  124508, 2011.
\newblock URL \url{http://stacks.iop.org/0034-4885/74/i=12/a=124508}.

\bibitem[Mazin et~al.(2008)Mazin, Singh, Johannes, and Du]{MazPRL1012008}
I.~I. Mazin, D.~J. Singh, M.~D. Johannes, and M.~H. Du.
\newblock Unconventional superconductivity with a sign reversal in the order
  parameter of {${\mathrm{LaFeAsO}}_{1\ensuremath{-}x}{\mathrm{F}}_{x}$}.
\newblock \emph{Phys. Rev. Lett.}, 101:\penalty0 057003, 2008.
\newblock \doi{10.1103/PhysRevLett.101.057003}.
\newblock URL \url{https://link.aps.org/doi/10.1103/PhysRevLett.101.057003}.

\bibitem[Chubukov et~al.(2008)Chubukov, Efremov, and Eremin]{ChuPRB782008}
A.~V. Chubukov, D.~V. Efremov, and I.~Eremin.
\newblock Magnetism, superconductivity, and pairing symmetry in iron-based
  superconductors.
\newblock \emph{Phys. Rev. B}, 78:\penalty0 134512, 2008.
\newblock \doi{10.1103/PhysRevB.78.134512}.
\newblock URL \url{https://link.aps.org/doi/10.1103/PhysRevB.78.134512}.

\bibitem[Cvetkovic and Tesanovic(2009)]{CveEPL852009}
V.~Cvetkovic and Z.~Tesanovic.
\newblock Multiband magnetism and superconductivity in {$\text{Fe}$}-based
  compounds.
\newblock \emph{EPL (Europhysics Letters)}, 85\penalty0 (3):\penalty0 37002,
  2009.
\newblock URL \url{http://stacks.iop.org/0295-5075/85/i=3/a=37002}.

\bibitem[Si and Abrahams(2008)]{SiQPRL1012008}
Qimiao Si and Elihu Abrahams.
\newblock Strong correlations and magnetic frustration in the high {${T}_{c}$}
  iron pnictides.
\newblock \emph{Phys. Rev. Lett.}, 101:\penalty0 076401, 2008.
\newblock \doi{10.1103/PhysRevLett.101.076401}.
\newblock URL \url{https://link.aps.org/doi/10.1103/PhysRevLett.101.076401}.

\bibitem[Castellanos-Gomez(2015)]{CasJPCL62015}
Andres Castellanos-Gomez.
\newblock Black phosphorus: Narrow gap, wide applications.
\newblock \emph{The Journal of Physical Chemistry Letters}, 6\penalty0
  (21):\penalty0 4280--4291, 2015.
\newblock \doi{10.1021/acs.jpclett.5b01686}.
\newblock URL \url{https://doi.org/10.1021/acs.jpclett.5b01686}.

\bibitem[Castellanos-Gomez et~al.(2013)Castellanos-Gomez, Rold\'an, Cappelluti,
  Buscema, Guinea, van~der Zant, and Steele]{CasNL132013}
Andres Castellanos-Gomez, Rafael Rold\'an, Emmanuele Cappelluti, Michele
  Buscema, Francisco Guinea, Herre S.~J. van~der Zant, and Gary~A. Steele.
\newblock Local strain engineering in atomically thin {$\text{MoS2}$}.
\newblock \emph{Nano Letters}, 13\penalty0 (11):\penalty0 5361--5366, 2013.
\newblock \doi{10.1021/nl402875m}.
\newblock URL \url{https://doi.org/10.1021/nl402875m}.

\bibitem[Sipos et~al.(2008)Sipos, Kusmartseva, Akrap, Berger, Forr{\'o}, and
  Tutis]{SipNM72008}
B.~Sipos, A.~F. Kusmartseva, A.~Akrap, H.~Berger, L.~Forr{\'o}, and E.~Tutis.
\newblock From {$\text{M}$}ott state to superconductivity in
  {$\text{1T-TaS2}$}.
\newblock \emph{Nature Materials}, 7:\penalty0 960, 2008.
\newblock URL \url{http://dx.doi.org/10.1038/nmat2318}.

\bibitem[Cao et~al.(2018{\natexlab{a}})Cao, Fatemi, Fang, Watanabe, Taniguchi,
  Kaxiras, and Jarillo-Herrero]{CaoN5562018_sc}
Yuan Cao, Valla Fatemi, Shiang Fang, Kenji Watanabe, Takashi Taniguchi,
  Efthimios Kaxiras, and Pablo Jarillo-Herrero.
\newblock Unconventional superconductivity in magic-angle graphene
  superlattices.
\newblock \emph{Nature}, 556:\penalty0 43, 2018{\natexlab{a}}.
\newblock URL \url{http://dx.doi.org/10.1038/nature26160}.

\bibitem[Cao et~al.(2018{\natexlab{b}})Cao, Fatemi, Demir, Fang, Tomarken, Luo,
  Sanchez-Yamagishi, Watanabe, Taniguchi, Kaxiras, Ashoori, and
  Jarillo-Herrero]{CaoN5562018_ins}
Yuan Cao, Valla Fatemi, Ahmet Demir, Shiang Fang, Spencer~L. Tomarken, Jason~Y.
  Luo, Javier~D. Sanchez-Yamagishi, Kenji Watanabe, Takashi Taniguchi,
  Efthimios Kaxiras, Ray~C. Ashoori, and Pablo Jarillo-Herrero.
\newblock Correlated insulator behaviour at half-filling in magic-angle
  graphene superlattices.
\newblock \emph{Nature}, 556:\penalty0 80, 2018{\natexlab{b}}.
\newblock URL \url{http://dx.doi.org/10.1038/nature26154}.

\bibitem[Reich et~al.(2002)Reich, Maultzsch, Thomsen, and
  Ordej\'on]{ReiPRB662002}
S.~Reich, J.~Maultzsch, C.~Thomsen, and P.~Ordej\'on.
\newblock Tight-binding description of graphene.
\newblock \emph{Phys. Rev. B}, 66:\penalty0 035412, 2002.
\newblock \doi{10.1103/PhysRevB.66.035412}.
\newblock URL \url{https://link.aps.org/doi/10.1103/PhysRevB.66.035412}.

\bibitem[Lopes~dos Santos et~al.(2007)Lopes~dos Santos, Peres, and
  Castro~Neto]{LopPRL992007}
J.~M.~B. Lopes~dos Santos, N.~M.~R. Peres, and A.~H. Castro~Neto.
\newblock Graphene bilayer with a twist: Electronic structure.
\newblock \emph{Phys. Rev. Lett.}, 99:\penalty0 256802, 2007.
\newblock \doi{10.1103/PhysRevLett.99.256802}.
\newblock URL \url{https://link.aps.org/doi/10.1103/PhysRevLett.99.256802}.

\bibitem[Su\'arez~Morell et~al.(2010)Su\'arez~Morell, Correa, Vargas, Pacheco,
  and Barticevic]{SuaPRB822010}
E.~Su\'arez~Morell, J.~D. Correa, P.~Vargas, M.~Pacheco, and Z.~Barticevic.
\newblock Flat bands in slightly twisted bilayer graphene: Tight-binding
  calculations.
\newblock \emph{Phys. Rev. B}, 82:\penalty0 121407, 2010.
\newblock \doi{10.1103/PhysRevB.82.121407}.
\newblock URL \url{https://link.aps.org/doi/10.1103/PhysRevB.82.121407}.

\bibitem[Bistritzer and MacDonald(2011)]{BisPNAS1082011}
Rafi Bistritzer and Allan~H. MacDonald.
\newblock Moir{\'e} bands in twisted double-layer graphene.
\newblock 108\penalty0 (30):\penalty0 12233--12237, 2011.
\newblock \doi{10.1073/pnas.1108174108}.
\newblock URL \url{http://www.pnas.org/content/108/30/12233}.

\bibitem[Cao et~al.(2016)Cao, Luo, Fatemi, Fang, Sanchez-Yamagishi, Watanabe,
  Taniguchi, Kaxiras, and Jarillo-Herrero]{CaoPRL1172016}
Y.~Cao, J.~Y. Luo, V.~Fatemi, S.~Fang, J.~D. Sanchez-Yamagishi, K.~Watanabe,
  T.~Taniguchi, E.~Kaxiras, and P.~Jarillo-Herrero.
\newblock Superlattice-induced insulating states and valley-protected orbits in
  twisted bilayer graphene.
\newblock \emph{Phys. Rev. Lett.}, 117:\penalty0 116804, 2016.
\newblock \doi{10.1103/PhysRevLett.117.116804}.
\newblock URL \url{https://link.aps.org/doi/10.1103/PhysRevLett.117.116804}.

\bibitem[Fang and Kaxiras(2016)]{FanPRB932016}
Shiang Fang and Efthimios Kaxiras.
\newblock Electronic structure theory of weakly interacting bilayers.
\newblock \emph{Phys. Rev. B}, 93:\penalty0 235153, 2016.
\newblock \doi{10.1103/PhysRevB.93.235153}.
\newblock URL \url{https://link.aps.org/doi/10.1103/PhysRevB.93.235153}.

\bibitem[Gonzalez-Arraga et~al.(2017)Gonzalez-Arraga, Lado, Guinea, and
  San-Jose]{GonPRL1192017}
Luis~A. Gonzalez-Arraga, J.~L. Lado, Francisco Guinea, and Pablo San-Jose.
\newblock Electrically controllable magnetism in twisted bilayer graphene.
\newblock \emph{Phys. Rev. Lett.}, 119:\penalty0 107201, 2017.
\newblock \doi{10.1103/PhysRevLett.119.107201}.
\newblock URL \url{https://link.aps.org/doi/10.1103/PhysRevLett.119.107201}.

\bibitem[Koshino et~al.(2018)Koshino, Yuan, Koretsune, Ochi, Kuroki, and
  Fu]{KosPRX82018}
Mikito Koshino, Noah F.~Q. Yuan, Takashi Koretsune, Masayuki Ochi, Kazuhiko
  Kuroki, and Liang Fu.
\newblock Maximally localized {$\text{W}$}annier orbitals and the extended
  {$\text{H}$}ubbard model for twisted bilayer graphene.
\newblock \emph{Phys. Rev. X}, 8:\penalty0 031087, 2018.
\newblock \doi{10.1103/PhysRevX.8.031087}.
\newblock URL \url{https://link.aps.org/doi/10.1103/PhysRevX.8.031087}.

\bibitem[Castro~Neto et~al.(2009)Castro~Neto, Guinea, Peres, Novoselov, and
  Geim]{CasRMP812009}
A.~H. Castro~Neto, F.~Guinea, N.~M.~R. Peres, K.~S. Novoselov, and A.~K. Geim.
\newblock The electronic properties of graphene.
\newblock \emph{Rev. Mod. Phys.}, 81:\penalty0 109--162, 2009.
\newblock URL \url{https://link.aps.org/doi/10.1103/RevModPhys.81.109}.

\bibitem[{Chen} et~al.(2018){Chen}, {Jiang}, {Wu}, {Lv}, {Li}, {Watanabe},
  {Taniguchi}, {Shi}, {Zhang}, and {Wang}]{Che1803.01985}
G.~{Chen}, L.~{Jiang}, S.~{Wu}, B.~{Lv}, H.~{Li}, K.~{Watanabe},
  T.~{Taniguchi}, Z.~{Shi}, Y.~{Zhang}, and F.~{Wang}.
\newblock Gate-tunable mott insulator in trilayer graphene-boron nitride
  moir\'e superlattice.
\newblock \emph{ArXiv e-prints}, 2018.
\newblock URL \url{https://arxiv.org/abs/1803.01985}.

\bibitem[Pizarro et~al.(2017)Pizarro, Calder\'on, Liu, Mu\~noz, and
  Bascones]{Pizarro1}
J.~M. Pizarro, M.~J. Calder\'on, J.~Liu, M.~C. Mu\~noz, and E.~Bascones.
\newblock Strong correlations and the search for high-{${T}_{c}$}
  superconductivity in chromium pnictides and chalcogenides.
\newblock \emph{Phys. Rev. B}, 95:\penalty0 075115, 2017.
\newblock \doi{10.1103/PhysRevB.95.075115}.
\newblock URL \url{https://link.aps.org/doi/10.1103/PhysRevB.95.075115}.

\bibitem[Pizarro and Bascones(2019)]{Pizarro2}
J.~M. Pizarro and E.~Bascones.
\newblock Strong electronic correlations and fermi surface reconstruction in
  the quasi-one-dimensional iron superconductor
  ${\mathrm{bafe}}_{2}{\mathrm{s}}_{3}$.
\newblock \emph{Phys. Rev. Materials}, 3:\penalty0 014801, 2019.
\newblock URL \url{https://link.aps.org/doi/10.1103/PhysRevMaterials.3.014801}.

\bibitem[Arita et~al.(2015)Arita, Ikeda, Sakai, and Suzuki]{AriPRB922015}
Ryotaro Arita, Hiroaki Ikeda, Shiro Sakai, and Michi-To Suzuki.
\newblock Ab initio downfolding study of the iron-based ladder superconductor
  {$BaFe_2S_3$}.
\newblock \emph{Phys. Rev. B}, 92:\penalty0 054515, 2015.
\newblock \doi{10.1103/PhysRevB.92.054515}.
\newblock URL \url{https://link.aps.org/doi/10.1103/PhysRevB.92.054515}.

\bibitem[{Pizarro} et~al.(2018){Pizarro}, {Calder{\'o}n}, and
  {Bascones}]{Pizarro3}
J.M. {Pizarro}, M.J. {Calder{\'o}n}, and E.~{Bascones}.
\newblock The nature of correlations in the insulating states of twisted
  bilayer graphene.
\newblock \emph{ArXiv e-prints}, 2018.

\bibitem[Terashima et~al.(2013)Terashima, Kurita, Kimata, Tomita, Tsuchiya,
  Imai, Sato, Kihou, Lee, Kito, Eisaki, Iyo, Saito, Fukazawa, Kohori, Harima,
  and Uji]{TerPRB872013}
Taichi Terashima, Nobuyuki Kurita, Motoi Kimata, Megumi Tomita, Satoshi
  Tsuchiya, Motoharu Imai, Akira Sato, Kunihiro Kihou, Chul-Ho Lee, Hijiri
  Kito, Hiroshi Eisaki, Akira Iyo, Taku Saito, Hideto Fukazawa, Yoh Kohori,
  Hisatomo Harima, and Shinya Uji.
\newblock Fermi surface in {$K{Fe}_{2}{As}_{2}$} determined via de
  {$\text{H}$}aas--van {$\text{A}$}lphen oscillation measurements.
\newblock \emph{Phys. Rev. B}, 87:\penalty0 224512, 2013.
\newblock \doi{10.1103/PhysRevB.87.224512}.
\newblock URL \url{https://link.aps.org/doi/10.1103/PhysRevB.87.224512}.

\bibitem[Hardy et~al.(2013)Hardy, B\"ohmer, Aoki, Burger, Wolf, Schweiss, Heid,
  Adelmann, Yao, Kotliar, Schmalian, and Meingast]{HarPRL1112013}
F.~Hardy, A.~E. B\"ohmer, D.~Aoki, P.~Burger, T.~Wolf, P.~Schweiss, R.~Heid,
  P.~Adelmann, Y.~X. Yao, G.~Kotliar, J.~Schmalian, and C.~Meingast.
\newblock Evidence of strong correlations and coherence-incoherence crossover
  in the iron pnictide superconductor {${\mathrm{KFe}}_{2}{\mathrm{As}}_{2}$}.
\newblock \emph{Phys. Rev. Lett.}, 111:\penalty0 027002, 2013.
\newblock \doi{10.1103/PhysRevLett.111.027002}.
\newblock URL \url{https://link.aps.org/doi/10.1103/PhysRevLett.111.027002}.

\bibitem[Nakajima et~al.(2014)Nakajima, Ishida, Tanaka, Kihou, Tomioka, Saito,
  Lee, Fukazawa, Kohori, Kakeshita, Iyo, Ito, Eisaki, and Uchida]{NakSC42014}
M.~Nakajima, S.~Ishida, T.~Tanaka, K.~Kihou, Y.~Tomioka, T.~Saito, C.~H. Lee,
  H.~Fukazawa, Y.~Kohori, T.~Kakeshita, A.~Iyo, T.~Ito, H.~Eisaki, and
  S.~Uchida.
\newblock Normal-state charge dynamics in doped {$BaFe2As2$}: Roles of doping
  and necessary ingredients for superconductivity.
\newblock \emph{Scientific Reports}, 4:\penalty0 5873 EP, 2014.
\newblock URL \url{http://dx.doi.org/10.1038/srep05873}.

\bibitem[Eilers et~al.(2016)Eilers, Grube, Zocco, Wolf, Merz, Schweiss, Heid,
  Eder, Yu, Zhu, Si, Shibauchi, and L\"ohneysen]{EilPRL1162016}
Felix Eilers, Kai Grube, Diego~A. Zocco, Thomas Wolf, Michael Merz, Peter
  Schweiss, Rolf Heid, Robert Eder, Rong Yu, Jian-Xin Zhu, Qimiao Si, Takasada
  Shibauchi, and Hilbert~v. L\"ohneysen.
\newblock Strain-driven approach to quantum criticality in
  {$A{\mathrm{Fe}}_{2}{\mathrm{As}}_{2}$} with {$A=\mathrm{K}$}, {$Rb$}, and
  {$Cs$}.
\newblock \emph{Phys. Rev. Lett.}, 116:\penalty0 237003, 2016.
\newblock \doi{10.1103/PhysRevLett.116.237003}.
\newblock URL \url{https://link.aps.org/doi/10.1103/PhysRevLett.116.237003}.

\bibitem[Hardy et~al.(2016)Hardy, B\"ohmer, de' Medici, Capone, Giovannetti,
  Eder, Wang, He, Wolf, Schweiss, Heid, Herbig, Adelmann, Fisher, and
  Meingast]{HarPRB942016}
F.~Hardy, A.~E. B\"ohmer, L.~de' Medici, M.~Capone, G.~Giovannetti, R.~Eder,
  L.~Wang, M.~He, T.~Wolf, P.~Schweiss, R.~Heid, A.~Herbig, P.~Adelmann, R.~A.
  Fisher, and C.~Meingast.
\newblock Strong correlations, strong coupling, and {$s$}-wave
  superconductivity in hole-doped {${\mathrm{BaFe}}_{2}{\mathrm{As}}_{2}$}
  single crystals.
\newblock \emph{Phys. Rev. B}, 94:\penalty0 205113, 2016.
\newblock \doi{10.1103/PhysRevB.94.205113}.
\newblock URL \url{https://link.aps.org/doi/10.1103/PhysRevB.94.205113}.

\bibitem[Hubbard and Flowers(1963)]{HubPRSLA2761963}
J.~Hubbard and B.H. Flowers.
\newblock Electron correlations in narrow energy bands.
\newblock \emph{Proceedings of the Royal Society of London A: Mathematical,
  Physical and Engineering Sciences}, 276\penalty0 (1365):\penalty0 238--257,
  1963.
\newblock \doi{10.1098/rspa.1963.0204}.
\newblock URL
  \url{http://rspa.royalsocietypublishing.org/content/276/1365/238}.

\bibitem[Hub(1964{\natexlab{a}})]{HubPRSLA2771964}
Electron correlations in narrow energy bands. {$\text{II}$}. the degenerate
  band case.
\newblock \emph{Proceedings of the Royal Society of London A: Mathematical,
  Physical and Engineering Sciences}, 277\penalty0 (1369):\penalty0 237--259,
  1964{\natexlab{a}}.
\newblock \doi{10.1098/rspa.1964.0019}.
\newblock URL
  \url{http://rspa.royalsocietypublishing.org/content/277/1369/237}.

\bibitem[Hub(1964{\natexlab{b}})]{HubPRSLA2811964}
Electron correlations in narrow energy bands {$\text{III}$}. an improved
  solution.
\newblock \emph{Proceedings of the Royal Society of London A: Mathematical,
  Physical and Engineering Sciences}, 281\penalty0 (1386):\penalty0 401--419,
  1964{\natexlab{b}}.
\newblock \doi{10.1098/rspa.1964.0190}.
\newblock URL
  \url{http://rspa.royalsocietypublishing.org/content/281/1386/401}.

\bibitem[Hub(1965)]{HubPRSLA2851965}
Electron correlations in narrow energy bands - {$\text{IV}$}. the atomic
  representation.
\newblock \emph{Proceedings of the Royal Society of London A: Mathematical,
  Physical and Engineering Sciences}, 285\penalty0 (1403):\penalty0 542--560,
  1965.
\newblock \doi{10.1098/rspa.1965.0124}.
\newblock URL
  \url{http://rspa.royalsocietypublishing.org/content/285/1403/542}.

\bibitem[Gunnarsson et~al.(1996)Gunnarsson, Koch, and Martin]{GunPRB541996}
Olle Gunnarsson, Erik Koch, and Richard~M. Martin.
\newblock Mott transition in degenerate {$\text{H}$}ubbard models: Application
  to doped fullerenes.
\newblock \emph{Phys. Rev. B}, 54:\penalty0 R11026--R11029, 1996.
\newblock \doi{10.1103/PhysRevB.54.R11026}.
\newblock URL \url{https://link.aps.org/doi/10.1103/PhysRevB.54.R11026}.

\bibitem[Florens et~al.(2002)Florens, Georges, Kotliar, and
  Parcollet]{FloPRB662002}
S.~Florens, A.~Georges, G.~Kotliar, and O.~Parcollet.
\newblock Mott transition at large orbital degeneracy: {$\text{D}$}ynamical
  mean-field theory.
\newblock \emph{Phys. Rev. B}, 66:\penalty0 205102, 2002.
\newblock \doi{10.1103/PhysRevB.66.205102}.
\newblock URL \url{https://link.aps.org/doi/10.1103/PhysRevB.66.205102}.

\bibitem[Bascones et~al.(2012)Bascones, Valenzuela, and
  Calder\'on]{BasPRB862012}
E.~Bascones, B.~Valenzuela, and M.~J. Calder\'on.
\newblock Orbital differentiation and the role of orbital ordering in the
  magnetic state of {$Fe$} superconductors.
\newblock \emph{Phys. Rev. B}, 86:\penalty0 174508, 2012.
\newblock \doi{10.1103/PhysRevB.86.174508}.
\newblock URL \url{https://link.aps.org/doi/10.1103/PhysRevB.86.174508}.

\bibitem[Calder\'on et~al.(2014)Calder\'on, Medici, Valenzuela, and
  Bascones]{CalPRB902014}
M.~J. Calder\'on, L.~de' Medici, B.~Valenzuela, and E.~Bascones.
\newblock Correlation, doping, and interband effects on the optical
  conductivity of iron superconductors.
\newblock \emph{Phys. Rev. B}, 90:\penalty0 115128, 2014.
\newblock \doi{10.1103/PhysRevB.90.115128}.
\newblock URL \url{https://link.aps.org/doi/10.1103/PhysRevB.90.115128}.

\bibitem[Rozenberg(1997)]{RozPRB551997}
Marcelo~J. Rozenberg.
\newblock Integer-filling metal-insulator transitions in the degenerate
  {$\text{H}$}ubbard model.
\newblock \emph{Phys. Rev. B}, 55:\penalty0 R4855--R4858, 1997.
\newblock \doi{10.1103/PhysRevB.55.R4855}.
\newblock URL \url{https://link.aps.org/doi/10.1103/PhysRevB.55.R4855}.

\bibitem[Yi et~al.(2017{\natexlab{a}})Yi, Zhang, Shen, and Lu]{YiMnpjQM22017}
Ming Yi, Yan Zhang, Zhi-Xun Shen, and Donghui Lu.
\newblock Role of the orbital degree of freedom in iron-based superconductors.
\newblock \emph{npj Quantum Materials}, 2\penalty0 (1):\penalty0 57,
  2017{\natexlab{a}}.
\newblock URL \url{https://doi.org/10.1038/s41535-017-0059-y}.

\bibitem[Yi et~al.(2015)Yi, Liu, Zhang, Yu, Zhu, Lee, Moore, Schmitt, Li,
  Riggs, Chu, Lv, Hu, Hashimoto, Mo, Hussain, Mao, Chu, Fisher, Si, Shen, and
  Lu]{YiMNC62015}
M.~Yi, Z.-K. Liu, Y.~Zhang, R.~Yu, J.-X. Zhu, J.~J. Lee, R.~G. Moore, F.~T.
  Schmitt, W.~Li, S.~C. Riggs, J.-H. Chu, B.~Lv, J.~Hu, M.~Hashimoto, S.-K. Mo,
  Z.~Hussain, Z.~Q. Mao, C.~W. Chu, I.~R. Fisher, Q.~Si, Z.-X. Shen, and D.~H.
  Lu.
\newblock Observation of universal strong orbital-dependent correlation effects
  in iron chalcogenides.
\newblock \emph{Nature Communications}, 6, Jul 2015.
\newblock URL \url{https://doi.org/10.1038/ncomms8777}.

\bibitem[Park et~al.(2013)Park, Mizoguchi, Kodama, ichi Shamoto, Otomo,
  Matsuishi, Kamiya, and Hosono]{ParIC522013}
Sang-Won Park, Hiroshi Mizoguchi, Katsuaki Kodama, Shin ichi Shamoto, Toshiya
  Otomo, Satoru Matsuishi, Toshio Kamiya, and Hideo Hosono.
\newblock Magnetic structure and electromagnetic properties of {$LnCrAsO$} with
  a {$ZrCuSiAs$}-type structure ({$Ln = La, Ce, Pr,$} and {$Nd$}).
\newblock \emph{Inorganic chemistry}, 52 23:\penalty0 13363--8, 2013.
\newblock URL \url{https://pubs.acs.org/doi/10.1021/ic401487q}.

\bibitem[Ramshaw et~al.(2015)Ramshaw, Sebastian, McDonald, Day, Tan, Zhu,
  Betts, Liang, Bonn, Hardy, and Harrison]{RamS62322015}
B.~J. Ramshaw, S.~E. Sebastian, R.~D. McDonald, James Day, B.~S. Tan, Z.~Zhu,
  J.~B. Betts, Ruixing Liang, D.~A. Bonn, W.~N. Hardy, and N.~Harrison.
\newblock Quasiparticle mass enhancement approaching optimal doping in a
  high-tc superconductor.
\newblock \emph{Science}, 348\penalty0 (6232):\penalty0 317--320, 2015.
\newblock URL \url{http://science.sciencemag.org/content/348/6232/317}.

\bibitem[Werner et~al.(2012{\natexlab{b}})Werner, Casula, Miyake, Aryasetiawan,
  Millis, and Biermann]{WernerNP82012}
Philipp Werner, Michele Casula, Takashi Miyake, Ferdi Aryasetiawan, Andrew~J.
  Millis, and Silke Biermann.
\newblock Satellites and large doping and temperature dependence of electronic
  properties in hole-doped {$BaFe_2As_2$}.
\newblock \emph{Nature Physics}, 8:\penalty0 331, 2012{\natexlab{b}}.
\newblock URL \url{http://dx.doi.org/10.1038/nphys2250}.

\bibitem[Misawa et~al.(2012)Misawa, Nakamura, and Imada]{MisPRL1082012}
Takahiro Misawa, Kazuma Nakamura, and Masatoshi Imada.
\newblock Ab initio evidence for strong correlation associated with
  {$\text{M}$}ott proximity in iron-based superconductors.
\newblock \emph{Phys. Rev. Lett.}, 108:\penalty0 177007, 2012.
\newblock \doi{10.1103/PhysRevLett.108.177007}.
\newblock URL \url{https://link.aps.org/doi/10.1103/PhysRevLett.108.177007}.

\bibitem[Bao et~al.(2015)Bao, Liu, Ma, Meng, Tang, Sun, Zhai, Jiang, Bai, Feng,
  Xu, and Cao]{BaoPRX52015}
Jin-Ke Bao, Ji-Yong Liu, Cong-Wei Ma, Zhi-Hao Meng, Zhang-Tu Tang, Yun-Lei Sun,
  Hui-Fei Zhai, Hao Jiang, Hua Bai, Chun-Mu Feng, Zhu-An Xu, and Guang-Han Cao.
\newblock Superconductivity in quasi-one-dimensional
  {${\mathrm{K}}_{2}{\mathrm{Cr}}_{3}{\mathrm{As}}_{3}$} with significant
  electron correlations.
\newblock \emph{Phys. Rev. X}, 5:\penalty0 011013, 2015.
\newblock \doi{10.1103/PhysRevX.5.011013}.
\newblock URL \url{https://link.aps.org/doi/10.1103/PhysRevX.5.011013}.

\bibitem[Simonson et~al.(2011)Simonson, Post, Marques, Smith, Khatib, Basov,
  and Aronson]{SimPRB842011}
J.~W. Simonson, K.~Post, C.~Marques, G.~Smith, O.~Khatib, D.~N. Basov, and
  M.~C. Aronson.
\newblock Gap states in insulating {$LaMnPO{}_{1\ensuremath{-}x}F{}_{x}$}
  ($x=0$--0.3).
\newblock \emph{Phys. Rev. B}, 84:\penalty0 165129, 2011.
\newblock \doi{10.1103/PhysRevB.84.165129}.
\newblock URL \url{https://link.aps.org/doi/10.1103/PhysRevB.84.165129}.

\bibitem[Satya et~al.(2011)Satya, Mani, Arulraj, Shekar, Vinod, Sundar, and
  Bharathi]{SatPRB842011}
A.~T. Satya, Awadhesh Mani, A.~Arulraj, N.~V.~Chandra Shekar, K.~Vinod, C.~S.
  Sundar, and A.~Bharathi.
\newblock Pressure-induced metallization of
  {$\text{Ba}$}{$\text{Mn}_2$}{$\text{As}_2$}.
\newblock \emph{Phys. Rev. B}, 84:\penalty0 180515, 2011.
\newblock \doi{10.1103/PhysRevB.84.180515}.
\newblock URL \url{https://link.aps.org/doi/10.1103/PhysRevB.84.180515}.

\bibitem[Simonson et~al.(2012)Simonson, Yin, Pezzoli, Guo, Liu, Post, Efimenko,
  Hollmann, Hu, Lin, Chen, Marques, Leyva, Smith, Lynn, Sun, Kotliar, Basov,
  Tjeng, and Aronson]{SimPNAS1092012}
J.~W. Simonson, Z.~P. Yin, M.~Pezzoli, J.~Guo, J.~Liu, K.~Post, A.~Efimenko,
  N.~Hollmann, Z.~Hu, H.-J. Lin, C.-T. Chen, C.~Marques, V.~Leyva, G.~Smith,
  J.~W. Lynn, L.~L. Sun, G.~Kotliar, D.~N. Basov, L.~H. Tjeng, and M.~C.
  Aronson.
\newblock From antiferromagnetic insulator to correlated metal in pressurized
  and doped {$LaMnPO$}.
\newblock \emph{Proceedings of the National Academy of Sciences}, 109\penalty0
  (27):\penalty0 E1815--E1819, 2012.
\newblock \doi{10.1073/pnas.1117366109}.
\newblock URL \url{http://www.pnas.org/content/109/27/E1815}.

\bibitem[Guo et~al.(2013)Guo, Simonson, Sun, Wu, Gao, Zhang, Gu, Kotliar,
  Aronson, and Zhao]{GuoSC32013}
Jing Guo, J.~W. Simonson, Liling Sun, Qi~Wu, Peiwen Gao, Chao Zhang, Dachun Gu,
  Gabriel Kotliar, Meigan Aronson, and Zhongxian Zhao.
\newblock Observation of antiferromagnetic order collapse in the pressurized
  insulator {$LaMnPO$}.
\newblock \emph{Scientific Reports}, 3:\penalty0 2555, 2013.
\newblock URL \url{http://dx.doi.org/10.1038/srep02555}.

\bibitem[Saparov et~al.(2013)Saparov, Singh, Garlea, and Sefat]{SapSC32013}
Bayrammurad Saparov, David~J. Singh, Vasile~O. Garlea, and Athena~S. Sefat.
\newblock Crystal, magnetic, and electronic structures, and properties of new
  {$BaMnPnF$} ({$Pn = As, Sb, Bi$}).
\newblock \emph{Scientific Reports}, 3:\penalty0 2154, 2013.
\newblock URL \url{http://dx.doi.org/10.1038/srep02154}.

\bibitem[{Beleanu} et~al.(2013){Beleanu}, {Kiss}, {Kreiner}, {K{\"o}hler},
  {M{\"u}chler}, {Schnelle}, {Burkhardt}, {Chadov}, {Medvediev}, {Ebke},
  {Cordier}, {Albert}, {Hoser}, {Bernardi}, {Larkin}, {Pr{\"o}pper}, {Boris},
  {Keimer}, and {Felser}]{Bel1307.6404}
A.~{Beleanu}, J.~{Kiss}, G.~{Kreiner}, C.~{K{\"o}hler}, L.~{M{\"u}chler},
  W.~{Schnelle}, U.~{Burkhardt}, S.~{Chadov}, S.~{Medvediev}, D.~{Ebke},
  G.~{Cordier}, B.~{Albert}, A.~{Hoser}, F.~{Bernardi}, T.~I. {Larkin},
  D.~{Pr{\"o}pper}, A.~V. {Boris}, B.~{Keimer}, and C.~{Felser}.
\newblock Large resistivity change and phase transition in {$LiMnAs$}.
\newblock \emph{ArXiv e-prints}, 2013.
\newblock URL \url{https://arxiv.org/abs/1307.6404}.

\bibitem[Lamsal et~al.(2013)Lamsal, Tucker, Heitmann, Kreyssig, Jesche, Pandey,
  Tian, McQueeney, Johnston, and Goldman]{LamPRB872013}
J.~Lamsal, G.~S. Tucker, T.~W. Heitmann, A.~Kreyssig, A.~Jesche, Abhishek
  Pandey, Wei Tian, R.~J. McQueeney, D.~C. Johnston, and A.~I. Goldman.
\newblock Persistence of local-moment antiferromagnetic order in
  {$Ba{}_{1\ensuremath{-}x}K{}_{x}Mn{}_{2}As{}_{2}$}.
\newblock \emph{Phys. Rev. B}, 87:\penalty0 144418, 2013.
\newblock \doi{10.1103/PhysRevB.87.144418}.
\newblock URL \url{https://link.aps.org/doi/10.1103/PhysRevB.87.144418}.

\bibitem[{McNally} et~al.(2014){McNally}, {Simonson}, {Post}, {Yin}, {Pezzoli},
  {Smith}, {Leyva}, {Marques}, {DeBeer-Schmitt}, {Kolesnikov}, {Zhao}, {Lynn},
  {Basov}, {Kotliar}, and {Aronson}]{McN1402.6312}
D.~E. {McNally}, J.~W. {Simonson}, K.~W. {Post}, Z.~P. {Yin}, M.~{Pezzoli},
  G.~J. {Smith}, V.~{Leyva}, C.~{Marques}, L.~{DeBeer-Schmitt}, A.~I.
  {Kolesnikov}, Y.~{Zhao}, J.~W. {Lynn}, D.~N. {Basov}, G.~{Kotliar}, and M.~C.
  {Aronson}.
\newblock Antiferromagnetic exchange, {$\text{H}$}und's coupling and the origin
  of the charge gap in {$LaMnPO$}.
\newblock \emph{ArXiv e-prints}, 2014.
\newblock URL \url{https://arxiv.org/abs/1402.6312}.

\bibitem[Dong et~al.(2014)Dong, Li, Huang, and Dagotto]{DonJAP1152014}
Shuai Dong, Wei Li, Xin Huang, and Elbio Dagotto.
\newblock First principles study of the magnetic properties of {$LaOMnAs$}.
\newblock \emph{Journal of Applied Physics}, 115\penalty0 (17):\penalty0
  17D723, 2014.
\newblock \doi{10.1063/1.4867757}.
\newblock URL \url{https://doi.org/10.1063/1.4867757}.

\bibitem[Wildman et~al.(2014)Wildman, Emery, and Mclaughlin]{WilPRB902014}
E.~J. Wildman, N.~Emery, and A.~C. Mclaughlin.
\newblock Electronic and magnetic properties of
  {${\mathrm{Nd}}_{1\ensuremath{-}x}{\mathrm{Sr}}_{x}\mathrm{MnAsO}$}
  oxyarsenides.
\newblock \emph{Phys. Rev. B}, 90:\penalty0 224413, 2014.
\newblock \doi{10.1103/PhysRevB.90.224413}.
\newblock URL \url{https://link.aps.org/doi/10.1103/PhysRevB.90.224413}.

\bibitem[Calder et~al.(2014)Calder, Saparov, Cao, Niedziela, Lumsden, Sefat,
  and Christianson]{CalPRB892014}
S.~Calder, B.~Saparov, H.~B. Cao, J.~L. Niedziela, M.~D. Lumsden, A.~S. Sefat,
  and A.~D. Christianson.
\newblock Magnetic structure and spin excitations in
  {${\text{BaMn}}_{2}{\text{Bi}}_{2}$}.
\newblock \emph{Phys. Rev. B}, 89:\penalty0 064417, 2014.
\newblock \doi{10.1103/PhysRevB.89.064417}.
\newblock URL \url{https://link.aps.org/doi/10.1103/PhysRevB.89.064417}.

\bibitem[Gu et~al.(2014)Gu, Dai, Le, Sun, Wu, Saparov, Guo, Gao, Zhang, Zhou,
  Zhang, Jin, Xiong, Li, Li, Li, Liu, Sefat, Hu, and Zhao]{GuDSC42014}
Dachun Gu, Xia Dai, Congcong Le, Liling Sun, Qi~Wu, Bayrammurad Saparov, Jing
  Guo, Peiwen Gao, Shan Zhang, Yazhou Zhou, Chao Zhang, Shifeng Jin, Lun Xiong,
  Rui Li, Yanchun Li, Xiaodong Li, Jing Liu, Athena~S. Sefat, Jiangping Hu, and
  Zhongxian Zhao.
\newblock Robust antiferromagnetism preventing superconductivity in pressurized
  {$(Ba_{0.61}K_{0.39})Mn_2Bi_2$}.
\newblock \emph{Scientific Reports}, 4:\penalty0 7342, 2014.
\newblock URL \url{http://dx.doi.org/10.1038/srep07342}.

\bibitem[Zhang et~al.(2015)Zhang, Tian, Peterson, Dennis, and
  Vaknin]{ZhaPRB912015}
Qiang Zhang, Wei Tian, Spencer~G. Peterson, Kevin~W. Dennis, and David Vaknin.
\newblock Spin reorientation and {$Ce$}--{$Mn$} coupling in antiferromagnetic
  oxypnictide {$CeMnAsO$}.
\newblock \emph{Phys. Rev. B}, 91:\penalty0 064418, 2015.
\newblock \doi{10.1103/PhysRevB.91.064418}.
\newblock URL \url{https://link.aps.org/doi/10.1103/PhysRevB.91.064418}.

\bibitem[Ueland et~al.(2015)Ueland, Pandey, Lee, Sapkota, Choi, Haskel,
  Rosenberg, Lang, Harmon, Johnston, Kreyssig, and Goldman]{UelPRL1142015}
B.~G. Ueland, Abhishek Pandey, Y.~Lee, A.~Sapkota, Y.~Choi, D.~Haskel, R.~A.
  Rosenberg, J.~C. Lang, B.~N. Harmon, D.~C. Johnston, A.~Kreyssig, and A.~I.
  Goldman.
\newblock Itinerant ferromagnetism in the {$As$} $4p$ conduction band of
  {${\mathrm{Ba}}_{0.6}{\mathrm{K}}_{0.4}{\mathrm{Mn}}_{2}{\mathrm{As}}_{2}$}
  identified by x-ray magnetic circular dichroism.
\newblock \emph{Phys. Rev. Lett.}, 114:\penalty0 217001, 2015.
\newblock \doi{10.1103/PhysRevLett.114.217001}.
\newblock URL \url{https://link.aps.org/doi/10.1103/PhysRevLett.114.217001}.

\bibitem[Pandey and Johnston(2015)]{PanPRB922015}
Abhishek Pandey and D.~C. Johnston.
\newblock
  {${\mathrm{Ba}}_{0.4}{\mathrm{Rb}}_{0.6}{\mathrm{Mn}}_{2}{\mathrm{As}}_{2}$}:
  A prototype half-metallic ferromagnet.
\newblock \emph{Phys. Rev. B}, 92:\penalty0 174401, Nov 2015.
\newblock \doi{10.1103/PhysRevB.92.174401}.
\newblock URL \url{https://link.aps.org/doi/10.1103/PhysRevB.92.174401}.

\bibitem[McNally et~al.(2015)McNally, Zellman, Yin, Post, He, Hao, Kotliar,
  Basov, Homes, and Aronson]{McNPRB922015}
D.~E. McNally, S.~Zellman, Z.~P. Yin, K.~W. Post, Hua He, K.~Hao, G.~Kotliar,
  D.~Basov, C.~C. Homes, and M.~C. Aronson.
\newblock From hund's insulator to fermi liquid: Optical spectroscopy study of
  {$K$} doping in {${\mathrm{BaMn}}_{2}{\mathrm{As}}_{2}$}.
\newblock \emph{Phys. Rev. B}, 92:\penalty0 115142, 2015.
\newblock \doi{10.1103/PhysRevB.92.115142}.
\newblock URL \url{https://link.aps.org/doi/10.1103/PhysRevB.92.115142}.

\bibitem[McGuire and Garlea(2016)]{McGPRB932016}
Michael~A. McGuire and V.~Ovidiu Garlea.
\newblock Short- and long-range magnetic order in {$LaMnAsO$}.
\newblock \emph{Phys. Rev. B}, 93:\penalty0 054404, 2016.
\newblock \doi{10.1103/PhysRevB.93.054404}.
\newblock URL \url{https://link.aps.org/doi/10.1103/PhysRevB.93.054404}.

\bibitem[Zhang et~al.(2016)Zhang, Kumar, Tian, Dennis, Goldman, and
  Vaknin]{ZhaPRB932016}
Qiang Zhang, C.~M.~N. Kumar, Wei Tian, Kevin~W. Dennis, Alan~I. Goldman, and
  David Vaknin.
\newblock Structure and magnetic properties of {$Ln\mathrm{MnSbO}$}
  ({$Ln=\mathrm{La}$} and {$Ce$}).
\newblock \emph{Phys. Rev. B}, 93:\penalty0 094413, 2016.
\newblock \doi{10.1103/PhysRevB.93.094413}.
\newblock URL \url{https://link.aps.org/doi/10.1103/PhysRevB.93.094413}.

\bibitem[Zingl et~al.(2016)Zingl, Assmann, Seth, Krivenko, and
  Aichhorn]{ZinPRB942016}
Manuel Zingl, Elias Assmann, Priyanka Seth, Igor Krivenko, and Markus Aichhorn.
\newblock Importance of effective dimensionality in manganese pnictides.
\newblock \emph{Phys. Rev. B}, 94:\penalty0 045130, 2016.
\newblock \doi{10.1103/PhysRevB.94.045130}.
\newblock URL \url{https://link.aps.org/doi/10.1103/PhysRevB.94.045130}.

\bibitem[Singh et~al.(2009)Singh, Sefat, McGuire, Sales, Mandrus, VanBebber,
  and Keppens]{SinPRB792009}
D.~J. Singh, A.~S. Sefat, M.~A. McGuire, B.~C. Sales, D.~Mandrus, L.~H.
  VanBebber, and V.~Keppens.
\newblock Itinerant antiferromagnetism in {${\text{BaCr}}_{2}{\text{As}}_{2}$}:
  Experimental characterization and electronic structure calculations.
\newblock \emph{Phys. Rev. B}, 79:\penalty0 094429, 2009.
\newblock \doi{10.1103/PhysRevB.79.094429}.
\newblock URL \url{https://link.aps.org/doi/10.1103/PhysRevB.79.094429}.

\bibitem[Ding et~al.(2009)Ding, Wang, and Ni]{DinSSC1492009}
Yi~Ding, Yanli Wang, and Jun Ni.
\newblock Electronic and magnetic properties of 3d transition-metal selenides
  from first principles.
\newblock \emph{Solid State Communications}, 149\penalty0 (13):\penalty0 505 --
  509, 2009.
\newblock \doi{https://doi.org/10.1016/j.ssc.2009.01.015}.
\newblock URL
  \url{http://www.sciencedirect.com/science/article/pii/S0038109809000295}.

\bibitem[Hu and Hu(2010)]{HuSJPCC1142010}
Shu-Jun Hu and Xiao Hu.
\newblock Half-metallic antiferromagnet bacrfeas2.
\newblock \emph{The Journal of Physical Chemistry C}, 114\penalty0
  (26):\penalty0 11614--11617, 2010.
\newblock \doi{10.1021/jp103328g}.
\newblock URL \url{https://doi.org/10.1021/jp103328g}.

\bibitem[{Hu} and {Hu}(2010)]{HuS1011.2576}
S.-J. {Hu} and X.~{Hu}.
\newblock Half-metallic antiferromagnet sheets in {$Sr_4M_2O_6CrFeAs_2$}
  ({$M=Sc, Cr$}) and their bulk form.
\newblock \emph{ArXiv e-prints}, 2010.
\newblock URL \url{https://arxiv.org/abs/1011.2576}.

\bibitem[Marty et~al.(2011)Marty, Christianson, Wang, Matsuda, Cao, VanBebber,
  Zarestky, Singh, Sefat, and Lumsden]{MarPRB832011}
K.~Marty, A.~D. Christianson, C.~H. Wang, M.~Matsuda, H.~Cao, L.~H. VanBebber,
  J.~L. Zarestky, D.~J. Singh, A.~S. Sefat, and M.~D. Lumsden.
\newblock Competing magnetic ground states in nonsuperconducting
  {$Ba({{\mathrm{Fe}}_{1\ensuremath{-}x}{\mathrm{Cr}}_{x}})_{2}{\mathrm{As}}_{2}$}
  as seen via neutron diffraction.
\newblock \emph{Phys. Rev. B}, 83:\penalty0 060509, 2011.
\newblock \doi{10.1103/PhysRevB.83.060509}.
\newblock URL \url{https://link.aps.org/doi/10.1103/PhysRevB.83.060509}.

\bibitem[Paramanik et~al.(2014)Paramanik, Prasad, Geibel, and
  Hossain]{ParPRB892014}
U.~B. Paramanik, R.~Prasad, C.~Geibel, and Z.~Hossain.
\newblock Itinerant and local-moment magnetism in
  {${\text{EuCr}}_{2}{\text{As}}_{2}$} single crystals.
\newblock \emph{Phys. Rev. B}, 89:\penalty0 144423, 2014.
\newblock \doi{10.1103/PhysRevB.89.144423}.
\newblock URL \url{https://link.aps.org/doi/10.1103/PhysRevB.89.144423}.

\bibitem[Jiang et~al.(2015)Jiang, Bao, Zhai, Tang, Sun, Liu, Wang, Bai, Xu, and
  Cao]{JiaPRB922015}
Hao Jiang, Jin-Ke Bao, Hui-Fei Zhai, Zhang-Tu Tang, Yun-Lei Sun, Yi~Liu,
  Zhi-Cheng Wang, Hua Bai, Zhu-An Xu, and Guang-Han Cao.
\newblock Physical properties and electronic structure of
  {${\mathrm{Sr}}_{2}{\mathrm{Cr}}_{3}{\mathrm{As}}_{2}{\mathrm{O}}_{2}$}
  containing {${\mathrm{CrO}}_{2}$} and {${\mathrm{Cr}}_{2}{\mathrm{As}}_{2}$}
  square-planar lattices.
\newblock \emph{Phys. Rev. B}, 92:\penalty0 205107, 2015.
\newblock \doi{10.1103/PhysRevB.92.205107}.
\newblock URL \url{https://link.aps.org/doi/10.1103/PhysRevB.92.205107}.

\bibitem[Slater and Koster(1954)]{SlaPR941954}
J.~C. Slater and G.~F. Koster.
\newblock Simplified {$\text{LCAO}$} method for the periodic potential problem.
\newblock \emph{Phys. Rev.}, 94:\penalty0 1498--1524, 1954.
\newblock \doi{10.1103/PhysRev.94.1498}.
\newblock URL \url{https://link.aps.org/doi/10.1103/PhysRev.94.1498}.

\bibitem[Takimoto et~al.(2004)Takimoto, Hotta, and Ueda]{TakPRB692004}
Tetsuya Takimoto, Takashi Hotta, and Kazuo Ueda.
\newblock Strong-coupling theory of superconductivity in a degenerate
  {$\text{H}$}ubbard model.
\newblock \emph{Phys. Rev. B}, 69:\penalty0 104504, 2004.
\newblock \doi{10.1103/PhysRevB.69.104504}.
\newblock URL \url{https://link.aps.org/doi/10.1103/PhysRevB.69.104504}.

\bibitem[Kubo(2007)]{KubPRB752007}
Katsunori Kubo.
\newblock Pairing symmetry in a two-orbital {$\text{H}$}ubbard model on a
  square lattice.
\newblock \emph{Phys. Rev. B}, 75:\penalty0 224509, 2007.
\newblock \doi{10.1103/PhysRevB.75.224509}.
\newblock URL \url{https://link.aps.org/doi/10.1103/PhysRevB.75.224509}.

\bibitem[Edelmann et~al.(2017)Edelmann, Sangiovanni, Capone, and de'
  Medici]{EdePRB952017}
Martin Edelmann, Giorgio Sangiovanni, Massimo Capone, and Luca de' Medici.
\newblock Chromium analogs of iron-based superconductors.
\newblock \emph{Phys. Rev. B}, 95:\penalty0 205118, 2017.
\newblock \doi{10.1103/PhysRevB.95.205118}.
\newblock URL \url{https://link.aps.org/doi/10.1103/PhysRevB.95.205118}.

\bibitem[Filsinger et~al.(2017)Filsinger, Schnelle, Adler, Fecher, Reehuis,
  Hoser, Hoffmann, Werner, Greenblatt, and Felser]{FilPRB952017}
Kai~A. Filsinger, Walter Schnelle, Peter Adler, Gerhard~H. Fecher, Manfred
  Reehuis, Andreas Hoser, Jens-Uwe Hoffmann, Peter Werner, Martha Greenblatt,
  and Claudia Felser.
\newblock Antiferromagnetic structure and electronic properties of
  {${\mathrm{BaCr}}_{2}{\mathrm{As}}_{2}$} and {${\mathrm{BaCrFeAs}}_{2}$}.
\newblock \emph{Phys. Rev. B}, 95:\penalty0 184414, May 2017.
\newblock \doi{10.1103/PhysRevB.95.184414}.
\newblock URL \url{https://link.aps.org/doi/10.1103/PhysRevB.95.184414}.

\bibitem[Nayak et~al.(2017)Nayak, Filsinger, Fecher, Chadov, Min{\'a}r, Rienks,
  B{\"u}chner, Parkin, Fink, and Felser]{NayPNAS1142017}
Jayita Nayak, Kai Filsinger, Gerhard~H. Fecher, Stanislav Chadov, J{\'a}n
  Min{\'a}r, Emile D.~L. Rienks, Bernd B{\"u}chner, Stuart~P. Parkin, J{\"o}rg
  Fink, and Claudia Felser.
\newblock Observation of a remarkable reduction of correlation effects in
  {$BaCr_2As_2$} by {$\text{ARPES}$}.
\newblock \emph{Proceedings of the National Academy of Sciences}, 114\penalty0
  (47):\penalty0 12425--12429, 2017.
\newblock \doi{10.1073/pnas.1702234114}.
\newblock URL \url{http://www.pnas.org/content/114/47/12425}.

\bibitem[Richard et~al.(2017)Richard, van Roekeghem, Lv, Qian, Kim, Hoesch, Hu,
  Sefat, Biermann, and Ding]{RicPRB952017}
P.~Richard, A.~van Roekeghem, B.~Q. Lv, Tian Qian, T.~K. Kim, M.~Hoesch, J.-P.
  Hu, Athena~S. Sefat, Silke Biermann, and H.~Ding.
\newblock Is {${\mathbf{BaCr}}_{2}{\mathbf{As}}_{2}$} symmetrical to
  {${\mathbf{BaFe}}_{2}{\mathbf{As}}_{2}$} with respect to half $3d$ shell
  filling?
\newblock \emph{Phys. Rev. B}, 95:\penalty0 184516, 2017.
\newblock \doi{10.1103/PhysRevB.95.184516}.
\newblock URL \url{https://link.aps.org/doi/10.1103/PhysRevB.95.184516}.

\bibitem[{Jishi} et~al.(2018){Jishi}, {Rodriguez}, {Haugan}, and
  {Susner}]{Jis1807.03041}
R.A. {Jishi}, J.P. {Rodriguez}, T.J. {Haugan}, and M.A. {Susner}.
\newblock Prediction of antiferromagnetism in chromium-phosphide confirmed by
  synthesis.
\newblock \emph{ArXiv e-prints}, 2018.
\newblock URL \url{https://arxiv.org/abs/1807.03041}.

\bibitem[Takahashi et~al.(2015)Takahashi, Sugimoto, Nambu, Yamauchi, Hirata,
  Kawakami, Avdeev, Matsubayashi, Du, Soeda, Nakano, Uwatoko, Ueda, Sato, and
  Ohgushi]{TakNM142015}
H.~Takahashi, A.~Sugimoto, Y.~Nambu, T.~Yamauchi, Y.~Hirata, T.~Kawakami,
  V.~Avdeev, K.~Matsubayashi, Ch. Du, F.and~Kawashima, H.~Soeda, S.~Nakano,
  Y.~Uwatoko, Y.~Ueda, T.J. Sato, and K.~Ohgushi.
\newblock Pressure-induced superconductivity in the iron-based ladder material
  {$BaFe_2S_3$}.
\newblock \emph{Nature Materials}, 14:\penalty0 1008, 2015.
\newblock \doi{10.1038/nmat4351}.
\newblock URL \url{https://www.nature.com/articles/nmat4351}.

\bibitem[Yamauchi et~al.(2015)Yamauchi, Hirata, Ueda, and
  Ohgushi]{YamPRL1152015}
Touru Yamauchi, Yasuyuki Hirata, Yutaka Ueda, and Kenya Ohgushi.
\newblock Pressure-induced {$\text{M}$}ott transition followed by a
  24-{$\text{K}$} superconducting phase in
  {${\mathrm{BaFe}}_{2}{\mathrm{S}}_{3}$}.
\newblock \emph{Phys. Rev. Lett.}, 115:\penalty0 246402, 2015.
\newblock \doi{10.1103/PhysRevLett.115.246402}.
\newblock URL \url{https://link.aps.org/doi/10.1103/PhysRevLett.115.246402}.

\bibitem[Ying et~al.(2017)Ying, Lei, Petrovic, Xiao, and
  Struzhkin]{YinPRB952017}
Jianjun Ying, Hechang Lei, Cedomir Petrovic, Yuming Xiao, and Viktor~V.
  Struzhkin.
\newblock Interplay of magnetism and superconductivity in the compressed
  {$Fe$}-ladder compound {${\mathrm{BaFe}}_{2}{\mathrm{Se}}_{3}$}.
\newblock \emph{Phys. Rev. B}, 95:\penalty0 241109, 2017.
\newblock \doi{10.1103/PhysRevB.95.241109}.
\newblock URL \url{https://link.aps.org/doi/10.1103/PhysRevB.95.241109}.

\bibitem[Caron et~al.(2011)Caron, Neilson, Miller, Llobet, and
  McQueen]{CarPRB842011}
J.~M. Caron, J.~R. Neilson, D.~C. Miller, A.~Llobet, and T.~M. McQueen.
\newblock Iron displacements and magnetoelastic coupling in the
  antiferromagnetic spin-ladder compound {$BaFe_{2}Se_{3}$}.
\newblock \emph{Phys. Rev. B}, 84:\penalty0 180409, 2011.
\newblock \doi{10.1103/PhysRevB.84.180409}.
\newblock URL \url{https://link.aps.org/doi/10.1103/PhysRevB.84.180409}.

\bibitem[Medvedev et~al.(2012)Medvedev, Nekrasov, and
  Sadovskii]{MedJETPL952012}
M.~V. Medvedev, I.~A. Nekrasov, and M.~V. Sadovskii.
\newblock Electronic and magnetic structure of a possible iron based
  superconductor {$BaFe_2Se_3$}.
\newblock \emph{JETP Letters}, 95\penalty0 (1):\penalty0 33--37, 2012.
\newblock \doi{10.1134/S0021364012010092}.
\newblock URL \url{https://doi.org/10.1134/S0021364012010092}.

\bibitem[Luo et~al.(2013)Luo, Nicholson, Rinc\'on, Liang, Riera, Alvarez, Wang,
  Ku, Samolyuk, Moreo, and Dagotto]{LuoPRB872013}
Qinlong Luo, Andrew Nicholson, Juli\'an Rinc\'on, Shuhua Liang, Jos\'e Riera,
  Gonzalo Alvarez, Limin Wang, Wei Ku, German~D. Samolyuk, Adriana Moreo, and
  Elbio Dagotto.
\newblock Magnetic states of the two-leg-ladder alkali metal iron selenides
  {$AFe_{2}Se_{3}$}.
\newblock \emph{Phys. Rev. B}, 87:\penalty0 024404, 2013.
\newblock \doi{10.1103/PhysRevB.87.024404}.
\newblock URL \url{https://link.aps.org/doi/10.1103/PhysRevB.87.024404}.

\bibitem[Chi et~al.(2016)Chi, Uwatoko, Cao, Hirata, Hashizume, Aoyama, and
  Ohgushi]{ChiPRL1172016}
Songxue Chi, Yoshiya Uwatoko, Huibo Cao, Yasuyuki Hirata, Kazuki Hashizume,
  Takuya Aoyama, and Kenya Ohgushi.
\newblock Magnetic precursor of the pressure-induced superconductivity in
  {$Fe$}-ladder compounds.
\newblock \emph{Phys. Rev. Lett.}, 117:\penalty0 047003, 2016.
\newblock \doi{10.1103/PhysRevLett.117.047003}.
\newblock URL \url{https://link.aps.org/doi/10.1103/PhysRevLett.117.047003}.

\bibitem[Wang et~al.(2017)Wang, Jin, Yi, Song, Jiang, Zhang, Sun, Luo,
  Christianson, Bourret-Courchesne, Lee, Yao, and Birgeneau]{WanPRB952017_123}
Meng Wang, S.~J. Jin, Ming Yi, Yu~Song, H.~C. Jiang, W.~L. Zhang, H.~L. Sun,
  H.~Q. Luo, A.~D. Christianson, E.~Bourret-Courchesne, D.~H. Lee, Dao-Xin Yao,
  and R.~J. Birgeneau.
\newblock Strong ferromagnetic exchange interaction under ambient pressure in
  {${\mathrm{BaFe}}_{2}{\mathrm{S}}_{3}$}.
\newblock \emph{Phys. Rev. B}, 95:\penalty0 060502, 2017.
\newblock \doi{10.1103/PhysRevB.95.060502}.
\newblock URL \url{https://link.aps.org/doi/10.1103/PhysRevB.95.060502}.

\bibitem[{Svitlyk} et~al.(2018){Svitlyk}, {Garbarino}, {Rosa}, {Pomjakushina},
  {Krzton-Maziopa}, {Conder}, {Nunez-Regueiro}, and {Mezouar}]{Svi1808.03952}
V.~{Svitlyk}, G.~{Garbarino}, A.~D. {Rosa}, E.~{Pomjakushina},
  A.~{Krzton-Maziopa}, K.~{Conder}, M.~{Nunez-Regueiro}, and M.~{Mezouar}.
\newblock High-pressure polymorphism of {$BaFe_2Se_3$}.
\newblock \emph{ArXiv e-prints}, 2018.
\newblock URL \url{https://arxiv.org/abs/1808.03952}.

\bibitem[Suzuki et~al.(2015)Suzuki, Arita, and Ikeda]{SuzPRB922015}
Michi-To Suzuki, Ryotaro Arita, and Hiroaki Ikeda.
\newblock First-principles study of magnetic properties in {$Fe$}-ladder
  compound {$\mathrm{BaFe}{}_{2}\mathrm{S}{}_{3}$}.
\newblock \emph{Phys. Rev. B}, 92:\penalty0 085116, 2015.
\newblock \doi{10.1103/PhysRevB.92.085116}.
\newblock URL \url{https://link.aps.org/doi/10.1103/PhysRevB.92.085116}.

\bibitem[Nambu et~al.(2012)Nambu, Ohgushi, Suzuki, Du, Avdeev, Uwatoko,
  Munakata, Fukazawa, Chi, Ueda, and Sato]{NamPRB852012}
Yusuke Nambu, Kenya Ohgushi, Shunpei Suzuki, Fei Du, Maxim Avdeev, Yoshiya
  Uwatoko, Koji Munakata, Hiroshi Fukazawa, Songxue Chi, Yutaka Ueda, and
  Taku~J. Sato.
\newblock Block magnetism coupled with local distortion in the iron-based
  spin-ladder compound {$BaFe_{2}Se_{3}$}.
\newblock \emph{Phys. Rev. B}, 85:\penalty0 064413, 2012.
\newblock \doi{10.1103/PhysRevB.85.064413}.
\newblock URL \url{https://link.aps.org/doi/10.1103/PhysRevB.85.064413}.

\bibitem[Caron et~al.(2012)Caron, Neilson, Miller, Arpino, Llobet, and
  McQueen]{CarPRB852012}
J.~M. Caron, J.~R. Neilson, D.~C. Miller, K.~Arpino, A.~Llobet, and T.~M.
  McQueen.
\newblock Orbital-selective magnetism in the spin-ladder iron selenides
  {$Ba_{1\ensuremath{-}x}K_{x}Fe_{2}Se_{3}$}.
\newblock \emph{Phys. Rev. B}, 85:\penalty0 180405, 2012.
\newblock \doi{10.1103/PhysRevB.85.180405}.
\newblock URL \url{https://link.aps.org/doi/10.1103/PhysRevB.85.180405}.

\bibitem[Hirata et~al.(2015)Hirata, Maki, Yamaura, Yamauchi, and
  Ohgushi]{HirPRB922015}
Yasuyuki Hirata, Sachiko Maki, Jun-ichi Yamaura, Touru Yamauchi, and Kenya
  Ohgushi.
\newblock Effects of stoichiometry and substitution in quasi-one-dimensional
  iron chalcogenide {${\mathrm{BaFe}}_{2}{\mathrm{S}}_{3}$}.
\newblock \emph{Phys. Rev. B}, 92:\penalty0 205109, 2015.
\newblock \doi{10.1103/PhysRevB.92.205109}.
\newblock URL \url{https://link.aps.org/doi/10.1103/PhysRevB.92.205109}.

\bibitem[Du et~al.(2012)Du, Ohgushi, Nambu, Kawakami, Avdeev, Hirata, Watanabe,
  Sato, and Ueda]{DuFPRB852012}
Fei Du, Kenya Ohgushi, Yusuke Nambu, Takateru Kawakami, Maxim Avdeev, Yasuyuki
  Hirata, Yoshitaka Watanabe, Taku~J Sato, and Yutaka Ueda.
\newblock Stripelike magnetism in a mixed-valence insulating state of the
  {$Fe$}-based ladder compound {$CsFe_{2}Se_{3}$}.
\newblock \emph{Phys. Rev. B}, 85:\penalty0 214436, 2012.
\newblock \doi{10.1103/PhysRevB.85.214436}.
\newblock URL \url{https://link.aps.org/doi/10.1103/PhysRevB.85.214436}.

\bibitem[Zhang et~al.(2018{\natexlab{a}})Zhang, Lin, Zhang, Dagotto, and
  Dong]{ZhaPRB972018}
Yang Zhang, Ling-Fang Lin, Jun-Jie Zhang, Elbio Dagotto, and Shuai Dong.
\newblock Sequential structural and antiferromagnetic transitions in
  {${\mathrm{BaFe}}_{2}{\mathrm{Se}}_{3}$} under pressure.
\newblock \emph{Phys. Rev. B}, 97:\penalty0 045119, 2018{\natexlab{a}}.
\newblock \doi{10.1103/PhysRevB.97.045119}.
\newblock URL \url{https://link.aps.org/doi/10.1103/PhysRevB.97.045119}.

\bibitem[Ootsuki et~al.(2015)Ootsuki, Saini, Du, Hirata, Ohgushi, Ueda, and
  Mizokawa]{OotPRB912015}
D.~Ootsuki, N.~L. Saini, F.~Du, Y.~Hirata, K.~Ohgushi, Y.~Ueda, and
  T.~Mizokawa.
\newblock Coexistence of localized and itinerant electrons in
  {${\mathrm{BaFe}}_{2}{X}_{3}$} {$(X=\mathrm{S}$} and {$Se$}) revealed by
  photoemission spectroscopy.
\newblock \emph{Phys. Rev. B}, 91:\penalty0 014505, 2015.
\newblock \doi{10.1103/PhysRevB.91.014505}.
\newblock URL \url{https://link.aps.org/doi/10.1103/PhysRevB.91.014505}.

\bibitem[Takubo et~al.(2017)Takubo, Yokoyama, Wadati, Iwasaki, Mizokawa, Boyko,
  Sutarto, He, Hashizume, Imaizumi, Aoyama, Imai, and Ohgushi]{TakPRB962017}
Kou Takubo, Yuichi Yokoyama, Hiroki Wadati, Shun Iwasaki, Takashi Mizokawa,
  Teak Boyko, Ronny Sutarto, Feizhou He, Kazuki Hashizume, Satoshi Imaizumi,
  Takuya Aoyama, Yoshinori Imai, and Kenya Ohgushi.
\newblock Orbital order and fluctuations in the two-leg ladder materials
  {${\mathrm{BaFe}}_{2}{X}_{3}$} ({$X=\mathrm{S}$} and {$Se$}) and
  {${\mathrm{CsFe}}_{2}{\mathrm{Se}}_{3}$}.
\newblock \emph{Phys. Rev. B}, 96:\penalty0 115157, 2017.
\newblock \doi{10.1103/PhysRevB.96.115157}.
\newblock URL \url{https://link.aps.org/doi/10.1103/PhysRevB.96.115157}.

\bibitem[Patel et~al.(2016)Patel, Nocera, Alvarez, Arita, Moreo, and
  Dagotto]{PatPRB942016}
Niravkumar~D. Patel, Alberto Nocera, Gonzalo Alvarez, Ryotaro Arita, Adriana
  Moreo, and Elbio Dagotto.
\newblock Magnetic properties and pairing tendencies of the iron-based
  superconducting ladder ${\mathrm{bafe}}_{2}{\mathrm{s}}_{3}$: Combined ab
  initio and density matrix renormalization group study.
\newblock \emph{Phys. Rev. B}, 94:\penalty0 075119, 2016.
\newblock URL \url{https://link.aps.org/doi/10.1103/PhysRevB.94.075119}.

\bibitem[Li et~al.(2016)Li, Kaushal, Wang, Tang, Alvarez, Nocera, Maier,
  Dagotto, and Johnston]{LiSPRB942016}
S.~Li, N.~Kaushal, Y.~Wang, Y.~Tang, G.~Alvarez, A.~Nocera, T.~A. Maier,
  E.~Dagotto, and S.~Johnston.
\newblock Nonlocal correlations in the orbital selective mott phase of a
  one-dimensional multiorbital hubbard model.
\newblock \emph{Phys. Rev. B}, 94:\penalty0 235126, 2016.
\newblock URL \url{https://link.aps.org/doi/10.1103/PhysRevB.94.235126}.

\bibitem[Fanfarillo et~al.(2017)Fanfarillo, Giovannetti, Capone, and
  Bascones]{FanPRB952017}
L.~Fanfarillo, G.~Giovannetti, M.~Capone, and E.~Bascones.
\newblock Nematicity at the hund's metal crossover in iron superconductors.
\newblock \emph{Phys. Rev. B}, 95:\penalty0 144511, 2017.
\newblock URL \url{https://link.aps.org/doi/10.1103/PhysRevB.95.144511}.

\bibitem[Jarillo-Herrero(2018)]{pablojarillo_talkAPS2018}
Pablo Jarillo-Herrero.
\newblock Magic-angle graphene superlattice.
\newblock APS March meeting 2018 talk, Mar 2018.
\newblock URL \url{https://www.youtube.com/watch?v=O2HVCjhuJlE\&t=}.

\bibitem[{Yankowitz} et~al.(2018){Yankowitz}, {Chen}, {Polshyn}, {Watanabe},
  {Taniguchi}, {Graf}, {Young}, and {Dean}]{Yan1808.07865}
M.~{Yankowitz}, S.~{Chen}, H.~{Polshyn}, K.~{Watanabe}, T.~{Taniguchi},
  D.~{Graf}, A.~F. {Young}, and C.R. {Dean}.
\newblock Tuning superconductivity in twisted bilayer graphene.
\newblock \emph{ArXiv e-prints}, 2018.
\newblock URL \url{https://arxiv.org/abs/1808.07865}.

\bibitem[Yuan and Fu(2018)]{YuaPRB982018}
Noah F.~Q. Yuan and Liang Fu.
\newblock Model for the metal-insulator transition in graphene superlattices
  and beyond.
\newblock \emph{Phys. Rev. B}, 98:\penalty0 045103, 2018.
\newblock \doi{10.1103/PhysRevB.98.045103}.
\newblock URL \url{https://link.aps.org/doi/10.1103/PhysRevB.98.045103}.

\bibitem[Po et~al.(2018)Po, Zou, Vishwanath, and Senthil]{PoHPRX82018}
Hoi~Chun Po, Liujun Zou, Ashvin Vishwanath, and T.~Senthil.
\newblock Origin of {$\text{M}$}ott insulating behavior and superconductivity
  in twisted bilayer graphene.
\newblock \emph{Phys. Rev. X}, 8:\penalty0 031089, 2018.
\newblock \doi{10.1103/PhysRevX.8.031089}.
\newblock URL \url{https://link.aps.org/doi/10.1103/PhysRevX.8.031089}.

\bibitem[Kang and Vafek(2018)]{KanPRX82018}
Jian Kang and Oskar Vafek.
\newblock Symmetry, maximally localized {$\text{W}$}annier states, and a
  low-energy model for twisted bilayer graphene narrow bands.
\newblock \emph{Phys. Rev. X}, 8:\penalty0 031088, 2018.
\newblock \doi{10.1103/PhysRevX.8.031088}.
\newblock URL \url{https://link.aps.org/doi/10.1103/PhysRevX.8.031088}.

\bibitem[{Rademaker} and {Mellado}(2018)]{Rad1805.05294}
L.~{Rademaker} and P.~{Mellado}.
\newblock Charge-transfer insulation in twisted bilayer graphene.
\newblock \emph{ArXiv e-prints}, 2018.
\newblock URL \url{https://arxiv.org/abs/1805.05294}.

\bibitem[Han et~al.(1998)Han, Jarrell, and Cox]{HanPRB581998}
J.~E. Han, M.~Jarrell, and D.~L. Cox.
\newblock Multiorbital {$\text{H}$}ubbard model in infinite dimensions:
  {$\text{Q}$}uantum {$\text{M}$}onte {$\text{C}$}arlo calculation.
\newblock \emph{Phys. Rev. B}, 58:\penalty0 R4199--R4202, 1998.
\newblock \doi{10.1103/PhysRevB.58.R4199}.
\newblock URL \url{https://link.aps.org/doi/10.1103/PhysRevB.58.R4199}.

\bibitem[de' Medici(2011)]{deMPRB832011}
Luca de' Medici.
\newblock Hund's coupling and its key role in tuning multiorbital correlations.
\newblock \emph{Phys. Rev. B}, 83:\penalty0 205112, 2011.
\newblock \doi{10.1103/PhysRevB.83.205112}.
\newblock URL \url{https://link.aps.org/doi/10.1103/PhysRevB.83.205112}.

\bibitem[Dodaro et~al.(2018)Dodaro, Kivelson, Schattner, Sun, and
  Wang]{DodPRB982018}
J.~F. Dodaro, S.~A. Kivelson, Y.~Schattner, X.~Q. Sun, and C.~Wang.
\newblock Phases of a phenomenological model of twisted bilayer graphene.
\newblock \emph{Phys. Rev. B}, 98:\penalty0 075154, 2018.
\newblock \doi{10.1103/PhysRevB.98.075154}.
\newblock URL \url{https://link.aps.org/doi/10.1103/PhysRevB.98.075154}.

\bibitem[{Liu} et~al.(2018){Liu}, {Zhang}, {Chen}, and {Yang}]{Liu1804.10009}
C.-C. {Liu}, L.-D. {Zhang}, W.-Q. {Chen}, and F.~{Yang}.
\newblock Chiral {$\text{SDW}$} and d + id superconductivity in the magic-angle
  twisted bilayer-graphene.
\newblock \emph{ArXiv e-prints}, 2018.
\newblock URL \url{https://arxiv.org/abs/1804.10009}.

\bibitem[Maier et~al.(2005)Maier, Jarrell, Pruschke, and Hettler]{MaiRMP772005}
Thomas Maier, Mark Jarrell, Thomas Pruschke, and Matthias~H. Hettler.
\newblock Quantum cluster theories.
\newblock \emph{Rev. Mod. Phys.}, 77:\penalty0 1027--1080, 2005.
\newblock \doi{10.1103/RevModPhys.77.1027}.
\newblock URL \url{https://link.aps.org/doi/10.1103/RevModPhys.77.1027}.

\bibitem[Poteryaev et~al.(2004)Poteryaev, Lichtenstein, and
  Kotliar]{PotPRL932004}
A.~I. Poteryaev, A.~I. Lichtenstein, and G.~Kotliar.
\newblock Nonlocal coulomb interactions and metal-insulator transition in
  {${\mathrm{T}\mathrm{i}}_{2}{\mathrm{O}}_{3}$}: A {$\text{C}$}luster
  {$\mathrm{L}\mathrm{D}\mathrm{A}+\mathrm{D}\mathrm{M}\mathrm{F}\mathrm{T}$}
  approach.
\newblock \emph{Phys. Rev. Lett.}, 93:\penalty0 086401, 2004.
\newblock \doi{10.1103/PhysRevLett.93.086401}.
\newblock URL \url{https://link.aps.org/doi/10.1103/PhysRevLett.93.086401}.

\bibitem[Biermann et~al.(2005)Biermann, Poteryaev, Lichtenstein, and
  Georges]{BiePRL942005}
S.~Biermann, A.~Poteryaev, A.~I. Lichtenstein, and A.~Georges.
\newblock Dynamical singlets and correlation-assisted {$\text{P}$}eierls
  transition in {${\mathrm{V}\mathrm{O}}_{2}$}.
\newblock \emph{Phys. Rev. Lett.}, 94:\penalty0 026404, 2005.
\newblock \doi{10.1103/PhysRevLett.94.026404}.
\newblock URL \url{https://link.aps.org/doi/10.1103/PhysRevLett.94.026404}.

\bibitem[Kita et~al.(2009)Kita, Ohashi, and Suga]{KitPRB792009}
Tomoko Kita, Takuma Ohashi, and Sei-ichiro Suga.
\newblock Spatial fluctuations of spin and orbital nature in the two-orbital
  {$\text{H}$}ubbard model.
\newblock \emph{Phys. Rev. B}, 79:\penalty0 245128, 2009.
\newblock \doi{10.1103/PhysRevB.79.245128}.
\newblock URL \url{https://link.aps.org/doi/10.1103/PhysRevB.79.245128}.

\bibitem[Nomura et~al.(2014)Nomura, Sakai, and Arita]{NomPRB892014}
Yusuke Nomura, Shiro Sakai, and Ryotaro Arita.
\newblock Multiorbital cluster dynamical mean-field theory with an improved
  continuous-time quantum {$\text{M}$}onte {$\text{C}$}arlo algorithm.
\newblock \emph{Phys. Rev. B}, 89:\penalty0 195146, 2014.
\newblock \doi{10.1103/PhysRevB.89.195146}.
\newblock URL \url{https://link.aps.org/doi/10.1103/PhysRevB.89.195146}.

\bibitem[Nomura et~al.(2015)Nomura, Sakai, and Arita]{NomPRB912015}
Yusuke Nomura, Shiro Sakai, and Ryotaro Arita.
\newblock Nonlocal correlations induced by {$\text{H}$}und's coupling: A
  cluster {$\text{DMFT}$} study.
\newblock \emph{Phys. Rev. B}, 91:\penalty0 235107, 2015.
\newblock \doi{10.1103/PhysRevB.91.235107}.
\newblock URL \url{https://link.aps.org/doi/10.1103/PhysRevB.91.235107}.

\bibitem[Liebsch and Wu(2013)]{LiePRB872013}
Ansgar Liebsch and Wei Wu.
\newblock Coulomb correlations in the honeycomb lattice: Role of translation
  symmetry.
\newblock \emph{Phys. Rev. B}, 87:\penalty0 205127, 2013.
\newblock \doi{10.1103/PhysRevB.87.205127}.
\newblock URL \url{https://link.aps.org/doi/10.1103/PhysRevB.87.205127}.

\bibitem[Li et~al.(2015)Li, He, and Lu]{LiQPRB922015}
Qing-Xiao Li, Rong-Qiang He, and Zhong-Yi Lu.
\newblock Correlated {$\text{D}$}irac semimetal by periodized cluster dynamical
  mean-field theory.
\newblock \emph{Phys. Rev. B}, 92:\penalty0 155127, 2015.
\newblock \doi{10.1103/PhysRevB.92.155127}.
\newblock URL \url{https://link.aps.org/doi/10.1103/PhysRevB.92.155127}.

\bibitem[Momoi and Kubo(1998)]{MomPRB581998}
Tsutomu Momoi and Kenn Kubo.
\newblock Ferromagnetism in the {$\text{H}$}ubbard model with orbital
  degeneracy in infinite dimensions.
\newblock \emph{Phys. Rev. B}, 58:\penalty0 R567--R570, 1998.
\newblock \doi{10.1103/PhysRevB.58.R567}.
\newblock URL \url{https://link.aps.org/doi/10.1103/PhysRevB.58.R567}.

\bibitem[Kim et~al.(2017)Kim, Penc, Nataf, and Mila]{KimPRB962017}
Francisco~H. Kim, Karlo Penc, Pierre Nataf, and Fr\'ed\'eric Mila.
\newblock Linear flavor-wave theory for fully antisymmetric {$SU(N)$}
  irreducible representations.
\newblock \emph{Phys. Rev. B}, 96:\penalty0 205142, 2017.
\newblock \doi{10.1103/PhysRevB.96.205142}.
\newblock URL \url{https://link.aps.org/doi/10.1103/PhysRevB.96.205142}.

\bibitem[De~Franco et~al.(2018)De~Franco, Tocchio, and Becca]{DeFPRB982018}
Caterina De~Franco, Luca~F. Tocchio, and Federico Becca.
\newblock Metal-insulator transitions, superconductivity, and magnetism in the
  two-band {$\text{H}$}ubbard model.
\newblock \emph{Phys. Rev. B}, 98:\penalty0 075117, 2018.
\newblock \doi{10.1103/PhysRevB.98.075117}.
\newblock URL \url{https://link.aps.org/doi/10.1103/PhysRevB.98.075117}.

\bibitem[Zhou et~al.(2016)Zhou, Wang, Meng, Wang, and Wu]{ZhoPRB932016}
Zhichao Zhou, Da~Wang, Zi~Yang Meng, Yu~Wang, and Congjun Wu.
\newblock Mott insulating states and quantum phase transitions of correlated
  {$\text{SU}(2N)$} {$\text{D}$}irac fermions.
\newblock \emph{Phys. Rev. B}, 93:\penalty0 245157, 2016.
\newblock \doi{10.1103/PhysRevB.93.245157}.
\newblock URL \url{https://link.aps.org/doi/10.1103/PhysRevB.93.245157}.

\bibitem[Corboz et~al.(2012)Corboz, Lajk\'o, L\"auchli, Penc, and
  Mila]{CorPRX22012}
Philippe Corboz, Mikl\'os Lajk\'o, Andreas~M. L\"auchli, Karlo Penc, and
  Fr\'ed\'eric Mila.
\newblock Spin-orbital quantum liquid on the honeycomb lattice.
\newblock \emph{Phys. Rev. X}, 2:\penalty0 041013, 2012.
\newblock \doi{10.1103/PhysRevX.2.041013}.
\newblock URL \url{https://link.aps.org/doi/10.1103/PhysRevX.2.041013}.

\bibitem[Jakab et~al.(2016)Jakab, Szirmai, Lewenstein, and
  Szirmai]{JakPRB932016}
D.~Jakab, E.~Szirmai, M.~Lewenstein, and G.~Szirmai.
\newblock Competing valence bond and symmetry-breaking {$\text{M}$}ott states
  of spin-$\frac{3}{2}$ fermions on a honeycomb lattice.
\newblock \emph{Phys. Rev. B}, 93:\penalty0 064434, 2016.
\newblock \doi{10.1103/PhysRevB.93.064434}.
\newblock URL \url{https://link.aps.org/doi/10.1103/PhysRevB.93.064434}.

\bibitem[Corboz et~al.(2011)Corboz, L\"auchli, Penc, Troyer, and
  Mila]{CorPRL1072011}
Philippe Corboz, Andreas~M. L\"auchli, Karlo Penc, Matthias Troyer, and
  Fr\'ed\'eric Mila.
\newblock Simultaneous dimerization and {$\text{SU}$}(4) symmetry breaking of
  4-color fermions on the square lattice.
\newblock \emph{Phys. Rev. Lett.}, 107:\penalty0 215301, 2011.
\newblock \doi{10.1103/PhysRevLett.107.215301}.
\newblock URL \url{https://link.aps.org/doi/10.1103/PhysRevLett.107.215301}.

\bibitem[{Po} et~al.(2018){Po}, {Zou}, {Senthil}, and
  {Vishwanath}]{PoH1808.02482}
H.C. {Po}, L.~{Zou}, T.~{Senthil}, and A.~{Vishwanath}.
\newblock Faithful tight-binding models and fragile topology of magic-angle
  bilayer graphene.
\newblock \emph{ArXiv e-prints}, 2018.
\newblock URL \url{https://arxiv.org/abs/1808.02482}.

\bibitem[Saito et~al.(2016)Saito, Nojima, and Iwasa]{SaiNRM22016}
Yu~Saito, Tsutomu Nojima, and Yoshihiro Iwasa.
\newblock Highly crystalline {$\text{2D}$} superconductors.
\newblock \emph{Nature Reviews Materials}, 2:\penalty0 16094, 2016.
\newblock URL \url{http://dx.doi.org/10.1038/natrevmats.2016.94}.

\bibitem[Efimkin and MacDonald(2018)]{EfiPRB982018}
Dmitry~K. Efimkin and Allan~H. MacDonald.
\newblock Helical network model for twisted bilayer graphene.
\newblock \emph{Phys. Rev. B}, 98:\penalty0 035404, 2018.
\newblock \doi{10.1103/PhysRevB.98.035404}.
\newblock URL \url{https://link.aps.org/doi/10.1103/PhysRevB.98.035404}.

\bibitem[Xu and Balents(2018)]{XuCPRL1212018}
Cenke Xu and Leon Balents.
\newblock Topological superconductivity in twisted multilayer graphene.
\newblock \emph{Phys. Rev. Lett.}, 121:\penalty0 087001, 2018.
\newblock \doi{10.1103/PhysRevLett.121.087001}.
\newblock URL \url{https://link.aps.org/doi/10.1103/PhysRevLett.121.087001}.

\bibitem[Guo et~al.(2018)Guo, Zhu, Feng, and Scalettar]{GuoPRB872018}
Huaiming Guo, Xingchuan Zhu, Shiping Feng, and Richard~T. Scalettar.
\newblock Pairing symmetry of interacting fermions on a twisted bilayer
  graphene superlattice.
\newblock \emph{Phys. Rev. B}, 97:\penalty0 235453, 2018.
\newblock \doi{10.1103/PhysRevB.97.235453}.
\newblock URL \url{https://link.aps.org/doi/10.1103/PhysRevB.97.235453}.

\bibitem[Moon and Koshino(2013)]{MooPRB872013}
Pilkyung Moon and Mikito Koshino.
\newblock Optical absorption in twisted bilayer graphene.
\newblock \emph{Phys. Rev. B}, 87:\penalty0 205404, 2013.
\newblock URL \url{https://link.aps.org/doi/10.1103/PhysRevB.87.205404}.

\bibitem[Nomura et~al.(2016)Nomura, Sakai, Capone, and Arita]{NomJPCM282016}
Yusuke Nomura, Shiro Sakai, Massimo Capone, and Ryotaro Arita.
\newblock Exotic s -wave superconductivity in alkali-doped fullerides.
\newblock \emph{Journal of Physics: Condensed Matter}, 28\penalty0
  (15):\penalty0 153001, 2016.
\newblock URL \url{http://stacks.iop.org/0953-8984/28/i=15/a=153001}.

\bibitem[Deng et~al.(2018)Deng, Yu, Song, Zhang, Wang, Sun, Yi, Wu, Wu, Zhu,
  Wang, Chen, and Zhang]{DenN5632018}
Yujun Deng, Yijun Yu, Yichen Song, Jingzhao Zhang, Nai~Zhou Wang, Zeyuan Sun,
  Yangfan Yi, Yi~Zheng Wu, Shiwei Wu, Junyi Zhu, Jing Wang, Xian~Hui Chen, and
  Yuanbo Zhang.
\newblock Gate-tunable room-temperature ferromagnetism in two-dimensional
  {$\text{Fe}_3$}{$\text{Ge}$}{$\text{Te}_2$}.
\newblock \emph{Nature}, 563\penalty0 (7729):\penalty0 94--99, 2018.
\newblock URL \url{https://doi.org/10.1038/s41586-018-0626-9}.

\bibitem[Fei et~al.(2018)Fei, Huang, Malinowski, Wang, Song, Sanchez, Yao,
  Xiao, Zhu, May, Wu, Cobden, Chu, and Xu]{FeiNM172018}
Zaiyao Fei, Bevin Huang, Paul Malinowski, Wenbo Wang, Tiancheng Song, Joshua
  Sanchez, Wang Yao, Di~Xiao, Xiaoyang Zhu, Andrew~F. May, Weida Wu, David~H.
  Cobden, Jiun-Haw Chu, and Xiaodong Xu.
\newblock Two-dimensional itinerant ferromagnetism in atomically thin
  {$\text{Fe}_3$}{$\text{Ge}$}{$\text{Te}_2$}.
\newblock \emph{Nature Materials}, 17\penalty0 (9):\penalty0 778--782, 2018.
\newblock URL \url{https://doi.org/10.1038/s41563-018-0149-7}.

\bibitem[Deiseroth et~al.(2006)Deiseroth, Aleksandrov, Reiner, Kienle, and
  Kremer]{DeiEJIC20062006}
Hans-J\"rg Deiseroth, Krasimir Aleksandrov, Christof Reiner, Lorenz Kienle, and
  Reinhard~K. Kremer.
\newblock {$\text{Fe}_3$}{$\text{Ge}$}{$\text{Te}_2$} and
  {$\text{Ni}_3$}{$\text{Ge}$}{$\text{Te}_2$} – two new layered
  transition-metal compounds: Crystal structures, {$\text{HRTEM}$}
  investigations, and magnetic and electrical properties.
\newblock \emph{European Journal of Inorganic Chemistry}, 2006\penalty0
  (8):\penalty0 1561--1567, 2006.
\newblock URL
  \url{https://onlinelibrary.wiley.com/doi/abs/10.1002/ejic.200501020}.

\bibitem[Chen et~al.(2013)Chen, Yang, Wang, Imai, Ohta, Michioka, Yoshimura,
  and Fang]{CheJPSJ822013}
Bin Chen, JinHu Yang, HangDong Wang, Masaki Imai, Hiroto Ohta, Chishiro
  Michioka, Kazuyoshi Yoshimura, and MingHu Fang.
\newblock Magnetic properties of layered itinerant electron ferromagnet
  {$\text{Fe}_3$}{$\text{Ge}$}{$\text{Te}_2$}.
\newblock \emph{Journal of the Physical Society of Japan}, 82\penalty0
  (12):\penalty0 124711, 2013.
\newblock URL \url{https://doi.org/10.7566/JPSJ.82.124711}.

\bibitem[Zhu et~al.(2016)Zhu, Janoschek, Chaves, Cezar, Durakiewicz, Ronning,
  Sassa, Mansson, Scott, Wakeham, Bauer, and Thompson]{ZhuPRB932016}
Jian-Xin Zhu, Marc Janoschek, D.~S. Chaves, J.~C. Cezar, Tomasz Durakiewicz,
  Filip Ronning, Yasmine Sassa, Martin Mansson, B.~L. Scott, N.~Wakeham,
  Eric~D. Bauer, and J.~D. Thompson.
\newblock Electronic correlation and magnetism in the ferromagnetic metal
  {${\mathrm{Fe}}_{3}{\mathrm{GeTe}}_{2}$}.
\newblock \emph{Phys. Rev. B}, 93:\penalty0 144404, 2016.
\newblock \doi{10.1103/PhysRevB.93.144404}.
\newblock URL \url{https://link.aps.org/doi/10.1103/PhysRevB.93.144404}.

\bibitem[Yi et~al.(2017{\natexlab{b}})Yi, Zhuang, Zou, Wu, Cao, Tang, Calder,
  Kent, Mandrus, and Gai]{Jie2DM42017}
Jieyu Yi, Houlong Zhuang, Qiang Zou, Zhiming Wu, Guixin Cao, Siwei Tang, S~A
  Calder, P~R~C Kent, David Mandrus, and Zheng Gai.
\newblock Competing antiferromagnetism in a quasi-2{$\text{D}$} itinerant
  ferromagnet: {$\text{Fe}_3$}{$\text{Ge}$}{$\text{Te}_2$}.
\newblock \emph{2D Materials}, 4\penalty0 (1):\penalty0 011005,
  2017{\natexlab{b}}.
\newblock URL \url{http://stacks.iop.org/2053-1583/4/i=1/a=011005}.

\bibitem[Zhang et~al.(2018{\natexlab{b}})Zhang, Lu, Zhu, Tan, Feng, Liu, Zhang,
  Chen, Liu, Luo, Xie, Luo, Zhang, and Lai]{ZhaSA42018}
Yun Zhang, Haiyan Lu, Xiegang Zhu, Shiyong Tan, Wei Feng, Qin Liu, Wen Zhang,
  Qiuyun Chen, Yi~Liu, Xuebing Luo, Donghua Xie, Lizhu Luo, Zhengjun Zhang, and
  Xinchun Lai.
\newblock Emergence of {$\text{K}$}ondo lattice behavior in a van der
  {$\text{W}$}aals itinerant ferromagnet,
  {$\text{Fe}_3$}{$\text{Ge}$}{$\text{Te}_2$}.
\newblock \emph{Science Advances}, 4\penalty0 (1), 2018{\natexlab{b}}.
\newblock URL \url{http://advances.sciencemag.org/content/4/1/eaao6791}.

\bibitem[Mermin and Wagner(1966)]{MerPRL171966}
N.~D. Mermin and H.~Wagner.
\newblock Absence of ferromagnetism or antiferromagnetism in one- or
  two-dimensional isotropic heisenberg models.
\newblock \emph{Phys. Rev. Lett.}, 17:\penalty0 1133--1136, 1966.
\newblock URL \url{https://link.aps.org/doi/10.1103/PhysRevLett.17.1133}.

\bibitem[B\"ohmer and Kreisel(2017)]{BohJPCM302017}
Anna~E B\"ohmer and Andreas Kreisel.
\newblock Nematicity, magnetism and superconductivity in {$\text{Fe}$
  $\text{Se}$}.
\newblock \emph{Journal of Physics: Condensed Matter}, 30\penalty0
  (2):\penalty0 023001, 2017.
\newblock URL
  \url{https://iopscience.iop.org/article/10.1088/1361-648X/aa9caa/meta}.

\bibitem[Guterding et~al.(2017)Guterding, Jeschke, Mazin, Glasbrenner,
  Bascones, and Valent\'{\i}]{GutPRL1182017}
Daniel Guterding, Harald~O. Jeschke, I.~I. Mazin, J.~K. Glasbrenner,
  E.~Bascones, and Roser Valent\'{\i}.
\newblock Nontrivial role of interlayer cation states in iron-based
  superconductors.
\newblock \emph{Phys. Rev. Lett.}, 118:\penalty0 017204, 2017.
\newblock URL \url{https://link.aps.org/doi/10.1103/PhysRevLett.118.017204}.

\bibitem[Mostofi et~al.(2014)Mostofi, Yates, Pizzi, Lee, Souza, Vanderbilt, and
  Marzari]{MosCPC1852014}
A.~A. Mostofi, J.~R. Yates, G.~Pizzi, Y.-S. Lee, I.~Souza, D.~Vanderbilt, and
  N.~Marzari.
\newblock An updated version of wannier90: A tool for obtaining
  maximally-localised wannier functions.
\newblock \emph{Comput. Phys. Commun.}, 185:\penalty0 2309, 2014.
\newblock \doi{10.1016/j.cpc.2014.05.003}.
\newblock URL \url{http://dx.doi.org/10.1016/j.cpc.2014.05.003}.

\bibitem[Hohenberg and Kohn(1964)]{HohPRB1361964}
P.~Hohenberg and W.~Kohn.
\newblock Inhomogeneous electron gas.
\newblock \emph{Phys. Rev.}, 136:\penalty0 B864--B871, 1964.
\newblock \doi{10.1103/PhysRev.136.B864}.
\newblock URL \url{https://link.aps.org/doi/10.1103/PhysRev.136.B864}.

\bibitem[Kohn and Sham(1965)]{KohPRB1401965}
W.~Kohn and L.~J. Sham.
\newblock Self-consistent equations including exchange and correlation effects.
\newblock \emph{Phys. Rev.}, 140:\penalty0 A1133--A1138, 1965.
\newblock \doi{10.1103/PhysRev.140.A1133}.
\newblock URL \url{https://link.aps.org/doi/10.1103/PhysRev.140.A1133}.

\bibitem[Ceperley and Alder(1980)]{CepPRL451980}
D.~M. Ceperley and B.~J. Alder.
\newblock Ground state of the electron gas by a stochastic method.
\newblock \emph{Phys. Rev. Lett.}, 45:\penalty0 566--569, 1980.
\newblock \doi{10.1103/PhysRevLett.45.566}.
\newblock URL \url{https://link.aps.org/doi/10.1103/PhysRevLett.45.566}.

\bibitem[Perdew and Wang(1992)]{PerPRB451992}
John~P. Perdew and Yue Wang.
\newblock Accurate and simple analytic representation of the electron-gas
  correlation energy.
\newblock \emph{Phys. Rev. B}, 45:\penalty0 13244--13249, 1992.
\newblock \doi{10.1103/PhysRevB.45.13244}.
\newblock URL \url{https://link.aps.org/doi/10.1103/PhysRevB.45.13244}.

\bibitem[Becke(1988)]{BecPRA381988}
A.~D. Becke.
\newblock Density-functional exchange-energy approximation with correct
  asymptotic behavior.
\newblock \emph{Phys. Rev. A}, 38:\penalty0 3098--3100, 1988.
\newblock \doi{10.1103/PhysRevA.38.3098}.
\newblock URL \url{https://link.aps.org/doi/10.1103/PhysRevA.38.3098}.

\bibitem[Wannier(1937)]{WanPR521937}
Gregory~H. Wannier.
\newblock The structure of electronic excitation levels in insulating crystals.
\newblock \emph{Phys. Rev.}, 52:\penalty0 191--197, 1937.
\newblock \doi{10.1103/PhysRev.52.191}.
\newblock URL \url{https://link.aps.org/doi/10.1103/PhysRev.52.191}.

\bibitem[Marzari and Vanderbilt(1997)]{MarPRB561997}
Nicola Marzari and David Vanderbilt.
\newblock Maximally localized generalized wannier functions for composite
  energy bands.
\newblock \emph{Phys. Rev. B}, 56:\penalty0 12847--12865, 1997.
\newblock \doi{10.1103/PhysRevB.56.12847}.
\newblock URL \url{https://link.aps.org/doi/10.1103/PhysRevB.56.12847}.

\bibitem[de'Medici()]{lucapriv}
Luca de'Medici.
\newblock \textit{Private communication}.

\bibitem[Broyden(1965)]{broyden}
C.~G. Broyden.
\newblock A class of methods for solving nonlinear simultaneous equations.
\newblock \emph{Math. Comp.}, 19:\penalty0 577--593, 1965.
\newblock URL
  \url{http://www.ams.org/journals/mcom/1965-19-092/S0025-5718-1965-0198670-6/home.html}.

\end{thebibliography}
\end{document}